\documentclass[a4paper,hidelinks, 12pt]{report}
\usepackage[utf8]{inputenc}
\usepackage{lmodern}
\usepackage{a4wide}
\usepackage{amsmath}
\usepackage{amsfonts}
\usepackage{amssymb}
\usepackage{cite}
\usepackage{xcolor}
\usepackage{xfrac}
\usepackage{soul}
\usepackage{mathtools}
\usepackage{hyperref}
\usepackage{enumitem}
\usepackage{physics}
\usepackage{multirow}
\usepackage{tikz}
\usetikzlibrary{arrows,positioning,backgrounds,calc,chains,decorations.pathmorphing,decorations.text,fit,patterns,positioning,shapes,shapes.geometric,trees,positioning}
\usepackage[section]{placeins}
\usepackage{listings}
\usepackage{etoolbox}
\usepackage{longtable}
\usepackage{threeparttable}
\usepackage{setspace}

\numberwithin{equation}{section}
\definecolor{MyOr}{RGB}{235,129,27}

\makeatletter
\let\TPT@hookin\@gobble
\let\TPT@hookarg\@gobble
\newcommand{\labeltext}[2]{%
  \@bsphack
  \csname phantomsection\endcsname 
  \def\@currentlabel{#1}{\label{#2}}%
  \@esphack
}
\makeatother

\newcommand*{\ctikz}[2][]{\hbox{\mathsurround=6pt$\vcenter{\hbox{\tikz[#1]{#2}}}$}}

\let\sillymacro\theequation 
\patchcmd\sillymacro{equation}{parentequation}{}{}

\allowdisplaybreaks

\def\FontLb{
  \usefont{T1}{phv}{b}{n}\fontsize{16pt}{16pt}\selectfont}

\def\FontMb{
  \usefont{T1}{phv}{b}{n}\fontsize{14pt}{14pt}\selectfont}
\def\FontSn{
  \usefont{T1}{phv}{m}{n}\fontsize{12pt}{12pt}\selectfont}

\begin{document}

\thispagestyle {empty}

\includegraphics[bb=9.5cm 11cm 0cm 0cm,scale=0.29]{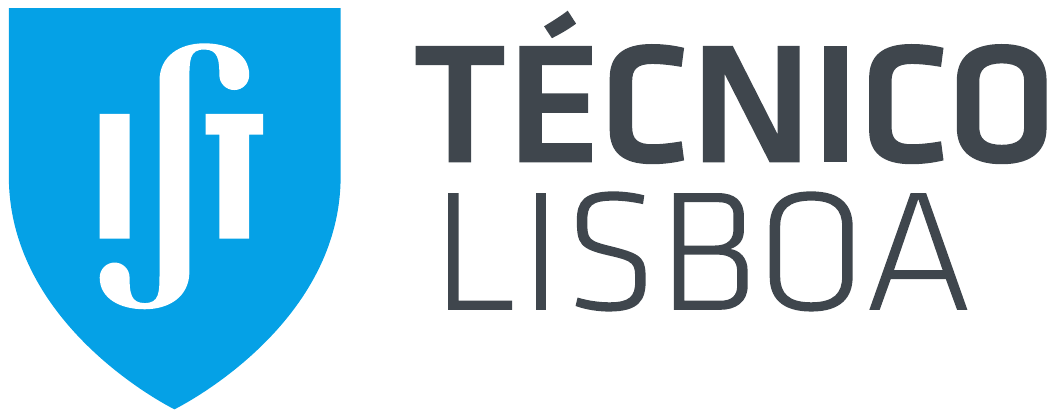}

\begin{center}

\begin{spacing}{1.5}
\vspace{2.5cm}

\vspace{1.0cm}
{\FontLb UNIVERSIDADE DE LISBOA} \\
{\FontLb INSTITUTO SUPERIOR TÉCNICO} \\

\vspace{2.0cm}
{\FontLb Dark Matter and CP Violation in Some\\ Symmetry-Constrained 3HDMs} \\

\vspace{2.6cm}
{\FontMb Anton Kun\v cinas} \\

\vspace{2.0cm}
{\FontSn %
\begin{tabular}{ll}
 \textbf{Supervisor}: & \textbf{Doctor} Maria Margarida Nesbitt Rebelo da Silva \\
 \textbf{Co-Supervisor}:  & \textbf{Doctor} Per Osland
\end{tabular} } \\

\vspace{2.0cm}
{\FontSn {\textbf{Thesis approved in a public defence for the award of the PhD degree in}}} \\
\vspace{0.3cm}
{\FontSn \textbf{Physics}} \\


\vspace{2.cm}
{\FontSn {\textbf{Funding Institution}}}\\
{\FontSn {Fundação para a Ciência e a Tecnologia (FCT)}}

\vspace*{\fill}
{\FontMb 2025}

\end{spacing}
\end{center}

\newpage
\thispagestyle {empty}

\includegraphics[bb=9.5cm 11cm 0cm 0cm,scale=0.29]{IST_A_CMYK_POS.pdf}

\begin{center}

\begin{spacing}{1.5}

\vspace{2.0cm}
{\FontLb UNIVERSIDADE DE LISBOA} \\
{\FontLb INSTITUTO SUPERIOR TÉCNICO} \\

\vspace{0.5cm}
{\FontLb Dark Matter and CP Violation in Some\\ Symmetry-Constrained 3HDMs} \\

\vspace{0.5cm}
{\FontMb Anton Kun\v cinas} \\

\vspace{1.5cm}
{\FontSn {\textbf{Jury final classification: Pass with Distinction}}}

\vspace{0.5cm}
{\FontSn {\textbf{Jury}}}\\
{\FontSn %
 \flushleft\textbf{Chairperson}: \textbf{Doctor} João Paulo Ferreira da Silva, Instituto Superior Técnico,\\ \hspace*{116pt} Universidade de Lisboa \\
 \flushleft\textbf{Members of the Committee:}\\
 \vspace{0.2cm}
 \hspace*{9.3pt}{\setlength{\tabcolsep}{1.75pt}\renewcommand{\arraystretch}{1}\begin{tabular}{lll}  
 & \textbf{Doctor}& Per Osland, Faculty of Science and Technology, University of Bergen,\\
 & & Norway\\
 & \textbf{Doctor}& Gustavo da Fonseca Castelo Branco, Instituto Superior Técnico,\\ 
 & & Universidade de Lisboa\\
 & \textbf{Doctor}& Bohdan Grządkowski, Faculty of Physics, University of Warsaw, Poland\\
 & \textbf{Doctor}& Geneviève Bélanger, Laboratoire d'Annecy-le-Vieux de Physique \\ 
 & & Théorique, France\\
 & \textbf{Doctor}& Filipe Rafael Joaquim, Instituto Superior Técnico, Universidade de Lisboa\\
 & \textbf{Doctor}& Joaquim Inácio da Silva Marcos, Instituto Superior Técnico,\\ 
 & & Universidade de Lisboa
\end{tabular} }} \\

\vspace{1cm}
{\flushleft{\FontSn {\textbf{Date of Thesis Submission:} 2025-02-07 }}\\
{\FontSn {\textbf{Date of Public Defence:} 2025-04-28 }}\\
{\FontSn {\textbf{Date of Thesis Acceptance:} 2025-06-09 }}\\}

\end{spacing}
\end{center}
\newpage

\pagestyle{plain}
\pagenumbering{roman}

\newpage
\thispagestyle{empty}
\vspace*{\fill}
\begin{center}
    \large \textit{This page intentionally left blank}
\end{center}
\vspace*{\fill}

\newpage

\section*{Abstract}
\addcontentsline{toc}{section}{Abstract}

It is now widely recognised that the Standard Model of Particle Physics is not the ultimate theory. Among the simplest extensions capable of addressing many of its shortcomings the multi-Higgs-doublet models. We focus on three-Higgs-doublet models, examining their potential to explain the dark matter puzzle and exploring whether they can introduce new sources of CP violation.

Multi-Higgs-doublet models can accommodate a dark matter candidate. An underlying symmetry not only ensures the candidate’s stability but also helps control the number of free parameters. We consider two scenarios. First, we explore dark matter candidates in the $S_3$-symmetric three-Higgs-doublet model. Notably, our findings reveal that the cases we discuss allow for different dark matter mass ranges compared to other candidates within three-Higgs-doublet models. Additionally, we investigate an alternative to conventional stabilisation via discrete symmetries by examining stabilisation through continuous symmetries. This alternative approach exhibits a unique characteristic---the emergence of mass-degenerate states.

We also explore the impact of CP violation in the $S_3$-symmetric three-Higgs-doublet model. In extended scalar frameworks, CP violation can arise either directly at the Lagrangian level or indirectly through the vacuum structure. Our analysis investigates how different representations of quarks under the $S_3$ group, combined with various vacuum configurations, lead to distinct phenomenological consequences, including the possibility of very light neutral scalars.

Overall, our extensive classification of promising scenarios for dark matter and CP violation lays a strong foundation for future research. Further detailed studies will be essential to fully explore our results. Our work paves the way for deeper insights into dark matter and CP violation within extended scalar sectors, underscores the rich complexity and potential of multi-Higgs models, and provides numerous results that can be readily utilised by model builders.

\vfill


\section*{Acknowledgements}
\addcontentsline{toc}{section}{Acknowledgements}

\vfil
First and foremost, a grandiose thank you to those who toiled alongside me, pouring countless hours into collaboration---your unwavering commitment, patience, and occasional bouts of despair (I am eager to spread credit among collaborators while remaining accountable for all blame) have truly shaped this work. Your dedication, like Virgil’s steady guidance, kept me from straying too far into the abyss. But let us not forget the real heroes: those who sacrificed even more, all in the name of the Almighty NHDMs! Your willingness to forgo sleep and sanity will not go unremembered (or at least, not immediately).

A special nod to the ever-watchful funding agencies for their infinite wisdom (or bureaucratic inertia),
\vspace{5pt}\newline
\centerline{\parbox{0.82\linewidth}{%
The work was partially supported by Funda\c c\~ ao para a  Ci\^ encia e a Tecnologia (FCT, Portugal) through the  projects CFTP-FCT Unit UIDB/00777/2020 and UIDP/00777/2020, CERN/FIS-PAR/0008/2019 and CERN/FIS-PAR/0002/2021, BL186/2024 IST-ID, which are partially funded through POCTI (FEDER), COMPETE, QREN and EU. Furthermore, the work has been supported by the FCT PhD fellowship with reference UI/BD/150735/2020. I would like to thank University of Bergen and Laboratoire d'Annecy-le-Vieux de Physique Théorique for hospitality, where collaboration visits took place.
}}\vspace{5pt}
\newline
whose benevolence ensured that I did not embark on an ill-advised pilgrimage through the three sacred Inferno circles of homelessness: the refrigerator box, the tent, and the car. Your merciful interventions kept me tethered to a semblance of stability---though, I must admit, the research might have been spicier had I been granted the immersive experience of urban nomadism.

\vfill

\section*{List of publications}
\addcontentsline{toc}{section}{List of publications}

\vfil
This thesis is based on the following publications, listed in chronological order:
\begin{itemize}
\item {\bf A.~Kun\v cinas}, P.~Osland and M.~N.~Rebelo, {\it $U(1)$-charged Dark Matter in three-Higgs-doublet models}, \href{https://link.springer.com/article/10.1007/JHEP11(2024)086}{JHEP \textbf{11} (2024), 086}, \href{https://arxiv.org/abs/2408.02728}{[2408.02728]};
\item {\bf A.~Kun\v cinas}, O.~M.~Ogreid, P.~Osland and M.~N.~Rebelo, {\it Complex $S_3$-symmetric 3HDM}, \href{https://link.springer.com/article/10.1007/JHEP07(2023)013}{JHEP \textbf{07} (2023), 013}, \href{https://arxiv.org/abs/2302.07210}{[2302.07210]};
\item {\bf A.~Kun\v cinas}, O.~M.~Ogreid, P.~Osland and M.~N.~Rebelo, {\it Dark matter in a CP-violating three-Higgs-doublet model with $S_3$ symmetry}, \href{https://journals.aps.org/prd/abstract/10.1103/PhysRevD.106.075002}{\textit{Phys. Rev. D} \textbf{106} (2022) 075002,}  \href{https://arxiv.org/abs/2204.05684}{[2204.05684]};
\item W.~Khater, {\bf A.~Kun\v cinas}, O.~M.~Ogreid, P.~Osland and M.~N.~Rebelo, {\it Dark matter in three-Higgs-doublet models with $S_3$ symmetry}, \href{https://link.springer.com/article/10.1007\%2FJHEP01\%282022\%29120}{JHEP \textbf{01} (2022), 120,} \href{https://arxiv.org/abs/2108.07026}{[2108.07026]}.
\end{itemize}
The research conducted by the author during his PhD studies has also led to the following contribution to conference proceedings:
\begin{itemize}
\item {\bf A.~Kun\v cinas}, {\it Status of dark matter in $S_3$-symmetric 3HDM}, \href{https://pos.sissa.it/436/}{PoS CORFU 2022 (2023) 32};
\item {\bf A.~Kun\v cinas}, O.~M.~Ogreid, P.~Osland and M.~N.~Rebelo, {\it Revisiting two dark matter candidates in $S_3$-symmetric three-Higgs-doublet models}, \href{https://pos.sissa.it/431/}{PoS DISCRETE 2022 (2023) 31,} \href{https://arxiv.org/abs/2301.12194}{[2301.12194]};
\item {\bf A.~Kun\v cinas}, O.~M.~Ogreid, P.~Osland and M.~N.~Rebelo, {\it Two dark matter candidates in three-Higgs-doublet models with $S_3$ symmetry}, \href{https://pos.sissa.it/405/}{PoS DISCRETE 2020-2021 (2022) 062,} \href{https://arxiv.org/abs/2204.08872}{[2204.08872]}.
\end{itemize}
\vfil

\tableofcontents
\cleardoublepage 



\chapter*{Abbreviations and notations}
\phantomsection
\addcontentsline{toc}{section}{Abbreviations and notations}

Abbreviations:
\begin{itemize}
\item \textbf{BBN} - Big Bang Nucleosynthesis;
\item \textbf{BSM} - Beyond the Standard Model;
\item \textbf{CDM} - Cold Dark Matter;
\item \textbf{CKM} - Cabibbo–Kobayashi–Maskawa;
\item \textbf{CMB} - Cosmic Microwave Background;
\item \textbf{C.L.} - Confidence Level;
\item \textbf{CP} - Charge Conjugation Parity;
\item \textbf{DM} - Dark Matter;
\item \textbf{EW} - Electroweak;
\item \textbf{FCNC} - Flavour-Changing Neutral Currents;
\item \textbf{FLRW} - Friedmann-Lemaître-Robertson-Walker;
\item \textbf{GCP} - General Charge Conjugation Parity;
\item \textbf{GIM} - Glashow–Iliopoulos–Maiani;
\item \textbf{h.c.} - Hermitian Conjugated;
\item \textbf{IDM} - Inert Doublet Model;
\item \textbf{LH} - Left-Handed;
\item \textbf{LHC} -  Large Hadron Collider;
\item \textbf{NFC} - Natural Flavour Conservation;
\item \textbf{NHDM} - $N$-Higgs-Doublet Model;
\item \textbf{NFW} - Navarro–Frenk–White;
\item \textbf{PDG} - Particle Data Group;
\item \textbf{PMNS} - Pontecorvo–Maki–Nakagawa–Sakata;
\item \textbf{QCD} - Quantum Chromodynamics;
\item \textbf{QED} - Quantum Electrodynamics;
\item \textbf{QFT} - Quantum Field Theory;
\item \textbf{RH} - Right-Handed;
\item \textbf{SM} - Standard Model;
\item \textbf{SSB} - Spontaneous Symmetry Breaking;
\item \textbf{vev} - Vacuum Expectation Value;
\item \textbf{WIMPs} - Weakly Interactive Massive Particles.
\end{itemize}

\hfill \break
List of symbols and shorthand notations:
\begin{itemize}
\item $\in$ - the ``element (member) of a set" symbol, \textit{e.g.}, $x \in X$ indicates that $x$ is included in the set of $X$; $x$ belongs to $X$;
\item $\{a,\,b,\,c,\, \dots\}$ - indicates a set of elements;
\item $\forall$ - the ``for all" symbol, \textit{e.g.}, $\forall\,\{a,\,b,\,c\}\in \mathcal{G}$ is understood as for all elements of the set $\{a,\,b,\,c\}$ contained in $\mathcal{G}$;
\item $[a;\,b ]$ - indicates range from $a$ to $b$;
\item $\mathcal{I}_n$ - an identity matrix of $\mathrm{dim}(\mathcal{I}_n)=n$;
\item $\mathcal{G} = \left\langle S | R \right\rangle$ - presentation of a group $\mathcal{G}$;
\item $\cong$ - symbol denoting isomorphism between two groups;
\item $|\mathcal{G}|$ - dimension of a group, $\dim(\mathcal{G})$;
\item $\delta_{ij}$ - the Kronecker delta;
\item $\mathcal{H} \leq \mathcal{G}$ - indicates that $\mathcal{H}$ is a subgroup of $\mathcal{G}$;
\item $\mathcal{N}\triangleleft \mathcal{G}$  - indicates that $\mathcal{N}$ is a normal subgroup of $\mathcal{G}$;
\item $\mathcal{G}/ \mathcal{N}$ - a quotient group;
\item $\mathcal{N} \rtimes\mathcal{H}$ - a semidirect group;
\item $h_{ij} \equiv h_i^\dagger h_j$ - is defined as an $SU(2)$ singlet scalar;
\item $	a \in \mathbb{R}$ - $a$ is a real quantity;
\item $	a \in \mathbb{C}$ - $a$ is a complex quantity.
\end{itemize}

\newpage
\hfill \break
Note:
\begin{itemize}
\item By Lagrangian we mean Lagrangian density;
\item Natural, ``God-given", units are used: $\hbar=c=1$;
\item We use the $``\to"$ symbol to identify transformations from one basis into another, for example, $A_i \to A_i^\prime = C_{ij} A_j$, usually dropping the intermediate $A_i^\prime$ quantity.
\end{itemize}

\chapter*{Preface}
\phantomsection
\addcontentsline{toc}{section}{Preface}

It is widely recognised that, despite its many successes, the Standard Model of Particle Physics is not the final theory. While the Standard Model has provided outstanding predictions and answered many key questions, it leaves several important issues unresolved. Among these are the origin of dark matter, the mechanism behind neutrino oscillations, the observed matter-antimatter asymmetry in the Universe, and various flavour anomalies. These are some of the shortcomings that have motivated the search for physics beyond the Standard Model.

One of the simplest and most natural extensions of the Standard Model involves enlarging the scalar sector. In the Standard Model, the Higgs mechanism is realised with a single Higgs doublet, which leads to one physical Higgs boson-a particle discovered at the LHC in 2012. Although there are so far no significant hints that the scalar sector of the Standard Model could be non-minimal, multi-Higgs-doublet models, particularly those with two or three Higgs doublets, are well motivated because they can address several of the Standard Model’s deficiencies. As more scalar doublets are added, the complexity of the model increases, leading to a rapid growth in the number of free parameters. This can, in turn, weaken the model’s ability to predict accurately, as it becomes fairly simple to match a large amount of data effortlessly.

To control this proliferation of parameters, symmetries are imposed on the scalar potential. Symmetries not only reduce the number of independent parameters by relating or eliminating certain couplings, but they can also provide a mechanism to stabilise potential dark matter candidates. In models with two or more scalar doublets, certain symmetries can give rise to vacua in which one or more vacuum expectation values vanish. If the symmetry that enforces vanishing vacuum expectation values is not broken, then that symmetry can be extended to the fermions, preventing interactions between the associated scalar fields and the fermions. This mechanism has been explored in detail in models such as the Inert Doublet Model and various multi-Higgs extensions.

A particularly interesting class of models is based on three-Higgs-doublet models with an underlying $S_3$ symmetry. The $S_3$ group, representing the permutation symmetry of three objects, has been used to constrain the scalar potential and explain the patterns in the Yukawa sector. In the $S_3$-symmetric three-Higgs-doublet model, the scalar potential can have several distinct vacuum configurations as a result of the minimisation process; we refer to these as implementations. Some of these implementations involve one or two vanishing vacuum expectation values. When a remnant of the $S_3$ symmetry survives spontaneous symmetry breaking, typically manifesting itself as a $\mathbb{Z}_2$ symmetry, the stability of the dark matter candidate is ensured. For instance, in the numerically analysed implementations, only one of the three Higgs doublets remains inert (with a vanishing vacuum expectation value), while the other two doublets are active, behaving similarly to a 2HDM in many respects. In this framework, the way fermions couple to the scalar sector is controlled by the $S_3$ symmetry. Notably, in these implementations there is no CP violation in the scalar sector unless additional phases are introduced. The phenomenology of this model is very rich.

In the context of dark matter, a detailed numerical analysis was performed for two implementations of the $S_3$-symmetric three-Higgs-doublet model: R-II-1a, an implementation with a real vacuum, and C-III-a, an implementation with a complex vacuum. For example, in the C-III-a model, after applying a series of theoretical and experimental constraints, a viable dark matter mass region was identified between approximately 28.9--41.9 GeV. This is significantly different from the typical mass range considered in the Inert Doublet Model or other three-Higgs-doublet models with inert doublets. These light states might be inaccessible through collider searches and indirect detection experiments. However, depending on the dark matter halo distribution profile applied to the indirect dark matter detection constraints, this model may be completely ruled out. Another interesting aspect present in both these implementations is the lack of heavy dark matter candidates---typically heavier than 500 GeV in the Inert Doublet Model. This is caused by the constrained parameter space of the $S_3$-symmetric three-Higgs-doublet model, which renders some of the scalar interactions proportional to masses rather than being free parameters.

Another aspect of extended scalar sectors is their potential to address CP violation. CP violation is a critical ingredient for explaining the observed matter-antimatter asymmetry in the Universe. In models with more than one Higgs doublet, CP violation can appear either explicitly in the Lagrangian via complex couplings or spontaneously through the vacuum structure. Spontaneous CP violation occurs when the Lagrangian is CP-symmetric but the vacuum is not, which requires that not all vacuum expectation values are real. In the multi-Higgs models, the existence of a basis where all couplings are real is sufficient, but not always necessary, to ensure explicit CP conservation; deviations from this scenario can then lead to CP violation either explicitly or spontaneously.

Within the $S_3$-symmetric three-Higgs-doublet model framework, various vacuum configurations have been identified that allow for both explicit and spontaneous CP violation. Detailed analyses have been performed, classifying the vacuum structures and determining which of these give rise to CP-violating phenomena in the scalar sector. For example, some implementations yield spontaneous CP violation when all quartic couplings are real, while other models require the introduction of complex couplings to achieve explicit CP violation. A particular model (referred to as C-V, having the most general vacuum configuration) stands out as especially promising. In this case, the complex Cabibbo–Kobayashi–Maskawa matrix is generated by phases arising from the vacuum expectation values, and there are no unwanted massless scalars present---a problem that other promising implementations in the context of the $S_3$-symmetric three-Higgs-doublet model suffer from. Moreover, there exists a viable parameter space where light neutral scalars (with masses of order $\mathcal{O}(\text{MeV})$) can be accommodated. Such light states, if they couple appropriately to the fermionic sector, might evade detection in current experiments yet have observable consequences in specific decay channels or cosmological observations.

In addition to discrete symmetries, the possibility of stabilising dark matter candidates via continuous symmetries has also been explored. Continuous symmetries tend to impose strict relations among the parameters of the scalar potential, often leading to the appearance of mass-degenerate neutral states. For example, in a three-Higgs-doublet model with an unbroken $U(1) \times U(1)$ symmetry, the case of the most general real scalar potential, the model features a multi-component dark matter sector with two independent mass scales. The numerical analysis of this model revealed a broad range of allowed dark matter masses, from about 45 GeV up to 2000 GeV. Interestingly, the decays of the heavier inert states into the lighter ones (two dark scalar sectors can interact, but the lightest particle of the heavy sector does not decay into particles of the lighter inert sector), can be adjusted, potentially altering the predicted relic density.

The extensive classification of both discrete and continuous symmetry models in the context of three-Higgs-doublet models has revealed that, although different symmetries might lead to scalar potentials with the same number of free parameters, they often produce different coupling patterns and mass relations among physical states. This means that, experimentally, it is possible to distinguish among these models by examining the specific patterns of interactions and the detailed structure of the scalar mass spectrum.

We start the discussion by introducing the Standard Model In Chapter~\ref{Ch:SM_Introduction}. It is assumed that the reader is familiar with some basic concepts of Quantum Mechanics and Quantum Field Theory. A brief discussion covers the particle content of the Standard Model and several interactions are discussed, though our primary focus is on Higgs physics. Then, in Chapter~\ref{Ch:Cosmology} we try to motivate why one should consider dark matter. Next, in Chapter~\ref{Ch:Group_Theory} we outline some basic concepts of Group Theory. No extensive knowledge of Group Theory is required from the reader. We outline some basic concepts and then try to present how one can apply groups to their favourite model. Then, in Chapter~\ref{Ch:2HDM} we discuss the two-Higgs-doublet model, which serves as an introduction to the three-Higgs-doublet model, presented in Chapter~\ref{Ch:S3_3HDM}. Chapter~\ref{Ch:S3_3HDM} marks the beginning of the original work. First, we discuss different vacuum configurations of the $S_3$-symmetric three-Higgs-doublet model based on several different aspects: explicit and spontaneous CP violation, presence of continuous symmetries, extension of the $S_3$ symmetry to fermions, presence of dark matter candidates. After identifying all possible implementations, in Chapter~\ref{Ch:DM_3HDM} we analyse two implementations of dark matter, aiming to determine whether they can accommodate a dark matter candidate. Finally, in Chapter~\ref{Ch:U1_3HDM} we cover continuous symmetries in the three-Higgs-doublet model. First, we try to check what possible continuous symmetries are realisable. Then, we check which of these models could accommodate a dark matter candidate while preserving the underlying symmetry. Finally, we perform a numerical check of the $U(1) \times U(1)$-symmetric three-Higgs-doublet model.

\chapter{Introduction to the Standard Model}\label{Ch:SM_Introduction}
\pagenumbering{arabic}

\vspace*{-7pt}
Up to these days the Standard Model (SM), or to be more precise extended by the right-handed neutrinos to allow for the neutrino masses, though there is no consensus on how this should be done, is in decent agreement with a significant amount of experimental data. Even though the SM has succeeded in predicting several experimental outcomes, it has several limitations and fails to address certain key aspects of the Universe and describe a handful of unexplained physical phenomena (some examples are provided in Section~\ref{Sec:BSM}). Therefore, the SM is not generally viewed as a complete theory of the fundamental interactions. The SM proved to be an efficient tool in describing different particle interactions like electromagnetic, weak, and strong forces, though not gravity, as well as classifying all of the known elementary particles, with the discovery of the Higgs boson being the major triumph of the SM~\cite{ATLAS:2012yve,CMS:2012qbp}.

Without delving into the well-known figures who laid the foundations of Quantum Mechanics and Special Relativity, the first major step towards the Quantum Field Theory (QFT) was done by P.A.M. Dirac, when he attempted to quantise the electromagnetic field~\cite{Dirac:1927dy,Dirac:1928hu}. He formulated the Quantum Electrodynamics (QED) theory, which describes the electromagnetic force interactions between the charged particles. Re-interpretation of the Dirac equation due to the negative-energy state hypothesised existence of anti-particles~\cite{Dirac:1930ek,Dirac:1931kp}, see also Ref.~\cite{Weyl:1929}. In a short period the first anti-particle---positron was (officially) discovered in 1932 by C. D. Anderson in cosmic rays~\cite{Anderson:1933mb}.

The next big step was done by C.-N. Yang and R. Mills. In 1954 they tried to explain strong interactions by extending the concept of the gauge theory to the non-Abelian groups~\cite{Yang:1954ek}, based on the special unitary groups $SU(n)$, which is in contrast to the simpler Abelian gauge theories. In the Yang–Mills theory the gauge group is represented by the Lie groups (a smooth manifold). One of the key elements of the non-Abelian gauge theories is that the gauge bosons can interact with one another. The Yang-Mills theory was crucial for the successful description of the combined model of the electromagnetic and weak interactions~\cite{Glashow:1959wxa,Salam:1959zz,Glashow:1961tr}. It should be noted that symmetries play a fundamental role in Elementary Particle Physics due to the fact that they help simplify the mathematical description of different interactions, dictate the properties of particles, and guide the formulation of physical laws.

In 1964 three independent groups proposed different approaches to how the mass terms can arise in the gauge-invariant models as a result of Spontaneous Symmetry Breaking (SSB). These groups were: F. Englert and R. Brout \cite{Englert:1964et}, P. Higgs \cite{Higgs:1964pj} and G. Guralnik, C. R. Hagen, T. Kibble \cite{Guralnik:1964eu}. Later the Higgs mechanism was incorporated into the Electroweak (EW) theory by S. Weinberg \cite{Weinberg:1967tq} and A. Salam \cite{Salam:1968rm}.

Let us now move on to the discussion of strong interactions. There was a lot of struggle to explain the binding force of nuclei until H. Yukawa proposed a theory in 1935~\cite{Yukawa:1935xg}. He suggested that the strong force between nucleons is mediated by a particle, which turned out to be pion~\cite{Lattes:1947mw}. In the following decades it became clear that the force responsible for binding protons and neutrons (and other hadrons) must operate at a much deeper level---at the level of quarks. The quark model was proposed by M. Gell-Mann~\cite{Gell-Mann:1964ewy} and G. Zweig~\cite{Zweig:1964ruk,Zweig:1964jf} in 1964. The model proposed that hadrons were not ``elementary" particles, but rather composed of bound states of quarks and anti-quarks. In 1973 the concept of colour as the source of the ``strong field" was developed into the theory of Quantum Chromodynamics (QCD)~\cite{Fritzsch:1973pi}. It was further formalised by the discovery of asymptotic freedom (strong force becomes weaker as quarks and gluons approach each other at high energies) by D. Gross, F. Wilczek~\cite{Gross:1973id} and D. Politzer~\cite{Politzer:1973fx}. Contrary, due to the colour confinement the strength of the strong force increases with an increase of a distance and it confines quarks and gluons into composite particles.

In the following chapter a brief introduction to the SM is presented. Some historical remarks are presented in Ref.~\cite{Iliopoulos:2025fhr}. A more detailed overview can be found in the textbooks~\cite{Cheng:1984vwu,Langacker:2010zza,Pal:2014xrq} and the Particle Data Group (PDG)~\cite{ParticleDataGroup:2024cfk}, on which this chapter is based. A basic understanding of Quantum Mechanics is assumed. It should be noted that, since our primary focus is on (extended) Higgs physics, a substantial portion of the discussions on the SM and QFT are omitted. For example, we do not cover strong interactions or the gauge boson self-interactions. Furthermore, we do not address the origin of neutrino mass and will assume, as was done in the early days of the SM, that neutrinos are massless.

\vspace*{12pt}
\section{Elementary particles}

The SM, including strong interactions, describes seventeen (the number of elementary particles depends on the context in which they are classified, $e.g.$, the number of gluons, whether one includes anti-particles, different colours of quarks) elementary particles and their interactions. The dynamics of the SM depends on nineteen parameters, values of which are established by experiment. These parameters are: nine charged fermion masses (without neutrinos) and the Higgs boson mass, four parameters to describe the Cabibbo-Kobayashi-Maskawa matrix (CKM), three gauge couplings which correspond to the three fundamental forces described by gauge theories, the vacuum expectation value (vev) and the QCD vacuum angle. The origin of most of these parameters is unknown and the SM does not provide any concrete answer. Apart from that, it is generally accepted that more parameters are needed, but not within the context of the SM. For example, depending on the nature of neutrinos, additional parameters should be considered in extended theories.

In the current view, all of the observable matter consists of twelve (taking into account flavours, but not colour charges or antiparticles) fundamental quantum fields with a half integer spin, which are elementary particles and are called fermions, and twelve corresponding anti-particles. These elementary particles can be categorised by their flavour, forming three distinct generations. Flavour states refer to the type of particle but do not specify whether the particle is massive. Abusing the notation, the physical fermions are presented in Table~\ref{SM_fer}.

{\renewcommand{\arraystretch}{1.3}
\begin{table}[htb]
\caption{Generations of physical matter of the SM.}
\label{SM_fer}
\begin{center}
\begin{tabular}{c|c|c|c|}
\cline{2-4}
\multicolumn{1}{c|}{}                    & \multicolumn{1}{c|}{Generation I} & \multicolumn{1}{c|}{Generation II} & \multicolumn{1}{c|}{Generation III} \\ \hline
\multicolumn{1}{|c|}{\multirow{2}{*}{Leptons}} &      $e$                 &          $\mu$             &        $\tau$               \\
\multicolumn{1}{|c|}{}           &      $\nu_e$                 &          $\nu_\mu$             &        $\nu_\tau$               \\ \cline{1-1} \hline
\multicolumn{1}{|c|}{\multirow{2}{*}{Quarks}} &       $d$                &     $s$                  &      $b$                 \\
\multicolumn{1}{|c|}{}            &       $u$                &      $c$                 &       $t$                \\ \cline{1-1} \hline 
\end{tabular}\vspace*{-9pt}
\end{center}
\end{table}}

Leptons can be split into two different sub-types: charged leptons (electron $e$, muon $\mu$, tau $\tau$) and neutral leptons (electron neutrino $\nu_e$,  muon neutrino $\nu_\mu$, tau neutrino $\nu_\tau$). Quarks can be split into two sub-types: up-type (up $u$, charm $c$, top $t$) and down-type (down $d$, strange $s$, bottom $b$) quarks. Quarks in contrast to leptons have a colour charge (this adds up to thirty six different quarks) and participate in strong interactions. Due to the colour confinement quarks are strongly bound together. In nature quarks are bound together to form composite particles---hadrons. Hadrons containing an odd number of valence quarks (at least three) are called baryons, while those with an even number of valence quarks are called mesons.

Another group of particles are called bosons. The gauge bosons, spin-one particles, are associated with the gauge fields that mediate interactions between the matter fields. Each fundamental interaction is associated with different gauge bosons, and they are responsible for transmitting the fundamental forces between particles. The SM describes three fundamental interactions---electromagnetic, weak, and strong. Photons are responsible for the electromagnetic interactions among the electrically charged particles, and are described by the QED theory. The $W^\pm$ and $Z$ bosons are responsible for mediating weak interactions among particles of different flavours. While $W^\pm$ interact only with the left-handed (LH) fermions and the right-handed (RH) anti-fermions, the $Z$ boson interacts with the LH particles and RH antiparticles. In total there are 8 gluons $g$, which are responsible for strong interactions. The strong interaction is only possible among particles with colour charge. All these interactions in the mathematical sense correspond to different gauge groups: the electromagnetic gauge symmetry is given by the $U(1)$ group, the weak gauge symmetry is given by the $SU(2)_L$ symmetry and strong interactions are described by the $SU(3)_C$ gauge group. All in all, the SM is described by the internal gauge symmetry:
\begin{equation}
\mathcal{G}_\mathrm{SM} \equiv SU(3)_C \times SU(2)_L \times U(1)_Y.
\end{equation}
In general, these mathematical objects can be considered as space-time-dependent transformations, under which they preserve physical quantities. For example, volume of a sphere does not change if it is rotated by an arbitrary angle or shifted in one of the planes.

Apart from the spin-one gauge bosons, there are spin-one bosons. These are referred to as scalar bosons. So far, only one elementary scalar boson has been discovered, and it is the Higgs boson. The Higgs boson also participates in the above mentioned weak interactions. 

\begin{figure}[htb]
\begin{center}
\includegraphics[scale=1.4]{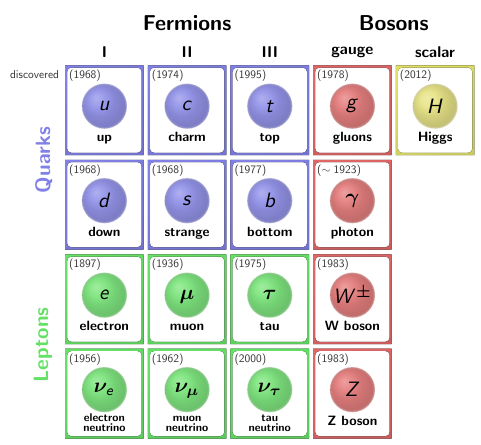}
\end{center}
\vspace*{-3mm}
\caption{In the physical basis of the SM of Particle Physics there are twelve (not including anti-particles and three different colours) fundamental fermions and five (counting the eight gluons and the two $W^\pm$ bosons as single entries) fundamental bosons. A familiar figure with atypical numbers; instead of the physical masses, electrical charges and spins, the discovery dates are provided. The up-to-date properties of the listed particles can be found in the PDG listings~\cite{ParticleDataGroup:2024cfk}. }\label{Fig:SM_particles}
\end{figure}

The elementary particles of the SM are presented in Figure~\ref{Fig:SM_particles}.

There are two types of symmetries: local and global. Under the global symmetry object is transformed uniformly at every point of the space-time. Conservation laws arise from global symmetries; in a more general sense global symmetries transform states of a system without altering its physical properties. In contrast, the local symmetry transformations vary depending on the base manifold. This means that local transformations act on a field differently, but smoothly, at each space-time point. Local symmetries are especially important because they often lead to the introduction of gauge fields. Under the gauge symmetry certain physical quantities can be transformed in a way that does not change the physical content of the theory. For example, in electromagnetism, the electromagnetic potential can be shifted by a function of space-time, the gauge function without affecting the physical observables, such as the electric and magnetic fields.

\section{Understanding Lagrangian density}

Before we proceed let us make a comment regarding the Lagrangian field theory. The Lagrangian is important because it provides a powerful framework for analysing the dynamics of physical systems. In classical mechanics Lagrangian is a function which summarises the dynamics of a particular system in terms of its generalised coordinates, velocities, and time. The Lagrangian for a system of particles is defined as:
\begin{equation}
L = T - V.
\end{equation}
The Lagrangian encapsulates the difference between the system's kinetic $T$ and potential energy $V$. By applying the Euler-Lagrange equation to the Lagrangian one can derive the equations of motion of the system,
\begin{equation}
\frac{d}{d t} \frac{\partial L}{\partial \dot{q}}-\frac{\partial L}{\partial q}=0,
\end{equation}
where $q$ denotes a set of generalised coordinates and the dotted variable denotes differentiation with respect to time, $\dot q = dx/dt$.

The Lagrangian formulation of mechanics is based on the principle of least action or Hamilton's principle. The principle of the least action states that the path a particle follows is the one for which the action is stationary, \textit{i.e.}, extremum or a saddle point. The action is a functional which is defined as the integral over time of the Lagrangian:
\begin{equation}
S = \int L dt.
\end{equation}

In the Lagrangian field theory formalism, one analyses fields as functions of the space-time. Equivalent to classical mechanics, where the Lagrangian encodes the dynamics of a particle, in field theory such quantity is encoded by the Lagrangian density $\mathcal{L}$ of the field configuration. In this case the action is expressed as:
\begin{equation}
S = \int L dt = \int  \int \mathcal{L} d^3 x dt.
\end{equation}
In short, the Lagrangian is the integral of the Lagrangian density over all space. Notice that the four-volume is a Lorentz invariant quantity, whereas it is not invariant with respect to a single temporal coordinate.

For the Lagrange density, the Euler-Lagrange equation is given by:
\begin{equation}
\frac{\partial \mathcal{L}}{\partial \varphi}-\partial_\mu\left(\frac{\partial \mathcal{L}}{\partial\left(\partial_\mu \varphi\right)}\right)=0.
\end{equation}

As can be realised, in the field theory we are interested in the Lagrangian density and not the classical Lagrangian, since $\mathcal{L}$ and not $L$ is a well-behaved quantity under the Lorentz transformations. For simplicity we shall refer to the Lagrangian density $\mathcal{L}$ simply as the Lagrangian.

\section{Quantum electrodynamics}

Electromagnetism is a well-known example of a gauge theory. Moreover, the first interaction which utilised the principles of a gauge invariance was QED. In QED the invariance of the Lagrangian under the local transformations plays a fundamental role. The exchange of photons mediates electromagnetic interactions. The QED interactions are described by the Abelian gauge theory with the symmetry group $U(1)$. The $U(1)$ group represents complex numbers with a modulus of one, $z=e^{i \theta}$. From the geometrical point of view it can be described as the unit circle in the complex plane.

The Lagrangian for Maxwell's equations in absence of sources (there are no external charges or currents) is given by
\begin{equation}\label{Eq:L_Maxwell_no_src}
\mathcal{L}_M = -\frac{1}{4} F_{\mu\nu} F^{\mu\nu},
\end{equation}
where $F_{\mu\nu}$ is the electromagnetic field strength tensor:
\begin{equation}
F_{\mu\nu} = \partial_\mu A_\nu - \partial_\nu A_\mu,
\end{equation}
and $A_\mu$ is the covariant four-potential of the electromagnetic field. The indices $\mu$ and $\nu$ run over the space-time dimensions. The $\mathcal{L}_M $ represents the kinetic energy of the electromagnetic field and it is a Lorentz-invariant quantity.

The equations of motion for a complex (bispinor) field $\psi$ with mass $m$ are given by the relativistic Dirac equations:
\begin{subequations}\label{Eq:complex_eq_mot}
\begin{align}
\left( i\gamma^\mu \partial_\mu -m \right)\psi = 0,\\
\quad \overline{\psi} \left( i\gamma^\mu \partial_\mu + m \right) = 0,
\end{align}
\end{subequations} 
where the field $\overline{\psi}$ is referred to as the Dirac adjoint, 
\begin{equation}
\overline{\psi} \equiv \psi^\dagger \gamma^0,
\end{equation}
and $\psi^\dagger$ is the Hermitian adjoint of the spinor $\psi$. Introduction of the Dirac adjoint is needed for the probability density conservation and Lorentz invariance. 

The Dirac $\gamma$ matrices play a key role in describing relativistic spin-1/2 particles. The Dirac $\gamma$ matrices are used to ensure that the Dirac equation is consistent with Lorentz invariance; to be more precise, a specific set of the $\gamma$ matrices corresponds to the matrix representation of the Clifford algebra,
\begin{equation}
\{\gamma^\mu, \gamma^\nu\} = \gamma^\mu \gamma^\nu + \gamma^\nu \gamma^\mu = 2 \eta^{\mu \nu},
\end{equation}
where $\eta^{\mu \nu}$ is the Minkowski metric with signature $(1,-1,-1,-1)$. The four contravariant $\gamma$ matrices in the Dirac (standard) representation are given by:
\begin{subequations}
\begin{align}
 \begin{split} &\gamma^0 = \begin{pmatrix}
1 & 0 & 0 & 0 \\
0 & 1 & 0 & 0 \\
0 & 0 & -1 & 0 \\
0 & 0 & 0 & -1
\end{pmatrix}, \quad
\gamma^1 = \begin{pmatrix}
0 & 0 & 0 & 1 \\
0 & 0 & 1 & 0 \\
0 & -1 & 0 & 0 \\
-1 & 0 & 0 & 0
\end{pmatrix},\\
&\gamma^2 = \begin{pmatrix}
0 & 0 & 0 & -i \\
0 & 0 & i & 0 \\
0 & i & 0 & 0 \\
-i & 0 & 0 & 0
\end{pmatrix}, \quad
\gamma^3 = \begin{pmatrix}
0 & 0 & 1 & 0 \\
0 & 0 & 0 & -1 \\
-1 & 0 & 0 & 0 \\
0 & 1 & 0 & 0
\end{pmatrix}, \end{split}\\
&\gamma^5 \equiv i \gamma^0 \gamma^1 \gamma^2 \gamma^3 =  \begin{pmatrix}
0 & 0 & 1 & 0 \\
0 & 0 & 0 & 1 \\
1 & 0 & 0 & 0 \\
0 & 1 & 0 & 0
\end{pmatrix}.\label{Eq:Gamma_5}
\end{align}
\end{subequations}
These can be written more compactly in terms of the Pauli $\sigma_i$ matrices,
\begin{equation}
\sigma_1 = \begin{pmatrix}
0 & 1 \\
1 & 0
\end{pmatrix}, \quad
\sigma_2 = \begin{pmatrix}
0 & -i \\
i & 0
\end{pmatrix}, \quad
\sigma_3 = \begin{pmatrix}
1 & 0 \\
0 & -1
\end{pmatrix},
\end{equation}
as:
\begin{equation}
\gamma^0 = \sigma_3 \otimes \mathcal{I}_2, ~ \gamma^j = i \sigma_2 \otimes \sigma_j,
\end{equation}
where $\mathcal{I}_n$ is the identity matrix of dimension $n$.

Some of the Dirac $\gamma$ matrices properties are:
\begin{equation}\label{Eq:gamma_properties}
\begin{aligned}
& (\gamma^0)^\dagger = \gamma^0,\qquad (\gamma^5)^\dagger = \gamma^5,\\
& (\gamma^5)^2 = 1,\qquad~ \left\lbrace \gamma^5, \gamma^\mu \right\rbrace = 0,\\
& (\gamma^\mu)^\dagger=\gamma^0\gamma^\mu\gamma^0=-\gamma^\mu,~\mathrm{for}~\mu\neq0.
\end{aligned}
\end{equation}

One can derive the Lagrangian for the Dirac fields by plugging eqs.~\eqref{Eq:complex_eq_mot} into the Euler-Lagrange equation. The Lagrangian is then given by:
\begin{equation}\label{Eq:Lagrangian_Dirac}
\mathcal{L}_D = \overline{\psi} \left( i\gamma^\mu \partial_\mu - m \right)\psi.
\end{equation}

The Dirac Lagrangian of eq.~\eqref{Eq:Lagrangian_Dirac} is invariant under the global $U(1)$ gauge transformation:
\begin{equation}\label{Eq:U1g_psi}
\psi  \rightarrow \psi^\prime = e^{iQ\xi} \psi, 
\end{equation}
where $Q$ is the charge operator acting on the bispinors as:
\begin{equation}
Q \begin{pmatrix}
\psi \\
\overline{\psi}
\end{pmatrix} = \begin{pmatrix}
+\psi \\
-\overline{\psi}
\end{pmatrix},
\end{equation}
and $\xi$ is an arbitrary real parameter independent of the space-time. For simplicity we shall drop the equality sign and the primes, so that transformations will be written as $A_i \to C_{ij} A_j$.

Let us consider an infinitesimal transformation,
\begin{equation}
e^{i\xi}=1+i\xi+\mathcal{O}(\xi^2),
\end{equation}
under which $\psi$ transforms as:
\begin{equation}
\begin{aligned}
\psi &\rightarrow \psi + i\xi\psi,\\
\overline{\psi} & \rightarrow \overline{\psi}-i\xi \overline{\psi}.
\end{aligned}
\end{equation}

Now, let us consider that the $\xi$ parameter is space-time dependent, \textit{i.e.}, a function of $x^\mu$. In this case the transformations are given by:
\begin{equation}
\begin{aligned}
\psi(x^\mu) &\rightarrow \psi(x^\mu) + i\xi(x^\mu)\psi(x^\mu), \\
\overline{\psi}(x^\mu) & \rightarrow \overline{\psi}(x^\mu) -i\xi(x^\mu)\overline{\psi}(x^\mu).
\end{aligned}
\end{equation}
The Dirac Lagrangian of eq.~\eqref{Eq:Lagrangian_Dirac} is not invariant under these transformations,
\begin{equation}
\left[ \overline{\psi}(x^\mu) -i\xi(x^\mu)\overline{\psi}(x^\mu)  \right] i\gamma^\mu\partial_\mu \left[ \psi(x^\mu) +i\xi(x^\mu)\psi(x^\mu)  \right] \neq i \overline \psi \gamma^\mu \partial_\mu \psi.
\end{equation}
Then the underlying $U(1)$ symmetry of eq.~\eqref{Eq:U1g_psi} is broken. This can be fixed by introducing a gauge fixing field $A_\mu$. Under the gauge transformation we have:
\begin{equation}
-eQA_\mu \rightarrow -eQ A_\mu + \partial_\mu \xi(x^\mu),
\end{equation}
so that
\begin{equation}
\delta \left( -e\overline{\psi}\gamma_\mu A_\mu\psi \right) = \delta \mathcal{L}_D.
\end{equation}
From here it follows that the gauged Dirac Lagrangian becomes
\begin{equation}
\mathcal{L}_D = \overline{\psi} \left[ i\gamma^\mu \left( \partial_\mu + ieQA_\mu \right) - m \right]\psi,
\end{equation}
where $e$ is the electric charge and $A_\mu$ is the electromagnetic four-potential (the photon field). 

One can simplify the above Lagrangian by introducing the covariant derivative
\begin{equation}
D_\mu = \partial_\mu + ieQA_\mu.
\end{equation}
Then, interactions of fermions with the electromagnetic field can be described by the following Lagrangian:
\begin{equation}\label{Eq:Dirac_Lagrangian_final}
\mathcal{L}_D = \overline{\psi} \left( i\gamma^\mu D_\mu - m \right)\psi.
\end{equation}

The Maxwell's Lagrangian~\eqref{Eq:L_Maxwell_no_src} and the Dirac Lagrangian \eqref{Eq:Dirac_Lagrangian_final} can be combined to describe the dynamics of the charged fermions, \textit{e.g.}, electrons and positrons, interacting with the electromagnetic field (photons):
\begin{equation}\label{Eq:L_QED}
\mathcal{L}_\text{QED} = - \frac{1}{4}F_{\mu\nu} F^{\mu\nu} + \overline{\psi} \left( i\gamma^\mu D_\mu - m \right) \psi.
\end{equation}

From the QED Lagrangian of eq.~\eqref{Eq:L_QED} one can note that photons interact directly only with the electrically charged particles, \textit{e.g.}, charged leptons, quarks and $W^\pm$ bosons. Having no electric charge, neutrinos do not couple to the electromagnetic field in the way other fermions do.

\section{Weak interactions}

The weak force is one of the known fundamental interactions of Nature. This force is responsible for the stochastic radioactive $\beta$-decay of atoms. This process can happen when particles, either elementary or composite, exchange virtual (since masses of the $W^\pm$ and $Z$ bosons are higher than of the typical particles in such interactions) weak bosons. The weak interaction is responsible for flavour changes of fermions (via the $W^\pm$ bosons). The weak interaction is mediated by the weak bosons, also referred to as the intermediate vector bosons, $W^\pm$ and $Z$. There are two differently charged $W^\pm$ boson, which are each other's anti-particle, while the $Z$ boson is electrically neutral and is its own anti-particle. All the weak bosons are spin-one particles. These particles are heavy: $m_{W^\pm}\approx 80$ GeV and $m_Z \approx 91$ GeV. Such high masses limit the range of the weak force.

In the early 1950s the strangeness quantum number~\cite{Pais:1952zz,Gell-Mann:1953hzm,Nakano:1953zz} (existence of the strange-quark will be predicted only a decade later~\cite{Gell-Mann:1964ewy,Zweig:1964ruk,Zweig:1964jf}) was introduced to explain why some heavy hadrons were easily created and decayed slower than expected. The strangeness quantum number was assumed to be preserved in the creation of such particles, but violated in their decays. It was found that the two charged mesons would decay to:
\begin{subequations}
\begin{align}
\theta^+ \to{}& \pi^+ + \pi^0,\\
\tau^+ \to{}& \pi^+ + \pi^+ + \pi^-.
\end{align}
\end{subequations}
With more precise measurements no difference was found between these two mesons, \textit{i.e.}, both $\theta^+$ and $\tau^+$ had indistinguishable masses and lifetimes. Yet, they differed in the final parity states, and were thought to be two distinct particles. This confusion was dubbed the $\theta{-}\tau$ puzzle and was resolved only after the discovery of parity violation in weak interactions; mesons decay through weak interactions. These days it is well understood that strangeness is conserved in strong and electromagnetic interactions but it is violated by the weak interactions. Nowadays, the $\theta^+$ and $\tau^+$ states are understood to be a single $K^+$ state.

The weak interaction is of particular interest since it is the only fundamental interaction that violates the parity symmetry (P), physical processes under the mirror reflection are not preserved, and also the charge conjugation parity (CP) symmetry, laws of physics are not identical if a particle is replaced by its anti-particle (C symmetry) and the spatial coordinates are flipped over (P symmetry). There was considerable experimental verification for parity invariance in both strong and electromagnetic interactions, but no experiments were conducted to test the parity properties of weak interactions. In 1956 C.-N. Yang and T.-D. Lee suggested that the weak interaction might violate parity conservation; so that the laws of physics are not ambidextrous. A year later C.-S. Wu and collaborators~\cite{Wu:1957my} discovered such violation by studying the nuclei of cobalt as it underwent the $\beta$-decay, ${_{27}^{60}\mathrm{Co}} \to {_{28}^{60}\mathrm{Ni}} + e + \bar{\nu_e} + 2 \gamma$. It was discovered that the majority of the observed electron, and hence the parity violation was established, were emitted in a direction opposite to that of the nuclear spin. Within a short time parity violation was also observed in the decays of muons~\cite{Garwin:1957hc,Friedman:1957mz}. 

Later, the $V{-}A$ (vector minus axial vector, originating due to the terms $\bar \psi \gamma^\mu \psi$ and $\bar \psi \gamma^\mu \gamma^5 \psi$) theory~\cite{Feynman:1958ty,Sudarshan:1958vf} was proposed based on the idea that the weak interactions violate parity. The key aspect of the theory is that the the weak interaction acts only on the L) particles. In short, the $V{-}A$ theory was a huge step in understanding the chiral nature of the weak force and the structure of  weak interactions.

Considering the spatial symmetry violations, and experimental proof, only the LH particles and the RH anti-particles can interact with the charged currents. This shall be described in more detail in Section~\ref{Eq:Chiral_Theory}. To describe how particles interact under the weak force, the quantum number of the weak isospin $T_3$ was introduced. Under the weak force quarks never decay into the state with the same weak isospin quantum number.

All the RH fermions and the LH anti-particles have a zero weak isospin value, while the LH particles and the RH anti-particles have a half-integer weak isospin value. The values of the LH isospins can be found in Table~\ref{Tab:weakisosp}. In order to get the weak isospin for the RH anti-particles one needs to multiply the values in Table~\ref{Tab:weakisosp} by minus one.

{\renewcommand{\arraystretch}{1.3}
\begin{table}[htb]
\caption{The weak isospin of the LH fermions in the SM. The RH anti-particles have opposite weak isospin. The RH particles and the LH anti-particles have a weak isospin equal to zero.}\label{Tab:weakisosp}
\begin{center}
\begin{tabular}{|ll|ll|ll|} \hline\hline
\multicolumn{2}{|l|}{Generation I} & \multicolumn{2}{|l|}{Generation II} & \multicolumn{2}{|l|}{Generation III} \\ 
\hline
\;$e^0$   & $-1/2$ & $\mu^0$     & $-1/2$ & $\tau^0$     & $-1/2$\\
\;$\nu_e^0$ & $+1/2$ & $\nu_{\mu}^0$ & $+1/2$ & $\nu_{\tau}^0$ & $+1/2$\\ \hline
\;$d^0$     & $-1/2$ & $s^0$         & $-1/2$ & $b^0$          & $-1/2$\\
\;$u^0$     & $+1/2$ & $c^0$         & $+1/2$ & $t^0$          & $+1/2$ \\ \hline\hline
\end{tabular}\vspace*{-9pt}
\end{center}
\end{table}}

Taking into consideration weak interactions, the particles from Table~\ref{SM_fer} can be written as the LH $SU(2)$ isospin doublets:
\begin{equation}\label{Eq:fermions_doublets}
\begin{aligned}
&(L^{0}_L)_{I}=\begin{pmatrix}
\nu_e^0\\
e^0
\end{pmatrix}_L,\qquad
&(L^{0}_L)_{II}=\begin{pmatrix}
\nu_\mu^0\\
\mu^0
\end{pmatrix}_L,\qquad
&(L^{0}_L)_{III}=\begin{pmatrix}
\nu_\tau^0\\
\tau^0
\end{pmatrix}_L,\\
&(Q^{0}_L)_{I}=\begin{pmatrix}
u^0\\d^0
\end{pmatrix}_L, \qquad
&(Q^{0}_L)_{II}=\begin{pmatrix}
c^0\\s^0
\end{pmatrix}_L, \qquad
&(Q^{0}_L)_{III}=\begin{pmatrix}
t^0\\b^0
\end{pmatrix}_L,
\end{aligned}
\end{equation}
and the RH singlets:
\begin{equation}
\begin{aligned}
&e_R^0,~~ \mu_R^0, ~~\tau_R^0, ~\left(\nu_{eR}^0, ~~\nu_{\mu R}^0, ~~\nu_{\tau R}^0\right),\\
&d_R^0,~~ s_R^0, ~~b_R^0, ~~~u_R^0, ~~~c_R^0, ~~~~t_R^0.\\
\end{aligned}
\end{equation}
It should be noted that the RH neutrinos are hypothetical particles that have a good motivation to exist in many models based on extensions of the SM. We shall further simplify notation by dropping the generation indices, so that the LH $SU(2)$ doublets are:
\begin{equation}\label{Eq:F_SU2_L}
Q_L^0=\binom{U_L^0}{D_L^0}=\binom{\left(u_L^0~c_L^0~t_L^0\right)^{\mathrm{T}}}{\left(d_L^0~s_L^0~b_L^0\right)^{\mathrm{T}}}, \quad L_L^0=\binom{N_L^0}{E_L^0}=\binom{\left(\nu_{e\,L}^0~\nu_{\mu\,L}^0~\nu_{\tau\,L}^0\right)^{\mathrm{T}}}{\left(e_L^0~\mu_L^0~\tau_L^0\right)^{\mathrm{T}}},
\end{equation}
and the RH $SU(2)$ singlet are:
\begin{equation}\label{Eq:F_SU2_R}
U_R^0=\left(\begin{array}{c}
u_R^0 \\
c_R^0 \\
t_R^0
\end{array}\right), \quad D_R^0=\left(\begin{array}{c}
d_R^0 \\
s_R^0 \\
b_R^0
\end{array}\right), \quad E_R^0=\left(\begin{array}{c}
e_R^0 \\
\mu_R^0 \\
\tau_R^0
\end{array}\right).
\end{equation}
One should note that, in general, the fields introduced above are in the weak basis, which is different from the physical mass basis. In order to get mass eigenstates one has to diagonalise mass matrices.

The conjugated fields of eq.~\eqref{Eq:fermions_doublets} are:
\begin{equation}
\overline{\Psi^0_L}= \left(\, \overline{\psi_{\alpha_i}^0} ~~ \overline{\psi_{\beta_i}^0} \,\right)_L,
\end{equation}
where $\psi_{\alpha_i}^0=\left\lbrace \nu_i^0,\,u_i^0 \right\rbrace$ and $\psi_{\beta_i}^0=\left\lbrace l_i^0,\,d_i^0 \right\rbrace$. We use the following notation for different generations of fermions:
\begin{equation}
\begin{aligned}
\nu_i^0&=\left\lbrace \nu_e^0,~\nu_\mu^0,~\nu_\tau ^0\right\rbrace,\\
\ell_i^0&=\left\lbrace e^0,~\mu^0,~\tau^0 \right\rbrace,\\
u_i^0&=\left\lbrace u^0,~c^0,~t^0 \right\rbrace,\\
d_i^0&=\left\lbrace d^0,~s^0,~b^0 \right\rbrace.\\
\end{aligned}
\end{equation}

\section{Electroweak theory}\label{Sec:EW_theory}

The first major step in the unification of the fundamental forces came with the unification of the electromagnetic force and the weak nuclear force into the electroweak force. From the mathematical point of view, the EW unification turned out not to be a simple task. The hugest obstacle is that while photons are massless, the weak bosons are not. Though, it took nearly a decade to discover the weak bosons after the theoretical prediction by S. Glashow, S. Weinberg, A. Salam.

Taking a look back at the early Universe, all the fundamental forces are assumed to have been once unified (at extremely high energies). The four fundamental forces might be a different manifestations of a single underlying force. This is based on the idea that the laws of physics are simple at the high energy scale and that the variety of forces we observe at low energies is a consequence of symmetry breaking as the Universe cooled down.

Now, consider that we need to write a gauge theory for a massive spin-one boson. Although the solution is provided by Maxwell's equations, we need not to forget that the photon is massless. A simple addition of the mass term would not result in a well-behaved Lorentz-invariance, since boosting into the rest frame of the assumed massive particle would correspond to no specific direction preferred by the polarisation vector. This, in turn, points out that such particle has (at least) three independent states. For example, a photon has a linear and a circular polarisation, and both of these can be considered to be special cases of the  elliptical polarisation. A possible solution comes from the description of the electrodynamics of a superconductor~\cite{Ginzburg:1950sr}.

Suppose we have a $U(1)$ gauge theory (in conventional superconductors, the phase of a wave function can change continuously, which is associated with the conservation of the particle number) given by the Lagrangian:
\begin{equation}
\mathcal{L} = - \frac{1}{4} F_{\mu\nu} F^{\mu\nu} + |D_\mu \phi|^2 - V(\phi),
\end{equation}
where $\phi$ is a complex scalar field, with the ground state given by:
\begin{equation}
\left \langle 0 | \phi | 0  \right\rangle = \frac{v}{\sqrt{2}},
\end{equation}
being the expectation value of the scalar potential $V(\phi)$. Since the scalar potential exhibits the $U(1)$ continuous symmetry, we expect to have a manifold of degenerate ground states. The complex scalar state is given by:
\begin{equation}
\phi = \frac{1}{\sqrt{2}} \left( v + \eta + i \chi \right).
\end{equation}
The complex part of the $\phi$ is associated with the $U(1)$ symmetry breaking since it shifts the vacuum state. This $\chi$ field can be rotated away via a local gauge transformation,
\begin{equation}
\phi \to e^{-i Q \alpha} \phi.
\end{equation}
By expanding the kinetic part of the Lagrangian and evaluating it at the ground state we get
\begin{equation}
|D_\mu \phi|^2 \Big|_{\left \langle 0 | \phi | 0  \right\rangle} = \frac{1}{2} e^2 Q^2 v^2 A_\mu A^\mu,
\end{equation}
a massive field $A_{\mu}$. This effect is known as the Meissner effect~\cite{Meissner:1933ela} in Condensed-Matter Physics. This effect is analogous to the Abelian Higgs mechanism, which we shall discuss in Section~\ref{Sec:SM_Higgs}. All in all, we have two degrees of freedom coming from $A_\mu$ and an additional one, as we wanted, from the Nambu-Goldstone boson $\chi$. The Nambu–Goldstone bosons, more commonly referred to as Goldstone bosons, are particles that arise necessarily in models with spontaneous breaking of continuous symmetries~\cite{Nambu:1960tm,Goldstone:1961eq}. Note that for the SM we need SSB of the non-Abelian gauge symmetry. This will be discussed in Section~\ref{Sec:SM_Higgs}.

Before moving to the discussion of the SSB in the SM, let us discuss some other aspects of the EW theory. The EW theory is described by the $SU(2)_L \times U(1)_Y$ gauge group. The $U(1)$ group was presented earlier, while $SU(n)$ is a special unitary group represented by the dimension $n$ unitary matrices with determinant one. The $SU(2)$ group has the following properties:
\begin{equation}
SU (2) = \left\lbrace
\begin{pmatrix}
\alpha & - \beta^\ast\\
\beta & \alpha^\ast
\end{pmatrix}:~ \alpha, \beta \in \mathbb{C},~ \vert \alpha \vert^2 + \vert \beta \vert^2 = 1
\right\rbrace.
\end{equation}

The Lie algebra of $\mathfrak{su}(2)$ in the real basis is generated by the following matrices:
\begin{equation}
u_1 = 
\begin{pmatrix}
0 & i\\
i & 0
\end{pmatrix}, \quad 
u_2 = 
\begin{pmatrix}
0 & -1\\
1 & 0
\end{pmatrix}, \quad 
u_3 = 
\begin{pmatrix}
i & 0\\
0 & - i
\end{pmatrix},
\end{equation}
which are related to the Pauli matrices by
\begin{equation}\label{Eq:Relation_to_Pauli_sigma}
u_1 = i \sigma_1,~ u_2 = -i \sigma_2,~ u_3 = i \sigma_3.
\end{equation}
The $\mathfrak{su}(2)$ group exponentiates to $SU(2)$, $U = e^{\alpha_n u_n} = e^{i \alpha_n \sigma_n}$.

The weak hypercharge $Y_W$ is a quantum number describing the EW interactions. It relates the electric charge and the weak isospin. The weak hypercharge is the generator of the $U(1)$ component of the EW gauge group. A specific combination of the quantum numbers is preserved:
\begin{equation}
Q = T_3 + \frac{Y_W}{2}.
\end{equation}
Considering this relation, possible values of the weak hypercharge are presented in Table~\ref{Tab:TWHC}.

{\renewcommand{\arraystretch}{1.3}
\begin{table}[htb]
\caption{The weak hypercharge in the SM for the LH and the RH fermions.}\label{Tab:TWHC}
\begin{center}
\begin{tabular}{|ll|ll|ll|} \hline \hline
\multicolumn{2}{|l|}{Generation I} & \multicolumn{2}{|l|}{Generation II} & \multicolumn{2}{|l|}{Generation III} \\ 
\hline
\;$e_L^0$     & $-1$   & $\mu_L^0$       & $-1$   & $\tau_L^0$       & $-1$\\
\;$e_R^0$   & $-2$   & $\mu_R^0$     & $-2$   & $\tau_R^0$     & $-2$\\
\;$\nu_{e_L}$   & $-1$   & $\nu_{\mu_L}^0$   & $-1$   & $\nu_{\tau_L}^0$   & $-1$\\ \hline
\;$d_L^0$       & $+1/3$ & $s_L^0$           & $+1/3$ & $b_L^0$            & $+1/3$\\ 
\;$d_R^0$     & $-2/3$ & $s_R^0$         & $-2/3$ & $b_R^0$          & $-2/3$\\ 
\;$u_L^0$       & $+1/3$ & $c_L^0$           & $+1/3$ & $t_L^0$            & $+1/3$\\
\;$u_R^0$     & $+4/3$ & $c_R^0$         & $+4/3$ & $t_R^0$          & $+4/3$\\ \hline\hline
\end{tabular}\vspace*{-9pt}
\end{center}
\end{table}}

The EW Lagrangian is given by:
\begin{equation}
\mathcal{L}_{SU(2) \times U(1)_Y} = \mathcal{L}_{\text{gauge}} + \mathcal{L}_{\text{fermions}} + \left[ \mathcal{L}_{\text{Yukawa}} + \mathcal{L}_\text{Higgs} \right],
\end{equation}
where the part in brackets will be covered in Section~\ref{Sec:SM_Higgs}. As for now, we shall focus on the gauge interactions $\mathcal{L}_\text{gauge}$ and the fermionic interactions $\mathcal{L}_\text{fermions}$.

The gauge term is:
\begin{equation}
\mathcal{L}_\text{gauge} = -\frac{1}{4} A_i^{\mu\nu} A^i_{\mu\nu} - \frac{1}{4}B^{\mu\nu} B_{\mu\nu},
\end{equation}
where
\begin{subequations}
\begin{align}
A_{\mu\nu}^i &=  \partial_\mu A_\nu ^i - \partial_\nu A_\mu ^i + g \varepsilon^{ijk}A_\mu^j A_\nu^k,\\
B_{\mu\nu} &= \partial_\mu B_\nu - \partial_\nu B_\mu,
\end{align}
\end{subequations}
are the field strength tensors.

The gauge part of the $SU(2) \times U(1)_Y$ Lagrangian describes interactions between the three weak isospin $A_i$ gauge fields and the weak hypercharge $B$ gauge field. These gauge fields mix together. After the electroweak symmetry breaking, as shall be shown in eq.~\eqref{Eq:WZgamma_as_AB}, these fields will correspond to the physical $W^{\pm}$, $Z$ and $\gamma$ states.

The fermions kinetic term consists of
\begin{equation}
\mathcal{L}_\text{fermions} = \mathcal{L}_\text{leptons} + \mathcal{L}_\text{quarks},
\end{equation}
where
\begin{subequations}
\begin{align}
\mathcal{L}_\text{leptons} ={}&i\overline{L_j^0}\gamma^\mu D_\mu^L L_j^0 + i \overline{\ell_j^0} \gamma^\mu D_\mu^R \ell_j^0 + \left( i \overline{\nu_j^0} \gamma^\mu D_\mu^R \nu_j^0 \right), \\
\mathcal{L}_\text{quarks}={}&i\overline{Q_j^0}\gamma^\mu D_\mu^L Q_j^0 + i \overline{u_j^0} \gamma^\mu D_\mu^R u_j^0 + i \overline{d_j^0} \gamma^\mu D_\mu^R d_j^0.
\end{align}
\end{subequations}
The superscript ``0" indicates that fermions are in the weak basis and those are not the mass eigenstates. The covariant derivatives acting on the fermionic fields are:
\begin{subequations}
\begin{align}
&D_\mu^L L_i^0 = \left( \partial_\mu-\frac{ig}{2}\sigma_i A_\mu^i+\frac{ig^\prime}{2} B_\mu \right) L_i^0,\\
&D_\mu^L Q_i^0 = \left( \partial_\mu-\frac{ig}{2}\sigma_i A_\mu^i-\frac{ig^\prime}{6} B_\mu \right) Q_i^0,\\
&D_\mu^R \ell_i^0= \left( \partial_\mu + ig^\prime B_\mu \right) \ell_i^0, \\
&D_\mu^R u_i^0= \left( \partial_\mu - \frac{ig^\prime 2}{3} B_\mu \right)u_i^0, \\
&D_\mu^R d_i^0= \left( \partial_\mu + \frac{ig^\prime}{3} B_\mu \right)d_i^0,
\end{align}
\end{subequations}
where $g$ and $ g^\prime$ are the $SU(2)$ and the $U(1)$ gauge couplings respectively.

\begin{figure}[htb]
\begin{center}
\includegraphics[scale=0.2]{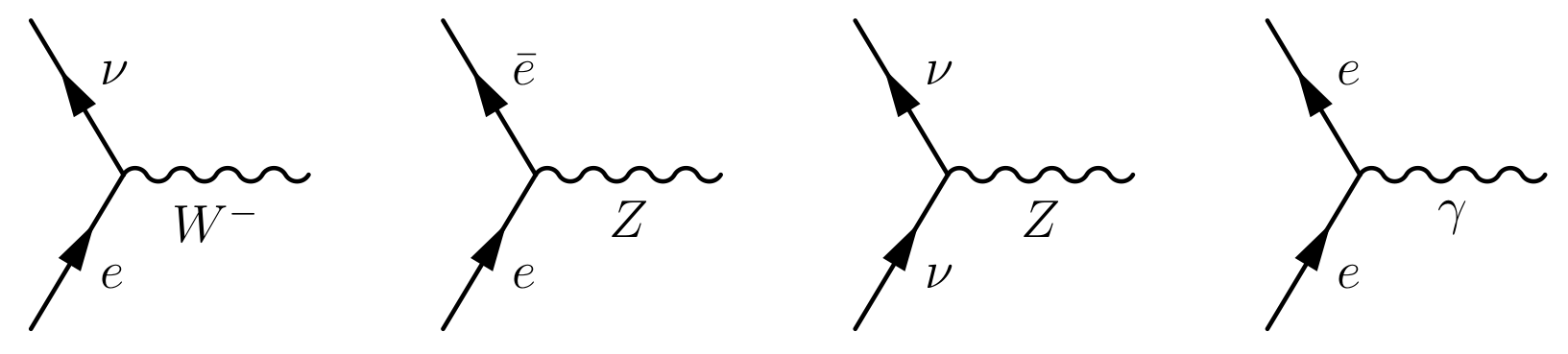}
\end{center}
\vspace*{-8mm}
\caption{Interactions between leptons and vector bosons are presented in terms of Feynman diagrams, which are representations of interactions among particles in QFT.}\label{Fig:WZgl}
\end{figure}

Possible interactions, in terms of the physical fields, of the first generation of leptons and the EW bosons at tree level are presented in Figure~\ref{Fig:WZgl}.

\section{Chiral theory}\label{Eq:Chiral_Theory}

The $\gamma^5$  matrix of eq.~\eqref{Eq:Gamma_5} is also called the chirality matrix. The $\gamma^5$ matrix of the Dirac representation is diagonalisable (change of a representation) by a unitary matrix $U$ via the transformation:
\begin{equation}
U\gamma^5 U^\dagger = \hat{\gamma}^5.
\end{equation}
It is easy to verify that the eigenvalues of $\gamma^5$ are $\pm1$. Consider that the LH $\psi_L$ and the RH $\psi_R$ fields are the eigenfunctions of $\gamma^5$:
\begin{equation}\label{Eq:gamma_eigenfunctions}
\gamma^5 \begin{pmatrix}
\psi_L \\ \psi_R
\end{pmatrix} = \begin{pmatrix}
-\psi_L \\ +\psi_R
\end{pmatrix}.
\end{equation}

Then, a spinor can be separated into the LH and the RH chiral components as:
\begin{equation}
\psi=\psi_L + \psi_R.
\end{equation}
Now, assume that the chirality projection  operator $P_i$ decomposes a spinor field into the LH and the RH components based on its chirality:
\begin{equation}\label{Eq:PLR}
\psi=\left( P_L + P_R \right)\psi=P_L \psi + P_R \psi = \psi_L + \psi_R.
\end{equation}
It follows that:
\begin{equation}\label{Eq:diffPLR}
\psi_L = P_L \psi,\quad\mathrm{and}\quad \psi_R = P_R \psi.
\end{equation}

How are the projection operators $P_L$ and $P_R$ defined? First of all, from eq.~\eqref{Eq:PLR} we can see that $P_L+P_R=1$. Secondly, using eqs.~\eqref{Eq:gamma_eigenfunctions} and  \eqref{Eq:diffPLR} we get:
\begin{subequations}
\begin{align}
& \psi_L = \frac{1}{2} \left( \psi_L + \psi_L + \psi_R - \psi_R \right) = \frac{1}{2} \psi - \frac{1}{2}\left( \gamma^5 \psi_L + \gamma^5 \psi_R \right) = \frac{1-\gamma^5}{2} \psi, \\
& \psi_R = \frac{1}{2} \left( \psi_L - \psi_L + \psi_R + \psi_R \right) = \frac{1}{2} \psi + \frac{1}{2}\left( \gamma^5 \psi_L + \gamma^5 \psi_R \right) = \frac{1+\gamma^5}{2} \psi.
\end{align}
\end{subequations}
We find that the projection operators are:
\begin{subequations}
\begin{align}
P_L ={}& \frac{1-\gamma^5}{2},\\
P_R ={}& \frac{1+\gamma^5}{2}.
\end{align}
\end{subequations}
The projection operators satisfy the following properties:
\begin{subequations}
\begin{align}
& P_{L,R}^\dagger = P_{L,R},\\ 
& P_L + P_R = \frac{1-\gamma^5+1+\gamma^5}{2} = 1,\\
&P_{L,R}^2=\frac{1\pm2\gamma^5+\left(\gamma^5\right)^2}{4}=\frac{1\pm\gamma^5}{2}=P_{L,R}, \label{Eq:PLR2_PLR}\\
&\left[ P_L, P_R \right] = \frac{1}{4} \left( 1 - \left( \gamma^5 \right)^2 - 1 + \left( \gamma^5 \right)^2 \right)=0, \\
&\left\lbrace P_L, P_R \right\rbrace = \frac{1}{4} \left( 1 - \left( \gamma^5 \right)^2 + 1 - \left( \gamma^5 \right)^2 \right)=0 .
\end{align}
\end{subequations}

Let us consider the Dirac Lagrangian of eq.~\eqref{Eq:Lagrangian_Dirac} by writing down different chiral states of the $\psi$ field:
\begin{equation}\label{prob_curr}
\begin{aligned}
\mathcal{L}_D &= \left( \overline{\psi_L} + \overline{\psi_R} \right) \left( i \gamma^\mu \partial_\mu - m \right) \left( \psi_L + \psi_R \right)\\
&=\overline{\psi_L} \left( i \gamma^\mu \partial_\mu - m \right) \psi_L 
 +\overline{\psi_L} \left( i \gamma^\mu \partial_\mu - m \right) \psi_R \\
 &~~~+\overline{\psi_R}\left( i \gamma^\mu \partial_\mu - m \right) \psi_L  
 +\overline{\psi_R} \left( i \gamma^\mu \partial_\mu - m \right) \psi_R,
\end{aligned}
\end{equation}
where the LH and the RH Dirac adjoint states can be written down as:
\begin{subequations}
\begin{align}
\overline{\psi_L} &= (\psi_L)^\dagger \gamma^0 = \psi^\dagger P_L \gamma^0 = \psi^\dagger \gamma^0 P_R = \overline{\psi}P_R,\\
\overline{\psi_R} &= (\psi_R)^\dagger \gamma^0 = \psi^\dagger P_R \gamma^0 = \psi^\dagger \gamma^0 P_L = \overline{\psi}P_L,
\end{align}
\end{subequations}
which are:
\begin{subequations}
\begin{align}
\overline{\psi_L} ={}& \overline{P_L\psi} = \overline{\psi}P_R,\\
\overline{\psi_R} ={}& \overline{P_R\psi} = \overline{\psi}P_L.
\end{align}
\end{subequations}
Taking into consideration the properties of the $\gamma$ matrices \eqref{Eq:gamma_properties} we get:
\begin{equation}
\overline{\psi}_L \gamma^\mu \psi_R = \overline{\psi}_R \gamma^\mu \psi_L =0.
\end{equation}

The Dirac Lagrangian in terms of the chiral fields $\psi_L$ and $\psi_R$ can be written as:
\begin{equation}
\mathcal{L}_D =
 i \overline{\psi_L} \gamma^\mu \partial_\mu \psi_L 
 +i \overline{\psi_R} \gamma^\mu \partial_\mu \psi_R - m (\overline{\psi_L} \psi_R +  \overline{\psi_R} \psi_L).
\end{equation}
As can be seen, the LH chiral fields couple only to the LH chiral fields, and the RH chiral fields couple only to the RH chiral fields. Moreover, the chirality states are space-time dependent since fields can fluctuate among the chiral states. 

The chiral fields $\psi_L$ and $\psi_R$ are also known as Weyl spinors. The Weyl spinor is constructed from two independent components. One of the possible options for the Weyl spinors is:
\begin{equation}
P_L  = \begin{pmatrix}
1 & 0 \\
0 & 0
\end{pmatrix}, \quad \mathrm{and} \quad P_R  = \begin{pmatrix}
0 & 0 \\
0 & 1
\end{pmatrix}.
\end{equation}
Thus the four-component spinor $\psi$ can be written as:
\begin{equation}
\psi = \begin{pmatrix}
\chi_R \\ \chi_L\end{pmatrix}, \text{ where }~ \psi_L = \begin{pmatrix}
0 \\ \chi_L 
\end{pmatrix}, ~ \psi_R = \begin{pmatrix}
\chi_R \\ 0
\end{pmatrix}.
\end{equation}
The $\chi_{L,R}$ states are the Weyl two-component fields.

The Dirac Lagrangian in terms of the Weyl fields is given by:
\begin{equation}
\mathcal{L}_D = \chi_L ^\dagger i \overline{\sigma}^\mu \partial_\mu \chi_L + \chi_R ^\dagger i \sigma^\mu \partial_\mu \chi_R - m \left( \chi_L^\dagger \chi_R + \chi_R^\dagger \chi_L \right),
\end{equation}
where
\begin{equation}
\sigma_\mu=\left( -i \sigma_1 \sigma_2 \sigma_3 ,\sigma_i \right), \quad  \overline{\sigma}^\mu = \left(  -i \sigma_1 \sigma_2 \sigma_3 ,-\sigma_i \right).
\end{equation}
In this case, the free field Dirac equations of motion are:
\begin{equation}
i\overline{\sigma}^\mu\partial_\mu \chi_L - m \chi_R = 0,\quad i\sigma^\mu \partial_\mu \chi_R - m \chi_L = 0.
\end{equation}

\section{The Standard Model Higgs sector}\label{Sec:SM_Higgs}

Now we shall turn our attention to some basic principles of the SM Higgs sector. In order to describe the Higgs mechanism, a complex scalar doublet is required, which we are already partly familiar with. We start with elemental mathematical concepts of scalar field theory. Apart from that, basic ideas of the SSB and the Higgs mechanism are presented. The SSB is one of the fundamental concepts of SM, without which several issues would arise. Finally we take a look at how the Higgs boson interacts with other particles.

The gauge bosons must be massless for the gauge theory to remain unbroken. This is the requirement for the QED and QCD. In order to implement massive gauge bosons $W^\pm,Z$ the theory should be extended to introduce symmetry breaking.  The simplest solution would be to introduce mass terms for the gauge bosons manually; however, this would violate the renormalisability of the theory. Obviously, renormalisable theories are more attractive since the infinities that arise in the process of quantisation (particle interactions) can be removed or absorbed into the re-defined (renormalised) parameters (masses and couplings).

\subsection{Scalar field theory}

The  scalar field is a mathematical object that at every point of the space-time is characterised by a single value and has no associated direction. For a well-behaved theory this physical quantity should be Lorentz invariant. In terms of the QFT scalar fields are associated with spin-zero quantum fields. The Higgs field is the only fundamental scalar quantum field which has been observed so far.

The Lagrangian of a real scalar field $\varphi$ is given by:
\begin{equation}
\mathcal{L}=\frac{1}{2}\left( \partial_\mu \varphi \right)\left( \partial^\mu \varphi \right) - \frac{1}{2}m^2\varphi^2 - \sum_{n=3}^{\infty} \frac{1}{n!} g_n \varphi^n,
\end{equation}
where the $n!$ factor is a symmetry factor due to identical interactions among the fields.

The field equation should satisfy the Klein-Gordon equation in some potential $V(\varphi)$:
\begin{equation}
\left( \partial_\mu \partial^\mu + m^2 \right) \varphi + \frac{\partial V (\varphi)}{\partial \varphi} = 0,
\end{equation}
notice that the mass term was extracted from the potential; in the most trivial case we would have $V (\varphi) \sim \frac{1}{2} m^2 \varphi^2$. The underlying symmetry will depend on the powers of $\varphi^n$. 

The simplest theory is given by the quartic interaction or the $\varphi^4$ theory. The $\varphi^4$ theory is renormalisable in four dimensions. The Lagrangian is invariant under the $\varphi \to - \varphi$ transformation, representing a $\mathbb{Z}_2$ symmetry. This symmetry is central in studying SSB. The Lagrangian for the $\varphi^4$ theory is then given by:
\begin{equation}
\mathcal{L} = \frac{1}{2} \left( \partial_\mu \varphi \partial^\mu \varphi \right) - \frac{1}{2} m^2 \varphi^2 - \frac{1}{4!} \lambda \varphi^4.
\end{equation}
  
Let us now take a look at a complex scalar field $\varphi \neq \varphi^\ast$. Assume that there exists such a field function $\varphi(x^\mu)$, which satisfies $\varphi(x^\mu)=\varphi(x^\mu)^\ast$. This is a real or a Hermitian field; notice that there is a complexified unitary map given by:
\begin{equation}\label{Eq:C_map_varphi}
\frac{1}{\sqrt{2}} \begin{pmatrix}
1 & i \\
1 & -i 
\end{pmatrix} \begin{pmatrix}
\varphi_1 \\
\varphi_2
\end{pmatrix} = \begin{pmatrix}
\varphi_1 + i \varphi_2 \\
\varphi_1 - i \varphi_2
\end{pmatrix} = \begin{pmatrix}
\varphi \\
\varphi^\ast
\end{pmatrix}.
\end{equation}

 The Lagrangian of a complex scalar field can be written as:
\begin{equation}
\mathcal{L}=(\partial_\mu\varphi)^\ast (\partial^\mu\varphi) - m^2 \varphi^\ast \varphi - V_\mathrm{int}(\varphi,\varphi^\ast),
\end{equation}
where the interaction term is:
\begin{equation}
V_\mathrm{int}(\varphi,\varphi^\ast)=\frac{\lambda}{4}\left( \varphi^\ast \varphi \right)^2 + \text{ (non-renormalisable terms) }.
\end{equation}

In general, it is possible to express the complex scalar field $\varphi$ in terms of two real fields, like in eq.~\eqref{Eq:C_map_varphi}:
\begin{subequations}\label{Eq:Complex_sc_f_decomp}
\begin{align}
\varphi ={}& \frac{1}{\sqrt{2}} \left( \varphi_1 + i\varphi_2 \right),\\
\varphi^\ast ={}& \frac{1}{\sqrt{2}} \left( \varphi_1 - i\varphi_2 \right).
\end{align}
\end{subequations}
One could note that this complex scalar field $\varphi$ can be treated as two independent scalar fields $\varphi_1$ and $\varphi_2$. These fields can be viewed as components of a two-dimensional vector. Under the global $U(1)$ transformation we have:
\begin{equation}
\begin{pmatrix}
\varphi_1' \\
\varphi_2' \\
\end{pmatrix} = \begin{pmatrix}
c_\theta & s_\theta\\
-s_\theta & c_\theta
\end{pmatrix} \begin{pmatrix}
\varphi_1 \\
\varphi_2
\end{pmatrix}.
\end{equation}

The most general renormalisable Lagrangian for a complex scalar field, expressed in terms of its components, is given by:
\begin{equation}
\mathcal{L}= \frac{1}{2}\left[ \left( \partial_\mu \varphi_1 \right)^2 +  \left( \partial_\mu \varphi_2 \right)^2 \right] - \frac{1}{2} m^2 \left( \varphi_1^2 + \varphi_2^2 \right) - \frac{\lambda}{4} \left( \varphi_1^2 + \varphi_2^2 \right)^2.
\end{equation}

\subsection{Spontaneous symmetry breaking}

The SSB of a Lagrangian occurs when the underlying symmetry of the system is not continuously invariant, \textit{e.g.}, the laws governing the system exhibit the symmetry, while the state of the system like the lowest energy state, vacuum, does not possess the same symmetry as encoded by the Lagrangian. This typically happens when the field acquires a non-zero vev, which breaks the symmetry of the theory. The SSB enables the existence of several vacuum states. 

The mass terms are not allowed in the Lagrangian for the gauge bosons and fermions as such Lagrangians would not be gauge invariant. On the other hand, the weak bosons are not massless. As a result, there should be some sort of mechanism through which bosons and fermions acquire their masses~\cite{Englert:1964et,Higgs:1964pj,Guralnik:1964eu}. In the SM the Higgs boson (scalar field) potential leads to SSB. The Lagrangian is symmetric under the electroweak gauge group $SU(2)_L \times U(1)_Y$, while the Higgs field vev, which is non-zero, spontaneously breaks the electroweak symmetry down to the electromagnetic symmetry $U(1)_Q$. During this spontaneous breaking the renormalisability of the theory is preserved~\cite{tHooft:1971qjg}.

Let us begin with a simple example. Assume a real scalar field. In this case, the Lagrangian is given by:
\begin{equation}
\mathcal{L} = \frac{1}{2} (\partial_{\mu} \phi )(\partial^{\mu} \phi) - V (\phi),
\end{equation}
where the scalar potential is:
\begin{equation}\label{Eq:V_Scalar_real_pot}
V (\phi) = \frac{1}{2} \mu^2 \phi^2 + \frac{1}{4} \lambda \phi^4.
\end{equation}
As in the $\varphi^4$ theory, the Lagrangian is invariant under the scalar field transformation $\phi \rightarrow - \phi$. In order to bound the scalar potential from below, the value of the coupling coefficient $\lambda$ should be positive, which is also know as the vacuum stability condition. Suppose that such condition is not satisfied, then asymptotically massive fields would let the system reach an unstable configuration, leading to an unphysical situation. 

Now, there are two different possibilities for the sign of the $\mu^2$ parameter, which should be understood as a value, and not a coefficient taken to the power of two. In the case of $\mu^2>0$ everything is trivial. When the vacuum of the potential, $v \equiv \left\langle \phi \right\rangle = \left\langle 0 | \phi | 0 \right\rangle$, satisfies a constant value, the minimum condition can be written as:
\begin{equation}\label{Eq:Min_V}
\frac{\partial V}{\partial \phi}\bigg|_{v}  = v \left( \mu^2 + \lambda v^2 \right) = 0.
\end{equation}
Here we differentiated the scalar potential of eq.~\eqref{Eq:V_Scalar_real_pot} with respect to the $\phi$ field and then evaluated the value at the minimum, $v$. The vacuum of such states corresponds to $v=0$. This Lagrangian represents a free particle characterised by a mass parameter $\mu^2$.

\begin{figure}[htb]
\begin{center}
\includegraphics[scale=1]{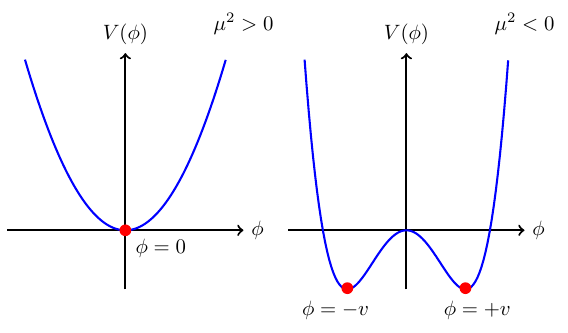}
\end{center}
\vspace*{-3mm}
\caption{The scalar potential for the case of $\mu^2>0$ (left) and for the case of $\mu^2 < 0$ (right) with minima specified.} \label{Fig:Higgs_pot}
\end{figure}

In the case of $\mu^2<0$ the previously mentioned point $v=0$ is no longer stable. Now we get two vacuum conditions by solving for the parentheses of eq.~\eqref{Eq:Min_V}):
\begin{equation}\label{Eq:Min_phi_pm}
\phi_0^{1,2} = \pm \sqrt{-\frac{\mu^2}{\lambda}} = \pm v.
\end{equation}
The scalar Lagrangian is invariant under $\phi \to - \phi$, and therefore there should be no difference in the choice of the vacuum state $\phi_0^1$ or $\phi_0^2$. The non-zero vev breaks the underlying $\mathbb{Z}_2$ symmetry. As it turns out, this is precisely what we wanted to achieve. Behaviour of the scalar potential is depicted in Figure~\ref{Fig:Higgs_pot}.

Let us consider perturbations around the minimum, which will be given by:
\begin{equation}
\phi=v+\eta.
\end{equation}
This can be also viewed as re-scaling of the $\phi$ field to represent the fact that the minimum is given by the $v$ constant.

By introducing such perturbation we introduce an excitation of the field, which turns out to be a physical particle. The scalar Lagrangian in terms of these fluctuations becomes:
\begin{equation}
\begin{aligned}
\mathcal{L} &= \frac{1}{2} \left[\partial_{\mu} (v+\eta) \right]\left[\partial^{\mu} (v+\eta)\right] - \frac{\mu^2}{2}\left( v + \eta \right)^2 - \frac{\lambda}{4}\left( v + \eta \right)^4\\
& \sim \frac{1}{2} \left(\partial_{\mu} \eta \right)^2 - \lambda v^2 \eta^2 - \lambda v \eta^3 - \frac{1}{4} \lambda \eta^4,
\end{aligned}
\end{equation}
where in the second line we dropped constant terms, which are those that do not depend on the $\eta$ field, and also utilised the fact that $\mu^2 = - \lambda v^2$, see eq.~\eqref{Eq:Min_phi_pm}. This Lagrangian describes the $\eta$ particle with a mass
\begin{equation}
m_{\eta} = \sqrt{2\lambda}v,
\end{equation}
along with cubic and quartic self-interactions. Notice that due to the cubic term the Lagrangian is no longer invariant under $\eta \to - \eta$.

Now, let us turn our attention to the complex scalar field, which we are already familiar with from eq.~\eqref{Eq:Complex_sc_f_decomp}. In this case the minimum of the scalar potential is given by 
\begin{equation}\label{Eq:Min_V_comp_phi}
\frac{\partial V}{\partial \phi_i}  = \phi_i \left[ \mu^2 + \lambda \left( \phi_1^2 + \phi_2^2 \right) \right] = 0.
\end{equation}
From here we get
\begin{equation}
\phi_1^2 + \phi_2^2 = -\frac{\mu^2}{\lambda},
\end{equation}
dismissing the other trivial solution, $\phi_i=0$.

Perturbations around the minima yield:
\begin{equation}
\phi = \frac{1}{\sqrt{2}} \left( \phi_1 + i \phi_2 \right) =  \frac{1}{\sqrt{2}} \left( v + \eta + i \chi \right),
\end{equation}
where fluctuations around the $\phi_2$ field are denoted by the $\chi$ excitation.

\begin{figure}[htb]
\begin{center}
\includegraphics[scale=1.4]{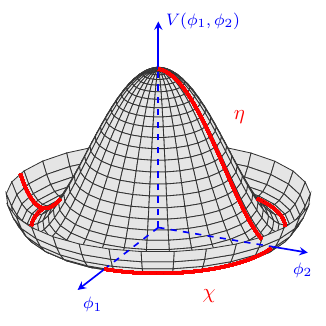}
\end{center}
\vspace*{-3mm}
\caption{Cartoon of the complex scalar potential with $\mu^2<0$ presented in terms of the $\phi_i$ fields in the Cartesian coordinate system and in terms of the $\eta$ and $\chi$ fields in the cylindrical coordinate system. Radial modes, $\eta$, are massive since there is resistance from the scalar potential, which is determined by its steepness, while rotational modes, $\chi$, are massless since those do not oscillate.}\label{Fig:Compl_fields_breaking}
\end{figure}

By expanding the scalar field in terms of the shifted fields we get that $\eta$ becomes massive, while the $\chi$ field remains massless. This amounts to the SSB of the $U(1)$ symmetry by rotating the complex fields $\phi_i$ to the mass basis of the $\eta{-}\chi$ space. The mass parameter is defined by the minima of the circle, which we chose to be $v$, see Figure~\ref{Fig:Compl_fields_breaking}. The massless particle is associated with the Goldstone boson, which we have already encountered in Section~\ref{Sec:EW_theory}.

In reality, the SM Higgs sector is described by the weak isospin doublet:
\begin{equation}
\Phi=\begin{pmatrix}
\phi^+ \\ \phi^0
\end{pmatrix} = \frac{1}{\sqrt{2}} \begin{pmatrix}
\phi_1 + i \phi_2 \\
\phi_3 + i \phi_4 \\
\end{pmatrix}.
\end{equation}

An identical technique, as in the case of the real scalar potential, can be applied to the complex scalar potential. In the complex case we are dealing with the four-dimensional space. It is possible to rotate vevs so that a single component of $\left\langle \phi_i \right\rangle$ is aligned with the direction of the vevs; the natural choice for such alignment would be the direction of either $\phi_1$ or $\phi_3$. Let us assume that there exists a non-zero vev, $v=\left\langle \phi_3  \right\rangle = \mathrm{Const}$, such that:
\begin{equation}
\phi_0 = \frac{1}{\sqrt{2}} \begin{pmatrix}
0 \\ v
\end{pmatrix}.
\end{equation}

By introducing fluctuations around the vev we get four different fields: one massive and three massless. The massive scalar particle corresponds to the SM Higgs boson. Three massless particles are the so-called Nambu-Goldstone bosons \cite{Nambu:1960tm,Goldstone:1961eq} or simply the Goldstone bosons. 

The Goldstone theorem states that if a generic continuous symmetry is spontaneously broken, then for each broken generator a massless scalar particle appears~\cite{Goldstone:1962es}. During the SSB of the $SU(2)_L \times U(1)_Y$ group three generators are broken: for the $SU(n)$ group there are $n^2-1$ generators and for the $U(n)$ group there are $n^2$ generators. These three Goldstone bosons correspond to the longitudinal polarization components of the weak bosons $W^\pm$ and $Z$. These Goldstone bosons are said to be ``eaten" to give the mass terms to the weak bosons.

We considered two cases, when $\mu^2 < 0$ and when $\mu^2 > 0$. One might be curious what happens at the point $\mu^2=0$. This case can not be treated in a classical way and one should consider higher-order corrections to the scalar potential. However, it was shown in Ref.~\cite{Coleman:1973jx} that in this case, with the zero mass term, SSB can still occur through the radiative corrections. It should be noted that this model has one less free parameter. 
 
\subsection{Higgs mechanism}

The main task of the Higgs mechanism is to explain why some of the elementary particles acquire mass terms while others do not. Without this mechanism no bosons would acquire masses. This mechanism assumes that there exists the Higgs field at every point of the space.

The SM Lagrangian for the Higgs field is:
\begin{equation}
\mathcal{L}_\text{Higgs} = \left( D^\mu \Phi \right)^\dagger \left( D_\mu \Phi \right) - V (\Phi),
\end{equation}
where the covariant derivative $D_\mu$ is given by:
\begin{equation}\label{Eq:dmuHiggs}
D_\mu=\partial_\mu + ig\frac{\sigma_i}{2}A_\mu^i + \frac{ig^\prime}{2} B_\mu.
\end{equation}
We do not cover the QCD theory. In general, the covariant derivative of the SM also combines the strong interactions via the $i\frac{g_s}{2} \lambda_a G_\mu^a$ term, where $\lambda_a$ are the Gell-Mann matrices and $g_s$ is the strong $SU(3)_C$ coupling.

The $SU(2)$ Higgs doublet can be parameterised as:
\begin{equation}
\Phi=e^{ \,\frac{i}{v} G^i \sigma_i }\begin{pmatrix}
0 \\  \frac{1}{\sqrt{2}}\left( v+h \right) \end{pmatrix},
\end{equation}
where $G^i$ are the Goldstone bosons. The Goldstone fields are unphysical and thus can be rotated away by an appropriate basis transformation. In the unitary gauge (a basis in which the Goldstone boson components are rotated away) the SM Higgs doublet is:
\begin{equation}
\Phi=\begin{pmatrix}
0 \\ \frac{1}{\sqrt{2}} \left( v+h \right)
\end{pmatrix}.
\end{equation}

Let us consider consider how the Higgs doublet transforms under the electroweak interactions. First, observe that under the $SU(2)_L$ generators of eq.~\eqref{Eq:Relation_to_Pauli_sigma} the Higgs doublet transforms as:
\begin{subequations}
\begin{align}
&\sigma_1 \Phi = \begin{pmatrix}
0 & 1 \\
1 & 0
\end{pmatrix} \begin{pmatrix}
0 \\ \frac{1}{\sqrt{2}} \left( v+h \right)
\end{pmatrix} =  \begin{pmatrix}
\frac{1}{\sqrt{2}}\left(v + h\right) \\
0
\end{pmatrix} \neq 0,\\
&\sigma_2 \Phi = \begin{pmatrix}
0 & -i \\
i & 0
\end{pmatrix} \begin{pmatrix}
0 \\ \frac{1}{\sqrt{2}} \left( v+h \right)
\end{pmatrix} = -i\begin{pmatrix}
 \frac{1}{\sqrt{2}} \left( v+h \right) \\ 0
\end{pmatrix} \neq 0,\\
&\sigma_3 \Phi = \begin{pmatrix}
1 & 0 \\
0 & -1
\end{pmatrix} \begin{pmatrix}
0 \\ \frac{1}{\sqrt{2}} \left( v+h \right)
\end{pmatrix} = -\begin{pmatrix}
0 \\ \frac{1}{\sqrt{2}} \left( v+h \right)
\end{pmatrix} \neq 0.
\end{align}
\end{subequations}
The Higgs doublet has a hypercharge of $Y=1$. Acting with the $U(1)_Y$ generator we get:
\begin{equation}
Y \Phi = (+1) \times \begin{pmatrix}
0 \\ \frac{1}{\sqrt{2}} \left( v+h \right)
\end{pmatrix} = \begin{pmatrix}
0 \\ \frac{1}{\sqrt{2}} \left( v+h \right)
\end{pmatrix} \neq 0.
\end{equation}

As can be observed, all generators of the $SU(2)_L \times U(1)_Y$ group are broken: acting with different generators does not leave the Higgs doublet invariant, $T_a \Phi \neq 0$. The hypercharge was carefully selected so that the lower component of the doublet does not carry an electric charge. 

A logical conclusion follows directly, the $U(1)_Q$ and $SU(3)_C$ generators are not broken. Therefore, under the Higgs mechanism generators are broken to the form:
\begin{equation}
SU(3)_C \times SU(2)_L \times U(1)_Y \rightarrow SU(3)_C \times U(1)_Q.
\end{equation}
The $SU(3)_C$ group is not broken since the $\Phi$ scalar doublet is affected by ``colour blindness". The $U(1)_Q$ symmetry will be discussed shortly.

\subsection{Interactions with bosons}

Let us consider interactions with photons. The $U(1)_Q$ generator, corresponding to the electric charge, leaves the vacuum invariant:
\begin{equation}
Q \left\langle \Phi \right\rangle =  \left( T_3 + \frac{Y}{2}  \right) \left\langle\Phi\right\rangle = \left[ \begin{pmatrix}
1/2 & 0 \\
0 & -1/2
\end{pmatrix} + \frac{1}{2} \begin{pmatrix}
1 & 0 \\
0 & 1
\end{pmatrix} \right] \begin{pmatrix}
0 \\ \sfrac{v}{\sqrt{2}}
\end{pmatrix} = 0.
\end{equation}
Therefore the photon remains massless, which is associated with the unbroken generator. An identical procedure can be applied to the eight gluons, which also remain massless.

The procedure of getting masses of the gauge bosons is straightforward. They are identified by evaluating the kinetic part of the Higgs Lagrangian at the minimum:

\begin{equation}
\begin{aligned}\label{Eq:Kinetic_part_Gen}
\left( D_\mu \Phi \right)^\dagger \left( D^\mu \Phi \right) \Big|_v & = \bigg| \left( \partial_\mu - ig\frac{\sigma_i}{2}A_\mu^i - \frac{ig^\prime}{2} B_\mu \right) \begin{pmatrix}
0 \\ \sfrac{v}{\sqrt{2}} \end{pmatrix}  \bigg| ^2 \\
&= \frac{v^2}{8} \bigg| \left( g \sigma^i A^i_\mu + g^\prime B_\mu \right) \begin{pmatrix}
0 \\ 1 
\end{pmatrix} \bigg| ^2 \\
&= \frac{v^2}{8} \bigg| \begin{pmatrix}
gA^1_\mu-igA_\mu^2 \\
-gA_\mu^3 + g^\prime B_\mu
\end{pmatrix} \bigg| ^2 \\
&= \frac{v^2}{8} \left( g^2 \left[ \left( A^1_\mu \right)^2 + \left( A^2_\mu \right)^2 \right] + \left[ gA_\mu^3-g^\prime B_\mu \right]^2 \right).
\end{aligned}
\end{equation} 
The mass eigenstates in terms of the gauge eigenstates are expressed as:
\begin{subequations}\label{Eq:WZgamma_as_AB}
\begin{align}
W_\mu^\pm &= \frac{A_\mu^1 \mp iA_\mu^2}{\sqrt{2}},\\
Z_\mu &= \frac{gA_\mu^3 - g^{\prime}B_\mu}{\sqrt{g^2+g^{\prime2}}},\\
A_\mu &= \frac{g^\prime A_\mu^3 + gB_\mu}{\sqrt{g^2+g^{\prime2}}}.
\end{align}
\end{subequations}
Then, via the term $W_\mu^+ W_\mu^-$ it easy to identify the mass of the $W^\pm$ boson as:
\begin{equation}
\frac{v^2 g^2}{8}  \left[ \left( A^1_\mu \right)^2 + \left( A^2_\mu \right)^2 \right] = \left(\frac{vg}{2}\right)^2  W_\mu^+ W_\mu^- = m_W^2 W_\mu^+ W_\mu^-.
\end{equation}
The remaining terms (the second square bracket of eq.~\eqref{Eq:Kinetic_part_Gen}) can be re-written as:
\begin{equation}
\frac{v^2}{8} \begin{pmatrix}
A_\mu^3 & B_\mu
\end{pmatrix} \begin{pmatrix}
g^2 & -gg^\prime \\
-gg^\prime & g^{\prime2}
\end{pmatrix} \begin{pmatrix}
A^{3\mu} \\
B^\mu
\end{pmatrix}.
\end{equation}
One can spot that one of the eigenvalues will be zero. In terms of the physical fields $Z_\mu$ and $A_\mu$, after diagonalising the mass matrix, we get:
\begin{subequations}
\begin{align}
m_Z & = \frac{v}{2} \sqrt{g^2+g^{\prime2}}, \\
m_A & = 0.
\end{align}
\end{subequations}

One can introduce the following relations between the gauge couplings,
\begin{subequations}
\begin{align}
\cos \theta_W &= \frac{g}{\sqrt{g^2+g^{\prime2}}},\\
\sin \theta_W &= \frac{g^\prime}{\sqrt{g^2+g^{\prime 2}}},\\
\tan \theta_W &= \frac{g^\prime}{g},
\end{align}
\end{subequations}
where $\theta_W$ is the weak mixing angle.

In terms of the $\theta_W$ mixing angle, the ratio of the masses of the gauge bosons can be expressed as:
\begin{equation}
\cos \theta_W = \frac{m_W}{m_Z}.
\end{equation}

So, we have identified masses of the gauge bosons, which can be thought of as couplings between the gauge fields and the vacuum of the Higgs boson. It should be stressed that the Higgs boson can also couple to the gauge fields. Interactions with the $W^\pm$ and $Z$ vector bosons are given by:
\begin{subequations}
\begin{align}
\mathcal{L}_{VVh} &= \left( \frac{g}{2 \cos \theta_W} m_Z Z_\mu Z^\mu + gm_W W_\mu^+ W^{\mu-} \right) h,\\
\mathcal{L}_{VVhh} & = \left( \frac{g^2}{8 \cos^2 \theta_W} Z_\mu Z^\mu +  \frac{g^2}{4} W_\mu^+ W^{\mu-} \right) hh.
\end{align}
\end{subequations}
Notice that the Higgs boson does not couple directly to photons. This should be of no surprise since we have seen that photons do not become massive, nor we want them to acquire masses. However, photons do couple to the Higgs doublet. In the provided calculations we assumed a unitary gauge. Should we have not done that, we would have gotten some additional interaction terms between the gauge bosons and the Goldstone bosons. 

\subsection{Interactions with fermions}\label{Sec:SM_LY}

We have identified the mass terms of the gauge bosons. The next step is to take a look at the fermionic sector. Everything would have been trivial provided that the $m\bar{\psi}\psi$ term was gauge invariant. Consider the following decomposition of the Dirac mass term:
\begin{equation}
\begin{aligned}
-\mathcal{L}_\text{Dirac}=&m_D\overline{\psi}\psi\\
=& m_D \left(\overline{\psi}_L+\overline{\psi}_R\right)\left(\psi_L+\psi_R\right)\\
=& m_D \left(\overline{\psi}P_R+\overline{\psi}P_L\right)\left(P_L\psi+P_R\psi\right)\\
=& m_D  \overline{\psi} \left(P_L^2 + 2 P_L P_R + P_R^2\right) \psi\\
=& m_D \left(  \overline{\psi}P_L \psi + \overline{\psi}P_R \psi \right)\\
=& m_D \left( \overline{\psi_L} \psi_R + \overline{\psi_R} \psi_L \right).
\end{aligned}
\end{equation}
It turns out that such a decomposition is not gauge invariant as gauge transformations of the LH and the RH fields are different. Therefore there should be another way to construct invariant mass terms for fermions. Let us consider how fermions acquire masses through the Higgs mechanism. 

Interactions between the Higgs doublet and fermions are given by the Yukawa Lagrangian:
\begin{equation}
\begin{aligned}
-\mathcal{L}_Y &= \overline{\psi_i}  Y_{ij} \phi \psi_j \\
&= \overline{\psi_{L_i}}Y_{ij} \phi\psi_{R_j} +  \overline{\psi_{R_i}}Y_{ij}^\dagger \phi^\dagger\psi_{L_J}\\
&= \overline{\psi_{L_i}}Y_{ij} \phi\psi_{R_j} + \mathrm{h.c.},
\end{aligned} 
\end{equation}
where $Y_{ij}$ are the so-called Yukawa couplings.

As an example, let us demonstrate how the down-type quarks interact with the Higgs doublet in the unitary gauge. For simplicity we shall assume that $Y_{ij} \in \mathbb{R}$ and also that there is a single generation of quarks:
\begin{equation}
\begin{aligned}\label{Eq:Ldb_phi_d}
-\mathcal{L}_Y^\text{down}&= \overline{Q_{L}^I} Y^d \Phi d_{R} +  \overline{d_{R}} (Y^d)^{\dagger} \Phi^\dagger Q_{L}^I\\
&=Y^d\left[ (\overline{u}~~\overline{d})_L\begin{pmatrix}
0 \\ \frac{1}{\sqrt{2}}\left(v+h\right)
\end{pmatrix} d_R + \overline{d_R} \left( 0  \quad \frac{1}{\sqrt{2}}\left(v+h\right) \right) \begin{pmatrix}
u \\ d
\end{pmatrix}_L \right]\\
& = \frac{Y^d \left( v+h \right)}{\sqrt{2}}\left[ \overline{d_L} d_R + \overline{d_R} d_L \right] + 0 \times \left[ \overline{u_L} d_R + \overline{d_R} u_L\right] \\
& = \frac{Y^d }{\sqrt{2}}v\overline{d}d+\frac{Y^d }{\sqrt{2}}h\overline{d}d\\
& = m_d \overline{d} d + \frac{m_d}{v}h\overline{d}d,
\end{aligned}
\end{equation}
where the first term is the mass of the down quark and the second term is the Higgs boson coupling to the down quark (generalised to an arbitrary fermion $f$):
\begin{equation}
g_{h\overline{f}f}=-i \frac{m_f}{v}.
\end{equation}
Also we utilised the fact that the Yukawa couplings can be expressed in terms of the mass parameters as:
\begin{equation}
Y^f=\sqrt{2}\frac{m_f}{v}.
\end{equation}

Following the above procedure one would succeed in deriving the mass terms for the down-type quarks and the charged leptons, but not for the other fermions. In order to generate the mass terms for the up-type quarks (and neutrinos) another term in the Lagrangian should be introduced. By intuition (notice where the upper component of the Higgs doublet appears in eq.~\eqref{Eq:Ldb_phi_d}), the term should be proportional to the inverted components of the Higgs doublet, \textit{i.e.}, the vev should appear as the top component. The conjugated Higgs doublet is given by:
\begin{equation}
\tilde{\Phi}\equiv -i \left( \Phi_i^\dagger \sigma_2 \right)^\mathrm{T} = i\sigma_2\Phi^\ast= \begin{pmatrix}
\frac{1}{\sqrt{2}}\left( v+h \right) \\
0
\end{pmatrix},
\end{equation}
which transforms the same way as $\Phi$ under $SU(2)_L$,
\begin{equation}
(i \sigma_2) u_i^\ast (i \sigma_2)^\dagger = u_i.
\end{equation}

Then, the Yukawa Lagrangian for the up-type quarks is:
\begin{equation}\label{Eq:Luhu}
\begin{aligned}
-\mathcal{L}_Y^\text{up} ={}&  \overline{Q_L^I} Y^u \tilde{\Phi} u_R +  \overline{u_R} (Y^u)^{\dagger} \tilde{\Phi}^\dagger Q_L^I\\
 ={}& Y^u \left[ (\overline{u}~~\overline{d})_L\begin{pmatrix}
 \frac{1}{\sqrt{2}}\left(v+h\right) \\ 0
\end{pmatrix} u_R + \overline{u_R} \left( \frac{1}{\sqrt{2}}\left(v+h\right) \quad  0\right) \begin{pmatrix}
u \\ d
\end{pmatrix}_L \right]\\
={}& \frac{Y^u \left( v+h \right)}{\sqrt{2}}\left[ \overline{u_L} u_R + \overline{u_R} u_L \right]\\
 ={}& m_u \overline{u} u + \frac{m_u}{v}h\overline{u}u,
\end{aligned}
\end{equation}

Combining the Yukawa Lagrangian for down-type quarks, given by eq.~\eqref{Eq:Ldb_phi_d}, with the Yukawa Lagrangian for the up-type quarks, given by eq.~\eqref{Eq:Luhu}, we get the full Yukawa Lagrangian for quarks:
\begin{equation}\label{Lquarks}
\begin{aligned}
-\mathcal{L}_Y^\text{quarks}=   \overline{Q_{L}^I} Y^d \Phi D_{R}^I +  \overline{D_{R}^I} (Y^d)^{\dagger} \Phi^\dagger Q_{L}^I + \overline{Q_L^I} Y^u \tilde{\Phi} U_R^I +  \overline{U_R^I} (Y^u)^{\dagger} \tilde{\Phi}^\dagger Q_L^I,
\end{aligned}
\end{equation}
generalised to three generations, see eqs.~\eqref{Eq:F_SU2_L} and \eqref{Eq:F_SU2_R}.

Masses of fermions are not predicted since the Yukawa couplings are free parameters in the SM. The Higgs mechanism only provides us with an answer how masses of fermions are generated, but not how and why the Yukawa couplings develop specific values.

In reality, the model is a bit more involved. First of all, there are three generations of fermions and not a single one as in the above discussed examples. Furthermore, there are no strict restrictions that the Yukawa couplings should be represented by real matrices. Apart from that, we assumed that the fermionic states are given as mass eigenstates (already diagonalised). In short, the Yukawa Lagrangian is of the form (for down-type quarks or charged leptons):
\begin{equation}
-\mathcal{L}_Y = \overline{\psi_{L_i}^0} Y_{ij} \Phi\psi_{R_j}^0 +  \overline{\psi_{R_i}^0} Y_{ij}^\dagger \Phi^\dagger\psi_{R_J}^0,
\end{equation}
where the superscript $``0"$ indicates that fermions are in the weak basis. This means that the corresponding fields transform according to the $SU(2)$ representation. Let us consider the unitary gauge once again:
\begin{equation}
-\mathcal{L}_Y^\text{up} = \overline{Q_{L_i}^0} \left( \mathcal{M}^u_{ij} + \frac{\mathcal{M}^u_{ij}}{v}h \right)u_{R_j}^0 + \mathrm{h.c.}, 
\end{equation}
where the mass matrix $\mathcal{M}^\xi_{ij}=(Y_{ij}^\xi v)/\sqrt{2}$ is not diagonal provided that the Yukawa couplings $Y_{ij}^\xi$ have off-diagonal elements. In order to get definite mass parameters for the particles one needs to diagonalise the  $\mathcal{M}^\xi_{ij}$ matrices. This can be done by introducing unitary transformation matrices $V_L$ and $V_R$ which satisfy,
\begin{equation}
V_{(L,R)}^{\xi\,\dagger}V_{(L,R)}^\xi = \mathcal{I}_{3}.
\end{equation}
The mass matrices are diagonalised via:
\begin{equation}
\hat{\mathcal{M}^\xi} \equiv V_L^{\xi\dagger} \mathcal{M}_{ij}^\xi V_R^\xi =  \begin{pmatrix}
m_1 & 0 & 0 \\
0 & m_2 & 0 \\
0 & 0 & m_3
\end{pmatrix},
\end{equation}
which is also equivalent to diagonalising the Yukawa matrices $Y_{ij}^\xi$ instead of $\mathcal{M}_{ij}^\xi$.

The fermion weak eigenstates can also be expressed as the mass eigenstates via:
\begin{equation}
\begin{aligned}
\psi_{({L,R)}_i} & = \left(V_{(L,R)_{ij}}^{\,\psi}\right)^\dagger \psi_{(L,R)_j}^{\,0}.
\end{aligned}
\end{equation}
Sometimes we shall drop the $``L"$ and $``R"$ indices of the diagonalisation matrices and refer to the LH diagonalisation matrix as $V$ and to the RH diagonalisation matrix as $U$,
\begin{subequations}
\begin{align}
u_L^0=V_u u_L, \quad u_R^0=U_u u_R \\
d_L^0=V_d d_L, \quad d_R^0=U_d d_R.
\end{align}
\end{subequations}

Diagonalisation of the fermion mass matrices also diagonalises the couplings to the Higgs boson; both $v$ and $h$ transform identically. Hence, the Yukawa couplings are diagonal and there are no Flavour-Changing Neutral Currents (FCNC) involving the Higgs boson---processes when a particle changes its flavour while interacting via a neutral current interaction. Absence of the FCNC is guaranteed if fermions, capable of mixing, couple to the same Higgs $n$-tuplet~\cite{Glashow:1976nt,Paschos:1976ay}.

Due to the mismatch between the weak and the mass eigenstates of different fermions there exists a mixing. All of the scalar-fermion interactions, consisting of the Yukawa matrices, can not be diagonalised simultaneously. For example, when the quark sector couples to the upper component  of the scalar doublets, $\phi^\pm$, after going into the mass eigenstates basis, we get two undefined matrices: $V_L^{u\,\dagger} V_L^d$ and $V_L^{d\,\dagger} V_L^u$. The quark mixing matrix provide information on the strength of the flavour-changing weak interactions, and is known as the CKM matrix~\cite{Cabibbo:1963yz,Kobayashi:1973fv}:
\begin{equation}\label{Eq:VCKM_def}
V_\mathrm{CKM} \equiv V_L^{u\,\dagger} V_L^d = 
\begin{pmatrix}
V_{ud} & V_{us} & V_{ub} \\
V_{cd} & V_{cs} & V_{cb} \\
V_{td} & V_{ts} & V_{tb} \\
\end{pmatrix}.
\end{equation}
It can be parameterised in terms of the Euler rotation angles and a complex $\delta$ phase~\cite{Chau:1984fp}. It should be noted that in the SM the CKM matrix is unitary; experimentally, the CKM matrix shows a small deviation of the first row from unitarity as discussed in Refs.~\cite{Belfatto:2019swo,Coutinho:2019aiy,Crivellin:2020lzu}. Also, historically, the CKM matrix was not derived from the Higgs mechanism but in the context of understanding why certain weak interactions involving neutral currents are rare and why the FCNC processes are suppressed in loop diagrams. This was done in the context of the Glashow–Iliopoulos–Maiani (GIM) mechanism~\cite{Glashow:1970gm}. It was later extended to three generations of quarks by M. Kobayashi and T. Maskawa.

In the SM there is no CKM-like matrix for leptons. However, it is well-established that neutrinos oscillate. As the SM assumes massless neutrinos, this behaviour was not originally included in the SM. However, the observation of neutrino oscillations provides direct evidence that neutrinos have small, nonzero masses, requiring an extension of the Standard Model. The Pontecorvo-Maki-Nakagawa-Sakata (PMNS) matrix~\cite{Pontecorvo:1967fh,Maki:1962mu}:
\begin{equation}
V_\mathrm{PMNS}=V_L^{\nu\,\dagger} V_L^l  = 
\begin{pmatrix}
V_{e1} & V_{e2} & V_{e3} \\
V_{\mu 1} & V_{\mu 2} & V_{\mu 3} \\
V_{\tau 1} & V_{\tau 2} & V_{\tau 3} \\
\end{pmatrix},
\end{equation}
is a unitary matrix that describes the mixing of neutrino flavours and the neutrino mass eigenstates. The violation of unitarity of the PMNS matrix, experimentally, is small~\cite{Esteban:2024eli}. The form of the parameterised PMNS matrix depends on whether neutrinos are of Dirac or Majorana origin.

\subsection{Decay rates}

The Higgs boson cannot be detected directly in particle detectors, at least not as electrons or photons are detected. The Higgs boson presence is worked out through its decay products and the overall energy signatures in the particle collision events. There are several possible decay processes. 

The decay rate refers to the probability per unit time that a particle will decay into other particles. It is essentially a measure of how quickly an unstable particle transforms into other particles due to the fundamental forces.  The decay rate $\Gamma$ is related to the lifetime $\tau$ of the particle via the following relation:
\begin{equation}
\Gamma = \frac{1}{\tau}.
\end{equation}
The total decay rate is the sum of all possible decay channels.

\begin{figure}[htb]
\begin{centering}
\begin{picture}(400,200)(0,0)
  \put(0,0){\includegraphics[width=220pt]{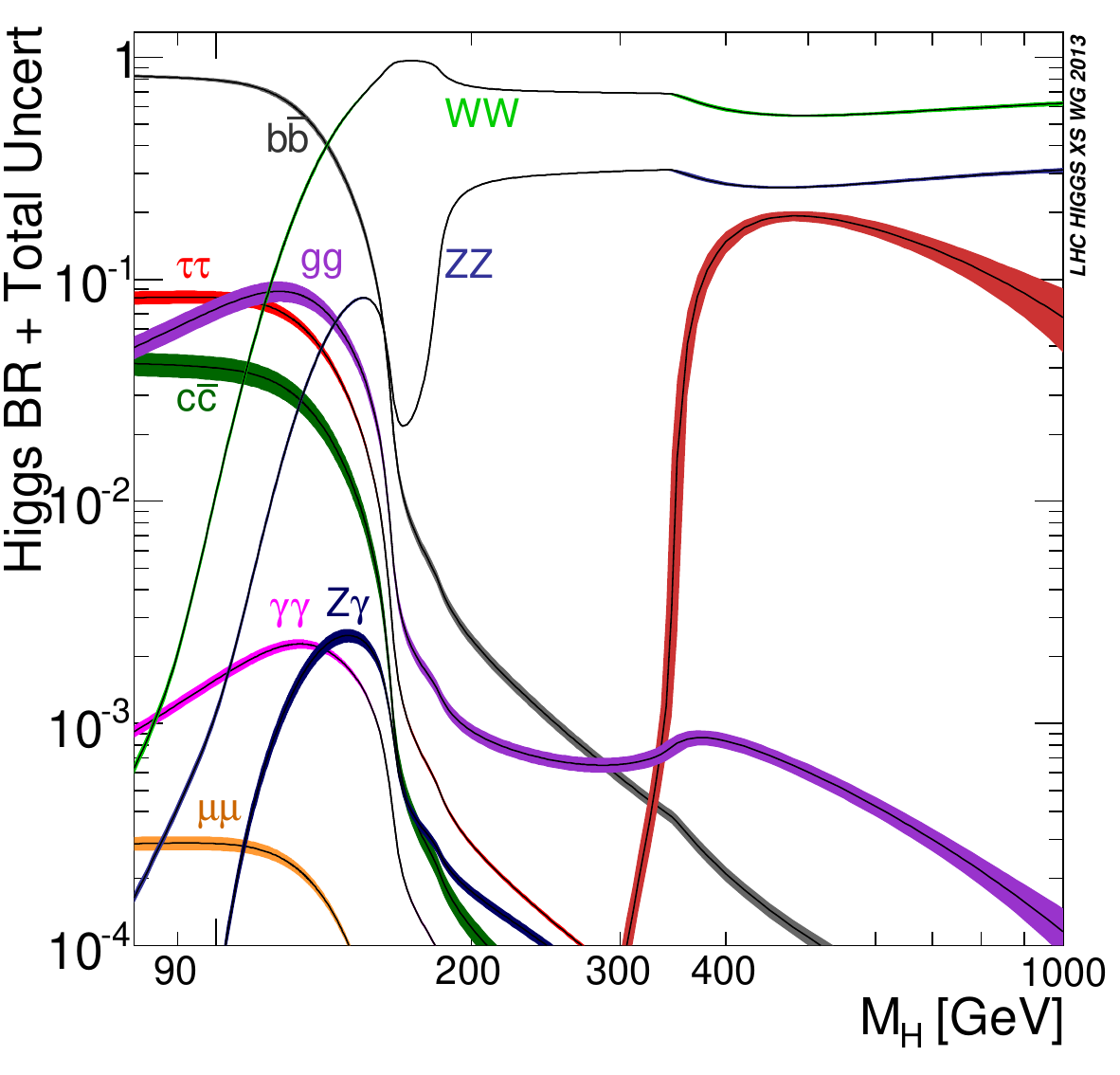}}
  \put(232,0){\includegraphics[width=220pt]{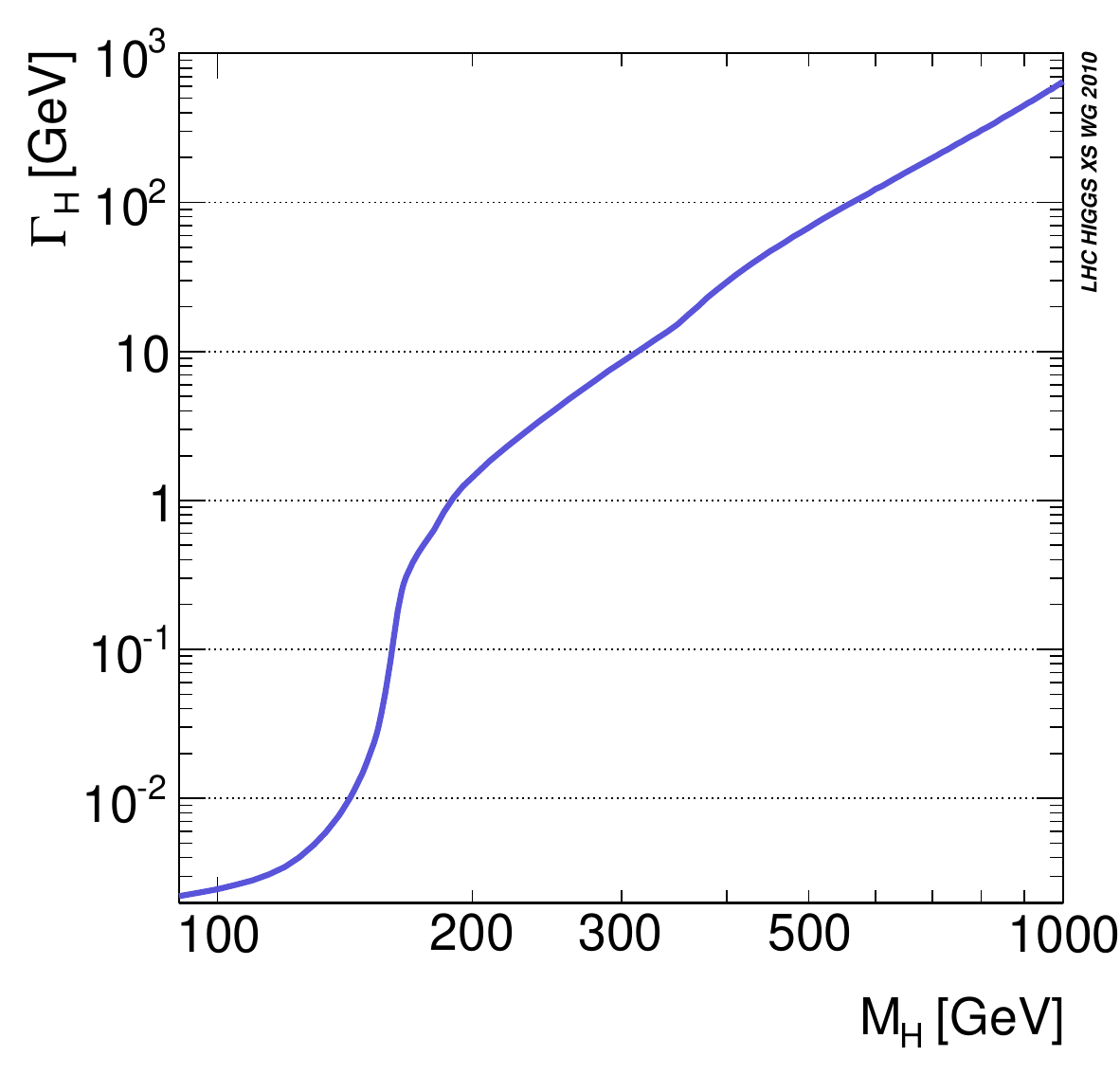}}
\end{picture}
\end{centering}
\caption{The SM Higgs boson branching ratios, and their uncertainties, and the total width as a function of the Higgs boson mass. Figures courtesy of the LHC Higgs Cross Section Working Group~\cite{LHCHiggsCrossSectionWorkingGroup:2013rie}.}\label{Fig:SMHiggsfig}
\end{figure}

Another useful parameter, when talking about the particle decays, is the branching ratio. The branching ratio refers to the fraction of a particular decay mode of a particle relative to the total number of the decay modes of that particle:
\begin{equation}
\mathrm{Br}_i=\frac{\Gamma_i}{\Gamma_{total}}.
\end{equation}
Possible decay channels of the Higgs boson are presented in Figure~\ref{Fig:SMHiggsfig}.

Let us take a look at the two body Higgs boson decay rates, detailed analysis can be found in Refs.~\cite{Gunion:1989we,Carena:2002es,Djouadi:2005gi,Denner:2011mq,LHCHiggsCrossSectionWorkingGroup:2016ypw,ParticleDataGroup:2024cfk}. One of the possible Higgs boson decays is into a pair of fermions. The decay width of such process is given by:
\begin{equation}
\Gamma_{h \rightarrow f \bar{f}} = \frac{N_C m_f^2 m_h}{8 \pi v^2} \left( 1 - \frac{4 m_f^2}{m_h^2} \right)^{\sfrac{3}{2}},
\end{equation}
where the colour number is $N_C=1$ for leptons or $N_C=3$ for quarks.

Another possible Higgs decay is into a pair of EW bosons. Then, the decay width of such process is given by:
\begin{equation}
\Gamma_{h \rightarrow VV^\ast} = \frac{m_V^4}{4 \pi m_h v^2} \sqrt{1 - \frac{4 m_V^2}{m_h^2}} \left( 3 + \frac{m_h^4}{4 m_V^4} - \frac{m_h^2}{m_V^2} \right).
\end{equation}
In the case of the $H \rightarrow ZZ$ decay there are two indistinguishable outgoing fields. Therefore the decay width should be multiplied by a factor of $1/2$. The Higgs boson is lighter than a pair of the EW bosons. Therefore, the Higgs boson can not  decay directly into a pair of the EW bosons. Due to the kinematic restrictions an off-shell gauge boson is produced.

The Higgs boson does not couple directly to neither photons nor gluons, rather it couples via loop interactions. Nevertheless, these channels are of significant importance at the Large Hadron Collider (LHC). Another decay channel mediated via loop diagrams is the Higgs boson decay to $Z\gamma$, evidence of which was discussed in Ref.~\cite{ATLAS:2023yqk}.

In the SM, production of the Higgs boson at the LHC is dominated by the gluon fusion process, contributing around eighty-ninety per cent of the total Higgs boson production events; in the lepton colliders, the dominant production mechanism would be associated $Z$ production (Higgstrahlung) or the $WW$ fusion process, depending on the total collision energy. The decay width to a pair of gluons is:
\begin{equation}
\Gamma_{h \rightarrow gg} = \frac{\alpha\alpha_s^2m_h^3}{8\pi^2\sin^2\theta_W m_W^2} \mathcal{J}(m_h^2,m_f^2),
\end{equation}
where $\mathcal{J}$ is a form factor. In this process, the QCD corrections play a significant role, with the QCD corrections of order sixty per cent~\cite{Spira:1995rr}. 

The Higgs di-photon decay at the lowest order is given by:
\begin{equation}
\Gamma\left( h\to\gamma\gamma \right) = \left| \mathcal{F} \right|^2 \left( \frac{\alpha}{4\pi} \right)^2 \frac{G_Fm_h^3}{8\sqrt{2}\pi},
\end{equation}
where
\begin{equation}
\mathcal{F} = \mathcal{F}_W(\beta_W) + \sum_{\mathrm{fermions}} N_CQ^2\mathcal{F}_f(\beta_f),
\end{equation}
and the corresponding contributions from the higher-order diagrams are:
\begin{subequations}
\begin{align}
&\mathcal{F}_W (\beta) = 2 + 3\beta + 3\beta\left( 2 - \beta \right)f(\beta),\\
&\mathcal{F}_f (\beta) = -2\beta\left[ 1 + \left( 1 - \beta \right)f(\beta) \right],\\
&f(\beta) = \begin{cases}
\arcsin^2\left( \beta^{-\sfrac{1}{2}} \right), & \beta\geq1,\\
-\frac{1}{4}\left[ \ln\left( \frac{1 + \sqrt{1-\beta}}{1 - \sqrt{1-\beta}} \right)\right], & \beta<1. \\
\end{cases}
\end{align}
\end{subequations}
The $\beta_{W,f}$ variables are the $W$ boson and the fermion (dominated by the top loop) mass relations to the Higgs boson mass:
\begin{equation}
\beta_W = \frac{4m_W^2}{m_h^2},\qquad \beta_f = \frac{4m_f^2}{m_h^2}.
\end{equation}

\begin{figure}[htb]
\begin{center}
\includegraphics[scale=0.225]{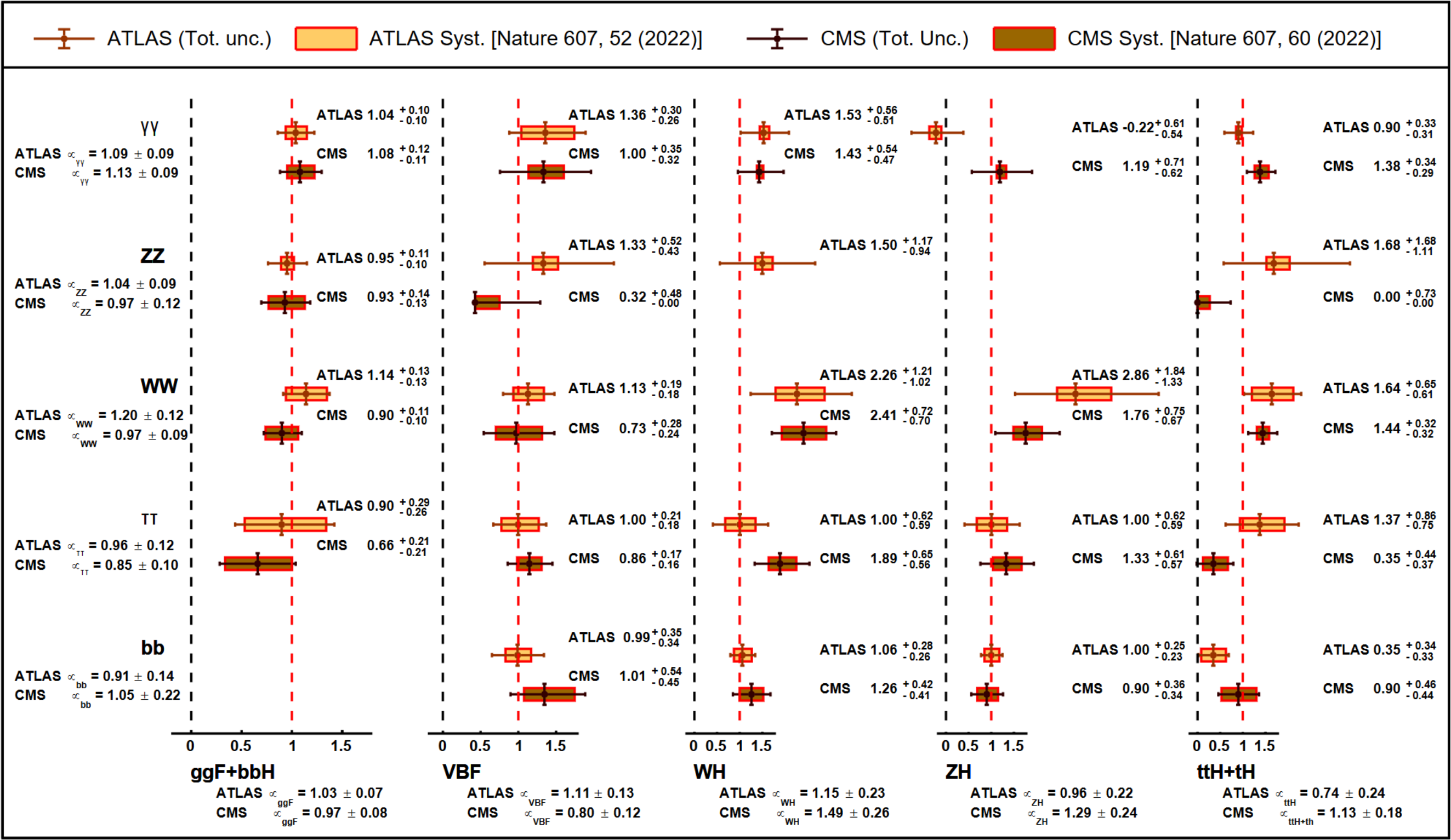}
\end{center}
\vspace*{-3mm}
\caption{Main production and decay channels of the measured Higgs boson state at the LHC. Figure courtesy of the PDG~\cite{ParticleDataGroup:2024cfk}, based on the ATLAS~\cite{ATLAS:2022vkf} and CMS~\cite{CMS:2022dwd} data.}\label{Fig:Higgs_exp_res}
\end{figure}

Comparison of the main production and decay channels, as measured at LHC by ATLAS and CMS, can be observed in Figure~\ref{Fig:Higgs_exp_res}. As of current, while some measurements do not coincide exactly with what the SM predicts for the Higgs boson, there is no significant deviation from the SM scenario. It should be noted the measurement of the Higgs boson self interactions at future particle collider facilities would be a more evident probe of the SM nature of the Higgs boson.

\subsection{Theoretical constraints}\label{Sec:SM_Theor_Constr}

In the SM, the Higgs boson mass is a free parameter. Provided that the perturbative framework of the SM holds, there are several theoretical constraints for the upper bound on the mass of the Higgs boson. When contributions from the lowest order Feynman diagrams are small, perturbative methods can be used. The idea of perturbative unitarity comes from the elastic scattering of the longitudinal components of the $W^\pm$ gauge boson, $W_L^+W_L^- \rightarrow W_L^+W_L^-$, amplitude of which grows with energy. Eventually this process violates the unitarity bound at energies $\Lambda \sim \text{TeV}$ in theories without the Higgs boson.  In the absence of the Higgs boson the considered elastic scattering would diverge quadratically. As a result, for the SM to remain unitary and renormalisable the Higgs boson should be introduced. 

Let us take a closer look at the unitarity bound. We start by considering the Jacob-Wick expansion~\cite{Jacob:1959at} of the Lorentz-invariant amplitude, assuming that the initial and the final particles have identical helicities:
\begin{equation}
\mathcal{M} = 16 \pi \sum_{l=0}^\infty \left(2l+1 \right) \mathcal{M}_l (|\overline{p}|) P_l (\cos \theta),
\end{equation}
where $\mathcal{M}_l$ is the partial-wave amplitude for the $l$'th wave and $P_l$ are the Legendre polynomials, depending on the $\theta$ scattering angle. The squared scattering amplitude, $|\mathcal{M}|^2$, is understood as the probability for scattering to occur, which can not be larger than unity, and hence the sum of the partial-waves is also bounded from above.

The partial-wave amplitude in terms of the Mandelstam variables was provided in Ref.~\cite{Balachandran:1968zza}:
\begin{equation}
\mathcal{M}_l(\overline{p}) = \frac{\sqrt{2}}{4i|\overline{p}|}\left( S_l-1 \right),
\end{equation}
where $S_l$ is a unitary matrix element and $\overline{p}$ is the momentum of the colliding particles in the centre-of-mass frame. In the high-energy limit we get:
\begin{equation}
|\overline{p}| \rightarrow \frac{1}{2}\sqrt{s}.
\end{equation}
The partial-wave amplitude can be decomposed into:
\begin{equation}
|\mathbb{I}\mathrm{m}(\mathcal{M}_l)| \geq |\mathcal{M}_l|^2 =  \mathbb{I}\mathrm{m}(\mathcal{M}_l) ^2 +  \mathbb{R}\mathrm{e}(\mathcal{M}_l) ^2.
\end{equation}
This, in turn, requires $|\mathcal{M}_l|\leq1$. There exists a much stronger constraint~\cite{Luscher:1988gc}:
\begin{equation}\label{Eq:Part_Re}
|\mathbb{R}\mathrm{e}(\mathcal{M}_l)|\leq \frac{1}{2}.
\end{equation}
The latter bound keeps results a bit farther away from the non-perturbative parameter space. As a matter of fact, there is no strong preference to which bound to use, and both bounds appear in literature. Alas, the looser $|\mathcal{M}_l|\leq1$ bound can be exploited as a vague glimpse of hope in some models for a wider range of the parameter space to survive.

The partial-wave amplitude can be written as:
\begin{equation}
\mathcal{M} = -  \frac{g^2 m_h^2}{2m_W^2}.
\end{equation}
From the $|\mathcal{M}_l|\leq1$ condition we get:
\begin{equation}
m_h^2 \leq \frac{4\pi \sqrt{2}}{G_F},
\end{equation}
where $G_F$ is the Fermi constant.

In the high-energy limit, using the Goldstone equivalence theorem~\cite{Cornwall:1974km,Vayonakis:1976vz,Lee:1977eg,Chanowitz:1985hj} (in the high-energy limit, the scattering amplitudes for the longitudinally polarised states of the massive gauge bosons correspond to the Goldstone counterparts), the Lagrangian for the quartic interactions in the basis of $\{w^-w^+,\, z^2,\,h^2,\,zh\}$ can be expressed as:
\begin{equation}
\mathcal{L} = -\frac{g^2 m_h^2}{8 m_W^2} \times \frac{1}{4}\left( 2w^-w^+ + z^2 + h^2 \right)^2,
\end{equation}
where the $w^\pm$ and $z$ fields should be interpreted as the longitudinal components, which are scalars. The scattering matrix in the same basis is then given by:
\begin{equation}
S = \begin{pmatrix}
  4 & \sqrt{2} & \sqrt{2} & 0 \\
  \sqrt{2} & 3 & 1 & 0 \\
  \sqrt{2} & 1 & 3 & 0 \\
  0  & 0 & 0 & 2 \\
 \end{pmatrix}.
\end{equation}
Only the first partial-wave amplitude is present, $\mathcal{M}^0$. There are three sources of the coefficients: the overall coefficients of the quartic couplings, every $\varphi^n$ should be normalised to $\varphi^2/\sqrt{2}$ due to the Bose-Einstein statistics, the symmetry factor $\varphi^n \to n!\,\varphi^n$ should be accounted for. Writing the $s$-matrix coefficients explicitly, in the basis of $\{w^-w^+,\, z^2,\,h^2,\,zh\}$ we have:
\begin{equation*}
\renewcommand\arraystretch{1.25}
  \underbrace{
\left(
\begin{array}{cccc}
 1 & \frac{1}{\sqrt{2}} & \frac{1}{\sqrt{2}} & 1 \\
 \frac{1}{\sqrt{2}} & \frac{1}{2} & \frac{1}{2} & \frac{1}{\sqrt{2}} \\
 \frac{1}{\sqrt{2}} & \frac{1}{2} & \frac{1}{2} & \frac{1}{\sqrt{2}} \\
 1 & \frac{1}{\sqrt{2}} & \frac{1}{\sqrt{2}} & 1 \\
\end{array}
\right)}_{\varphi^2 \to \varphi^2/\sqrt{2}}
  \underbrace{
\left(
\begin{array}{cccc}
 4 & 2 & 2 & 1 \\
 2 & 24 & 4 & 6 \\
 2 & 4 & 24 & 6 \\
 1 & 6 & 6 & 4 \\
\end{array}
\right)}_{\varphi^n \to n!\,\varphi^n} 
\underbrace{
\left(
\begin{array}{cccc}
 1 & 1 & 1 & 0 \\
 1 & \frac{1}{4} & \frac{1}{2} & 0 \\
 1 & \frac{1}{2} & \frac{1}{4} & 0 \\
 0 & 0 & 0 & \frac{1}{2} \\
\end{array}
\right)}_{\text{quartic terms of }\mathcal{L}}.
\end{equation*}
The largest eigenvalue of the above $S$-matrix has to be less or equal to the unit element, $\frac{G_F m_h^2}{4 \pi \sqrt{2}}\,\frac{3}{2}\leq 1$. Taking everything into consideration we can evaluate that the Higgs boson should not be heavier than:
\begin{equation}
m_h < \sqrt{\frac{8\pi\sqrt{2}}{3G_F}} \approx  1008~\mathrm{GeV},
\end{equation}
and is known as the Lee-Quigg-Thacker bound~\cite{Lee:1977yc}. On the other hand, the stronger restriction of eq.~\eqref{Eq:Part_Re} yields:
\begin{equation}
m_h < \sqrt{\frac{4\pi\sqrt{2}}{3G_F}} \approx  713~\mathrm{GeV}.
\end{equation}
If the Higgs boson was found to be heavier than the above established limit, perturbation theory would break down, and the model would cease to be renormalisable. This scenario is particularly plausible for incomplete models that rely on effective theories. It is important to note that we considered only the tree-level diagrams for the scattering processes. If higher-order processes were to contribute significantly, that would signal that the perturbation theory should likely fail.  

Another class of constraints comes from the oblique parameters. These parameters are important when there are contributions from new particles which affect the precision observables through vacuum polarisation loops, $e.g.$, new fermions or additional scalar tuplets. One of such parameters is the $\rho$ parameter, which shows the relative strength of the neutral and charged current weak interactions \cite{Ross:1975fq}. In the SM, the value of the $\rho$ parameter,
\begin{equation}
\rho = \frac{m_W^2}{m_Z^2 \cos^2 \theta_W},
\end{equation}
is predicted to be unity, and it is protected by the exact custodial symmetry (a symmetry that remains after SSB~\cite{Sikivie:1980hm}). This arises because the SM predicts a very specific relationship between the masses of the EW gauge bosons, which is a consequence of the EW symmetry breaking via the Higgs mechanism. Taking a step forward, the general expression for the tree-level $\rho$ with $N$ multiplets is \cite{Langacker:1980js}:
\begin{equation}
\rho = \sum_{i=1}^N  \frac{\left( (T_3)_i\left[ (T_3)_i + 1 \right] - \frac{Y_i^2}{4} \right) v_i}{\frac{1}{2} Y_i^2 v_i}.
\end{equation}

The other set of parameters we want to mention are the so-called Peskin–Takeuchi parameters~\cite{Peskin:1990zt,Peskin:1991sw}, also known as the electroweak oblique corrections, and can be viewed as Wilson coefficients of the higher-dimension operators. These parameters (usually only $S,~T$ and $U$ are evaluated) are used to characterise effects Beyond the Standard Model (BSM) physics, particularly in the context of the EW interactions. These parameters show possible new physics contribution to the EW radiative corrections; the vacuum polarization diagrams that contribute to the four fermion scattering processes. The Peskin-Takeuchi parameters are defined in such a way that at the reference point, that would be in the SM, they are zero. Parameters of the oblique corrections are the self energies of the $W^\pm,~Z,~\gamma$ bosons. The self energy of the vacuum polarisation amplitude is given by:
\begin{equation}
\Pi_{ij}(q^2) = \Pi_{ij} (0) + q^2 \Pi^\prime_{ij}(0),
\end{equation}
where there are four possible options for the self energies: $\Pi_{WW}$, $\Pi_{ZZ}$, $\Pi_{\gamma\gamma}$ and $\Pi_{Z\gamma}$. 

The Peskin-Takeuchi parameters are defined as:
\begin{subequations}
\begin{align}
\frac{\alpha\,S}{4 \sin^2 \theta_W \cos^2 \theta_W} ={}&  \Pi^\prime_{ZZ}(0) - \frac{\cos^2 \theta_W-\sin^2 \theta_W}{\cos \theta_W \sin \theta_W}\Pi^\prime_{Z\gamma}(0) - \Pi^\prime_{\gamma\gamma}(0),\\
\alpha\,T ={}& \frac{\Pi_{WW}(0)}{m_W^2} - \frac{\Pi_{ZZ}(0)}{m_Z^2},\\
\begin{split} \frac{\alpha\, U}{4\sin^2 \theta_W} ={}&  \Big[ \Pi^\prime_{WW} (0) - \cos^2 \theta_W \Pi^\prime_{ZZ}(0)\\ &~ - 2  \sin (2\theta_W) \Pi^\prime_{Z\gamma}(0) - \sin^2 \theta_W \Pi^\prime_{\gamma\gamma}(0) \Big], \end{split}\
\end{align}
\end{subequations}
where $\alpha$ is the fine structure constant. The above equations assume the $\overline{\text{MS}}$ renormalisation scheme~\cite{tHooft:1973mfk,Weinberg:1973xwm}.

The $S$ parameter shows the possible extension of the fermionic sector. It measures the symmetry between the number of different chirality fermions that carry weak isospin. The $T$ parameter is dependant on the difference between the loop corrections of the $W$ vacuum polarisation function and the $Z$ vacuum polarisation function. It measures discrepancy of the total weak isospin of a new model. Also, it is related to the previously mentioned $\rho$ parameter via,
\begin{equation}
\alpha \, T = \frac{\rho-1}{\rho}.
\end{equation}
Both the $S$ and $T$ parameters are sensitive to the mass of the Higgs boson. The last parameter $U$ gets a small contribution from the BSM physics. This is because the $U$ parameter is a dimension-eight operator.

\section{Physics beyond the Standard Model}\label{Sec:BSM}

Although it might sound that the SM successfully describes Nature, this is far from being so. Indeed the SM is a great physical model, which is capable of answering many important questions and made outstanding predictions. Acknowledging the fact that the SM is not a complete theory, it is worth taking a note that there is the BSM physics. There is sufficient evidence that there is a variety of phenomena, which the SM can not explain. Some of these are:
\begin{itemize}
\item Theory of gravity;
\item Dark Matter (DM) and dark energy;
\item Asymmetry of matter-antimatter;
\item Neutrino masses;
\item Anomalous magnetic moment of the muon;
\item Strong CP problem.
\end{itemize}

One of the hugest drawbacks of the SM is that it does not explain gravity. In classical General Relativity, gravity is not a force mediated by a particle, like the other forces in the quantum mechanical framework, but rather by a curvature of space-time caused by the energy and momentum of matter and radiation. An introduction of a graviton to the SM Lagrangian does not correspond to what is observed experimentally. So far the attempts to quantise gravity have been futile since gravity remains to be non-renormalisable in QFT and there are major issues at the Planck scale, at which quantum gravitational effects are expected to become significant.

Another huge drawback is that the SM explains roughly only five per cent of the mass-energy present in the Universe. Around twenty-seven per cent of the energy is the hypothetical DM, while the bulk of the energy should consist of the so-called dark energy. The DM is a postulated form of matter that does not interact with the electromagnetic radiation, indicating that it is likely electrically neutral, but exerts gravitational influence. Observations of galaxy rotation curves, gravitational lensing, and Cosmic Microwave Background (CMB) radiation suggest existence of DM. There are various candidates for DM in the BSM theories. Several candidates shall be mentioned in Chapter~\ref{Ch:Cosmology}. Talking about dark energy, the SM does not provide a satisfactory explanation for the observed acceleration of the expansion of the Universe.

As evident, the SM must be extended to address some of the described issues. A straightforward approach is to enlarge the SM Higgs sector by introducing additional tuplets.

\chapter{Basic cosmological principles}\label{Ch:Cosmology}

There is abundant evidence that the luminous (observable) matter is responsible only for a tiny fraction of all matter in the Universe, unless our understanding of general relativity is wrong and needs to be modified, which is not discussed here. The failure to apply Kepler’s third law to the observed rotation curves of the galaxies leads to the speculation that there exists a non-luminous matter component. The further studies lead to even more hints that there is indeed a significant amount of matter missing. In this Chapter we discuss basic cosmological ideas which motivate the scientific community to believe that there is DM. We shall mainly follow Ref.~\cite{Plehn:2017fdg} for mathematical aspects, though it should be noted there are several inconsistencies in derivations. 

Our description of the Universe can go back to physics of the Planck epoch, of energies around ${ 10^{19}}$ GeV, beyond which our comprehension of Physics miserably fails. General relativity proposes a gravitational singularity at higher energies. In the standard cosmological $\Lambda_\mathrm{CDM}$ model (where $\Lambda$ stands for the cosmological constant and CDM is the abbreviation for Cold Dark Matter), described by the Friedmann-Lemaître-Robertson-Walker (FLRW) metric, it is assumed that at temperatures around $T \sim 10^{16}$ GeV the SM gauge forces were unified into a yet unknown symmetry governed by the Grand Unified Theory; gravity was separated. Afterwards, depending on the inflationary model~\cite{Martin:2013tda} chosen, the Universe underwent a rapid expansion. With an immense amount of energy released, a hot dense ``quark soup"  spread out. As the strong force was separated from the other forces, a large amount of exotic particles was formed. At around $T\sim10^2$ GeV the Higgs field was formed, the underlying symmetry was broken and particles became massive. At $T\sim 10$ GeV Weakly Interactive Massive Particles (WIMPs)~\cite{Lee:1977ua} decoupled from the thermal bath. Then, at $T\sim 100$ MeV QCD transition happened and confinement of quarks and gluons resulted in the hadronic plasma. At around $T\sim 1$ MeV neutrinos froze out, photons no longer photodissociated light nuclei and primordial nucleosynthesis proceeded. Finally, light neutral atoms were formed and the Universe became transparent to photons.

We shall focus on WIMPs as DM candidates. Since, basically, ``nothing" is known about the origin of the hypothetical DM, numerous hypotheses exist regarding the composition of DM. There are several well-motivated candidates. For example, in the inflationary or the radiation-dominated phases of the early Universe extremely dense concentrations of subatomic matter could have undergone gravitational collapse, forming primordial black holes, which are possible DM candidates~\cite{Sasaki:2018dmp,Carr:2020xqk,Green:2020jor}. There are several candidates within supersymmetry, \textit{e.g.}, neutralino WIMPs and gravitinos~\cite{Moroi:1993mb,Feng:2004mt,Rychkov:2007uq,Ellwanger:2009dp}. Another possible DM candidate is sterile neutrino, which does not interact via the SM weak force~\cite{Lesgourgues:2006nd,Boyarsky:2009ix,Boyarsky:2018tvu,Bringmann:2022aim}. Yet another constituents of DM could be axions~\cite{Kim:2008hd,Jaeckel:2010ni,Arias:2012az,Marsh:2015xka,Irastorza:2018dyq,Ferreira:2020fam}. Or, perhaps, in the presence of extra dimensions the Kaluza-Klein excitations could serve as candidates for DM~\cite{Burnell:2005hm,Kong:2005hn,Hooper:2007qk,Belanger:2010yx}. However, there is always an option that we might have been fooled by gravity~\cite{Sanders:2002pf,Bekenstein:2004ne,Clifton:2011jh}, but these theories have their own problems. There are many more candidates, see Refs.~\cite{Bertone:2016nfn,Arcadi:2017kky,Roszkowski:2017nbc,Cirelli:2024ssz} for an overview. Also the question of the DM candidate is not only in its nature, but also the production mechanism~\cite{Carlson:1992fn,Moroi:1999zb,Lin:2000qq,McDonald:2001vt,Hall:2009bx,Petraki:2013wwa,Zurek:2013wia,Li:2013nal,Hochberg:2014dra,Co:2015pka,Kuflik:2015isi,Pappadopulo:2016pkp,Tulin:2017ara,Kramer:2020sbb,DAgnolo:2020mpt,Bringmann:2021tjr,Puetter:2022ucx}.

\section{Evolution of the Universe}\label{Sec.Universe_Evolution}

\begin{figure}[htb]
\begin{center}
\includegraphics[scale=0.42]{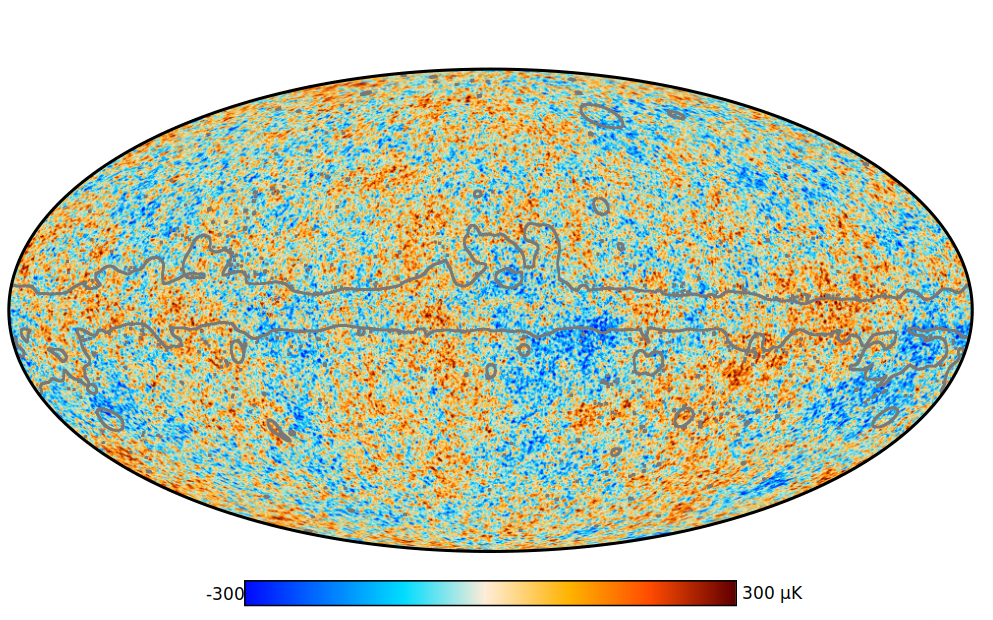}
\end{center}
\vspace*{-4mm}
\caption{ The 2018 Planck map of the temeprature anisotropies of the CMB, extracted using the spectral matching independent component analysis. The gray outline shows the extent of the confidence mask. Figure taken from Ref.~\cite{Planck_CMB_link}. }
\label{Fig:Planck_CMB}
\end{figure}

The geometry of spacetime is determined by the energy content of the Universe and is described by the Einstein field equations:
\begin{equation}
R_{\mu \nu}-\frac{1}{2} R g_{\mu \nu} \equiv G_{\mu \nu}=8 \pi G T_{\mu \nu}+\Lambda g_{\mu \nu},
\end{equation}
including the matter fields and the cosmological constant $\Lambda$, where $R_{\mu \nu}$ is the Ricci tensor, $R = g^{\mu \nu} R_{\mu \nu}$ is the Ricci scalar, $g_{\mu \nu}$ is the spacetime metric, and $T_{\mu \nu}$ is the energy-momentum tensor. This equation relates the geometry of spacetime to the matter and energy content of the Universe. To solve this equation intrinsic symmetries are conjectured. The Universe is assumed to be highly homogeneous on large scales~\cite{deBernardis:2000sbo,Hinshaw:2012aka,Aghanim:2018eyx} as seen from the CMB, see Figure~\ref{Fig:Planck_CMB}. 

Evolution of a homogeneous and isotropic Universe is described by the FLRW metric:
\begin{equation}
g_{\mu\nu} dx^\mu dx^\nu = d s^{2}=d t^{2}-a(t)^{2}\left(\frac{d r^{2}}{1-k r^{2}}+r^{2} d \theta^{2}+r^{2} \sin ^{2} \theta d \phi^{2}\right),
\end{equation}
where $a(t)$ is the scale factor and $k$ is the curvature with $k=\{-1,\,0,\,1\}$ for an open, a flat or a closed Universe respectively. The $\{r,\,\theta,\,\phi\}$ variables are the co-moving spatial coordinates. The FLRW metric is an exact solution to Einstein’s field equations describing a homogeneous and isotropic Universe.

The energy-momentum tensor depends on the energy density $\rho_t = T_{00}$ and the pressure $p_j = T_{jj}$.  By using the diagonal form of the stress-energy tensor, $T_\nu^\mu = \mathrm{diag}(\rho, \overline{-p})$, for a perfect fluid in its rest frame, we can extract equations describing dynamics of the Universe in a different form,
\begin{subequations}
\begin{align}
\ddot a &= - \frac{4 \pi G}{3} a \sum_i \left( \rho_i + 3 p_i \right),\\
H^2 &\equiv \left( \frac{\dot a }{a} \right)^2 = \frac{8 \pi G}{3} \sum_i \rho_i - \frac{k}{a^2},
\end{align}
\end{subequations}
which are known as the Friedmann equations, where $H$ is the Hubble parameter. These equations determine the dynamics of the Universe. It is convenient to re-write the Friedmann equations using the density parameter $\Omega$. The density parameters are averaged. The (mean) matter density is a fraction to the critical density
\begin{equation}\label{Eq:DensityFraction}
\Omega_\mathrm{M} = \frac{\rho_\mathrm{M}}{\rho_\mathrm{Crit}} = \frac{\rho_\mathrm{M}}{\left(  \frac{3H_0^2}{8\pi G}\right)},
\end{equation}
where $G \approx 6.67\times 10^{-11}$ N m$^2$/kg$^2$ is the gravitational constant and $H_0 \approx 70$ km/s/Mpc is the present-day Hubble parameter, which describes  expansion of the Universe. In terms of the averaged parameters, the first Friedmann equation is given by
\begin{equation}
1 = \sum_i \Omega_i - \frac{k}{a^2 H^2},
\end{equation}
where $\sum_i \Omega_i = \Omega_\mathrm{R} + \Omega_\mathrm{M} + \Omega_\Lambda$. The $\Omega_\mathrm{R}$ value represents the present day radiation density, which can be neglected, $\Omega_\mathrm{R} h^2 = \mathcal{O}(10^{-5})$~\cite{ParticleDataGroup:2024cfk}, where $h$ is the scaled Hubble constant, $h \approx 0.71$. The other parameters are discussed in more detail below.  With this in mind, extrapolation back in time results in a singularity in the past. Therefore, it is assumed that the Universe started as a hot and dense phase, the Big Bang. As time goes on, the Universe expands and cools down in the process. 

The $\Omega_\mathrm{M}$ parameters includes the mean fractional density of forms of matter, both baryonic (ordinary) and CDM. The matter density has been evaluated in a variety of ways: mass-to-light ratio~\cite{Bahcall:1995cy}, bispectrum  of redshift distortion~\cite{Verde:2001sf}, linear perturbation growth rate~\cite{Nesseris:2007pa}, baryon acoustic oscillations~\cite{Addison:2013haa}, galaxy–galaxy lensing~\cite{More:2014uva}. The current value is estimated to be $\Omega_\mathrm{M} \approx 0.315  \pm 0.007$~\cite{ParticleDataGroup:2024cfk}. Naively one could have thought that the value should be close to unity. Surprisingly, cosmological observations~\cite{Riess:1998cb,Perlmutter:1998np} seem to indicate that a large part of the Universe's energy, around seventy per cent, $\Omega_\Lambda \approx 0.685 \pm 0.007$~\cite{ParticleDataGroup:2024cfk}, is made of a component with a negative pressure which is called dark energy. Due to the fact that the dark energy component is so large, it causes acceleration of the expansion of the Universe.

Many observations~\cite{Zwicky:1933gu,Trimble:1987ee,Sofue:2000jx,Clowe:2006eq,Allen:2011zs,Salucci:2018hqu,Simon:2019nxf}, along with CMB, from different sources and at different scales point out that there is a non-baryonic matter component. It is possible to approximate the density of the CDM (in the $\Lambda_\mathrm{CDM}$ model), which is $\Omega_\mathrm{CDM}h^2 \approx 0.1200 \pm 0.0012$~\cite{ParticleDataGroup:2024cfk} with $h=0.674 \pm 0.005$. Modern cosmological models assume that the CDM candidate is nearly electromagnetically neutral~\cite{McDermott:2010pa,Dolgov:2013una}, is of a non-baryonic origin, preferably weakly interacting with ordinary matter, its self-interactions cannot be too strong~\cite{Randall:2007ph,Harvey:2015hha,Tulin:2017ara}, and sufficiently stable or long-lived. 

There are two online tools~\cite{WebCMBsim,WebCMBsim2} which demonstrate how the structure of the Universe changes based on $\Omega_i$ densities.

\section{Boltzmann equations}\label{Sec.Boltzmann_eq}

The most promising and theoretically well-studied DM candidates are WIMPs. Due to the weak interaction strength and stability, usually associated with some discrete symmetry, which forbids specific decays, WIMPs would yield a right abundance in the early Universe. WIMPs can be classified based on velocities when they decoupled from the thermal plasma. These are: cold (non-relativistic), warm, hot (relativistic) DM. The hot DM was relativistic when the DM particles fell out of thermal equilibrium. The typical mass of such candidates is of the eV order. Due to the large mean free path these particles did not form clumps. However, hot WIMPs are not compatible with some observed constraints of the average velocities~\cite{Tremaine:1979we,White:1984yj,Abazajian:2005xn,Viel:2013fqw,Schneider:2013wwa}. In this scenario structures form by fragmentation; first, large structures are formed and then disentangle into the smaller ones. In contrast, in the cold CDM theory clusters are formed smoothly. Under the gravitational force small clusters are formed. Then, those hierarchically grow into larger ones. This seems to be in good agreement with cosmological large-scale structures. Moreover, the CDM is preferred by the cosmic microwave background radiation. The typical mass range of the CDM candidate is in the GeV-TeV range. Finally, there is a possibility of  warm DM in the range of a few keV. During decoupling from the plasma it was relativistic. It was produced non-thermally. These particles would freeze-out before reaching thermal equilibrium. The warm DM is capable of solving several issues~\cite{Lovell:2011rd} which the CDM is not able to~\cite{Ghigna:1998vn,Moore:1999nt,Klypin:1999uc}, \textit{e.g.}, the so-called ``satellite problem", which points out that there is a large discrepancy between the simulated and observed subhalos. It should be noted that there are no decent known candidates within the SM.

WIMPs can be produced thermally and non-thermally. The thermal relics production depends on the rate of an interaction process and the Universe expansion, both of which are time-dependent processes. In the early stage of the Universe there was a thermal, to be more precise, chemical, equilibrium between annihilation and pair production of the SM and DM species. Then, with the Universe expanding it cooled down. The interaction rate of the SM-DM became inefficient and smaller than the Hubble parameter. At this point WIMPs decouple or ``freeze-out". Therefore, WIMPs are said to be thermally produced as those were in thermal equilibrium and eventually frozen-out. The relic abundance of the decoupled particle species from the thermal bath will eventually stay constant afterwards.  Another possibility is that WIMPs were produced non-thermally~\cite{Baer:2014eja}.

DM may be produced in a predictive manner as a thermal relic. After all, the early Universe is a simple place. Let us assume a hypothetical massive particle, which we shall denote as $\chi$. At high temperatures $\chi$ is in thermal equilibrium, as all of the particles. We shall assume that $\chi$ is thermally created and the observed relic density is governed by the freeze-out mechanism. Another assumption is that the hypothetical $\chi$ is massive enough, $m_\chi \sim $ GeV, which will guarantee that it will be  non-relativistic after decoupling from the thermal bath.

Assume a four-point interaction,
\begin{equation*}
\ctikz{ \node[text width=0.5cm] at (-1,0) {$\chi$};\draw [-] (-1,0.6) -- (-0.32,0.2); \draw [-] (-1,-0.6) -- (-0.32,-0.2); \fill[pattern=north east lines](0,0)circle(0.4);\draw(0,0)circle(0.4); \draw [-] (1,0.6) -- (0.32,0.2); \draw [-] (1,-0.6) -- (0.32,-0.2);\node[text width=0.5cm] at (0.85,0) {$f$};},
\end{equation*}
which can be read in several ways:
\begin{itemize}
\item $\chi$ annihilation, $\chi \bar\chi \to f \bar f$;
\item $\chi$ scattering of $f$, $\chi f \to \chi f$;
\item $\chi$ pair-production, $f \bar f \to \chi \bar\chi$;
\end{itemize}
The equilibrium abundance is maintained since the process is not one-way directed.

As the Universe expands, the plasma temperature drops down,
\begin{equation}
T = T_0 (1+z),
\end{equation}
where $T_0 \approx 2.7$ K is the present day CMB temperature, and $z$ is the redshift parameter defined as
\begin{equation}
z= \frac{\lambda_\mathrm{observed} - \lambda_\mathrm{emitted}}{ \lambda_\mathrm{emitted}},
\end{equation}
where $\lambda$ are wavelengths. Then, at some point the Universe  is cool enough, $T < m_\chi$, causing the equilibrium to decrease exponentially until it falls below the expansion rate, $\Gamma < H$. At this point collisions become highly unlikely. The current relic abundance can be evaluated by solving the Boltzmann equations, which describe how the number density $n(t)$ changes with time. Quantitatively, we need to solve
\begin{equation}
\frac{d n_{\chi}}{d t}+3 H n_{\chi} = -\left\langle\sigma_{\mathrm{ann}} v\right\rangle\left(n_{\chi}^{2}-n_{\chi-\mathrm{eq}}^{2}\right),
\end{equation}
where $\left\langle\sigma_{\mathrm{ann}} v\right\rangle$ is the thermally averaged annihilation cross section of $\chi\chi$ decaying into lighter particles times the relative velocity (in the non-relativistic case, $v \ll 1$, we need to distinguish velocities of the individual states), and $n_{\chi-\mathrm{eq}}$ is the number density in thermal equilibrium. The Boltzmann equation accounts for the expansion of the Universe (dilution from expansion), given by the left-hand side, and represents how the system acts with the number-changing interactions, given by the right-hand side. On the right-hand side, the first term, $ -\left\langle\sigma_{\mathrm{ann} }v \right\rangle n_{\chi}^{2}$, represents the annihilation of DM particles into SM particles, $\chi \chi \to f \bar f$. This process reduces the DM number density over time. The second term, $\left\langle\sigma_{\mathrm{ann}} v\right\rangle n_{\chi-\mathrm{eq}}^{2}$, represents the creation of DM particles from SM interactions. This is only possible when the Universe is hot enough for thermal equilibrium. In the case of $n_{\chi}=n_{\chi-\mathrm{eq}}$ annihilation and production balance each other. As a result, the change in density would be due to the expansion of the Universe. As the Universe expands and cools down, the production rate decreases, and the  annihilation rate starts to dominates, leading to freeze-out when interactions become too rare to maintain equilibrium.

The Boltzmann equations can be solved analytically, assuming some approximations. It is convenient to change variables from the time scale to the temperature scale via $t \to x = m/T$ and by replacing the number density by the co-moving number density, $n \to Y = n/s$, where $s$ is the entropy density. Then, the Boltzmann equation in terms of the new variables takes the form 
\begin{equation}
\frac{x}{Y_{\mathrm{eq}}} \frac{d Y}{d x}=-\frac{n_{\mathrm{eq}}\left\langle\sigma_\mathrm{ann} v\right\rangle}{H}\left(\frac{Y^{2}}{Y_{\mathrm{eq}}^{2}}-1\right).
\end{equation}
After the freeze-out, $Y$ approaches a certain number and eventually becomes a constant, determined by $\left\langle\sigma_\mathrm{ann} v\right\rangle$. Based on the cross section value, the higher its value is, the lower the relic density value becomes.

Let us consider a particle with mass $m^{-2}_\chi \sim \left\langle\sigma_\mathrm{ann} v\right\rangle$. The freeze out happens at the point when $n_\mathrm{eq} \left\langle\sigma_\mathrm{ann} v\right\rangle \sim H$. For a massive non-relativistic particle this translates into roughly $x \sim 30$. As the expansion rate is slow, the freeze out will occur later than at the temperature $T\sim m_\chi$, \textit{e.g.}, for $m_\chi \sim 300$ GeV the freeze-out temperature is around $T \sim 10$ GeV. The relic density can also be found governed by
\begin{equation}
\frac{\rho_\chi(T)}{\rho(T)} \sim  \frac{x}{M_\mathrm{Planck} m_\chi \left\langle\sigma_\mathrm{ann} v\right\rangle}.
\end{equation}
An intriguing fact is that a typical weak scale cross section and masses of around $m_\chi \sim \mathcal{O}$(10) GeV yield the predicted value of $\Omega h^2 \sim 0.1$. This fact is referred to as the WIMP miracle. 

In principle, the picture is far more complicated. There exist numerous computer codes like $\mathsf{DarkBit}$~\cite{Workgroup:2017lvb}, $\mathsf{DarkSUSY}$~\cite{Gondolo:2004sc}, $\mathsf{micrOMEGAs}$~\cite{Belanger:2004yn}, capable of solving the Boltzmann equations with great precision. The full picture involves taking into consideration a large number of processes. In general, there will be an additional contribution due to co-annihilations. In a typical scenario the dark sector would consist of more than a single (DM candidate) particle. Then, those particles could annihilate into a pair of SM particles, $\chi_1 \chi_2 \to f \bar f$. Apart from that, there could be a resonant annihilation, which happens via a particle $\eta$ when $m_\chi \approx m_\eta /2$, which would yield a lower relic abundance. 

\section{The Big Bang nucleosynthesis}\label{Sec.BBN}

The Big Bang Nucleosynthesis (BBN)~\cite{Kawasaki:2004yh,Kawasaki:2004qu,Jedamzik:2006xz} plays a crucial role in the DM production. Usefulness of BBN relies on the fact that the Universe is in thermodynamic equilibrium and that only nuclear reactions are out of equilibrium. The BBN is a period during the early Universe which took place between roughly 1 to 20 minutes after the Big Bang. During this period the temperature of the Universe was at $T \sim $ MeV. Abundance of the light elements was produced during BBN: D ($^2$H), $^3$He and $^4$He, $^7$Li. The heavier elements were created by the stellar nucleosynthesis~\cite{Hoyle:1954zz,Burbidge:1957vc}.

Let us discuss some main aspects and points of the BBN. The BBN is highly sensitive to the early radiation dominated era. The initial BBN conditions are considered to be $T\sim 1$ MeV, \textit{i.e.}, in the time scale $t\sim$1 s. At temperatures above $T \gg 1$ MeV there was a chemical equilibrium in the thermal bath between protons and neutrons, maintained by weak interactions,
\begin{align*}
n + \nu_e \leftrightarrows p + e^-, \\
n+ e^+ \leftrightarrows p + \bar{\nu}_e,\\
n \leftrightarrows p + e^- + \bar \nu_e.
\end{align*}
During this period, the Universe was dominated by relativistic $e^\pm$, $\nu_i$ and $\bar\nu_i$, and $\gamma$, which were in thermal equilibrium. The weak interaction rates were faster than the expansion of the Universe. As the temperature dropped down, the weak interaction rates fell out of equilibrium, neutrinos decoupled from the thermal bath and at some point later $e^\pm$ pairs annihilated while heating the photon gas. This caused the neutron-proton ratio to freeze, with the ratio
\begin{equation}
\frac{n_{n}}{n_{p}}=e^{-\frac{m_{n}-m_{p}}{T}} = e^{- \frac{1.293 \text{ MeV}}{T}} \approx \frac{1}{6}.
\end{equation}
This ratio is sensitive to known physical interactions as the neutron-proton mass difference is determined by strong and electromagnetic interactions, and the freeze-out temperature depends on both gravitational and weak interactions. Furthermore, neutrons decayed though
\begin{equation*}
n \to p e^- \bar{\nu}_e.
\end{equation*}
The $\beta$-decay reduced the neutron fraction to $n/p \sim 1/7$ before neutrons became bound in deuterium. However, due to a high number of photons deuterium photodissociated
\begin{equation*}
\mathrm{D} + \gamma \to n + p.
\end{equation*}
It is crucial to realise that any spike in the ambient background of the mass-energy density will result in a higher freeze-out temperature.

At temperatures around $T \sim 0.1$ MeV, nuclei could begin to form without being ripped apart. Regardless of the reaction chain, nearly all neutrons fused to $^4$He. This happened as $^4$He has the highest binding energy per nucleon. Due to the absence of stable nuclei with mass number 5 or 8 and the large Coulomb barrier, heavier nuclei did not form. Therefore, we can estimate the relative abundance of $^4$He by weight, with a simple argument that
\begin{equation}
Y_p = \frac{2 \frac{n_n}{n_p}}{1+\frac{n_n}{n_p}} \approx 0.25.
\end{equation}

The rates of the aforementioned reactions strongly depend on the density of baryons, which is given by the baryon-to-photon ratio
\begin{equation}
\eta = \frac{\eta_b}{\eta_\gamma},
\end{equation}
which is provided in the form of $\eta_{10}= \eta \times 10^{10}$ by PDG and is $\eta_{10} = 5.8-6.5$~\cite{ParticleDataGroup:2024cfk}. This value can also be expressed as $\Omega_\mathrm{B}h^2 \approx \eta_{10}/274 = (0.021-0.024)$~\cite{ParticleDataGroup:2024cfk}.

At the beginning, the rates of the $^4$He production were not high enough to reach the nuclear statistical equilibrium, as the abundance is low and the Coulomb suppression is significant. At this stage only a tiny amount of $^7$Li could have been synthesised. Obviously, for larger values of $\eta_{10}$ the nuclei formation is more efficient. This explains why the presence of decaying particles during the BBN would significantly disrupt the nuclear formation chains. This will depend on the context of the decaying particles if those are long-lived or not, charged or neutral. All in all, the BBN constrains the relic density of late particles. 

\begin{figure}[htb]
\begin{center}
\includegraphics[scale=0.35]{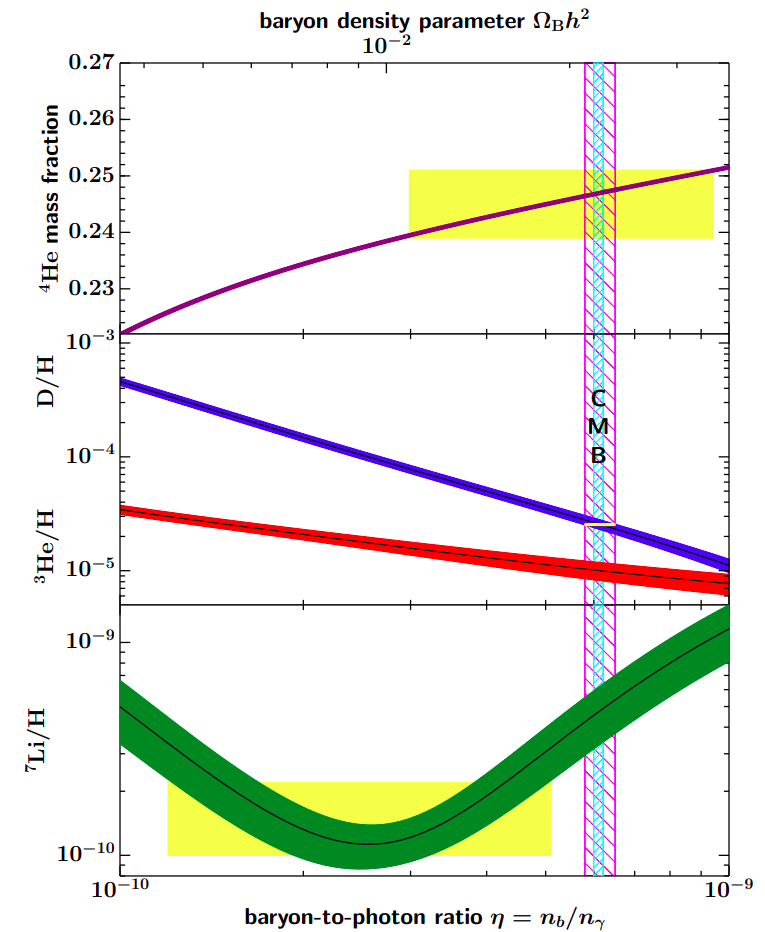}
\end{center}
\vspace*{-4mm}
\caption{ The primordial abundances of $^4$He, D, $^3$He, and $^7$Li as predicted by the standard model of the BBN---the bands show the 95\% CL range~\cite{Fields:2019pfx}. Boxes indicate the observed light element abundances. The narrow blue vertical band indicates the CMB measure of the cosmic baryon density, while the wider red band indicates the BBN D+$^4$He concordance range (both at 95\% CL). Figure taken from Ref.~\cite{ParticleDataGroup:2024cfk}. }
\label{Fig:BaryonToPhoton}
\end{figure}

One of the amazing predictions of the standard BBN is that once the nuclear reactions are known it is easy to compute the light element abundances, which is given in terms of a single parameter $\eta_{10}$, see Figure~\ref{Fig:BaryonToPhoton} for the Schramm plot, showing the primordial element abundances as a function of the baryon-to-photon ratio. A crucial test is whether the value of the baryon-to-photon ratio fits the CMB observation. An amazing agreement between the BBN and the CMB is the D/H abundance, which is the blue line in Figure~\ref{Fig:BaryonToPhoton}. On the other hand, there is a slight issue when it comes to the ratio of $^7$Li/H, which is known as the lithium problem~\cite{Coc:2003ce,Angulo:2005mi,Cyburt:2008kw,Fields:2011zzb}.

\chapter{Some aspects of Group Theory}\label{Ch:Group_Theory}

In this chapter we shall outline some basic concepts of Group Theory. We start by defining which set of elements can be characterised from the mathematical point of view as groups. As we are predominantly interested in constructing invariants under some group $\mathcal{G}$, we shall primarily focus on concepts necessary to understand the basic approach to the construction of such invariants. In this chapter we shall follow several books~\cite{joshi1997elements,jones2020groups,James_Liebeck_2001} on Group Theory.

\section{Generalities}

A non-empty set (object) $\mathcal{G}$ combined with an operation is called a group if it satisfies four key properties (axioms):
\begin{itemize}
\item Identity \\
Every group includes an identity element $e$, such that the identity element of a group is unique. Suppose on the contrary: two elements $e$ and $e^\prime$ are both identities. However, then $e e^\prime = e$ and $e^\prime e = e$, and hence $e = e^\prime$.

\item Inverse \\
For $\forall\,a\in \mathcal{G}$ there is a unique inverse element, such that $a\,a^{-1}=a^{-1}a=e$. Assume that there are two inverse elements $\bar a$ and $\tilde{a}$ of an element $a$. Then, $\bar a = e \bar a = (\tilde a a) \bar a = \tilde a (a \bar a) = \tilde a e = \tilde a$. As a result, $\bar a = \tilde a$.

\item Associativity\\
The multiplication is associative, \textit{i.e}, $(ab)c=a(bc),~\forall\,\{a,\,b,\,c\}\in \mathcal{G}$.

\item Closure \\
Any product of two group elements should result in a group element, \textit{i.e.}, if $a \in \mathcal{G}$ and $b \in \mathcal{G}$, then $\{ab,\,ba\} \in \mathcal{G}$.

\end{itemize}

As an example let us consider the set of integers $\mathbb{Z}$ under addition $(\mathbb{Z},\,+)$ (the group ``multiplication" in this case is addition of numbers):
\begin{itemize}

\item Identity \\
The identity elements is 0; $a + 0 = a,~ \forall a \in \mathbb{Z}$.

\item Inverse \\
The inverse of any integer $a$ is $-a$ since $a + (-a) =0$.

\item Associativity\\
Addition of integers is associative, $(a + b) + c = a + (b + c), ~ \forall \{a,\,b,\,c\} \in \mathbb{Z}$.

\item Closure \\
The sum of any two integers is an integer, $a + b =c,~ c\in \mathbb{Z}$.

\end{itemize}
As can be seen, $(\mathbb{Z},\,+)$ can be identified as a group, since it satisfies all of the group properties.

A group is called an Abelian group if all elements of a group commute, \textit{i.e.}, $a b = b a$ for any two elements of $\mathcal{G}$.

\section[\texorpdfstring{$S_3$}{S3} as a permutation group]{\boldmath$S_3$ as a permutation group} \label{Sec:S3_perm}

\begin{figure}[htb]
\begin{center}
\includegraphics[scale=1.0]{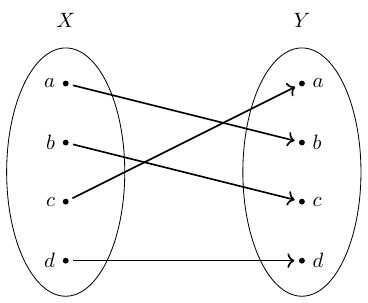}
\end{center}
\vspace*{-4mm}
\caption{A bijective function, $f:\,X \to Y$, as a permutation of a particular set $X$ into another set $Y$. For example, $f(a)=b$.}
\label{Fig:bijections_as_perm}
\end{figure}

Let us discuss a specific type of groups---symmetric groups, denoted as $S_n$, where $n$ refers to the number of elements (this will later be referred to as ``order") in a set that is being permuted (rearrangement of elements of a particular set). The operation of the $S_n$ group is the composition of permutations: permutations are applied in a sequence one after another. Each element of the $S_n$ group can be written as a product of cyclic permutations with disjoint (a particular element of a group permutes the object only once) cycles. All elements (permutations) of $S_n$ are bijection (one-to-one correspondence between two sets; an injective function maps each element of the co-domain, \textit{i.e.}, set of destination of a function,  from at most one element of the domain, while the surjective function maps each element of the co-domain to at least one element of the domain) from a set to itself. An example of a particular permutation presented as bijection from $X$ to $Y$ is demonstrated in Figure~\ref{Fig:bijections_as_perm}.

%
%
%

The set of elements in $S_n$ is the set of all possible permutations of $n$ distinct objects. All possible permutations of the $S_3$ group acting on a vector $v = (x,\, y,\, z)$ are:
\begin{equation}\label{Eq:S3_as_permutations}
\begin{aligned}
e =& \begin{pmatrix}
 1 & 0 & 0 \\
 0 & 1 & 0 \\
 0 & 0 & 1 \\
\end{pmatrix} = \mathcal{I}_3,\\
a =& \begin{pmatrix}
 0 & 1 & 0 \\
 1 & 0 & 0 \\
 0 & 0 & 1 \\
\end{pmatrix}, \quad b = \begin{pmatrix}
 1 & 0 & 0 \\
 0 & 0 & 1 \\
 0 & 1 & 0 \\
\end{pmatrix}, \quad c = \begin{pmatrix}
 0 & 0 & 1 \\
 0 & 1 & 0 \\
 1 & 0 & 0 \\
\end{pmatrix},\\
d =& \begin{pmatrix}
 0 & 1 & 0 \\
 0 & 0 & 1 \\
 1 & 0 & 0 \\
\end{pmatrix}, \quad f = \begin{pmatrix}
 0 & 0 & 1 \\
 1 & 0 & 0 \\
 0 & 1 & 0 \\
\end{pmatrix}.
\end{aligned}
\end{equation}
For example acting with $a$ on $v$ we have:
\begin{equation}
\begin{pmatrix}
 0 & 1 & 0 \\
 1 & 0 & 0 \\
 0 & 0 & 1 \\
\end{pmatrix} \begin{pmatrix}
x \\
y \\
z
\end{pmatrix} = \begin{pmatrix}
y \\
x \\
z
\end{pmatrix},
\end{equation}
where the first two elements were swapped, $x \leftrightarrow y$. The Cauchy two-line notation represents a permutation by listing the elements of a transformation in the first row and their corresponding images in the second row directly below them, $e.g.$,
\begin{equation}
a= \begin{pmatrix}
1 & 2 & 3 \\
2 & 1 & 3
\end{pmatrix}.
\end{equation}
This can be further simplified to a cycle notation, represented by listing its elements in cycles. Each closed cycle begins with an element, followed by its image, and then the image of a previous image, and so on, until returning to the initial starting element. For example, for $a$ we have two cycles $1\to 2 \to 1$ and $3 \to 3$. In the cycle notation this corresponds to $(12)(3)$. Elements that are mapped to themselves, $(3)$, are often omitted.

The permutation matrices of eq.~\eqref{Eq:S3_as_permutations} can be presented in the cyclic notation as:
\begin{equation}\label{Eq:S3_cycles}
\begin{aligned}
& e = \begin{pmatrix}
1 & 2 & 3 \\
1 & 2 & 3
\end{pmatrix} = (1)(2)(3),\\
& a= \begin{pmatrix}
1 & 2 & 3 \\
2 & 1 & 3
\end{pmatrix}=(12)(3),~
b= \begin{pmatrix}
1 & 2 & 3 \\
1 & 3 & 2
\end{pmatrix}=(1)(23),~
c= \begin{pmatrix}
1 & 2 & 3 \\
3 & 2 & 1
\end{pmatrix}=(13)(2),\\
& d = \begin{pmatrix}
1 & 2 & 3 \\
2 & 3 & 1
\end{pmatrix}=(123),~
f = \begin{pmatrix}
1 & 3 & 2 \\
3 & 2 & 1
\end{pmatrix}=(132).
\end{aligned}
\end{equation}

It can be easily verified that there are no other possible permutations acting on $v$. Therefore, we can define the $S_3$ group as a set of the elements:
\begin{equation}\label{Eq:S3_all_elem}
S_3 = \{e,\,a,\,b,\,c,\,d,\,f\},
\end{equation}
which consists of six elements. In Group Theory, the order of a finite group (a group may also contain an infinite number of elements) is the number of the elements from which it is formed. Therefore, the order of the $S_3$ group is six. In general, the number of all possible permutations (the total number of ways to arrange $n$ elements) can be expressed in terms of a factorial. The number of elements, or equivalently the order, of the $S_n$ group is $n!$.

The $S_3$ group is the simplest non-Abelian group. This can be verified, for instance, by applying two permutations in an opposite order,
\begin{equation}\label{Eq:ab_neq_ba}
\begin{aligned}
a\,b = \begin{pmatrix}
0 & 0 & 1 \\
1 & 0 & 0 \\
0 & 1 & 0
\end{pmatrix} &= f, \quad
b\,a = \begin{pmatrix}
0 & 1 & 0 \\
0 & 0 & 1 \\
1 & 0 & 0
\end{pmatrix} = d,\\
& \quad a\,b \neq b\,a.
\end{aligned}
\end{equation}
In a more general sense, one could construct a table that would show the result of applying the group operation to each pair of elements of a particular group; arranging all the possible products. Such construction is called a Cayley table. The construction of such tables may prove to be extremely useful for small groups, when tackling specific tasks. The Cayley table can help visualise the behaviour of the group operation and several properties of the group, \textit{e.g.}, commutativity or closure. The Cayley table for the $S_3$ group is provided in Table~\ref{Table:Caylely_Tab_S3}.

{{\renewcommand{\arraystretch}{1.3}
\begin{table}[htb]
\caption{The Cayley table for the $S_3$ group under the multiplication operation $(\cdot)$. It can be seen that the entries of the table are not symmetric, and hence we can deduce that $S_3$ is a non-Abelian group.}
\label{Table:Caylely_Tab_S3}
\begin{center}
\begin{tabular}{c||cccccc} \hline\hline
$S_3$ & $e$ & $a$ & $b$ & $c$ & $d$ & $f$ \\ \hline \hline
$e$ & $e$ & $a$ & $b$ & $c$ & $d$ & $f$ \\
$a$ & $a$ & $e$ & $f$ & $d$ & $c$ & $b$ \\
$b$ & $b$ & $d$ & $e$ & $f$ & $a$ & $c$ \\
$c$ & $c$ & $f$ & $d$ & $e$ & $b$ & $a$ \\
$d$ & $d$ & $b$ & $c$ & $a$ & $f$ & $e$ \\
$f$ & $f$ & $c$ & $a$ & $b$ & $e$ & $d$ \\ \hline \hline
\end{tabular}\vspace*{-9pt}
\end{center}
\end{table}}

While the presentation of the $S_3$ group provided in eq.~\eqref{Eq:S3_all_elem} might seem reasonable and understandable, for groups of a higher order it would become quite challenging to try to write down the entire set of elements. A group $\mathcal{G}$ can be presented in terms of a generating set. A generating set of a group $\mathcal{G}$ is a subset of the group set, like the one of eq.~\eqref{Eq:S3_all_elem}, such that all elements of a particular $\mathcal{G}$ can be expressed by combining elements of the subset using the group operation and its inverses. In short, one can look for a subset of elements that can ``generate" the entire group through repeated application of the group operation.

Now, consider the discussed $S_3$ group. We have already seen that by considering a particular, $\{a,\,b\}$, subset of $S_3$ it is possible to generate two other elements, namely $\{f,\, d\}$, see eq.~\eqref{Eq:ab_neq_ba}. The final element can be generated via $a\,b\,a = b\,a\,b =c$. One could present $S_3$ as a group generated by the elements $a$ and $b$, which would be the minimal generating set. However, notice that in eq.~\eqref{Eq:S3_cycles} there are several different patterns of cycles, to be more precise compare the structure of the element in lines two and three, those are $(xy)(z)$ and $(xyz)$. Maybe it is more natural to present the $S_3$ group in terms of two elements, one from each of those lines? Since this section serves as an introduction to another topic, we shall omit a more advanced discussion of the generating sets. But, consider that we pick $\{a,\,d\}$. Then, by multiplying the elements we get
\begin{equation}
d^2 = f, ~ d\,a = b, ~ d^2\, a = c.
\end{equation}
Therefore, we can re-write eq.~\eqref{Eq:S3_all_elem} as 
\begin{equation}\label{Eq:S3_a_d_presentation}
S_3 = \{e,\,a,\,da,\,d^2a,\,d,\,d^2\},
\end{equation}
while preserving the order of the occurrence of the elements.

A presentation of a group is one method of describing a group using a set of generators (a minimial set of elements) and writing every element of a particular set as a product of relations of these generators. In a shorthand notation a $\mathcal{G}$ group can be described as 
\begin{equation}
\mathcal{G} = \left\langle S | R \right\rangle,
\end{equation}
where $S=\{g_1,\, g_2,\, \dots,\, g_n\}$ is a set of $n$ generators (like $\{a,\,d\}$ for the case of $S_3$) and $R$ is a set of relations among these generators. For the $S_3$ group we can write the presentation
\begin{equation}\label{Eq:S3_presentation}
S_3 = \left\langle a,\,d~|~d^3=a^2=e,\, ada = d^2\right\rangle,
\end{equation}
which follows from eq.~\eqref{Eq:S3_a_d_presentation}. Usually there are several ways of presenting the same group; the group presentation is not unique. The Tietze transformations and the Gröbner basis algorithms are standard methods for simplifying and modifying presentations of groups.

Let us mention some other presentations of groups. For example, for a cyclic group $\mathbb{Z}_n$ we have,
\begin{equation}
\mathbb{Z}_n = \left\langle g~|~g^n=e\right\rangle,
\end{equation}
or for a direct product of two cyclic groups its presentation can be expressed as
\begin{equation}
\mathbb{Z}_m \times \mathbb{Z}_n  = \left\langle x,\, y~|~x^m = y^n = e,\, xy=yx \right\rangle,
\end{equation}
or for a semidirect product (two groups do not commute and, as a result, the second group ``acts" on the first group) of two groups we have
\begin{equation}
\mathbb{Z}_m \rtimes \mathbb{Z}_n  = \left\langle x,\, y~|~x^m = y^n = e,\, x^{-1} y x = y^k \right\rangle,~\mathrm{gcd}(n,k)=1,
\end{equation}
where gcd(n,k) is the greatest common divisor function.

As a side note, and a fun fact, there exist infinite finitely presented groups~\cite{sapir2007grouptheoryproblems} (groups that are defined by a finite set of generators and a finite set of relations, but are infinite in size).

Let us now consider how two of the $S_3$ elements act on the $v$ vector,
\begin{subequations}
\begin{align}
a\,v =& \begin{pmatrix}
 0 & 1 & 0 \\
 1 & 0 & 0 \\
 0 & 0 & 1 \\
\end{pmatrix} \begin{pmatrix}
x \\
y \\
z
\end{pmatrix} = \begin{pmatrix}
y \\
x \\
z
\end{pmatrix},\\
d\,v =& \begin{pmatrix}
 0 & 1 & 0 \\
 0 & 0 & 1 \\
 1 & 0 & 0 \\
\end{pmatrix} \begin{pmatrix}
x \\
y \\
z
\end{pmatrix} = \begin{pmatrix}
y \\
z \\
x
\end{pmatrix}.
\end{align}
\end{subequations}
It is trivial to notice that the action of neither $a$ nor $d$ leaves the vector $v$ invariant. However, when acting with generators of a particular group on a vector space, we want to find some orthogonal combinations of the vector space (a non-orthogonal vector would indicate that it could be decomposed into the other orthogonal vectors) that would be intact when acted on by a specific group. In the example provided above we can guess that the invariant combinations under any element of the $S_3$ group can be expressed in terms of polynomials of degree $n$ following a simple rule:
\begin{equation}
P^{(n)}(i,\,j,\,k) \equiv \sum_{i,j,k} C_{ijk} x^i y^j z^k, \quad i+j+k=n,~ i \geq j \geq k \geq 0,
\end{equation}
where we assume that the order of the elements $\{x,\, y,\, z\}$ does not matter and we consider positive definite powers of the polynomials, and $C_{ijk}$ are some coefficients, taking values of $\{0,\,1\}$ in this example. The invariant combinations under the $S_3$ group are:
\begin{subequations}\label{Eq:S3_Polynom_powers}
\begin{align}
n=1~:~& x + y +z,\\
n=2~:~& \{x^2 + y^2 + z^2,\, xy + yz + zx\},\\
n=3~:~& \{xyz,\, x^2 y + y^2 z + z^2 x + x^2 z + z^2 y + y^2 x\}, \\
\begin{split}n=4~:~& \{x y z (x + y + z),\, x^4 + y^4 + z^4, \\
& ~ x^2 y^2 + x^2 z^2 + y^2 z^2,\,x^3 y+x^3 z+x y^3+x z^3+y^3 z+y z^3\}, 
\end{split}
\\ \nonumber
\dots
\end{align}
\end{subequations}
These polynomials were constructed by picking a specific $P^{(n)}(i,\,j,\,k)$ and acting with all elements of $S_3$ (not the minimal set) one at a time. Therefore, the above combinations are simultaneously invariant under all permutations, see eq.~\eqref{Eq:S3_as_permutations}, of the $S_3$ group.

Consider another example. Let us consider a polynomial of degree two,
\begin{equation}
P(x,y,z) = v A v^\mathrm{T} +  B v ^\mathrm{T} + C,
\end{equation}
where
\begin{equation}
A = \frac{1}{2} \begin{pmatrix}
2A_{11} & A_{12} & A_{13} \\
A_{12} & 2A_{22} & A_{23} \\
A_{13} & A_{23} & 2A_{33}
\end{pmatrix}, \quad B= \left( B_1~B_2~B_3 \right),
\end{equation}
and $C$ is a constant coefficient, since it is not multiplied by $v$. We need to relate elements of $A$ and $B$ in a way such that $P(x,y,z)$ would be invariant under all possible transformations of the underlying symmetry.

The $v$ vector transforms under a three-dimensional matrix representation as
\begin{equation}
v_i^\prime = D(g)_{ij} v_j.
\end{equation}
By taking the generator $a$ we see that the coefficients transform as
\begin{equation}
\begin{pmatrix}
A_{11} \\
A_{12} \\
A_{13} \\
A_{22} \\
A_{23} \\
A_{33} \\
B_1 \\
B_2 \\
B_3 \\
C
\end{pmatrix}^\prime = \begin{pmatrix}
 0 & 0 & 0 & 1 & 0 & 0 & 0 & 0 & 0 & 0 \\
 0 & 1 & 0 & 0 & 0 & 0 & 0 & 0 & 0 & 0 \\
 0 & 0 & 0 & 0 & 1 & 0 & 0 & 0 & 0 & 0 \\
 1 & 0 & 0 & 0 & 0 & 0 & 0 & 0 & 0 & 0 \\
 0 & 0 & 1 & 0 & 0 & 0 & 0 & 0 & 0 & 0 \\
 0 & 0 & 0 & 0 & 0 & 1 & 0 & 0 & 0 & 0 \\
 0 & 0 & 0 & 0 & 0 & 0 & 0 & 1 & 0 & 0 \\
 0 & 0 & 0 & 0 & 0 & 0 & 1 & 0 & 0 & 0 \\
 0 & 0 & 0 & 0 & 0 & 0 & 0 & 0 & 1 & 0 \\
 0 & 0 & 0 & 0 & 0 & 0 & 0 & 0 & 0 & 1 \\
\end{pmatrix} \begin{pmatrix}
A_{11} \\
A_{12} \\
A_{13} \\
A_{22} \\
A_{23} \\
A_{33} \\
B_1 \\
B_2 \\
B_3 \\
C
\end{pmatrix}.
\end{equation}
By comparing the primed and unprimed coefficients, we can deduce that the sum of these should be an invariant quantity. For example, we see that $A_{11}$ becomes $A_{22}$, while $A_{22}$ transforms into $A_{11}$. Therefore, $A_{11} + A_{22}$ under the action of the generator $a$ remains invariant. As a result, invariance under $S_3$ forces:
\begin{equation}
A = \frac{1}{2} \begin{pmatrix}
2A_{ii} & A_{ij} & A_{ij} \\
A_{ij} & 2A_{ii} & A_{ij} \\
A_{ij} & A_{ij} & 2A_{ii}
\end{pmatrix}, \quad B= \left( B_i~B_i~B_i \right),
\end{equation}
using a shorthand notation by dropping the summation indices, $e.g.$, $A_{ij} = A_{12} + A_{23} + A_{31}$. At the end of the day, we got that in total there are three invariant coefficients given by $B_i$ and $A_{ii},\,A_{ij}$. As might have been expected, this does coincide with the results provided in eq.~\eqref{Eq:S3_Polynom_powers}.

There is yet another approach which relies on Representation Theory. Consider the eigenvalues and eigenvectors of the generators $a$ and $b$. For the generator $a$ we have
the following eigensystem:
\begin{equation}
\begin{array}{c}
\text{eigenvalues} \to\\
\text{eigenvectors} \to \\
\end{array}
\left(
\begin{array}{ccc}
 -1 & 1 & 1 \\
 \frac{1}{\sqrt{2}}(-1,1,0) ~~& (0,0,1) ~~& \frac{1}{\sqrt{2}}(1,1,0) \\
\end{array}
\right),
\end{equation}
where in the upper row eigenvalues are listed, each paired with its corresponding eigenvector directly below.

For $b$, the eigensystem is given by:
\begin{equation}
\left(
\begin{array}{ccc}
 e^{\frac{-2 i \pi }{3}} & 1 & e^{\frac{2 i \pi }{3}} \\
 \frac{1}{\sqrt{3}}\left(e^{-\frac{2 i \pi }{3}},e^{\frac{2 i \pi }{3}},1\right) ~~& \frac{1}{\sqrt{3}}(1,1,1) ~~& \frac{1}{\sqrt{3}}\left(e^{\frac{2 i \pi }{3}},e^{-\frac{2 i \pi }{3}},1\right) \\
\end{array}
\right).
\end{equation}

It is easy to notice that the eigenvectors corresponding to the value of ``1" (comparing $a$ and $b$) are not mutually orthogonal, $e.g.$,
\begin{equation}
\frac{1}{\sqrt{3}}(1,1,1) \cdot (0,0,1) = \frac{1}{\sqrt{3}} \neq 0.
\end{equation}
Actually, generators of \textit{reducible representations} share at least a single eigenvector. This statement arises from the fact that in a reducible representation there exists  an invariant subspace that is invariant under the action of the group elements. If a particular representation is reducible, it can be decomposed into a unique direct sum of irreducible representations. 

By picking a unitary matrix,
\begin{equation}\label{Eq:S3_red_U_rot_tribi}
U = 
\left(
\begin{array}{ccc}
 \frac{1}{\sqrt{2}} & -\frac{1}{\sqrt{2}} & 0 \\
 \frac{1}{\sqrt{6}} & \frac{1}{\sqrt{6}} & -\sqrt{\frac{2}{3}} \\
 \frac{1}{\sqrt{3}} & \frac{1}{\sqrt{3}} & \frac{1}{\sqrt{3}} \\
\end{array}
\right),
\end{equation}
we can make a transformation into a new basis via
\begin{equation}
g_i^\prime = U g_i U^\dagger,
\end{equation}
so that the generators of eq.~\eqref{Eq:S3_as_permutations} become
\begin{equation}\label{Eq:S3_as_eqtr}
\begin{aligned}
e^\prime =& \begin{pmatrix}
 1 & 0 & 0 \\
 0 & 1 & 0 \\
 0 & 0 & 1 \\
\end{pmatrix},\\
a^\prime =& \begin{pmatrix}
 -1 & 0 & 0 \\
 0 & 1 & 0 \\
 0 & 0 & 1 \\
\end{pmatrix}, \quad b^\prime = \begin{pmatrix}
 \frac{1}{2} & \frac{\sqrt{3}}{2} & 0 \\
 \frac{\sqrt{3}}{2} & -\frac{1}{2} & 0 \\
 0 & 0 & 1 \\
\end{pmatrix}, \quad c^\prime = \begin{pmatrix}
 \frac{1}{2} & -\frac{\sqrt{3}}{2} & 0 \\
 -\frac{\sqrt{3}}{2} & -\frac{1}{2} & 0 \\
 0 & 0 & 1 \\
\end{pmatrix},\\
d^\prime =& \begin{pmatrix}
 -\frac{1}{2} & \frac{\sqrt{3}}{2} & 0 \\
 -\frac{\sqrt{3}}{2} & -\frac{1}{2} & 0 \\
 0 & 0 & 1 \\
\end{pmatrix}, \quad f^\prime = \begin{pmatrix}
 -\frac{1}{2} & -\frac{\sqrt{3}}{2} & 0 \\
 \frac{\sqrt{3}}{2} & -\frac{1}{2} & 0 \\
 0 & 0 & 1 \\
\end{pmatrix}.
\end{aligned}
\end{equation}
Notice that all generators are simultaneously block diagonalised. In the new basis the generators look like rotations and reflections. Indeed, we could define the reflection operator,
\begin{equation}
\hat{\mathcal{F}}(\theta) = \begin{pmatrix}
-\cos \theta & \sin \theta \\
\sin \theta & \cos \theta
\end{pmatrix},
\end{equation}
and the rotation operator,
\begin{equation}
\hat{\mathcal{R}}(\theta) = \begin{pmatrix}
\cos \theta & \sin \theta \\
-\sin \theta & \cos \theta
\end{pmatrix}.
\end{equation}
Then, for instance, the $S_3$ generators can be expressed in terms of these operators as,
\begin{equation}\label{Eq:S3_as_RF}
\begin{aligned}
e^\prime = \hat{\mathcal{R}}\left(0\right), \quad d = \hat{\mathcal{R}}\left(\frac{2\pi}{3}\right), \quad f = \hat{\mathcal{R}}\left(-\frac{2\pi}{3}\right),\\
a^\prime = \hat{\mathcal{F}}\left(0\right), \quad b^\prime = \hat{\mathcal{F}}\left(\frac{2\pi}{3}\right), \quad c^\prime = \hat{\mathcal{F}}\left(-\frac{2\pi}{3}\right).
\end{aligned}
\end{equation}

\section[\texorpdfstring{$S_3$}{S3} as a group of an equilateral triangle]{\boldmath$S_3$ as a group of an equilateral triangle}\label{Sec:S3_as_equilat_tri}

Consider an equilateral triangle with vertices as in Figure~\ref{Fig:Eq_Tr}.

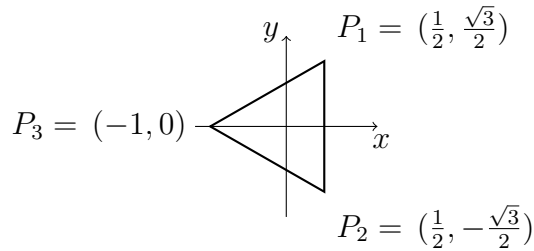
\begin{figure}[htb]
 \centering
\begin{tikzpicture}[
  triangle/.style={
   regular polygon,
   regular polygon sides=3,
   minimum size=2cm,
   draw,
   thick,
   label=corner 1:{$P_3=\,(-1,0)~$},
   label=corner 2:{$P_2=\,(\frac{1}{2},-\frac{\sqrt{3}}{2}) $},
   label=corner 3:{$P_1=\,(\frac{1}{2},\frac{\sqrt{3}}{2}) $}}
]
\node [triangle,rotate=90] (a) {};
\draw [thin , ->] (0,-1.2) -- (0,1.2);
\node at (-0.2,1.25) (nodeyi) {$y$};
\draw [thin , ->] (-1.2,0)--(1.2,0);
\node at (1.25,-0.2) (nodexi) {$x$};
\end{tikzpicture}
\caption{An equilateral triangle with its vertices provided, $P_i = (x_i,\,y_i)$.}\label{Fig:Eq_Tr} 
\end{figure}

There are several possible symmetry operations: two rotation (or three, counting the identity element) and three reflections. These are schematically visualised in Figure~\ref{Fig:Sym_of_eq_tr}. Let us define rotations by the generators $r_i$ and reflections by the generators $s_j$. 
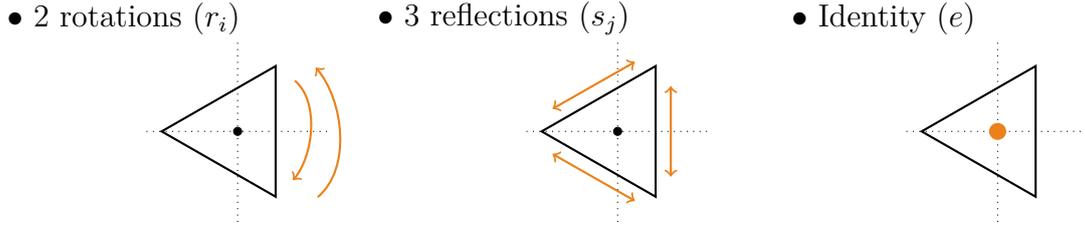
\begin{figure}[htb]
\centering
\begin{tikzpicture}
[
  triangle/.style={
   regular polygon,
   regular polygon sides=3,
   minimum size=2cm,
   thick,
   draw}
]
\node at (-1.5,1.5) { $\bullet$ 2 rotations ($r_i$)};
\node [triangle,rotate=90] (a) {};
\draw[dotted] (0,-1.2)--(0,1.2);
\draw[dotted] (-1.2,0)--(1.2,0);
\draw[fill] (0,0) circle  [radius=1.5pt];
\path [MyOr,bend right,style={<-,shorten >=0.2cm,shorten <=0.2cm},thick,out=-40,in=230] (0.6,-0.8) edge (0.6,0.8);
\path [MyOr,bend right,style={->,shorten >=0.2cm,shorten <=0.2cm},thick,out=-50,in=220] (0.9,-1) edge (0.9,1);
\node at (3.5,1.5) { $\bullet$ 3 reflections ($s_j$)};
\node [triangle,rotate=90] at (5,0) {};
\draw[dotted] (5,-1.2)--(5,1.2);
\draw[dotted] (3.8,0)--(6.2,0);
\draw[fill] (5,0) circle  [radius=1.5pt];
\path [MyOr,style={<->, thick, shorten >=0.2cm,shorten <=0.2cm}] (5.7,-0.8) edge (5.7,0.8);
\path [MyOr,style={<->, thick, shorten >=0.2cm,shorten <=0.2cm}] (3.97,0.2) edge (5.4,1.02);
\path [MyOr,style={<->, thick, shorten >=0.2cm,shorten <=0.2cm}] (3.97,-0.2) edge (5.4,-1.02);
\node at (8.5,1.5) {$\bullet$ Identity ($e$)};
\node [triangle,rotate=90] at (10,0) {};
\draw[dotted] (10,-1.2)--(10,1.2);
\draw[dotted] (8.8,0)--(11.2,0);
\draw[MyOr,fill] (10,0) circle  [radius=3pt];
 \end{tikzpicture}
 \caption{Depiction of symmetries of an equilateral triangle.}\label{Fig:Sym_of_eq_tr}
\end{figure}

A dihedral group $D_n$ is the group of symmetries of a regular $n$-gon (polygon). For the case of an equilateral triangle there are six elements,
\begin{equation}\label{Eq:set_of_D3}
D_3 = \{ e, \, r_1,\, r_2,\, s_1,\, s_2,\, s_3\}.
\end{equation}
The minimal set of $D_3$ consists of two elements $\{r,\,s\}$.

Let us check how different transformations, which leave the equilateral triangle of Figure~\ref{Fig:Eq_Tr} invariant, look like. Consider a rotation around the $x$-axis, which we shall denote to be a two-dimensional $A$ transformation. Acting on the vertices of the equilateral triangle with $A$ we get:
\begin{equation}
A\,P_1 = P_2, \quad A\,P_2 = P_1, \quad A\,P_3 = P_3,
\end{equation}
where $P_i$ are the vertices of Figure~\ref{Fig:Eq_Tr}. Spelling it out, we need to solve for $A$:
\begin{subequations}
\begin{align}
& A\,P_1 = P_2~ \left\{
\begin{aligned}
&A_{11} \frac{1}{2} + A_{12} \frac{\sqrt{3}}{2} = \frac{1}{2},\\
&A_{21} \frac{1}{2} + A_{22} \frac{\sqrt{3}}{2} = -\frac{\sqrt{3}}{2},
\end{aligned}\quad
\right. \\
& A\,P_2 = P_1~ \left\{
\begin{aligned}
&A_{11} \frac{1}{2} + A_{12} \left( -\frac{\sqrt{3}}{2} \right) = \frac{1}{2},\\
&A_{21} \frac{1}{2} + A_{22} \left( -\frac{\sqrt{3}}{2} \right) = \frac{\sqrt{3}}{2},
\end{aligned}\quad
\right. \\
& A\,P_3 = P_3~ \left\{
\begin{aligned}
&A_{11} (-1) + A_{12}\,\left(0\right) = -1,\\
&A_{21} (-1) + A_{22}\,\left(0\right) = 0.
\end{aligned}\quad
\right.
\end{align}
\end{subequations}
There is a single solution given by:
\begin{equation}
A = \begin{pmatrix}
1 & 0 \\
0 & -1
\end{pmatrix}.
\end{equation}

Then, there are two other reflections, which are:
\begin{equation}
B= \frac{1}{2}\begin{pmatrix}
-1 & \sqrt{3} \\
\sqrt{3} & 1
\end{pmatrix}, \quad
C= \frac{1}{2}\begin{pmatrix}
-1 & -\sqrt{3} \\
-\sqrt{3} & 1
\end{pmatrix},
\end{equation}
Apart from that it is possible to rotate the equilateral triangle by
\begin{equation}
D= \frac{1}{2}\begin{pmatrix}
-1 & -\sqrt{3}\\
\sqrt{3} & -1
\end{pmatrix}, \quad
F= \frac{1}{2}\begin{pmatrix}
-1 & \sqrt{3} \\
-\sqrt{3} & -1
\end{pmatrix}.
\end{equation}

Now, assume that we got $r=D$ and $s=A$; transformations of different origin, one rotations and one reflection. The presentation of $D_3$, see eq.~\eqref{Eq:set_of_D3}, is then:
\begin{equation}
D_3 = \left\langle  r,\, s~|~r^3 = s^2 =e,\, srs =r^2 \right\rangle.
\end{equation}
This looks exactly like that of the $S_3$ group, see eq.~\eqref{Eq:S3_presentation}. Even though these two groups were defined differently, they have a structure-preserving map (bijective map) between themselves. Since these two are structurally identical, we call them isomorphic, $S_3 \cong D_3$. Another interesting aspect to consider is that we defined $D_3$ to act on the vertices of the equilateral triangle, however we could have written all of the generators acting on a six-dimensional vector, which would correspond to the two components of $x_i$ and $y_i$ multiplied by three vertices. For example, by acting with $A$ we expect the following reflection,
\begin{equation}
\begin{pmatrix}
x_1 \\
x_2 \\
x_3 \\
y_1 \\
y_2 \\
y_3
\end{pmatrix}^\prime = \begin{pmatrix}
 0 & 1 & 0 & 0 & 0 & 0 \\
 1 & 0 & 0 & 0 & 0 & 0 \\
 0 & 0 & 1 & 0 & 0 & 0 \\
 0 & 0 & 0 & 0 & 1 & 0 \\
 0 & 0 & 0 & 1 & 0 & 0 \\
 0 & 0 & 0 & 0 & 0 & 1 \\
\end{pmatrix} \begin{pmatrix}
x_1 \\
x_2 \\
x_3 \\
y_1 \\
y_2 \\
y_3
\end{pmatrix},
\end{equation}
which corresponds to $\mathcal{I}_2 \otimes a$ of eq.~\eqref{Eq:S3_as_permutations}.

We could also have assumed that the equilateral triangle of Figure~\ref{Fig:Eq_Tr} was inclined into an additional $z$ dimension. This way the generators would be represented by three-dimensional matrices. However, it is trivial to deduce from a geometrical perspective that the equilateral triangle is a two-dimensional figure. If we were given only the transformation matrices in the $(x,\,y,\,z)$ space, we would have noticed that some of the transformation matrices would share common eigenvectors. Hence, such transformations would be reducible.

\section{Quotient and semidirect groups}

Some subsets of elements of groups can form groups themselves. If a considered subset $\mathcal{H}$ respects the same operation as the original group and all of the group properties (definitions) are satisfied, then such subset forms a subgroup, usually denoted as $\mathcal{H} \leq \mathcal{G}$. For the case of $S_3$, in terms of eq.~\eqref{Eq:S3_as_permutations}, the subgroups are:
\begin{equation}
\{e\},~ \{e,\,a\},~ \{e,\,b\},~ \{e,\,c\},~ \{e,\,d,\,f\}.
\end{equation}
The Lagrange theorem states that if $\mathcal{H}$ is a subgroup of a finite $\mathcal{G}$, then $|\mathcal{H}|$ divides $|\mathcal{G}|$.

A group is called normal, denoted as $\mathcal{N}\triangleleft \mathcal{G}$, if it remains unchanged under conjugation ($g n g^{-1}$) by all elements of the group $\mathcal{G}$,
\begin{equation}
g n g^{-1} \in \mathcal{N}\leq \mathcal{G},\quad \forall g \in \mathcal{G},~\forall n \in \mathcal{N}.
\end{equation} 
An equivalent condition for a subgroup $\mathcal{H}$ to be called normal would be that left,
\begin{equation}
g \mathcal{H}=\{g h \mid h \in \mathcal{H}\},
\end{equation}
and right cosets,
\begin{equation}
\mathcal{H} g=\{h g \mid h \in \mathcal{H}\},
\end{equation}
are equal, $g\mathcal{H} = \mathcal{H}g$.

This interpretation allows to define a quotient group,
\begin{equation}
\mathcal{G}/\mathcal{N} = \{g \mathcal{N} \mid g \in \mathcal{G}\},
\end{equation}
which is a set of all left cosets ($g \mathcal{N}$) of $\mathcal{N}$ in $\mathcal{G}$. To be more precise, if we expanded the set of $\mathcal{G}/\mathcal{N} = \{ g \mathcal{N} \}$, note that $\mathcal{N}$ is a set of generators itself, we would have gotten the initial set given by $\mathcal{G}$, $\{ g \mathcal{N} \} = \{g_1 \mathcal{N},\, \dots, g_i \mathcal{N}\} = \{g_1 \{ n_1,\, \dots,\, n_j\},\, \dots, g_i \{ n_1,\, \dots,\, n_j\}\} = \{g_1,\, \dots,\, g_i\} = \mathcal{G}$.

Consider the $S_3$ group. The normal subgroup of $S_3$ is $\mathcal{N} = \{e,\, d,\, d^2 \}$, which can be identified as $A_3 \cong \mathbb{Z}_3$. In general we have that $A_n \triangleleft  S_n$. The alternating group $A_n$ is the group of even permutations, while $S_n$ is a group of all permutations. From the Lagrange theorem we can deduce that the order of the quotient group should be
\begin{equation}
|S_n/A_n|= \frac{n!}{n!/2} = 2,
\end{equation}
which can be identified as that of $\mathbb{Z}_2$. In the case of $S_3$ we have
\begin{equation}
S_3 / \mathbb{Z}_3  = \{ \mathcal{N}, a \mathcal{N} \},~~\mathcal{N} = \{e,\, d,\, d^2 \}.
\end{equation}
By expanding the set $\{ \mathcal{N}, a \mathcal{N} \}$ all elements of the $S_3$ group are restored.

Since $S_3 / \mathbb{Z}_3  \cong \mathbb{Z}_2$, can the $S_3$ group be presented as a product of $\mathbb{Z}_2 \times \mathbb{Z}_3$? 

In the case of the direct product, $\mathbb{Z}_2 \times \mathbb{Z}_3$ it is assumed that the two groups commute. Given,
\begin{equation}
\mathbb{Z}_2 = \left\langle a~|~a^2=e\right\rangle, \quad \mathbb{Z}_3 = \left\langle b~|~b^3=e\right\rangle,
\end{equation} 
we get that $(ab)^n = a^n b^n$ due to the commutation of the two generators. Then we could introduce a new generator $c=ab$. In the new basis, it can be verified that,
\begin{equation}
c^n : \{ab,\, b^2,\, a^3,\, b,\, ab^2,\,e \},\quad \text{ for } n=\{1,\dots,6\}.
\end{equation}
Therefore we would get $\mathbb{Z}_2 \times \mathbb{Z}_3 \cong \mathbb{Z}_6$, which is an Abelian group, and cannot be isomorphic to the non-Abelian $S_3$. 

On the other hand, a semidirect product of these groups is $\mathbb{Z}_3 \rtimes \mathbb{Z}_2 \cong S_3$, since $ab \neq ba$. However, a definition of a semidirect product is more involved, for our discussion it is sufficient to note that one of the groups is a normal subgroup ($\mathbb{Z}_3$), while the other group acts non-trivially ($\mathbb{Z}_2$). 

\section{Representations and characters}

In Group Theory, a representation of the group $\mathcal{G}$ is a way of associating and describing the elements of $\mathcal{G}$ in terms of bijective linear transformations (or matrices) of a vector space. In general, a representation of a group ($D(g),~g \in \mathcal{G}$) is a homomorphism from $\mathcal{G}$ to the group of general invertible linear transformations ($GL(V)$) on a vector space ($V$),
\begin{equation}
D~:~ \mathcal{G} \rightarrow GL(V),
\end{equation}
such that
\begin{equation}
D(g_1 g_2) = D(g_1) D(g_2)~~\forall~g_i \in \mathcal{G},
\end{equation}
and each is invertible,
\begin{equation}
D(g^{-1}) = D(g)^{-1}.
\end{equation}
In the most general case $ GL(V)$ is a group $GL(n,\,\mathbb{C})$ of dimension $n$ of non-singular complex matrices.

For example, consider the $S_2 \cong \mathbb{Z}_2$ group with the representation
\begin{equation}
D(e) = \mathcal{I}_2, \quad D(g) = \begin{pmatrix}
0 & 1 \\
1 & 0
\end{pmatrix}.
\end{equation}
We get,
\begin{subequations}
\begin{align}
D(g) ={}& D(e) D(g) = D(g) D(e),\\
D(e) ={}& D(g) D(g).
\end{align}
\end{subequations}
In this trivial example the generators $e$ and $g$ are mapped (injective function) to distinct linear transformations (or matrices) in the representation. This way the entire group structure is preserved; no information is lost. Such representations are called faithful. The representation $D(g)$ is non-diagonal with the eigensystem:
\begin{equation}
\left(
\begin{array}{ccc}
 -1 & 1 \\
 \left(-1,\,1\right) ~~ & \left(1,\,1\right) \\
\end{array}
\right).
\end{equation}
The eigenvectors show that this representation can be decomposed into two irreducible components given by a symmetric subspace, $(1,\,1)$, and by an antisymmetric subspace, $\left(-1,\,1\right)$. Both subspaces are invariant under the action of $S_2$. Therefore, the original representation is reducible since it can be split into these two independent subspaces. As will soon be shown, one could deduce if a particular representation is reducible by checking the inner product of the non-trivial representations. The inner product of each of the eigenvectors of $D(g)$ by itself is 2. Any result different from unity indicates that the matrix representation in question is reducible. 

In order to construct a representation of $\mathcal{G}$ we need to assign each distinct operation a unique element. Let us define a bijective conjugation operation as
\begin{equation}\label{Eq:Conj_Cl_def}
b = g \,a\,g^{-1} \quad \forall~g \in \mathcal{G}\text{ and }\{a,\,b\} \in \mathcal{G}.
\end{equation}

At this point it may become obvious why the elements of eqs.~\eqref{Eq:S3_as_permutations}, or eqs.~\eqref{Eq:S3_as_eqtr}, or eqs.~\eqref{Eq:S3_as_RF} were written in separate lines or combined together. Let us consider the $S_3$ group as it was presented in eq.~\eqref{Eq:S3_a_d_presentation}, $S_3 = \{e,\,a,\,da,\,d^2a,\,d,\,d^2\}$. For $S_3$ we got the following conjugacy classes (different sets obeying eq.~\eqref{Eq:Conj_Cl_def}):
\begin{equation}
\begin{aligned}
\mathcal{C}_1&:~\{e\},\\
\mathcal{C}_2&:~\{a,\,da,\,d^2a\},\\
\mathcal{C}_3&:~\{d,\, d^2\}.
\end{aligned}
\end{equation}

Each of the conjugacy classes corresponds to a different transformation operator, \textit{e.g.,} rotations or reflections. Consider the conjugacy class $\mathcal{C}_3:\,\{d,\,d^2\}$, which corresponds to rotations. We got,
\begin{equation}
(d)^n = \prod_n \left( g (d^2) g^{-1} \right)= g (d^2)^n g^{-1}.
\end{equation}
Assuming that $(d^2)^n = e$, we can conclude that all elements of the same conjugacy class have the same order; the smallest positive integer $n$ such that $g^n=e$. The solution of the above equation is $n=3x$ for $x \in \mathbb{Z}$; though it makes sense to restrict solutions to positive integers.

We are already aware that the permutation matrices of the $S_3$ group in the triplet representation can be written down in block-diagonal forms, which were presented in eq.~\eqref{Eq:S3_as_eqtr}.

Consider an arbitrary finite group. Since this group is finite it should have a finite number of irreducible representations. Here is an analogy: an $n$-dimensional vector can be expressed (at most) in terms of $n$ different orthonormal vectors. The task of decomposing a reducible triplet of the $S_3$ group was somewhat equivalent to identifying an orthogonal basis. Apart from that we already know that there are several conjugacy classes, the generators of which are alike in their structure. Therefore, we can deduce that the number of irreducible representations of a finite group equals the number of its conjugacy classes. If this is true then (at least) the orthogonality relations should hold and the complete set of irreducible representations should span the entire space of all possible representations of a given group.  

Let us start by writing the conjugacy classes of the $S_3$ group in a table-like format:
\begin{equation}\label{Eq:S3_const_ch_tab}
\begin{tabular}{l|c|c|c} \hline \hline
 &   $\mathrm{dim}(\mathcal{C}_1)=1,~ \mathcal{C}_1$ & $\mathrm{dim}(\mathcal{C}_2)=3,~\mathcal{C}_2$ & $\mathrm{dim}(\mathcal{C}_3)=2,~ \mathcal{C}_3$\\ \hline \hline
$D_1$ & 1 & 1 & 1 \\ \hline
$D_2$ & $e$& $\{a,\,da,\,d^2a\}$ & $\{d,\,d^2\}$ \\ \hline \hline
\end{tabular}
\end{equation}
Here we split the reducible triplet ($D_\mathrm{R}$) into the block-diagonal parts as
\begin{equation}
D_\mathrm{R} = \mathrm{diag}(D_2,\, D_1) = D_2 \oplus D_1.
\end{equation}
It is important to highlight that the reduction of a reducible representation into its irreducible components is unique. One should keep in mind that every $\mathcal{G}$ has an identity/trivial representation, in which every element of the group is mapped to the identity transformation in the representation space, \textit{i.e}, no transformation is performed. The trivial representation is typically one-dimensional.

In general, sets like $\{d,\,d^2\}$ and $\{a,\,da,\,d^2a\}$ provide little information (talking about eq.~\eqref{Eq:S3_const_ch_tab}) about the structure of the underlying group. Writing down a single generator from either list is also not particularly helpful, since the conjugacy classes do not form subgroups. Rather, a very useful quantity is called the character of the group representation. The character of a group representation $D(g),~g \in \mathcal{G}$ is defined as the trace of the corresponding representation,
\begin{equation}
 \chi^{(D_n)}(g) = \mathrm{tr}\left(D_n(g)\right).
\end{equation} 

Using the character function, the above table can be written as
\begin{equation}\label{Eq:S3_const_ch_tab_2}
\begin{tabular}{l|c|c|c} \hline \hline
 &   \begin{tabular}[l]{@{}c@{}} $\mathcal{C}_1$ \\ $\{e\}$ \end{tabular} & \begin{tabular}[l]{@{}c@{}} $3\mathcal{C}_2$ \\ $\{a,\,da,\,d^2a\}$ \end{tabular} & \begin{tabular}[l]{@{}c@{}} $2\mathcal{C}_3$ \\ $\{d,\,d^2\}$ \end{tabular}\\ \hline \hline
$\chi^{(D_1)}$ & 1 & 1 & 1 \\ \hline
$\chi^{(D_2)}$ & $2$ & 0 & -1\\ \hline \hline
\end{tabular}
\end{equation}

As stated above, the number of irreducible representations should equal the number of conjugacy classes. Therefore, one irreducible representation is missing. Another useful statement in finding irreducible representations is that the sum of the squares of the dimensions of the irreducible representations ($d_i^2$) is equal to the group order ($|\mathcal{G}|$),
\begin{equation}\label{Eq:d2_reps}
\sum_{i=1}^n d_i^2 = |\mathcal{G}|,~ d_i \leq d_{i+1}.
\end{equation}
Since dimensions are positive integers we can put an upper limit by evaluating
\begin{equation}
d_n^2 \leq |\mathcal{G}| - (n-1).
\end{equation}
Besides, the degree of the irreducible representations divides the order of the group $\mathcal{G}$.

The dimensions of the irreducible representations can be read directly from the column represented by the conjugacy class $\mathcal{C}_1$ since it corresponds to the identity element. In the case of the $S_3$ group we have three conjugacy classes, and so we need to solve
\begin{equation}\label{Eq_S3_rep_di2_sum}
d_1^2 + d_2^2 + d_3^2 = |S_3| = 6.
\end{equation}
There is a single solution with $d_i = \{1,\,1,\,2 \}$, which indicates that we are missing another representation of dimension one.

There are two orthogonality theorems, which are essential for the construction, or verification, of the character tables. The first orthogonality theorem (for rows of the
character table) can be expressed in terms of the relations,
\begin{equation}
\sum_k \chi^{\left(D_i\right)}(\mathcal{C}_k) \chi^{\left(D_{j}\right)}(\mathcal{C}_k)^\ast N_k= |\mathcal{G}| \delta_{ij},
\end{equation}
where $\delta_{ij}$ is the Kronecker delta and $N_k$ represents the number of elements of the conjugacy class $\mathcal{C}_k$. Mind that some entries in the character table can be complex quantities.

The second orthogonality theorem (for columns of the character table) is given by a set of relations,
\begin{equation}
\sum_k \chi^{\left(D_k\right)}\left(\mathcal{C}_i\right)\chi^{\left(D_k\right)}\left(\mathcal{C}_{j}\right)^\ast N_j= |\mathcal{G}| \delta_{ij},
\end{equation}

Let us identify the last missing representation of eq.~\eqref{Eq:S3_const_ch_tab_2}. From eq.~\eqref{Eq_S3_rep_di2_sum} we already know that it should be of dimension one. Suppose that the missing row in \eqref{Eq:S3_const_ch_tab_2} is given by $D_{1^\prime}:(1,\,x,\,y)$, with $x$ and $y$ to be determined. Then, using the orthogonality relations we get:
\begin{subequations}
\begin{align}
(D_1,D_{1^\prime}):~ &(1)(1)(1)+(1)(x)(3)+(1)(y)(2) =  1 + 3x + 2y =0, \\
(D_2,D_{1^\prime}):~ &(2)(1)(1)+(0)(x)(3)+(-1)(y)(2) = 2 - 2y = 0, \\
(\mathcal{C}_1,\mathcal{C}_2):~ &3\left[(1)(1)+(1)(x)+(2)(0)\right] = 1 + x = 0, \\
(\mathcal{C}_1,\mathcal{C}_3):~ &2\left[(1)(1)+(1)(y)+(2)(-1) \right] = -1 + y  = 0, \\
(\mathcal{C}_2,\mathcal{C}_3):~ &2\left[(1)(1)+(x)(y)+(0)(-1) \right] = 1 + xy = 0,
\end{align}
\end{subequations}
the solution of which is $x=-1,~y=1$. As a result, the character table for the $S_3$ group is given in Table~\ref{Tab:S3_char}.

{{\renewcommand{\arraystretch}{1}
\begin{table}[h]
\caption{The character table of the $S_3$ group.}
\label{Tab:S3_char}
\begin{center}
\begin{tabular}{l|c|c|c} \hline \hline
$S_3$ &   \begin{tabular}[l]{@{}c@{}} $\mathcal{C}_1$ \\ $\{e\}$ \end{tabular} & \begin{tabular}[l]{@{}c@{}} $3\mathcal{C}_2$ \\ $\{a,\,da,\,d^2a\}$ \end{tabular} & \begin{tabular}[l]{@{}c@{}} $2\mathcal{C}_3$ \\ $\{d,\,d^2\}$ \end{tabular}\\ \hline \hline
$\chi^{(D_1)}$ & 1 & 1 & 1 \\ \hline
$\chi^{(D_1^\prime)}$ & 1 & -1 & 1 \\ \hline
$\chi^{(D_2)}$ & $2$ & 0 & -1\\ \hline \hline
\end{tabular}
\end{center}
\end{table}}

The sign representation, $D_{1^\prime}$, in the context of the $S_3$ group can be easily understood by considering a homomorphism from the permutation group to a quotient (factor) group. All of the $S_3$ permutations are even or odd, or as it was shown in eq.~\eqref{Eq:S3_as_RF} can be expressed in terms of only two operators. So, the sign representation can be viewed as a homomorphism to the number of transpositions. The two-cycles, $\{(12),\, (13),\,(23)\}$, of eqs.~\eqref{Eq:S3_cycles} are odd, $i.e.$, there is a single swap of two elements. Therefore, these elements are assigned the value of ``-1".

Now, consider that we are given some representation $D_\mathrm{R}$ of a particular group and we need to identify whether it is reducible, and if it is reducible, which representations it is reducible to. Remember that each reducible representation is given by a unique combination of irreducible representations,
\begin{equation}\label{Eq:DR_reducable_oplus}
D_\mathrm{R} = \sideset{}{}\bigoplus_{i \in \mathcal{K}} \,a_i D_i,
\end{equation}
where the direct sum is over the index set of $\mathcal{K}$ (for the case of $S_3:~\mathcal{K}=\{1,\, 1^\prime,\, 2\}$) and the $a_i$ coefficients denote the number of times the irreducible representation appears in the decomposition. These coefficients can be found by solving
\begin{equation}
a_i = \frac{1}{|\mathcal{G}|} \sum_j \chi^{(D_\mathrm{R})}(\mathcal{C}_j) \chi^{(D_i)}(\mathcal{C}_j)^\ast.
\end{equation}

For example, in the case of the reducible triplets of the $S_3$ group, see eqs.~\eqref{Eq:S3_as_permutations}, a vector composed from the characters is given by $D_\mathrm{R}:(3,\,1,\,0)$. Then,
\begin{subequations}\label{Eq:S3_decomp_examp}
\begin{align}
a_1 ={}& \frac{1}{6}\left[ (1)(3)(1) + (3)(1)(1) + (2)(0)(1) \right] = 1,\\
a_{1^\prime} ={}& \frac{1}{6}\left[ (1)(3)(1) + (3)(1)(-1) + (2)(0)(1) \right] = 0,\\
a_2 ={}& \frac{1}{6}\left[ (1)(3)(2) + (3)(1)(0) + (2)(0)(-1) \right] = 1.
\end{align}
\end{subequations}
Therefore, $D_\mathrm{R} = D_1 \oplus D_2$.

One thing to note is that several non-isomorphic groups can share the same character table. This occurs because the character table does not encode the entire structure of a particular group, but rather only partial information about the conjugacy classes and irreducible representations. For example, both $D_4$ and $Q_8$ share the same character table. For the non-Abelian groups, with the same number of conjugacy classes, the constraints on the dimensions of the representations, see eq.~\eqref{Eq:d2_reps}, might be way too restrictive. Finite groups, which have the same character table, with order less than 2000 were analysed in Ref.~\cite{cocke2019databasegroupsequivalentcharacter}.

Character tables of many different groups are available in different formats. Systems for computational algebra can be used to obtain such tables. For example, a popular system for Group Theory is $\mathtt{GAP}$~\cite{GAP4}. By running the code:
\begin{lstlisting}
 G := SymmetricGroup(3);
CT := CharacterTable(G);
Display(CT);
\end{lstlisting}
we get the following output:
\begin{lstlisting}
     2  1  1  .
     3  1  .  1

       1a 2a 3a
    2P 1a 1a 3a
    3P 1a 2a 1a

X.1     1 -1  1
X.2     2  . -1
X.3     1  1  1
\end{lstlisting}
In the above output the character table is presented in the last three lines, ``X.i". The ``." in the listings indicates ``0". When compared to Table~\ref{Tab:S3_char} we can see that characters of representations are presented in a different order. The conjugacy classes other than the one containing the identity element are specified by entries ``2P" and ``3P", displaying dimensionalities.

We can implement a non-optimised function (for real representations) in $\mathtt{GAP}$ to find the decomposition of a particular representation by utilising the following function (though, it should be noted that there is a built-in function $\mathtt{ConstituentsOfCharacter()}$, the output of which is harder to read):
\begin{lstlisting}
DecomposeChars:= function(G, rep)
 local CT,ch,i,ScPr,Lch,res;
 CT := CharacterTable(G);
 ch:=Irr(CT);
 p:=Character(G,rep);
 Lch:=Length(ch);
 res:=[];
 for i in [1..Lch] do
   ScPr := ScalarProduct(p,ch[i]);
   if ScPr <> 0  then
     Add(res, JoinStringsWithSeparator([ScPr," ","X.",i],""));
   fi;
 od;
 Display(JoinStringsWithSeparator(res," + "));
end;
\end{lstlisting}
Then, we can evaluate the vector of characters, $D_\mathrm{R}:(3,\,1,\,0)$, as in the example provided by eqs.~\eqref{Eq:S3_decomp_examp}, by running:
\begin{lstlisting}
DecomposeChars(SymmetricGroup(3), [3,1,0]);
\end{lstlisting}
Notice that the character table for $\mathtt{SymmetricGroup(3)}$ is provided in a different order than in Table~\ref{Tab:S3_char}.

\section{Construction of invariants}

Now, suppose we need to decompose a tensor product of two irreducible representations, $D_\mathrm{R} = D_i \otimes D_j$, and
\begin{equation}
\chi^{\left(D_\mathrm{R}\right)}(\mathcal{C}_k) =  \chi^{\left(D_i\right)}(\mathcal{C}_k) \chi^{\left(D_{j}\right)}(\mathcal{C}_k),
\end{equation}
which amounts to multiplying characters of the two relevant representations.

We need to solve (also applicable to complex representations)
\begin{equation}
\mathcal{J} a_i = \begin{pmatrix}
\chi^{\left(D_\mathrm{R}\right)}(\mathcal{C}_1)\\
\chi^{\left(D_\mathrm{R}\right)}(\mathcal{C}_2) \\
\chi^{\left(D_\mathrm{R}\right)}(\mathcal{C}_3)
\end{pmatrix},
\end{equation}
where $a_i$ are the coefficients of eq.~\eqref{Eq:DR_reducable_oplus}, which will correspond to the number of irreducible representations contained within the investigated representation. The transposed matrix
\begin{equation}
\mathcal{J}^\mathrm{T} = \begin{pmatrix}
1 & 1 & 1 \\
1 & -1 & 1 \\
2 & 0 & -1
\end{pmatrix},
\end{equation}
is given by the entries of the character table of $S_3$, see Table~\ref{Tab:S3_char}.

 Below we list several possible combinations:
\begin{subequations}
\begin{align}
& \mathbf{1} \otimes \mathbf{1}:~ \mathcal{J} a_i =\begin{pmatrix}
1 \\
1 \\
1 \\
\end{pmatrix},\text{ results in } a_i=(1,\,0,\,0), \\
&\mathbf{1} \otimes \mathbf{1^\prime}:~ \mathcal{J} a_i =\begin{pmatrix}
1 \\
-1 \\
1 \\
\end{pmatrix},\text{ results in } a_i=(0,\,1,\,0), \\
&\mathbf{1} \otimes \mathbf{2}:~ \mathcal{J} a_i =\begin{pmatrix}
2 \\
0 \\
-1 \\
\end{pmatrix},\text{ results in } a_i=(0,\,0,\,1), \\
&\mathbf{1^\prime} \otimes \mathbf{1^\prime}:~ \mathcal{J} a_i =\begin{pmatrix}
1 \\
1 \\
1 \\
\end{pmatrix},\text{ results in } a_i=(1,\,0,\,0), \\
&\mathbf{1^\prime} \otimes \mathbf{2}:~ \mathcal{J} a_i =\begin{pmatrix}
2 \\
0 \\
-1 \\
\end{pmatrix},\text{ results in } a_i=(0,\,0,\,1), \\
&\mathbf{2} \otimes \mathbf{2}:~ \mathcal{J} a_i =\begin{pmatrix}
2 \\
0 \\
1 \\
\end{pmatrix},\text{ results in } a_i=(1,\,1,\,1), \\
&\mathbf{2} \otimes \mathbf{2} \otimes \mathbf{2} \otimes \mathbf{2}:~ \mathcal{J} a_i =\begin{pmatrix}
16 \\
0 \\
1 \\
\end{pmatrix},\text{ results in } a_i=(3,\,3,\,5) \label{Eq:2t2t2t2},
\end{align}
\end{subequations}
where, for simplicity, we defined representations in terms of integers in bold font, $\mathbf{n} \equiv D_{n}$.

We can also utilise the $\mathtt{DecomposeChars(\mathcal{G}, [\chi^{(D_i)}])}$ function, presented in the previous section, for the decomposition of the tensor products. For example, for the tensor product $\mathbf{2} \otimes \mathbf{2} \otimes \mathbf{2} \otimes \mathbf{2}$ we evaluate:
\begin{lstlisting}
 G := SymmetricGroup(3);
CT := CharacterTable(G);
ch := Irr(CT);
DecomposeChars(SymmetricGroup(3),ch[2]*ch[2]*ch[2]*ch[2]);
\end{lstlisting}
The output of which is:
\begin{lstlisting}
3 X.1 + 5 X.2 + 3 X.3
\end{lstlisting}
It shows how $\mathbf{2} \otimes \mathbf{2} \otimes \mathbf{2} \otimes \mathbf{2}$ can be decomposed into different representations, ``X.i",  multiplied by the overall coefficients (3, 5, 3). Note that $\mathsf{GAP}$ does not fix the order of representations in a particular way, and hence the output looks different from eq.~\eqref{Eq:2t2t2t2}. One has to refer to $\mathtt{CharacterTable(G)}$ to see which representations are given by ``X.i".

To summarise, we got the following tensor products:
\begin{subequations}\label{Eq:S3_diff_tensor_pr}
\begin{align}
\mathbf{ 1 } \otimes \text{any} &= \text{any},\\
\mathbf { 1^\prime }  \otimes \mathbf { 1^\prime } &= \mathbf { 1 },\\
\mathbf { 1^\prime } \otimes \mathbf { 2 } &= \mathbf { 2 },\\ 
\mathbf { 2 } \otimes \mathbf { 2 } &= \mathbf { 1 } \oplus \mathbf { 1^\prime } \oplus \mathbf { 2 }.
\end{align}
\end{subequations}

The next step is to find corresponding invariants, which will be associated with singlets (trivial representations), $\mathbf{1}$.

Let us consider the following $S_3$ generators, see eq.~\eqref{Eq:S3_as_eqtr}:
\begin{equation}
\begin{aligned}
e = \mathcal{I}_2, \quad
a =  \begin{pmatrix}
-1 & 0 \\
0 & 1
\end{pmatrix}, \quad
d = \frac{1}{2}\begin{pmatrix}
-1 & \sqrt{3} \\
-\sqrt{3} & -1
\end{pmatrix}.
\end{aligned}
\end{equation}

Assume that we got an $\mathbb{R}^4$ space with $\{S,~A,~D_{1,2}\}$, where those indicate a singlet \mbox{$S \equiv \mathbf{1}$}, a pseudosinglet $A\equiv \mathbf{1}$, and a doublet $D_{1,2}\equiv \mathbf{2}$. We got the following transformation rules under the elements $\{a,\,d\}$:
\begin{equation}
\begin{aligned}
&S\xrightarrow{\{a,\,d\}}S,\\
&A\xrightarrow{a}-A,~A\xrightarrow{d}A,\\
&\begin{pmatrix}
D_1 \\
D_2
\end{pmatrix} \xrightarrow{a} \begin{pmatrix}
-D_1 \\
D_2
\end{pmatrix},~ \begin{pmatrix}
D_1 \\
D_2
\end{pmatrix} \xrightarrow{d} \frac{1}{2}\begin{pmatrix}
-D_1 +\sqrt{3} D_2 \\
-\sqrt{3} D_1 - D_2
\end{pmatrix},
\end{aligned}
\end{equation}
which could be understood by considering the character table of the $S_3$ group presented in Table~\ref{Tab:S3_char}. Next, consider the tensor product between two different doublets, which we shall denote $X$ and $Y$,
\begin{equation}\label{S3_D_under_a}
XY^{(a)}_i \equiv (D_2(a)X) \otimes (D_2(a)Y) = \begin{pmatrix}
 1 & 0 & 0 & 0 \\
 0 & -1 & 0 & 0 \\
 0 & 0 & -1 & 0 \\
 0 & 0 & 0 & 1 \\
\end{pmatrix} \begin{pmatrix}
x_1 y_1 \\
x_1 y_2 \\
x_2 y_1 \\
x_2 y_2
\end{pmatrix} = \begin{pmatrix}
x_1 y_1 \\
-x_1 y_2 \\
-x_2 y_1 \\
x_2 y_2
\end{pmatrix},
\end{equation}
where $i$ is the running index indicating the row. Apart from that,
\begin{equation}
\begin{aligned}
XY^{(d)}_i \equiv (D_2(d)X) \otimes (D_2(d)Y) ={}& \begin{pmatrix}
 \frac{1}{4} & -\frac{\sqrt{3}}{4} & -\frac{\sqrt{3}}{4} & \frac{3}{4} \\
 \frac{\sqrt{3}}{4} & \frac{1}{4} & -\frac{3}{4} & -\frac{\sqrt{3}}{4} \\
 \frac{\sqrt{3}}{4} & -\frac{3}{4} & \frac{1}{4} & -\frac{\sqrt{3}}{4} \\
 \frac{3}{4} & \frac{\sqrt{3}}{4} & \frac{\sqrt{3}}{4} & \frac{1}{4} \\
\end{pmatrix} \begin{pmatrix}
x_1 y_1 \\
x_1 y_2 \\
x_2 y_1 \\
x_2 y_2
\end{pmatrix}\\ 
={}& \frac{1}{4}\begin{pmatrix}
x_1 y_1 - \sqrt{3} x_1 y_2  - \sqrt{3} x_2 y_1 + 3 x_2 y_2\\
 \sqrt{3}x_1 y_1 + x_1 y_2 - 3 x_2 y_1 - \sqrt{3} x_2 y_2 \\
\sqrt{3} x_1 y_1 -3 x_1 y_2 + x_2 y_1 - \sqrt{3} x_2 y_2\\
 3 x_1 y_1 + \sqrt{3} x_1 y_2 + \sqrt{3}  x_2 y_1 + x_2 y_2
\end{pmatrix},
\end{aligned}
\end{equation}
while under the singlet and pseudosinglet representations the tensor product is invariant; notice that for $A\xrightarrow{a}-A$ we get $(-A)(-A)$ and that $\mathbf{1^\prime}\otimes \mathbf{1^\prime} = \mathbf{1}$. Therefore, after finding the invariant combinations of doublets we could conclude that those should also be invariant under the other representations.

From the tensor product $XY^{(a)}$ of eq.~\eqref{S3_D_under_a} we can notice that the $XY^{(a)}_{1}$ and $XY^{(a)}_{4}$ remain invariant while $XY^{(a)}_{2}$ and $XY^{(a)}_{3}$ change their signs. We could conjecture that these combinations are invariant, which is true:
\begin{subequations}
\begin{align}
\mathbf{1}:\quad & XY^{(a)}_{1} + XY^{(a)}_{4} = XY^{(d)}_{1} + XY^{(d)}_{4}, \\
\mathbf{1^\prime}:\quad &-\left( XY^{(a)}_{2} - XY^{(a)}_{3} \right) = XY^{(d)}_{2} - XY^{(d)}_{3}.
\end{align}
\end{subequations}
Although there is some ``guessing" involved, it is relatively easy to build the dimension-one invariants if some of the generators are presented in a diagonal form, like $a$. 

So far we got that
\begin{equation}
\begin{pmatrix}
1 & 0 & 0 & 1 \\
0 & 1 & -1 & 0 \\
\multicolumn{4}{c}{\dots} \\
\multicolumn{4}{c}{\dots} \\
\end{pmatrix} \begin{pmatrix}
x_1 y_1 \\
x_1 y_2 \\
x_2 y_1 \\
x_2 y_2
\end{pmatrix} = \begin{pmatrix}
x_1 y_1 + x_2 y_2 \\
x_1 y_2 - x_2 y_1 \\
\dots \\
\dots
\end{pmatrix}.
\end{equation}
Since different irreducible representations should be orthogonal to each other, we could write the remaining combinations as
\begin{equation}
\begin{pmatrix}
1 & 0 & 0 & 1 \\
0 & 1 & -1 & 0 \\
a_1 & 0 & 0 & -a_1 \\
0 & a_2 & a_2 & 0 \\
\end{pmatrix} \begin{pmatrix}
x_1 y_1 \\
x_1 y_2 \\
x_2 y_1 \\
x_2 y_2
\end{pmatrix} = \begin{pmatrix}
x_1 y_1 + x_2 y_2 \\
x_1 y_2 - x_2 y_1 \\
a_1 x_1 y_1 - a_1 x_2 y_2 \\
a_2 x_1 y_2 + a_2 x_2 y_1
\end{pmatrix},
\end{equation}
where 
\begin{equation}
\begin{pmatrix}
a_1 x_1 y_1 - a_1 x_2 y_2 \\
a_2 x_1 y_2 + a_2 x_2 y_1
\end{pmatrix}
\end{equation}
should transform as $\mathbf{2}$, see eq.~\eqref{Eq:S3_diff_tensor_pr}. We need to find an invariant combination that would satisfy  
\begin{subequations}
\begin{align}
a_{ij}^k XY^{(a)}  ={}& a (a_{ij}^k x_i y_j), \\
a_{ij}^k XY^{(d)}  ={}& d (a_{ij}^k x_i y_j),
\end{align}
\end{subequations}
where we have slightly abused the dimensionality of the $a_{ij}^k$ coefficient on the right side, due to the mapping of $(i,\,j)\to i^\prime$, which is given by $f(i,j)=i+j+|i-1|-1$. The ``$k$" variable stands for the upper/lower components of a new irreducible doublet. By solving the system of equations we get
\begin{subequations}
\begin{align}
\begin{pmatrix}
XY^{(a)}_2 + XY^{(a)}_3 \\
XY^{(a)}_1 - XY^{(a)}_4
\end{pmatrix} ={}& D_2(a) \begin{pmatrix}
x_1 y_2 + x_2 y_1 \\
x_1 y_1 - x_2 y_2
\end{pmatrix}, \\
\begin{pmatrix}
XY^{(d)}_2 + XY^{(d)}_3 \\
XY^{(d)}_1 - XY^{(d)}_4
\end{pmatrix} ={}& D_2(d) \begin{pmatrix}
x_1 y_2 + x_2 y_1 \\
x_1 y_1 - x_2 y_2
\end{pmatrix}.
\end{align}
\end{subequations}

Now we need to consider the products of the one-dimensional representations, which shall be denoted $z$, with the two-dimensional representation, $X$. For the trivial representation we have,
\begin{subequations}
\begin{align}
Xz^{(a)}_i \equiv (D_2(a)X) \otimes (D_1(a)z) ={}& \begin{pmatrix}
-1 & 0 \\
 0 & 1\\
\end{pmatrix} \begin{pmatrix}
x_1 z \\
x_2 z 
\end{pmatrix} = \begin{pmatrix}
-x_1 z \\
x_2 z
\end{pmatrix},\\
Xz^{(d)}_i \equiv (D_2(d)X) \otimes (D_1(d)z) ={}& \frac{1}{2}\begin{pmatrix}
-1 & \sqrt{3} \\
 -\sqrt{3} & -1\\
\end{pmatrix} \begin{pmatrix}
x_1 z \\
x_2 z 
\end{pmatrix} = \frac{1}{2}\begin{pmatrix}
  -x_1 z_2 + \sqrt{3} x_2 z\\
 -\sqrt{3} x_1 z-x_2 z_2 \\
\end{pmatrix}.
\end{align}
\end{subequations}
Then, the invariant combination is
\begin{equation}
\mathbf{2} = \begin{pmatrix}
x_1 z \\
x_2 z
\end{pmatrix}.
\end{equation}

The final tensor product is between a pseudosinglet and a doublet (remember that $D_{1^\prime}(a)=-1$, see Table~\ref{Tab:S3_char}; a one-dimensional representation coincides with the character of the group representation),
\begin{subequations}
\begin{align}
Xz^{(a)}_i \equiv (D_2(a)X) \otimes (D_{1^\prime}(a)z) ={}& \begin{pmatrix}
1 & 0 \\
 0 & -1\\
\end{pmatrix} \begin{pmatrix}
x_1 z \\
x_2 z 
\end{pmatrix} = \begin{pmatrix}
-x_1 z \\
x_2 z
\end{pmatrix},\\
Xz^{(d)}_i \equiv (D_2(d)X) \otimes (D_{1^\prime}(d)z) ={}& \frac{1}{2}\begin{pmatrix}
-1 & \sqrt{3} \\
 -\sqrt{3} & -1\\
\end{pmatrix} \begin{pmatrix}
x_1 z \\
x_2 z 
\end{pmatrix} = \frac{1}{2}\begin{pmatrix}
  -x_1 z_2 + \sqrt{3} x_2 z\\
 -\sqrt{3} x_1 z-x_2 z_2 \\
\end{pmatrix}.
\end{align}
\end{subequations}
Then, we find that
\begin{subequations}
\begin{align}
\begin{pmatrix}
-Xz^{(a)}_2 \\
Xz^{(a)}_1
\end{pmatrix} ={}& D_2(a) \begin{pmatrix}
-x_2 z \\
x_1 z
\end{pmatrix}, \\
\begin{pmatrix}
-Xz^{(d)}_2 \\
Xz^{(d)}_1
\end{pmatrix} ={}& D_2(d) \begin{pmatrix}
-x_2 z \\
x_1 z
\end{pmatrix}.
\end{align}
\end{subequations}
Notice that in the first line we want a new doublet to be invariant under the action of $D_2(a)$, $i.e.$, ``it" does not hold information about being constructed out of the pseudosinglet representation.

We are left with the tensor products between the two one-dimensional representations. 

To sum up, for the $S_3$ group we have the following tensor products, see eq.~\eqref{Eq:S3_diff_tensor_pr}, projected onto the vector space,
\begin{subequations}\label{S3_real_rep_tensor_proj}
\begin{align}
(x)_\mathbf{1} \otimes (y)_\mathbf{1} &= (xy)_\mathbf{1},\\
(x)_\mathbf{1} \otimes (y)_\mathbf{1^\prime} &= (xy)_\mathbf{1^\prime},\\
\begin{pmatrix}
x_1 \\
x_2
\end{pmatrix}_\mathbf{2} \otimes (y)_\mathbf{1} &= \begin{pmatrix}
x_1 y \\
x_2 y
\end{pmatrix}_\mathbf{2},\\
(x)_\mathbf { 1^\prime }  \otimes (y)_\mathbf { 1^\prime } &= (xy)_\mathbf { 1 },\\
\begin{pmatrix}
x_1 \\
x_2
\end{pmatrix}_\mathbf{2}  \otimes (y)_\mathbf { 1^\prime } &= \begin{pmatrix}
- x_2 y \\
x_1 y
\end{pmatrix}_\mathbf{2},\\
\begin{pmatrix}
x_1 \\
x_2
\end{pmatrix}_\mathbf{2}\otimes \begin{pmatrix}
y_1 \\
y_2
\end{pmatrix}_\mathbf{2} &= \left(x_1 y_1+x_2 y_2\right)_{\mathbf{1}} \oplus\left(x_1 y_2-x_2 y_1\right)_{\mathbf{1}^{\prime}}\oplus\begin{pmatrix}
x_1 y_2+x_2 y_1 \\
x_1 y_1-x_2 y_2
\end{pmatrix}_\mathbf {2 } \label{Eq:S3_2x2_decomp_xy}.
\end{align}
\end{subequations}

As can be seen, projection of the tensor products onto the vector space is quite a tedious process. Tensor products for some discrete group are presented in Ref.~\cite{Ishimori:2010au}. 

For the sake of completeness, let us consider the complex $S_3$ basis. One possibility to simplify calculations is to look for a basis in which generators of dimensionality two or higher would be presented in terms of diagonal or anti-diagonal matrices. On the other hand, in the real basis of $S_3$ we did not have to consider cases where representations transformed as complex quantities. 

The generators of the real basis $g_i^{\mathbb{R}}$ and of the complex basis $g_i^{\mathbb{C}}$ are connected via a unitary transformation:
\begin{equation}
U g_i^{\mathbb{R}} U^\dagger = g_i^{\mathbb{C}},
\end{equation}
where
\begin{equation}
U= \frac{1}{\sqrt{2}} \begin{pmatrix}
1 & i \\
1 & -i
\end{pmatrix}.
\end{equation}
Then we get,
\begin{equation}\label{Eq:S3_complex_rep_gen}
\begin{aligned}
U\{e,a,da,d^2a,d,d^2\}U^\dagger = \bigg\{
& \mathcal{I}_2,\, \left(
\begin{array}{cc}
 0 & -1 \\
 -1 & 0 \\
\end{array}
\right),\,\left(
\begin{array}{cc}
 0 & -\omega ^2 \\
 -\omega  & 0 \\
\end{array}
\right),\,\\ 
&\left(
\begin{array}{cc}
 0 & -\omega  \\
 -\omega ^2 & 0 \\
\end{array}
\right),\,\left(
\begin{array}{cc}
 \omega ^2 & 0 \\
 0 & \omega  \\
\end{array}
\right),\,\left(
\begin{array}{cc}
 \omega  & 0 \\
 0 & \omega ^2 \\
\end{array}
\right)
\bigg\},
\end{aligned}
\end{equation}
where $\omega = e^{2 i \pi/3}$. Note that $\omega + \omega^2 = -1$. For the tensor product of two doublets $X$ and $Y$ we have
\begin{equation}\label{S3_D2xD2_comp}
XY^{(a)}_i \equiv (D_2(a)X) \otimes (D_2(a)Y) = \begin{pmatrix}
 0 & 0 & 0 & 1 \\
 0 & 0 & 1 & 0 \\
 0 & 1 & 0 & 0 \\
 1 & 0 & 0 & 0 \\
\end{pmatrix} \begin{pmatrix}
x_1 y_1 \\
x_1 y_2 \\
x_2 y_1 \\
x_2 y_2
\end{pmatrix} = \begin{pmatrix}
x_2 y_2 \\
x_2 y_1 \\
x_1 y_2 \\
x_1 y_1
\end{pmatrix}.
\end{equation}
With a slight abuse of notation (dropping the ``$\dagger$" symbol from the components) we have,
\begin{equation}
\begin{aligned}
& (D_2(a)X) \otimes (D_2(a)Y) = (D_2(a)X) \otimes (D_2(a^\dagger)Y^\dagger) \\ & = (D_2(a^\dagger)X^\dagger) \otimes (D_2(a)Y)= (D_2(a^\dagger)X^\dagger) \otimes (D_2(a^\dagger)Y^\dagger).
\end{aligned}
\end{equation}
While the tensor products under $b$ are:
\begin{subequations}
\begin{align}
(D_2(b)X) \otimes (D_2(b)Y) ={}& \mathrm{diag}\left( \omega,\, 1,\, 1,\, \omega^2\right) (X Y)_i,\\
(D_2(b^\dagger)X^\dagger) \otimes (D_2(b)Y) ={}& \mathrm{diag}\left(1,\, \omega^2,\, \omega,\, 1 \right) (X^\dagger Y)_i,\\
(D_2(b)X) \otimes (D_2(b^\dagger)Y^\dagger) ={}& \mathrm{diag}\left(1,\, \omega,\, \omega^2,\, 1  \right) (X Y^\dagger)_i,\\
(D_2(b^\dagger)X^\dagger) \otimes (D_2(b^\dagger)Y^\dagger) ={}& \mathrm{diag}\left( \omega^2,\, 1,\, 1,\, \omega\right) (X^\dagger Y^\dagger)_i.
\end{align}
\end{subequations}
By comparing equations that are related by conjugation we can relate $(\omega^2)^\ast = \omega$, as expected. Therefore, it is sufficient to know how the first two tensor products of the above equation transform.

Following an identical procedure to the case of the real $S_3$ representations we get:
\begin{subequations}
\begin{align}
\mathbf{1}:~& x_1 y_2 + x_2 y_1, \quad x_1^\dagger y_1 + x_2^\dagger y_2,\\
\mathbf{1^\prime}:~& x_1 y_2 - x_2 y_1, \quad x_1^\dagger y_1 - x_2^\dagger y_2,\\
\mathbf{2}:~&\begin{pmatrix}
x_2 y_2 \\
x_1 y_1
\end{pmatrix}, \quad \begin{pmatrix}
x_1^\dagger y_2 \\
x_2^\dagger y_1
\end{pmatrix}.
\end{align}
\end{subequations}
The other doublet (arising from $\mathbf{1^\prime} \otimes \mathbf{2}$) is:
\begin{equation}
\mathbf{2}:~\begin{pmatrix}
-x_1 z \\
x_2 z
\end{pmatrix}, \quad \begin{pmatrix}
-x_1^\dagger z \\
x_2^\dagger z
\end{pmatrix}.
\end{equation}

\section{Moving towards invariant scalar potentials}\label{Sec:S3_4HDM_constr}

Now, we shall try to derive an $S_3$-symmetric scalar potential. It is convenient to start by considering the $SU(2)$ singlets of the scalar fields,
\begin{equation}\label{Eq:hij_def}
h_{ij} \equiv h_i^\dagger h_j,
\end{equation}
with the scalar potential given by,
\begin{equation}
V = \mu_{ij}^2 h_{ij} + \lambda_{ijkl} h_{ij} h_{kl} + \mathrm{h.c.},
\end{equation}
where $\mu_{ij}^2$ and $\lambda_{ijkl}$ are some coefficients. Due to the Hermiticity of the scalar potential not all of these couplings are free. For the time being, not much physical insight is needed to understand the structure of the scalar potential; consider it to be a specific polynomial. For the curious reader, from a physical perspective, we are considering the $N$-Higgs-Doublet Models (NHDMs), which shall be discussed in Chapter~\ref{Ch:2HDM}. It should be noted that the summation sign $\sum$ in the above scalar potential was omitted; summation is implied over the subindices. To simplify notation, we shall assume that all coefficients are real, but it can be easily generalised to complex ones.

We need to identify how to form singlets under the $S_3$ group, acting on $h_{ij}$ and $h_{ij} h_{kl}$. We are already aware how to form singlets, $e.g.$, see eqs.~\eqref{S3_real_rep_tensor_proj}. All in all, regardless of whether we consider the real or the complex $S_3$ representation, we could generate the following $S_3$ structures from the tensor products of two components:
\begin{equation}
\begin{aligned}
&\mathbf{\mathbf{1}}:~[\mathbf{2}\otimes\mathbf{2}]_\mathbf{1},\,[\mathbf{1}\otimes\mathbf{1}]_\mathbf{1},\,[\mathbf{1}^\prime \otimes \mathbf{1}^\prime]_\mathbf{1},\\
&\mathbf{\mathbf{1}^\prime}:~[\mathbf{2}\otimes\mathbf{2}]_{\mathbf{1}^\prime},\,[\mathbf{1}\otimes\mathbf{1}^\prime]_{\mathbf{1}^\prime},\,[\mathbf{1}^\prime\otimes\mathbf{1}]_{\mathbf{1}^\prime},\\
&\mathbf{\mathbf{2}}:~[\mathbf{2}\otimes\mathbf{2}]_\mathbf{2},\,[\mathbf{1}\otimes\mathbf{2}]_\mathbf{2},\,[\mathbf{2}\otimes\mathbf{1}]_\mathbf{2},\,[\mathbf{1}^\prime \otimes\mathbf{2}]_\mathbf{2},\,[\mathbf{2}\otimes\mathbf{1}^\prime]_\mathbf{2}.
\end{aligned}
\end{equation}
Then, the $S_3$ singlets involving only two tensors can be read directly from the first line of the above equation:
\begin{equation}\label{Eq:V2_S3_vsim}
V \sim [\mathbf{2}\otimes\mathbf{2}]_\mathbf{1} +  [\mathbf{1}\otimes\mathbf{1}]_\mathbf{1} +  [\mathbf{1}^\prime \otimes \mathbf{1}^\prime]_\mathbf{1}.
\end{equation}
For the time being we shall not multiply different components by coefficients $\mu^2$ and $\lambda$. The sequence of tensor products, separated by ``+", should be understood as unique invariant elements.

When four tensors are considered, it is impossible to have a single $\mathbf{2}$, as one of the four tensors. Also, there should be an even number of $\mathbf{1^\prime}$ to form a trivial singlet. Taking into account all possible tuples, there are the following distinct quartic elements present:
\begin{equation}
\allowdisplaybreaks
\begin{aligned}
V \sim{}& ~[\mathbf{2}\otimes\mathbf{2}]_\mathbf{1}\otimes[\mathbf{2}\otimes\mathbf{2}]_\mathbf{1} + [\mathbf{2}\otimes\mathbf{2}]_{\mathbf{\mathbf{1}^\prime}}\otimes[\mathbf{2}\otimes\mathbf{2}]_{\mathbf{\mathbf{1}^\prime}} + [\mathbf{2}\otimes\mathbf{2}]_\mathbf{2}\otimes[\mathbf{2}\otimes\mathbf{2}]_\mathbf{2}\\
&+ \left( [\mathbf{2}\otimes\mathbf{2}]_\mathbf{2}\otimes[\mathbf{1}\otimes\mathbf{2}]_\mathbf{2}  + \xleftrightarrow{\mathrm{sym}} \right) +  \left( [\mathbf{2}\otimes\mathbf{2}]_\mathbf{2}\otimes[\mathbf{\mathbf{1}^\prime}\otimes\mathbf{2}]_\mathbf{2}  + \xleftrightarrow{\mathrm{sym}} \right)\\
&+ [\mathbf{2}\otimes\mathbf{2}]_\mathbf{1}\otimes[\mathbf{1}\otimes\mathbf{1}]_\mathbf{1} + [\mathbf{2}\otimes\mathbf{2}]_\mathbf{1}\otimes[\mathbf{\mathbf{1}^\prime}\otimes\mathbf{\mathbf{1}^\prime}]_\mathbf{1}\\
&+ [\mathbf{1}\otimes\mathbf{2}]_\mathbf{2}\otimes[\mathbf{2}\otimes\mathbf{1}]_\mathbf{2} + [\mathbf{\mathbf{1}^\prime}\otimes\mathbf{2}]_\mathbf{2}\otimes[\mathbf{2}\otimes\mathbf{\mathbf{1}^\prime}]_\mathbf{2}\\
&+ \left( [\mathbf{1}\otimes\mathbf{2}]_\mathbf{2}\otimes[\mathbf{1}\otimes\mathbf{2}]_\mathbf{2}  + \xleftrightarrow{\mathrm{sym}}\right)+ \left( [\mathbf{\mathbf{1}^\prime}\otimes\mathbf{2}]_\mathbf{2}\otimes[\mathbf{\mathbf{1}^\prime}\otimes\mathbf{2}]_\mathbf{2} + \xleftrightarrow{\mathrm{sym}}\right)\\
&+ [\mathbf{1}\otimes\mathbf{1}]_\mathbf{1}\otimes[\mathbf{1}\otimes\mathbf{1}]_\mathbf{1} + [\mathbf{\mathbf{1}^\prime}\otimes\mathbf{\mathbf{1}^\prime}]_\mathbf{1}\otimes[\mathbf{\mathbf{1}^\prime}\otimes\mathbf{\mathbf{1}^\prime}]_\mathbf{1}\\
&+ [\mathbf{1}\otimes\mathbf{1}]_\mathbf{1}\otimes[\mathbf{\mathbf{1}^\prime}\otimes\mathbf{\mathbf{1}^\prime}]_\mathbf{1} + [\mathbf{1}\otimes\mathbf{\mathbf{1}^\prime}]_{\mathbf{\mathbf{1}^\prime}}\otimes[\mathbf{\mathbf{1}^\prime}\otimes\mathbf{1}]_{\mathbf{\mathbf{1}^\prime}} + \left( [\mathbf{1} \otimes \mathbf{\mathbf{1}^\prime}]_{\mathbf{\mathbf{1}^\prime}} \otimes [\mathbf{1} \otimes \mathbf{\mathbf{1}^\prime}]_{\mathbf{\mathbf{1}^\prime}} + \xleftrightarrow{\mathrm{sym}} \right)\\
&+ \left( [\mathbf{2} \otimes \mathbf{2}]_{\mathbf{\mathbf{1}^\prime}} \otimes [\mathbf{1} \otimes \mathbf{\mathbf{1}^\prime}]_{\mathbf{\mathbf{1}^\prime}} + \xleftrightarrow{\mathrm{sym}} \right)\\
&+ \left\lbrace \left( [\mathbf{1} \otimes \mathbf{2}]_\mathbf{2} \otimes [\mathbf{\mathbf{1}^\prime} \otimes \mathbf{2}]_\mathbf{2} + \xleftrightarrow{\mathrm{sym}} \right) + \left( [\mathbf{1} \otimes \mathbf{2}]_\mathbf{2} \otimes [\mathbf{2} \otimes \mathbf{\mathbf{1}^\prime}]_\mathbf{2} + \xleftrightarrow{\mathrm{sym}} \right) \right\rbrace,
\end{aligned}
\end{equation}
where the $\xleftrightarrow{\mathrm{sym}}$ symbol stands for:
\begin{equation}
\begin{aligned}
~[\mathbf{2}\otimes\mathbf{2}]_\mathbf{2}\otimes[\mathbf{\mathbf{1}^\prime}\otimes\mathbf{2}]_\mathbf{2} + \xleftrightarrow{\mathrm{sym}}\,= [\mathbf{2}\otimes\mathbf{2}]_\mathbf{2}\otimes[\mathbf{\mathbf{1}^\prime}\otimes\mathbf{2}]_\mathbf{2} + [\mathbf{2}\otimes\mathbf{2}]_\mathbf{2}\otimes[\mathbf{2}\otimes\mathbf{\mathbf{1}^\prime}]_\mathbf{2},
\end{aligned}
\end{equation}
which is equivalent to the ``h.c." part, but is not automatically satisfied by considering the tensor products on their own.

We would like to note that the tensor products are associative.

Let us assign the $SU(2)$ doublets to the following representations of the $S_3$ group:
\begin{equation}
\mathbf{1}:\, h_s , \quad \mathbf{1}^\prime:\, h_a, \quad \mathbf{2}:\,   \begin{pmatrix}
h_1 \\
h_2
\end{pmatrix}.
\end{equation}
To be more precise, we are looking at the 4HDM.

The most general $S_3$-symmetric scalar potential can be written as:
\begin{equation}\label{Eq:S3_4HDM_parts}
V = V_2 + V_4^{D} +  V_4^{A} + V_4^{S} + V_4^{AS}.
\end{equation}
The bilinear part is given by:
\begin{equation}\label{Eq:S3_4HDM_V2}
V_2 = \mu_{SS}^2 h_{SS}  + {\mu}_{AA}^2 h_{AA} + \mu_{11}^2 \left( h_{11} + h_{22} \right),
\end{equation}
which can be read from eq.~\eqref{Eq:V2_S3_vsim}.

The remaining quartic parts are:
\begin{subequations}\label{Eq:S3_4HDM_V4}
\begin{align}
\begin{split}
V_4^D={}&\,\lambda_1[\mathbf{2}\otimes\mathbf{2}]_\mathbf{1}\otimes[\mathbf{2}\otimes\mathbf{2}]_\mathbf{1} + \lambda_2[\mathbf{2}\otimes\mathbf{2}]_{\mathbf{\mathbf{1}^\prime}}\otimes[\mathbf{2}\otimes\mathbf{2}]_{\mathbf{\mathbf{1}^\prime}} + \lambda_3[\mathbf{2}\otimes\mathbf{2}]_\mathbf{2}\otimes[\mathbf{2}\otimes\mathbf{2}]_\mathbf{2}\\
={}&\,\lambda_1 \left( h_{11} + h_{22} \right)^2 + \lambda_2 \left( h_{12} - h_{21} \right)^2+ \lambda_3 \left[ \left( h_{11} - h_{22} \right)^2 + \left( h_{12} + h_{21} \right)^2 \right]
\end{split}\\
\begin{split}
V_4^S={}&\,\lambda_4\left( [\mathbf{2}\otimes\mathbf{2}]_\mathbf{2}\otimes[\mathbf{1}\otimes\mathbf{2}]_\mathbf{2}  + \mathrm{h.c.} \right) + \lambda_5 [\mathbf{2}\otimes\mathbf{2}]_\mathbf{1}\otimes[\mathbf{1}\otimes\mathbf{1}]_\mathbf{1} \\
&+ \lambda_6 [\mathbf{1}\otimes\mathbf{2}]_\mathbf{2}\otimes[\mathbf{2}\otimes\mathbf{1}]_\mathbf{2}+ \lambda_7 \left( [\mathbf{1}\otimes\mathbf{2}]_\mathbf{2}\otimes[\mathbf{1}\otimes\mathbf{2}]_\mathbf{2}  +\mathrm{h.c.}  \right)\\
&+ \lambda_8[\mathbf{1}\otimes\mathbf{1}]_\mathbf{1}\otimes[\mathbf{1}\otimes\mathbf{1}]_\mathbf{1}\\
&={}\lambda_4 \left[ h_{S1} \left( h_{12} + h_{21} \right) + h_{S2} \left( h_{11} - h_{22} \right) +  \mathrm{h.c.}\right]+ \lambda_5  h_{SS}  \left( h_{11} + h_{22} \right)\\
& + \lambda_6 \left( h_{1S}  h_{S1} + h_{2S} h_{S2} \right)+ \lambda_7 \left( h_{S1}^2 + h_{S2}^2 +  \mathrm{h.c.} \right)\\
& + \lambda_8 h_{SS}^2,
\end{split} \label{Eq:S3_3HDM_S3}\\
\begin{split}
V_4^A={}&\,\tilde\lambda_4\left( [\mathbf{2}\otimes\mathbf{2}]_\mathbf{2}\otimes[\mathbf{\mathbf{1}^\prime}\otimes\mathbf{2}]_\mathbf{2}  + \mathrm{h.c.} \right) + \tilde\lambda_5 [\mathbf{2}\otimes\mathbf{2}]_\mathbf{1}\otimes[\mathbf{\mathbf{1}^\prime}\otimes\mathbf{\mathbf{1}^\prime}]_\mathbf{1}\\
& + \tilde\lambda_6 [\mathbf{\mathbf{1}^\prime}\otimes\mathbf{2}]_\mathbf{2}\otimes[\mathbf{2}\otimes\mathbf{\mathbf{1}^\prime}]_\mathbf{2}+ \tilde\lambda_7 \left( [\mathbf{\mathbf{1}^\prime}\otimes\mathbf{2}]_\mathbf{2}\otimes[\mathbf{\mathbf{1}^\prime}\otimes\mathbf{2}]_\mathbf{2}  +\mathrm{h.c.}  \right)\\
& + \tilde\lambda_8[\mathbf{\mathbf{1}^\prime}\otimes\mathbf{\mathbf{1}^\prime}]_\mathbf{1}\otimes[\mathbf{\mathbf{1}^\prime}\otimes\mathbf{\mathbf{1}^\prime}]_\mathbf{1}\\
={}&\,\tilde\lambda_4 \left[ h_{A2} \left( h_{12} + h_{21} \right) - h_{A1} \left( h_{11} - h_{22} \right) +  \mathrm{h.c.}\right]+ \tilde\lambda_5  h_{AA} \left( h_{11} + h_{22} \right)\\
& + \tilde\lambda_6 \left( h_{1A} h_{A1} + h_{2A} h_{A2} \right)+ \tilde\lambda_7 \left( h_{A1}^2 + h_{A2}^2 +  \mathrm{h.c.} \right)\\
& + \tilde\lambda_8 h_{AA}^2.
\end{split}\\
\begin{split}
V_4^{AS} ={}&\,\lambda_9[\mathbf{1}\otimes\mathbf{1}]_\mathbf{1}\otimes[\mathbf{\mathbf{1}^\prime}\otimes\mathbf{\mathbf{1}^\prime}]_\mathbf{1} + \lambda_{10}[\mathbf{1}\otimes\mathbf{\mathbf{1}^\prime}]_{\mathbf{\mathbf{1}^\prime}}\otimes[\mathbf{\mathbf{1}^\prime}\otimes\mathbf{1}]_{\mathbf{\mathbf{1}^\prime}}\\
& + \lambda_{11}\left( [\mathbf{1} \otimes \mathbf{\mathbf{1}^\prime}]_{\mathbf{\mathbf{1}^\prime}} \otimes [\mathbf{1} \otimes \mathbf{\mathbf{1}^\prime}]_{\mathbf{\mathbf{1}^\prime}} + \mathrm{h.c.} \right)+ \lambda_{12}\left( [\mathbf{2} \otimes \mathbf{2}]_{\mathbf{\mathbf{1}^\prime}} \otimes [\mathbf{1} \otimes \mathbf{\mathbf{1}^\prime}]_{\mathbf{\mathbf{1}^\prime}} + \mathrm{h.c.} \right)\\
&+  \lambda_{13}\left( [\mathbf{1} \otimes \mathbf{2}]_\mathbf{2} \otimes [\mathbf{\mathbf{1}^\prime} \otimes \mathbf{2}]_\mathbf{2} + \mathrm{h.c.} \right) + \lambda_{14}\left( [\mathbf{1} \otimes \mathbf{2}]_\mathbf{2} \otimes [\mathbf{2} \otimes \mathbf{\mathbf{1}^\prime}]_\mathbf{2} + \mathrm{h.c.} \right) \\
={}&\,\lambda_9 h_{SS}h_{AA} + \lambda_{10} h_{SA}h_{AS} \\
&+ \lambda_{11} \left(h_{SA}^2 + \mathrm{h.c.}\right)+\lambda_{12}\left[ h_{SA} \left( h_{12} - h_{21} \right) + \mathrm{h.c.}\right]\\
&+\lambda_{13} \left( h_{S1}h_{A2} - h_{S2}h_{A1} + \mathrm{h.c.} \right) + \lambda_{14} \left( h_{S1}h_{2A} - h_{S2}h_{1A} + \mathrm{h.c.} \right).
\end{split}
\end{align}
\end{subequations}

There were attempts to construct an $S_3$-symmetric (and $S_3 \times \mathbb{Z}_2$-symmetric) scalar potential in Refs.~\cite{Espinoza:2018itz,Espinoza:2020qyf}, utilising all three representations of the $S_3$ group. However, they are missing some terms. For a discussion of possible symmetries of 4HDM see Refs.~\cite{Shao:2023oxt,Shao:2024ibu}.

The whole procedure of finding the $S_3$ singlets can be simplified by utilising the projection operator (which is used to decompose a reducible representation to the irreducible ones),
\begin{equation}\label{Eq:Proj_Oper}
\mathcal{P}_a = \frac{|D_\mathrm{R}(g)|}{|\mathcal{G}|} \sum_{g} \chi^{(D_a)}(g)^\ast D_\mathrm{R}(g),
\end{equation}
where summation is over all generators of the reducible representation, and the $a$ index indicates onto which of the irreducible representations $D_\mathrm{R}$ is decomposed. This method is sometimes called reduction by idempotents since the projection operator $\mathcal{P}_a$ provides an idempotent matrix (multiplying the matrix by itself yields itself).

For example, for the tensor product of two doublets we have:
\begin{subequations}
\begin{align}
\mathcal{P}_\mathbf{1} ={}& \frac{1}{6} \left[ (1)\,\mathcal{I}_4 + (1) \left(
\begin{array}{cccc}
 \frac{3}{2} & 0 & 0 & \frac{3}{2} \\
 0 & -\frac{3}{2} & \frac{3}{2} & 0 \\
 0 & \frac{3}{2} & -\frac{3}{2} & 0 \\
 \frac{3}{2} & 0 & 0 & \frac{3}{2} \\
\end{array}
\right) + (1) \left(
\begin{array}{cccc}
 \frac{1}{2} & 0 & 0 & \frac{3}{2} \\
 0 & \frac{1}{2} & -\frac{3}{2} & 0 \\
 0 & -\frac{3}{2} & \frac{1}{2} & 0 \\
 \frac{3}{2} & 0 & 0 & \frac{1}{2} \\
\end{array}
\right) \right],\\
\mathcal{P}_{\mathbf{1^\prime}} ={}& \frac{1}{6} \left[ (1)\,\mathcal{I}_4 + (-1) \left(
\begin{array}{cccc}
 \frac{3}{2} & 0 & 0 & \frac{3}{2} \\
 0 & -\frac{3}{2} & \frac{3}{2} & 0 \\
 0 & \frac{3}{2} & -\frac{3}{2} & 0 \\
 \frac{3}{2} & 0 & 0 & \frac{3}{2} \\
\end{array}
\right) + (1) \left(
\begin{array}{cccc}
 \frac{1}{2} & 0 & 0 & \frac{3}{2} \\
 0 & \frac{1}{2} & -\frac{3}{2} & 0 \\
 0 & -\frac{3}{2} & \frac{1}{2} & 0 \\
 \frac{3}{2} & 0 & 0 & \frac{1}{2} \\
\end{array}
\right) \right],\\
\mathcal{P}_\mathbf{2} ={}& \frac{2}{6} \left[ (1)\,\mathcal{I}_4 + (0) \left(
\begin{array}{cccc}
 \frac{3}{2} & 0 & 0 & \frac{3}{2} \\
 0 & -\frac{3}{2} & \frac{3}{2} & 0 \\
 0 & \frac{3}{2} & -\frac{3}{2} & 0 \\
 \frac{3}{2} & 0 & 0 & \frac{3}{2} \\
\end{array}
\right) + (-1) \left(
\begin{array}{cccc}
 \frac{1}{2} & 0 & 0 & \frac{3}{2} \\
 0 & \frac{1}{2} & -\frac{3}{2} & 0 \\
 0 & -\frac{3}{2} & \frac{1}{2} & 0 \\
 \frac{3}{2} & 0 & 0 & \frac{1}{2} \\
\end{array}
\right) \right].
\end{align}
\end{subequations}
These matrices are not orthogonal. After orthogonalising them and projecting onto the vector space of $X$ and $Y$ we get:
\begin{subequations}
\begin{align}
\mathbf{1}:&~\frac{1}{\sqrt{2}}\left(
\begin{array}{cccc}
 1 & 0 & 0 & 1 \\
 0 & 0 & 0 & 0 \\
 0 & 0 & 0 & 0 \\
 0 & 0 & 0 & 0 \\
\end{array}
\right) \begin{pmatrix}
x_1 y_1 \\
x_1 y_2 \\
x_2 y_1 \\
x_2 y_2
\end{pmatrix} \sim{} x_1 y_1 + x_2 y_2,\\
\mathbf{1^\prime}:&~\frac{1}{\sqrt{2}}\left(
\begin{array}{cccc}
 0 & 0 & 0 & 0 \\
 0 & 1 & -1 & 0 \\
 0 & 0 & 0 & 0 \\
 0 & 0 & 0 & 0 \\
\end{array}
\right) \begin{pmatrix}
x_1 y_1 \\
x_1 y_2 \\
x_2 y_1 \\
x_2 y_2
\end{pmatrix} \sim{} x_1 y_2 - x_2 y_1,\\
\mathbf{2}:&~\frac{1}{\sqrt{2}}\left(
\begin{array}{cccc}
 1 & 0 & 0 & -1 \\
 0 & 1 & 1 & 0 \\
 0 & 0 & 0 & 0 \\
 0 & 0 & 0 & 0 \\
\end{array}
\right) \begin{pmatrix}
x_1 y_1 \\
x_1 y_2 \\
x_2 y_1 \\
x_2 y_2
\end{pmatrix} \sim{} \begin{pmatrix}
x_1 y_2 + x_2 y_1 \\
x_1 y_1 - x_2 y_2
\end{pmatrix},
\end{align}
\end{subequations}
which is consistent with eq.~\eqref{Eq:S3_2x2_decomp_xy} up to the factor of $1/\sqrt{2}$.

This method can be generalised to get singlets from different combinations of tensor products. For example, in order to get the scalar potential of eq.~\eqref{Eq:S3_4HDM_parts}, we need to construct four-dimensional representations, $D_\mathrm{R} = \mathbf{1} \oplus \mathbf{1^\prime} \oplus \mathbf{2}$, and  tensor products $V \otimes V$ and $V \otimes V \otimes V \otimes V$, where $V= \left( h_s,\, h_a,\, h_1,\, h_2 \right)^\mathrm{T}$. For the bilinear part one gets a vector of $\mathrm{dim}(V \otimes V)=16$ with three (after orthogonalisation) non-zero components, which are identical to those of eq.~\eqref{Eq:S3_4HDM_V2}. For the quartic part we have $\mathrm{dim}(V \otimes V\otimes V\otimes V)=256$ of which 160 are zero before orthogonalisation and 213 are zero after orthogonalisation. This method depends significantly on computational tools, whereas the previously discussed approach could be tackled by pen and paper. On the other hand, for the projection operator/reduction by idempotents one does not need to know how the tensor products transform, like in eq.~\eqref{S3_real_rep_tensor_proj}. But one would have to invest more time when re-writing the scalar potential in a more readable form due to the nature of the method. 

The feasibility of a particular method depends on dimensions and the assumed structure. Yet another technique for the construction of finite group invariants was covered in Refs.~\cite{Sloane:77,Patera:78}.

Let us consider an $S_3$-symmetric potential constructed out of two different doublets, $D_\mathrm{R} = \Phi_\mathbf{2} \oplus \Psi_\mathbf{2}$, where $\Phi=(h_1\,h_2)^\mathrm{T}$ and $\Psi=(h_3\,h_4)^\mathrm{T}$. It should be noted that as a result of applying symmetries, the resulting scalar potential might be symmetric under a larger group. For this specific representation, utilising the projection operator method we get:
\begin{equation}
V = V_2 + V_4^\mathbb{R} + V_4^\mathbb{C},
\end{equation}
where
\begin{subequations}
\begin{align}
V_2 ={}& \mu_{1}^2 (h_{11} + h_{22}) + \mu_{2}^2 (h_{33} + h_{44}) + \mu_{3}^2 (h_{13} + h_{24}) ,\\
\begin{split}V_4^\mathbb{R} ={}&  \lambda_1 (h_{11} + h_{22})^2 + \lambda_2 (h_{33} + h_{44})^2 + \lambda_3 (h_{12} - h_{21})^2 + \lambda_4 (h_{34} - h_{43})^2\\
& + \lambda_5 (h_{11} + h_{22})(h_{33} + h_{44}) + \lambda_6 (h_{13} + h_{24})(h_{31} + h_{42}) \\
& + \lambda_7 (h_{12} - h_{21})(h_{34} - h_{43}) + \lambda_8 (h_{14} - h_{23})(h_{32} - h_{41}) \\
& + \lambda_9 \left[ (h_{12} + h_{21})^2 +  (h_{11} - h_{22})^2 \right] + \lambda_{10} \left[ (h_{34} + h_{43})^2 +  (h_{33} - h_{44})^2 \right]\\
& + \lambda_{11} \left[ (h_{12} + h_{21})(h_{34}+h_{43}) + (h_{11}-h_{22})(h_{33}-h_{44})\right]\\
& + \lambda_{12} \left[ (h_{14} + h_{23})(h_{32}+h_{41}) + (h_{13}-h_{24})(h_{31}-h_{42})\right],
\end{split}\\
\begin{split}V_4^\mathbb{C} ={}&  \lambda_{13} (h_{13} + h_{24})^2 + \lambda_{14} (h_{14} - h_{23})^2 + \lambda_{15} \left[ (h_{14} + h_{23})^2 + (h_{13}-h_{24})^2 \right]\\
& + \lambda_{16} (h_{11} + h_{22})(h_{13} + h_{24}) + \lambda_{17} (h_{13} + h_{24})(h_{33} + h_{44})\\
& + \lambda_{18} (h_{12} - h_{21})(h_{14} - h_{23}) + \lambda_{19} (h_{14} - h_{23})(h_{34} - h_{43})\\
& + \lambda_{20} \left[ (h_{12} + h_{21})(h_{14}+h_{23}) + (h_{11}-h_{22})(h_{13}-h_{24})\right]\\
& + \lambda_{21} \left[ (h_{14} + h_{23})(h_{34}+h_{43}) + (h_{13}-h_{24})(h_{33}-h_{44})\right] + \mathrm{h.c.}
\end{split}
\end{align}
\end{subequations}

\section[Comparing different \texorpdfstring{$S_3$}{S3}-symmetric potentials]{Comparing different \boldmath$S_3$-symmetric potentials }\label{Sec:Different_S3_constr}

It is quite obvious that Physics should be basis independent, and in our case different methods should yield identical scalar potentials invariant under some group $\mathcal{G}$. So far, we have not tried to construct the $S_3$-symmetric scalar potential by applying the naive method, see the discussion of constructing invariants out of the polynomials in Section~\ref{Sec:S3_perm}. 

Assume that we want the scalar potential to be explicitly symmetric under permutations of eq.~\eqref{Eq:S3_as_permutations}, which would correspond to a reducible representation. In this case, the general approach would be to construct the most general (out of all tuples) scalar potential without any restrictions on the $\mu$ or $\lambda$ couplings. For example, the bilinear part would correspond to:
\begin{equation}
V_2 = \mu_{ii}^2 h_{ii} + \mu_{ij}^2 h_{ij} + \mathrm{h.c.}
\end{equation}
Next, one would have to go through all of the $S_3$ generators $g_i$ and solve for $\mu^2$ so that
\begin{equation}
V_2(h_1,\,h_2,\,h_3)- V_2^\prime(g_i\{h_1,\,h_2,\,h_3\} )=0\quad \forall g_i \in S_3,
\end{equation}
where $V_2^\prime$ stands for the bilinear part in the new basis.
For example, by picking 
\begin{equation}
d= \begin{pmatrix}
 0 & 1 & 0 \\
 0 & 0 & 1 \\
 1 & 0 & 0 \\
\end{pmatrix},
\end{equation}
of eq.~\eqref{Eq:S3_as_permutations} we get that the transformations rules for the $SU(2)$ singlets are:
\begin{equation}
\begin{aligned}
& h_{11} \to h_{22}, \quad h_{22} \to h_{33}, \quad h_{33} \to h_{11},\\
& h_{12} \to h_{23}, \quad h_{23} \to h_{31},\quad h_{31} \to h_{12},\\
& h_{13} \to h_{21}, \quad h_{21} \to h_{32},\quad h_{32} \to h_{13}.
\end{aligned}
\end{equation}
In general, the transformation rule is give by:
\begin{equation}
h_{ij} \to \mathcal{R}_{ik}^\ast \mathcal{R}_{jl} h_{kl},
\end{equation}
where the matrix $\mathcal{R}$ defines the set of transformation rules, \textit{i.e.}, $\mathcal{R} = g$, and summation is implied over indices $k$ and $l$.

Then, the solution to
\begin{equation}
\begin{aligned}
V_2- V_2^\prime ={}& (\mu_{11}^2 - \mu_{33}^2) h_{11} + (\mu_{22}^2 - \mu_{11}^2) h_{22} + (\mu_{33}^2 - \mu_{22}^2) h_{33},\\
& +(\mu_{12}^2 - \mu_{13}^2) h_{12} + (\mu_{23}^2 - \mu_{12}^2) h_{23} + (\mu_{13}^2 - \mu_{23}^2) h_{31},\\
& +(\mu_{13}^2 - \mu_{23}^2) h_{13} + (\mu_{12}^2 - \mu_{13}^2) h_{21} + (\mu_{23}^2 - \mu_{21}^2) h_{32}\\
={}& 0,
\end{aligned}
\end{equation}
is given by:
\begin{subequations}
\begin{align}
& \mu_{11}^2 = \mu_{22}^2 = \mu_{33}^2.\\
& \mu_{12}^2 = \mu_{23}^2 = \mu_{13}^2.
\end{align}
\end{subequations}
This procedure is identical for the quartic part of the potential.

In the reducible-triplet framework the $S_3$-symmetric scalar potential was written down by Derman~\cite{Derman:1978rx,Derman:1979nf},
\begin{subequations}\label{Eq:S3_Derman_redtr}
\begin{align}
V_2 &= -\lambda  \sum_i \phi_{ii} + \frac{1}{2} \gamma \sum_{i<j}  \left( \phi_{ij} + \mathrm{h.c.}\right), \\
\begin{split}V_4 &=  A \sum_i \phi_{ii}^2+\sum_{i<j} \left[ C \phi_{ii}  \phi_{jj}  + \bar{C}\phi_{ij}\phi_{ji} + \frac{1}{2}D\left( \phi_{ij}^2 + \mathrm{h.c.} \right)
\right] +\frac{1}{2}\sum_{i\neq j} \left(   E_1 \phi_{ii}  \phi_{ij}  + \mathrm{h.c.}\right) \\
&\quad + \frac{1}{2}\sum_{i \neq j \neq k \neq i ,j<k} \left(  E_2 \phi_{ij} \phi_{ki} +  E_3 \phi_{ii}  \phi_{kj} + E_4  \phi_{ij} \phi_{ik} + \mathrm{h.c.}
\right).\end{split}
\end{align}
\end{subequations}
Notice that due to the underlying $S_3$ symmetry, not all of the coefficients can be complex; there are two complex couplings, $E_1$ and $E_4$, which, for simplicity, can be written as
\begin{equation}
E_i \to e^{i \theta_i} E_i,\quad i=\{1,4\}.
\end{equation}
Although the notation is slightly different, it should not take much effort to understand it. The $SU(2)$ singlets are written as $\phi_{ij}$, which was done for an easier comparison and the justification will be given soon.

The above scalar potential is symmetric under all permutations of the three fields. In Section~\ref{Sec:S3_as_equilat_tri} it was discussed that $S_3 \cong D_3$. Let us now build the scalar potential in the irreducible representation. The first thing to note is that in the character table for the $S_3$ group, see Table~\ref{Tab:S3_char}, there is no irreducible triplet representation (we want to have a triplet representation in order to compare the scalar potentials) and we have already come across the reducible quadruplet representation in Section~\ref{Sec:S3_4HDM_constr}. There are only two non-trivial possibilities to build a reducible triplet representation:
\begin{subequations}
\begin{align}
D_\mathbf{3} = D_\mathbf{1} \oplus D_\mathbf{2} \equiv h_s \oplus \begin{pmatrix}
h_1 \\
h_2
\end{pmatrix}:& \quad a = \begin{pmatrix}
 1 & 0 & 0 \\
 0 & -1 & 0 \\
 0 & 0 & 1 \\
\end{pmatrix},\qquad d = \begin{pmatrix}
 1 & 0 & 0\\
 0 &  -\frac{1}{2} & \frac{\sqrt{3}}{2} \\
 0 &  -\frac{\sqrt{3}}{2} & -\frac{1}{2}
\end{pmatrix},\\
D_\mathbf{3^\prime} = D_\mathbf{1^\prime} \oplus D_\mathbf{2} \equiv h_a \oplus \begin{pmatrix}
h_1 \\
h_2
\end{pmatrix}:& \quad a = \begin{pmatrix}
 -1 & 0 & 0 \\
 0 & -1 & 0 \\
 0 & 0 & 1 \\
\end{pmatrix},\quad d = \begin{pmatrix}
 1 & 0 & 0\\
 0 &  -\frac{1}{2} & \frac{\sqrt{3}}{2} \\
 0 &  -\frac{\sqrt{3}}{2} & -\frac{1}{2}
\end{pmatrix},
\end{align}
\end{subequations}
where the only difference is in the first element of the generator $a$. For the $D_\mathbf{3}$ representation the invariant scalar potential is:
\begin{subequations}\label{Eq:S3_pot_s12}
\begin{align}
V_2^S ={}& \mu_{SS}^2 h_{SS} + \mu_{11}^2 \left( h_{11} + h_{22} \right),\\
\begin{split}
V_4^S={}&\,\lambda_1 \left( h_{11} + h_{22} \right)^2 + \lambda_2 \left( h_{12} - h_{21} \right)^2+ \lambda_3 \left[ \left( h_{11} - h_{22} \right)^2 + \left( h_{12} + h_{21} \right)^2 \right] \\
&+ \left\lbrace \lambda_4 \left[ h_{S1} \left( h_{12} + h_{21} \right) + h_{S2} \left( h_{11} - h_{22} \right)\right] +  \mathrm{h.c.} \right\rbrace + \lambda_5  h_{SS}  \left( h_{11} + h_{22} \right)\\
& + \lambda_6 \left( h_{1S}  h_{S1} + h_{2S} h_{S2} \right) +  \left[\lambda_7 \left( h_{S1}^2 + h_{S2}^2\right) +  \mathrm{h.c.}\right] + \lambda_8 h_{SS}^2,
\end{split}
\end{align}
\end{subequations}
while for the $D_\mathbf{3^\prime}$ representation the invariant scalar potential is:
\begin{subequations}\label{Eq:S3_pot_a12}
\begin{align}
V_2^A ={}& {\mu}_{AA}^2 h_{AA} + \mu_{11}^2 \left( h_{11} + h_{22} \right),\\
\begin{split}
V_4^A={}&\,\lambda_1 \left( h_{11} + h_{22} \right)^2 + \lambda_2 \left( h_{12} - h_{21} \right)^2+ \lambda_3 \left[ \left( h_{11} - h_{22} \right)^2 + \left( h_{12} + h_{21} \right)^2 \right] \\
&+ \left\lbrace \lambda_4 \left[ h_{A2} \left( h_{12} + h_{21} \right) - h_{A1} \left( h_{11} - h_{22} \right)\right] +  \mathrm{h.c.}\right\rbrace + \lambda_5  h_{AA} \left( h_{11} + h_{22} \right)\\
& + \lambda_6 \left( h_{1A} h_{A1} + h_{2A} h_{A2} \right)+ \left[\lambda_7 \left( h_{A1}^2 + h_{A2}^2\right) +  \mathrm{h.c.}\right]  + \lambda_8 h_{AA}^2.
\end{split}
\end{align}
\end{subequations}
Notice that both of these are consistent with eqs.~\eqref{Eq:S3_4HDM_V2} and \eqref{Eq:S3_4HDM_V4}.

The only difference between these scalar potentials, apart from re-labeling the indices, is the $\lambda_4$ term:
\begin{subequations}\label{Eq:Relating_S3S_S3A}
\begin{align}
D_\mathbf{3}\,:& \quad \lambda_4 \left[ h_{S1} \left( h_{12} + h_{21} \right) + h_{S2} \left( h_{11} - h_{22} \right) +  \mathrm{h.c.}\right],\\
D_\mathbf{3^\prime}:& \quad \lambda_4 \left[ h_{A2} \left( h_{12} + h_{21} \right) - h_{A1} \left( h_{11} - h_{22} \right) +  \mathrm{h.c.}\right].
\end{align}
\end{subequations}
Then, the transformation rules to go from one reducible triplet representation to the other one are given by:
\begin{equation}
h_s \leftrightarrow h_a, \quad h_1 \leftrightarrow h_2.
\end{equation}
The only term sensitive to the $h_1 \leftrightarrow h_2$ permutations is $\lambda_4$. As a spoiler, to be discussed in Section~\ref{Sec:S3_br_cont_symm}, removal of the $\lambda_4$ term increases the overall symmetry to that of $O(2)$.

The scalar potentials of eq.~\eqref{Eq:S3_pot_s12} and eq.~\eqref{Eq:S3_Derman_redtr} are related via:
\begin{subequations}
\begin{alignat}{3}
& \textbf{2:}\quad{}&\begin{pmatrix}
h_1 \\ h_2 
\end{pmatrix} &= \begin{pmatrix}
\frac{1}{\sqrt{2}} \left( \phi_1 - \phi_2 \right)\\
\frac{1}{\sqrt{6}} \left( \phi_1 + \phi_2 - 2\phi_3 \right)
\end{pmatrix},\\
&\textbf{1:} {}&h_S &= \frac{1}{\sqrt{3}} \left( \phi_1 + \phi_2 + \phi_3 \right).
\end{alignat}
\end{subequations}
Notice that the transformation matrix is exactly the one given by eq.~\eqref{Eq:S3_red_U_rot_tribi}, which is not a coincidence: this unitary matrix was utilised to bring the reducible triplet representation (composed of permutations of three objects) to a block-diagonal form. Then, obviously the transformation laws from one basis of the scalar potential to the other one would also depend on that same matrix.

Using the above transformations we can relate couplings in different representations:
\begin{subequations}
\begin{align}
\mu_0^2 ={}& \gamma - \lambda,\\
\mu_1^2 ={}& - \left( \frac{1}{2} \gamma + \lambda\right),\\
\lambda_1 ={}& \frac{1}{12}\left( 4A + 4C + \bar{C} + D - 4 E_1 \cos \theta_1 + E_2 - 2E_3 + E_4 \cos \theta_4 \right),\\
\lambda_2 ={}&\frac{1}{4}\left( -\bar{C} + D + E_2 - E_4 \cos \theta_4 \right),\\
\lambda_3 ={}& \frac{1}{12} \left( 2 A - C + 2 \bar{C} + 2 D - 2 E_1 \cos \theta_1 - E_2 + 2 E_3 - E_4 \cos \theta_4 \right),\\
\begin{split}\lambda_4 ={}& \frac{1}{6 \sqrt{2}} \Big( 4 A - 2 C - 2 \bar{C} - 2 D - E_1 \cos \theta_1  + E_2 + E_3 +  E_4 \cos \theta_4\\&\hspace{50pt}- 3 i \left[  E_1 \sin \theta_1 - E_4 \sin \theta_4 \right] \Big),\end{split}\\
\lambda_5 ={}& \frac{1}{6}\left( 4 A +4 C -2\bar{C} - 2 D + 2 E_1 \cos \theta_1 - 2 E_2 + E_3 - 2 E_4 \cos \theta_4 \right),\\
\lambda_6 ={}& \frac{1}{6}\left( 4 A - 2 C + 4 \bar{C} - 2 D + 2 E_1 \cos \theta_1 + E_2 - 2 E_3 - 2 E_4 \cos \theta_4 \right),\\
\begin{split} \lambda_7 ={}& \frac{1}{12} \Big( 4 A - 2 C - 2 \bar{C} + 4 D + 2 E_1 \cos \theta_1  - 2 E_2 - 2 E_3 +  E_4 \cos \theta_4\\&\hspace{40pt}- 3 i \left[ 2 E_1 \sin \theta_1 + E_4 \sin \theta_4 \right] \Big),\end{split}\\
\lambda_8 ={}& \frac{1}{3} \left( A + C + \bar{C} + D + 2  E_1 \cos \theta_1 + E_2 + E_3 +  E_4 \cos \theta_4\right),
\end{align}
\end{subequations}
where $\{\lambda_4, \, \lambda_7\} \in \mathbb{C}$. The complex $\lambda_i$ couplings could be written down as $\lambda_i = \lambda_i^\mathrm{R} + i \lambda_i ^ \mathrm{I}$. Then, $\lambda_i^\mathrm{R} = f_1\left(A,\, C,\, \bar C,\, D,\, E_1,\, E_2,\, E_3,\, E_4\right)$ and $i\lambda_i^\mathrm{I} = if_2\left(E_1,\, E_4\right)$.

What about the transformation laws from the reducible $D_\mathbf{3^\prime}$ triplet basis? By utilising the transformation of eq.~\eqref{Eq:S3_red_U_rot_tribi} we get that in the new basis the $S_3$ generators are given by:
\begin{equation}
a= \frac{1}{3}\begin{pmatrix}
 1 & -2 & -2 \\
 -2 & -2 & 1 \\
 -2 & 1 & -2 \\
\end{pmatrix}, \quad d= \begin{pmatrix}
 0 & 1 & 0 \\
 0 & 0 & 1 \\
 1 & 0 & 0 \\
\end{pmatrix},
\end{equation}
while first going from $D_\mathbf{3^\prime}$ to $D_\mathbf{3}$ by interchanging $h_1 \leftrightarrow h_2$ (not a symmetry operation) yields:
\begin{equation}
a= \begin{pmatrix}
 -1 & 0 & 0 \\
 0 & 0 & -1 \\
 0 & -1 & 0 \\
\end{pmatrix}, \quad d= \begin{pmatrix}
 0 & 0 & 1 \\
 1 & 0 & 0 \\
 0 & 1 & 0 \\
\end{pmatrix}.
\end{equation}
So, by utilisng different basis transformations we can see how the invariant scalar potentials are connected.

The only basis we have not covered is the complex $S_3$ representation, see eq.~\eqref{Eq:S3_complex_rep_gen}. By requiring invariance of $SU(2)$ singlets under the complex $S_3$ generators we get:
\begin{subequations}\label{Eq:S3_pot_C_s12}
\begin{align}
V_2^S ={}& \mu_{SS}^2 h_{SS} + \mu_{11}^2 \left( h_{11} + h_{22} \right),\\
\begin{split}
V_4^S={}&\,\lambda_1 \left( h_{11}^2 + h_{22}^2 \right) + \lambda_2 h_{12} h_{21} + \lambda_3 h_{11} h_{22}  + \left[\lambda_4   \left( h_{12} h_{1S} - h_{21} h_{2S} \right)  + \mathrm{h.c.} \right]\\
&+ \lambda_5  h_{SS}  \left( h_{11} + h_{22} \right) + \lambda_6 \left( h_{1S}  h_{S1} + h_{2S} h_{S2} \right)+ \left(\lambda_7  h_{1S} h_{2S}  +  \mathrm{h.c.} \right)\\
& + \lambda_8 h_{SS}^2,
\end{split}\label{Eq:S3_pot_C_s12_V4}
\end{align}
\end{subequations}
or
\begin{subequations}\label{Eq:S3_pot_C_a12}
\begin{align}
V_2^A ={}& \mu_{AA}^2 h_{AA} + \mu_{11}^2 \left( h_{11} + h_{22} \right),\\
\begin{split}
V_4^A={}&\,\lambda_1 \left( h_{11}^2 + h_{22}^2 \right) + \lambda_2 h_{12} h_{21} + \lambda_3 h_{11} h_{22}  + \left[\lambda_4   \left(h_{12} h_{1A} +h_{21} h_{2A} \right)  + \mathrm{h.c.} \right]\\
&+ \lambda_5  h_{AA}  \left( h_{11} + h_{22} \right) + \lambda_6 \left( h_{1A}  h_{A1} + h_{2A} h_{A2} \right)+ \left(\lambda_7  h_{1A} h_{2A}  +  \mathrm{h.c.} \right)\\
& + \lambda_8 h_{AA}^2.
\end{split}
\end{align}
\end{subequations}

In order to get the invariant scalar potential by utilising the projection operator of eq.~\eqref{Eq:Proj_Oper} it is essential to realise that the triplet representations are no longer real. For convenience, one can think of the projected representation onto the conjugated vector space as of an anti-triplet, $\overline D_\mathrm{R} = D_\mathrm{R} ^\ast$. Then  one needs to analyse the following tensor products:
\begin{equation}
\overline D_\mathrm{R} \otimes D_\mathrm{R} ~\text{  and  }~ \overline D_\mathrm{R} \otimes D_\mathrm{R} \otimes \overline D_\mathrm{R} \otimes D_\mathrm{R}.\end{equation}
For example, for the complex representation $D_\mathbf{3} = (h_1\, h_2)^\mathrm{T} \oplus h_S$ of $S_3$, the latter projection of the tensor product yields the following vector of $S_3$ singlets:
\begin{equation}
\mathcal{P}_\mathbf{1} \begin{pmatrix}
h_{11} h_{11} \\
h_{11} h_{12} \\
h_{11} h_{1S} \\
\dots \\
h_{S1} h_{SS} \\
h_{S2} h_{SS} \\
h_{SS} h_{SS}
\end{pmatrix} \sim \begin{pmatrix}
h_{11}^2 + h_{22}^2 \\
h_{11} h_{22} \\
(h_{11} + h_{22}) h_{SS} \\
h_{12} h_{1S} - h_{21} h_{2S} \\
h_{12} h_{21} \\
h_{12} h_{S2} - h_{21} h_{S1} \\
h_{1s} h_{2S} \\
h_{1S} h_{S1} + h_{2S} h_{S2} \\
h_{21} h_{S1} - h_{12} h_{S2} \\
h_{S1} h_{S2} \\
h_{SS}^2
\end{pmatrix},
\end{equation}
where, for simplicity, we dropped the numerical indices, since those can be later absorbed by coefficients. Mind that in order to get simpler analytic expressions $\mathcal{P}_\mathbf{1}$ should be orthogonalised.

This vector should then be compared to the quartic scalar potential of eq.~\eqref{Eq:S3_pot_C_s12_V4}. When comparing, only the $\lambda_4$ term might cause some issues. Assume some dummy couplings $c_i$. Then we have:
\begin{equation}
\begin{aligned}
{}&c_1 ( h_{12} h_{1S} - h_{21} h_{2S}  ) + c_2 (h_{12} h_{S2} - h_{21} h_{S1}) + c_3 ( h_{21} h_{S1} - h_{12} h_{S2}  ) = \\
&\quad =  \left[ h_{12} h_{1S} ( c_1 ) - h_{21} h_{2S} (c_1) \right]  +  \left[ h_{21} h_{S1} (c_3 - c_2)  - h_{12} h_{S2} (c_3 - c_2) \right].
\end{aligned}
\end{equation}
Due to the hermiticity of the scalar potential one needs to require $c_3 - c_2 = c_1^\ast$.

At the end of the day, as expected, different methods of constructing invariants might have yielded different forms, but those, as was shown, are all connected by unitary transformations.

\chapter{The two-Higgs-doublet model}\label{Ch:2HDM}

Despite the fact that the properties of the 125 GeV state discovered in 2012~\cite{ATLAS:2012yve,CMS:2012qbp} are in experimental agreement with the SM Higgs predictions~\cite{CMS:2022dwd,ATLAS:2022vkf}, there are compelling reasons to consider multi-Higgs extensions of the SM. One of the most scrutinised BSM theories is the two-Higgs-doublet model (2HDM)~\cite{Lee:1973iz,Gunion:1989we,Branco:2011iw}, where a second $SU(2)$ scalar doublet is added to the SM case. In this Chapter we shall focus on the 2HDM and try to motivate why it might be beneficial to extend the scalar EW sector. As a matter of fact, there are several advantages and motivations to extend the scalar sector. From the mathematical point of view such extension would be considered to be one of the simplest BSM theories due to the nature of scalars. Apart from that, in many BSM models additional scalar fields arise naturally, \textit{e.g.} in Supersymmetry~\cite{Martin:1997ns,Djouadi:2005gj}. From the physical point of view, such extension could tackle several shortcomings of the SM, though not all at once and simultaneously. For example, in contrast to the SM, the 2HDM can accommodate the DM candidate if the additional scalar doublet is stabilised by an underlying symmetry~\cite{Deshpande:1977rw,Silveira:1985rk,Barbieri:2006dq,LopezHonorez:2006gr}. Another aspect, which the 2HDM can address, is the deficiency of CP violation sources in the SM. The new source of CP violation can be expressed in terms of spontaneous CP violation~\cite{Lee:1973iz,Branco:1980sz,Branco:1985aq} or explicit CP violation~\cite{Weinberg:1990me,Pilaftsis:1999qt}. In contrast to the SM~\cite{Gavela:1993ts,Farakos:1994kx,Gavela:1994dt}, CP violation in the 2HDM can contribute to the EW baryogenesis~\cite{Turok:1990zg,Funakubo:1993jg,Joyce:1994zt,Cline:1995dg,Cline:1996mga,Fromme:2006cm}.

Symmetries play an essential role in the construction of the QFT models. The earliest encounter with symmetries, in the framework of 2HDMs, was in the context of the Peccei-Quinn $U(1)_{PQ}$ theory~\cite{Peccei:1977hh,Peccei:1977ur,Weinberg:1977ma,Wilczek:1977pj} and in the context of the $\mathbb{Z}_2$ stabilisation of the 2HDM, which is known as the Inert Doublet Model (IDM)~\cite{Deshpande:1977rw}, which can address a plethora of phenomena~\cite{Tao:1996vb,Barbieri:2000gf,Ma:2006km,Barbieri:2006dq,Ma:2006fn,Casas:2006bd,Hambye:2007vf}. A considerable effort was put into identifying possible symmetries of the 2HDM scalar sector from different perspectives~\cite{Pomarol:1993mu,Ivanov:2005hg,Ivanov:2006yq,Gerard:2007kn,Ivanov:2007de,Ferreira:2009wh,Ferreira:2010yh,Battye:2011jj,Pilaftsis:2011ed,BhupalDev:2014bir,Pilaftsis:2016erj,Haber:2018iwr,Darvishi:2019dbh,Bento:2020jei,Ferreira:2020ana,Ferreira:2022gjh,Ferreira:2023dke,Doring:2024kdg,Trautner:2025yxz}. It should be noted that though all these symmetries hold at low energies, they are expected to be broken at high energies~\cite{Kallosh:1995hi,Banks:2010zn,Harlow:2018jwu,Harlow:2018tng}, \textit{e.g.}, by quantum (whatever it might be) gravity. Apart from that, the symmetry-stabilised 2HDMs may be required to be broken at low energies attributed to unphysical results, like predicting massless physical scalars or predicting wrong properties of fermions, possibly in the context of expanding the underlying symmetry to the Yukawa sector.

\section{The scalar potential}

In the 2HDM, a second Higgs doublet with the same quantum numbers as the first one is added, \textit{i.e.}, it transforms in the $(\mathbf{1},\mathbf{2},\mathbf{1})$ representation  under the $\mathcal{G}_{SM}$ group. The 2HDM consists of two indistinguishable, for the time being, $SU(2)$ scalar doublets,
\begin{equation}\label{Eq:NHDM_doublets_ud}
h_i = \begin{pmatrix}
h_i^+ \\
h_i^0
\end{pmatrix},
\end{equation}
for $i=\{1,2\}$, which is the Higgs flavour index. The corresponding conjugated fields are given by:
\begin{equation}
\tilde{h}_i \equiv i\sigma_2 h^{\ast}_i=\begin{pmatrix}
~~(h_i^0)^\ast\\ -h_i
\end{pmatrix}.
\end{equation}

The most general renormalisable scalar potential of the 2HDM would be a polynomial of all possible 2-tuples (bilinear terms) and 4-tuples (quartic terms). We present the scalar potential in terms of the $SU(2)$ singlets $h_{ij}=h_i^\dagger h_j$, which were defined in eq.~\eqref{Eq:hij_def}, for easier typing and to save space. The only restriction is that these combinations should be orthonormal. Due to multiple possibilities of bases choices there arises a problem of the invariant parameters. All physically observable parameters must be basis independent. The most general $SU(2)_L \times U(1)_Y$ gauge-invariant 2HDM scalar potential may be written as~\cite{Haber:1993an,Wu:1994ja,Davidson:2005cw} (for easier understanding the overall numerical coefficients were absorbed into the couplings, so there is a difference with the provided literature):
\begin{equation}\label{Eq:V_2HDM_generic}
\begin{aligned}
V ={}& \mu_{11}^2 h_{11} + \mu_{22}^2 h_{22} + \left( \mu_{12}^{2} h_{12} + \mathrm{h.c.} \right)\\
&+ \lambda_1 h_{11}^2 + \lambda_2 h_{22}^2 + \lambda_3 h_{11} h_{22} + \lambda_4 h_{12}h_{21}\\
&+ \left(\lambda_5 h_{12}^2 + \lambda_6 h_{11}h_{12} + \lambda_7 h_{12}h_{22} + \mathrm{h.c.} \right),
\end{aligned}
\end{equation}
where $\{m_{11}^2,\,m_{22}^2,\,\lambda_1,\,\lambda_2,\,\lambda_3,\,\lambda_4\} \in \mathbb{R}$ and, in general, $\{m_{12}^2,\,\lambda_5,\,\lambda_6,\,\lambda_7\} \in \mathbb{C}$. The latter complex parameters can give rise to CP violation in the Higgs sector. Presence of the complex parameters is necessary but not sufficient for the CP violation in the scalar sector.  In total, there are six plus eight real parameters, which adds up to fourteen parameters. However, it should be noted that due to the freedom of the $SU(2)$ basis change, not all of theses degrees have physical significance, and it is possible to absorb three parameters, resulting in eleven free physical parameters.

A model with just eleven free parameters comes at an expensive computational cost. A numerical scan with fixed absolute parameter values, taking into account both signs, $\pm a$, would result in $2^{11}=2\,048$ data points, while a scan over two different values would yield $4^{11} = 4\,194\,304$ data points. When moving to the general NHDMs, the scalar potential allows for even more terms. In order to write the renormalisable scalar potential in a compact way, one can introduce a set of linearly independent operators:
\begin{equation}\label{Eq:oper}
\Phi = \left\lbrace  h_{ii},\,\frac{1}{2}\left( h_{ij} + h_{ji} \right),\,-\frac{i}{2}\left( h_{ij} - h_{ji} \right) \right\rbrace.
\end{equation}
While constructing the NHDM scalar potential from the above operators, one would not be double counting the complex parameters.

The number of free real parameters in the NHDM is given by~\cite{Olaussen:2010aq}:
\begin{equation}\label{Eq:NHDM_param_tot}
N_{tot}  = \frac{1}{2} N^2 (N^2 + 3),
\end{equation}
where $N$ is the number of the $SU(2)$ doublets. Substituting $N=2$ we get fourteen real parameters for the 2HDM and for the 3HDM the number of parameters increases to fifty-four. By utilising the $SU(N)$ transformation, and rotating away redundant degrees, the number of free parameters of eq.~\eqref{Eq:NHDM_param_tot} is reduced by $N^2-1$ to:
\begin{equation}
N_{ind} = \frac{1}{2} (N^2 + N^4 + 2),
\end{equation}
while for the real scalar potential the number of independent free parameters is:
\begin{equation}
N_{\mathbb{R}{-}ind} = \frac{1}{4} N (3N + N^3 + 4).
\end{equation}

It should be obvious that the predictability of the NHDMs is quickly lost once more free parameters are introduced. That is why symmetries play an important role---not only invariance of the scalar potential under a symmetry leads to specific physics, but also the underlying symmetries introduce a ``natural" way to control the total number of free parameters.

Before moving on to the scalar potential, let us mention several different notations. We have already seen one in eq.~\eqref{Eq:V_2HDM_generic}, which lists all terms explicitly. This might not be the most optimal from the perspective of the available terms or the properties studied. An alternative compact form was presented in Refs.~\cite{Botella:1994cs,Branco:1999fs} to study basis invariants and transformations:
\begin{equation}
V = \mu_{ij}^2 \,h_{ij} + \lambda_{ijkl} \, h_{ij}h_{kl},
\end{equation}
where Hermiticity of the potential implies:
\begin{subequations}
\begin{align}
\mu_{ij}^2 ={}& (\mu_{ji}^2)^\ast,\\
\lambda_{ijkl} = \lambda_{klij} ={}& \lambda_{jilk}^\ast = \lambda_{lkji}^\ast.
\end{align}
\end{subequations}
As a rule of thumb, one should always be careful with the summation of indices, not to double count them, and the overall signs. For example, since $\lambda_{ijkl} = \lambda_{klij}$ we could have introduced a factor of $1/2$ or have restricted the summation over indices.

Yet another way of presenting the scalar potential relies on $SU(2)$ being the double-cover of the $SO(3)$ rotation matrices,
\begin{equation}
\phi:~ SU(2) \to SO(3),
\end{equation}
the homomorphism of which may be presented in terms of
\begin{equation}
\mathcal{R}(U)_{i j}=\frac{1}{2} \operatorname{Tr}\left(\sigma_i U^\dagger \sigma_j U\right),
\end{equation}
where $\mathcal{R}\in SO(3)$ and $U\in SU(2)$. In the adjoint representation of $SU(2)$ the combinations of the scalar doublets,
\begin{subequations}\label{Eq:2HDM_bilinear}
\begin{align}
r_0 ={}& h_{11} + h_{22},\\
r_1 ={}& \mathbb{R}\mathrm{e}(h_{12}),~  r_2 ={} \mathbb{I}\mathrm{m}(h_{12}),~  r_3 ={} h_{11} - h_{22},
\end{align}
\end{subequations}
transform as a singlet and a triplet, $r_i = h_a^\dagger (\sigma_i)_{ab} h_b$. Such mapping is only valid for the 2HDM, \textit{e.g.}, see Refs.~\cite{Nishi:2006tg,Nishi:2007nh,Ivanov:2010ww,Maniatis:2014oza}. In terms of the real-valued bilinears, the 2HDM scalar potential can be expressed as:
\begin{equation}
\begin{aligned}
V ={}& M_0 r_0+M_i r_i+\Lambda_0 r_0^2+L_i r_0 r_i+\Lambda_{i j} r_i r_j\\
  ={}& M_\mu r_\mu + L_{\mu\nu} r_\mu r_\nu.
\end{aligned}
\end{equation}
Notice that in this notation, the scalar potential is quadratic, rather than quartic. The bilinear formalism was initially introduced for the studies of the minima~\cite{Velhinho:1994np} and was further developed in Refs.~\cite{Nagel:2004sw,Ivanov:2005hg,Maniatis:2006fs,Ivanov:2006yq,Maniatis:2006jd,Maniatis:2007vn,Ivanov:2007de,Nishi:2007dv}.

\section{Counting physical states}\label{Sec:Phys_Spectrum_2HDM}

By introducing a second doublet, as in eq.~\eqref{Eq:NHDM_doublets_ud}, there is no clear way to distinguish two doublets, since both $h_1$ and $h_2$ are interchangeable, $h_1 \leftrightarrow h_2$. In the SM, the key role of the Higgs mechanism was SSB. Now, in general, the $SU(2)$ scalar doublet can be decomposed into the fields:
\begin{equation}\label{Eq:Def_doublet}
h_i = \begin{pmatrix}
h_i^+ \\
\frac{1}{\sqrt{2}} (v_i + \eta_i + i \chi_i)
\end{pmatrix}.
\end{equation}
In the SM, only the $\eta$ field was shown to be physical, which is the Higgs boson. In the 2HDM there are now eight fields (depending on how one counts the upper component). As in the SM case, three fields are ``eaten" to give masses to the weak bosons. Hence we are left with five fields. One of these fields should be associated with the Higgs boson. What are the remaining fields? As we shall see, in the 2HDM there is a charged physical scalar and three neutral physical scalars, one of which is the SM Higgs boson.

Assume that the two doublets develop non-zero vevs:
\begin{equation}\label{Eq:2HDM_vevs_N}
\left\langle h_1 \right\rangle = \frac{1}{\sqrt{2}} \begin{pmatrix}
0 \\ v_1
\end{pmatrix}, \quad \left\langle h_2 \right\rangle = \frac{1}{\sqrt{2}} \begin{pmatrix}
0 \\ v_2
\end{pmatrix},
\end{equation}
which we shall assume to be real.

The Higgs field has a vev of around 246 GeV. This value is fixed by the Fermi coupling constant; we shall not consider time variations of the vev~\cite{Dixit:1987at,Kujat:1999rk,Uzan:2002vq,Yoo:2002vw,Casadio:2007ip,Fung:2021wbz}. Any departure from this value, or small perturbation, would be ``visible" by experiments. Therefore, we need to require that
\begin{equation}
v_1^2 + v_2^2 = v^2 \approx 246^2~ \text{GeV}^2.
\end{equation}
The vevs can be parameterised in terms of the $\beta$ angle as
\begin{equation}
v_1^2 + v_2^2 = v^2 (\cos^2\beta + \sin^2 \beta).
\end{equation}
By inverting, we get:
\begin{equation}\label{Eq:Tan_b}
\tan \beta = \frac{v \sin \beta}{v  \cos \beta} = \frac{v_2}{v_1}.
\end{equation}
Without loss of generality, the $\beta$ angle can be fixed to the first quadrant~\cite{Carena:2002es}.

The choice of vevs in eq.~\eqref{Eq:2HDM_vevs_N} is not unique. These can be chosen to be complex,
\begin{equation}\label{Eq:2HDM_vev_C}
\left\langle h_1 \right\rangle = \frac{1}{\sqrt{2}} \begin{pmatrix}
0 \\ v_1 
\end{pmatrix}, \quad \left\langle h_2 \right\rangle = \frac{1}{\sqrt{2}} \begin{pmatrix}
0 \\ v_2 e^{i \theta}
\end{pmatrix},
\end{equation}
where due to the overall $U(1)_Y$ symmetry one of the phases was rotated away. Although there is a complex parameter, in general, this does not guarantee CP violation. Complex vevs are of interest and shall be considered in the forthcoming chapters.

Yet another possibility would be to introduce a charge-breaking vacuum,
\begin{equation}
\left\langle h_1 \right\rangle = \frac{1}{\sqrt{2}} \begin{pmatrix}
0 \\ v_1
\end{pmatrix}, \quad \left\langle h_2 \right\rangle = \frac{1}{\sqrt{2}} \begin{pmatrix}
\alpha \\ v_2
\end{pmatrix}.
\end{equation}
Due to the presence of a non-zero vev $\alpha$, the electrical charge conservation will be broken, which would lead to the photon being promoted to a massive state. We shall omit discussion of these vevs.

The first step towards the identification of the physical spectrum of the 2HDM would be to consider the extrema of the scalar potential. Let us consider the scalar potential of eq.~\eqref{Eq:V_2HDM_generic} with vacuum given by eq.~\eqref{Eq:2HDM_vevs_N}. For simplicity, assume that all couplings are real, $\{\mu_{ij}^2, \lambda_{ijkl}\} \in \mathbb{R}$. In order to get the minimum conditions of the scalar potential we need to evaluate the following derivatives:
\begin{equation}
 \frac{\partial V}{\partial \xi_i}  \bigg |_{v} = 0
\end{equation}
at the vacuum level, by sending all of the scalar fields of eq.~\eqref{Eq:Def_doublet} to zero. Here, $\xi_i$ can be chosen to be a set of real fields, $\eta_i$ or $\chi_i$, or the vevs, $v_i$. Based on the form of the scalar potential it may be beneficial to consider a mixture of derivatives. By differentiating the scalar potential with respect to the vevs we get:
\begin{subequations}
\begin{align}
\begin{split}
 \frac{\partial V}{\partial v_1}  \bigg |_{v} = {}& \mu_{11}^2 v_1 + \mu_{12}^2 v_2 + \lambda_1 v_1^3 \\
& + \frac{1}{2} \left[ 3 \lambda_6 v_1^2 + \left( \lambda_3 + \lambda_4 + 2 \lambda_5 \right) v_1 v_2 + \lambda_7 v_2^2 \right] v_2
\end{split}\\
\begin{split}
 \frac{\partial V}{\partial v_2}  \bigg |_{v} = {}& \mu_{22}^2 v_2 + \mu_{12}^2 v_1 + \lambda_2 v_2^3 \\
& + \frac{1}{2} \left[ \lambda_6 v_1^2 + \left( \lambda_3 + \lambda_4 + 2 \lambda_5 \right) v_1 v_2 + 3\lambda_7 v_2^2 \right] v_1.
\end{split}
\end{align}
\end{subequations}
It is common to solve the minimisation conditions in terms of the bilinear coefficients, $\mu_{ij}^2$. The trivial solution $v_1 = v_2 = 0$ is of no particular interest. Assuming that neither $v_1$ nor $v_2$ vanishes, the minimisation conditions are:
\begin{subequations}
\begin{align}
\mu_{11}^2 ={}& - \mu_{12}^2 \frac{v_2}{v_1}- \lambda_1 v_1^2 - \left[ 3 \lambda_6 v_1^2 + \left( \lambda_3 + \lambda_4 + 2 \lambda_5 \right) v_1 v_2 + \lambda_7 v_2^2 \right] \frac{v_2}{2 v_1},\\
\mu_{22}^2 ={}& - \mu_{12}^2 \frac{v_1}{v_2}- \lambda_2 v_2^2 - \left[ \lambda_6  v_1^2 + \left( \lambda_3 + \lambda_4 + 2 \lambda_5 \right) v_1 v_2 + 3\lambda_7 v_2^2 \right] \frac{v_1}{2 v_2}.
\end{align}
\end{subequations}
Another possible solution is given by an implementation when a single vev vanishes. For the case of $v_2=0$, the minimisation conditions are given by:
\begin{subequations}
\begin{align}
\mu_{11}^2 ={}& - \lambda_1 v_1^2,\\
\mu_{12}^2 ={}& - \frac{1}{2} \lambda_6 v_1^2.
\end{align}
\end{subequations}
The case of $v_1=0$ is equivalent to the previous case  by interchanging the relevant indices of the couplings and substituting $v_1 \leftrightarrow v_2$.

To sum up, there are three different vacuum configurations:
\begin{subequations}
\begin{align*}
&\text{Case A: } (v_1,\, v_2) = (0,\,0),\\
&\text{Case B: }  (v_1,\, v_2) = (v,\,0) \text{ or } (v_1,\, v_2) = (0,\,v),\\
&\text{Case C: }  (v_1,\, v_2).
\end{align*}
\end{subequations}

In general, $h_i$ are not the physical scalar doublets, unless the scalar states are aligned with the mass eigenstates, \textit{i.e.}, the mass-squared matrices, which we shall discuss in detail in a moment, are diagonal. When we talk about the physical scalar states, we mean the states that are associated with the mass eigenstates. However, one should be careful with this interpretation since not all physical processes are described by the mass eigenstates, \textit{e.g.}, neutrino oscillations.

In order to get masses of the scalar particles we need to evaluate the mass-squared matrix,
\begin{equation}
\mathcal{M}^2 =  \frac{\partial V}{\partial \xi_i \xi_j}  \bigg |_{v},
\end{equation}
after minimising the scalar potential. Terms which contribute to $\mathcal{M}^2$ are proportional to $\mu_{ij}^2 \xi_i \xi_j$ and $\lambda_{ijkl} v_i v_j \xi_k \xi_l$. The mass-squared matrices for Case B and Case C are block-diagonal in the basis of $\{h_i^+,\, \eta_i,\, \chi_i\}$,
\begin{equation}
\mathcal{M}^2 = \mathcal{M}^2_{h^\pm} \oplus \mathcal{M}^2_\eta \oplus \mathcal{M}^2_\chi = \mathrm{diag}\left( \mathcal{M}^2_{h^\pm},\,  \mathcal{M}^2_\eta,\, \mathcal{M}^2_\chi \right) .
\end{equation}
In general, there is mixing between $\mathcal{M}^2_\eta$ and $\mathcal{M}^2_\chi$ if coefficients are allowed to be complex.

In Case B, $(v,0)$, the mass-squared matrices are given by:
\begin{subequations}\label{Eq:M2_2HDM_CaseB}
\begin{align}
\mathcal{M}^2_{h^\pm} ={}& \begin{pmatrix}
0 & 0 \\
0 & \mu_{22}^2 + \frac{1}{2} \lambda_3 v^2
\end{pmatrix},\\
\mathcal{M}^2_{\eta} ={}& \begin{pmatrix}
2 \lambda_1 v^2 & \lambda_6 v^2 \\
\lambda_6 v^2 & \mu_{22}^2 + \frac{1}{2}\left( \lambda_3 + \lambda_4 + 2 \lambda_5 \right) v^2
\end{pmatrix},\label{Eq:M2_CaseB_Meta}\\
\mathcal{M}^2_{\chi} ={}& \begin{pmatrix}
0 & 0 \\
0 & \mu_{22}^2 + \frac{1}{2}\left( \lambda_3 + \lambda_4 - 2 \lambda_5 \right) v^2
\end{pmatrix},
\end{align}
\end{subequations}
while for Case C those are:
\begin{subequations}\label{Eq:M2_2HDM_CaseC}
\begin{align}
\mathcal{M}^2_{h^\pm} ={}& \begin{pmatrix}
-\frac{v_2}{2 v_1}(\mathcal{M}^2_{h^\pm})_{11} & \frac{1}{2}(\mathcal{M}^2_{h^\pm})_{11} \\
\frac{1}{2}(\mathcal{M}^2_{h^\pm})_{11} & -\frac{v_1}{2 v_2}(\mathcal{M}^2_{h^\pm})_{11}
\end{pmatrix},\\
\mathcal{M}^2_{\eta} ={}& \begin{pmatrix}
(\mathcal{M}^2_{\eta})_{11} & (\mathcal{M}^2_{\eta})_{12} \\
(\mathcal{M}^2_{\eta})_{12} & (\mathcal{M}^2_{\eta})_{22}
\end{pmatrix},\\
\mathcal{M}^2_{\chi} ={}& \begin{pmatrix}
- \frac{v_2}{2 v_1}(\mathcal{M}^2_{\chi})_{11} & \frac{1}{2}(\mathcal{M}^2_{\chi})_{11} \\
\frac{1}{2}(\mathcal{M}^2_{\chi})_{11} & \frac{- v_1}{2 v_2}(\mathcal{M}^2_{\chi})_{11}
\end{pmatrix},
\end{align}
\end{subequations}
where 
\begin{subequations}
\begin{align}
(\mathcal{M}^2_{h^\pm})_{11} = {}& 2 \mu_{12}^2 + \lambda_6 v_1^2 + \left( \lambda_4 + 2 \lambda_5 \right) v_1 v_2 + \lambda_7 v_2^2,\\
(\mathcal{M}^2_{\eta})_{11} = {} & \frac{1}{2 v_1} \left( -2 \mu_{12}^2 v_2 + 4 \lambda_1 v_1^3 + 3 \lambda_6 v_1^2 v_2 - \lambda_7 v_2^2 \right),\\
(\mathcal{M}^2_{\eta})_{12} = {} & \mu_{12}^2 + \frac{3}{2} \lambda_6 v_1^2 + \left( \lambda_3 + \lambda_4 + 2 \lambda_5\right) v_1 v_2 + \frac{3}{2} \lambda_7 v_2^2 ,\\
(\mathcal{M}^2_{\eta})_{22} = {} & \frac{1}{2 v_2} \left( -2 \mu_{12}^2 v_1 + 4 \lambda_2 v_2^3 - \lambda_6 v_1^3 + 3 \lambda_7 v_1 v_2^2 \right),\\
(\mathcal{M}^2_{\chi})_{11} = {} & 2 \mu_{12}^2 + \lambda_6 v_1^2 + 4 \lambda_5 v_1 v_2 + \lambda_7 v_2^2.
\end{align}
\end{subequations}
 
The form of the mass-squared matrices suggests that the scalar states are not in the physical basis. In order to get the mass-squared parameters, the relevant $\mathcal{M}_\xi^2$ need to be diagonalised.

For the real-valued scalar potential we have
\begin{equation}
\mathcal{M}_\xi^2 = \begin{pmatrix}
A & B \\
B & C
\end{pmatrix},
\end{equation}
which needs to be diagonalised. The eigenvalues of the above $\mathcal{M}_\xi^2$ are the mass-squared parameters
\begin{equation}
m_{\Xi_i}^2=\frac{1}{2}\left( A + C \pm \sqrt{\left( C - A \right)^2 + 4B^2} \right),
\end{equation}
where $\Xi_i$ are linear combinations of $\xi_i$.

The $\mathcal{M}_\xi^2$ matrix can be diagonalised by a unitary matrix $U$,
\begin{equation}
U \mathcal{M}_\xi^2  U^\dagger = \hat {\mathcal{M}_\Xi^2} = m_{\Xi_1}^2 \oplus m_{\Xi_2}^2,
\end{equation}
In the case of the real-valued $\mathcal{M}_\xi^2$ it suffices to assume that $U$ is a rotation matrix in the two-dimensional Euclidean space, \textit{e.g.},
\begin{equation}
\mathcal{R}_\theta = \begin{pmatrix}
\cos \theta & \sin \theta \\
-\sin \theta & \cos \theta
\end{pmatrix}.
\end{equation}
Then we have:
\begin{equation}
\begin{aligned}
\mathcal{R}_\theta \mathcal{M}_\xi^2 \mathcal{R}_\theta^\mathrm{T} ={}& 
\begin{pmatrix}
A \cos^2 \theta + B \sin(2 \theta) + C \sin^2 \theta & - \frac{1}{2}\left( A - C \right) \sin (2 \theta) + B \cos (2 \theta) \\
- \frac{1}{2}\left( A - C \right) \sin (2 \theta) + B \cos (2 \theta) & A \sin^2 \theta - B \sin (2 \theta) + C \cos^2 \theta
\end{pmatrix}\\
={}& \begin{pmatrix}
m_{\Xi_1}^2 & 0 \\
0 & m_{\Xi_2}^2
\end{pmatrix}.
\end{aligned}
\end{equation}
Solving for the $\theta$ angle yields
\begin{equation}
\tan (2 \theta ) = \frac{2 B }{A - C},
\end{equation}
assuming no specific ordering of the $m_{\Xi_i}^2$ masses for a free $\theta$ angle, \textit{i.e.}, either of the $\Xi_i$ states can be the lightest.

We started by writing down the mass-squared matrix for the $\xi_i$ states, which are not the mass eigenstates due to form of the mass-squared matrix $\mathcal{M}^2_\xi$. The two states can be combined into a doublet,
\begin{equation}
\xi = \begin{pmatrix}
\xi_1 \\
\xi_2
\end{pmatrix}.
\end{equation}
Since we know that $\mathcal{R}_\theta$ diagonalises $\mathcal{M}^2_\xi$, we can also relate the mass eigenstates $\Xi_i$ to the initial $\xi_i$ ones via
\begin{equation}
\Xi = \begin{pmatrix}
\Xi_1 \\
\Xi_2
\end{pmatrix}  = \mathcal{R}_\theta \xi = \begin{pmatrix}
\cos \theta & \sin \theta \\
-\sin \theta & \cos \theta
\end{pmatrix} \begin{pmatrix}
\xi_1 \\
\xi_2
\end{pmatrix} = \begin{pmatrix}
\xi_1 \cos \theta  + \xi_2 \sin \theta  \\
-\xi_1 \sin \theta  + \xi_2  \cos \theta 
\end{pmatrix},
\end{equation}
which can be inverted:
\begin{equation}
 \begin{pmatrix}
\xi_1 \\
\xi_2
\end{pmatrix} = \begin{pmatrix}
\Xi_1 \cos \theta  -\Xi_2 \sin \theta  \\
\Xi_1 \sin \theta  + \Xi_2  \cos \theta 
\end{pmatrix}.
\end{equation}

Validity of the diagonalisation procedure and identification of the states can be checked by verifying that
\begin{equation}
\begin{aligned}
\hat{\mathcal{M}_\xi^2} ={}& \mathcal{R}_\theta \mathcal{M}_\xi^2 \mathcal{R}_\theta^\dagger,\\
\Xi^\dagger \hat{\mathcal{M}_\xi^2} \Xi ={}&  \Xi^\dagger \mathcal{R}_\theta \mathcal{M}_\xi^2 \mathcal{R}_\theta^\dagger \Xi,\\
\Xi^\dagger \hat{\mathcal{M}_\xi^2} \Xi ={}&  \xi^\dagger  \mathcal{M}_\xi^2 \xi,
\end{aligned}
\end{equation} 
holds. Notice that on the left side we have the diagonalised mass-squared matrix, composed out of the eigenvalues, along with the $\Xi$ states, while on the right side we can identify the initial $\xi_i$ states and the non-diagonal $\mathcal{M}_\xi^2$.

Following an identical procedure for eqs.~\eqref{Eq:M2_2HDM_CaseB} and \eqref{Eq:M2_2HDM_CaseC} we can identify the physical content of the 2HDM.

For Case B, see eq.~\eqref{Eq:M2_2HDM_CaseB}, the mass eigenstates are:
\begin{subequations}\label{Eq:CaseB_Phys_States}
\begin{align}
G^+ ={} & h_1^+,\quad G^- = h_1^-,\\
H^+ ={} & h_2^+,\quad H^- = h_2^-,\\
h   ={} & \eta_1 \cos \alpha + \eta_2 \sin \alpha,\\
H   ={} & -\eta_1 \sin \alpha + \eta_2 \cos \alpha,\\
G^0 ={} & \chi_1,\\
A   ={} & \chi_2,
\end{align}
\end{subequations}
where we had to diagonalise only the $\eta_i$ states, see eq.~\eqref{Eq:M2_CaseB_Meta}. By comparing the elements of $\mathcal{M}^2_\xi$ of eq.~\eqref{Eq:M2_2HDM_CaseB} we can identify the mass-squared parameters to be
\begin{subequations}
\begin{align}
m_{G^\pm}^2 ={} & 0 ,\\
m_{H^\pm}^2 ={} & \mu_{22}^2 + \frac{1}{2} \lambda_3 v^2,\\
m_{h}^2     ={} & \frac{1}{4}\left[ 2 \mu_{22}^2 + \left( 4 \lambda_1 + \lambda_3 + \lambda_4 + 2 \lambda_5 \right) v^2 - \Delta \right],\\
m_{H}^2     ={} & \frac{1}{4}\left[ 2 \mu_{22}^2 + \left( 4 \lambda_1 + \lambda_3 + \lambda_4 + 2 \lambda_5 \right) v^2 + \Delta \right],\\
m_{G^0}^2   ={} & 0 ,\\
m_{A}^2     ={} & \mu_{22}^2 + \frac{1}{2}\left( \lambda_3 + \lambda_4 - 2 \lambda_5 \right) v^2,
\end{align}
\end{subequations}
where
\begin{equation}
\begin{aligned}
\Delta ={}& 4 (\mu_{22}^2)^2 + 4 \mu_{22}^2 \left( -4 \lambda_1 + \lambda_3 + \lambda_4 + 2 \lambda_5 \right) v^2 \\
& + \left[ \left( -4 \lambda_1 + \lambda_3 + \lambda_4 + 2 \lambda_5 \right)^2 + 16 \lambda_6^2 \right] v^4.
\end{aligned}
\end{equation}
In comparison to the SM Higgs sector we can spot several additional scalar states. First of all, there is a pair of physical charged scalars, $H^\pm$, which were absent in the SM. Apart from that there are three neutral states $h$, $H$ and $A$. Let us identify the CP properties of these states. We start by re-writing the $SU(2)$ scalar doublet of eq.~\eqref{Eq:Def_doublet} in terms of the physical fields, see eq.~\eqref{Eq:CaseB_Phys_States}:
\begin{subequations}
\begin{align}\label{Eq:CaseB_h1_ME}
h_1 ={}& \begin{pmatrix}
G^+ \\
\frac{1}{\sqrt{2}}\left( v + h \cos \alpha - H \sin \alpha + i G^0 \right)
\end{pmatrix},\\
h_2 ={}&  \begin{pmatrix}
H^+ \\
\frac{1}{\sqrt{2}}\left( h \sin \alpha + H \cos \alpha + i A \right)
\end{pmatrix}.\label{Eq:CaseB_h2_ME}
\end{align}
\end{subequations}
We can spot that the Goldstone bosons transform together with the vev, which is something we expect to see.

Inserting the doublets in terms of the mass eigenstates into the kinetic Lagrangian,
\begin{equation}
\mathcal{L}_K = (D_\mu h_1)^\dagger (D^\mu h_1) + (D_\mu h_2)^\dagger (D^\mu h_2),
\end{equation}
where the covariant derivative is
\begin{equation}\label{Eq:D_mu_general}
D_\mu \begin{pmatrix}
h^+_i \\ h^0_i
\end{pmatrix} = \begin{pmatrix}
\left[ \partial_\mu +\frac{i g}{2} \frac{\mathrm{c}_{2w}}{\mathrm{c}_w} Z_\mu + ie A_\mu \right] h^+_i + \frac{ig}{\sqrt{2}} W_\mu^+ h^0_i \\
\frac{ig}{\sqrt{2}} W_\mu^- h^+_i + \left[ \partial_\mu - \frac{ig}{2\mathrm{c}_w}Z_\mu \right] h_i^0
\end{pmatrix},
\end{equation}
yields:
\begin{subequations}\label{Eq:vK}
\begin{align}\label{Eq:LVVH_CaseB}
\begin{split}
\mathcal{L}_{VVH} =& \left( \frac{g}{2 \cos \theta_W}m_ZZ_\mu Z^\mu + g m_W W_\mu^+ W^{\mu-} \right) \left( h \cos \alpha - H \sin \alpha \right),
 \end{split}\\
\begin{split}\label{Eq:LVHH_CaseB}
\mathcal{L}_{VHH} =& -\frac{ g}{2 \cos \theta_W}Z^\mu \left( - \sin \alpha~h \overset\leftrightarrow{\partial_\mu} A + \cos \alpha ~ H \overset\leftrightarrow{\partial_\mu} A \right)\\
& - \frac{g}{2}\bigg\{ i W_\mu^+ \left( - \sin \alpha ~ H^- \overset\leftrightarrow{\partial^\mu} h   
+ \cos \alpha ~ H^-\overset\leftrightarrow{\partial^\mu}H + i H^- \overset\leftrightarrow{\partial^\mu} A \right) + \mathrm{h.c.} \bigg\}\\
& + \left( i e A^\mu + \frac{i g}{2} \frac{\cos (2\theta_W)}{\cos \theta_W} Z^\mu \right) \left( H^+ \overset\leftrightarrow{\partial_\mu} H^-  \right),
\end{split}\\
\begin{split}\label{Eq:LVVHH_CaseB}
\mathcal{L}_{VVHH} =& \left( \frac{g^2}{8 \cos^2\theta_W}Z_\mu Z^\mu + \frac{g^2}{4} W_\mu^+ W^{\mu-} \right) \left( h^2 + H^2 +A^2\right)\\
& + \bigg\{ \left( \frac{e g}{2} A^\mu W_\mu^+ - \frac{g^2}{2} \frac{\sin^2\theta_W}{\cos \theta_W}Z^\mu W_\mu^+ \right)\\ 
& \hspace{30pt} \times \left(  \sin \alpha ~ hH^- - \cos \alpha ~ HH^- + iAH^- \right) + \mathrm{h.c.} \bigg\}\\
&+ \left( e^2 A_\mu A^\mu + e g \frac{\cos (2\theta_W)}{\cos \theta_W}A_\mu Z^\mu + \frac{g^2}{4} \frac{\cos^2(2\theta_W)}{\cos^2\theta_W}Z_\mu Z^\mu + \frac{g^2}{2} W_\mu^- W^{\mu +} \right) \\
&\hspace{15pt}\times\left( h^-h^+ + H^-H^+ \right),
\end{split}
\end{align}
\end{subequations}
where we dropped couplings involving the Goldstone bosons, \textit{i.e.}, evaluated the kinetic Lagrangian in the unitary gauge.

The Feynman rules for the interactions of the scalars $\varphi_i$ and the gauge bosons $V_j$ are:
\begin{equation}
\begin{aligned}
\varphi_i V_j V_k & = i\, S\, g \left( \varphi_i V_j V_k \right) g^{\mu\nu},\\
\varphi_i \overset\leftrightarrow{\partial} \varphi_j V_k & = S \,g \left( \varphi_i \varphi_j V_k \right) \left( p_j - p_i \right)^\mu \mathrm{,\,for \,\, all \,\, momenta \,\, ingoing,}\\
\varphi_i \varphi_j V_k V_l & = i\, S\, g \left( \varphi_i \varphi_j V_k V_l \right) g^{\mu\nu},
\end{aligned}
\end{equation}
where $S$ is the symmetry factor $S=\Pi_i n_i!$, for $i$ identical particles of species $n$; the differently charged states count as different species. The  $g(\dots)$ are coefficients extracted from eqs.~\ref{Eq:vK}, and $p_i$ are the incoming four-momenta.

By observing eq.~\eqref{Eq:LVVH_CaseB} we can deduce that both $h$ and $H$ act as CP-even states, since those couple to $Z_\mu Z^\mu$. By checking the first line of eq.~\eqref{Eq:LVHH_CaseB} we can notice that there is a non-zero $Z h A$ vertex, which, according to the previous finding should indicate that the $Z A$ pair is CP-even, and hence $A$ is CP-odd. This observation is according to the form of the $SU(2)$ doublet in eq.~\eqref{Eq:CaseB_h2_ME} due to the presence of the imaginary unit. In some instances of the NHDMs, identification of the CP nature of the neutral states, according to the the above scheme, is more involved since those might be CP-indefinite.

To sum up, we figured out that in the 2HDM there are three neutral physical scalars. Both $h$ and $H$ turned out to be CP-even states while the $A$ scalar is a CP-odd state. The $A$ state is often referred to as a pseudoscalar, since it couples with $\gamma_5$ to fermions, which we shall discuss in Section~\ref{Sec:2HDM_Yukawa_L}. There are no physical pseudoscalars present in the SM, apart from the neutral Goldstone boson. Therefore, both the charged and the pseudoscalar states introduce different decay modes and branching ratios. An outline of possible processes can be found in Refs.~\cite{Gunion:1989we,Aoki:2009ha} and in the public code $\mathsf{2HDMC}$~\cite{Eriksson:2009ws}.

We are left only with the task to figure out how the Higgs boson looks like in the 2HDM. Let us turn our attention to eq.~\eqref{Eq:CaseB_h1_ME}. From the SM Higgs mechanism we know that the Higgs boson transforms together with the vev. Governed by this, we can postulate that the SM Higgs boson should be $h_{SM} = h \cos \alpha - H \sin \alpha$. While the 125 GeV state closely resembles the SM Higgs boson~\cite{CMS:2022dwd,ATLAS:2022vkf}, there is no firm exclusion of the NHDMs, and models with extended EW scalar sector thrive. In the context of the NHDMs, one tries to incorporate the 125 GeV into their model. One of the disadvantages of the NHDMs is that with so many free parameters of a model, in most instances, it is a tough task to completely rule out the available parameter space, based on the current experimental constraints. We shall refer to the 125 GeV state as the SM-like Higgs boson, along with requiring that its couplings to the SM sector are in line with the experimental results~\cite{ParticleDataGroup:2024cfk}.

\section{Basis transformations}

So far we have covered Case B, what about Case C? For Case C, see eq.~\eqref{Eq:M2_2HDM_CaseC}, the mass eigenstates are:
\begin{subequations}
\begin{align}
G^+ ={} & h_1^+ \cos \gamma  + h_2^+ \sin \gamma ,\quad G^- = (G^+)^\dagger,\\
H^+ ={} & - h_1^+ \sin \gamma  + h_2^+ \cos \gamma ,\quad H^- = (H^+)^\dagger,\\
h   ={} & \eta_1 \cos \alpha + \eta_2 \sin \alpha,\\
H   ={} & -\eta_1 \sin \alpha + \eta_2 \cos \alpha,\\
G^0 ={} & \chi_1 \cos \delta  + \chi_2 \sin \delta,\\
A   ={} &-\chi_1 \sin \delta  + \chi_2 \cos \delta.
\end{align}
\end{subequations}

Both $\mathcal{M}^2_{h^\pm}$ and $\mathcal{M}^2_{\chi}$ of eq.~\eqref{Eq:M2_2HDM_CaseC} can be presented as
\begin{equation}
\mathcal{M}^2_i = \frac{1}{2}(\mathcal{M}^2_i)_{11} \begin{pmatrix}
-\frac{v_2}{v_1} & 1 \\
1 & -\frac{v_1}{v_2}
\end{pmatrix}.
\end{equation}
Therefore, $\mathcal{M}^2_{h^\pm}$ and $\mathcal{M}^2_{\chi}$ can be diagonalised simultaneously. Suppose that the diagonalisation matrix is given by
\begin{equation}\label{Eq:R_CaseC_TanB}
\mathcal{R}= \frac{1}{v} \begin{pmatrix}
v_1 & v_2\\
-v_2 & v_1
\end{pmatrix},
\end{equation}
rather than by a rotation in a two-dimensional plane. This transformation yields,
\begin{equation}
\begin{aligned}
\hat{\mathcal{M}_i^2} ={}& \frac{1}{2}(\mathcal{M}^2_i)_{11} \frac{1}{v} \begin{pmatrix}
v_1 & v_2\\
-v_2 & v_1
\end{pmatrix}
\begin{pmatrix}
-\frac{v_2}{v_1} & 1 \\
1 & -\frac{v_1}{v_2}
\end{pmatrix}
\frac{1}{v} \begin{pmatrix}
v_1 & -v_2\\
v_2 & v_1
\end{pmatrix}\\
={}& \frac{1}{2}(\mathcal{M}^2_i)_{11} \begin{pmatrix}
0 & 0 \\
0 & -\frac{v^2}{v_1 v_2}
\end{pmatrix}.
\end{aligned}
\end{equation}

By parameterising the diagonalisation matrix of eq.~\eqref{Eq:R_CaseC_TanB} in terms of the $\beta$ angle of eq.~\eqref{Eq:Tan_b} we get:
\begin{equation}
\mathcal{R}_\beta = \begin{pmatrix}
\cos \beta & \sin \beta \\
-\sin \beta & \cos \beta
\end{pmatrix}.
\end{equation}
Acting with $\mathcal{R}_\beta$ on the vevs yields:
\begin{equation}
\mathcal{R}_\beta \begin{pmatrix}
v_1 \\
v_2
\end{pmatrix} = \begin{pmatrix}
\frac{v_1^2 + v_2^2}{v} \\
0
\end{pmatrix} = \begin{pmatrix}
v \\
0
\end{pmatrix}.
\end{equation}

By following the above procedure we rotated the $SU(2)$ scalar doublets into the so-called Higgs basis. The Higgs basis rotation is a transformation in the scalar field space that results in a basis where only one of the scalar doublets obtains a vev. As a result,
\begin{equation}
\left\langle H_1 \right\rangle = \frac{1}{\sqrt{2}} \begin{pmatrix}
0 \\ v
\end{pmatrix}, \quad \left\langle H_2 \right\rangle = \frac{1}{\sqrt{2}} \begin{pmatrix}
0 \\ 0
\end{pmatrix},
\end{equation}
where $H_i$ are the $SU(2)$ scalar doublets in the Higgs basis, $H_i = (\mathcal{R}_\beta)_{ij} h_j$. 

In a general sense, allowing for a complex vevs, $(v \cos \beta,\, v \sin \beta e^{i \theta})$, as in eq.~\eqref{Eq:2HDM_vev_C}, the Higgs basis transformation is given by:
\begin{equation}
\mathcal{R}_{HB} = \frac{1}{v} \begin{pmatrix}
v_1 & v_2 e^{-i \theta} \\
-v_2 & v_1 e^{-i \theta} \\
\end{pmatrix} = \begin{pmatrix}
\cos \beta & \sin \beta\,e^{-i \theta}\\
-\sin \beta & \cos \beta\,e^{-i \theta}
\end{pmatrix}.
\end{equation}
Due to the fact that we are now dealing with complex vevs, the relation $v_1^2 + v_2^2 = v^2$  becomes $|v_1|^2 + |v_2|^2 = v^2$.

The Higgs basis is not uniquely defined as $H_2$ can be rotated by an arbitrary phase,
\begin{equation}
 H_2 \to e^{i \vartheta} H_2.
\end{equation} 
As a result, the scalar doublets $H_i$ in the Higgs basis can be expressed in terms of the original $h_i$ fields as:
\begin{subequations}\label{Eq:2HDM_gen_HB_tr}
\begin{align}
H_1 ={}& \cos \beta\, h_1  + e^{-i \theta}\sin \beta\, h_2,\\
H_2 ={}& e^{i \vartheta} \left( -\sin \beta\, h_1  + e^{-i \theta} \cos \beta \, h_2 \right).
\end{align}
\end{subequations}

It is always possible to redefine the scalar doublets by an arbitrary unitary transformation. Applying a global $U$ transformation to the doublets results in:
\begin{equation}
h_i \to U_{ij} \varphi_j, \qquad h_{i}^{\dagger} \to \varphi_{j}^{\dagger}U_{ji}^{\ast},
\end{equation}
where $\varphi_i$ are the scalar doublets in the new basis. 
  
Then, the $SU(2)$ singlets transforms as:
\begin{equation}
h_{ij} = h_i^\dagger h_j \to \left( U_{ik} \varphi_k \right)^\dagger \left( U_{jl} \varphi_l \right) = \varphi_k^\dagger U_{ki}^\ast U_{jl} \varphi_l = U_{ki}^\ast U_{jl} \varphi_{kl}.
\end{equation}

Let us check what happens with the bilinear part of the scalar potential of eq.~\eqref{Eq:V_2HDM_generic} after applying the above procedure. Multiplying the above term by the bilinear coupling we have:
\begin{equation}
\begin{aligned}
\mu_{ij}^2 h_{ij} \to U_{ki}^\ast \mu_{ij}^2 U_{jl} \varphi_{kl} ={}& U_{1i}^\ast \mu_{ij}^2 U_{j1} \varphi_{11} + U_{2i}^\ast \mu_{ij}^2 U_{j2} \varphi_{22} \\
& + U_{1i}^\ast \mu_{ij}^2 U_{j2} \varphi_{12} + U_{2i}^\ast \mu_{ij}^2 U_{j1} \varphi_{21},
\end{aligned}
\end{equation}
where terms in the second line are interrelated due to the Hermiticity of the scalar potential. So, in the new basis, assuming the most general transformation, each term splits into four new terms. We can preserve the form of the initial scalar potential by collecting coefficients for common $h_{ij}$ terms. The bilinear part of the scalar potential transforms as
\begin{equation}\label{Eq:V2_basis_change}
\begin{aligned}
V_2 ={}& \mu_{11}^2 h_{11} + \mu_{22}^2 h_{22} + \left( \mu_{12}^2 h_{12} + \mathrm{h.c.} \right)\\
 \to {}& \left( U_{11}^\ast \mu_{11}^2 U_{11}  + U_{11}^\ast \mu_{12}^2 U_{21} + U_{12}^\ast (\mu_{12}^2)^\ast U_{11} + U_{12}^\ast \mu_{22}^2 U_{21}\right) \varphi_{11} \\
 & + \left( U_{21}^\ast \mu_{11}^2 U_{12}  + U_{21}^\ast \mu_{12}^2 U_{22} + U_{22}^\ast (\mu_{12}^2)^\ast U_{12} + U_{22}^\ast \mu_{22}^2 U_{22}\right) \varphi_{22}\\
 & + \left[ \left( U_{11}^\ast \mu_{11}^2 U_{12}  + U_{11}^\ast \mu_{12}^2 U_{22} + U_{12}^\ast (\mu_{12}^2)^\ast U_{12} + U_{12}^\ast \mu_{22}^2 U_{22}\right) \varphi_{12} + \mathrm{h.c.} \right]\\
\equiv{}& m_{11}^2 \varphi_{11} + m_{22}^2 \varphi_{22} + \left( m_{12}^2 \varphi_{12} + \mathrm{h.c.} \right).
\end{aligned}
\end{equation}
There are some things to note. First of all, transformations into another basis can produce new pairs of $\varphi_{ij}$, \textit{e.g.}, requiring $\mu_{12}^2=0$ does not automatically result in no terms multiplied by the $\varphi_{12}$ $SU(2)$ singlet. However, terms multiplied by $\varphi_{12}$ will be a linear combination of $m_{11}^2$ and $m_{22}^2$ in the $\varphi$ basis, \textit{i.e.}, no new degrees of freedom are introduced. In some situations, given an arbitrary scalar potential, a change of a basis might be beneficial since in the other basis redundant couplings might become obvious. Secondly, the Hermiticity of the scalar potential is preserved, provided that the basis change is given by a unitary transformation matrix, since $x + x^\ast = 2 \mathbb{R}\mathrm{e}(x)$. For example, in the first line of eq.~\eqref{Eq:V2_basis_change} the second and third terms might be complex, however
\begin{equation}
(U_{11}^\ast \mu_{12}^2 U_{21})^\dagger = U_{12}^\ast (\mu_{12}^2)^\ast U_{11},
\end{equation}
and therefore $U_{11}^\ast \mu_{12}^2 U_{21} + U_{12}^\ast (\mu_{12}^2)^\ast U_{11} \in \mathbb{R}$. However, the term multiplied by $\varphi_{12}$ depends on $U_{11}^\ast \mu_{12}^2 U_{22} + U_{12}^\ast (\mu_{12}^2)^\ast U_{12} \in \mathbb{C}$ for $U \in \mathbb{R}$. 

We have outlined the procedure for a basis transformation for the bilinear part $V_2$. This was done for simplicity. The same reasoning can be applied to the whole scalar potential. For further reading one can check Refs.~\cite{Davidson:2005cw,Haber:2006ue} for the basis transformations.

Now, after discussing how basis transformations look like, we can turn our attention back to the Higgs basis transformation. In the Higgs basis, utilising eq.~\eqref{Eq:2HDM_gen_HB_tr}, the scalar potential can be written as,
\begin{equation}
\begin{aligned}
V={} & Y_1 H_{11} +Y_2 H_{22} + \left( Y_3 e^{-i \vartheta} H_{12} + \mathrm{ h.c. }\right) \\
& + Z_1 H_{11}^2 + Z_1 H_{22}^2 + Z_3 H_{11} H_{22} + Z_4H_{12}H_{21} \\
& +\left(  Z_5 e^{-2 i \vartheta} H_{12}^2 + Z_6 e^{-i \vartheta} H_{11}H_{12} + Z_7 e^{-i \vartheta} H_{12} H_{22} + \mathrm{ h.c. }\right),
\end{aligned}
\end{equation}
which, basically, coincides with the scalar potential of eq.~\eqref{Eq:V_2HDM_generic}. Then, in the Higgs basis, solution Case C, $(v_1,\, v_2)$, coincides with solution Case B, $(v,\, 0)$, in the initial basis. 

\section{The Yukawa Lagrangian}\label{Sec:2HDM_Yukawa_L}

In contrast to the SM Yukawa Lagrangian, the 2HDM (and most NHDMs) Yukawa Lagrangian allows for some interesting, yet undesirable, properties. The main difference is that in general 2HDM allows for tree-level FCNC; in the SM diagonalisation of the fermionic mass matrices automatically resulted in no presence of tree-level FCNC. The FCNC are strongly constrained by experiments, and hence different symmetries are applied to avoid these unobserved interactions. Nevertheless, FCNC are phenomenologically accepted as long as those are severely suppressed within the framework of the studied model. On the other hand, it is interesting to note that in the 2HDM both doublets can acquire vevs. This might be the key to the problem of the fermion masses. It is unknown why there are three generations of fermions and masses of different families do not coincide, $e.g.$, $m_t/m_b \approx 41$. One of the possible solutions could be that the down-type quarks couple to one scalar doublet while the up-type type quarks couple to the other scalar doublet.

The most general gauge invariant 2HDM Yukawa Lagrangian, excluding the RH $\nu$, is:
\begin{equation}\label{T3YL}
\begin{aligned}
-\mathcal{L}_Y ={}& \overline{Q_L^0} Y_1^u  \tilde{h}_1 U_R^0 + \overline{Q_L^0} Y_1^d  h_1 D_R^0 + \overline{L_L^0} Y_1^e  h_1 E_R^0\\
& + \overline{Q_L^0}  Y_2^u \tilde{h}_2 U_R^0 + \overline{Q_L^0} Y_2^d  h_2 D_R^0 + \overline{L_L^0}  Y_2^e h_2 E_R^0 + \mathrm{h.c.}
\end{aligned}
\end{equation}
As in the SM Yukawa Lagrangian, the superscript ``0" indicates that fermions are in the weak basis. Then, for the $(v_1,\,v_2)$ vacuum configuration the mass matrix of fermions is
\begin{equation}\label{Eq:Gen_Mf_2HDM}
\mathcal{M}_f = \frac{1}{\sqrt{2}} \left( Y_1^f v_1 + Y_2^f v_2 \right).
\end{equation}
It is diagonalisable via
\begin{equation}\label{Eq:Diagonalising_Mf}
\hat{\mathcal{M}_f} = V_f^\dagger \mathcal{M}_f U_f.
\end{equation}
Expressing $Y_i^f$ in terms of the fermion mass matrix $\hat{\mathcal{M}}_f$ does not automatically diagonalise the other Yukawa coupling $Y_j^f$, leading to the emergence of FCNCs. A naive approach would be to align the Yukawa couplings, $Y_1^f \sim Y_2^f$, so that the Yukawa couplings associated with different scalar doublets would be diagonalisable simultaneously~\cite{Pich:2009sp,Ferreira:2010xe,Bijnens:2011gd,Celis:2013rcs}.

It might not be straightforward to diagonalise the most general $\mathcal{M}_f$. Assuming that the diagonalised fermionic mass matrix is real, we can construct:
\begin{equation}
\hat{\mathcal{M}}_f^2 = \hat{\mathcal{M}_f} \hat{\mathcal{M}}_f^\dagger = \left( V_f^\dagger \mathcal{M}_f U_f\right) \left( U_f^\dagger \mathcal{M}_f^\dagger V_f\right) = V_f^\dagger  \mathcal{M}_f \mathcal{M}_f^\dagger V_f,
\end{equation}
and likewise for $\hat{\mathcal{M}}_f^\dagger \hat{\mathcal{M}_f}$. Then we can define the Hermitian mass-squared matrices to be:
\begin{subequations}\label{Eq:M_Herm_f}
\begin{align}
\mathcal{H}_f = \mathcal{M}_f \mathcal{M}_f^\dagger = V_f \hat{\mathcal{M}}_f^2 V_f^\dagger,\\
\mathcal{H}_f^\dagger =  \mathcal{M}_f^\dagger \mathcal{M}_f= U_f \hat{\mathcal{M}}_f^2 U_f^\dagger.
\end{align}
\end{subequations}
By solving for the LH diagonalisation matrix, the RH one can be obtained via:
\begin{equation}
U_f = \mathcal{M}_f^\dagger V_f \hat{\mathcal{M}_f}.
\end{equation}

The natural way to deal with the tree-level FCNC is to introduce symmetries in a way that the criteria of Refs.~\cite{Glashow:1976nt,Paschos:1976ay} would be satisfied. By observing the mass matrix of eq.~\eqref{Eq:Gen_Mf_2HDM} we can deduce that the underlying symmetry should act in an opposite way on the scalar doublets. For example, by introducing the $\mathbb{Z}_2$ symmetry, acting non-trivially only on the first $SU(2)$ scalar doublet, $h_1 \to - h_1$, we would decouple it from the fermions, since neither of those would have a non-trivial charge under $\mathbb{Z}_2$. This model is known as Type-I 2HDM. In the Type-II model, which happens naturally in the Supersymmetry framework, the $\mathbb{Z}_2$ symmetry is extended to the RH down-type quarks and charged leptons, $h_1 \to - h_1,~ D_R \to - D_R,~ E_R \to - E_R$. This particular choice of charges forces these fields to appear together. As a result, $h_1$ couples to $D$ and $E$, while $h_2$ couples to $U$. The $\mathbb{Z}_2$ charges of the Type-I model combined with $E_R \to -E_R$ result in the Type-X/lepton-specific model, in which the RH leptons couple to $h_1$. While the Type-II with opposite charges for $E_R$ yields the Type-Y/flipped model. These models are summarised in Table~\ref{Tab:2HDM_NFC}. The Type-III model is reserved for the general Yukawa Lagrangian, without the $\mathbb{Z}_2$ symmetry.

{\renewcommand{\arraystretch}{1.3}
\begin{table}[htb]
\caption{ Models of Natural Flavour Conservation (NFC) as a result of the underlying $\mathbb{Z}_2$ symmetry in the 2HDM. Couplings of the RH fermions to different scalar doublets are presented. We do not consider the RH neutrinos.}
\label{Tab:2HDM_NFC}
\begin{center}
\begin{tabular}{|c|c|c|c|c|c|c|c|}\hline\hline 
$\mathbb{Z}_2$ & $h_1$ & $h_2$ & $Q_L$ & $L_L$ & $U_R$ & $D_R$ & $E_R$ \\ \hline 
Type-I    & + & - & + & + & - & - & - \\
Type-II   & + & - & + & + & - & + & +  \\
Type-X    & + & - & + & + & - & - & +  \\ 
Type-Y    & + & - & + & + & - & + & -  \\ 
\hline\hline
\end{tabular}\vspace*{-9pt}
\end{center}
\end{table}}

To make it easier to understand how the Yukawa Lagrangian looks like we provide two examples: the Type-I 2HDM Yukawa Lagrangian is
\begin{equation}
-\mathcal{L}_Y =  \overline{Q_L^0}  Y_2^u \tilde{h}_2 U_R^0 + \overline{Q_L^0} Y_2^d  h_2 D_R^0 + \overline{L_L^0}  Y_2^e h_2 E_R^0 + \mathrm{h.c.},
\end{equation}
while for Type-II 2HDM we have
\begin{equation}
-\mathcal{L}_Y = \overline{Q_L^0}  Y_2^u \tilde{h}_2 U_R^0 + \overline{Q_L^0} Y_1^d  h_1 D_R^0 + \overline{L_L^0} Y_1^e  h_1 E_R^0  + \mathrm{h.c.}
\end{equation}

In the 2HDM we have two new types of particles---the charged scalar and the pseudoscalar, which couple to fermions differently from the the discussed CP-even state in Section~\ref{Sec:SM_LY}. Since we did not cover how the Goldstone bosons coupled to the fermions in the SM, let us take a look at a simplified Yukawa Lagrangian. The procedure can then be generalised to the 2HDM Yukawa Lagrangian.

First, we shall consider interactions of the charged scalar, $h^\pm$. Assume the following Lagrangian:
\begin{equation}
\begin{aligned}
- \mathcal{L}_Y ={}&  \overline{Q_L^0}Y^u \tilde h U_R^0 + \overline{Q_L^0} Y^d h D_R^0 + \mathrm{h.c.}\\
\sim{}& \overline{D_L^0}Y^u (-h^-) U_R^0 + \overline{U_L^0} Y^d h^+ D_R^0 + \mathrm{h.c.} \\
={}& h^+ \left[ \overline{U_L^0} Y^d D_R^0 -  \overline{U_R^0} (Y^u)^\dagger D_L^0 \right] + h^- \left[ \dots \right] \\
={}& h^+ \left[ \overline{U_L} V_u^\dagger Y^d U_d D_R -  \overline{U_R} U_u^\dagger (Y^u)^\dagger V_d D_L \right] + \mathrm{h.c.}\\
={}& h^+ \left[ \overline{U_L} V_u^\dagger \left( V_d V_d^\dagger \right) Y^d U_d D_R -  \overline{U_R} U_u^\dagger (Y^u)^\dagger \left( V_u V_u^\dagger \right) V_d D_L \right] + \mathrm{h.c.}
\end{aligned}
\end{equation}
We expanded the scalar potential by accounting only for the terms proportional to the charged scalar $h^\pm$. Note that $\tilde{h} \sim (0 ~ -h^-)^\mathrm{T}$. Then, we can freely insert the $V_f V_f^\dagger$ matrix since is corresponds to the identity matrix. This procedure helps us diagonalise the mass terms, see eq.~\eqref{Eq:Diagonalising_Mf}, as well as identify the CKM matrix, $V_\mathrm{CKM} = V_u^\dagger V_d$, see eq.~\eqref{Eq:VCKM_def}. Now, recall that $P_{L,R}^2 = P_{L,R}$, see eq.~\eqref{Eq:PLR2_PLR}. Considering all these points, the above Lagrangian can be written as
\begin{equation}
-\mathcal{L}_Y = h^+ \overline{U} \left[ \frac{\sqrt{2}}{v} V_\mathrm{CKM}  \hat{\mathcal{M}_d} P_R - \frac{\sqrt{2}}{v}  \hat{\mathcal{M}_u} V_\mathrm{CKM} P_L  \right] D + \mathrm{h.c.}
\end{equation}
This is how the charged scalars interact with fermions.

Moving on, assume that we want to check interactions of a scalar, which is multiplied by the imaginary unit, $i\,a^0$. The Yukawa Lagrangian is then,
\begin{equation}
\begin{aligned}
- \mathcal{L}_Y ={}&  \overline{Q_L^0}Y^u \tilde h U_R^0 + \overline{Q_L^0} Y^d h D_R^0 + \mathrm{h.c.}\\
\sim{}& \overline{U_L^0}Y^u (-i\,a^0) U_R^0 + \overline{D_L^0} Y^d (i\,a^0) D_R^0 + \mathrm{h.c.} \\
={}& a \overline{U} \left[ -i P_R \frac{1}{v} \hat{\mathcal{M}_u} + i P_L \frac{1}{v} \hat{\mathcal{M}_u}\right] U + a \overline{D} \left[ i P_R \frac{1}{v} \hat{\mathcal{M}_d} - i P_L \frac{1}{v} \hat{\mathcal{M}_d}\right] D \\
={}& -i \frac{\hat{\mathcal{M}_u}}{v} a \overline U \gamma_5 U  + i \frac{\hat{\mathcal{M}_d}}{v} a \overline D \gamma_5 D. 
\end{aligned}
\end{equation}
Presence of the $\gamma_5$ matrix indicates that $a^0$ is a pseudoscalar, as was discussed at the end of Section~\ref{Sec:Phys_Spectrum_2HDM}. In some cases scalars can couple as CP-indefinite states, $i.e.$, with a coupling $\sim C_1 \bar f f + i C_2 \bar f \gamma_5 f$, where $C_i$ are some coefficients.

Assume that the $SU(2)$ doublets in terms of the mass eigenstates are given by
\begin{subequations}
\begin{align}
h_1=\begin{pmatrix}
G^{+} \cos \beta - H^{+} \sin \beta \\ \frac{1}{\sqrt{2}}\left( v \cos \beta + h \cos \alpha - H \sin \alpha + i \left(G^0 \cos \beta-A \sin \beta\right) \right)
\end{pmatrix},\\
h_2=\begin{pmatrix}
G^{+} \sin \beta + H^{+} \cos \beta \\ \frac{1}{\sqrt{2}}\left( v \sin \beta + h \sin \alpha + H \cos \alpha + i \left(G^0 \sin \beta + A \cos \beta\right) \right)
\end{pmatrix}.
\end{align}
\end{subequations}
Following the procedure of Ref.~\cite{Davidson:2005cw}, by expanding the Yukawa Lagrangian we get the fermionic mass matrix, see eq.~\eqref{Eq:Gen_Mf_2HDM}:
\begin{equation}
\mathcal{M}_f = \frac{v}{\sqrt{2}} \left( Y_1^f \cos \beta + Y_2^f \sin \beta \right) \equiv \frac{v}{\sqrt{2}}\,\kappa^f,
\end{equation}
parameterised in terms of $\kappa^f$. Apart from that, we can define an orthogonal combination, $\kappa^f \perp \rho^f$,
\begin{equation}
\rho^f \equiv - Y_1^f \sin \beta + Y_2^f \cos \beta.
\end{equation}
By going into the Higgs basis, $H_i$, the Yukawa Lagrangian expressed in terms of $\kappa^f$ and $\rho^f$ becomes:
\begin{equation}\label{Eq:2HDM_LY_Gen_HB}
\begin{aligned}
-\mathcal{L}_Y ={}& \overline{Q_L^0} \kappa^u  \tilde{H}_1 U_R^0 + \overline{Q_L^0} \kappa^d  H_1 D_R^0 + \overline{L_L^0} \kappa^e  H_1 E_R^0\\
& + \overline{Q_L^0} \rho^u \tilde{H}_2 U_R^0 + \overline{Q_L^0} \rho^d  H_2 D_R^0 + \overline{L_L^0}  \rho^e H_2 E_R^0 + \mathrm{h.c.}
\end{aligned}
\end{equation}
Expanding the Yukawa Lagrangian in the model-independent form yields the following interactions of the physical scalars:
\begin{equation}
\begin{split}\label{Eq:General_Ly_2HDM}
-\mathcal{L}_{Y} \sim {} & \frac{1}{\sqrt{2}} h \overline{U}\left[\kappa^u \cos (\beta-\alpha)-\rho^u \sin (\beta-\alpha)\right] U \\
& + \frac{1}{\sqrt{2}} H \overline{U}\left[\kappa^u \sin (\beta-\alpha)+\rho^u \cos (\beta-\alpha)\right] U  -\frac{i}{\sqrt{2}} A \overline{U} \gamma_5 \rho^u U \\
& + \frac{1}{\sqrt{2}} h \overline{D}\left[\kappa^d \cos (\beta-\alpha)-\rho^d \sin (\beta-\alpha)\right] D \\
& + \frac{1}{\sqrt{2}} H \overline{D}\left[\kappa^d \sin (\beta-\alpha)+\rho^d \cos (\beta-\alpha)\right] D + \frac{i}{\sqrt{2}} A \overline{D} \gamma_5 \rho^d D \\
& + \frac{1}{\sqrt{2}} h \overline{E}\left[\kappa^e \cos (\beta-\alpha)-\rho^e \sin (\beta-\alpha)\right] E \\
& + \frac{1}{\sqrt{2}} H \overline{E}\left[\kappa^e \sin (\beta-\alpha)+\rho^e \cos (\beta-\alpha)\right] E + \frac{i}{\sqrt{2}} A \overline{E} \gamma_5 \rho^e E \\
& + \left\{\overline{U}\left(V_{\mathrm{CKM}} \rho^d P_R-\rho^u V_{\mathrm{CKM}} P_L\right) D H^{+} + \overline{N} \rho^e P_R E H^{+} + \mathrm{h.c. }\right\} .
\end{split}
\end{equation}
We recall that in the Higgs basis the vev is aligned with a single scalar doublet. As a result we have,
\begin{equation}
\kappa^f = \frac{\sqrt{2}}{v} \hat{\mathcal{M}_f}.
\end{equation}
Relations between the Yukawa couplings are summarised in Table~\ref{Tab:2HDM_NFC_couplings}. For a slightly alternative parameterisation consider Ref.~\cite{Aoki:2009ha}.

{\renewcommand{\arraystretch}{1.3}
\begin{table}[htb]
\caption{ Rotated Yukawa couplings in the 2HDM of NFC. The $\mathbb{Z}_2$ charges are assigned according to Table~\ref{Tab:2HDM_NFC}. These values are to be inserted in the model-independent Yukawa Lagrangian of eq.~\eqref{Eq:General_Ly_2HDM}.}
\label{Tab:2HDM_NFC_couplings}
\begin{center}
\begin{tabular}{|c|c|c|c|}\hline\hline 
          & $\rho^u$ & $\rho^d$ & $\rho^e$  \\ \hline 
Type-I    & $\kappa^u \cot \beta$ & $\kappa^d \cot \beta$ & $\kappa^e \cot \beta$ \\
Type-II   & $\kappa^u \cot \beta$ & $-\kappa^d \tan \beta$ & $-\kappa^e \tan \beta$ \\
Type-X    & $\kappa^u \cot \beta$ & $-\kappa^d \tan \beta$ & $\kappa^e \cot \beta$ \\
Type-Y    & $\kappa^u \cot \beta$ & $\kappa^d \cot \beta$ & $-\kappa^e \tan \beta$ \\
\hline\hline
\end{tabular}\vspace*{-9pt}
\end{center}
\end{table}}

\section{Symmetries of the 2HDM}

In Chapter~\ref{Sec:Different_S3_constr} we got acquainted with the construction of the $S_3$-symmetric 3HDM. Now, we turn our attention to possible symmetries of the 2HDM. The first thing to note is that under unitary transformations the gauge-kinetic terms remain invariant. Therefore, the task of identifying possible symmetries of the 2HDM reduces to checking the finite sub-groups of at most $U(2)$. It is known that the scalar potential is invariant under the global $U(1)$ symmetry, $h_{ij} \rightarrow e^{i\left( \theta_j - \theta_i \right)} h_{ij}$ for $h_i \to e^{i \theta_i} h_i$, assuming that both doublets transform with the same phase. Actually, $SU(n) \times U(1)$ is homomorphic to $U(n)$. Let us embark upon the topic of symmetries in the 2HDM by verifying this statement in the context of the scalar potential.

The $U(2)$-invariant matrix can be parameterised in terms of
\begin{equation}
\mathcal{R}_{U(2)} = \begin{pmatrix}
 e^{i (\alpha + \beta +\frac{ \phi }{2})} \cos \theta  & e^{i (\alpha - \beta +\frac{ \phi }{2})}  \sin \theta \\
 -e^{i (-\alpha + \beta +\frac{ \phi }{2})} \sin \theta  & e^{i (-\alpha - \beta +\frac{ \phi }{2})} \cos \theta
\end{pmatrix},
\end{equation}
while the $SU(2)$-invariant matrix can be parameterised in terms of
\begin{equation}
\mathcal{R}_{SU(2)} = \begin{pmatrix}
 e^{i (\alpha +\beta )} \cos \theta & i e^{i (\alpha -\beta )} \sin \theta \\
 i e^{i (-\alpha +\beta )} \sin \theta & e^{i (-\alpha -\beta )} \cos \theta \\
\end{pmatrix}.
\end{equation}
In both cases, as expected, the scalar potential is identical:
\begin{equation}
V = \mu_a^2 \left( h_{11} + h_{22} \right) + \lambda_a \left( h_{11} + h_{22} \right)^2 + \lambda_b \left(  h_{12}h_{21} - h_{11} h_{22}\right).
\end{equation}
Actually, in this case we have $SU(2)/ \mathbb{Z}_2  \cong SO(3)$.

We are already familiar with the CP transformation, under which
\begin{equation}
h_i (t, \overline{x}) \to (h_i)_{CP} (t, \overline{x}) =  h_i^\ast (t, -\overline{x}).
\end{equation}
The scalar doublets can be combined into a reducible space $h \oplus h^\ast$, where $h=(h_1~ h_2)^\mathrm{T}$. In this space symmetries can be represented by:
\begin{subequations}
\begin{align}
\begin{pmatrix}
h \\
h^\ast
\end{pmatrix} = \begin{pmatrix}
A & 0 \\
0 & A^\ast
\end{pmatrix}\begin{pmatrix}
h \\
h^\ast
\end{pmatrix},\\
\begin{pmatrix}
h \\
h^\ast
\end{pmatrix} = \begin{pmatrix}
0 & B \\
B^\ast & 0
\end{pmatrix}\begin{pmatrix}
h \\
h^\ast
\end{pmatrix}.
\end{align}
\end{subequations}
For example, we could have:
\begin{subequations}
\begin{align}
A: ~ \begin{pmatrix}
h_1 \\
h_2
\end{pmatrix} \to \begin{pmatrix}
e^{-i \theta} & 0 \\
0 & e^{i \theta}
\end{pmatrix} \begin{pmatrix}
h_1 \\
h_2
\end{pmatrix}, \quad \begin{pmatrix}
h_1^\ast \\
h_2^\ast
\end{pmatrix} \to \begin{pmatrix}
e^{i \theta} & 0 \\
0 & e^{-i \theta}
\end{pmatrix} \begin{pmatrix}
h_1^\ast \\
h_2^\ast
\end{pmatrix},\\
B: ~ \begin{pmatrix}
h_1 \\
h_2
\end{pmatrix} \to \begin{pmatrix}
e^{-i \theta} & 0 \\
0 & e^{i \theta}
\end{pmatrix} \begin{pmatrix}
h_1^\ast \\
h_2^\ast
\end{pmatrix}, \quad \begin{pmatrix}
h_1^\ast \\
h_2^\ast
\end{pmatrix} \to \begin{pmatrix}
e^{i \theta} & 0 \\
0 & e^{-i \theta}
\end{pmatrix} \begin{pmatrix}
h_1 \\
h_2
\end{pmatrix}.
\end{align}
\end{subequations}
These are two categories of global symmetry transformations that can be imposed on the 2HDM scalar potential, that leave the kinetic Lagrangian part invariant. Symmetries similar to the $A$ transformation are known as the Higgs-family/flavour symmetries, while symmetries similar to the $B$ transformation are known as the general CP (GCP) symmetries~\cite{Lee:1966ik,Ecker:1981wv,Ecker:1987qp,Neufeld:1987wa}.

We have omitted discussion of the symmetries of the Yukawa Lagrangian. Since there are three families of fermions, one needs to consider at most finite sub-groups of $U(3)$. While the Yukawa Lagrangian may be invariant under a specific symmetry, the same symmetry might not be realisable within the framework of the 2HDM scalar potential. Suppose that the Yukawa Lagrangian is invariant under a discrete symmetry $\mathcal{G}$. Requiring that the scalar potential should also be invariant under $\mathcal{G}$, might automatically yield that the scalar potential becomes invariant under another continuous symmetry. For some symmetries of the Yukawa Lagrangian this gets generalised to NHDMs.

In the previous section we saw that the $\mathbb{Z}_2$ symmetry results in NFC. Suppose we want to expand the $\mathbb{Z}_2$ symmetry to the scalar potential. Due to the overall $U(1)$ symmetry, there are two possibilities to assign the $\mathbb{Z}_2$ charges to the scalar doublets, \mbox{$\mathbb{Z}_2:~ h_1 \to \pm h_1,~ h_2 \to \mp h_2$}. However, the assignment of the $\mathbb{Z}_2$ charges affects the structure of the Yukawa Lagrangian.

The $\mathbb{Z}_2$-symmetric scalar potential has no odd numbers of identical doublets,
\begin{equation}\label{Eq:V_2HDM_Z2}
\begin{aligned}
V ={}& \mu_{11}^2 h_{11} + \mu_{22}^2 h_{22} + \lambda_1 h_{11}^2 + \lambda_2 h_{22}^2 + \lambda_3 h_{11} h_{22} + \lambda_4 h_{12}h_{21} + \left(\lambda_5 h_{12}^2 + \mathrm{h.c.} \right)
\end{aligned}
\end{equation}
Provided that the vacuum does not break the underlying $\mathbb{Z}_2$ symmetry, we get the IDM. It shall be discussed in Section~\ref{Sec:IDM}.

Now, let us try to construct a scalar potential invariant under $\mathbb{Z}_3$. Choosing any two different $\mathbb{Z}_3$ charges from $\{1, e^{-i2\pi/3}, e^{i2\pi/3}\}$ does not alter the structure of the scalar potential, since there is only one relative phase between the two doublets, but does impact the form of the Yukawa Lagrangian, see Ref.~\cite{Ferreira:2025ymc}. The $\mathbb{Z}_3$-invariant scalar potential is:
\begin{equation}\label{Eq:V_2HDM_U1}
\begin{aligned}
V ={}& \mu_{11}^2 h_{11} + \mu_{22}^2 h_{22} + \lambda_1 h_{11}^2 + \lambda_2 h_{22}^2 + \lambda_3 h_{11} h_{22} + \lambda_4 h_{12}h_{21}.
\end{aligned}
\end{equation}
By maximally breaking the $\mathbb{Z}_3$ symmetry, assuming that the vacuum is given by non-vanishing values of $(v_1,\, v_2)$, we shall get two massless neutral states $m_{\chi_1}^2 = m_{\chi_2}^2 =0.$ In most cases, presence of additional massless states is the consequence of the Goldstone theorem, signaling that there is an additional continuous symmetry broken, apart from the $SU(2)_L \times U(1)_Y$ symmetry of the SM. If we take a closer look at the scalar potential, we may spot that the different $SU(2)$ doublets, no ordering is assumed, always appear in pairs of $h_i^\dagger h_i = |h_i|^2$. Therefore, we have:
\begin{equation}
h_i \to e^{i \theta_i}h_i, \quad h_i^\dagger h_i \to e^{-i \theta_i} h_i^\dagger e^{i \theta_i} h_i = h_i^\dagger h_i,
\end{equation}
which can be identified as an additional $U(1)$ symmetry. As expected, requiring
\begin{equation}
U(1):~ h_1 \to e^{\pm i \theta} h_1, ~ h_2 \to e^{\mp i \theta} h_2,
\end{equation}
results in the scalar potential of eq.~\eqref{Eq:V_2HDM_U1}. The above charges were related to a common one via the global $U(1)$ symmetry. Obviously, such approach is not systematic when trying to identify continuous symmetries. The basis-independent conditions for detecting continuous symmetries in the bilinear formalism were discussed in Refs.~\cite{Ivanov:2012fp}.

In case of the 2HDM, the underlying $U(1)$ symmetry is larger, and hence $\mathbb{Z}_3$ is considered to be non-realisable; due to the overall $U(1)$ invariance of the 2HDM we are interested in the relative phase between the two scalar doublets. Actually, the identified $U(1)$ symmetry corresponds to the $U(1)_{PQ}$ Peccei–Quinn symmetry~\cite{Peccei:1977hh}. Any larger $\mathbb{Z}_n$, $n \geq 3 $, symmetries, expressed in terms of the $n$'th roots of unity,  will unavoidably result in the 2HDM invariant under the $U(1)$ symmetry.

The three discussed symmetries summarise possible realisable Higgs-family groups. Moving to the GCP symmetries, the simplest one is $\mathrm{CP1}:~h_1 \to h_1^\ast,~ h_2 \to h_2^\ast$. In this case all parameters of the scalar potential are forced to be real. In the bilinear space this symmetry corresponds to a specific $\mathbb{Z}_2$ symmetry. Apart from this GCP, there are two other symmetries: CP2 ($\mathbb{Z}_2 \times \mathbb{Z}_2 \times \mathbb{Z}_2$) and CP3 ($\mathbb{Z}_2 \times O(2)$).

To sum up, there are six known global realisable symmetries of the 2HDM scalar potential. Using the bilinear formalism it is simple to  confirm this statement~\cite{Ivanov:2005hg,Ivanov:2006yq}. As it happens, the number of possible symmetries is larger if one considers the (accidental) custodial symmetry group~\cite{Battye:2011jj,Pilaftsis:2011ed,Darvishi:2019dbh} or the renormalisation group invariant ``GOOFy" symmetries~\cite{Ferreira:2023dke,Trautner:2025yxz}. The specific relations among the parameters of the 2HDM scalar potential, as a result of applying different symmetries are presented in Table~\ref{Table:Diff_Cases}.

{{\renewcommand{\arraystretch}{1.13}
\setlength\LTcapwidth{\linewidth}
\begin{center}
\begin{longtable}[htb]{|c|c|l|} 
\caption{ Symmetries of the 2HDM. In the second column the numbers of independent bilinear plus quartic couplings are provided. In the third column specific relations among the parameters are shown. The table is split into three blocks: in the first block Higgs-family and GCP symmetries are presented, in the second block---accidental custodial symmetries, in the third block---renormalisation group invariant ``GOOFy" symmetries. A peculiar observation is that each of these blocks consists of exactly seven entries.}
\label{Table:Diff_Cases} \\ \hline\hline
\begin{tabular}[l]{@{}c@{}} Underlying \\ symmetry \end{tabular} & \begin{tabular}[l]{@{}c@{}} Independent \\ couplings \end{tabular} & \begin{tabular}[l]{@{}c@{}} Allowed couplings and  necessary relations \end{tabular} \\ \hline  
\multicolumn{3}{c}{}\\[-1.2em]
\hline
- & 3 + 7 & $\mu_{11}^2,\, \mu_{22}^2,\, \mu_{12}^2,  ~~ \lambda_1,\, \lambda_2,\, \lambda_3,\, \lambda_4,\, \lambda_5,\, \lambda_6,\, \lambda_7$ \\ \hline 
$\mathbb{Z}_2$ & 2 + 5 & $\mu_{11}^2,\, \mu_{22}^2,  ~~ \lambda_1,\, \lambda_2,\, \lambda_3,\, \lambda_4,\, \lambda_5$ \\ \hline 
$U(1)$ & 2 + 4 & $\mu_{11}^2,\, \mu_{22}^2,  ~~ \lambda_1,\, \lambda_2,\, \lambda_3,\, \lambda_4$ \\ \hline 
$SO(3)$ & 1 + 2 & $\mu_{11}^2=\mu_{22}^2,  ~~ \lambda_1=\lambda_2,\, \lambda_3,\, \lambda_4=2 \lambda_1 - \lambda_3$ \\ \hline 
CP1 & 3 + 7 & \begin{tabular}[l]{@{}l@{}}$\mu_{11}^2,\, \mu_{22}^2,\, \mathbb{R}\mathrm{e}(\mu_{12}^2),$\\ $\lambda_1,\, \lambda_2,\, \lambda_3,\, \lambda_4,\, \mathbb{R}\mathrm{e}(\lambda_5),\, \mathbb{R}\mathrm{e}(\lambda_6),\, \mathbb{R}\mathrm{e}(\lambda_7)$\end{tabular} \\ \hline 
CP2 & 1 + 4 & $\mu_{11}^2=\mu_{22}^2,  ~~ \lambda_1=\lambda_2,\, \lambda_3,\, \lambda_4,\, \mathbb{R}\mathrm{e}(\lambda_5)$ \\ \hline 
CP3 & 1 + 3 & \begin{tabular}[l]{@{}l@{}}$\mu_{11}^2=\mu_{22}^2$\\$\lambda_1=\lambda_2,\, \lambda_3,\, \lambda_4,\, 2\mathbb{R}\mathrm{e}(\lambda_5) = 2 \lambda_1 - \lambda_3 - \lambda_4$\end{tabular} \\ \hline  
\multicolumn{3}{c}{}\\[-1em]
\hline
$Sp(2)_{h_1+h_2}$ & 3 + 6 & \begin{tabular}[l]{@{}l@{}}$\mu_{11}^2,\, \mu_{22}^2,\, \mathbb{R}\mathrm{e}(\mu_{12}^2),$\\$\lambda_1,\, \lambda_2,\, \lambda_3,\, \lambda_4=2\mathbb{R}\mathrm{e}(\lambda_5),\, \mathbb{R}\mathrm{e}(\lambda_6),\, \mathbb{R}\mathrm{e}(\lambda_7)$\end{tabular} \\ \hline 
$(\mathrm{CP1}\rtimes S_2) \times Sp(2)_{h_1+h_2}$ & 2 + 4 & \begin{tabular}[l]{@{}l@{}}$\mu_{11}^2=\mu_{22}^2,\, \mathbb{R}\mathrm{e}(\mu_{12}^2),$\\$\lambda_1=\lambda_2,\, \lambda_3,\, \lambda_4=2\mathbb{R}\mathrm{e}(\lambda_5),\, \mathbb{R}\mathrm{e}(\lambda_6)=\mathbb{R}\mathrm{e}(\lambda_7)$\end{tabular} \\ \hline 
$(S_2 \rtimes \mathbb{Z}_2) \times Sp(2)_{h_1+h_2}$ & 1 + 3 & $\mu_{11}^2=\mu_{22}^2,  ~~ \lambda_1=\lambda_2,\, \lambda_3,\, \pm\lambda_4=2\mathbb{R}\mathrm{e}(\lambda_5)$ \\ \hline 
$U(1) \times Sp(2)_{h_1h_2}$ & 1 + 2 & $\mu_{11}^2=\mu_{22}^2,  ~~ 2\lambda_1=2\lambda_2=\lambda_3,\, \lambda_4$ \\ \hline 
$Sp(2)_{h_1} \times Sp(2)_{h_2}$ & 2 + 3 & $\mu_{11}^2,\, \mu_{22}^2,  ~~ \lambda_1,\, \lambda_2,\, \lambda_3$ \\ \hline 
$S_2 \times Sp(2)_{h_1} \times Sp(2)_{h_2}$ & 1 + 2 & $\mu_{11}^2 = \mu_{22}^2,  ~~ \lambda_1 = \lambda_2,\, \lambda_3$ \\ \hline 
$Sp(4)$ & 1 + 1 & $\mu_{11}^2=\mu_{22}^2,  ~~ 2\lambda_1=2\lambda_2=\lambda_3$ \\ \hline  
\multicolumn{3}{c}{}\\[-1em]
\hline
$r_0$ & 2 + 4 & $\mu_{11}^2=-\mu_{22}^2,\, \mu_{12}^2,  ~~ \lambda_1=\lambda_2,\, \lambda_3,\, \lambda_4,\, \mathbb{R}\mathrm{e}(\lambda_5)$ \\ \hline 
0$\mathbb{Z}_2$ & 1 + 4 & $\mu_{11}^2=-\mu_{22}^2,  ~~ \lambda_1=\lambda_2,\, \lambda_3,\, \lambda_4,\, \mathbb{R}\mathrm{e}(\lambda_5)$ \\ \hline 
0$U(1)$ & 1 + 3 & $\mu_{11}^2=-\mu_{22}^2,  ~~ \lambda_1=\lambda_2,\, \lambda_3,\, \lambda_4$ \\ \hline 
0$SO(3)$ & 0 + 2 & $\lambda_1=\lambda_2,\, \lambda_3,\, \lambda_4=2 \lambda_1 - \lambda_3$ \\ \hline 
0CP1 & 2 + 4 & \begin{tabular}[l]{@{}l@{}}$\mu_{11}^2=-\mu_{22}^2,\, \mathbb{R}\mathrm{e}(\mu_{12}^2),$\\$\lambda_1=\lambda_2,\, \lambda_3,\, \lambda_4,\, \mathbb{R}\mathrm{e}(\lambda_5)$\end{tabular} \\ \hline 
0CP2 & 0 + 4 & $\lambda_1=\lambda_2,\, \lambda_3,\, \lambda_4,\, \mathbb{R}\mathrm{e}(\lambda_5)$ \\ \hline 
0CP3 & 0 + 3 & $\lambda_1=\lambda_2,\, \lambda_3,\, \lambda_4,\, 2\mathbb{R}\mathrm{e}(\lambda_5) = 2 \lambda_1 - \lambda_3 - \lambda_4$ \\ \hline  \hline 

\end{longtable}
\end{center}}

\section{The Inert Doublet Model}\label{Sec:IDM}

The final topic we want to cover in terms of the 2HDM is the presence of DM in the IDM. The IDM corresponds to an unbroken $\mathbb{Z}_2$-symmetric 2HDM, see eq.~\eqref{Eq:V_2HDM_Z2}. For definiteness, assume that the $\mathbb{Z}_2$ symmetry acts non-trivially on the second doublet, $h_2 \to - h_2$. For the vacuum to preserve the underlying symmetry it needs to satisfy \mbox{$\left\langle h_2 \right\rangle \to - \left\langle h_2 \right\rangle.$} This indicates that we are looking at the $(v,\,0)$ vacuum configuration. Then, the IDM resembles to the earlier discussed Case B with some further restrictions on the scalar potential. In the 2HDM, the scalar potential with a $\mathbb{Z}_2$-symmetry can be expressed using seven parameters. A common parameterisation of this potential involves setting $\mu_{12}^2 = 0$, and additionally, $\lambda_5$ can be chosen to be real, while $\lambda_6 = \lambda_7 = 0$.

The second scalar doublet is referred to as an inert/dark doublet. This indicates that its components do not interact with the SM fermions at tree level. Apart from this, inert doublets do not participate in the EW symmetry breaking, since the vev is forced to be zero by the underlying $\mathbb{Z}_2$. The lightest neutral component, either the state associated with $\eta$ or with $\chi$, of the IDM is considered to be the DM candidate, classified as WIMP. Actually, it is the vacuum invariance under the $\mathbb{Z}_2$ symmetry that guarantees stability of the DM candidate.

Consider Table~\ref{Tab:2HDM_NFC}. By requiring that $h_2$ is the only doublet that receives the ``-" $\mathbb{Z}_2$ charge, while all other fields transform trivially under $\mathbb{Z}_2$, it should be clear that $h_2$ will not couple to the SM fermions. However, this does not prevent particles constituting $h_2$ from decoupling from the SM fermions. Decoupling is guaranteed if there is no mixing between the scalar states. Now, by checking the mass-squared matrices of Case B, see eq.~\eqref{Eq:M2_2HDM_CaseB}, we can notice that only $\mathcal{M}_\eta^2$ has off-diagonal elements, proportional to $\lambda_6$. We recall that this coupling is not allowed by the underlying $\mathbb{Z}_2$ symmetry. As a result, we verified that there are no tree level couplings between the inert doublet and the SM fermions.

What about the gauge couplings? The kinetic part of the Lagrangian was expanded in eq.~\eqref{Eq:D_mu_general}. In the IDM we have $\alpha=0$ (no mixing between doublets) due to $\lambda_6=0$. Then, there are no trilinear couplings involving two gauge bosons and the inert scalar doublet, see eq.~\eqref{Eq:LVVH_CaseB}. This indicates that the lightest particle of the inert sector can not decay into two off-shell $Z$ bosons, with a subsequent decay into a final state of four fermions. On the other hand, the trilinear couplings involving two scalars and a single gauge boson do not vanish, see eq.~\eqref{Eq:LVHH_CaseB}. These vertices are $\{Z\eta_2\chi_2,\, W^\pm h_2^\mp\eta_2,\, W^\pm h_2^\mp\chi_2\}$. The $Z\eta_2\chi_2$ vertex indicates that the pair of $\eta_2 \chi_2$ scalars is CP-odd, and hence the two neutral scalars must be of opposite CP parities. However, due to the freedom of the $h_2 \to i h_2$ transformation, since $v_2=0$, there is an unambiguity which of the states is CP-even and which is CP-odd.

For simplicity, assume that the $\eta_2$ state is the lightest particle of the inert doublet; or else the lightest scalar could be $\chi_2$. Several decay processes are possible depending on the masses of the inert scalars. For $m_{\eta_2} + m_{\chi_2} > m_Z$ the dominant decay of $\chi_2$ would be into $\eta_2 \bar{f}f$. For $m_{\eta_2} + m_{h^\pm_2} > m_{W^\pm}$ the dominant decay of $h^\pm_2$ would be into $W^\mp \eta_2$ or $W^\mp\chi_2\to  W^\mp Z \eta_2$, followed by a decay of the gauge bosons into fermions. In these cases there is no enhancement of the widths of the SM particles. On the other hand, if $m_{\eta_2} + m_{\chi_2} < m_Z$ or $m_{\eta_2} + m_{h^\pm_2} < m_{W^\pm}$, widths of the gauge bosons would be enhanced due to the fact that the gauge bosons could decay into DM. Such processes are highly constrained by the LEP searches. A conservative bound on the light charged scalars is usually assumed to be $m_{H^\pm} \geq 80\text{ GeV}$~\cite{Pierce:2007ut,Arbey:2017gmh}. Apart from that, measurements of the $Z$ widths at LEP forbid decays of the $Z$ boson into the lighter scalars.

This leaves us to the discussion of the scalar interactions, which we did not mention in the previous sections. To get the trilinear and quartic interactions one can expand the scalar potential in terms of the mass eigenstates and then pick relevant terms, not forgetting to multiply those by a symmetry factor. An equivalent procedure would be to evaluate derivative,
\begin{equation}
g(\xi_i \xi_j \xi_k \xi_l) = -i \frac{\partial V}{\partial \xi_i  \partial \xi_j  \partial \xi_k \partial \xi_l}.
\end{equation}
When interested in a potential DM candidate, one should worry about the presence of trilinear couplings involving a single inert particle and two active sector scalars. In this case the DM candidate could decay via active scalars to a final state of four fermions. It should be noted that presence of such couplings is not strictly forbidden, but it will result in fine-tuning of the lifetime of the DM candidate so that it would exceed the age of the Universe, and also it would lead to mixing between the two scalar sectors. As it happens, due to the underlying $\mathbb{Z}_2$ symmetry there are no such trilinear vertices present in the IDM. Presence of these would require minuscule numerical values of $\lambda_6$. The $\lambda_7$ coupling would induce trilinear interactions between the inert and active sectors.

One thing we did not mention is that although $\lambda_5$ can be complex, by rotating the inert doublet $h_2 \to e^{-i/2 \arg(\lambda_5) }h_2$ we can absorb the phase of the $\lambda_5$ coupling; yet it is possible to have CP violation once a soft symmetry breaking (violation of the underlying symmetry by the bilinear terms) coupling $\mu_{12}^2$ is introduced~\cite{Ginzburg:2002wt,Khater:2003wq,ElKaffas:2006gdt,Grzadkowski:2009iz,Arhrib:2010ju}. As a consequence, the overall sign of $\lambda_5$ is irrelevant. However, the absolute value of the $\lambda_5$ coupling is a physical parameter.

The masses of the IDM are given by :
\begin{subequations}
\begin{align}
m_{\eta_1}^2     ={} & 2  \lambda_1 v^2,\\
m_{h_2^\pm}^2 ={} & \mu_{22}^2 + \frac{1}{2} \lambda_3 v^2,\\
m_{\eta_2}^2     ={} & \mu_{22}^2 + \frac{1}{2}\left( \lambda_3 + \lambda_4 + 2 \lambda_5 \right) v^2,\\
m_{\chi_2}^2     ={} & \mu_{22}^2 + \frac{1}{2}\left( \lambda_3 + \lambda_4 - 2 \lambda_5 \right) v^2.
\end{align}
\end{subequations}
Inverting the mass-squared parameters (for the numerical scans it is convenient to use the mass-squared parameters and physical scalar couplings as input parameters) and solving for the couplings yields:
\begin{subequations}
\begin{align}
\mu_{22}^2 ={}& m_{h_2^\pm}^2 - \frac{v^2 \lambda_3}{2},\\
\lambda_1 = {}&  \frac{m_{\eta_1}^2}{2 v^2},\\
\lambda_4 = {}& \frac{m_{\eta_2}^2 + m_{\chi_2}^2-2 m_{h_2^\pm}^2}{v^2},\\
\lambda_5 = {}& \frac{m_{\chi_2}^2 - m_{\eta_2}^2}{2 v^2} \label{Eq:l5_2HDM}.
\end{align}
\end{subequations}
While the sign of $\lambda_5$ would be swapping $\eta_2 \leftrightarrow \chi_2$ states, the absolute value determines the mass splitting of the neutral inert states.

Let us discuss the cosmological consequences of the scalar DM. Prior to discussing the IDM we shall mention some simplified scalar DM models. A DM candidate can be accommodated by non-trivial $SU(2)$ scalar tuples with minimal quantum numbers. These are commonly referred to as minimal DM models~\cite{Cirelli:2005uq,Cirelli:2007xd,Cirelli:2008id,Cirelli:2009uv,Hambye:2009pw,Cai:2012kt,Earl:2013jsa,Garcia-Cely:2015dda,DelNobile:2015bqo}. The simplest extension of the SM, which could accommodate DM is obtained by adding a scalar singlet~\cite{Silveira:1985rk,McDonald:1993ex,Burgess:2000yq,Patt:2006fw,Barger:2007im,Andreas:2008xy}, rather than a doublet. The singlet model does not allow for a charged inert scalar, and if the SM model is extended by a real scalar singlet, a single additional neutral state will be present. The singlet model is also stabilised by an explicit $\mathbb{Z}_2$ symmetry, to prevent specific decay channels. The parameter space of the singlet models was studied in Refs.~\cite{Cline:2013gha,Feng:2014vea,Beniwal:2015sdl,Cuoco:2016jqt,He:2016mls,Casas:2017jjg,Athron:2017kgt}. In general, two DM regions were identified: a low-mass region 55--63~GeV and a high-mass region above around 100-500~GeV, depending on the implementation. The scalar singlet model stabilised by $\mathbb{Z}_3$ was considered in Refs.~\cite{Belanger:2012zr,Ko:2014nha}. 

The IDM was studied extensively a decade ago \cite{Majumdar:2006nt,LopezHonorez:2006gr,Gustafsson:2007pc,Hambye:2007vf,Cao:2007rm,Lundstrom:2008ai,Agrawal:2008xz,Nezri:2009jd,Hambye:2009pw,Dolle:2009fn,Arina:2009um,Dolle:2009ft,Honorez:2010re,Miao:2010rg,LopezHonorez:2010tb,Sokolowska:2011sb}. Two DM mass regions were identified: a low-/intermediate-mass region, in the region of around 5 to 100~GeV (the upper bound depends on the mass of the SM Higgs boson and could be extend up to 160~GeV for the 500~GeV Higgs boson), and a high-mass region, beyond about 530~GeV. Above 80~GeV annihilation to a pair of gauge bosons in the early Universe becomes very fast and the relic DM density would be undersaturated. In addition, DM annihilations into a pair of SM Higgs bosons is feasible. These decays are controlled by the portal $\lambda_L\equiv \lambda_3 + \lambda_4 + 2\lambda_5$  coupling, sometimes abbreviated as $\lambda_{345}$. For small $\lambda_L$ values the overall effect is negligible. For sufficiently heavy DM particles, above 500~GeV, the annihilation rate drops and the relic density parameter becomes compatible with the Planck measurements~\cite{Planck:2018vyg}.

For small mass splittings between the inert states, coannihilation effects get stronger. By controlling the $\lambda_L$ coupling, a relic density in agreement with observations can be reached. For fixed values of mass splittings, as the masses of the inert states become heavier, the absolute value of the $\lambda_L$ portal coupling should also be increased to counter overabundance of DM. On the other hand, for a fixed $\lambda_L$ value, if the mass splittings are increased, the relic density drops. The total relic density depends not only on the absolute value of the $\lambda_L$ parameter, but also on its sign. Interplay of the portal coupling and mass splittings of the inert states may result in an acceptable relic density for DM candidates as heavy as several TeV.

After the Higgs boson discovery, there was some interest in constraining the IDM via the Higgs boson di-photon decays~\cite{Swiezewska:2012eh,Goudelis:2013uca,Krawczyk:2013jta,Arhrib:2013ela}. Neither astroparticle nor collider constraints are consistent with the light, below 45--50~GeV, IDM DM masses~{\cite{Ilnicka:2015jba,Diaz:2015pyv}. The low-/intermediate-mass region, around 55--74~GeV, is consistent with the experimental observations~\cite{Belyaev:2016lok}. Another consistent region, above some 500~GeV, is the high-mass region~\cite{Belyaev:2016lok}. The IDM region of 74--500~GeV is viable provided that additional particles constitute the inert sector. More recently, Ref.~\cite{Kalinowski:2018ylg} confirmed the mentioned DM mass regions. Apart from that they provided a set of benchmarks for the $e^+e^-$ studies. Some further collider-related topics were investigated in Refs.~\cite{Aoki:2013lhm,Ilnicka:2015jba,Hashemi:2015swh,Kalinowski:2018kdn,Dercks:2018wch,Braathen:2024lyl}.

\begin{figure}[htb]
\begin{center}
\includegraphics[scale=0.2]{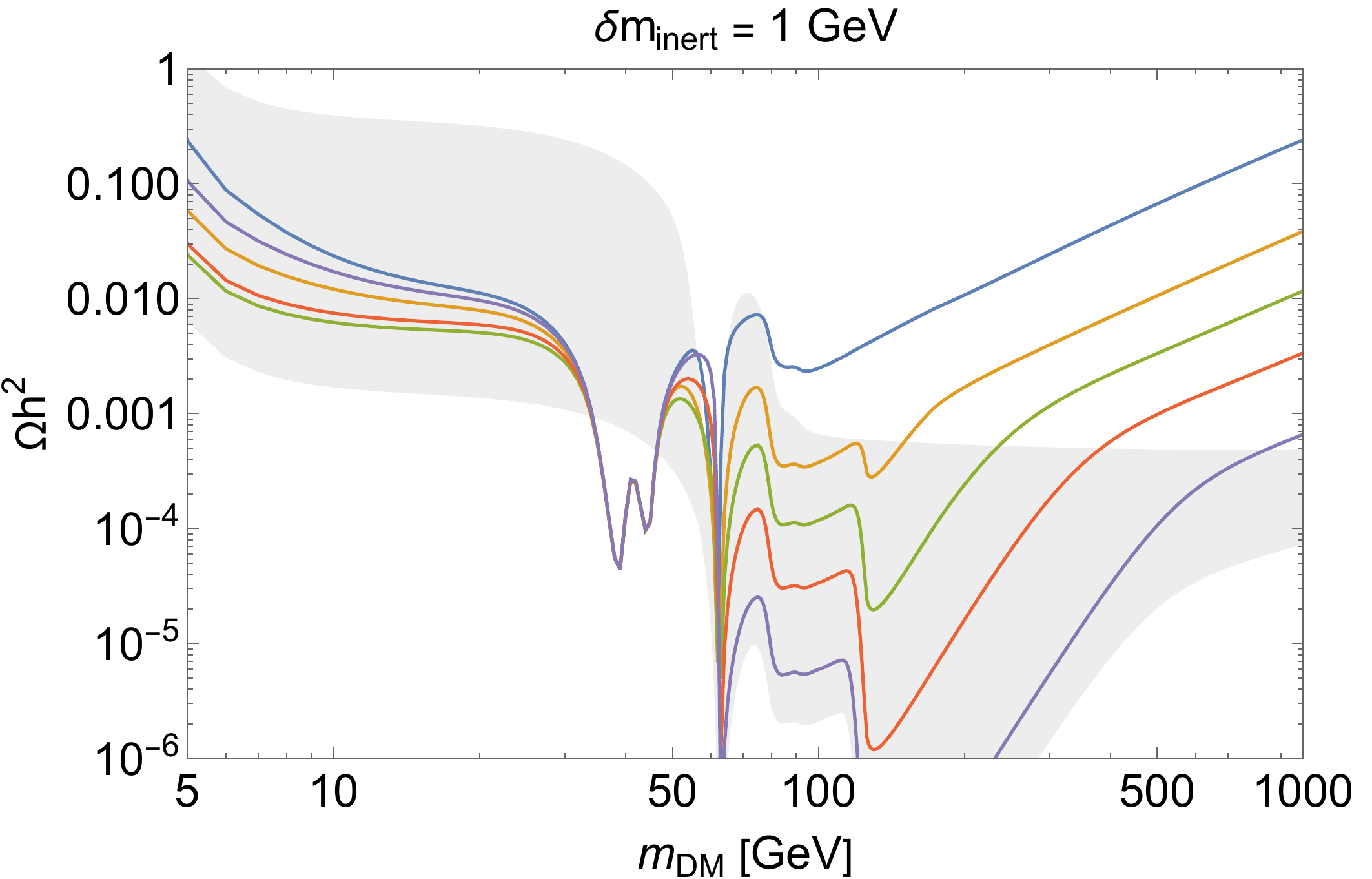}
\includegraphics[scale=0.2]{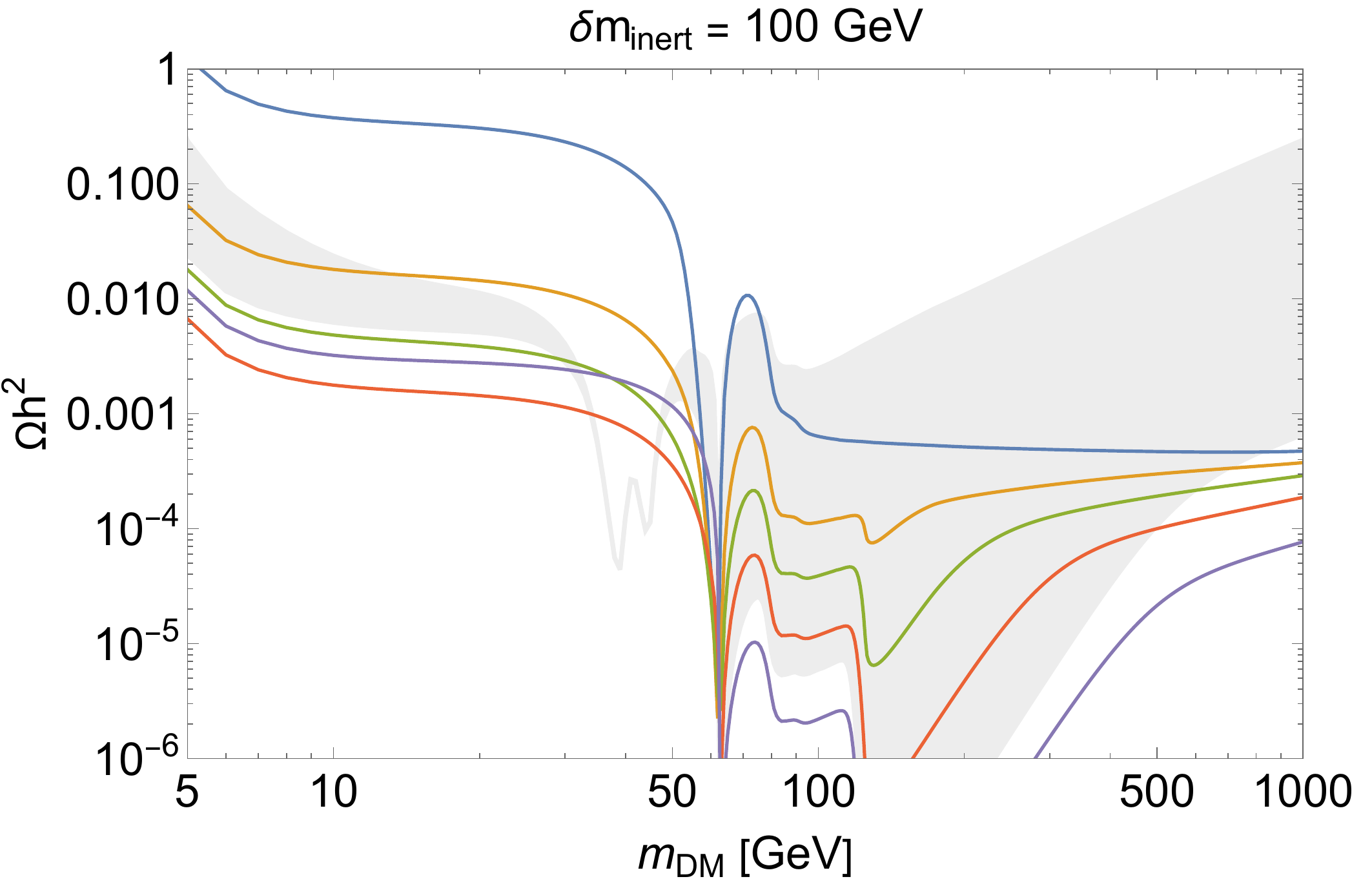}
\includegraphics[scale=0.15]{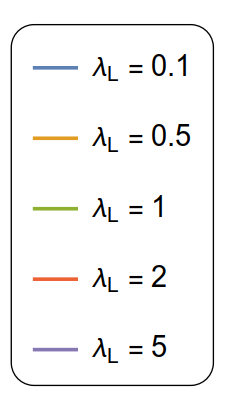}
\end{center}
\vspace*{-4mm}
\caption{ A cartoon of the relic density as a function of the DM mass in the IDM. Different lines show behaviour of the relic density for specific $\lambda_L$ values. The grey region represents the area enclosed between the lines of the other plot. It is assumed that masses of the inert scalars are given by $m_\text{DM} + \delta m_\text{inert}$. Left plot: the mass splitting between the inert scalars is 1 GeV. Right plot: the mass splitting is 100 GeV.}
\label{Fig:mass-ranges}
\end{figure}

A typical behaviour (effects depend on numerical values of the model) of the relic density as a function of the DM mass is presented in Figure~\ref{Fig:mass-ranges}.

\newpage
Here, we present some explanation of what phenomena might occur in different regions:
\begin{itemize}
\item
For the DM masses below $50~\text{GeV}$, DM would annihilate too fast via the off-shell Higgs boson, which would further decay to lighter fermions. For 5~GeV, the leading contribution is from decays into $b$-quarks. This region is highly constrained by the direct DM detection experiments XENONnT~\cite{XENON:2020kmp,XENON:2022ltv,XENON:2023cxc} and LUX-ZEPLIN~\cite{LZ:2021xov,LZ:2022lsv,LZ:2024zvo}.
\item
For small mass splittings between the inert scalars, resonant coannihilation into gauge bosons occurs for the DM masses of 40--45~GeV. These are fixed by the gauge couplings. Another resonant coannihilation occurs due to annihilation into the Higgs boson. Contrary to coannihilations to gauge bosons, coannihilation into the Higgs boson depends on the value of the portal coupling $\lambda_L$. The relevant dips can be seen in Figure~\ref{Fig:mass-ranges}, blue line. 
\item
The 50--80~GeV region is consistent with the total relic density.
\item
In the region around 80--125~GeV DM annihilates too fast via a pair of gauge or Higgs bosons. The latter channel depends on the portal coupling.
\item
In the region around 125--500~GeV DM annihilates too fast via two on-/off-shell gauge bosons. Depending on the $\lambda_L$ value, annihilation into a pair of Higgs bosons can become significant. Now, annihilations via loop contributions into $t$-quarks are possible.
\item
The region above $500~\text{GeV}$ is compatible with the observed relic density. An important mechanism is due to the coupling to longitudinal vector bosons~\cite{Hambye:2009pw}. For heavy DM, to counteract the increase of the total relic density, large values of $\lambda_L$ are required. Eventually, at some tens of TeV, interactions would become non-perturbative as a large value of $\lambda_L$ would be required. Coannihilations between the inert scalars become important in the limit of mass degeneracy.
\end{itemize}

Finally, let us mention presence of the DM candidates in other 2HDMs. For this, we need to refer back to Table~\ref{Table:Diff_Cases}. For a truly stable DM we need to require that there are no $\mu_{12}^2$ and $\lambda_6$ couplings present, while $\lambda_7$ does not result in the undesirable behaviour. For example, the $U(1)$-symmetric model, which can be represented in terms of six parameters, since it requires $\lambda_5=0$, resembles the IDM. In this context, when a parameter is removed, it implies a degeneracy in the model, meaning that the removal of the parameter leads to the vanishing of some physical observable, such as a mass or coupling. This simplification reflects a situation where the model becomes less flexible, and certain interactions or particle masses are tied together or constrained~\cite{Ferreira:2020ana}.  In the context of the $U(1)$ model, this indicates that the exact $U(1)$ symmetry at the EW scale would imply mass degeneracy between the two states with opposite CP parities $m_\chi = m_\eta$~\cite{Haber:2018iwr}, see eq.~\eqref{Eq:l5_2HDM}. In other words, by removing the parameter, two real scalars are combined into a complex scalar field, which is associated with a conserved $U(1)$ symmetry charge. The mass degeneracies can be problematic for the direct DM searches~\cite{Barbieri:2006dq,Hambye:2009pw,Arina:2009um,Escudero:2016gzx}. Models with a small mass splitting between the inert neutral states were considered in Refs.~\cite{Alves:2016bib,Jueid:2020rek}.

Another possibility is the spontaneous breaking of the underlying $\mathbb{Z}_2$ symmetry, where both vevs are non-zero. In this case, only one neutral state becomes decoupled from the gauge bosons and the charged scalars. In the context of the $U(1)$-symmetric 2HDM, if the vacuum breaks the $U(1)$ symmetry, a massless inert state arises, which is identified as the Peccei-Quinn axion \cite{Peccei:1977hh,Peccei:1977ur}.

Another aspect to consider is that the presence of a DM candidate might not be apparent. For example, it is known that $S_2 \cong \mathbb{Z}_2$. In the context of the 2HDM we have a permutation $h_1 \leftrightarrow h_2$ symmetry. These two symmetries are related by the basis transformation:
\begin{equation}\label{Eq:Rot_tr_S2_to_Z2}
\begin{pmatrix}
h_1 \\
h_2 
\end{pmatrix} = \begin{pmatrix}
\cos \theta & \sin \theta \\
-\sin \theta & \cos \theta
\end{pmatrix} \begin{pmatrix}
h_1^\prime \\
h_2^\prime
\end{pmatrix},
\end{equation}
where $\theta= \pm \pi/4$. Starting from the $S_2$ symmetry, we get
\begin{equation}\label{Eq:Rot_S2_to_Z2}
\begin{pmatrix}
\cos (\pm \frac{\pi}{4}) & \sin (\pm \frac{\pi}{4})\\
-\sin (\pm \frac{\pi}{4}) & \cos (\pm \frac{\pi}{4})
\end{pmatrix}
\begin{pmatrix}
0 & 1 \\
1 & 0  
\end{pmatrix}
\begin{pmatrix}
\cos (\pm \frac{\pi}{4}) & \sin (\pm \frac{\pi}{4}) \\
-\sin (\pm \frac{\pi}{4}) & \cos (\pm \frac{\pi}{4})
\end{pmatrix}^\dagger = \begin{pmatrix}
\pm 1 & 0 \\
0 & \mp 1 
\end{pmatrix}.
\end{equation}
In the new basis vevs are related, $1/2\,(v,\, \mp v)$. As a result, from the theoretical point of view, by only looking at the scalar potential along with the vacuum configuration, it might not always be apparent that there is a DM candidate present. A simple solution would be to go into the Higgs basis and check how the mass eigenstates look like.

\chapter{Vacua of the \texorpdfstring{\boldmath$S_3$}{S3}-symmetric three-Higgs-doublet model}\label{Ch:S3_3HDM}

After getting familiarised with the 2HDM in the previous chapter, we turn our attention to the three-Higgs-doublet model (3HDM). Why one might be interested in extending the scalar sector by yet another $SU(2)$ scalar doublet? A naive answer could be that there are three gauge groups, $\mathcal{G}_{SM}$, in the SM along with three fermionic families, so why one should not assume that there could be three scalar doublets? Although partially intended as a joke, there might genuinely be such a preference in Nature. For example, there is an empirical Koide relationship between the masses of the charged leptons~\cite{Koide:1982ax,Koide:1983qe}, which might be a numerological gimmick. Given the advancements in science over the past century, we should always be prepared to embrace the unexpected.

On a more serious note, there are several challenges that the 2HDM cannot address simultaneously. For example, presence of a DM candidate, in the IDM, forces the scalar potential to become real, and hence, IDM contains no additional source of CP violation with respect to the SM. Explaining the observed baryon asymmetry of the Universe necessitates new sources of CP violation~\cite{Sakharov:1967dj,Kuzmin:1985mm,Gavela:1993ts,Huet:1994jb,Gavela:1994dt}. A softly broken $\mathbb{Z}_2$ model reintroduces CP violation; however, it no longer provides a viable DM candidate unless heavy fine-tuning is applied to ensure compatibility with the lifetime of the Universe. On the other hand, additional sources of CP violation would become negligible if $\mu_{12}^2 \approx 0$, see eq.~\eqref{Eq:V_2HDM_generic}.  Furthermore, there are by now several experimental anomalies in the flavour sector possibly hinting at the existence of new physics~\cite{Graverini:2018riw,Belle-II:2018jsg}. 

Despite the rich phenomenology of the 2HDM, models incorporating three or more Higgs doublets have been attracting considerable attention in the literature. With three doublets it is possible to have both a DM candidate and new sources of CP violation in both active or inert sectors~\cite{Grzadkowski:2009bt,Grzadkowski:2010au,Cordero-Cid:2016krd,Azevedo:2018fmj,Cordero-Cid:2018man,Ivanov:2018srm,Kuncinas:2022whn,Biermann:2022meg}. A simple solution would be to have two non-inert doublets along with one inert doublet~\cite{Grzadkowski:2009bt,Grzadkowski:2010au,Osland:2013sla}. As in the case of the 2HDM, depending on the structure of the vacuum, FCNCs might be present. Within the 3HDM framework, it is possible to introduce an additional source of CP violation in the scalar sector while simultaneously maintaining natural flavour conservation~\cite{Weinberg:1976hu,Deshpande:1976yp,Gatto:1979mr,Branco:1979pv,Branco:1980sz}. It has been observed that, in many cases, imposing additional symmetries on the NHDMs eliminates the possibility of CP violation in the scalar sector. Various examples of different symmetries have been explored in the context of 3HDM, where CP violation in the scalar sector can occur, including $A_4$ and $S_4$~\cite{deAdelhartToorop:2010jxh,Degee:2012sk,GonzalezFelipe:2013xok}, $\Delta(27)$~\cite{Branco:1983tn,deMedeirosVarzielas:2011zw,deMedeirosVarzielas:2012ylr,Ma:2013xqa,Nishi:2013jqa,Fallbacher:2015rea}, exotic CP4~\cite{Ivanov:2015mwl,Ferreira:2017tvy,Ivanov:2018ime,Ivanov:2021pnr}.

In the previous chapter we discussed that the predictability of models with multiple doublets diminishes rapidly due to the number of free parameters of the scalar potential. Therefore, it is crucial to constrain these by requiring invariance under symmetries~\cite{Ferreira:2008zy,Ivanov:2011ae,Ivanov:2012ry,Ivanov:2012fp,Keus:2013hya,Ivanov:2014doa,Pilaftsis:2016erj,deMedeirosVarzielas:2019rrp,Darvishi:2019dbh,Bree:2024edl,Kuncinas:2024zjq,Doring:2024kdg}. In this chapter we focus on the $S_3$-symmetric 3HDM, which was studied by several authors in the past, starting in the late 70s by Pakvasa and Sugawara~\cite{Pakvasa:1977in} in the irreducible representation of $S_3$, and Derman and Tsao \cite{Derman:1978rx,Derman:1979nf} in the reducible representation of $S_3$. The main aspect of these papers was to try to explain the structure of the Yukawa Lagrangian. Drawing from several historical lessons, we devoted a substantial portion of time to discussing various approaches to constructing the $S_3$-symmetric scalar potential in Chapter~\ref{Ch:Group_Theory}. We shall focus on the irreducible $S_3$ representation.  We analyse different implementations, as a result of different vacuum configurations, and try to characterise these, further expanding the classification of Ref.~\cite{Emmanuel-Costa:2016vej}. In Ref.~\cite{Emmanuel-Costa:2016vej} a real scalar potential was studied. In Ref.~\cite{Kuncinas:2020wrn} we examined to which continuous symmetries $S_3$ can be broken into as a result of the minimisation conditions. Spontaneously broken continuous symmetries lead to the emergence of unwanted massless states, which might be promoted to massive ones by the introduction of soft breaking terms. These were also classified in Ref.~\cite{Kuncinas:2020wrn}. With real couplings, the case of Ref.~\cite{Emmanuel-Costa:2016vej}, CP is explicitly conserved; spontaneous CP violation is possible in some cases. In Ref.~\cite{Kuncinas:2023ycz} we allowed for complex couplings. In the presence of complex couplings, some of the CP-conserving vacuum structures of the real potential correspond to regions of parameter space that permit explicit CP violation. Finally, in Ref.~\cite{Khater:2021wcx} we classified implementations based on their ability to accommodate a DM candidate.

This chapter is based on articles~\cite{Kuncinas:2020wrn,Khater:2021wcx,Kuncinas:2023ycz}.

\section{The scalar potential}

We have already constructed the $S_3$-symmetric scalar potential in Section~\ref{Sec:Different_S3_constr}. For convenience, we shall start by repeating several aspects here.

The $S_3$ group is a non-Abelian group. It can be presented as a permutation group of three objects, in the case of 3HDM: permutation of the three Higgs doublets $\lbrace\phi_1,~\phi_2,~\phi_3 \rbrace$, where
\begin{equation}\label{Eq:phi_dub}
\phi_i = \begin{pmatrix}
\varphi_i^+ \\
\frac{1}{\sqrt{2}} \left( \rho_i + \tilde \eta_i + i \tilde \chi_i \right)
\end{pmatrix},~\mathrm{for}~i=\{1,\,2,\,3\},
\end{equation}
where $\varphi_i^+$ is a complex field, $\tilde{\eta}_i$ and $\tilde{\chi}_i$ are real fields, and, in general, the $\rho_i$ vacuum can be complex.

The $S_3$ has two one-dimensional irreducible representations $\textbf{1}$ and  $\textbf{1}^\prime$ and a two-dimensional doublet irreducible representation $\textbf{2}$. We shall consider the following representations:
\begin{equation}\label{Eq:S312S}
\begin{aligned}
& \textbf{2:}\qquad \begin{pmatrix}
h_1 \\ h_2 
\end{pmatrix} = \begin{pmatrix}
\frac{1}{\sqrt{2}} \left( \phi_1 - \phi_2 \right)\\
\frac{1}{\sqrt{6}} \left( \phi_1 + \phi_2 - 2\phi_3 \right)
\end{pmatrix},\\
& \textbf{1:}\qquad~~~~~ h_S = \frac{1}{\sqrt{3}} \left( \phi_1 + \phi_2 + \phi_3 \right),
\end{aligned}
\end{equation}
where $S$ in $h_S$ indicates that the $SU(2)$ scalar doublet transforms as an $S_3$ $s$inglet. 

In the new basis, the $h_i$ fields are defined as:
\begin{equation}\label{Eq:h_dub}
\begin{aligned}
& h_i = \begin{pmatrix}
h_i^+ \\
\frac{1}{\sqrt{2}} \left( v_i + \eta_i + i \chi_i \right)\end{pmatrix},~~i=\{1,\,2\},\\
& h_S = \begin{pmatrix}
h_S^+ \\
\frac{1}{\sqrt{2}} \left( v_S + \eta_S + i \chi_S \right)\end{pmatrix},
\end{aligned}
\end{equation}
where the vevs $v_i$ and $v_S$ can be complex. If CP is not broken explicitly, complex vevs could result in spontaneous CP violation. We shall write complex vevs as
\begin{equation}
(v_1,\, v_2,\,v_S ) \to (\hat v_1 e^{i \sigma_1},\, \hat v_2 e^{i \sigma_2},\, \hat v_S e^{i \sigma_S}),
\end{equation}
following the notation of Ref.~\cite{Emmanuel-Costa:2016vej}. The hatted vevs, $\hat v_i$, represent absolute values. Due to the overall $U(1)$ symmetry of the scalar potential, without loss of generality, we can rotate one of the phases away, $h_i \to e^{-i \sigma_j} h_i$. We shall assume $\sigma_S=0$.

Transformation, see eq.~\eqref{Eq:S3_red_U_rot_tribi}, from the basis of eq.~\eqref{Eq:phi_dub} to the basis of eq.~\eqref{Eq:h_dub} is given by:
\begin{equation}\label{h_i expand}
\begin{pmatrix}
h_1 \\ h_2 \\ h_S
\end{pmatrix} = \begin{pmatrix}
\frac{1}{\sqrt{2}} & -\frac{1}{\sqrt{2}} & 0 \\
\frac{1}{\sqrt{6}} & \frac{1}{\sqrt{6}} & -\frac{2}{\sqrt{6}} \\
\frac{1}{\sqrt{3}} & \frac{1}{\sqrt{3}} & \frac{1}{\sqrt{3}}
\end{pmatrix} \begin{pmatrix}
\phi_1 \\ \phi_2 \\ \phi_3
\end{pmatrix},
\end{equation}
or equivalently:
\begin{equation}
\begin{pmatrix}
\phi_1 \\ \phi_2 \\ \phi_3
\end{pmatrix} = \begin{pmatrix}
\frac{1}{\sqrt{2}} & \frac{1}{\sqrt{6}} & \frac{1}{\sqrt{3}} \\
-\frac{1}{\sqrt{2}} & \frac{1}{\sqrt{6}} & \frac{1}{\sqrt{3}} \\
0 & -\frac{2}{\sqrt{3}} & \frac{1}{\sqrt{3}}
\end{pmatrix} \begin{pmatrix}
h_1 \\ h_2 \\ h_S
\end{pmatrix}.
\end{equation}

The most general renormalisable $S_3$-invariant potential can be written as~\cite{Pakvasa:1977in,Kubo:2004ps,Teshima:2012cg,Das:2014fea}:
\begin{equation}\label{Eq:V_S3_3HDM}
\begin{aligned}
V ={}& V_2 + V_4,\\
V_2 ={}& \mu_{11}^2 \left( h_{11} + h_{22} \right) + \mu_{SS}^2 h_{SS},\\
V_4 ={}& \lambda_1 \left( h_{11} + h_{22} \right)^2 + \lambda_2 \left( h_{12} - h_{21} \right)^2+ \lambda_3 \left[ \left( h_{11} - h_{22} \right)^2 + \left( h_{12} + h_{21} \right)^2 \right] \\
&+ \left\{ \lambda_4 \left[ h_{S1} \left( h_{12} + h_{21} \right) + h_{S2} \left( h_{11} - h_{22} \right)\right] +  \mathrm{h.c.} \right\}+ \lambda_5  h_{SS}  \left( h_{11} + h_{22} \right)\\
& + \lambda_6 \left( h_{1S}  h_{S1} + h_{2S} h_{S2} \right)+ \left\{ \lambda_7 \left( h_{S1}^2 + h_{S2}^2  \right) +  \mathrm{h.c.} \right\} + \lambda_8 h_{SS}^2.
\end{aligned}
\end{equation}
Here, both $\lambda_4$ and $\lambda_7$ can be complex. The scalar potential was derived earlier in eqs.~\ref{Eq:S3_pot_s12}. The scalar potential in the reducible basis was presented by Derman in Refs.~\cite{Derman:1978rx,Derman:1979nf}. It was derived in eq.~\eqref{Eq:S3_Derman_redtr}.

Another possibility, for the scalar doublets transforming in the $\mathbf{1}^\prime \oplus \mathbf{2}$ representation, was presented in eq.~\eqref{Eq:S3_pot_a12}. In the case of the pseudosinglet-doublet representation there is no unitary transformation into the defining representation of $S_3$. Such representation would yield an equivalent scalar potential up to the $\lambda_4$ coupling, see eq.~\eqref{Eq:Relating_S3S_S3A}:
\begin{subequations}
\begin{align}
\mathbf{1} \oplus \mathbf{2}:~& \left\lbrace \lambda_4 \left[ h_{S1} \left( h_{12} + h_{21} \right) + h_{S2} \left( h_{11} - h_{22} \right)\right] +  \mathrm{h.c.} \right\rbrace, \\
\mathbf{1}^\prime \oplus \mathbf{2}:~& \left\lbrace \lambda_4 \left[ h_{A2} \left( h_{12} + h_{21} \right) - h_{A1} \left( h_{11} - h_{22} \right)\right] +  \mathrm{h.c.}  \right\rbrace.
\end{align}
\end{subequations}
These two are connected by a substitution of $h_s \leftrightarrow h_a$ along with an interchange of $h_1 \leftrightarrow h_2$. As a result, the two representations are equivalent. However, the Yukawa Lagrangians would not be equivalent. We shall not consider the pseudosinglet-doublet representation.

Actually, there is a redundant coupling in the basis of eq.~\eqref{Eq:V_S3_3HDM}. Since we shall be interested in CP violation, we need to define a suitable basis for the scalar potential. The most general approach would rely on the fact that both, or either, of the couplings $\{\lambda_4,\, \lambda_7\} \in \mathbb{C}$. As we shall discuss, it is sufficient to consider when either $\lambda_4$ or $\lambda_7$, but not both, acquires a non-vanishing phase. It is convenient to take $\lambda_4$ complex and $\lambda_7$ real.

Consider the following basis rotation of the $SU(2)$ doublets,
\begin{equation}\label{Eq:h_ph_theta}
h_i = e^{i \theta_i} \varphi_i, \quad i=\{1,2\}.
\end{equation}
We do not need to consider transformations of $h_S$ since we work in the basis of $v_S \in \mathbb{R}$. Several couplings are sensitive to such rotations: $\{ \lambda_2+\lambda_3,\, \lambda_4,\,\lambda_7\}$. As mentioned earlier, only $\{\lambda_4,\, \lambda_7\} \in \mathbb{C}$, these can be parameterised in the complex polar notation as
\begin{equation}\label{Eq:l4l7_alpha}
\lambda_i = e^{i \alpha_i}|\lambda_i|, \quad i=\{4,7\}. 
\end{equation}
In total, there are three sets of phases: the $\sigma_i$ vev phases, the $\theta_i$ phases, describing a basis change, and the $\alpha_i$ phases of $\lambda_4$ and $\lambda_7$.

In the $\phi$ basis of eq.~\eqref{Eq:h_ph_theta}, $V_4$ can be split into two parts,
\begin{equation}
V_4=V_4^0+V_4^\text{phase},
\end{equation}
where we are primarily interested in the phase sensitive part:
\begin{align}\label{Eq:V4C_explicit_phases-0}
\begin{split}  V_4^\text{phase} &= 
e^{-2i(\theta_1 - \theta_2)} \left( \lambda_2 + \lambda_3 \right) \varphi_{12}^2\\
&\quad + |\lambda_4| \bigg\{ e^{i(2 \theta_1 - \theta_2 + \alpha_4)} \varphi_{S1} \varphi_{21}  + e^{i(\theta_2 + \alpha_4)}\left[ \varphi_{S1} \varphi_{12} + \varphi_{S2}\left( \varphi_{11} - \varphi_{22} \right)  \right]  \bigg\} \\
&\quad + |\lambda_7| \left[e^{i(2 \theta_1 + \alpha_7)} \varphi_{S1}^2 + e^{i(2 \theta_2 + \alpha_7)} \varphi_{S2}^2\right] + \mathrm{h.c.}
\end{split}
\end{align}
It is interesting to note that in the $h$ basis $(\lambda_2 + \lambda_3) \in \mathbb{R}$, while in the $\phi$ basis we have $(\lambda_2 + \lambda_3) \in \mathbb{C}$. We recall that the $\theta_i$ phases come from the basis transformations, and hence without loss of generality, we can restore $(\lambda_2 + \lambda_3) \in \mathbb{R}$ in the $\phi$ basis by choosing $\theta_1=\theta_2\equiv \theta$, while still allowing for $\{\lambda_4,\, \lambda_7\} \in \mathbb{C}$. This choice results in,
\begin{equation}
\begin{aligned} \label{Eq:V4C_explicit_phases}
\left(V_4^\text{phase}\right)^\prime ={}& |\lambda_4| e^{i(\theta + \alpha_4)} \bigg\{ \phi_{S1} \phi_{21}  + \left[ \phi_{S1} \phi_{12} + \phi_{S2}\left( \phi_{11} - \phi_{22} \right)  \right]  \bigg\}\\ 
&+|\lambda_7| e^{i(2 \theta + \alpha_7)} \left[ \phi_{S1}^2 + \phi_{S2}^2 \right] +\mathrm{h.c.}
\end{aligned}
\end{equation}
Furthermore, we may rotate one of the $\alpha_i$ phases away via the $\theta$ phase. In consistency with Ref.~\cite{Kuncinas:2023ycz}, we choose the $\lambda_7$ coupling to become real; the minimisation conditions of different $S_3$ implementations were checked and it was concluded that $\lambda_7 \in \mathbb{R}$ is a more convenient choice then $\lambda_4 \in \mathbb{R}$. Then, this choice results in $2 \theta + \alpha_7 = 0 \mod \pi$. We shall assume that
\begin{equation}\label{Eq:lambda_4_explicit_l4R+l4I}
\lambda_4 \to e^{i \alpha_4}|\lambda_4| \equiv \lambda_4^\mathrm{R} + i \lambda_4^\mathrm{I}.
\end{equation}
Since the scalar potential is not invariant under a phase rotation of eq.~\eqref{Eq:h_ph_theta}, we need to consider both real and complex vacua, along with $\lambda_4 \in \mathbb{C}$. This concludes our discussion, showing that the potential of eq.~\eqref{Eq:V_S3_3HDM} contains one redundant coupling, $\mathbb{I}\mathrm{m}(\lambda_7)=0$.

\section{Different vacuum implementations}\label{Sec:SectionRealComplexVEV}

The unique characteristic of different implementations within the $S_3$-symmetric 3HDMs is a choice of different vacuum configurations. Different vevs yield different minimisation conditions, which can result in different symmetries, and different mass-squared matrices, which result in unique mass eigenstates, further altering the physical couplings. The first step is to minimise the scalar potential. We are interested in characterising different vevs and a way around is to consider derivatives with respect to vevs, which are independent. 

For the minimisation conditions we consider a set of the first-order derivatives with respect to the to the neutral fields $\eta_i$ and $\chi_i$. Assume a general phase rotation of the $S_3$ doublet~\eqref{Eq:h_ph_theta} along with complex coupling as was done eq.~\eqref{Eq:l4l7_alpha}. In this case we can require that vevs are real, $(v_1,\,v_2,\,v_S)$, due to the phase transformations of the $S_3$ doublets, given by $\theta_i$. The minimisation conditions are derived by requiring that the first-order derivatives vanish:
\begin{subequations}
\begin{align}
\begin{split} \dfrac{\partial V}{\partial \eta_1} \Bigg|_v ={}& \mu_{11}^2   v_1 
+\lambda_1   v_1 \left(   v_1^2 +   v_2^2 \right)
-2 \lambda_2 \sin^2(\theta_1-\theta_2)   v_1   v_2^2\\&
+\lambda_3 \left[   v_1^3 + \cos(2\theta_1-2\theta_2)   v_1   v_2^2 \right]\\&
+\lambda_4 \left[ \cos(2 \theta_1 - \theta_2 + \alpha_4)  + 2 \cos(\theta_2 + \alpha_4)\right]   v_1   v_2   v_S\\&
+\frac{1}{2} \left( \lambda_5 + \lambda_6 \right)  v_1   v_S^2 + \lambda_7 \cos(2\theta_1 + \alpha_7)  v_1  v_S^2,
\end{split}\\
\begin{split} \dfrac{\partial V}{\partial \eta_2} \Bigg|_v ={}& \mu_{11}^2 v_2 
+\lambda_1   v_2 \left(   v_1^2 +   v_2^2 \right)
-2 \lambda_2 \sin^2(\theta_1-\theta_2)   v_1^2   v_2\\&
+\lambda_3 \left[   v_2^3 + \cos(2\theta_1-2\theta_2)   v_1^2   v_2 \right]\\&
+\frac{1}{2} \lambda_4 \left\{ \left[ \cos(2 \theta_1 - \theta_2 + \alpha_4)  + 2 \cos(\theta_2 + \alpha_4)\right]v_1^2 - 3 \cos(\theta_2+\alpha_4) v_2^2 \right\}  v_S\\&
+\frac{1}{2} \left( \lambda_5 + \lambda_6 \right)  v_2   v_S^2 + \lambda_7 \cos(2\theta_2 + \alpha_7)  v_2  v_S^2,
\end{split}\\
\begin{split} \dfrac{\partial V}{\partial \eta_S} \Bigg|_v ={}&   \mu_{SS}^2   v_S
+ \frac{1}{2} \lambda_4 \left\{ \left[ \cos(2 \theta_1 -\theta_2+\alpha_4) + 2 \cos(\theta_2+\alpha_4) \right] v_1^2  v_2 - \cos(\theta_2+\alpha_4)  v_2^3 \right\}\\&
+ \frac{1}{2} \left( \lambda_5 +\lambda_6 \right) \left(   v_1^2 +   v_2^2 \right)   v_S
+ \lambda_7 \left[ \cos(2 \theta_1+\alpha_7) v_1^2 + \cos(2 \theta_2+\alpha_7) v_2^2  \right]  v_S\\&
+ \lambda_8   v_S^3    ,\end{split}\\
\begin{split} \dfrac{\partial V}{\partial \chi_1} \Bigg|_v ={}&  
- \lambda_2 \sin(2\theta_1-2\theta_2) v_1  v_2^2
-\lambda_3 \sin(2\theta_1-2 \theta_2) v_1  v_2^2\\&
-\lambda_4 \sin(2\theta_1 -\theta_2+\alpha_4)   v_1   v_2   v_S
- \lambda_7 \sin(2\theta_1+\alpha_7)   v_1   v_S^2     ,\end{split}\\
\begin{split} \dfrac{\partial V}{\partial \chi_2} \Bigg|_v ={}&  
\lambda_2 \sin(2\theta_1-2\theta_2) v_1^2  v_2
+\lambda_3 \sin(2\theta_1-2 \theta_2) v_1^2  v_2\\&
+\frac{1}{2}\lambda_4 \left[ \sin(2\theta_1 -\theta_2+\alpha_4) v_1^2 - \sin(\theta_2+\alpha_4)\left( 2 v_1^2 - v_2^2 \right)\right]v_S\\&
- \lambda_7 \sin(2\theta_2+\alpha_7)   v_2   v_S^2     ,\end{split}\\
\begin{split} \dfrac{\partial V}{\partial \chi_S} \Bigg|_v ={}& 
\frac{1}{2} \lambda_4 \left\{ \left[ \sin(2 \theta_1 -\theta_2+\alpha_4) + 2 \sin(\theta_2+\alpha_4) \right] v_1^2 v_2 - \sin(\theta_2+\alpha_4) v_2^3 \right\}\\&
+ \lambda_7 \left[ \sin(2 \theta_1+\alpha_7) v_1^2 + \sin(2 \theta_2+\alpha_7) v_2^2  \right]  v_S.\end{split}
\end{align}
\end{subequations}

For real vacua, no $\theta_i$ phases, the first-order derivatives get simplified to:
\begin{subequations}
\begin{align}
 \dfrac{\partial V}{\partial \eta_1} \Bigg|_v ={}& \frac{1}{2} v_1 \left[ 2 \mu_{11}^2 + 2 \left(\lambda_1 + \lambda_3\right)\left(v_1^2+v_2^2\right) + 6 \lambda_4^\mathrm{R} v_2 v_S + \left(\lambda_5 + \lambda_6 + 2 \lambda_7\right) v_S^2\right],\\
\begin{split} \dfrac{\partial V}{\partial \eta_2} \Bigg|_v ={}& \frac{1}{2} \left[ 2 \mu_{11}^2 v_2 + 2 \left( \lambda_1 + \lambda_3 \right)\left( v_1^2 + v_2^2\right)v_2   +3 \lambda_4^\mathrm{R}\left(v_1^2 - v_2^2\right)v_S \right.\\& \hspace{15pt}+ \left.\left( \lambda_5 + \lambda_6 + 2 \lambda_7 \right)v_2 v_S^2 \right],\end{split}\\
\begin{split} \dfrac{\partial V}{\partial \eta_S} \Bigg|_v ={}& \frac{1}{2} \left[ 2 \mu_{SS}^2 v_S + \lambda_4^\mathrm{R} \left( 3 v_1^2 v_2 - v_2^3 \right)   + \left( \lambda_5 + \lambda_6 + 2 \lambda_7 \right)\left(v_1^2 + v_2^2\right)v_S + 2 \lambda_8 v_S^3\right],\end{split}\\
 \dfrac{\partial V}{\partial \chi_1} \Bigg|_v ={}& - \lambda_4^\mathrm{I} v_1 v_2 v_S,\label{Eq:DV_Dchi1}\\
 \dfrac{\partial V}{\partial \chi_2} \Bigg|_v ={}& -\frac{1}{2} \lambda_4^\mathrm{I} \left(v_1^2-v_2^2\right) v_S,\label{Eq:DV_Dchi2}\\
 \dfrac{\partial V}{\partial \chi_S} \Bigg|_v ={}& \frac{1}{2}\lambda_4^\mathrm{I}\left(3 v_1^2 - v_2^2\right)v_2.\label{Eq:DV_Dchi3}
\end{align}
\end{subequations}

The most general complex vacuum with $\lambda_4 \in \mathbb{C}$ results in the following derivatives:
\begin{subequations} \label{Eq:Derivatives_Cvev}
\begin{align}
\begin{split} \dfrac{\partial V}{\partial \eta_1} \Bigg|_v ={}& \mu_{11}^2 \cos\sigma_1 \hat v_1 
+\lambda_1 \cos\sigma_1 \hat v_1 \left( \hat v_1^2 + \hat v_2^2 \right)
+2 \lambda_2 \sin(\sigma_1-\sigma_2)\sin\sigma_2 \hat v_1 \hat v_2^2\\&
+\lambda_3 \left[ \cos\sigma_1 \hat v_1^3 + \cos(\sigma_1-2\sigma_2) \hat v_1 \hat v_2^2 \right]\\&
+\lambda_4^\mathrm{R} \left[ 3 \cos\sigma_1 \cos\sigma_2 + \sin\sigma_1 \sin\sigma_2 \right] \hat v_1 \hat v_2 \hat v_S\\&
-\lambda_4^\mathrm{I} \sin(\sigma_1 +\sigma_2) \hat v_1 \hat v_2 \hat v_S
+\frac{1}{2} \left( \lambda_5 + \lambda_6 + 2 \lambda_7 \right) \cos\sigma_1 \hat v_1 \hat v_S^2,
\end{split}\\
\begin{split} \dfrac{\partial V}{\partial \eta_2} \Bigg|_v ={}&  \mu_{11}^2 \cos\sigma_2 \hat v_2
+ \lambda_1 \cos\sigma_2 \hat v_2 \left( \hat v_1^2 + \hat v_2^2 \right)
- 2 \lambda_2 \sin\sigma_1 \sin(\sigma_1-\sigma_2) \hat v_1^2 \hat v_2\\&
+ \lambda_3 \left[ \cos(2 \sigma_1 -\sigma_2) \hat v_1^2 \hat v_2 + \cos\sigma_2 \hat v_2^3 \right]\\&
+ \frac{1}{2} \lambda_4^\mathrm{R} \left[ \left( 2 + \cos2 \sigma_1 \right)\hat v_1^2 - \left( 2 + \cos2 \sigma_2 \right) \hat v_2^2 \right] \hat v_S\\&
- \lambda_4^\mathrm{I} \left( \cos\sigma_1 \sin\sigma_1 \hat v_1^2 - \cos\sigma_2\sin\sigma_2 \hat v_2^2 \right) \hat v_S\\&
+ \frac{1}{2} \left( \lambda_5 +\lambda_6 +2 \lambda_7 \right) \cos\sigma_2 \hat v_2 \hat v_S^2,\end{split}\\
\begin{split} \dfrac{\partial V}{\partial \eta_S} \Bigg|_v ={}&   \mu_{SS}^2 \hat v_S
+ \frac{1}{2} \lambda_4^\mathrm{R} \left\lbrace \left[ \cos(2 \sigma_1 -\sigma_2) + 2 \cos\sigma_2 \right] \hat v_1^2 \hat v_2 - \cos\sigma_2 \hat v_2^3 \right\rbrace\\&
- \frac{1}{2} \lambda_4^\mathrm{I} \left\lbrace \left[ \sin(2 \sigma_1 -\sigma_2) + 2 \sin\sigma_2 \right] \hat v_1^2 \hat v_2 - \sin\sigma_2 \hat v_2^3 \right\rbrace\\&
+ \frac{1}{2} \left( \lambda_5 +\lambda_6 \right) \left( \hat v_1^2 + \hat v_2^2 \right) \hat v_S
+ \lambda_7 \left( \cos2 \sigma_1 \hat v_1^2 + \cos2 \sigma_2 \hat v_2^2  \right) \hat v_S + \lambda_8 \hat v_S^3    ,\end{split}\\
\begin{split} \dfrac{\partial V}{\partial \chi_1} \Bigg|_v ={}&  \mu_{11}^2 \sin\sigma_1 \hat v_1 
+\lambda_1 \sin\sigma_1 \hat v_1 \left( \hat v_1^2 + \hat v_2^2 \right)
- 2 \lambda_2 \sin(\sigma_1-\sigma_2)\cos\sigma_2 \hat v_1 \hat v_2^2\\&
+\lambda_3 \left[ \sin\sigma_1 \hat v_1^3 - \sin(\sigma_1-2 \sigma_2) \hat v_1 \hat v_2^2 \right]
+\lambda_4^\mathrm{R} \sin(\sigma_1 +\sigma_2) \hat v_1 \hat v_2 \hat v_S\\&
-\lambda_4^\mathrm{I} \left[ 3 \sin\sigma_1 \sin\sigma_2 + \cos\sigma_1 \cos\sigma_2 \right] \hat v_1 \hat v_2 \hat v_S\\&
+\frac{1}{2} \left( \lambda_5 + \lambda_6 - 2 \lambda_7 \right) \sin\sigma_1 \hat v_1 \hat v_S^2,\end{split}\\
\begin{split} \dfrac{\partial V}{\partial \chi_2} \Bigg|_v ={}& \mu_{11}^2 \sin\sigma_2 \hat v_2 
+\lambda_1 \sin\sigma_2 \hat v_2 \left( \hat v_1^2 + \hat v_2^2 \right)
+2 \lambda_2 \sin(\sigma_1-\sigma_2)\cos\sigma_1 \hat v_1^2 \hat v_2\\&
+\lambda_3 \left[ \sin(2 \sigma_1- \sigma_2) \hat v_1^2 \hat v_2 + \sin\sigma_2 \hat v_2^3 \right]\\&
+\lambda_4^\mathrm{R} \left[ \cos\sigma_1 \sin\sigma_1 \hat v_1^2 - \cos\sigma_2 \sin\sigma_2 \hat v_2^2 \right]\hat v_S \\&
- \frac{1}{2} \lambda_4^\mathrm{I} \left[ \left( 2 - \cos2 \sigma_1 \right) \hat v_1^2 - \left( 2 - \cos2 \sigma_2 \right)\hat v_2^2   \right] \hat v_S\\&
+\frac{1}{2} \left( \lambda_5 + \lambda_6 - 2 \lambda_7 \right) \sin\sigma_2 \hat v_2 \hat v_S^2,\end{split}\\
\begin{split} \dfrac{\partial V}{\partial \chi_S} \Bigg|_v ={}&
\frac{1}{2} \lambda_4^\mathrm{R} \left\lbrace \left[ \sin(2 \sigma_1 -\sigma_2) + 2 \sin\sigma_2 \right] \hat v_1^2 \hat v_2 - \sin\sigma_2 \hat v_2^3 \right\rbrace\\&
+ \frac{1}{2} \lambda_4^\mathrm{I} \left\lbrace \left[ \cos(2 \sigma_1 -\sigma_2) + 2 \cos\sigma_2 \right] \hat v_1^2 \hat v_2 - \cos\sigma_2 \hat v_2^3 \right\rbrace\\&
+ \lambda_7 \left( \sin2 \sigma_1 \hat v_1^2 + \sin2 \sigma_2 \hat v_2^2  \right) \hat v_S.\end{split}
\end{align}
\end{subequations}

\subsection{Real vacua with real couplings}\label{Sec:Cases_vR_lR}

Let us begin by reviewing possible real vacuum configurations, assuming that the couplings of the scalar potential are real. The minimisation conditions result in:
\begin{subequations}\label{Eq:mu_min}
\begin{align}
\mu_{11}^2 &= -\left(\lambda_1 +\lambda_3 \right)\left( v_1^2 + v_2^2 \right)- \frac{1}{2} v_S \left[ 6 \lambda_4 v_2 + \left( \lambda_5 + \lambda_6 + 2 \lambda_7 \right)v_S \right],\label{Eq:mu_min1}\\
\mu_{11}^2 &= -\left(\lambda_1 +\lambda_3 \right)\left( v_1^2 + v_2^2 \right)- \frac{1}{2} \frac{v_S}{v_2} \left[ 3 \lambda_4 \left( v_1^2 - v_2^2 \right)+ \left( \lambda_5 + \lambda_6 + 2 \lambda_7 \right)v_2v_S \right],\label{Eq:mu_min2}\\
\mu_{SS}^2 &=\frac{1}{2v_S}\left[ \lambda_4 \left( -3 v_1^2 v_2 + v_2^3 \right) -   \left( \lambda_5 + \lambda_6 + 2\lambda_7 \right)\left( v_1^2 + v_2^2 \right)v_S + 2\lambda_8 v_S^2 \right].\label{Eq:mu_min3}
\end{align}
\end{subequations}
Additional conditions should be taken into account since eq.~\eqref{Eq:mu_min1} and eq.~\eqref{Eq:mu_min2} are not equal to one another for general vevs. For self-consistency,
\begin{equation}\label{Eq:SelfConsis1}
 \lambda_4 v_S \left( v_1^2 - 3 v_2^2 \right)=0
\end{equation}
needs to be satisfied. From the self-consistency condition we can see that there are several possible solutions: $\lambda_4=0$, or $v_S=0$, or $v_1 =\pm \sqrt{3}v_2$. If one of the vevs acquires a zero value, the corresponding derivative should vanish. For example, picking $v_1=0$ automatically satisfies the self-consistency condition since the $\mu_{11}^2$ coupling becomes uniquely defined. The choice of $v_2=0$ leads to additional minimisation conditions since derivative with respect to $\eta_2$ does not vanish. In the case of $v_S=0$ we need to solve:
\begin{equation}
\lambda_4 v_2 \left( 3 v_1^2 - v_2^2 \right)=0.
\end{equation}
Regardless that $v_S=0$ is a viable solution to eq.~\eqref{Eq:SelfConsis1}, it should be supplemented by: $\lambda_4=0$ or $v_2=0$ or $v_2 =\pm \sqrt{3}v_1$. The only non-trivial case when two of the vevs vanish is when $v_1=v_S=0$. In this case we need to solve:
\begin{equation}
\lambda_4 v_2^3=0.
\end{equation}

As can be observed, various implementations are possible based on different choices of vevs. Before presenting different vacuum configurations let us note that the classification is based on and adopts the notation of Ref.~\cite{Emmanuel-Costa:2016vej}. For example, consider the ``R-I-2b" vacuum configuration. Here, ``R" stands for a real vacuum configuration. The Roman numeral shows the total number of constraints. The last combination of numeral and letter is used to distinguish different configurations in the same category.

Real vacuum configurations are presented in a format of a name, vevs $( v_1,\,v_2,\,v_S )$ and the minimisation conditions:
{\begin{itemize}
\setlength\itemsep{1.2em}
\item \textbf{R-0}, $(0,\,0,\,0 )$.
\item \textbf{R-I-1}, $(0,\,0,\,v )$: \hspace*{5pt}$\mu_{SS}^2 = - \lambda_8 v^2.$
\item \textbf{R-I-2a}, $(v,\,0,\,0 )$: \hspace*{5pt}$\mu_{11}^2 = - \left( \lambda_1 + \lambda_3 \right) v^2.$
\item \textbf{R-I-2b}, $(v_1,\,-\sqrt{3} v_1,\,0 )$: \hspace*{5pt}$\mu_{11}^2 = - \left( \lambda_1 + \lambda_3 \right) v^2.$
\item \textbf{R-I-2c}, $(v_1,\,\sqrt{3} v_1,\,0 )$: \hspace*{5pt}$\mu_{11}^2 = - \left( \lambda_1 + \lambda_3 \right) v^2.$
\item \textbf{R-II-1a}, $(0,\,v_2,\,v_S )$: \hspace*{5pt}\begin{tabular}[l]{@{}l@{}}$\mu_{SS}^2 = \frac{1}{2}\lambda_4 \frac{v_2^3}{v_S}-\frac{1}{2}\left( \lambda_5 + \lambda_6 + 2\lambda_7 \right)v_2^2-\lambda_8 v_S^2,$\\ $\mu_{11}^2 = - \left( \lambda_1 + \lambda_3 \right) v_2^2 + \frac{3}{2} \lambda_4 v_2 v_S - \frac{1}{2}\left( \lambda_5 + \lambda_6 + 2\lambda_7 \right)v_S^2.$\end{tabular}
\item \textbf{R-II-1b}, $(- \sqrt{3} v_2,\,v_2,\,v_S )$: \begin{tabular}[l]{@{}l@{}}$\mu_{SS}^2 = -4\lambda_4 \frac{v_2^3}{v_S}-2\left( \lambda_5 + \lambda_6 + 2\lambda_7 \right)v_2^2-\lambda_8 v_S^2,$\\ $\mu_{11}^2 = - 4\left( \lambda_1 + \lambda_3 \right) v_2^2 - 3\lambda_4 v_2 v_S - \frac{1}{2}\left( \lambda_5 + \lambda_6 + 2\lambda_7 \right)v_S^2.$\end{tabular}
\item \textbf{R-II-1c}, $(\sqrt{3} v_2,\,v_2,,\,v_S )$:\hspace*{5pt} \begin{tabular}[l]{@{}l@{}}$\mu_{SS}^2 = -4\lambda_4 \frac{v_2^3}{v_S}-2\left( \lambda_5 + \lambda_6 + 2\lambda_7 \right)v_2^2-\lambda_8 v_S^2,$\\ $\mu_{11}^2 = - 4\left( \lambda_1 + \lambda_3 \right) v_2^2 - 3\lambda_4 v_2 v_S - \frac{1}{2}\left( \lambda_5 + \lambda_6 + 2\lambda_7 \right)v_S^2.$\end{tabular}
\item \textbf{R-II-2}, $(0,\,v,\,0 )$: \hspace*{5pt}\begin{tabular}[l]{@{}l@{}}$\mu_{11}^2=-\left( \lambda_1 + \lambda_3 \right)v^2,$\\ $\lambda_4=0.$\end{tabular}
\item \textbf{R-II-3}, $(v_1,\,v_2,\,0 )$: \hspace*{5pt}\begin{tabular}[l]{@{}l@{}}$\mu_{11}^2=-\left( \lambda_1 + \lambda_3 \right)v^2,$\\ $\lambda_4=0.$\end{tabular}
\item \textbf{R-III}, $(v_1,\,v_2,\,v_S )$: \hspace*{5pt}\begin{tabular}[l]{@{}l@{}} $\mu_{SS}^2= -\frac{1}{2}\left( \lambda_5 + \lambda_6 + 2 \lambda_7 \right) \left( v_1^2 + v_2^2 \right)-\lambda_8 v_S^2,$ \\$\mu_{11}^2=-\left( \lambda_1 + \lambda_3 \right) \left( v_1^2 + v_2^2 \right) - \frac{1}{2} \left( \lambda_5 + \lambda_6 + 2 \lambda_7 \right)v_S^2,$\\ $\lambda_4=0.$\end{tabular}
\end{itemize}}

\subsection{Complex vacua with real couplings}

After checking different real vacuum configurations in the previous section, we shall consider complex vacuum $( \hat{v}_1 e^{i \sigma_1},\,\hat{v}_2 e^{i \sigma_2},\,\hat{v}_S)$ with real couplings. Due to the fact that vevs are required to be complex, we need to consider five first-order derivatives.

We start by analysing the most general vev, $( \hat{v}_1 e^{i \sigma_1},\,\hat{v}_2 e^{i \sigma_2},\,\hat{v}_S)$. In this case, the self-consistency condition is given by:
\begin{equation}\label{Eq:SelfConsisC2}
\begin{aligned}
&-2\left( \lambda_2 + \lambda_3 \right) \hat{v}_2 \left( \hat{v}_1^2 - \hat{v}_2^2  \right) \sin^2\left( \sigma_1 - \sigma_2 \right)\\
& +  \frac{1}{2} \lambda_4 \hat{v}_S \left[ \cos \sigma_2  \left( 2\hat{v}_1^2 - 7\hat{v}_2 \right) + \cos\left( 2\sigma_1 - \sigma_2 \right) \left( \hat{v}_1^2 - 2\hat{v}_2 \right) \right]\\
&  - \lambda_7  \hat{v}_2 \hat{v}_S^2 \left[ \cos (2\sigma_1) - \cos(2\sigma_2) \right]=0.
\end{aligned}
\end{equation}
Apart from that,
\begin{subequations}
\begin{align}
\begin{split}
 & \left( \lambda_2 + \lambda_3 \right)\hat{v}_1^2\hat{v}_2^2 \sin\left(  2\sigma_1 - 2\sigma_2\right) + \lambda_4 \hat{v}_1^2 \hat{v}_2 \hat{v}_S \sin\left( 2\sigma_1 - \sigma_2 \right)\\
 & \hspace{190pt} +\lambda_7 \hat{v}_1^2 \hat{v}_S^2 \sin(2\sigma_1)=0,
\end{split}\\
\begin{split}
 &\left( \lambda_2 + \lambda_3 \right)\hat{v}_1^2\hat{v}_2^2 \sin \left(  2\sigma_1 - 2\sigma_2\right) + \frac{1}{2} \lambda_4 \hat{v}_2 \hat{v}_S \left[ \hat{v}_1^2 \sin\left( 2\sigma_1 - \sigma_2 \right) - \left( 2\hat{v}_1^2 - \hat{v}_2^2 \right)\sin \sigma_2 \right]\\
 &  \hspace{190pt} - \lambda_7 \hat{v}_2^2 \hat{v}_S^2 \sin(2\sigma_2)=0,
\end{split}
\end{align}
\end{subequations}
need to be satisfied.

Different complex implementations (now, ``C" indicates complex vacuum configurations) are:
{\begin{itemize}
\setlength\itemsep{1.2em}
\item \textbf{C-I-a}, $(\hat{v}_1,\,\pm i\hat{v}_1,\,0)$: \hspace*{5pt}$\mu_{11}^2 = - \frac{1}{2}\left( \lambda_1 - \lambda_2 \right) v^2.$
\item \textbf{C-III-a}, $(0,\,\hat{v}_2 e^{i\sigma_2},\,\hat{v}_S)$: \hspace*{5pt}\begin{tabular}[l]{@{}l@{}} $\mu_{SS}^2 = -\frac{1}{2}\left( \lambda_5 + \lambda_6 - 2\lambda_7 \right)\hat{v}_2^2 - \lambda_8 \hat{v}_S^2,$  \\$ \mu_{11}^2 = - \left( \lambda_1 + \lambda_3\right)\hat{v}_2^2 - \frac{1}{2}\left( \lambda_5 + \lambda_6 - 2\lambda_7 - 8 \cos^2 \sigma_2 \lambda_7 \right)\hat{v}_S^2,$\\$\lambda_4 = \frac{4 \cos \sigma_2\hat{v}_S}{\hat{v}_2}\lambda_7.$
\end{tabular}
\item \textbf{C-III-b}, $(\pm i \hat{v}_1,\,0,\,\hat{v}_S)$: \hspace*{5pt}\begin{tabular}[l]{@{}l@{}} $\mu_{SS}^2 = -\frac{1}{2}\left( \lambda_5 + \lambda_6 - 2\lambda_7 \right)\hat{v}_1^2 - \lambda_8 \hat{v}_S^2,$ \\$ \mu_{11}^2 = - \left( \lambda_1 + \lambda_3\right)\hat{v}_1^2 - \frac{1}{2}\left( \lambda_5 + \lambda_6 - 2\lambda_7 \right)\hat{v}_S^2,$\\$\lambda_4 = 0.$
\end{tabular}
\item \textbf{C-III-c}, $(\hat{v}_1 e^{i\sigma_1},\,\hat{v}_2 e^{i\sigma_2},\,0)$: \hspace*{5pt}\begin{tabular}[l]{@{}l@{}} $\mu_{11}^2 = -\left( \lambda_1 + \lambda_3 \right)\left( \hat{v}_1^2 + \hat{v}_2^2 \right),$ \\ $\lambda_2 + \lambda_3 = 0,$ \\ $\lambda_4=0.$\end{tabular}
\item \textbf{C-III-d}, $(\pm i\hat{v}_1,\,\hat{v}_2,\,\hat{v}_S)$:\\ \hspace*{40pt}\begin{tabular}[l]{@{}l@{}}
$\mu_{SS}^2 = \left( \lambda_2 + \lambda_3 \right)\frac{\left( \hat{v}_1^2 - \hat{v}_2^2 \right)^2}{\hat{v}_S^2} - \frac{\left( \hat{v}_1^2 - \hat{v}_2^2 \right)\left( \hat{v}_1^2 - 3\hat{v}_2^2 \right)}{4\hat{v}_2 \hat{v}_S}\lambda_4-\frac{1}{2}\left( \lambda_5 + \lambda_6 \right)\left( \hat{v}_1^2 + \hat{v}_2^2 \right) - \lambda_8 \hat{v}_S^2,$\\
$\mu_{11}^2 = - \left( \lambda_1 - \lambda_2 \right)\left( \hat{v}_1^2 + \hat{v}_2^2 \right) -  \frac{\hat{v}_S\left( \hat{v}_1^2 - \hat{v}_2^2 \right)}{4\hat{v}_2}\lambda_4 - \frac{1}{2}\left( \lambda_5 + \lambda_6 \right)\hat{v}_S^2,$\\
$\lambda_7 = \frac{ \left( \hat{v}_1^2 - \hat{v}_2^2 \right)}{\hat{v}_S^2}\left( \lambda_2 + \lambda_3 \right) - \frac{\left( \hat{v}_1^2 - 5\hat{v}_2^2 \right)}{4\hat{v}_2\hat{v}_S}\lambda_4.$ 
\end{tabular}
\item \textbf{C-III-e}, $(\pm i\hat{v}_1,-\hat{v}_2,\,\hat{v}_S)$:\\ \hspace*{40pt} \begin{tabular}[l]{@{}l@{}} 
$\mu_{SS}^2 = \left( \lambda_2 + \lambda_3 \right)\frac{\left( \hat{v}_1^2 - \hat{v}_2^2 \right)^2}{\hat{v}_S^2} + \frac{\left( \hat{v}_1^2 - \hat{v}_2^2 \right)\left( \hat{v}_1^2 - 3\hat{v}_2^2 \right)}{4\hat{v}_2 \hat{v}_S}\lambda_4 -\frac{1}{2}\left( \lambda_5 + \lambda_6 \right)\left( \hat{v}_1^2 + \hat{v}_2^2 \right) - \lambda_8 \hat{v}_S^2,$\\
$\mu_{11}^2 = - \left( \lambda_1 - \lambda_2 \right)\left( \hat{v}_1^2 + \hat{v}_2^2 \right) +  \frac{\hat{v}_S\left( \hat{v}_1^2 - \hat{v}_2^2 \right)}{4\hat{v}_2}\lambda_4 - \frac{1}{2}\left( \lambda_5 + \lambda_6 \right)\hat{v}_S^2,$\\
$\lambda_7 = \frac{ \left( \hat{v}_1^2 - \hat{v}_2^2 \right)}{\hat{v}_S^2}\left( \lambda_2 + \lambda_3 \right) + \frac{\left( \hat{v}_1^2 - 5\hat{v}_2^2 \right)}{4\hat{v}_2\hat{v}_S}\lambda_4.$
\end{tabular}
\item \textbf{C-III-f}, $(\pm i\hat{v}_1,\,i\hat{v}_2,\,\hat{v}_S)$: \hspace*{5pt}\begin{tabular}[l]{@{}l@{}} 
$\mu_{SS}^2 = -\frac{1}{2} \left( \lambda_5 + \lambda_6 -2\lambda_7 \right)\left( \hat{v}_1^2 + \hat{v}_2^2 \right) - \lambda_8 \hat{v}_S^2,$\\ 
$\mu_{11}^2 = - \left( \lambda_1 + \lambda_3 \right) \left( \hat{v}_1^2 + \hat{v}_2^2 \right) - \frac{1}{2} \left( \lambda_5 + \lambda_6 - 2\lambda_7 \right)\hat{v}_S^2,$\\
$\lambda_4 =0.$
\end{tabular}
\item \textbf{C-III-g}, $(\pm i\hat{v}_1,\,-i\hat{v}_2,\,\hat{v}_S)$: \hspace*{5pt}\begin{tabular}[l]{@{}l@{}} 
$\mu_{SS}^2 = -\frac{1}{2} \left( \lambda_5 + \lambda_6 -2\lambda_7 \right)\left( \hat{v}_1^2 + \hat{v}_2^2 \right) - \lambda_8 \hat{v}_S^2,$\\ 
$\mu_{11}^2 = - \left( \lambda_1 + \lambda_3 \right) \left( \hat{v}_1^2 + \hat{v}_2^2 \right) - \frac{1}{2} \left( \lambda_5 + \lambda_6 - 2\lambda_7 \right)\hat{v}_S^2,$\\
$\lambda_4 =0.$
\end{tabular}
\item \textbf{C-III-h}, $(\sqrt{3} \hat{v}_2 e^{i \sigma_2},\,\pm \hat{v}_2 e^{i \sigma_2},\,\hat{v}_S)$:\\ \hspace*{120pt} \begin{tabular}[l]{@{}l@{}} 
$\mu_{SS}^2 = -2\left( \lambda_5 + \lambda_6 -2\lambda_7 \right)\hat{v}_2^2- \lambda_8 \hat{v}_S^2,$\\ 
$\mu_{11}^2 = - 4\left( \lambda_1 + \lambda_3 \right)\hat{v}_2^2 - \frac{1}{2} \left( \lambda_5 + \lambda_6 - 2\lambda_7  - 8 \cos^2 \sigma_2 \lambda_7\right)\hat{v}_S^2,$\\
$\lambda_4 = \mp \frac{2 \cos \sigma_2 \hat{v}_S}{\hat{v}_2} \lambda_7.$
\end{tabular}
\item \textbf{C-III-i}, $( \sqrt{\frac{3(1+\tan^2 \sigma_1)}{1+9 \tan^2 \sigma_1 }} \hat{v}_2 e^{i \sigma_1},\, \pm \hat{v}_2 e^{-i \arctan(3 \tan \sigma_1)},\,\hat{v}_S )$:\\ \hspace*{55pt} {\renewcommand{\arraystretch}{1.2}\begin{tabular}[l]{@{}l@{}} 
$\mu_{SS}^2 = \frac{16 \left( 1 - 3 \tan^2 \sigma_1 \right)^2}{\left( 1 + 9 \tan^2 \sigma_1 \right)^2} \left( \lambda_2 + \lambda_3 \right)\frac{\hat{v}_2^4}{\hat{v}_S^2} \pm \frac{6\left( 1 - \tan^2 \sigma_1\right)\left( 1-3\tan^2 \sigma_1\right)}{ \left( 1 + 9 \tan^2 \sigma_1\right)^{\sfrac{3}{2}}}\lambda_4\frac{\hat{v}_2^3}{\hat{v}_S}$\\ \hspace{35pt}$-\frac{2\left( 1 + 3 \tan^2 \sigma_1\right)}{\left( 1 + 9 \tan^2 \sigma_1\right)}\left( \lambda_5 + \lambda_6 \right) \hat{v}_2^2 - \lambda_8 \hat{v}_S^2,$\\
$\mu_{11}^2 = - \frac{4 \left( 1 + 3 \tan^2 \sigma_1\right)}{\left( 1 + 9 \tan^2 \sigma_1 \right)}\left( \lambda_1 - \lambda_2 \right)\hat{v}_2^2 \mp \frac{\left( 1 - 3 \tan^2 \sigma_1\right)}{2 \sqrt{1 + 9 \tan^2 \sigma_1 }} \lambda_4 \hat{v}_2 \hat{v}_S -\frac{1}{2}\left( \lambda_5 + \lambda_6 \right)\hat{v}_S^2,$\\
$\lambda_7 = - \frac{4 \left( 1 - 3 \tan^2 \sigma_1\right)}{\left( 1 + 9 \tan^2 \sigma_1\right)}\left( \lambda_2 + \lambda_3 \right)\frac{\hat{v}_2^2}{\hat{v}_S^2} \mp \frac{\left( 5 - 3 \tan^2 \sigma_1\right)}{2 \sqrt{1 + 9 \tan^2 \sigma_1}}\lambda_4 \frac{\hat{v}_2}{\hat{v}_S}.$
\end{tabular}}
\item \textbf{C-IV-a}, $(\hat{v}_1 e^{i\sigma_1},\,0,\,\hat{v}_S)$: \hspace*{5pt}\begin{tabular}[l]{@{}l@{}} 
$\mu_{SS}^2 = -\frac{1}{2}\left( \lambda_5 + \lambda_6 \right)\hat{v}_1^2 - \lambda_8 \hat{v}_S^2,$\\
$\mu_{11}^2 = - \left( \lambda_1 + \lambda_3 \right)\hat{v}_1^2 - \frac{1}{2}\left( \lambda_5 + \lambda_6 \right)\hat{v}_S^2,$\\
$\lambda_4 = 0,$\\
$\lambda_7 = 0.$
\end{tabular}
\item \textbf{C-IV-b}, $(\hat{v}_1,\,\pm i\hat{v}_2,\,\hat{v}_S)$: \hspace*{5pt}\begin{tabular}[l]{@{}l@{}} 
$\mu_{SS}^2 = \left( \lambda_2 + \lambda_3 \right)\frac{\left( \hat{v}_1^2 - \hat{v}_2^2 \right)^2}{\hat{v}_S^2}- \frac{1}{2}\left( \lambda_5 + \lambda_6 \right)\left( \hat{v}_1^2 + \hat{v}_2^2 \right) - \lambda_8 \hat{v}_S^2,$\\
$\mu_{11}^2 = - \left( \lambda_1 - \lambda_2  \right)\left( \hat{v}_1^2 + \hat{v}_2^2 \right)- \frac{1}{2}\left( \lambda_5 + \lambda_6 \right)\hat{v}_S^2,$\\
$\lambda_4 = 0,$\\
$\lambda_7 = -\frac{\left( \hat{v}_1^2 - \hat{v}_2^2 \right)}{\hat{v}_S^2} \left( \lambda_2 + \lambda_3 \right).$
\end{tabular}
\newpage
\item \textbf{C-IV-c}, $(\sqrt{1 + 2 \cos^2 \sigma_2}\hat{v}_2,\,\hat{v}_2 e^{i \sigma_2},\,\hat{v}_S)$:\\ \hspace*{10pt}
  \begin{tabular}[l]{@{}l@{}}
$\mu_{SS}^2 = 2 \cos^2 \sigma_2\left( 1 + \cos^2 \sigma_2 \right)\left( \lambda_2 + \lambda_3 \right)\frac{\hat{v}_2^4}{\hat{v}_S^2} - \left( 1 + \cos^2 \sigma_2 \right)\left( \lambda_5 + \lambda_6 \right)\hat{v}_2^2 - \lambda_8 \hat{v}_S^2,$\\
$\mu_{11}^2 = -\left[ 2 \left( 1 + \cos^2 \sigma_2 \right)\lambda_1 - \left( 2 + 3 \cos^2 \sigma_2 \right)\lambda_2 - \cos^2 \sigma_2\lambda_3 \right]\hat{v}_2^2$
$-\frac{1}{2}\left( \lambda_5 + \lambda_6 \right)\hat{v}_S^2,$\\
$\lambda_4 = -2\cos \sigma_2\left( \lambda_2 + \lambda_3 \right)\frac{\hat{v}_2}{\hat{v}_S},$\\
$\lambda_7 =\cos^2 \sigma_2\left( \lambda_2 + \lambda_3 \right)\frac{\hat{v}_2^2}{\hat{v}_S^2}.$
\end{tabular}
\item \textbf{C-IV-d}, $(\hat{v}_1 e^{i\sigma_1},\,\pm \hat{v}_2 e^{i\sigma_1},\,\hat{v}_S)$: \hspace*{5pt}\begin{tabular}[l]{@{}l@{}} 
$\mu_{SS}^2 = -\frac{1}{2}\left( \lambda_5 + \lambda_6 \right)\left( \hat{v}_1^2 + \hat{v}_2^2 \right) - \lambda_8 \hat{v}_S^2,$\\
$\mu_{11}^2 = - \left( \lambda_1 + \lambda_3 \right)\left( \hat{v}_1^2 + \hat{v}_2^2 \right)- \frac{1}{2}\left( \lambda_5 + \lambda_6 \right)\hat{v}_S^2,$\\
$\lambda_4 = 0,$\\
$\lambda_7 = 0.$
\end{tabular}
\item \textbf{C-IV-e}, $(\sqrt{-\frac{\sin (2\sigma_2)}{\sin (2\sigma_1)}} \hat{v}_2 e^{i\sigma_1},\,\hat{v}_2 e^{i \sigma_2},\,\hat{v}_S)$:\\ \hspace*{60pt}
 \begin{tabular}[l]{@{}l@{}} 
$\mu_{SS}^2 = \frac{\sin^2(2 \sigma_1 - 2\sigma_2)}{\sin^2 \sigma_1} \left( \lambda_2 + \lambda_3 \right) \frac{\hat{v}_2^4}{\hat{v}_S^2}-\frac{1}{2} \left( 1 - \frac{\sin (2\sigma_2)}{\sin (2\sigma_1)} \right)\left( \lambda_5 + \lambda_6 \right)\hat{v}_2^2 - \lambda_8 \hat{v}_S^2,$\\
$\mu_{11}^2 = - \left( 1 - \frac{\sin (2\sigma_2)}{\sin (2\sigma_1)} \right) \left( \lambda_1 - \lambda_2 \right)\hat{v}_2^2 - \frac{1}{2} \left( \lambda_5 + \lambda_6 \right)\hat{v}_S^2,$\\
$\lambda_4 = 0,$\\
$\lambda_7 = -\frac{\sin (2 \sigma_1 - 2 \sigma_2)}{\sin (2\sigma_1)}\left( \lambda_2 + \lambda_3 \right)\frac{\hat{v}_2^2}{\hat{v}_S^2}.$
\end{tabular}
\item \textbf{C-IV-f}, $(\sqrt{2 + \frac{\cos(\sigma_1 - 2\sigma_2)}{\cos \sigma_1}} \hat{v}_2 e^{i\sigma_1},\,\hat{v}_2 e^{i \sigma_2},\,\hat{v}_S)$:\\ \hspace*{90pt}
 {\renewcommand{\arraystretch}{1.2}\begin{tabular}[l]{@{}l@{}} 
$\mu_{SS}^2 = -\frac{\left( \cos(\sigma_1 - 2\sigma_2) + 3 \cos \sigma_1\right)\cos (\sigma_2 - \sigma_1)}{2 \cos^2 \sigma_1} \lambda_4 \frac{\hat{v}_2^3}{\hat{v}_S}$\\ \hspace{30pt}
$-\frac{\cos(\sigma_1 - 2\sigma_2) + 3 \cos \sigma_1}{2\cos \sigma_1}\left( \lambda_5 + \lambda_6 \right)\hat{v}_2^2 - \lambda_8 \hat{v}_S^2,$\\
$\mu_{11}^2 = -\frac{ \cos(\sigma_1 - 2\sigma_2) + 3 \cos \sigma_1}{\cos \sigma_1}\left( \lambda_1 + \lambda_3 \right)\hat{v}_2^2$\\ \hspace{30pt}
$-\frac{3\cos (2\sigma_1) + 2\cos (2\sigma_1 - 2\sigma_2) + \cos (2\sigma_2) +4 }{4 \cos (\sigma_1 - \sigma_2)\cos \sigma_1}\lambda_4 \hat{v}_2 \hat{v}_S - \frac{1}{2}\left( \lambda_5 + \lambda_6 \right)\hat{v}_S^2,$\\
$\lambda_4 = -2\frac{\cos (\sigma_2-\sigma_1)}{\cos \sigma_1}\left( \lambda_2 + \lambda_3 \right)\frac{\hat{v}_2}{\hat{v}_S},$\\
$\lambda_7 =\frac{\cos (\sigma_2-\sigma_1)^2}{\cos \sigma_1^2}\left( \lambda_2 + \lambda_3 \right)\frac{\hat{v}_2^2}{\hat{v}_S^2}.$
\end{tabular}}
\item \textbf{C-V}, $(\hat{v}_1 e^{i\sigma_1},\,\hat{v}_2 e^{i\sigma_2},\,\hat{v}_S)$: \hspace*{5pt} \begin{tabular}[l]{@{}l@{}} 
$\mu_{SS}^2 = -\frac{1}{2}\left( \lambda_5 + \lambda_6 \right)\left( \hat{v}_1^2 + \hat{v}_2^2 \right) - \lambda_8 \hat{v}_S^2,$\\
$\mu_{11}^2 = - \left( \lambda_1 + \lambda_3 \right)\left( \hat{v}_1^2 + \hat{v}_2^2 \right)- \frac{1}{2}\left( \lambda_5 + \lambda_6 \right)\hat{v}_S^2,$\\
$\lambda_2 + \lambda_3 = 0,$\\
$\lambda_4 = 0,$\\
$\lambda_7 = 0.$
\end{tabular}
\end{itemize}

Since we allow for complex vacua, this prompts the question of whether CP can be spontaneously violated~\cite{Lee:1973iz}. One can check if vevs obey the following relation~\cite{Branco:1983tn}:
\begin{equation}
U_{ij} \left\langle 0 | h_j | 0 \right\rangle^\ast = \left\langle 0 | h_i | 0 \right\rangle,
\end{equation}
for $U$ being the (CP) symmetry of the scalar potential. There is spontaneous CP violation if the above equation is not satisfied. In Ref.~\cite{Emmanuel-Costa:2016vej} it was shown that out of all (seventeen) complex vacua, only four cases allow for spontaneous CP violation. These are:
\begin{itemize}
\item \textbf{C-III-a}, $(0,\,\hat v_2 e^{i \sigma_2},\, \hat v_S)$;
\item \textbf{C-III-h}, $(\sqrt{3} \hat v_2 e^{i \sigma_2},\, \pm \hat v_2 e^{i \sigma_2} ,\, \hat v_S)$;
\item \textbf{C-IV-c}, $\left(\sqrt{1 + 2 \cos^2 \sigma_2} \hat v_2 ,\, \hat v_2 e^{i \sigma_2} ,\, \hat v_S\right)$;
\item \textbf{C-IV-f}, $\left(\sqrt{2 + \frac{\cos(\sigma_1 - 2\sigma_2)}{\cos \sigma_1}} \hat v_2 e^{i \sigma_1},\, \hat v_2 e^{i \sigma_2} ,\, \hat v_S\right)$.
\end{itemize}

\subsection{Real vacua with complex couplings}

After discussing vacua with real couplings, we shall cover both real and complex vacuum configurations with a complex $\lambda_4$ coupling. Now, apart from spontaneous CP violation it is possible to have explicit CP violation. Vacuum configurations with complex couplings were classified in Ref.~\cite{Kuncinas:2023ycz}.

We start by considering real vacua with complex couplings. By checking the first-order derivatives~\eqref{Eq:DV_Dchi1}--\eqref{Eq:DV_Dchi3} we can identify that some of the real vacuum configurations will result in $\lambda_4^\mathrm{I} = 0$. These are:
\begin{itemize}
\item \textbf{R-II-1a}, \hspace{5pt} $(0,\, v_2,\, v_S)$;
\item \textbf{R-II-1b,c}, \hspace{5pt}$(\mp \sqrt{3} v_2,\, v_2,\, v_S)$;
\item \textbf{R-II-2}, \hspace{5pt}$(0,\, v_2,\, 0)$;
\item \textbf{R-II-3},\hspace{5pt} $(v_1,\, v_2,\, 0)$;
\item \textbf{R-III}, \hspace{5pt}$(v_1,\, v_2,\, v_S)$.
\end{itemize}
These implementations coincide with the ones presented in Section~\ref{Sec:Cases_vR_lR}, and are of no particular interest. Apart from these, there are the following models:
\begin{itemize}
\item  \textbf{R-I-1}, $(0,\,0,\, v)$: \hspace{5pt}$\mu_{SS}^2=- \lambda_8 v^2$;
\item \textbf{R-I-2a}, $(v,\, 0,\, 0)$: \hspace{5pt}$\mu_{11}^2=-(\lambda_1+\lambda_3)v^2$;
\item \textbf{R-I-2b, R-I-2c}, $( v_1,\, \pm \sqrt{3} v_1,\, 0)$:\hspace{5pt} $\mu_{11}^2=-(\lambda_1+\lambda_3)v^2$.
\end{itemize}
These models result in explicit CP violation. This will be discussed in Section~\ref{Sec:CPV_cond}.

\subsection{Complex vacua with complex couplings}\label{Sec:Diff_Im_Models}

Finally, we have to consider complex vevs with complex couplings. As in several cases of real vacua with complex couplings, there are implementations which force $\lambda_4^\mathrm{I}=0$:
\begin{itemize}
\item    \textbf{C-III-b}, $(\pm i \hat v_1 ,\, 0 ,\, \hat v_S)$;
\item    \textbf{C-III-c}, $(\hat v_1 e^{i \sigma_1},\, \hat v_2 e^{i \sigma_2} ,\, 0)$;
\item    \textbf{C-III-d}, $(\pm i \hat v_1  ,\,  \hat v_2 ,\, \hat v_S)$;
\item    \textbf{C-III-e}, $(\pm i \hat v_1  ,\, - \hat v_2 ,\, \hat v_S)$;
\item    \textbf{C-III-f}, $(\pm i \hat v_1 ,\, i \hat v_2 ,\, \hat v_S)$;
\item    \textbf{C-III-g}, $(\pm i \hat v_1 ,\, -i \hat v_2 ,\, \hat v_S)$;
\item    \textbf{C-III-i}, $\left(\sqrt{\frac{3(1 + \tan^2 \sigma_1)}{1+9\tan^2 \sigma_1}} \hat v_2 e^{i \sigma_1} ,\, \pm \hat v_2 e^{-i \arctan(3 \tan \sigma_1)} ,\, \hat v_S\right)$;
\item    \textbf{C-IV-a}, $\left(\hat v_1 e^{i \sigma_1} ,\, 0 ,\, \hat v_S\right)$;
\item    \textbf{C-IV-c}, $(\sqrt{1 + 2 \cos^2 \sigma_2}\hat{v}_2,\,\hat{v}_2 e^{i \sigma_2},\,\hat{v}_S)$;
\item    \textbf{C-IV-d}, $\left(\hat v_1 e^{i \sigma_1} ,\, \pm \hat v_2 e^{i \sigma_1} ,\, \hat v_S\right)$;
\item    \textbf{C-IV-e}, $\left(\sqrt{-\frac{\sin (2 \sigma_2)}{\sin (2\sigma_1)}} \hat v_2 e^{i \sigma_1} ,\, \hat v_2 e^{i \sigma_2} ,\, \hat v_S\right)$;
\item    \textbf{C-IV-f}, $(\sqrt{2 + \frac{\cos(\sigma_1 - 2\sigma_2)}{\cos \sigma_1}} \hat{v}_2 e^{i\sigma_1},\,\hat{v}_2 e^{i \sigma_2},\,\hat{v}_S)$.
\end{itemize}

Apart from that, there are several implementations which result in $\lambda_4^\mathrm{R}=0$, though $\lambda_4^\mathrm{I}$ is left as a free parameter. As shall be shown in Section~\ref{Sec:CPV_cond}, there is no explicit CP violation present in these implementations. The vacuum configurations with $\lambda_4^\mathrm{R}=0$ are:
\begin{itemize}
\item \textbf{C-IV-b},  $\left(\hat v_1 ,\, \pm i \hat v_2 ,\, \hat v_S\right)$\\
It was pointed out in Ref.~\cite{Emmanuel-Costa:2016vej} that there is no spontaneous CP violation in C-IV-b.

\item \textbf{C-IV-e}{\boldmath$^\prime$}, $(\pm \hat v_1e^{-i\arctan(3\tan\sigma_2)},\, \sqrt{\frac{3(1+\tan^2\sigma_2)}{1+9\tan^2\sigma_2}}\hat v_1e^{i\sigma_2},\,\hat v_S)$\\
This is an unidentified vacuum, which seems to resemble the C-III-i vacuum, with a change of the $ v_1 \leftrightarrow  v_2$ entries. In reality, it is C-IV-e with $\hat v_2 = \sqrt{- \sin(2\sigma_1) / \sin(2\sigma_2)}$ and $\sigma_1 = - \arctan\left(3 \tan \sigma_2\right)$, or \mbox{$\sigma_1 = - \arctan\left(3 \tan \sigma_2\right)+\pi$} for $-v_1$. With $\lambda_4 \in \mathbb{C}$ the minimisation conditions differ from the C-IV-e implementation. This model does not violate CP explicitly, however there could be spontaneous CP violation. This implementation yields a real scalar potential and a real vacuum by going into another basis:
\begin{equation}
\begin{pmatrix}
	h_1\\
	h_2\\
	h_S
\end{pmatrix}
=
\begin{pmatrix}
	e^{-i\gamma_0}\cos\alpha & e^{-i(\gamma_0+\gamma_2)}\sin\alpha & 0\\
	e^{-i(\gamma_0-\gamma_1)}\sin\alpha & -e^{-i(\gamma_0-\gamma_1+\gamma_2)}\cos\alpha & 0\\
	0 & 0 & 1
\end{pmatrix}
\begin{pmatrix}
	h_1^\prime\\
	h_2^\prime\\
	h_S^\prime
\end{pmatrix},
\end{equation}
with
\begin{subequations}
\begin{align}
\gamma_0&=\arctan\left(\frac{
\sqrt{3}(1+2\cos2\alpha)}
{\sqrt{-1-2\cos4\alpha}}
\right),\\
\gamma_1&=-\arctan\left(\frac{\sqrt{-1-2\cos4\alpha}}
{\sqrt{3}\cos2\alpha}
\right),\\
\gamma_2&=\arctan\left(\frac{
	\sqrt{3}}
{\sqrt{-1-2\cos4\alpha}}
\right),
\end{align}
\end{subequations}
where
\begin{subequations}
\begin{align}
\begin{array}{l}
\alpha=\frac{5\pi}{24}\text{ if } \cos\sigma<0,\vspace*{5pt}\\
\alpha=\frac{7\pi}{24} \text{ if } \cos\sigma>0,
\end{array} \text{ for } ( \hat v_1e^{-i\arctan(3\tan\sigma_2)}, \sqrt{\frac{3(1+\tan^2\sigma_2)}{1+9\tan^2\sigma_2}}\hat v_1e^{i\sigma_2},\hat v_S),\\
\begin{array}{l}
\alpha=\frac{5\pi}{24}\text{ if } \cos\sigma>0,\vspace*{5pt}\\
\alpha=\frac{7\pi}{24} \text{ if } \cos\sigma<0,
\end{array} \text{ for } (-\hat v_1e^{-i\arctan(3\tan\sigma_2)}, \sqrt{\frac{3(1+\tan^2\sigma_2)}{1+9\tan^2\sigma_2}}\hat v_1e^{i\sigma_2},\hat v_S).
\end{align}
\end{subequations}
These solutions in terms of $\alpha$ are not unique since there are several continuous regions which yield rotations into a real basis.
\end{itemize}

Apart from the discussed vacuum configurations, there are several which result in CP violation:
\begin{itemize}
\setlength\itemsep{1.2em}
\item \textbf{C-I-a}, $(\hat v_1,\,\pm i \hat v_1,\,0)$: \hspace*{5pt} $\mu_{11}^2 = -2 \left( \lambda_1 - \lambda_2 \right) \hat v_1^2$.
\item \textbf{C-III-a}, $(0,\,\hat v_2 e^{i \sigma_2},\, \hat v_S)$:\\ \hspace*{75pt}
\begin{tabular}[l]{@{}l@{}} 
$\mu_{SS}^{2} =-\frac{1}{2} \left( \lambda_5 + \lambda_6 - 2 \lambda_7 \right)\hat{v}_{2}^{2}-\lambda_{8} \hat{v}_{S}^{2} - \lambda_4^\mathrm{I} \frac{\hat v_2^3}{2 \sin\sigma_2 \hat v_S},$ \\
$\mu_{11}^{2}=-\left(\lambda_{1}+\lambda_{3}\right) \hat{v}_{2}^{2}-\frac{1}{2}\left[ \lambda_5 + \lambda_6 - \lambda_7 \left( 2 + 8\cos^{2} \sigma_{2}\right) \right] \hat{v}_{S}^{2} - \frac{3}{2} \lambda_4^\mathrm{I} \frac{\hat v_2 \hat v_S}{\sin\sigma_2},$ \\
$\lambda_4^\mathrm{R} =- \lambda_4^\mathrm{I} \cot \sigma_2 +4\lambda_7\frac{ \cos \sigma_{2} \hat{v}_{S}}{\hat{v}_{2}}.$
\end{tabular}
\item \textbf{C-III-h}, $(\sqrt{3} \hat v_2 e^{i \sigma_2},\, \pm \hat v_2 e^{i \sigma_2} ,\, \hat v_S)$:\\ \hspace*{65pt}
\begin{tabular}[l]{@{}l@{}} 
$\mu_{SS}^{2}=- 2\left(\lambda_5 + \lambda_6 - 2 \lambda_7\right)\hat{v}_{2}^{2}-\lambda_{8} \hat{v}_{S}^{2}  \pm 4 \lambda_4^\mathrm{I} \frac{\hat v_2^3}{\sin\sigma_2 \hat v_S},$ \\
$\mu_{11}^{2}=-4\left(\lambda_{1}+\lambda_{3}\right) \hat{v}_{2}^{2}-\frac{1}{2}\left[\lambda_5 + \lambda_6 - 2\lambda_{7}\left( 3 + 2 \cos 2\sigma_{2} \right) \right] \hat{v}_{S}^{2} \pm 3 \lambda_4^\mathrm{I} \frac{\hat v_2 \hat v_S}{\sin\sigma_2},$\\
$\lambda_4^\mathrm{R}= \mp 2 \lambda_7\frac{\cos\sigma_{2} \hat{v}_{S}}{\hat{v}_{2}}  - \lambda_4^\mathrm{I} \cot\sigma_2.$
\end{tabular}
\item \textbf{C-IV-g}, $(\hat v_1e^{i\sigma_1},\, \pm i \hat v_1e^{i\sigma_1},\, \hat v_S)$: \hspace*{5pt}\begin{tabular}[l]{@{}l@{}} 
$\mu_{SS}^2 = - (\lambda_5+\lambda_6)\hat v_1^2-\lambda _8 \hat v_S^2,$\\
$\mu_{11}^2 =  -2\left( \lambda _1- \lambda _2\right)\hat v_1^2-\frac{1}{2} (\lambda_5 + \lambda_6) \hat v_S^2,$ \\
$\lambda_4^\mathrm{R} =\pm \frac{\sin3\sigma_1 \hat v_S}{\hat v_1}\lambda_7,$\\ 
$\lambda_4^\mathrm{I} =\pm \frac{\cos3\sigma_1 \hat v_S}{\hat v_1}\lambda_7.$
\end{tabular}\vspace*{10pt}\\
This implementation is unique to $\lambda_4 \in \mathbb{C}$.\\
The neutral mass-squared has two negative eigenvalues.

\item \textbf{C-V}, $\left(\hat v_1 e^{i \sigma_1} ,\, \hat v_2 e^{i \sigma_2} ,\, \hat v_S\right)$:
\begin{subequations}
\begin{align*}
\begin{split}
\mu_{0}^{2}&= - \frac{1}{2} \left( \lambda_5 + \lambda_6\right) \left( \hat{v}_1^2 + \hat{v}_2^2 \right)+ \lambda_4^\mathrm{I} \frac{\hat{v}_2}{\hat{v}_S} \frac{C_1}{C_2} - \lambda_8 \hat{v}_S^2
\end{split},\\
\mu_{1}^{2}&=-\left(\lambda_{1}-\lambda_{2}\right)\left(\hat{v}_{1}^{2}+\hat{v}_{2}^{2}\right) - \lambda_4^\mathrm{I}\frac{\hat{v}_S}{\hat{v}_2} \frac{C_3}{C_4}-\frac{1}{2}\left(\lambda_{5}+\lambda_{6}\right) \hat{v}_{S}^{2},\\
\lambda_{2}+\lambda_{3}&= \lambda_4^\mathrm{I} \frac{\hat{v}_S}{\hat{v}_2} \frac{C_5}{C_6},\\
\lambda_4^\mathrm{R}&= \lambda_4^\mathrm{I} \frac{ \sin \sigma_1 \hat{v}_1^2 - \left[ 2 \sin \sigma_1 - \sin\left( \sigma_1 - 2 \sigma_2 \right) \right] \hat{v}_2^2 }{\cos \sigma_1 \hat{v}_1^2 - \left[ 2 \cos \sigma_1 + \cos \left( \sigma_1 - 2 \sigma_2 \right) \right] \hat{v}_2^2},\\
\lambda_7&= - \lambda_4^\mathrm{I} \frac{\hat{v}_2}{\hat{v}_S} \frac{C_7}{C_2},
\end{align*}
\end{subequations}
where
\begin{subequations}
\begin{align*}
\begin{split}
C_1 &= \cos\left( \sigma_1 - 3 \sigma_2 \right) \hat{v}_2^6 - \left[ \cos\left( 3 \sigma_1 - 5 \sigma_2 \right)  + 8 \cos\left( \sigma_1 - \sigma_2 \right) \cos 2 \sigma_2 \right] \hat{v}_1^2 \hat{v}_2^4\\
&\quad+ \cos\left( 3 \sigma_1 - \sigma_2 \right) \hat{v}_1^2 \left( 2 \hat{v}_1^4 - 5 \hat{v}_1^2 \hat{v}_2^2 + 2 \hat{v}_2^4 \right) + \cos\left( \sigma_1 + \sigma_2 \right) \left( \hat{v}_1^6 - 2 \hat{v}_1^4 \hat{v}_2^2 + 2 \hat{v}_2^6 \right),
\end{split}\\
\begin{split}
C_2 &= 2 \left\lbrace \cos \sigma_1 \hat{v}_1^2  - \left[2 \cos \sigma_1 + \cos \left( \sigma_1 - 2 \sigma_2 \right)\right]  \hat{v}_2^2 \right\rbrace \left\lbrace \sin 2 \sigma_1  \hat{v}_1^2 + \sin 2 \sigma_2  \hat{v}_2^2 \right\rbrace,
\end{split}\\
C_3&= \left\lbrace  \hat v_1^2 - \left[ 2 + \cos\left( 2 \sigma_1 - 2 \sigma_2 \right) \right] \hat v_2^2 \right\rbrace^2 + \sin^2\left( 2\sigma_1 - 2 \sigma_2\right) \hat v_2^4,\\
C_4&= 4 \sin\left( \sigma_1 - \sigma_2 \right) \left\lbrace \cos \sigma_1 \hat{v}_1^2 - \left[ 2 \cos \sigma_1 + \cos \left( \sigma_1 - 2 \sigma_2 \right) \right]\hat{v}_2^2 \right\rbrace,\\
\begin{split}
C_5 &= 2 \cos  2 \sigma_1  \sin \left( 2 \sigma_1 - 2 \sigma_2 \right) \hat{v}_1^2 \hat{v}_2^2 + 2 \left[ \sin 2 \sigma_2  - \sin \left( 2 \sigma_1 - 4 \sigma_4 \right) \right] \hat{v}_2^4\\
&\quad + \sin  2 \sigma_1  \left( \hat{v}_1^4 - 6 \hat{v}_1^2 \hat{v}_2^2 + 5 \hat{v}_2^4 \right),
\end{split}\\
C_6 &= C_4 \left[ \sin 2 \sigma_1  \hat{v}_1^2 + \sin 2 \sigma_2  \hat{v}_2^2 \right],\\
C_7 &= \cos\left( \sigma_1 - \sigma_2 \right) \left\lbrace 3 \hat{v}_1^4 - \left[ 8 + 2 \cos \left( 2 \sigma_1 - 2 \sigma_2 \right) \right] \hat{v}_1^2 \hat{v}_2^2 + 3 \hat{v}_2^4 \right\rbrace.
\end{align*}
\end{subequations}

\end{itemize}

\section{Explicit CP violation}\label{Sec:CPV_cond}

After listing all possible vacuum configurations, we need to verify claims about (explicit) CP violation in implementations with complex couplings. There are two possibilities for explicit CP violation: soft CP violation (CP violating phases present in the scalar potential can be confined to the bilinear terms via basis changes) and hard CP violation (CP violating phases present in the quartic part of the scalar potential cannot be removed by a basis change). It was shown that there exist models where phases present in the scalar potential cannot be rotated away via a basis change, yet the scalar potential is CP invariant. A well-known example of such models is the CP4 model~\cite{Ivanov:2015mwl,Ferreira:2017tvy,Ivanov:2018ime,Ivanov:2021pnr,Zhao:2023hws,Liu:2024aew}. We are not interested in the irremovable phases of the CP4 model since such phases do not violate CP.

Consider that the $S_3$-symmetric scalar potential is presented in the following form:
\begin{subequations}
	\begin{align}
		V_2={}&Y_{ij}  h_{ij} ,\\
		V_4={}&\frac{1}{2}Z_{ijkl} h_{ij} h_{kl},
	\end{align}
\end{subequations}
where the non-zero elements of the $Y$- and $Z$-tensors are
\begin{subequations}
\begin{align}
 Y_{11}=Y_{22}=\mu_{11}^2, ~\quad Y_{33}=\mu_{SS}^2,
\end{align}
and
\begin{align}
		&\begin{aligned}
			&Z_{1111}=Z_{2222}=2\lambda_1+2\lambda_3,\\
			&Z_{1122}=Z_{2211}=2\lambda_1-2\lambda_3,\\
			&Z_{1221}=Z_{2112}=-2\lambda_2+2\lambda_3,\\
			&Z_{1212}=Z_{2121}=2\lambda_2+2\lambda_3,\\
		\end{aligned}\hspace{20pt}
		\begin{aligned}
			&Z_{3333}=2\lambda_8,\\
			&Z_{1133}=Z_{2233}=Z_{3311}=Z_{3322}=\lambda_5,\\
			&Z_{1331}=Z_{2332}=Z_{3113}=Z_{3223}=\lambda_6,\\
			&Z_{1313}=Z_{2323}=Z_{3131}=Z_{3232}=2\lambda_7,\\
		\end{aligned}\\
		&Z_{1123}=Z_{1213}=Z_{1312}=Z_{1321}=Z_{2113}=Z_{2311}=-Z_{2223}=-Z_{2322}=\lambda_4^\mathrm{R} - i \lambda_4^\mathrm{I}, \nonumber \\
		&Z_{1132}=Z_{1231}=Z_{2131}=Z_{3112}=Z_{3121}=Z_{3211}=-Z_{2232}=-Z_{3222}=\lambda_4^\mathrm{R} + i \lambda_4^\mathrm{I}. \nonumber
\end{align}
\end{subequations}

In this form it is more efficient to establish quantities that are invariant under basis changes. We shall utilise the technique of constructing CP-odd invariants from the $Y$-/$Z$-tensors as outlined in Refs.~\cite{Mendez:1991gp,Lavoura:1994fv,Botella:1994cs,Branco:2005em,Davidson:2005cw,Gunion:2005ja,Haber:2006ue}. It should be noted that this is not a unique approach and some other methods (in some cases more elegant) were presented in Refs.~\cite{Nishi:2006tg,deMedeirosVarzielas:2016rii,Ogreid:2017alh,Trautner:2018ipq,Ivanov:2019kyh}.

We shall define the following CP-odd invariants, see Ref.~\cite{Kuncinas:2023ycz}:
\begin{subequations}\label{Eq:I6_inv}
\begin{align}
\mathrm{I}_{5Z}^{(1)}={}& \mathbb{I}\mathrm{m} \left[Z_{aabc}Z_{dbef}Z_{cghe}Z_{idgh}Z_{fijj}\right],\\
\mathrm{I}_{5Z}^{(2)}={}& \mathbb{I}\mathrm{m} \left[Z_{abbc}Z_{daef}Z_{cghe}Z_{idgh}Z_{fjji}\right],\\
\mathrm{I}_{6Z}^{(1)}={}& \mathbb{I}\mathrm{m}\left[Z_{abcd}Z_{baef}Z_{gchi}Z_{djke}Z_{fkil}Z_{jglh}\right],\\
\mathrm{I}_{6Z}^{(2)}={}& \mathbb{I}\mathrm{m}\left[Z_{abcd}Z_{baef}Z_{gchi}Z_{dejk}Z_{fhkl}Z_{lgij}\right],\\
\mathrm{I}_{7Z}={}& \mathbb{I}\mathrm{m}\left[Z_{abcd}Z_{eafc}Z_{bgdh}Z_{iejk}Z_{gflm}Z_{hlkn}Z_{minj}\right],\\
\stepcounter{parentequation}
\gdef\theparentequation{\sillymacro}
\setcounter{equation}{-1}
\mathrm{I}_{2Y3Z}={}& \mathbb{I}\mathrm{m} \left[Z_{abcd}Z_{befg}Z_{dchf}Y_{ga}Y_{eh} \right] \label{Eq:I2Y3Z_inv}.
\end{align}
\end{subequations}
These are sufficient but not uniquely defined invariants.

From the mathematical point of view we are constructing knots of different orders, which have to be closed, with vertices presented by the $Z$-tensors, the $Y$-tensors correspond to altering the connecting lines between the $Z$ nodes; consider an example in Figure~\ref{Fig:Knots}. Then, one has to take into account different orderings of the subindices of $Z_{ijkl}$. The only requirement is that each of the indices should appear once and only once in an odd and in an even position. For instance, $Z_{aaab}Z_{bcdd}$ (there are three $a$ and one $c$) and $Z_{abcd}Z_{cdba}$ ($c$ and $d$ appear in identical positions) are wrong combinations. This is a computationally heavy task. These invariants are summed over their subindices. In particular, for order five invariants there are 28 different knots, yielding 226 different combinations in terms of the $Z$-tensors. While it is relatively straightforward to identify various arrangements, the main challenge arises when summing over the $Y$- and $Z$-tensors.

\begin{figure}[htb]
\begin{center}
\includegraphics[scale=0.18]{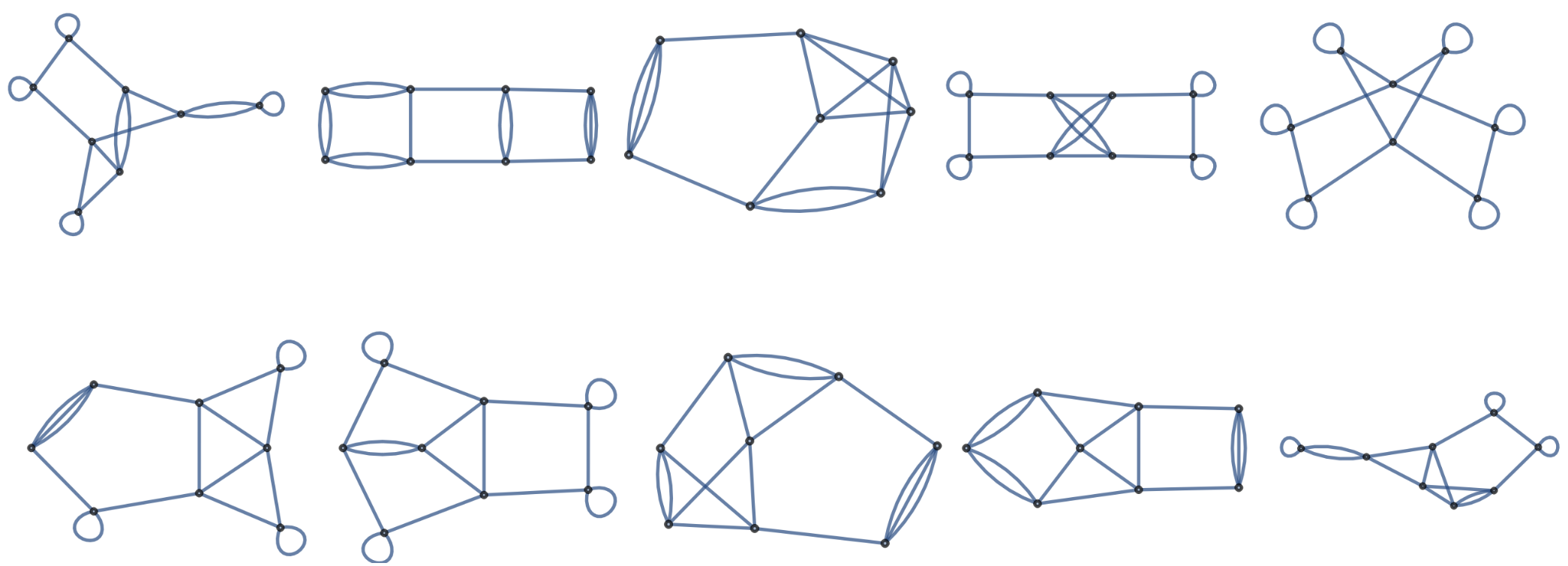}
\end{center}
\vspace*{-4mm}
\caption{ Several knots representing invariants of order eight, $Z^8$.}
\label{Fig:Knots}
\end{figure}

Let us take a closer look. By expanding the $\mathrm{I}_{5Z}$ invariants of eq.~\eqref{Eq:I6_inv} we get:
\begin{subequations}
\begin{align}
\begin{split}
\mathrm{I}_{5Z}^{(1)}={}& 16 \lambda_4^\mathrm{R} \lambda_4^\mathrm{I} \lambda_7 (4\lambda_1-\lambda_5-2\lambda_8)^2,
\end{split}\\
\begin{split}
\mathrm{I}_{5Z}^{(2)}={}& 16 \lambda_4^\mathrm{R} \lambda_4^\mathrm{I} \lambda_7(2\lambda_1-2\lambda_2+4\lambda_3-\lambda_6-2\lambda_8)^2.
\end{split}
\end{align}
\end{subequations}
We are interested in solutions when these invariants vanish. This can happen when:
\begin{itemize}
\item {\bf Solution 0:} $\lambda_4^\mathrm{I}=0$.\\
This is a trivial case of no explicit CP violation. The spontaneous CP violation was already covered in Ref.~\cite{Emmanuel-Costa:2016vej}.
\item {\bf Solution 1:} $\lambda_4^\mathrm{R}=0$.\\
It is possible to transform the scalar potential into the real basis via $h_2\to i h_2$. We may at most have spontaneous CP violation.  
\item {\bf Solution 2:} $\lambda_7=0$.\\
By going back to the most general phase-sensitive part of the scalar potential, given by eq.~\eqref{Eq:V4C_explicit_phases}, we see that the choice of $\theta=-\alpha_4$ results in a real potential. Apart from that, since $\lambda_4$ is the only complex coupling, it can be ``absorbed" via $ h_S  \to e^{i \arg(\lambda_4)}  h_S^\prime $. As in the above cases, CP can only be violated spontaneously.
\item {\bf Solution 3:} $\lambda_5=2(2\lambda_1-\lambda_8)$, $\lambda_6=2(\lambda_1-\lambda_2+2\lambda_3-\lambda_8)$ and $\lambda_4^\mathrm{R} \lambda_4^\mathrm{I} \lambda_7 \neq 0$.\\
These constraints are not sufficient to make the quartic part of the scalar potential CP-invariant. It is necessary to examine the higher-order invariants.
\end{itemize}

In Ref.~\cite{Kuncinas:2023ycz} we established that the $V_4$ part of the $S_3$-symmetric scalar potential explicitly conserves CP if and only if $\mathrm{I}_{5Z}^{(1)}=\mathrm{I}_{5Z}^{(2)}=\mathrm{I}_{6Z}^{(1)}=\mathrm{I}_{6Z}^{(2)}=\mathrm{I}_{7Z}=0$. Vanishing of the $Z$-tensor invariants is satisfied by one of the following solutions: 
\begin{itemize}
\item {\bf Solutions 0--2}: discussed above;
\item {\bf Solution 3 (\boldmath$\lambda_4^\mathrm{R} \lambda_4^\mathrm{I} \lambda_7  \neq  0$)}:
\begin{gather*}
\begin{aligned}
\left( \lambda_4^\mathrm{R} \right)^2={}& -\frac{\left[  (\lambda_2 + \lambda_3)-\lambda_7\right] \left[ 2(\lambda_2 + \lambda_3)+\lambda_7\right]^2}{\lambda_7},\\
\left( \lambda_4^\mathrm{I} \right)^2={}& \frac{\left[ (\lambda_2 + \lambda_3)+\lambda_7\right] \left[2(\lambda_2 + \lambda_3)-\lambda_7\right]^2}{\lambda_7},
\end{aligned}\hspace{50pt}
\begin{aligned}
\lambda_5 ={}& 2 \left( \lambda_1 + \lambda_2 \right),\\
\lambda_6 ={}& 4 \lambda_3,\\
\lambda_8 ={}& \lambda_1 - \lambda_2,
\end{aligned}
\end{gather*}
with
\begin{equation} \label{Eq:lam4-mod}
|\lambda_4|^2=(\lambda_4^\mathrm{R})^2+(\lambda_4^\mathrm{I})^2=2\lambda_7^2.
\end{equation}
\end{itemize}

Now we need to prove that the quartic part of the scalar potential is CP-invariant provided that the constraints of Solution 3 are satisfied. The phase-sensitive part of $V_4$, see eq.~\eqref{Eq:V4C_explicit_phases-0}, is given by
\begin{equation}
\begin{aligned}
V_4^\mathrm{phase} ={} &
\frac{\lambda_2 + \lambda_3}{2}\Big[(h_{11} - h_{22}^2 + 2h_{12}^2 + 2h_{21}^2 + 2 h_{SS}(h_{11} + h_{22})\\
&\hspace*{45pt}-h_{SS}^2 + 4 h_{1S} h_{S1} + 4h_{2S}h_{S2}\Big]\\
&+ \left\lbrace \lambda_4\left[ h_{S1}(h_{12} + h_{21}) + h_{S2} (h_{11} - h_{22}) \right] + \mathrm{h.c.} \right\rbrace \\
&+ \left\lbrace\lambda_7 \left[ h_{S1}^2 + h_{S2}^2\right] +\mathrm{h.c.}\right\rbrace.
\end{aligned}
\end{equation}
Note that Solution 3 fixes $\lambda_4$ in terms of  $(\lambda_2 + \lambda_3)$ and $\lambda_7$. We do not need to consider the whole plane of $(\lambda_2 + \lambda_3){-}\lambda_7$ for the analysis of the CP conditions. This is because by definition $(\lambda_4^\mathrm{R})^2$ and $(\lambda_4^\mathrm{I})^2$  must be necessarily positive. The relevant regions consist of $(\lambda_2 + \lambda_3)=0$ (no restrictions on $\lambda_7$ apart from $\lambda_7 \neq 0$) as well as the region given by $\lambda_7^2-(\lambda_2 + \lambda_3)^2>0$, except for the two lines given by $\lambda_7=\pm (\lambda_2 + \lambda_3)$. 

For $(\lambda_2 + \lambda_3)=0$, the basis transformation
\begin{equation}
\begin{pmatrix}
h_1 \\
h_2 \\
h_S
\end{pmatrix}
= \begin{pmatrix}
1 & 0 & 0 \\
0 & e^{i \theta} \cos \alpha & i e^{i \theta} \sin \alpha \\
0 & \sin \alpha & -i \cos \alpha 
\end{pmatrix} 
\begin{pmatrix}
h_1^\prime \\
h_2^\prime \\
h_S^\prime
\end{pmatrix},
\end{equation}
where
\begin{equation}
\theta ={} - \frac{\pi}{4},\qquad
\alpha ={} -\frac{1}{2} \arctan \sqrt{2},
\end{equation}
forces all couplings to become real.

For the remaining plane of $(\lambda_2 + \lambda_3){-}\lambda_7$, the basis transformation into a basis with real coefficients is given by:
\begin{equation}
\begin{pmatrix}
h_1 \\
h_2 \\
h_S
\end{pmatrix}  = 
\begin{pmatrix}
1 & 0 & 0 \\
0 & e^{i \theta} \cos \alpha &  e^{i \left( \theta + \phi \right)} \sin \alpha \\
0 & \sin \alpha & -e^{i \phi} \cos \alpha 
\end{pmatrix} 
\begin{pmatrix}
h_1^\prime \\
h_2^\prime \\
h_S^\prime
\end{pmatrix},
\end{equation}
where
\begin{subequations}
\begin{align}
\alpha &= \arctan\left(\frac{(\lambda_2 + \lambda_3)\sin(2\theta)}{\lambda_4^\mathrm{I}\cos\theta-\lambda_4^\mathrm{R}\sin\theta}\right),\\
\phi &= \arctan\left( \frac{\lambda_4^\mathrm{I} \cos \theta + \lambda_4^\mathrm{R} \sin \theta }{(\lambda_2 + \lambda_3)\sin\left(2 \alpha\right) + \left( \lambda_4^\mathrm{R} \cos \theta - \lambda_4^\mathrm{I} \sin \theta \right)\cos\left(2 \alpha\right) } \right),
\end{align}
\end{subequations}
while the general form of the $\theta$ angle is given by
\begin{equation}
2 \theta = \pm \arctan\left(\frac{\sqrt{3}\lambda_7+2\sqrt{\lambda_7^2-(\lambda_2 + \lambda_3)^2}}{(\lambda_2 + \lambda_3)}\right) + \left\lbrace \begin{array}{ll}
& \pi \text{ when } (\lambda_2 + \lambda_3)\lambda_7<0,\\
& 0 \text{ else}.
\end{array}\right.
\end{equation}

At the end of the day, this verifies that Solution 3 results in a CP conserving $V_4$. Furthermore, it was checked that the CP-odd invariants of order eight do not introduce any new constraints, which verifies that it is sufficient to analyse $Z$-tensors of order up to seven.

Next, we need to prove that $\mathrm{I}_{5Z}^{(1)}=\mathrm{I}_{5Z}^{(2)}=\mathrm{I}_{6Z}^{(1)}=\mathrm{I}_{6Z}^{(2)}=\mathrm{I}_{7Z}=\mathrm{I}_{2Y3Z}=\mathrm{I}_{2Y3Z}=0$ implies a CP invariant scalar potential; not just the quartic part. Comparing to the above conditions, we now need to account for the vanishing of $\mathrm{I}_{2Y3Z}$,
\begin{equation}\label{Eq:Mu_Eq_Req}
\mathrm{I}_{2Y3Z}= 16 \lambda_4^\mathrm{R} \lambda_4^\mathrm{I} \lambda_7(\mu_{SS}^2-\mu_{11}^2)^2.
\end{equation}
Solutions~0--2 force all six invariants to vanish simultaneously. On the other hand, Solution 3 is modified with one additional constraint, namely $\mu_{11}^2=\mu_{SS}^2$:
\begin{itemize}
\item {\bf Solution 3$^\prime$ (\boldmath$\lambda_4^\mathrm{R} \lambda_4^\mathrm{I} \lambda_7  \neq  0$): }
\begin{gather*}
\begin{aligned}
\mu_{11}^2 ={}& \mu_{SS}^2,\\
\left( \lambda_4^\mathrm{R} \right)^2={}& -\frac{\left[  (\lambda_2 + \lambda_3)-\lambda_7\right] \left[ 2(\lambda_2 + \lambda_3)+\lambda_7\right]^2}{\lambda_7},\\
\left( \lambda_4^\mathrm{I} \right)^2={}& \frac{\left[ (\lambda_2 + \lambda_3)+\lambda_7\right] \left[2(\lambda_2 + \lambda_3)-\lambda_7\right]^2}{\lambda_7},
\end{aligned}\hspace{50pt}
\begin{aligned}
&\\
\lambda_5 ={}& 2 \left( \lambda_1 + \lambda_2 \right),\\
\lambda_6 ={}& 4 \lambda_3,\\
\lambda_8 ={}& \lambda_1 - \lambda_2.
\end{aligned}
\end{gather*}
\end{itemize}
For Solutions 0--2, the basis changes to a real quartic part of the scalar potential were represented by phase rotations of doublets, leaving the bilinear part unaffected. Solution 3$^\prime$ needs more elaboration. For Solution 3$^\prime$, the constraint $\mu_{11}^2=\mu_{SS}^2$ yields the $V_2$ part invariant under the $SU(3)$ transformations. Apart from that, Solution 3$^\prime$ enlarges the $S_3$-symmetric potential to $\Delta(54)$. This can be confirmed by carrying out a basis rotation
\begin{equation}
\begin{pmatrix}
h_1 \\
h_2 \\
h_S
\end{pmatrix} = \frac{1}{\sqrt{2}}\begin{pmatrix}
1 & -i & 0 \\
-i & 1 & 0 \\
0 & 0 & \sqrt{2}
\end{pmatrix} \begin{pmatrix}
\phi_1 \\
\phi_2 \\
\phi_3 
\end{pmatrix}.
\end{equation}
In the $\phi$ basis the scalar potential is:
\begin{subequations}\label{Eq:Delta54_Pot}
\begin{align}
V_2  ={}& \mu_{11}^2(\phi_{11} + \phi_{22} + \phi_{33}),\\
\begin{split}
V_4 ={}& \frac{1}{3}\left( 3 \lambda_1 + \lambda_2 \right)\left(\phi_{11}+\phi_{22}+\phi_{33}\right)^2 +4\lambda_3\left(\left|\phi_{12}\right|^2+\left|\phi_{23}\right|^2+\left|\phi_{31}\right|^2\right)  \\
 & - \frac{4 \lambda_2}{3}\left[\phi_{11}^2 + \phi_{22}^2 + \phi_{33}^2 -\phi_{11}\phi_{22}-\phi_{22}\phi_{33}-\phi_{33}\phi_{11}\right]\\
&+\left\{ 2i \lambda_7\,\phi_{13}\phi_{23}+ \sqrt{2}\lambda_4\,\phi_{21}\phi_{31}- i\sqrt{2}\lambda_4\,\phi_{32}\phi_{12} + \mathrm{h.c.} \right\}.
\end{split} \raisetag{3.5\baselineskip}
\end{align}
\end{subequations}
The potential has the structure of the $\Delta(54)$-symmetric scalar potential. This can be verified by checking eqs.~(52) and (53) of Ref.~\cite{deMedeirosVarzielas:2019rrp}. 

It is important to highlight that the aforementioned conclusions are specific to the $S_3$-symmetric 3HDM. For the general 3HDM, the complete set of CP-odd invariants required for explicit CP conservation remains unidentified. Apart from that, the above approach addresses only the CP properties of the scalar potential and not those of the vacuum. Spontaneous CP violation may still occur.

\section{Identifying continuous symmetries}\label{Sec:S3_br_cont_symm}

Continuous symmetries broken by vevs result in the emergence of massless scalars~\cite{Nambu:1960tm,Goldstone:1961eq,Goldstone:1962es}. These states are ruled out by experiments. In Ref.~\cite{Kuncinas:2020wrn} it was pointed out that several interesting vacua lead to the existence of massless scalars. These states arise from the spontaneous breaking of accidental continuous symmetries, which emerge due to constraints imposed by the minimisation conditions, mainly due to relations between the $\lambda$ couplings. In most cases massless states can be promoted to massive ones through the introduction of soft symmetry-breaking terms. Identifying additional symmetries in a basis-independent way is an involved routine. In Ref.~\cite{deMedeirosVarzielas:2019rrp} a technique for identifying these in terms of the 3HDMs was outlined.

Let us consider the neutral mass-squared matrices of the 3HDM. The general neutral mass-squared matrix $\mathcal{M}^2_0$ is of dimension six. The determinant of this matrix is a product of all its eigenvalues:
\begin{equation}
\det\left( \mathcal{M}^2_0 \right)= \prod _{i=1}^6 m_{H_i^0}^2,
\end{equation}
where $m_{H_i^0}^2$ are the mass-squared parameters of the scalars. Due to the existence of the neutral Goldstone boson, $G^0$, it is easy to conclude that the determinant will yield zero. 

We can find the eigenvalues by solving the characteristic equation:
\begin{equation}\label{Eq:det_eigenvalues_char}
\begin{aligned}
\det\left( \mathcal{M}^2_0 - \lambda\, \mathcal{I}_6 \right) &= \left( m_{H_1^0}^2 - \lambda \right)\left( m_{H_2^0}^2 - \lambda \right)...\left( m_{H_6^0}^2 - \lambda \right)\\
&= \sum_{i=0}^6 (-1)^i\lambda^i c_i,
\end{aligned}
\end{equation}
where $c_i$ are exponential Bell polynomials which satisfy the Cayley-Hamiltonian theorem:
\begin{subequations}
\begin{align}
c_0 &= \det\left( \mathcal{M}^2_0 \right) ,\\
\begin{split}
c_1 &= \frac{1}{120} \Bigg[ \tr\left( \mathcal{M}^2_0 \right)^5 - 10 \tr\left( \mathcal{M}^2_0 \right)^3 \tr\left( (\mathcal{M}^2_0)^2 \right)+ 15 \tr\left( \mathcal{M}^2_0 \right) \tr\left( (\mathcal{M}^2_0)^2 \right)^2\\ &\hspace{35pt}  + 20 \tr\left( \mathcal{M}^2_0 \right)^2 \tr\left( (\mathcal{M}^2_0)^3 \right) -20 \tr\left( (\mathcal{M}^2_0)^2 \right)\tr\left( (\mathcal{M}^2_0)^3 \right) \\ &\hspace{35pt}- 30 \tr\left( \mathcal{M}^2_0 \right) \tr\left( (\mathcal{M}^2_0)^4 \right) +24 \tr\left( (\mathcal{M}^2_0)^5 \right) \Bigg],
\end{split}\\
\begin{split}
c_2 &= \frac{1}{24}\left[ \tr\left( \mathcal{M}^2_0 \right)^4 - 6 \tr\left( \mathcal{M}^2_0 \right)^2 \tr\left( (\mathcal{M}^2_0)^2 \right)   + 3 \tr\left( (\mathcal{M}^2_0)^2 \right)^2 \right. \\ &\left. \hspace{35pt}+ 8 \tr\left( \mathcal{M}^2_0 \right) \tr\left( (\mathcal{M}^2_0)^3 \right) - 6 \tr\left( (\mathcal{M}^2_0)^4 \right)\right],\end{split}\\
c_3 &= \frac{1}{6} \left[ \tr\left( \mathcal{M}^2_0 \right)^3 - 3 \tr\left( \mathcal{M}^2_0 \right)\tr \left( (\mathcal{M}^2_0)^2 \right) + 2\tr\left( (\mathcal{M}^2_0)^3 \right) \right],\\
c_4 &= \frac{1}{2} \left[ \tr\left( \mathcal{M}^2_0 \right)^2 - \tr\left( (\mathcal{M}^2_0)^2 \right) \right],\\
c_5 &= \tr\left( \mathcal{M}^2_0 \right),\\
c_6 &= 1.
\end{align}
\end{subequations}

Due to the presence of the neutral Goldstone boson we have $c_0=0$. Of particular interest are the coefficients $c_{1..5}$, as the highest-order polynomial among them, satisfying $c_i=0$, indicates the presence of additional $i$ massless scalars.

A special role is played by the $\lambda_4$ couplings. For $\lambda_4\neq0$, in addition to $S_3$, there are two independent symmetries: the global $U(1)$ symmetry and the $\mathbb{Z}_2$ symmetry, under which $h_1 \to - h_1$. In some cases the minimisation conditions force $\lambda_4=0$. In this case the scalar potential becomes invariant under the larger $O(2)$ continuous symmetry~\cite{Das:2014fea}. The $O(2)$ symmetry is realised among the $h_1$ and $h_2$ fields,
\begin{equation}
\begin{pmatrix}
h_1^\prime \\
h_2^\prime
\end{pmatrix} = \begin{pmatrix}
\cos \theta & - \sin \theta \\
\sin\theta & \cos \theta
\end{pmatrix} \begin{pmatrix}
h_1 \\
h_2
\end{pmatrix}.
\end{equation}
The $O(2)$-symmetric 3HDM is also symmetric under the sign changes of either $h_2$ or $h_S$.

If, along with $\lambda_4=0$, it is required that $\lambda_7=0$, the underlying symmetry is increased to $O(2) \times U(1)$. Finally, if in addition to the previous conditions  $\lambda_2 + \lambda_3 = 0$ is imposed, the underlying symmetry is enhanced to $SU(2)$. If $\lambda_4=0$ is satisfied only along with $\lambda_2 + \lambda_3 = 0$, namely the C-III-c case, the underlying symmetry is still $O(2)$. As a result the $S_3$-symmetric scalar potential, due to the minimisation conditions, can be increased to the following continuous symmetries:
\begin{itemize}
\item {\boldmath$O(2)$}: $\lambda_4=0$. If spontaneously broken, an additional neutral massless states appears;
\item {\boldmath$O(2)\times U(1)$}: $\lambda_4=\lambda_7=0$. If spontaneously broken, two additional neutral massless states appears;
\item {\boldmath$SU(2)$}: $\lambda_4=\lambda_7 = \lambda_2 + \lambda_3  = 0$. If spontaneously broken, three additional neutral massless states appears.
\end{itemize}
Different cases are sumamrised in Table~\ref{Table:S3-real-vacua}.

{\renewcommand{\arraystretch}{1.1}
\begin{table}[htb]
\caption{ Massless states and mass degeneracies of different implementations of the unbroken $S_3$-symmetric 3HDM. In the second column relevant minimisation conditions are displayed which, in most cases, lead to additional continuous symmetries. The $L$ function indicates that a linear expression in its arguments vanishes. In the broken models only neutral massless states arise. The numbers are displayed by not taking into account the neutral would-be Goldstone boson. In the last column degeneracies refer to massive pairs, where the first entry in the parenthesis refers to the charged sector and the second to the neutral sector. }
\label{Table:S3-real-vacua}
\begin{center}
\begin{tabular}{|c|c|c|c|c|c|c|}
\hline\hline
Vacuum & \begin{tabular}[l]{@{}c@{}} Minimisation \\ conditions ($\lambda$) \end{tabular} & Symmetry & \begin{tabular}[l]{@{}c@{}} Massless \\ neutral states \end{tabular} & \begin{tabular}[l]{@{}c@{}} Mass \\ degeneracies \end{tabular}  \\ \hline \hline
R-I-1 &   & &  & (1, 2) \\ \hline
R-II-2 & $\lambda_4=0$   & $O(2)$ & 1 &  \\ \hline
R-II-3 & $\lambda_4=0$  & $O(2)$ & 1 &   \\ \hline
R-III & $\lambda_4=0$  & $O(2)$ & 1 &   \\ \hline\hline
C-I-a &   & &  & (none, 2) \\ \hline
C-III-b & $\lambda_4=0$ &  $O(2)$ & 1 &  \\ \hline
C-III-c &  $\lambda_4= \lambda_2 + \lambda_3 = 0$ & $O(2)$ & 2 &  \\ \hline
C-III-f,g & $\lambda_4=0$ & $O(2)$ & 1 &  \\ \hline
C-IV-a & $\lambda_4= \lambda_7 = 0$ &  $O(2)\times U(1)$ & 2 &  \\ \hline
C-IV-b & $\lambda_4=0$ & $O(2)$ & 1 &  \\ \hline
C-IV-c & \begin{tabular}[l]{@{}c@{}} $L(\lambda_2+\lambda_3,\lambda_4)$ \\ $L(\lambda_2+\lambda_3,\lambda_7)$ \end{tabular} &   & 1 &  \\ \hline
C-IV-d & $\lambda_4= \lambda_7 = 0$ & $O(2)\times U(1)$ & 2 &  \\ \hline
C-IV-e & $\lambda_4=0$ & $O(2)$ & 1 &  \\ \hline
C-IV-f & \begin{tabular}[l]{@{}c@{}} $L(\lambda_2+\lambda_3,\lambda_4)$ \\ $L(\lambda_2+\lambda_3,\lambda_7)$ \end{tabular} &  & 1 &  \\ \hline
C-V & \begin{tabular}[l]{@{}c@{}} $\lambda_4= \lambda_2 + \lambda_3$ \\ $=\lambda_7= 0$ \end{tabular} & $SU(2)$ & 3 &  \\ \hline \hline
\end{tabular}\vspace*{-9pt}
\end{center}
\end{table}}

In some instances it is possible to promote massless states to massive ones by introducing soft symmetry breaking terms:
\begin{align}\label{Eq:VSoftGenericBasis}
V_2^\prime &=\mu_{22}^2 \left( h_{11} - h_{22} \right) + \frac{1}{2} \left\lbrace  \mu_{12}^2 h_{12} + \mu_{S1}^2 h_{S1} + \mu_{S2}^2 h_{S2} + \mathrm{h.c.}\right\rbrace.
\end{align}

When adding soft terms, we do not need to consider all the different vacuum configurations (real and complex) and the various combinations of soft terms. Actually, we shall assume that all soft symmetry-breaking terms are introduced simultaneously. This results in six bilinear terms but only five parameters come from the vevs: three absolute values and two phases. Since there are more independent bilinear terms than the number of degrees of freedom of the vacuum, the minimisation conditions will not constrain or relate the vevs. Consequently, we only need to focus on which components of the vevs vanish. Following this approach, we classified different implementations in Ref.~\cite{Kuncinas:2020wrn}.

\subsection{Real vacua}

For real vacua we have:
\begin{itemize}
\setlength\itemsep{1.2em}
\item {\boldmath$(0,\,0,\,v_S)$}: \hspace*{5pt}\begin{tabular}[l]{@{}l@{}} $\mu_{SS}^2= -\lambda _8 v_S^2,$ \\ $\mu_{S1}^2=\mu_{S2}^2=0$. \end{tabular}
\item {\boldmath$(0,\,v,\,0)$}: \hspace*{5pt}\begin{tabular}[l]{@{}l@{}} $\mu_{11}^2=\mu_{22}^2 - \left(\lambda _1+\lambda _3\right) v^2,$\\ $\mu_{12}^2=0,$\\ $\mu_{S2}^2=v^2\lambda_4.$ \end{tabular}
\item {\boldmath$(v,\,0,\,0)$}: \hspace*{5pt}\begin{tabular}[l]{@{}l@{}} $\mu_{11}^2=-\mu_{22}^2- \left(\lambda _1+\lambda _3\right) v^2,$\\ $\mu_{12}^2=\mu_{S1}^2=0.$\end{tabular}
\item {\boldmath$(0,\,v_2,\,v_S)$}: \hspace*{5pt}\begin{tabular}[l]{@{}l@{}} $\mu_{SS}^2=-\frac{1}{2}\mu_{S2}^2\frac{ v_2}{v_S}+ \frac{1}{2}\lambda _4\frac{ v_2^3}{v_S}
-\frac{1}{2} (\lambda_5 + \lambda_6 + 2 \lambda_7) v_2^2-\lambda _8 v_S^2,$\\ $\mu_{11}^2=\mu_{22}^2-\frac{1}{2}\mu_{S2}^2\frac{v_S}{v_2} -\left( \lambda _1+ \lambda _3\right) v_2^2+\frac{3}{2} \lambda _4 v_2 v_S-\frac{1}{2} (\lambda_5 + \lambda_6 + 2 \lambda_7) v_S^2,$\\ $\mu_{12}^2=-\mu_{S1}^2\frac{v_S}{v_2}.$ \end{tabular}
\item {\boldmath$(v_1,\,0,\,v_S)$}: \hspace*{5pt}\begin{tabular}[l]{@{}l@{}} $\mu_{SS}^2=-\frac{1}{2}\mu_{S1}^2\frac{v_1}{v_S} -\frac{1}{2} (\lambda_5 + \lambda_6 + 2 \lambda_7) v_1^2-\lambda _8 v_S^2,$\\ $\mu_{11}^2=-\mu_{22}^2-\frac{1}{2}\mu_{S1}^2\frac{v_S}{v_1} - \left( \lambda _1+ \lambda _3\right) v_1^2-\frac{1}{2} (\lambda_5 + \lambda_6 + 2 \lambda_7) v_S^2,$\\ $\mu_{12}^2=-\mu_{S2}^2\frac{v_S}{v_1}-3\lambda_4v_1v_S.$ \end{tabular}
\item {\boldmath$(v_1,\,v_2,\,0)$}: \hspace*{5pt}\begin{tabular}[l]{@{}l@{}} $\mu_{11}^2=-\mu_{12}^2\frac{v_1^2+v_2^2}{4v_1v_2} -\left(\lambda _1+\lambda _3\right)(v_1^2+v_2^2),$\\ $\mu_{22}^2=\mu_{12}^2\frac{v_1^2-v_2^2}{4v_1v_2},$\\ $\mu_{S1}^2=\left[-\mu_{S2}^2+(v_2^2-3v_1^2)\lambda_4\right]\frac{v_2}{v_1}.$\end{tabular}
\item {\boldmath$(v_1,\,v_2,\,v_S)$}: \\ \hspace*{5pt}\begin{tabular}[l]{@{}l@{}} $\mu_{SS}^2=-\frac{1}{2}\mu_{S1}^2\frac{v_1}{v_S}-\frac{1}{2}\mu_{S2}^2\frac{v_2}{v_S}-\frac{1}{2}\lambda_4\frac{v_2(3v_1^2-v_2^2)}{v_S} -\frac{1}{2} (\lambda_5 + \lambda_6 + 2 \lambda_7)( v_1^2+ v_2^2)-\lambda _8 v_S^2,$ \\ $\mu_{11}^2=-\mu_{12}^2\frac{v_1^2+v_2^2}{4v_1v_2} -\frac{v_S}{4v_1}\mu_{S1}^2-\frac{v_S}{4v_2}\mu_{S2}^2-\frac{3}{4}\lambda_4\frac{(v_1^2+v_2^2)v_S}{v_2}$\\ $\hspace{27pt}- \left( \lambda _1+ \lambda _3\right)( v_1^2+ v_2^2)-\frac{1}{2} (\lambda_5 + \lambda_6 + 2 \lambda_7) v_S^2,$ \\ $\mu_{22}^2=\mu_{12}^2\frac{v_1^2-v_2^2}{4v_1v_2}-\frac{v_S}{4v_1}\mu_{S1}^2+\frac{v_S}{4v_2}\mu_{S2}^2-\frac{3}{4}\lambda_4\frac{v_S(3v_2^2-v_1^2)}{v_2}.$ \end{tabular}
\end{itemize}
As shown in the list above, not all soft symmetry-breaking terms survive. For instance, in the case of $(0,\,0,\,v)$ vacuum, the $\mu_{S1}^2$ and $\mu_{S2}^2$ terms vanish after minimising the scalar potential. In the absence of soft symmetry-breaking terms, there is mass degeneracy among the charged physical scalars, as well as two pairs of mass-degenerate neutral scalars. However, if either $\mu_{12}^2$ or $\mu_{22}^2$ is present, there is no mass degeneracy. 

By studying the above vacuum configurations, we can also hypothesise under which conditions the unwanted massless states can be promoted to massive ones. For example, for vacuum $(0,\,v,\,0)$, R-II-2-like, we can notice that the $\lambda_4=0$ condition is ``absorbed" by the $\mu_{S2}^2 = v^2 \lambda_4$ minimisation condition. In this case we still have one neutral massless state. The massless state is a consequence of relating $\mu_{S2}^2$ and $\lambda_4$. The presence of either $\mu_{22}^2$ or $\mu_{S1}^2$ bilinear term is necessary to ensure that no unwanted massless states appear.

Finally, in some cases, \textit{e.g.}, $(0,\,v_2,\,v_S)$, we can notice that the soft terms $\mu_{12}^2$ and $\mu_{S1}^2$ are mutually proportional. This indicates that they have to coexist, \textit{i.e.}, vanishing of either term requires vanishing of the other term.

\subsection{Complex vacua}

For complex vacua with soft terms we have the following implementations:
\begin{itemize}
\setlength\itemsep{1.2em}
\item {\boldmath$ (0,\hat v_2e^{i\sigma_2},\hat v_S) $}: \hspace*{5pt}\begin{tabular}[l]{@{}l@{}} $\mu_{SS}^2=- \frac{1}{2} (\lambda_5 + \lambda_6 - 2 \lambda_7) \hat{v}_2^2-\lambda _8 \hat{v}_S^2,$\\
$\mu_{11}^2=\mu_{22}^2-\left(\lambda _1+\lambda _3\right) \hat{v}_2^2+\cos\sigma_2\hat v_2\hat v_S\lambda_4- \frac{1}{2} (\lambda_5 + \lambda_6 - 2 \lambda_7)\hat{v}_S^2,$ \\
$\mu_{S2}^2=\hat v_2(\hat v_2\lambda_4-4\cos\sigma_2\hat v_S\lambda_7),$ \\
$\mu_{12}^2=\mu_{S1}^2=0.$ \end{tabular}
\item {\boldmath$(\hat v_1e^{i\sigma_1},0,\hat v_S)$}: \hspace*{5pt}\begin{tabular}[l]{@{}l@{}} $\mu_{SS}^2=-\frac{1}{2} (\lambda_5 + \lambda_6 - 2 \lambda_7) \hat{v}_1^2-\lambda _8 \hat{v}_S^2,$\\
$\mu_{11}^2=-\mu_{22}^2-\left( \lambda _1+ \lambda _3\right) \hat{v}_1^2-\frac{1}{2} (\lambda_5 + \lambda_6 - 2 \lambda_7) \hat{v}_S^2,$ \\
$\mu_{12}^2=-2\cos\sigma_1\hat v_1\hat v_S\lambda_4,$ \\ 
$\mu_{S1}^2=-4\cos\sigma_1\hat v_1\hat v_S\lambda_7,$ \\
$\mu_{S2}^2=-\lambda_4\hat v_1^2.$ \end{tabular}
\item {\boldmath$(\hat v_1 e^{i\sigma_1},\hat v_2e^{i\sigma_2},0)$}: \hspace*{5pt}\begin{tabular}[l]{@{}l@{}} $\mu_{11}^2=-(\lambda_1-\lambda_2)(\hat{v}_1^2+\hat{v}_2^2),$\\
$\mu_{22}^2=-(\hat v_1^2-\hat v_2^2)(\lambda_2+\lambda_3),$ \\
$\mu_{12}^2=-4\hat v_1\hat v_2\cos(\sigma_2-\sigma_1)(\lambda_2+\lambda_3),$\\
$\mu_{S1}^2=-2\hat v_1\hat v_2\cos(\sigma_2-\sigma_1)\lambda_4,$\\
$\mu_{S2}^2=-(\hat v_1^2-\hat v_2^2)\lambda_4.$ \end{tabular}
\item {\boldmath$(\hat v_1 e^{i\sigma_1},\hat v_2e^{i\sigma_2},\hat v_S)$} provided $\sin(\sigma_2-\sigma_1)\neq0$:\\ {\renewcommand{\arraystretch}{1.2} \hspace*{5pt}\begin{tabular}[l]{@{}l@{}} $\mu_{11}^2=\{2\hat v_S^2(\sin^2\sigma_1\hat w_1^2 + \sin^2\sigma_2\hat w_2^2)\mu_{SS}^2-4\hat v_1^2\hat v_2^2(\hat v_1^2+\hat v_2^2)\sin^2(\sigma_2-\sigma_1)(\lambda_1-\lambda_2)$ \\ 
\hspace{40pt}$-\hat v_S^2[2\hat v_1^2\hat v_2^2\sin^2(\sigma_2-\sigma_1)
-(\hat v_1^2+\hat v_2^2)(\sin^2\sigma_1\hat w_1^2 + \sin^2\sigma_2\hat w_2^2)](\lambda_5 + \lambda_6 - 2 \lambda_7)$\\
\hspace{40pt}$+2\hat v_S^4(\sin^2\sigma_1\hat w_1^2 + \sin^2\sigma_2\hat w_2^2)\lambda_8\}/[4\hat v_1^2\hat v_2^2\sin^2(\sigma_2-\sigma_1)],$ \\ 
$\mu_{22}^2=-\{2\hat v_S^2(\sin^2\sigma_1\hat w_1^2 - \sin^2\sigma_2\hat w_2^2)\mu_{SS}^2+4\hat v_1^2\hat v_2^2(\hat v_1^2
-\hat v_2^2)\sin^2(\sigma_2-\sigma_1)(\lambda_2+\lambda_3) $ \\ 
\hspace{45pt}$+4\cos\sigma_2\hat v_1^2\hat v_2^3\hat v_S\sin^2(\sigma_2-\sigma_1)\lambda_4$\\
\hspace{45pt}$+\hat v_S^2(\sin^2\sigma_1\hat w_1^2 - \sin^2\sigma_2\hat w_2^2)[(\hat v_1^2+\hat v_2^2)(\lambda_5 + \lambda_6 - 2 \lambda_7) +2\hat v_S^2\lambda_8]\}/$\\
\hspace{65pt}$[4\hat v_1^2\hat v_2^2\sin^2(\sigma_2-\sigma_1)],$ \\ 
$\mu_{12}^2=-\{2\sin\sigma_1\sin\sigma_2\hat v_S^2\mu_{SS}^2 
+4\hat v_1^2\hat v_2^2\cos (\sigma_2-\sigma_1) \sin^2 (\sigma_2-\sigma_1)(\lambda_2+\lambda_3)$\\
\hspace{45pt}$+2\cos\sigma_1\hat v_1^2\hat v_2\hat v_S\sin^2(\sigma_2-\sigma_1)\lambda_4
+\sin\sigma_1\sin\sigma_2(\hat v_1^2+\hat v_2^2)\hat v_S^2(\lambda_5 + \lambda_6 - 2 \lambda_7)$ \\ 
\hspace{45pt}$+2\sin\sigma_1\sin\sigma_2\hat v_S^4\lambda_8
\}/[\hat v_1\hat v_2\sin^2(\sigma_2-\sigma_1)],$ \\
$\mu_{S1}^2=-\{2\sin\sigma_2\hat v_S\mu_{SS}^2
+\hat v_1^2\hat v_2\sin[2(\sigma_2-\sigma_1)]\lambda_4
+\sin\sigma_2\hat v_S(\hat v_1^2+\hat v_2^2)(\lambda_5 + \lambda_6 - 2 \lambda_7)$ \\
\hspace{45pt}$+4\cos\sigma_1\hat v_1^2\hat v_S\sin(\sigma_2-\sigma_1)\lambda_7
+2\sin\sigma_2\hat v_S^3\lambda_8
\}/[\hat v_1\sin(\sigma_2-\sigma_1)],$ \\
 $\mu_{S2}^2=\{2\sin\sigma_1 \hat v_S\mu_{SS}^2
+\hat v_2(\hat v_2^2-\hat v_1^2)\sin(\sigma_2-\sigma_1)\lambda_4$\\
\hspace{40pt}$+\sin\sigma_1\hat v_S(\hat v_1^2+\hat v_2^2)(\lambda_5 + \lambda_6 - 2 \lambda_7) $ \\
\hspace{40pt}$-4\cos\sigma_2\hat v_2^2\hat v_S\sin(\sigma_2-\sigma_1)\lambda_7
+2\sin\sigma_1\hat v_S^3\lambda_8
\}/[\hat v_2\sin(\sigma_2-\sigma_1)].$ \end{tabular}}
\item {\boldmath$(\hat v_1 e^{i\sigma},\pm\hat v_2e^{i\sigma},\hat v_S)$} provided $\sin\sigma\neq0$: \\ {\renewcommand{\arraystretch}{1.2} \hspace*{5pt}\begin{tabular}[l]{@{}l@{}} $\mu_{SS}^2  =-\frac{1}{2}  \left(\hat{v}_1^2+\hat{v}_2^2\right)(\lambda_5 + \lambda_6 - 2 \lambda_7) -\lambda_8 \hat{v}_S^2,$\\
	$\mu_{11}^2  = \frac{-1}{4 \hat{v}_1 \hat{v}_2} \{ 4\hat v_1\hat v_2(\hat v_1^2+\hat v_2^2)(\lambda_1+\lambda_3)$\\
	\hspace{60pt}$+2 \hat{v}_1 \hat{v}_S \left[\left(\hat{v}_1^2+\hat{v}_2^2\right) \cos\sigma\lambda_4
	+\hat{v}_2 \hat{v}_S(\lambda_5 + \lambda_6 - 2 \lambda_7)\right] \pm(\hat v_1^2+\hat v_2^2)\mu_{12}^2 \},$\\
	$\mu_{22}^2  = \frac{1}{4 \hat{v}_1 \hat{v}_2} \{ 2 \lambda_4 \hat{v}_1 \hat{v}_S \left(\hat{v}_1^2-3 \hat{v}_2^2\right)  \cos\sigma\pm\left(\hat{v}_1^2-\hat{v}_2^2\right)\mu_{12}^2  \},$\\
	$\mu_{S1}^2  = \mp2 \hat{v}_1 (\lambda_4 \hat{v}_2+2 \lambda_7 \hat{v}_S \cos\sigma),$\\
	$\mu_{S2}^2  = \lambda_4 \left(\hat{v}_2^2-\hat{v}_1^2\right)-4 \lambda_7 \hat{v}_2 \hat{v}_S \cos\sigma.$ \end{tabular}}
\end{itemize}

\section{Yukawa Lagrangian}\label{Sec:SecScalarFermionInter}

So far we have identified different implementations in the $S_3$-symmetric potential based on real and complex vevs,  with and without complex coefficients. This was done in Section~\ref{Sec:SectionRealComplexVEV}. Then, since some of the models have complex coefficients, we had to test whether these survive or there is a basis in which they become real. This was done in Section~\ref{Sec:CPV_cond}. Finally, in Section~\ref{Sec:S3_br_cont_symm} we checked if the minimisation conditions could result in additional symmetries, namely continuous ones. Spontaneous breaking of these symmetries results in unwanted massless states. The next possible classification is to check the Yukawa Lagrangian: if it is possible to generate correct masses of fermions and reproduce their mixing, provided that the $S_3$ symmetry is expanded to fermions. We shall assume that neutrinos are massless.

\subsection[Different \texorpdfstring{$S_3$}{S3} representations]{Different \boldmath$S_3$ representations}

Since the scalar potential is $S_3$-symmetric, we need to consider that fermions also transform under $S_3$. The $S_3$ symmetric Yukawa Lagrangian in the context of 3HDMs was previously studied by several authors~\cite{Kubo:2003iw,Teshima:2005bk,Mondragon:2007nk,Mondragon:2007af,Teshima:2012cg,GonzalezCanales:2012blg,Ma:2013zca,GonzalezCanales:2013pdx,Das:2015sca,Cruz:2017add}. However, no systematic approach was discussed: different implementations of vevs, as a consequence of the minimisation conditions were not considered. There are multiple ways to group fermions into the $S_3$ multiplets. The most trivial approach relies on transforming all fermions as singlets or pseudosinglets provided that $v_S \neq 0$. We recall that $\left[\mathbf{1} \otimes \mathbf{1} \otimes \mathbf{1}\right]_\mathbf{1}$ and $\left[\mathbf{1^\prime} \otimes \mathbf{1} \otimes \mathbf{1^\prime}\right]_\mathbf{1}$, see eqs.~\eqref{Eq:S3_diff_tensor_pr}. When fermions transform as singlets under $S_3$, the fermion mass matrices are given by:
\begin{subequations} \label{Eq.FMMws}
\begin{align}
\mathcal{M}_u={}&\frac{1}{\sqrt{2}} Y^u w^*_S,\\
\mathcal{M}_d={}&\frac{1}{\sqrt{2}} Y^d w_S,
\end{align}
\end{subequations}
where the Yukawa couplings, which can be complex, and are not constrained by the $S_3$ symmetry. In this sense, such Yukawa Lagrangian is identical to the SM case. It is always possible to generate correct fermionic masses and the CKM matrix. 

Another option involves assigning non-trivial $S_3$ charges to fermions by requiring that those transform under the singlet-doublet representation as
\begin{equation}
\begin{aligned}
&\mathbf{2}: \hspace{10pt} \begin{pmatrix}
\overline{Q}_1 \\ \overline{Q}_2
\end{pmatrix}_{\!\!L}, \,\,
\begin{pmatrix}
u_1 \\ u_2
\end{pmatrix}_{\!\!R}, \,\,
\begin{pmatrix}
d_1 \\ d_2
\end{pmatrix}_{\!\!R}, \,\,
\begin{pmatrix}
h_1 \\ h_2
\end{pmatrix},\\
&\mathbf{1}: \hspace{20pt}\overline{Q}_{3L}, \,\, u_{3R}, \,\, d_{3R}, \,\, h_S.
\end{aligned}
\end{equation}

Now, consider possible Yukawa terms. For an easier derivation of the $S_3$ singlets we list results from eq.~\eqref{Eq:S3_diff_tensor_pr}:
\begin{subequations}
\begin{align*}
\mathbf{ 1 } \otimes \text{any} &= \text{any},\\
\mathbf { 1^\prime }  \otimes \mathbf { 1^\prime } &= \mathbf { 1 },\\
\mathbf { 1^\prime } \otimes \mathbf { 2 } &= \mathbf { 2 },\\ 
\mathbf { 2 } \otimes \mathbf { 2 } &= \mathbf { 1 } \oplus \mathbf { 1^\prime } \oplus \mathbf { 2 }.
\end{align*}
\end{subequations}
For example, if we want to couple $(h_S)_\mathbf{ 1 } $ to fermions, there are two possibilities: it has to couple to two $\mathbf{ 1 } $ or to two $\mathbf{ 2 } $. No other combination will result in an $S_3$ singlet. If we want to write a term $\overline{f}_\mathbf{ 2 } (h_S)_\mathbf{ 1 } f_\mathbf{ 2 }$, it has to be multiplied by a single Yukawa coupling; after all, it should be viewed as a unique $S_3$ singlet. Considering the down-type quark sector, we get~\cite{Kubo:2004ps}:
\begin{subequations}\label{Eq: YukawaSinglets}
\begin{align}
&y_1^d: \,\,\, \overline{ Q_{1L}^0}h_S d_{1R}^{\,0}+\overline{Q_{2L}^0}h_S d_{2R}^{\,0},  \\
&y_2^d: \,\,\, \left(\overline{ Q_{1L}^0}h_2 + \overline{ Q_{2L}^{0}} h_1\right)d_{1R}^{\,0} + \left( \overline{Q_{1L}^0} h_1-\overline{Q_{2L}^0}h_2 \right)d_{2R}^{\,0}, \\
&y_3^d: \,\,\, \overline{ Q_{3L}^0}h_S d_{3R}^{\,0}, \\
&y_4^d: \,\,\, \left(\overline{ Q_{1L}^0}h_1+\overline{ Q_{2L}^0}h_2\right)d_{3R}^{\,0}, \\
&y_5^d: \,\,\, \overline{ Q_{3L}^0}\left(h_1 d_{1R}^{\,0}+h_2 d_{2R}^{\,0}\right),
\end{align}
\end{subequations}
where $y_i^d$ are the Yukawa couplings.

As a result, the quark sector of the Yukawa Lagrangian is given by:
\begin{subequations}
\begin{align}
&\begin{split}-\mathcal{L}_Y^d ={}& y_1^d \left[ \overline{ Q_{1L}^0}h_S d_{1R}^{\,0}+\overline{ Q_{2L}^0}h_S d_{2R}^{\,0} \right] + y_2^d \left[ (\overline{ Q_{1L}^0}h_2+\overline{ Q_{2L}^0}h_1)d_{1R}^{\,0} +(\overline{ Q_{1L}^0}h_1-\overline{ Q_{2L}^0}h_2)d_{2R}^{\,0} \right] \\
&\, + y_3^d \overline{ Q_{3L}^0}h_S d_{3R}^{\,0} + y_4^d \left[ (\overline{ Q_{1L}^0}h_1+\overline{ Q_{2L}^0}h_2)d_{3R}^{\,0} \right] + y_5^d \overline{ Q_{3L}^0}(h_1 d_{1R}^{\,0}+h_2 d_{2R}^{\,0}),
\end{split}\\
&\begin{split}-\mathcal{L}_Y^u ={}& y_1^u \left[ \overline{ Q_{1L}^0}\tilde{h}_S u_{1R}^{\,0}+\overline{ Q_{2L}^0}\tilde{h}_S u_{2R}^{\,0} \right] + y_2^u \left[ (\overline{ Q_{1L}^0}\tilde{h}_2+\overline{ Q_{2L}^0}\tilde{h}_1)u_{1R}^{\,0} +(\overline{ Q_{1L}^0}\tilde{h}_1-\overline{ Q_{2L}^0}\tilde{h}_2)u_{2R}^{\,0} \right] \\
&\, + y_3^u \overline{ Q_{3L}^0}\tilde{h}_S u_{3R}^{\,0} + y_4^u \left[ (\overline{ Q_{1L}^0}\tilde{h}_1+\overline{ Q_{2L}^0}\tilde{h}_2)u_{3R}^{\,0} \right] + y_5^u \overline{ Q_{3L}^0}(\tilde{h}_1 u_{1R}^{\,0}+\tilde{h}_2 u_{2R}^{\,0}).
\end{split}
\end{align}
\end{subequations}
Note that we can extract the Yukawa couplings to the front since those are coefficients. However, writing them in terms of the fermionic mass matrix yields:
\begin{subequations}\label{Eq:GeneralFM}
\begin{align}
\mathcal{M}_u= \frac{1}{\sqrt{2}}
\begin{pmatrix}
y_1^u v_S^\ast+y_2^u v_2^\ast & y_2^u v_1^\ast & y_4^u v_1^\ast \\
y_2^u v_1^\ast & y_1^u v_S^\ast-y_2^u v_2^\ast & y_4^u v_2^\ast \\
y_5^u v_1^\ast & y_5^u v_2^\ast & y_3^u v_S^\ast
\end{pmatrix},\\
\mathcal{M}_d= \frac{1}{\sqrt{2}}
\begin{pmatrix}
y_1^dv_S+y_2^dv_2 & y_2^d v_1 & y_4^d v_1 \\
y_2^d v_1 & y_1^d v_S-y_2^d v_2 & y_4^d v_2 \\
y_5^d v_1 & y_5^d v_2 & y_3^d v_S
\end{pmatrix}\label{Eq:SS_rep_D}.
\end{align}
\end{subequations}
Decomposition of the mass matrices into different parts yields:
\begin{subequations}\label{Eq:DifferentMDoublets}
\begin{align}
\mathcal{M}_d  = \begin{pmatrix}
0 & y_2 & y_4 \\
y_2 & 0 & 0 \\
y_5 & 0 & 0
\end{pmatrix}v_1 + \begin{pmatrix}
y_2 & 0 & 0 \\
0 & -y_2 & y_4 \\
0 & y_5 & 0
\end{pmatrix} v_2 + \begin{pmatrix}
y_1 & 0 & 0 \\
0 & y_1 & 0 \\
0 & 0 & y_3
\end{pmatrix} v_S.
\end{align}
\end{subequations}
In the above form it is straightforward to consider which of the $S_3$ implementations result in unrealistic cases. For example, models with a single active $SU(2)$ doublet result in unrealistic  eigenvalues: for R-I-2a the first generation becomes massless, for R-II-2 one of the eigenvalues is negative, for R-I-1 two states are mass-degenerate.

Apart from assigning the fermionic fields to the singlet-doublet representation, one could consider a pseudosinglet-doublet representation. Different patterns start to emerge. 
This is an important moment to emphasise that the analysis is conducted within the singlet-doublet representation of the scalar sector. If we had considered a pseudosinglet-doublet representation, it would have been necessary to include at least one fermionic field transforming in the $S_3$ pseudosinglet representation, since $\mathbf { 1^\prime }  \otimes \mathbf { 1^\prime } = \mathbf { 1 }$. Back to the pseudosinglet-doublet representation of fermions, it is possible to assign  $\mathbf{1^\prime}$ to fermions in the following way:
\begin{subequations}\label{Eq:Other_rep_d}
\begin{align}
\mathbf{1^\prime}:Q_3,\,d_{3R}:\qquad  &\mathcal{M}_d=
\frac{1}{\sqrt{2}}\begin{pmatrix}
y_1^dv_S+y_2^dv_2 & y_2^d v_1 & y_4^d v_2 \\
y_2^d v_1 & y_1^d v_S-y_2^d v_2 & -y_4^d v_1 \\
y_5^d v_2 & -y_5^d v_1 & y_3^d v_S
\end{pmatrix},\\
\mathbf{1^\prime}:Q_3,~\mathbf{1}:\,d_{3R}:\qquad  &\mathcal{M}_d=
\frac{1}{\sqrt{2}}\begin{pmatrix}
y_1^dv_S+y_2^dv_2 & y_2^d v_1 & y_4^d v_1 \\
y_2^d v_1 & y_1^d v_S-y_2^d v_2 & y_4^d v_2 \\
y_5^d v_2 & -y_5^d v_1 & 0
\end{pmatrix},\label{Eq:MD_ps_s}\\
\mathbf{1}:Q_3,~\mathbf{1^\prime}:\,d_{3R}:\qquad  &\mathcal{M}_d=
\frac{1}{\sqrt{2}}\begin{pmatrix}
y_1^dv_S+y_2^dv_2 & y_2^d v_1 & y_4^d v_2 \\
y_2^d v_1 & y_1^d v_S-y_2^d v_2 & -y_4^d v_1 \\
y_5^d v_1 & y_5^d v_2 & 0
\end{pmatrix}.\label{Eq:MD_s_ps}
\end{align}
\end{subequations}
The latter two representations, eq.~\eqref{Eq:MD_ps_s} and eq.~\eqref{Eq:MD_s_ps}, are equivalent. Without loss of generality, we can re-define $\{y_2^d,\, y_5^d\} \to - \{y_2^d,\, y_5^d\}$ of eq.~\eqref{Eq:MD_s_ps}. Next, we can rotate $\mathcal{M}_d$ by
\begin{equation}
U =  \begin{pmatrix}
0 & -1 & 0 \\
1 & 0 & 0 \\
0 & 0 & 1
\end{pmatrix}.
\end{equation}
In the new basis $\mathcal{M}_d$ of eq.~\eqref{Eq:MD_s_ps} becomes $\mathcal{M}_d$ of eq.~\eqref{Eq:MD_ps_s}. An interesting feature of the mixed representations is that the number of Yukawa couplings decreases from five to four, $y_3^d=0$.

Apart from that it is possible to assign different chiral states to different representations:
\begin{equation}
\mathbf{1}:Q_3,~\mathbf{1^\prime}:\,u_{3R},\,\,d_{3R}, \qquad 
\mathbf{1^\prime}:Q_3,~\mathbf{1}:\,u_{3R},\,\,d_{3R}.
\end{equation}
Moreover, we could also assign the up-quarks and down-quarks to different representations:
\begin{equation}
\begin{aligned}[c]
\mathbf{1}:Q_3,~\mathbf{1}:\,u_{3R},~\mathbf{1^\prime}:\,d_{3R},\\
\mathbf{1}:Q_3,~\mathbf{1^\prime}:\,u_{3R},~\mathbf{1}:\,d_{3R},
\end{aligned}
\qquad
\begin{aligned}[c]
\mathbf{1^\prime}:Q_3,~\mathbf{1}:\,u_{3R},~\mathbf{1^\prime}:\,d_{3R},\\
\mathbf{1^\prime}:Q_3,~\mathbf{1^\prime}:\,u_{3R},~\mathbf{1}:\,d_{3R}.
\end{aligned}
\end{equation}
In total, this brings us to eight different possibilities to construct the Yukawa Lagrangian due to different $S_3$ charges. These are: four combinations in the down-type sector, eqs.~(\ref{Eq:SS_rep_D}), (\ref{Eq:Other_rep_d}) times two different possibilities for the up-type quarks. Due to the fact that two representations yield identical mass matrices, $(\mathbf{1}:Q_3,\,\mathbf{1^\prime}:\,u_{3R},\,d_{3R})$ and $(\mathbf{1^\prime}:Q_3,\,\mathbf{1}:\,u_{3R},\,d_{3R})$, the number of independent solutions is reduced by one.

\subsection{Analysing different vacua}

After discussing different structures of the Yukawa Lagrangian, we can proceed to check whether any of the vacuum configurations could result in interesting structures. We restrict discussion to implementations with CP violation, cases which could explain the structure of the CKM matrix, and require that the discussed cases do not result in massless states, exact $S_3$ symmetry is assumed. This limits our discussion to the following models:
\begin{itemize}
\item \textbf{R-I-1}, $(0,\,0,\, v_S)$\\
Since only $v_S \neq 0$, this limits our possibilities to the discussion of trivial transformations of fermions. The complex CKM parameters would then be generated by complex Yukawa couplings. This case is of no particular interest.

\item \textbf{R-I-2a},  $(v_1,\,0,\, 0)$, and \textbf{R-I-2b,c},  $(v_1,\,\pm \sqrt{3} v_1,\, 0)$

Grouping fermions into identical singlet or pseudosinglet representations results in $\det\left(\mathcal{M}_f\right)=0$, $i.e.$, there are massless fermions. However, grouping fermions into mixed representations of singlets and pseudosinglets yields $\det\left(\mathcal{M}_f\right) \neq 0$. Nonetheless, when diagonalised, the form of the mass matrices predicts that two of the fermions are nearly mass degenerate, which is not true.

\item \textbf{C-I-a},  $(\hat v_1,\,\pm i \hat v_1 ,\, 0)$

The Hermitian mass-squared matrix, see eq.~\eqref{Eq:M_Herm_f}, yields:
\begin{equation}
\mathcal{H}_d = \mathcal{M}_d \mathcal{M}_d^\dagger = \frac{\hat v_1^2}{2}\begin{pmatrix}
|y_4^d|^2 + 2 |y_2^d|^2 & \pm i \left( 2|y_2^d|^2 -|y_4^d|^2 \right) & 0 \\
\mp i \left( 2|y_2^d|^2 -|y_4^d|^2 \right) &  |y_4^d|^2 + 2 |y_2^d|^2 & 0 \\
0 & 0 & 2|y_5^d|^2
\end{pmatrix},
\end{equation}
indicating that the CKM matrix is unrealistic.

\item \textbf{C-III-a}, $(0,\,\hat{v}_2 e^{i\sigma_2},\,\hat{v}_S)$

After transforming the mass matrix, it can be written as
\begin{equation}
\mathcal{M}_d = \frac{1}{\sqrt{2}}\begin{pmatrix}
e^{i \sigma_2} y_2^d \hat v_2 + y_1^d \hat v_S & 0 & 0 \\
0 & -e^{i \sigma_2} y_2^d \hat v_2 +  y_1^d \hat v_S &  e^{i \sigma_2} y_4^d \hat v_2 \\
0 & e^{i \sigma_2} y_5^d \hat v_2 &  y_3^d \hat v_S
\end{pmatrix},
\end{equation}
indicating that the CKM matrix is unrealistic. A choice of mixed representations results  in $\det\left(\mathcal{M}_f\right)=0$. Therefore, the only choice is to assume that fermions transform trivially under $S_3$.

\item \textbf{C-III-h}, $(\sqrt{3} \hat{v}_2 e^{i \sigma_2},\,\pm \hat{v}_2 e^{i \sigma_2},\,\hat{v}_S)$

In the singlet-doublet representation, by utilisng
\begin{equation}
U = \begin{pmatrix}
\cos \theta & \sin \theta & 0 \\
-\sin \theta & \cos \theta & 0 \\
0 & 0 & 1
\end{pmatrix},
\end{equation}
with $\theta = - \pi/3$, we get the down-type mass matrix
\begin{equation}
U \mathcal{M}_d U^\mathrm{T} = \frac{1}{\sqrt{2}} \begin{pmatrix}
- 2 e^{i \sigma_2} y_2^d \hat w_2 + y_1^d \hat w_S & 0 & 0 \\
0 & 2 e^{i \sigma_2} y_2^d \hat w_2 + y_1^d \hat w_S & 2 e^{i \sigma_2} y_4^d \hat w_2 \\
0 & 2 e^{i \sigma_2} y_5^d \hat w_2 & y_3^d \hat w_S
\end{pmatrix}.
\end{equation}
In the case of the pseudosinglet representation, by choosing $\theta = - 5\pi/6$ and re-defining $y_2^d \to -y_2^d$, we get the same mass matrix. As in the previous cases, the block-diagonal form suggests that the CKM matrix is unrealistic. Mixed representations yield $\det\left(\mathcal{M}_f\right)=0$. The only feasible option is to adopt a trivial Yukawa sector.
\end{itemize}

Although it is possible to assign non-trivial charges to fermions in several of the above listed examples, none of these choices yield accurate experimental predictions. Therefore, one has to assume that all fermions are singlets under $S_3$, which, in some cases is not possible due to $v_S = 0$.

The remaining cases (C-IV-c, C-IV-f, C-IV-g, C-V) require further discussion, as they have to be explored numerically due to the structure of the vacuum. A comprehensive and systematic study of these implementations, encompassing both the scalar and Yukawa sectors, is beyond the scope of the work. Instead, we rely on a simplified tree-level analysis, disregarding the leptonic sector, to determine whether a specific model can be ruled out. We fit several parameters, adopting a 3-$\sigma$ tolerance of values taken from the PDG~\cite{ParticleDataGroup:2024cfk}:
\begin{itemize}
\item Masses of the up-type and down-type quarks;
\item The absolute values, arguments of the unitarity triangle ($\alpha,\,\sin 2\beta,\,\gamma$) and independent measure of CP violation ($J$)~\cite{Jarlskog:1985ht,Dunietz:1985uy,Bernabeu:1986fc} of the CKM matrix;
\end{itemize}
For the C-V model, we perform several additional checks:
\begin{itemize}
\item Interactions of the SM-like Higgs boson with fermions. We assume the Higgs boson signal strength in the $b$-quark channel~\cite{CMS:2020zge,ATLAS:2020fcp,ATLAS:2020jwz} as a reference point and apply the corresponding limits to other channels;
\item Suppressed scalar mediated FCNC~\cite{CMS:2021gfa,ATLAS:2022gzn};
\item CP properties of the SM-like Higgs boson~\cite{ATLAS:2020ior,CMS:2022dbt};
\item Upper limit on the decay of the $t$-quark into lighter charged scalars when decays are not kinematically suppressed~\cite{CMS:2020osd,ATLAS:2021zyv};
\end{itemize}
Additional checks are conducted only for the C-V implementation, as it requires input from the scalar sector. Specifically, determining the interaction strength of the SM-like Higgs boson with fermions necessitates knowledge of the scalar mass eigenstates. The other models (C-IV-c, C-IV-f, C-IV-g) cannot produce a realistic scalar content unless the $S_3$ symmetry is softly broken. We do not consider the scalar sectors of these.

Fitting the discussed constraints is performed by taking vevs as inputs and scanning over ten free Yukawa couplings ($y_i^u,\,y_i^d$) in the range $\{|y_i^u|,\,|y_i^d|\} \leq \sqrt{4\pi}$. One might worry about hitting the Landau pole. Actually, only two of the Yukawa couplings (one for the up-type quarks and one for the down-type quarks) are typically within the range of $0.1 \leq\{|y_i^u|,\,|y_i^d|\} \leq 1.5$. Other Yukawa couplings are of order $\mathcal{O}(10^{-10}) - \mathcal{O}(10^{-3})$. Additionally, a valid fit can also be achieved when two of the ten Yukawa couplings are set to zero. However, this scenario provides less flexibility in fitting the scalar-fermion interactions. Partially due to the fact that the numerical scan was not performed in the scalar sector (additional limits on vevs) it was possible to achieve good fits in the fermionic sector with only eight different Yukawa couplings. For the discussion of the numerical approach consider Ref.~\cite{Kuncinas:2023ycz}.

\newpage
The remaining models are:
\begin{itemize}

\item \textbf{C-IV-c}, $(\sqrt{1 + 2 \cos^2 \sigma_2}\hat{v}_2,\,\hat{v}_2 e^{i \sigma_2},\,\hat{v}_S)$

For fermions transforming under $(\mathbf{1}:Q_3,~\mathbf{1^\prime}:\,u_{3R},\,\,d_{3R})$ or $(\mathbf{1^\prime}:Q_3,~\mathbf{1}:\,u_{3R},\,\,d_{3R})$ representations, assuming real Yukawa couplings, the CKM matrix is not in agreement with the experimental observations. Therefore, one has to assume complex Yukawa couplings. In the non-mixed representations it is sufficient to assume real Yukawa couplings.

\item \textbf{C-IV-f}, $(\sqrt{2 + \frac{\cos(\sigma_1 - 2\sigma_2)}{\cos \sigma_1}} \hat{v}_2 e^{i\sigma_1},\,\hat{v}_2 e^{i \sigma_2},\,\hat{v}_S)$

Due to the fact that the C-IV-c implementation is contained within C-IV-f with $\sigma_1=0$ results do not differ. Though there is more freedom due to the $\sigma_1$ phase. One still has to require complex Yukawa couplings in the mixed representations.

\item \textbf{C-IV-g}, $(\hat v_1e^{i\sigma_1},\, \pm i \hat v_1e^{i\sigma_1},\, \hat v_S)$

While the vacuum configuration is totally different from C-IV-c, there are still three degrees of freedom coming from vevs. The results of fitting the Yukawa sector are identical to those of C-IV-c

This model is only possible when $\lambda_4^\mathrm{I} \neq 0$. There are negative mass-squared scalars present. Introduction of soft symmetry breaking terms could possibly produce non-negative eigenvalues.

\item \textbf{ C-V}, $(\hat{v}_1 e^{i\sigma_1},\,\hat{v}_2 e^{i\sigma_2},\,\hat{v}_S)$

The three previous implementations result in massless states (C-IV-c, C-IV-f) or negative mass-squared terms (C-IV-g). Contrary, the scalar sector of C-V is well-behaved. In view of this, C-V was further scanned based on:
\begin{itemize}
\item Perturbativity, stability and unitarity constraints, see Section~\ref{Sec:Constraints_3HDM_gen};
\item LEP constraints. We set a conservative lower bound for the charged scalars, $m_{\varphi_i^\pm} \geq 70\text{ GeV}$~\cite{Pierce:2007ut,Arbey:2017gmh}. Decays of the $W^\pm$ and $Z$ into a pair of scalars are kinematically suppressed~\cite{Schael:2013ita};
\item One of the neutral scalars has to be $h \approx 125$~GeV;
\item Decay of the SM-like Higgs boson into lighter scalars is assumed to be within $\mathrm{Br(h\to \text{BSM}) \lesssim 0.1}$. This constraint is approximated by fixing the total width of the SM-like Higgs boson to be $\Gamma_h^\mathrm{tot} = 5.6\,\text{MeV}$. This includes SM-like scalar-gauge and scalar-fermion couplings;
\item SM-like scalar-gauge couplings;
\item Electroweak oblique parameters, see Section~\ref{Sec:SectionSTU};
\end{itemize}
Some constraints were not considered: di-photon partial width, $B$-physics, the electric dipole moment. 

For C-V we can parameterise vevs in terms of two angles:
\begin{equation}
\hat v_1 = v \sin \alpha \cos \beta, \quad \hat v_2 = v \cos \alpha \cos \beta, \quad \hat v_S = v \sin \beta.
\end{equation}
The available parameter space of the parameterised vevs, after applying the discussed constraints, is displayed in Figure ~\ref{Fig:C-V_VEVs}. The central point of the left panel corresponds to $\hat v_1^2=\hat v_2^2=v^2/4$ and $\hat v_S^2=v^2/2$. A striking feature of the right panel is the presence of alternating bands of allowed and excluded regions, $\sigma_2\simeq\sigma_1\pm n\pi$. When both complex phases approach zero (the limit of real vevs) we get that: $\lambda_4^\mathrm{R} \to 0$, both $(\lambda_2 + \lambda_3)$ and $\lambda_7$ being periodic in $\sigma_2$, with a period of $\pi$, diverge and break the perturbativity limit. This does not happen in the real C-V implementation due to the fact that the minimisation conditions are no longer interconnected by the $\lambda_4^\mathrm{I}$ coupling, or rather they are forced to vanish.

\begin{figure}[htb]
\begin{center}
\includegraphics[scale=0.275]{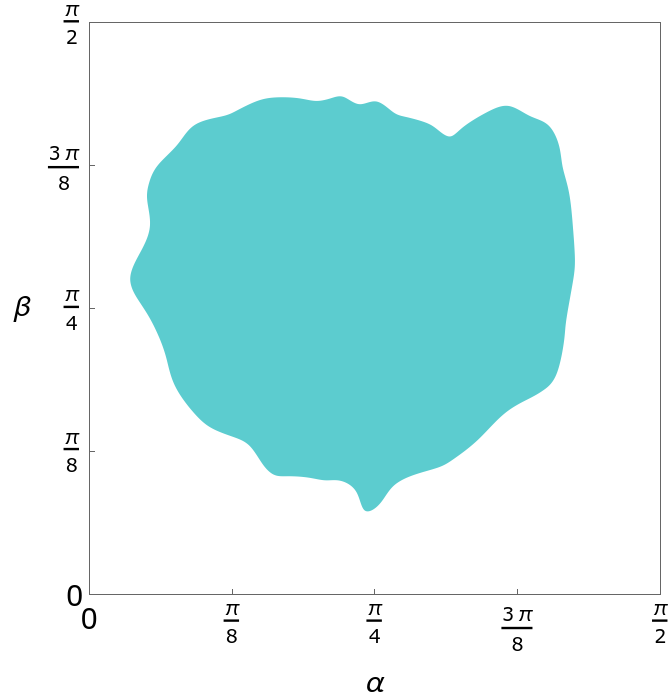}\hspace*{10pt}
\includegraphics[scale=0.275]{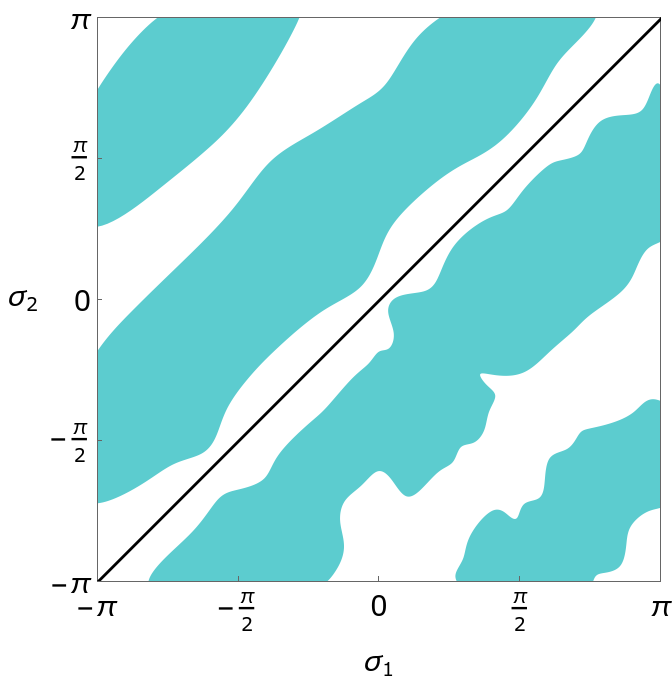}
\end{center}
\caption{Scatter plots of vevs in the C-V implementation with $\lambda_4^\mathrm{I}\neq0$. Left: the absolute values of vevs parameterised in terms of two angles. Right: phases of the vevs. Figures taken from Ref.~\cite{Kuncinas:2023ycz}.}
\label{Fig:C-V_VEVs}
\end{figure}

\begin{figure}[htb]
\begin{center}
\includegraphics[scale=0.275]{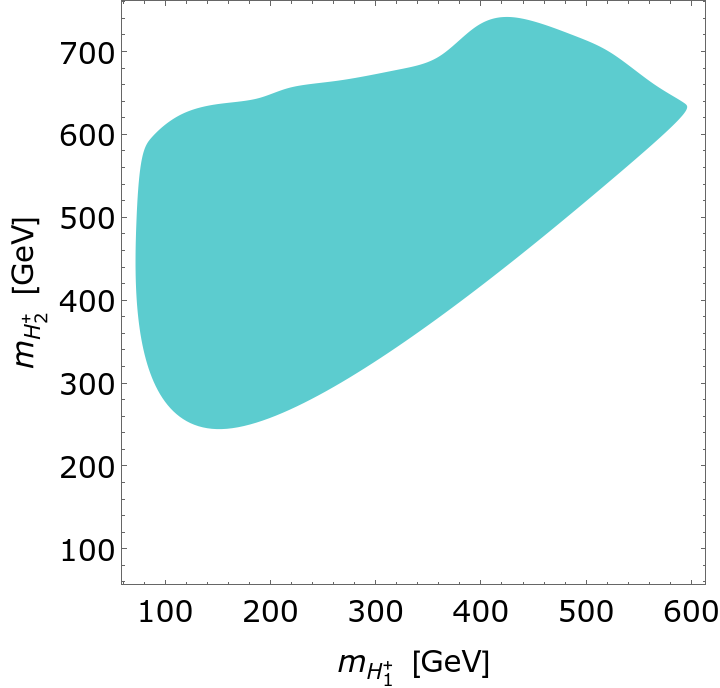}
\includegraphics[scale=0.275]{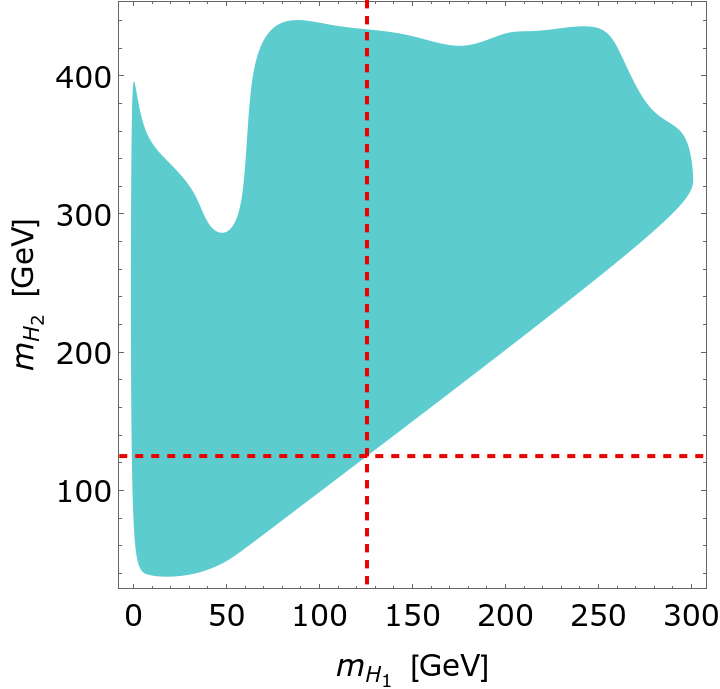}
\includegraphics[scale=0.275]{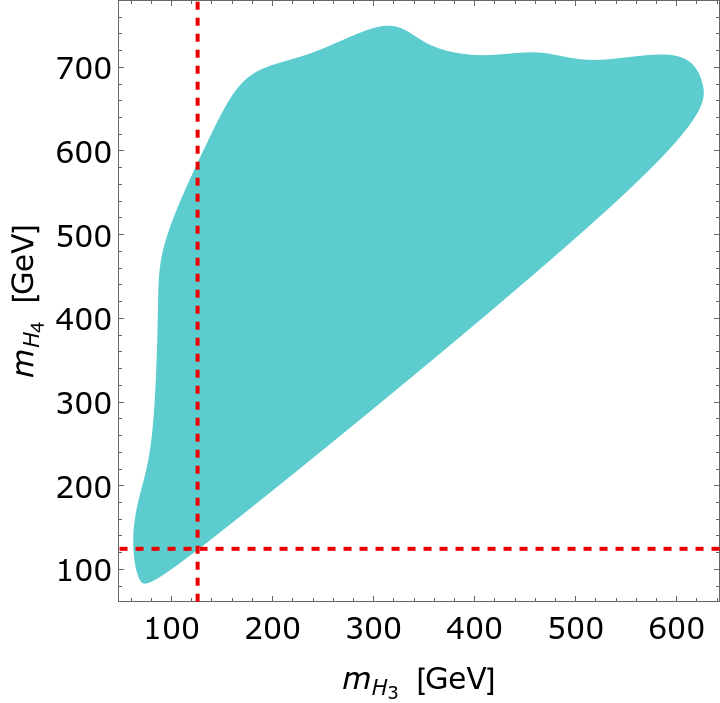}
\end{center}
\caption{Scatter plots of masses that satisfy constraints in the C-V implementation with $\lambda_4^\mathrm{I}\neq0$. Left: the charged sector, $H_i^\pm$. Middle and right: the active sector, $H_i$. In the neutral sector the dashed red line indicates a 125 GeV state. Possible locations of the SM-like $h$ are indicated by the dashed lines. Figures taken from Ref.~\cite{Kuncinas:2023ycz}.}
\label{Fig:C-V_masses}
\end{figure}

The complex C-V implementation accommodates two charged scalars, $H_i^\pm$ and five neutral physical scalars ($h,H_1, H_2, H_3, H_4$) which all mix. Mixing of these results in CP-indefinite states. The $H_i$ scalars are assumed to be ordered as $m_{H_i} \leq m_{H_{i+1}}$. The $h$ state must be compatible with the SM-like Higgs boson. There is no strict requirement that the SM-like Higgs boson has to be the lightest. The only requirement is that it has to be within the mass tolerance of the 125 GeV state. The mass-scatter plots are presented in Figure~\ref{Fig:C-V_masses}. For different mass hierarchies, the dashed line indicates the 125 GeV state. Presence of the red line for $m_{H_4}$ suggests that the SM-like Higgs boson could be the heaviest among the neutral states. We acknowledge that some regions of the parameter space remain unconstrained due to the incomplete consideration of all LHC-related constraints in Ref.~\cite{Kuncinas:2023ycz}.

It is of interest to consider light scalars~\cite{Fox:2017uwr,Cepeda:2021rql,Biekotter:2022jyr,Robens:2022zgk,Plantey:2022jdg}. One of the striking features of the discussed implementation is the presence of light states $m_{H_1} = \mathcal{O}(\text{MeV})$, not excluded by the considered constraints. There is sufficient freedom to suppress decays of the SM-like Higgs boson to such light states, $g(h H_1 H_1) \sim \mathcal{O}(10^{-6})$. These light scalars arise for small $\lambda_4$ and $\lambda_7$ values. A simple assumption might be that this case becomes similar to C-V with $\lambda_4^\mathrm{I}=0$. However, in that case we have $\lambda_2 + \lambda_3 =0$ due to the minimisation conditions, while for C-V with $\lambda_4^\mathrm{I} \neq 0$ we have $0.1 < \lambda_2 + \lambda_3 < 3$. Apart from that, C-V with $\lambda_4^\mathrm{I}=0$ is enlarged to $SU(3)$, and there are three unwanted Goldstone states present.

Interactions of the neutral $H_i$ scalars with fermions are left unconstrained. Unlike in the SM case, where $g(h \bar f  f) = - i\,m_{f}/v$, due to the CP-indefinite nature of the C-V scalars, the scalar-fermion couplings are more involved. In general, we have:
\begin{equation}
g(H_i \bar{f}_j f_k) =  \bar{f}_j \left( V_f^\dagger \mathcal{M}_{(i,f)}^\mathrm{int} U_f P_R + U_f^\dagger \left(\mathcal{M}_{(i,f)}^\mathrm{int}\right)^\dagger V_f  P_L  \right)_{jk} f_k,
\end{equation}
where $\mathcal{M}_f^\mathrm{int}$ represents interaction matrices, which, when rotated, are given by:
\begin{equation} \label{Eq:MintDiag}
\tilde{\mathcal{M}}_{(i,f)}^\mathrm{int} = V_f ^\dagger \mathcal{M}_i^\mathrm{int} U_f.
\end{equation}
Therefore,
\begin{equation}
g(H_i \bar{f}_j f_k) =  \bar{f}_j \left( \tilde{\mathcal{M}}_{(i,f)}^\mathrm{int} P_R + \left( \tilde{\mathcal{M}}_{(i,f)}^\mathrm{int}\right)^\dagger P_L  \right)_{jk} f_k.
\end{equation}
Interaction of $H_i \bar{f}_j f_k$ are given by:
\begin{equation}
\frac{1}{2} \left[ \left( \tilde{\mathcal{M}}_{jk}  + (\tilde{\mathcal{M}}_{kj})^* \right) + \gamma_5 \left( \tilde{\mathcal{M}}_{jk}  - (\tilde{\mathcal{M}}_{kj})^* \right) \right],
\end{equation}
where $\tilde{\mathcal{M}}_{jk}$ are $\left(\tilde{\mathcal{M}}_{(i,f)}^\mathrm{int}\right)_{jk}$. It should be of no surprise that FCNCs are present. For the diagonal interactions $H_i \bar{f}_j f_j$ we get:
\begin{equation}
\tilde{\mathcal{M}}_{jj} \gamma_5 + \mathbb{R}\mathrm{e}(\tilde{\mathcal{M}}_{jj}) \left(1 - \gamma_5 \right).
\end{equation}
In principle, it is possible to generate scalar-fermion interactions with the diagonal elements of $\tilde{\mathcal{M}}_{(i,f)}^\mathrm{int}$ suppressed over the off-diagonal ones, resulting in $g(H_i \bar f_j f_k) \gg g(H_i \bar f_j f_j)$, or that $g(H_i \bar f_j f_k)=0$, even for $j=k$. Due to significant freedom in the Yukawa Lagrangian, it is possible that the $\mathcal{O}(\text{MeV})$ scalars of C-V would escape the detector. As a consequence, the loop-induced observables, $e.g.$, di-photon signals, $B$-physics, electric dipole moment, $(g-2)_\mu$, should be considered to constrain the parameter space. It should be noted that in the context of the C-V implementation it was sufficient to assume real Yukawa couplings. With complex Yukawa couplings, depending on under which of the $S_3$ representations fermions transform, from eight to ten additional degrees of freedom could be introduced in the quark sector.
\end{itemize}

\newpage

\subsection{Discussing different cases}

Let us briefly discuss implementations which we did not consider; these do no result in CP violation of the scalar sector. Introduction of soft symmetry-breaking terms, see Section~\ref{Sec:S3_br_cont_symm}, could provide additional interesting cases from the perspective of the Yuakwa Lagrangian. When complex soft terms are included, the ability to rotate away their phases depends on which terms are considered. Full analysis of the softly broken cases was not considered in Ref.~\cite{Kuncinas:2023ycz} since such cases would introduce many possibilities: there could be a single soft term or all four present. In total, a few hundred different cases could become possible. In Table~\ref{Table:SSB} we list structures of the Yukawa Lagrangian, provided there are CP sources in the scalar sector.

One of the interesting cases is C-IV-b, $\left( \hat v_1,\, \pm i \hat v_2 ,\, \hat v_S\right)$. By performing an analysis of the Yukawa sector alone, we found that it is possible to generate the CKM matrix by means of a single imaginary unit, \textit{i.e.}, non-trivial phases were not required.

{{\renewcommand{\arraystretch}{1.15}
\begin{table}[htb]
\caption{ Summary of different CP violating implementations in the $S_3$-symmetric 3HDM which require soft symmetry-breaking terms for CP violation in the scalar sector. In the third column we present different realisable structures of the Yukawa Lagrangian, where ``-" entries indicate unrealistic predictions. Implementations with complex vevs, $Y_f \in \mathbb{C}$ indicates that one is required to introduce complex Yukawa couplings to generate the CKM matrix. In other cases (apart from the implementations with real vevs) it is sufficient to have real Yukawa couplings.}
\label{Table:SSB}
\begin{center}
\begin{tabular}{|c|c|c|}\hline\hline
Vacuum & vevs & $\mathcal{L}_Y$ \\ \hline\hline
R-II-1a &  $\left( 0,\,v_2,\,v_S \right)$ & trivial\\ \hline
R-II-1b,c &  $\left( v_1,\, \pm v_1/\sqrt{3},\, v_S \right)$ & trivial  \\ \hline
R-II-2 &  $\left( 0,\, v,\,0 \right)$ & - \\ \hline
R-II-3 & $\left( v_1,\, v_2,\, 0 \right)$ & any except for trivial \\ \hline
R-III &  $\left( v_1,\, v_2,\, v_S \right)$ & any \\ \hline
R-III-s & $\left( v_1,\,0,\, v_S \right)$ & any \\ \hline\hline
C-III-b & $\left( \pm i v_1,\, 0 ,\, \hat v_S \right)$ & any with $Y_f \in \mathbb{C}$  \\ \hline
C-III-c & $\left( \hat v_1 e^{i \sigma_1} ,\, \hat v_2 e^{i \sigma_2} ,\, 0\right)$ & any except for trivial \\ \hline
C-III-d & $\left( \pm i \hat v_1,\, \hat v_2 ,\, \hat v_S\right)$ & any with $Y_f \in \mathbb{C}$  \\ \hline
C-III-e & $\left( \pm i \hat v_1,\, -\hat v_2 ,\, \hat v_S\right)$ & any with $Y_f \in \mathbb{C}$  \\ \hline
C-III-f & $\left( \pm i \hat v_1,\, i \hat v_2 ,\, \hat v_S\right)$ &  any \\ \hline
C-III-g & $\left( \pm i \hat v_1,\, -i \hat v_2  ,\, \hat v_S\right)$ & any \\ \hline
C-III-i & $\left( \sqrt{\frac{3 \left( 1 + \tan^2 \sigma_1 \right)}{1 + 9 \tan^2 \sigma_1}} \hat v_2 e^{i \sigma_1} ,\, \pm \hat v_2 e^{-i \arctan(3 \tan \sigma_1)} ,\, \hat v_S\right)$ & any \\ \hline
C-IV-a & $\left( \hat v_1 e^{i \sigma_1} ,\, 0 ,\, \hat v_S\right)$ &  any \\ \hline
C-IV-b & $\left( \hat v_1,\, \pm i \hat v_2 ,\, \hat v_S\right)$ &  any \\ \hline
C-IV-d & $\left( \hat v_1 e^{i \sigma_1},\, \pm \hat v_2 e^{i \sigma_1} ,\, \hat v_S\right)$ & any \\ \hline
C-IV-e & $\left( \sqrt{-\frac{\sin 2 \sigma_2}{\sin 2 \sigma_1}} \hat v_2 e^{i \sigma_1},\, \hat v_2 e^{i \sigma_2} ,\, \hat v_S\right)$ & any \\ \hline
\end{tabular}
\end{center}
\end{table}}

For implementations that involve CP violation in the scalar sector, results of different representations of fermions under $S_3$ are summarised in Table~\ref{Table:CPV_models_list}.  In total, there are four implementations with $\lambda_4^\mathrm{I}=0$ that induce spontaneous CP violation due to the phases of the vevs~\cite{Emmanuel-Costa:2016vej}: C-III-a, C-III-h, C-IV-c, C-IV-f. The vacuum structures of C-III-a and C-III-h require fermions to be trivially charged under $S_3$. Then, C-IV-c and C-IV-f with different representations of fermions could, in principle, yield correct results. It is interesting to note that neither C-IV-c nor C-IV-f survives when $\lambda_4 \in \mathbb{C}$. It is important to note that this fact (applicable also to C-V) does not contradict results of Ref.~\cite{Branco:2015bfb}, where it was conjectured that whenever a symmetry of the scalar potential prevents explicit CP violation, it also prevents spontaneous CP violation. The crucial difference is that here we consider specific directions of vevs. Although both C-IV-c and C-IV-f yield good numerical fits in the Yukawa Lagrangian, both implementations suffer from unwanted massless states. To sum up, for $\lambda_4 \in \mathbb{R}$ we have to consider C-III-a or C-III-h and that fermions transforms trivially under $S_3$, which is of no particular interest in the context of the Yukawa Lagrangian.

{{\renewcommand{\arraystretch}{1.15}
\begin{table}[htb]
\caption{Summary of different CP violating implementations in the $S_3$-symmetric 3HDM. In the first column we specify whether the scalar potential can be complex. The CPV column indicates cases with spontaneous CP violation, after Ref.~\cite{Emmanuel-Costa:2016vej} with $\lambda_4^\mathrm{I}=0$, and explicit CP violation, $\lambda_4^\mathrm{I} \neq 0$. In the final column we present different realisable structures of the Yukawa Lagrangian, where ``-" entries indicate unrealistic predictions.}
\label{Table:CPV_models_list}
\begin{center}
\begin{threeparttable}
\begin{tabular}{|c|c|c|c|c|}\hline\hline
$V$ & Vacuum & vevs & CPV & $\mathcal{L}_Y$ \\ \hline\hline
$\mathbb{C}$ & R-I-1 & $(0,\,0,\,v)$ & explicit & trivial \\ \hline
$\mathbb{C}$ & R-I-2a & $(v,\,0,\,0)$ & explicit &  - \\ \hline
$\mathbb{C}$ & R-I-2b,c & $(v_1,\, \pm \sqrt{3} v_1,\,0)$ & explicit & - \\ \hline\hline
$\mathbb{C}$ & C-I-a & $(\hat v_1,\,\pm i \hat v_1,\,0)$ & explicit &  - \\ \hline
\begin{tabular}[l]{@{}c@{}} $\mathbb{C}$ \\ $\mathbb{R}$ \end{tabular} & C-III-a & $(0,\,\hat v_2 e^{i \sigma_2},\, \hat v_S)$ &\begin{tabular}[c]{@{}c@{}} explicit\\ spontaneous \end{tabular} & trivial \\ \hline
\begin{tabular}[l]{@{}c@{}} $\mathbb{C}$\\ $\mathbb{R}$ \end{tabular} & C-III-h & $(\sqrt{3} \hat v_2 e^{i \sigma_2},\, \pm \hat v_2 e^{i \sigma_2} ,\, \hat v_S)$ &\begin{tabular}[c]{@{}c@{}} explicit\\ spontaneous \end{tabular} & trivial \\ \hline
$\mathbb{R}^\alpha$ & C-IV-c & $\left(\sqrt{1 + 2 \cos^2 \sigma_2} \hat v_2 ,\, \hat v_2 e^{i \sigma_2} ,\, \hat v_S\right)$ & spontaneous & any \\ \hline
$\mathbb{R}^\alpha$ & C-IV-f & $\left(\sqrt{2 + \frac{\cos(\sigma_1 - 2\sigma_2)}{\cos \sigma_1}} \hat v_2 e^{i \sigma_1},\, \hat v_2 e^{i \sigma_2} ,\, \hat v_S\right)$ &  spontaneous & any \\ \hline
$\mathbb{C}^\beta$ & C-IV-g & $(\hat v_1e^{i\sigma_1},\, \pm i \hat v_1e^{i\sigma_1},\, \hat v_S)$ & explicit  & any \\ \hline
$\mathbb{C}$ & C-V & $\left(\hat v_1 e^{i \sigma_1} ,\, \hat v_2 e^{i \sigma_2} ,\, \hat v_S\right)$ & explicit & any \\ \hline
\end{tabular}
  \footnotesize
  \begin{tablenotes}
  \item[$\alpha$] In C-IV-c and C-IV-f there are unwanted massless scalars present.
  \item[$\beta$] In C-IV-g there are two negative mass-squared parameters.
  \end{tablenotes}
  \end{threeparttable}
\end{center}
\end{table}}

Allowing for complex couplings significantly alters the situation; there is more freedom to implement CP violation. If we consider real vacua, R-I-1, R-I-2a and R-I-2b,c can result in explicit CP violation. Out of these, only R-I-1 could produce a realistic Yukawa sector, but it has to be a trivial one. Therefore, we have to consider explicit CP violation in implementations with complex vevs: C-I-a, C-III-a, C-III-h, C-IV-g, C-V. Out of these, only C-I-a fails to describe the observed fermionic content correctly. Then, both C-III-a and C-III-h require a trivial Yukawa sector. This narrows the list of interesting cases down to C-IV-g and C-V. The C-IV-g implementation does not appear in the classification of Ref.~\cite{Emmanuel-Costa:2016vej} since it does not survive when $\lambda_4^\mathrm{I}=0$.  This case might seem interesting as it is capable of generating the experimentally observed CKM matrix with just a single phase. Unfortunately, this case is ruled out due to unrealistic scalar masses. As a consequence, only C-V, the most general vacuum configuration, survives. Although it is the most general vacuum, this does not indicate that all other implementations can be derived from C-V. As can be observed from Figure~\ref{Fig:C-V_VEVs}, there exists a viable region of parameter space where at least one of the phases is close to zero. A remarkable feature of C-V is its potential to accommodate light neutral scalars, which could, in principle, escape detectors. A more detailed analysis of C-V is yet to be performed elsewhere.

\section{Dark Matter candidates}\label{Sec:SectionDM}

The final criterion for categorising different $S_3$ vacua is the potential to accommodate a DM candidate. Our approach is to identify vacua with at least one vanishing vev. We recall that for a doublet to accommodate a DM candidate, it must have a vanishing vev in the representation of a symmetry under which it gets stabilised; otherwise, it would decay via gauge couplings, $SW^+W^-$, $SZZ$, or into fermions.  In Table~\ref{Table:S3-vacua} we list vacua with at least one vanishing vev. Whenever $\lambda_4=0$, the scalar potential acquires an additional continuous symmetry, as was discussed in Section~\ref{Sec:S3_br_cont_symm}.

{\renewcommand{\arraystretch}{1.3}
\begin{table}[htb]\footnotesize
\caption{Different implementations of the $S_3$-symmetric potential that might accommodate DM due to a vanishing vev. In the last column remnant symmetries are displayed. The symmetry $h_1\to-h_1$ is present in the irreducible representation. Then, there are several remnant symmetries, which are explicit in the defining representation, see the scalar potential of eq.~\eqref{Eq:S3_Derman_redtr}. These are: $S_2$, $S_3$, $\mathbb{Z}_3$. }
\label{Table:S3-vacua}
\begin{center}
\begin{tabular}{|c|c|c|c|c|c|}
\hline\hline
Vacuum & vevs & \begin{tabular}[c]{@{}c@{}} Are massless \\ states present? \end{tabular} & Symmetry \\
\hline
\hline
R-I-1 & $(0,\,0,\,v)$ &  & $S_3$, $h_1\to -h_1$ \\
\hline
R-I-2a & $(v,\,0,\,0)$ &   & $S_2$ \\
\hline
R-I-2b,2c & $(v_1,\,\pm\sqrt3 v_1,\,0)$ &   & $S_2$ \\
\hline
R-II-1a & $(0,\,v_2,\,v_S)$  &  & $S_2$, $h_1\to-h_1$\\
\hline
R-II-2 & $(0,\,v,\,0)$ & yes & $S_2$, $h_1\to -h_1, h_S\to -h_S$ \\
\hline
R-II-3 & $(v_1,\,v_2,\,0)$ & yes & $h_S\to-h_S$  \\
\hline
R-III-s & $(v_1,\,0,\,v_S)$ & yes & $h_2\to-h_2$  \\
\hline
C-I-a & $(\hat v_1,\,\pm i \hat v_1,\,0)$ &  & cyclic $\mathbb{Z}_3$ \\
\hline
C-III-a & $(0,\, \hat v_2e^{i\sigma_2},\,\hat v_S)$ & & $S_2$, $h_1\to-h_1$ \\
\hline
C-III-b & $(\pm i \hat v_1,\,0,\,\hat v_S)$ & yes & $h_2\to-h_2$  \\
\hline
C-III-c & $(\hat v_1e^{i\sigma_1},\,\hat v_2e^{i\sigma_2},\,0)$ & yes & $h_S\to-h_S$  \\
\hline
C-IV-a & $(\hat v_1 e^{i\sigma_1},\,0,\,\hat v_S)$ & yes & $h_2\to-h_2$ \\
\hline
\hline
\end{tabular}
\end{center}
\end{table}}

In Ref.~\cite{Adulpravitchai:2011ei}, a two scalar singlets extended model was considered, with the DM candidate stabilised by $S_3$. The scalar extended models differ from the doublet extended models by: there are no charged scalars and no interactions with gauge bosons. Principally, due to the absence of the latter, it is much easier to match relic abundance. Here, we do not consider that the DM candidate should be stabilised specifically by the $S_3$ group, but rather consider which of the earlier mentioned implementations result in a plausible DM candidate, \textit{i.e.}, we check for remnants of $S_3$; consider the last column of Table~\ref{Table:S3-vacua}. 

All of the $SU(2)$ scalar doublets in the $S_3$-symmetric scalar potential appear in pairs with themselves, except for terms multiplied by the $\lambda_4$ coupling. Provided that the DM candidate resides in $h_1$, it is automatically satisfied by $\mathbb{Z}_2$, we recall \mbox{$S_3 \cong \mathbb{Z}_3 \rtimes \mathbb{Z}_2$}. An equivalent stabilisation condition would be to have an inert pair of $h_2$ and $h_S$. 

Below, we summarise the pros and cons of different vacua in the $S_3$-symmetric 3HDM:
\begin{itemize}
\item
\textbf{R-I-1: $(0,\,0,\,v)$}\\
The DM candidate is assumed to reside in $h_1$. There are three pairs of mass-degenerate states: one charged pair and two neutral. Due to the presence of the $\lambda_4$ term, there are different scalar couplings present among the mass-degenerate states. Actually, taking into account loop contributions, mass degeneracy can be lifted.


\item
\textbf{R-I-2a: $(v,\,0,\,0)$}\\
While the $\mathbb{Z}_2$ of $h_1$ is not preserved by a non vanishing vev, the $\mathbb{Z}_2$ symmetry is preserved for ${(h_2,h_S) \to -(h_2,h_S)}$. However, regardless of the fermion representations, the Yukawa Lagrangian will always predict massless states.

\item
\textbf{R-I-2b,2c: $(v_1,\,\pm\sqrt3 v_1,\,0)$}\\
Although there is a vanishing vev, mixing is present in the mass-squared matrix since $\lambda_4 \neq 0$, and thus, the DM candidate is not stabilised. The $\mathcal{L}_Y$ is unrealistic.

\item
\textbf{R-II-1a: $(0,\, v_2,\, v_S)$}\\
This case yields a viable DM candidate, provided that all fermions transform trivially under $S_3$. Other representations of fermions predict a block-diagonal structure of the CKM matrix. This implementation was studied in Refs.~\cite{Khater:2021wcx,Kuncinas:2023hpf}.

\item
\textbf{R-II-2: $(0,\,v,\,0)$}\\
The underlying symmetry is increased to $O(2)$ since $\lambda_4=0$. One has to introduce soft terms: $\nu_{12}^2$ does not survive, $\nu_{S2}^2$ does not promote the massless states to massive ones, $\mu_{22}^2$ and $\nu_{S1}^2$ are free parameters. Regardless, $\mathcal{L}_Y$ is unrealistic.

\item
\textbf{R-II-3: $(v_1,\,v_2,\,0)$}\\
Once again, it is required that $\lambda_4=0$. Not to spoil the form of the vevs (presence of some soft terms forces relations among the vevs), one needs to assume that both $\mu_{22}^2$ and $\nu_{12}^2$ are added. However, the Yukawa sector yields unrealistic results.

\item
\textbf{R-III-s: \boldmath$(v_1,\,0,\,v_S)$}\\
This is a special case of R-III with $v_2=0$. This case could result in a viable DM candidate provided the $S_3$ symmetry is softly broken by $\mu_{22}^2$ and/or $\nu_{S1}^2$.

\item
\textbf{C-I-a: \boldmath$(\hat v_1,\,\pm i \hat v_1,\,0)$}\\
In this case there is mixing of the $h_S$ doublet with other doublets. Stabilisation of DM requires $\lambda_4=0$ to be imposed. In turn, the $O(2)$ symmetry is spontaneously broken. No soft terms survive the minimisation conditions. Even if one could produce a model with a stable DM candidate, the Yukawa sector is unrealistic.

\item
\textbf{C-III-a: \boldmath$(0,\,\hat v_2e^{i\sigma_2},\,\hat v_S)$}\\
Considering a trivial Yukawa sector, this implementation could provide a viable DM candidate. This implementation was studied in Refs.~\cite{Kuncinas:2022whn,Kuncinas:2023hpf}.

\item
\textbf{C-III-b: \boldmath$(\pm i \hat v_1,\,0,\,\hat v_S)$}\\
After introducing soft symmetry breaking term $\mu_{22}^2$, this case results in no massless particles and could accommodate a DM candidate.  

\item
\textbf{C-III-c: \boldmath$(\hat v_1e^{i\sigma_1},\,\hat v_2e^{i\sigma_2},\,0)$}\\
In this case there are two massless states: one due to the spontaneously broken $O(2)$ and the other one is an accidental massless state. Based on which soft terms are introduced, the vevs get altered~\cite{Kuncinas:2020wrn}:  $\mu_{22}^2$ results in $(\pm i \hat{v}_1, \hat{v}_2, 0)$, $\nu_{12}^2$ yields in  ${(\hat{v} e^{i \sigma /2}, \hat{v} e^{-i \sigma /2}, 0)}$, $\mu_{22}^2-\nu_{12}^2$ forces $(\hat{v}_1e^{i\sigma_1},\hat{v}_2e^{i\sigma_2},0)$. Any, except for a trivial representation, since $\hat v_S =0$, may result in a realistic Yukawa Lagrangian.

\item
\textbf{C-IV-a: \boldmath$(\hat v_1e^{i\sigma_1},\, 0,\, \hat v_S)$}\\
The DM candidate is stabilised by the $\mathbb{Z}_2$ symmetry, which is a remnant of $O(2)$. As in the previous cases, presence of $\lambda_4=0$ signals that there are massless states present. To fix it, one can introduce either $\mu_{22}^2$ or $\nu_{S1}^2$. The inclusion of the $\mu_{22}^2$ term modifies the structure of the vacuum, C-III-b: $(i \hat v_1, 0, \hat v_S)$.

\end{itemize}

There are several implementations which could accommodate a DM candidate:
\begin{itemize}
\item R-I-1/ R-I-1-$\mu_2^2$: $\mathrm{DM}\sim h_1$ or $(h_1,\,h_2)$;
\item R-II-1a: $\mathrm{DM}\sim h_1$;
\item R-III-s-$(\mu_2^2, \nu_{01}^2)$: $\mathrm{DM}\sim h_2$;
\item C-III-a: $\mathrm{DM}\sim h_1$;
\item C-III-b-$\mu_2^2$: $\mathrm{DM}\sim h_2$;
\item C-III-c-$(\mu_2^2, \nu_{12}^2)$: $\mathrm{DM}\sim h_S$;
\item C-IV-a-$(\mu_2^2, \nu_{01}^2)$: $\mathrm{DM}\sim h_2$.
\end{itemize}
Despite several implementations provided in the list above, only three  do not require soft symmetry-breaking terms and feature a realistic Yukawa sector, though fermions are forced to transform as singlets under the $S_3$ group. These are R-I-1, R-II-1a, C-III-a. We shall discuss these implementations in Chapter~\ref{Ch:DM_3HDM}.

\section{Overview of implementations}

In the previous sections we considered the $S_3$-symmetric 3HDM implementations from different angles: real and complex vevs, real and complex couplings, additional continuous symmetries, soft symmetry breaking terms, explicit and spontaneous CP violation, the Yukawa Lagrangian, DM candidates. For convenience, in Table~\ref{Table:Diff_Cases_Sum} we summarise different implementations.

{{\renewcommand{\arraystretch}{1.2}
\setlength\LTcapwidth{\linewidth}
\begin{center}
\begin{longtable}[htb]{|c|c|c|c|c|c|c|} 
\caption{ Different vacua of the $S_3$-symmetric 3HDM with $S_3$ extended to $\mathcal{L}_Y$. In the fourth column, continuous symmetries are shown when identified. In the $\mathcal{L}_Y$ column, $\mathbf{1}$ represents a trivial sector,  $(\mathbf{1},\,\mathbf{1}^\prime,\,\mathbf{2})$ indicates different representations, $Y_f\in \mathbb{C}$ specifies that complex couplings should be considered. In the last column, DM candidates are highlighted, where ``$\mu_{ij}^2$" necessitates presence of the soft symmetry-breaking terms. }
\label{Table:Diff_Cases_Sum} \\ \hline\hline
Vacuum & $\lambda_4 = 0$ & $\lambda_4^\mathrm{I} \neq 0$ & \begin{tabular}[l]{@{}c@{}} Unrealistic \\ masses \end{tabular} & CP violation & $\mathcal{L}_Y$ & DM \\ \hline\hline
R-I-1     &   &   &    & explicit &$\mathbf{1}$  & $\checkmark$                \\ \hline
R-I-2a    &   &   &    & explicit & -   &                 \\ \hline
R-I-2b,c  &   &   &    & explicit & -   &                 \\ \hline
R-II-1a   &   & 0 &    &     &$\mathbf{1}$  & $\checkmark$                \\ \hline
R-II-1b,c &   & 0 &    &     &$\mathbf{1}$  &                 \\ \hline
R-II-2    & 0 & 0 & $O(2)$ &     & -   &                 \\ \hline
R-II-3    & 0 & 0 & $O(2)$ &     &$(\mathbf{1},\,\mathbf{1}^\prime,\,\mathbf{2})$ &                 \\ \hline
R-III     & 0 & 0 & $O(2)$ &     &$(\mathbf{1},\,\mathbf{1}^\prime,\,\mathbf{2})$ & \begin{tabular}[l]{@{}c@{}} $\mu_{ij}^2$   \\ $(v_2=0)$ \end{tabular} \\ \hline\hline

C-I-a     &   &                                                         &                                                                        & explicit       & -   &    \\ \hline
C-III-a   &   &                                                         &                                                                        & \begin{tabular}[c]{@{}c@{}} explicit\\ spontaneous \end{tabular} &$\mathbf{1}$  & $\checkmark$   \\ \hline
C-III-b   & 0 & 0                                                       & $O(2)$                                                                 &           & \begin{tabular}[c]{@{}c@{}}$(\mathbf{1},\,\mathbf{1}^\prime,\,\mathbf{2}),$\\ $Y_f\in \mathbb{C}$ \end{tabular}   & $\mu_{ij}^2$   \\ \hline
C-III-c   & 0 & 0                                                       & $O(2)$                                                                 &           &$(\mathbf{1},\,\mathbf{1}^\prime,\,\mathbf{2})$ & $\mu_{ij}^2$  \\ \hline
C-III-d,e &   & 0                                                       &                                                                        &           & \begin{tabular}[c]{@{}c@{}}$(\mathbf{1},\,\mathbf{1}^\prime,\,\mathbf{2}),$\\ $Y_f\in \mathbb{C}$ \end{tabular}  &    \\ \hline
C-III-f   & 0 & 0                                                       & $O(2)$                                                                 &           &$(\mathbf{1},\,\mathbf{1}^\prime,\,\mathbf{2})$ &    \\ \hline
C-III-g   & 0 & 0                                                       & $O(2)$                                                                 &           &$(\mathbf{1},\,\mathbf{1}^\prime,\,\mathbf{2})$ &    \\ \hline
C-III-h   &   &                                                         &                                                                        & \begin{tabular}[c]{@{}c@{}} explicit\\ spontaneous \end{tabular} &$\mathbf{1}$  &    \\ \hline
C-III-i   &    & 0                                                       &                                                                        &           &$(\mathbf{1},\,\mathbf{1}^\prime,\,\mathbf{2})$ &    \\ \hline
C-IV-a    &  0 & 0                                                       & $O(2) \times U(1)$                                                     &           &$(\mathbf{1},\,\mathbf{1}^\prime,\,\mathbf{2})$ & $\mu_{ij}^2$  \\ \hline
C-IV-b    &  0 & 0                                                       & $O(2)$                                                                 &           &$(\mathbf{1},\,\mathbf{1}^\prime,\,\mathbf{2})$ &    \\ \hline
C-IV-c    &    & 0                                                       & $\checkmark$                                                                     & spontaneous     &$(\mathbf{1},\,\mathbf{1}^\prime,\,\mathbf{2})$ &    \\ \hline
C-IV-d    &  0 & 0                                                       & $O(2) \times U(1)$                                                     &           &$(\mathbf{1},\,\mathbf{1}^\prime,\,\mathbf{2})$ &    \\ \hline
C-IV-e    &  0 & 0                                                       & $O(2)$                                                                 &           &$(\mathbf{1},\,\mathbf{1}^\prime,\,\mathbf{2})$ &    \\ \hline
C-IV-f    &    & 0                                                       & $\checkmark$                                                                       & spontaneous     &$(\mathbf{1},\,\mathbf{1}^\prime,\,\mathbf{2})$ &    \\ \hline
C-IV-g    & \multicolumn{2}{c|}{only $\lambda_4^\mathrm{I} \neq 0$}      & $\checkmark$                                                                       & explicit      &$(\mathbf{1},\,\mathbf{1}^\prime,\,\mathbf{2})$ &    \\ \hline
C-V       &  \multicolumn{2}{c|}{\begin{tabular}[c]{@{}c@{}} $\lambda_4 = 0$ \\  $\lambda_4^\mathrm{I} \neq 0$\end{tabular}} & \begin{tabular}[c]{@{}c@{}}$SU(2)$\\ - \end{tabular} & \begin{tabular}[c]{@{}c@{}} - \\ explicit  \end{tabular}    &$(\mathbf{1},\,\mathbf{1}^\prime,\,\mathbf{2})$ &    \\ \hline

\end{longtable}
\end{center}}

\section{Navigating the parameter space}\label{Sec:Constraints_3HDM_gen}

In Section~\ref{Sec:SecScalarFermionInter} we mentioned a range of constraints in the context of the numerical analysis of the scalar sector of the C-V implementation. We shall discuss these constraints. The theoretical constraints consist of several checks: unitarity, perturbativity, stability; these can be abbreviated as UPS (whoops). Some of the constraints can be verified simply by understanding the structure of the scalar potential (due to the underlying symmetries) and knowing the numerical values of the quartic couplings. However, for a full analysis of the theoretical constraints (specifically for the perturbativity constraint), one needs to know how the mass eigenstates of scalars look like. Apart from these we mention how the Peskin-Takeuchi/ electroweak oblique parameters are evaluated in the NHMDs. In order to evaluate these, it is essential to determine the form of the scalar mass eigenstates.

\subsection{Perturbativity}

Perturbativity constraints refer to the limitations placed on the parameters of the scalar sector to ensure that the theory remains consistent with the perturbation theory. Specifically, these constraints arise from the requirement that couplings of the scalar potential should not be too large, otherwise higher-order corrections might become leading in the perturbation expansion. The soft perturbativity limit is established by imposing constraints on the quartic couplings:
\begin{equation}\label{Eq:lamabsperturb}
|\lambda_i| \leq \lambda_\mathrm{max}.
\end{equation}
The conservative choice is to limit the absolute value by $4 \pi$. A smaller value $\lambda_\mathrm{max} = 2 \pi$ was adopted in Ref.~\cite{Nierste:1995zx}. We consider that $\lambda_\mathrm{max} = 4 \pi$. Successfully passing one consistency check does not necessarily ensure that the other perturbativity check will be satisfied.

\subsection{Stability}

Another requirement is stability (boundedness from below) of the scalar potential~\cite{Deshpande:1977rw,Sher:1988mj}. This constraint was previously mentioned in the context of the SM Higgs boson. In short, for model to be stable, the scalar potential must be bounded from below in all directions in field space. For large field values, the scalar potential should not asymptotically tend to negative infinity. Such behaviour would correspond to an unstable model, resulting in an unphysical vacuum. This implies that the scalar potential should remain positive in all directions of field space for large field values, \textit{i.e.}, for $h_i$ approaching infinity. This is the fundamental constraint, as it ensures the existence of a stable minimum.

In the context of the $S_3$-symmetric model, necessary, but not sufficient, stability conditions were derived in Ref.~\cite{Das:2014fea}:
\begin{subequations}\label{Eq:GenStabilityCond}
\begin{align}
\lambda_1 >& ~0 ,\\
\lambda_8 >& ~0 ,\\
\lambda_1 + \lambda_3 >& ~0,\\
2\lambda_1 + \left( \lambda_3 - \lambda_2 \right) >& ~|\lambda_2 + \lambda_3|,\\
\lambda_5 + 2 \sqrt{\lambda_8 \left( \lambda_1 + \lambda_3 \right)} >& ~0,\\
\lambda_5 + \lambda_6 + 2 \sqrt{\lambda_8 \left( \lambda_1 + \lambda_3 \right)} >& ~2|\lambda_7|,\\
\lambda_1 + \lambda_3 + \lambda_5 + \lambda_6 + 2\lambda_7 + \lambda_8 >& ~2 |\lambda_4|.
\end{align}
\end{subequations}
In Ref.~\cite{Emmanuel-Costa:2016vej}, following the approach of Refs.~\cite{ElKaffas:2006gdt,Grzadkowski:2009bt} (see also Ref.~\cite{Maniatis:2014oza} for the bilinear approach), it was shown that, although stability conditions for the $S_3$-symmetric potential are involved, when $S_3$ is broken down to $O(2)$, due to $\lambda_4=0$, they can be expressed in explicit form. We emphasise that the above constraints are necessary but not sufficient.

We can parameterise the $SU(2)$ scalar doublets in terms of the spinor components as:
\begin{equation}\label{Eq:StabilityGeneralDoublets}
h_i = || h_i || \hat{h}_i, \quad i = \{1, 2, S \},
\end{equation}
where $|| h_i ||$ denotes the norm and $\hat h_i$ stands for a unit spinor. From the above parameteresiation it should be obvious that the complex product between the spinors relies on six degrees of freedom~\cite{Emmanuel-Costa:2016vej},
\begin{subequations}\label{Eq:Wrong_Six_Spinors}
\begin{align}
\hat{h}_2^\dagger \hat{h}_1 &= \rho_3 e^{i\theta_3},\\
\hat{h}_S^\dagger \hat{h}_2 &= \rho_1 e^{i\theta_1},\\
\hat{h}_1^\dagger \hat{h}_S &= \rho_2 e^{i\theta_2}.
\end{align}
\end{subequations}

It was pointed out in section III-C of Ref.~\cite{Faro:2019vcd} that these six variables are not mutually independent, and hence the positivity conditions would lead to an over-constrained $\lambda$ parameter space. This indicates that a lower value than the true minimum could be reached. A possible way around is to parameterise the norms of the spinors $|| h_i ||$ in terms of the spherical coordinates:
\begin{equation}
|| h_1 || = r \cos \gamma \sin \theta,  \qquad || h_2 || = r \sin \gamma \sin \theta, \qquad || h_S || = r \cos \theta,
\end{equation}
and the unit spinors in terms of
\begin{equation}
\hat{h}_1 = \begin{pmatrix}
0 \\ 1
\end{pmatrix}, \qquad \hat{h}_2 = \begin{pmatrix}
\sin \alpha_2 \\ \cos \alpha_2 \, e^{i \beta_2}
\end{pmatrix},\qquad \hat{h}_S = e^{i \delta} 
\begin{pmatrix}
\sin \alpha_3 \\
\cos \alpha_3 \, e^{i \beta_3}
\end{pmatrix}.
\end{equation}
Notice that the unit spinors are parameterised in terms of five angles, rather than six in eq.~\eqref{Eq:Wrong_Six_Spinors}.

Regardless, due to the freedom of the $\lambda_4$ coupling, the stability conditions are quite complex. Our approach involves first checking the necessary stability constraints presented in eqs.~\eqref{Eq:GenStabilityCond} and if those are satisfied, a numerical check is performed. For the latter one we utilise the $\mathsf{Mathematica}$ function $\mathsf{NMinimize}$ by
employing several algorithms.

To perform the numerical check, we need to evaluate the minimum of the quartic part of the scalar potential. In the new basis, we have:
\begin{equation}\label{Eq:Vpr_Min_Cond}
V_4^\prime = \sum_{i=1}^{8} \lambda_i A_i \geq 0,\quad \forall ~\{ \gamma,\,\theta,\,\alpha_2,\,\alpha_3,\,\beta_2,\,\beta_3,\, \delta\},
\end{equation}
where
\begin{subequations}
\begin{align}
A_1 &= \sin ^4 \theta, \\
A_2 &= -4 \cos^2 \alpha_2 \, \sin^2 \beta_2 \, \sin^2 \gamma \, \cos^2 \gamma \, \sin^4 \theta \,,\\
A_3 &= \sin ^4\theta \, \left[2 \sin ^2\gamma \, \cos ^2\gamma \, \left(\cos ^2\alpha_2 \, \cos (2 \beta_2)-\sin ^2\alpha_2\right)+\sin ^4\gamma+\cos ^4\gamma \right],\\
\begin{split}
A_4 &= -2 \sin \gamma \, \sin ^3\theta \, \cos \theta \, \Big\{-\sin \alpha_2 \, \sin \alpha_3 \, \cos (2 \gamma ) \cos \delta\\
&\hspace{70pt}+ \cos \alpha_2 \, \cos \alpha_3\sin ^2\gamma \, \cos (\beta_2-\beta_3-\delta )\\
&\hspace{70pt}-\cos \alpha_2 \, \cos \alpha_3 \cos ^2\gamma \, [2 \cos (\beta_2-\beta_3-\delta )+\cos (\beta_2+\beta_3+\delta )]
\Big\},
\end{split}\\
A_5 &= \sin ^2\theta \, \cos ^2\theta,\\
\begin{split}
A_6 &= \sin ^2\theta \, \cos ^2\theta \, \Big\{ \cos ^2\alpha_3 \, \cos ^2\gamma + \cos ^2\alpha_2 \, \cos ^2\alpha_3\sin ^2\gamma \\
&\hspace{50pt} +\sin ^2\gamma \, \left[\sin \alpha_3 \, \sin (2 \alpha_2) \cos \alpha_3 \, \cos (\beta_2-\beta_3)+\sin ^2\alpha_2 \, \sin^2\alpha_3 \right]\Big\},
\end{split}\\
\begin{split}
A_7 &= 2 \sin ^2\theta \, \cos ^2\theta \, \Big\{ \cos ^2\alpha_3 \, \cos ^2\gamma \, \cos (2\beta_3+2\delta )\\
&+\sin ^2\gamma \cos ^2\alpha_2 \, \cos ^2\alpha_3 \, \cos (2\beta_2-2\beta_3-2\delta)\\
&+\sin \alpha_3 \, \sin ^2\gamma \left[\sin (2 \alpha_2) \cos \alpha_3 \, \cos (\beta_2-\beta_3-2 \delta )+\sin ^2\alpha_2 \, \sin \alpha_3 \, \cos (2 \delta )\right]\Big\},
\end{split}\\
A_8 &= \cos ^4 \theta.
\end{align}
\end{subequations}

The validity of the conditions presented in eq.~\eqref{Eq:GenStabilityCond} is verified in Table~\ref{Table:ReproducedAnalyticalStability}. 

{\renewcommand{\arraystretch}{1.3}
\begin{table}[htb]
\caption{Reproduction of the necessary, but not sufficient, stability conditions from Ref.~\cite{Das:2014fea} in terms of the parameterisation given by~\eqref{Eq:Vpr_Min_Cond}.}
\label{Table:ReproducedAnalyticalStability}
\begin{center}
\begin{tabular}{|c|c|c|}
\hline\hline
Conditions & $V_4^\prime  > 0$ & \begin{tabular}[c]{@{}l@{}}  Conditions of Ref.~\cite{Das:2014fea} \end{tabular} \\ \hline\hline
\begin{tabular}[c]{@{}c@{}} $\gamma=\frac{\pi}{4},$ \\ $\theta = \alpha_2 = \frac{\pi}{2}$\end{tabular} & $\lambda_1$ & (4a) \\ \hline
$\theta=0$ & $\lambda_8$ & (4b) \\ \hline
\begin{tabular}[c]{@{}c@{}} $\gamma=\frac{\pi}{4},\, \theta= \frac{\pi}{2},$ \\ $\alpha_2=0,\, \beta_2=\{0, \frac{\pi}{2}\}$\end{tabular} & \begin{tabular}[c]{@{}c@{}} $\lambda_1+\lambda_3$ \\ $\lambda_1-\lambda_2$\end{tabular} & (4c) and (4d) \\ \hline
\begin{tabular}[c]{@{}c@{}} $\gamma = 0, \, \delta = 0,$ \\
$\tan \theta = \sqrt{\frac{\lambda_8}{\sqrt{\left( \lambda_1 + \lambda_3 \right)\lambda_8}}},$ \\
$\alpha_3 = \{0, \frac{\pi}{2}\},\,\beta_3 = \{0, \frac{\pi}{2}\}$ \end{tabular}  & \begin{tabular}[c]{@{}c@{}} $\lambda_5 + \mathrm{min}\left(0,\lambda_6 - 2\left| \lambda_7 \right|\right)$ \\ $+ 2 \sqrt{\left(\lambda_1+\lambda_3\right)\lambda_8}$\end{tabular} & (4e) and (4f) \\ \hline
\begin{tabular}[c]{@{}c@{}} $\theta= \frac{\pi}{4}, \, \gamma = \frac{\pi}{2},$\\ $\alpha_2 = \alpha_3 = \frac{\pi}{2},\,\delta=\{\pi,2\pi\}$ \end{tabular}  & \begin{tabular}[c]{@{}c@{}} $\lambda_1 + \lambda_3 \pm 2\lambda_4+ \lambda_5$ \\  $+ \lambda_6 + 2\lambda_7 + \lambda_8$\end{tabular} &  (4g)\\
\hline\hline
\end{tabular}
\end{center}
\end{table}}

\subsection{Unitarity}

The unitarity condition is assessed by ensuring that the absolute values of the eigenvalues of the scattering matrix remain within a specified limit. In our scan we assume \mbox{$|\Lambda_i| \leq 16\pi$}~\cite{Lee:1977eg}. Some authors prefer a more severe bound $|\Lambda_i| \leq 8\pi$~\cite{Luscher:1988gc, Marciano:1989ns}. The unitarity conditions for the real $S_3$-symmetric scalar potential were derived in Ref.~\cite{Das:2014fea}. Due to the $\lambda_4^\mathrm{I}$ coupling, additional off-diagonal mixing terms will arise in the $S$-matrix. In both real and complex $S_3$ model there are common eigenvalues of the $S$-matrix:
\begin{subequations}
\begin{align}
e_1 \,(b_6) & = \lambda_5 - \lambda_6,\\
e_2 \,(b_3) & = 2 \left( \lambda_1 - 5 \lambda_2 - 2 \lambda_3 \right),\\
e_3 \,(b_4,b_5) & = 2 \left( \lambda_1 \pm \lambda_2 - 2 \lambda_3 \right),\\
e_4 \,(a_2^\pm) & = \lambda_1+\lambda_2+2 \lambda_3+\lambda_8\pm \sqrt{(\lambda_1+\lambda_2+2 \lambda_3-\lambda_8)^2+8 \lambda_7^2},\\
e_5 \,(a_3^\pm)& = \lambda_1-\lambda_2+2 \lambda_3+\lambda_8\pm \sqrt{(-\lambda_1+\lambda_2-2 \lambda_3+\lambda_8)^2+2 \lambda_6^2},\\
e_6 \,(a_5^\pm)& = 5 \lambda_1-\lambda_2+2 \lambda_3+3 \lambda_8\pm \sqrt{(-5 \lambda_1+\lambda_2-2 \lambda_3+3 \lambda_8)^2+2 (2 \lambda_5+\lambda_6)^2},\\
e_7 \,(a_1^\pm)& = \frac{1}{2} \left(2 \lambda_1-2 \lambda_2+\lambda_5+\lambda_6\pm \sqrt{(-2 \lambda_1+2 \lambda_2+\lambda_5+\lambda_6)^2+16 |\lambda_4|^2 }\right).
\end{align}
\end{subequations}
For simplicity, we presented in parenthesis  the eigenvalues in terms of the Ref.~\cite{Das:2014fea} notation, $\{a_{1,2,3,4}^\pm,\, b_{3,4,5,6}\}$. Apart from these, there are additional eigenvalues which differ due to the $\lambda_4 \in \mathbb{C}$ coupling. Rather than presenting them analytically, it is more straightforward to provide the relevant portions of the scattering matrices, which are block-diagonal:
\begin{subequations}\label{Eq:S_eigen_to_solve}
\begin{align}
\mathcal{N}_{55} ={}& \begin{pmatrix}
2 \left( \lambda_1 + \lambda_2 \right) & -2i \lambda_4^\mathrm{I} & 2\lambda_4^\mathrm{R}\\
2i\lambda_4^\mathrm{I} & \lambda_5 - 2 \lambda_7 & 0 \\
2\lambda_4^\mathrm{R} & 0 & \lambda_5 + 2 \lambda_7
\end{pmatrix},\\
\mathcal{N}_{66} ={}& \begin{pmatrix}
2 \left( \lambda_1 + \lambda_2 + 4 \lambda_3 \right) & -6i \lambda_4^\mathrm{I} & 6\lambda_4^\mathrm{R}\\
6i\lambda_4^\mathrm{I} & \lambda_5 + 2 \lambda_6 - 6 \lambda_7 & 0 \\
6\lambda_4^\mathrm{R} & 0 & \lambda_5 + 2 \lambda_6 + 6 \lambda_7
\end{pmatrix}.
\end{align}
\end{subequations}
In the limit of $\lambda_4^\mathrm{I} = 0$ these become: $\mathcal{N}_{55}\,(b_2,\, a_4^\pm)$ and $\mathcal{N}_{66}\,(b_1,\, a_6^\pm)$. Next, we proceed with the derivation of the results.

The unitarity constraint was previously discussed in Section~\ref{Sec:SM_Theor_Constr}. The process of finding the unitarity constraint is then generalised to three doublets. It suffices to analyse the $S$-matrix consisting of the $2 \to 2$ scalar eigenstate processes.

The $S$-matrix consists of the two-particle states $\Psi_i^n$, $S_{i}=\left\langle  \Psi_i^n \,|\, \Psi_i^n\right\rangle$ . These states can be directly identified from the quartic terms in the scalar potential, which are expressed in terms of $SU(2)$ doublets:
\begin{equation}
h_i = \begin{pmatrix}
w_i^+ \\
n_i
\end{pmatrix}.
\end{equation}
Notice that we do not need to account for the $1/\sqrt{2}$ factor in the lower component. Since we do not expand the neutral component in terms of the scalar fields, it is important to note that $n_i \in \mathbb{C}$.

Given the computational complexity of determining the eigenvalues of the most general scattering matrix, it is worthwhile to explore a basis in which the scattering matrix is block-diagonal. This approach is reasonable since not all 2-body scattering processes are possible, as they are constrained by the $S_3$-symmetric potential and discrete symmetries such as CP.

The eigenvalues of the $S$-matrix can be expressed in terms of the eigenvalues of each sub block-diagonal matrix:
\begin{equation}
\det \left( S-\lambda\, \mathcal{I}\right)= \det \left( S_1-\lambda\, \mathcal{I}\right) \times \cdots \times \det \left( S_n-\lambda\, \mathcal{I}\right).
\end{equation}
Due to the conservation of the electric charge in the scattering processes, it is straightforward to split the $S$-matrix into the following block-diagonal parts:
\begin{equation}
S= \mathrm{diag}\left( S^0,\, S^+,\, S^{++} \right),
\end{equation}
where the superscripts indicate charges of the incoming/outgoing two-body states.

Some of the off-diagonal matrix entries are symmetric, making it advantageous to apply a unitary transformation,
\begin{equation}
U = \frac{1}{\sqrt{2}} \begin{pmatrix}
1 & -1 \\
1 & 1
\end{pmatrix},
\end{equation}
which shall be useful to remove several off-diagonal entries.

We begin by examining the neutral two-body interactions. Utilising the $\mathbb{Z}_2$ symmetry of $h_1$, the states can be divided into three distinct blocks:
\begin{subequations}
\begin{align}
\Psi^0_1 &= \Big\{ \left.| n_1 n_1\right\rangle,\, \left.| n_2 n_2\right\rangle,\, \left.| n_S n_S\right\rangle,\, \left.| n_2 n_S\right\rangle \Big\},\\
\Psi^0_2 &= \Big\{ \left.| n_1 n_2\right\rangle,\, \left.| n_1 n_S\right\rangle \Big\},\\
\Psi^0_3 &= \Big\{  \left.| w^+_1 w^-_1 \right\rangle,\ \left.| w^+_2 w^-_2 \right\rangle,\ \left.| w^+_2 w^-_S \right\rangle,\ \left.| w^+_S w^-_2 \right\rangle,\ \left.| n_1 n_1^\ast\right\rangle,  \\
&\qquad \left.| n_2 n_2^\ast \right\rangle,\ \left.| n_2 n_S^\ast \right\rangle,\ \left.| n_S n_2^\ast \right\rangle,\ \left.| w^+_S w^-_S \right\rangle,\ \left.| n_S n_S^\ast \right\rangle \Big\},\\
\Psi^0_4 &= \Big\{  \left.| w^+_1 w^-_2 \right\rangle,\ \left.| w^+_2 w^-_1 \right\rangle,\ \left.| w^+_1 w^-_S \right\rangle,\ \left.| w^+_S w^-_1 \right\rangle,\\
&\qquad  \left.| n_1 n_2^\ast \right\rangle,\ \left.| n_2 n_1^\ast \right\rangle,\ \left.| n_1 n_S^\ast \right\rangle,\ \left.| n_S n_1^\ast \right\rangle \Big\}.
\end{align}
\end{subequations}

The first two block forms of $S^0$ are given by:
\begin{subequations}
\begin{align}
S_1^0 & = \begin{pmatrix}
2 \left( \lambda_1 + \lambda_3 \right) & 2 \left( \lambda_2 + \lambda_3 \right) & 2 \lambda_7 & \sqrt{2}\left( \lambda_4^\mathrm{R} + i \lambda_4^\mathrm{I} \right) \\
2 \left( \lambda_2 + \lambda_3 \right) & 2 \left( \lambda_1 + \lambda_3 \right) & 2 \lambda_7 & -\sqrt{2}\left( \lambda_4^\mathrm{R} + i \lambda_4^\mathrm{I} \right) \\
2 \lambda_7 & 2 \lambda_7 & 2 \lambda_8 & 0 \\
\sqrt{2}\left( \lambda_4^\mathrm{R} - i \lambda_4^\mathrm{I} \right) & \sqrt{2}\left(-\lambda_4^\mathrm{R} + i \lambda_4^\mathrm{I} \right)  & 0 & \lambda_5 + \lambda_6
\end{pmatrix},\\
S_2^0 &= \begin{pmatrix}
2 \left( \lambda_1 - \lambda_2 \right) & 2 \left( \lambda_4^\mathrm{R} + i \lambda_4^\mathrm{I} \right) \\
2 \left( \lambda_4^\mathrm{R} - i \lambda_4^\mathrm{I} \right) & \lambda_5 + \lambda_6
\end{pmatrix}.
\end{align}
\end{subequations}

Next, we rotate the relevant block of $\Psi^0_3$ by
\begin{equation}
\mathcal{P} \mathcal{R}\,S_3^0\,\mathcal{R}^\dagger \mathcal{P}^\mathrm{T} = \mathrm{diag} \left( \begin{pmatrix}
\mathcal{N}_{11} & \mathcal{N}_{12} \\
\mathcal{N}_{12} & \mathcal{N}_{11}
\end{pmatrix},\,\mathcal{N}_{33} \right),
\end{equation} 
where
\begin{subequations}
\begin{align}
\mathcal{R}={}&\mathcal{I}_5 \otimes U,\\
\mathcal{P}={}& \begin{pmatrix}
 1 & 0 & 0 & 0 & 0 & 0 & 0 & 0 & 0 & 0 \\
 0 & 0 & 1 & 0 & 0 & 0 & 0 & 0 & 0 & 0 \\
 0 & 0 & 0 & 1 & 0 & 0 & 0 & 0 & 0 & 0 \\
 0 & 0 & 0 & 0 & 1 & 0 & 0 & 0 & 0 & 0 \\
 0 & 0 & 0 & 0 & 0 & 0 & 1 & 0 & 0 & 0 \\
 0 & 0 & 0 & 0 & 0 & 0 & 0 & 1 & 0 & 0 \\
 0 & 1 & 0 & 0 & 0 & 0 & 0 & 0 & 0 & 0 \\
 0 & 0 & 0 & 0 & 0 & 1 & 0 & 0 & 0 & 0 \\
 0 & 0 & 0 & 0 & 0 & 0 & 0 & 0 & 1 & 0 \\
 0 & 0 & 0 & 0 & 0 & 0 & 0 & 0 & 0 & 1 
\end{pmatrix}.
\end{align}
\end{subequations}
In the new basis, the elements of $S_3^0$ are
\begin{subequations}\label{Eq:N11_N12_N33}
\begin{align}
\mathcal{N}_{11}=&{}\begin{pmatrix}
2\left( \lambda_1 + \lambda_2 + 2 \lambda_3 \right) & - 4i \lambda_4^\mathrm{I} &  4 \lambda_4^\mathrm{R}\\
4i \lambda_4^\mathrm{I} & \lambda_5 + \lambda_6 - 4 \lambda_7 & 0 \\
4 \lambda_4^\mathrm{R} & 0 & \lambda_5 + \lambda_6 + 4 \lambda_7
\end{pmatrix},\\
\mathcal{N}_{12}=&{}\begin{pmatrix}
4 \lambda_3 & - 2i \lambda_4^\mathrm{I} &  2 \lambda_4^\mathrm{R}\\
2i \lambda_4^\mathrm{I} & \lambda_6 - 2 \lambda_7 & 0 \\
2 \lambda_4^\mathrm{R} & 0 & \lambda_6 + 2 \lambda_7
\end{pmatrix},\\
\mathcal{N}_{33}=&{}\begin{pmatrix}
6 \lambda_1 - 2 \lambda_2 + 4 \lambda_3 & 4 \lambda_1 & \lambda_6 & 2\lambda_5 + \lambda_6\\
4 \lambda_1 & 6 \lambda_1 - 2 \lambda_2 + 4 \lambda_3 & -\lambda_6 & 2 \lambda_5 +\lambda_6\\
\lambda_6 & -\lambda_6 & 2 \lambda_8 & 0 \\
2\lambda_5 + \lambda_6 & 2\lambda_5 + \lambda_6 & 0 & 6 \lambda_8
\end{pmatrix}.
\end{align}
\end{subequations}
From the above matrices it should be easy to spot that the $\lambda_4^\mathrm{I}$ coupling will enter in the eigenvalues. In the limit of $\lambda_4^\mathrm{I}=0$, the eigenvalues correspond to $\{a_4^\pm,\, a_6^\pm,\, b_1,\, b_2\}$. These are the sole eigenvalues that differ between the models with real and complex coefficients.

The final part of the neutral $S$-matrix corresponds to scattering of the $\Psi^0_4$ states:
\begin{equation}
\mathcal{P} \mathcal{R}\,S_4^0\,\mathcal{R}^\dagger \mathcal{P}^\mathrm{T} = \mathrm{diag} \left(
\mathcal{N}_{44},\, \begin{pmatrix}
\mathcal{N}_{11} & \mathcal{N}_{12} \\
\mathcal{N}_{12} & \mathcal{N}_{11}
\end{pmatrix} \right),
\end{equation} 
where the basis transformations are given by:
\begin{subequations}
\begin{align}
\mathcal{R}={}&\mathcal{I}_4 \otimes U,\\
\mathcal{P}={}& \begin{pmatrix}
 1 & 0 & 0 & 0 & 0 & 0 & 0 & 0 \\
 0 & 0 & 0 & 0 & 1 & 0 & 0 & 0 \\
 0 & 1 & 0 & 0 & 0 & 0 & 0 & 0 \\
 0 & 0 & 1 & 0 & 0 & 0 & 0 & 0 \\
 0 & 0 & 0 & 1 & 0 & 0 & 0 & 0 \\
 0 & 0 & 0 & 0 & 0 & 1 & 0 & 0 \\
 0 & 0 & 0 & 0 & 0 & 0 & 1 & 0 \\
 0 & 0 & 0 & 0 & 0 & 0 & 0 & 1 
\end{pmatrix}.
\end{align}
\end{subequations}
Then,
\begin{equation}
\mathcal{N}_{44} = \begin{pmatrix}
2 \lambda_1 - 6 \lambda_2 - 4 \lambda_3 & -4 \lambda_2 \\
-4 \lambda_2 & 2 \lambda_1 - 6 \lambda_2 - 4 \lambda_3
\end{pmatrix}.
\end{equation}

Next, we move to the the singly-charged two-body states. The scattering matrix is block-diagonal in the basis of:
\begin{subequations}
\begin{align}
\Psi^+_1 & = \Big\{ \left.| n_1 w^+_1 \right\rangle,\ \left.| n_2 w^+_2 \right\rangle,\ \left.| n_S w^+_S \right\rangle,\ \left.| n_S w^+_2 \right\rangle,\ \left.| n_2 w^+_S \right\rangle \Big\},\\
\Psi^+_2 &= \Big\{ \left.| n_1 w^+_2 \right\rangle,\ \left.| n_2 w^+_1 \right\rangle,\ \left.| n_1 w^+_S \right\rangle,\ \left.| n_S w^+_1 \right\rangle \Big\},\\
\Psi^+_3 &= \Big\{ \left.| n_1^\ast w^+_1 \right\rangle,\ \left.| n_2^\ast w^+_2 \right\rangle,\ \left.| n_S^\ast w^+_S \right\rangle,\ \left.| n_S^\ast w^+_2 \right\rangle,\ \left.| n_2^\ast w^+_S \right\rangle \Big\},\\
\Psi^+_4 &= \Big\{ \left.| n_1^\ast w^+_2 \right\rangle,\ \left.| n_2^\ast w^+_1 \right\rangle,\ \left.| n_1^\ast w^+_S \right\rangle,\ \left.| n_S^\ast w^+_1 \right\rangle \Big\}.
\end{align}
\end{subequations}

The first piece of the $S$-matrix, corresponding to the $\Psi^+_1$ states, can be rotated by
\begin{subequations}
\begin{align}
\mathcal{R}={}& \mathrm{diag} (U,\,1,\,U),\\
\mathcal{P}={}& \begin{pmatrix}
1 & 0 & 0 & 0 & 0\\
0 & 0 & 0 & 0 & 1\\
0 & 0 & 0 & 1 & 0\\
0 & 0 & 1 & 0 & 0\\
0 & 1 & 0 & 0 & 0 
\end{pmatrix},
\end{align}
\end{subequations}
which yields
\begin{equation}
\mathcal{P} \mathcal{R}\,S_1^+\,\mathcal{R}^\dagger \mathcal{P}^\mathrm{T} = \begin{pmatrix}
2 \left( \lambda_1 - \lambda_2 \right) & 2 \left( \lambda_4^\mathrm{R} + i \lambda_4^\mathrm{I} \right) & 0 & 0 & 0\\
2\left( \lambda_4^\mathrm{R} - i \lambda_4^\mathrm{I} \right) & \lambda_5 + \lambda_6 & 0 & 0 & 0\\
0 & 0 & \lambda_5 - \lambda_6 & 0 & 0\\
0 & 0 & 0 & 2\lambda_8 & 2 \sqrt{2} \lambda_7 \\
0 & 0 & 0 & 2\sqrt{2} \lambda_7 & 2\left( \lambda_1 + \lambda_2 + 2 \lambda_3 \right)
\end{pmatrix}.
\end{equation}

The $S_2^+$ part can be rotated by
\begin{subequations}
\begin{align}
\mathcal{R}={}& \mathcal{I}_2 \otimes U,\\
\mathcal{P}={}& \begin{pmatrix}
1 & 0 & 0 & 0 \\
0 & 0 & 1 & 0 \\
0 & 1 & 0 & 0 \\
0 & 0 & 0 & 1 \\
\end{pmatrix},
\end{align}
\end{subequations}
yielding
\begin{equation}
\mathcal{P} \mathcal{R}\,S_2^+\,\mathcal{R}^\dagger \mathcal{P}^\mathrm{T} = \begin{pmatrix}
2\left( \lambda_1 + \lambda_2 - 2 \lambda_3 \right) & 0 & 0 & 0 \\
0 & \lambda_5 - \lambda_6 & 0 & 0 \\
0 & 0 & 2 \left( \lambda_1 - \lambda_2 \right) & 2 \left( \lambda_4^\mathrm{R} + i \lambda_4^\mathrm{I} \right)\\
0 & 0 & 2\left( \lambda_4^\mathrm{R} - i \lambda_4^\mathrm{I} \right) & \lambda_5 + \lambda_6
\end{pmatrix}.
\end{equation}

For $S_3^+$ we have 
\begin{equation}
\mathcal{P} \mathcal{R}\,S_3^+\,\mathcal{R}^\dagger \mathcal{P}^\mathrm{T} = \mathrm{diag} \left( \begin{pmatrix}
2 \lambda_8 & \sqrt{2} \lambda_6 \\
\sqrt{2} \lambda_6 & 2\left( \lambda_1 - \lambda_2 + 2 \lambda_3 \right)
\end{pmatrix},\, \mathcal{N}_{55} \right),
\end{equation}
where
\begin{subequations}
\begin{align}
\mathcal{R}={}& \mathrm{diag} (U,\,1,\,U),\\
\mathcal{P}={}& \begin{pmatrix}
0 & 0 & 1 & 0 & 0\\
0 & 1 & 0 & 0 & 0\\
1 & 0 & 0 & 0 & 0\\
0 & 0 & 0 & 1 & 0\\
0 & 0 & 0 & 0 & 1 
\end{pmatrix},
\end{align}
\end{subequations}
and
\begin{equation}
\mathcal{N}_{55} = \begin{pmatrix}
2 \left( \lambda_1 + \lambda_2 \right) & -2i \lambda_4^\mathrm{I} & 2\lambda_4^\mathrm{R}\\
2i\lambda_4^\mathrm{I} & \lambda_5 - 2 \lambda_7 & 0 \\
2\lambda_4^\mathrm{R} & 0 & \lambda_5 + 2 \lambda_7
\end{pmatrix}.
\end{equation}
The eigenvalues of $\mathcal{N}_{55}$ in the limit of $\lambda_4^\mathrm{I}=0$ are $\{a_4^\pm,\, b_2 \}$. The eigenvalues of $\mathcal{N}_{55}$ correspond to the eigenvalues of the first block of $S_3^0$, see eq.~\eqref{Eq:N11_N12_N33}. This observation allows us to conclude that the matrix of eq.~\eqref{Eq:N11_N12_N33} can be split into two block-diagonal structures of dimension three, with one corresponding to $\mathcal{N}_{55}$. By analysing the structure of $\mathcal{N}_{55}$ and its eigenvalues, we can deduce that the other block should correspond to
\begin{equation}
\mathcal{N}_{66} = \begin{pmatrix}
2 \left( \lambda_1 + \lambda_2 + 4 \lambda_3 \right) & -6i \lambda_4^\mathrm{I} & 6\lambda_4^\mathrm{R}\\
6i\lambda_4^\mathrm{I} & \lambda_5 + 2 \lambda_6 - 6 \lambda_7 & 0 \\
6\lambda_4^\mathrm{R} & 0 & \lambda_5 + 2 \lambda_6 + 6 \lambda_7
\end{pmatrix}.
\end{equation}
Although the exact form of the rotation matrix matrices is not defined,
\begin{subequations}\label{Eq:New_S30_S40}
\begin{align}
\mathcal{R}_3 \,S_3^0 \,\mathcal{R}_3^\dagger ={}& \mathrm{diag} \left( \mathcal{N}_{33},\, \mathcal{N}_{55},\, \mathcal{N}_{66} \right),\\
\mathcal{R}_4 \,S_4^0 \,\mathcal{R}_4^\dagger ={}& \mathrm{diag} \left( \mathcal{N}_{44},\, \mathcal{N}_{55},\, \mathcal{N}_{66} \right),
\end{align}
\end{subequations} 
the eigenvalues of the $S$-matrix are the key focus when analysing the unitarity conditions.

Finally, the last block of the scattering matrix involving the singly-charged states is
\begin{equation}
\mathcal{R}\,S_4^+\,\mathcal{R}^\dagger = \mathrm{diag} \left( 2\left( \lambda_1 - \lambda_2 - 2 \lambda_3 \right),\, \mathcal{N}_{55}^\ast \right),
\end{equation}
where
\begin{equation}
\mathcal{R}= \mathcal{I}_2 \otimes U.
\end{equation}

In general, the part corresponding to the doubly-charged states of the $S$-matrix can be omitted since it produces redundant eigenvalues. However, for completeness, we list these states below:
\begin{subequations}
\begin{align}
\Psi^{++}_1 &= \Big\{ \left.| w^+_1 w^+_1 \right\rangle,\, \left.| w^+_2 w^+_2 \right\rangle,\, \left.| w^+_S w^+_S \right\rangle,\ \left.| w^+_2 w^+_S \right\rangle \Big\},\\
\Psi^{++}_2 &= \Big\{ \left.| w^+_1 w^+_2 \right\rangle,\, \left.| w^+_1 w^+_S \right\rangle \Big\}.
\end{align}
\end{subequations}
The relevant parts of the scattering matrices are:
\begin{subequations}
\begin{align}
S_1^{++} &= S_1^0,\\
S_2^{++} &= S_2^0.
\end{align}
\end{subequations}

\subsection{Electroweak oblique parameters}\label{Sec:SectionSTU}

Finally, we want to mention the electroweak oblique parameters. These were discussed in Section~\ref{Sec:SM_Theor_Constr}. In the context of extended scalar models, these parameters determine the extent to which the electroweak sector can be extended beyond the SM reference point. The PDG~\cite{ParticleDataGroup:2024cfk} provides the following values:
\begin{subequations}
\begin{align}
S &= ~~0.04 \pm 0.10, \\
T &= ~~0.01 \pm 0.12, \\
U &= -0.01 \pm 0.09.
\end{align}
\end{subequations}
Fixing $U=0$~\cite{Grinstein:1991cd}, motivated by the suppression factor $m_\mathrm{new}^2/m_Z^2$ compared to $S$ and $T$, yields~\cite{ParticleDataGroup:2024cfk} : 
\begin{subequations}
\begin{align}
S &= -0.05 \pm 0.07, \\
T &= ~~0.00 \pm 0.06.
\end{align}
\end{subequations}
The approach for deriving the electroweak-oblique parameters for the NHDM was presented in Refs.~\cite{Grimus:2007if,Grimus:2008nb}. For simplicity, suppose that we have the following mass eigenstates:
\begin{subequations}
\begin{equation}
\begin{pmatrix}
e^{i \sigma} h_1^0 \\
h_2^0 \\
e^{i \gamma} h_S^0
\end{pmatrix} =  \begin{pmatrix}
e^{i \sigma} \left( i G^0 + \sum_{i=1}^3 \mathcal{R}^0_{i1} H_i \right) \\
\sum_{i=1}^3 \left( \mathcal{R}_{i2}^0 + i \mathcal{R}_{i3}^0 \right) H_i \\
e^{i \gamma} \left(\varphi_1 + i \varphi_2\right) \\
\end{pmatrix} = V \begin{pmatrix}
G^0 \\
H_1 \\
H_2 \\
H_3 \\
\varphi_1 \\
\varphi_2 \\
\end{pmatrix},
\end{equation}
with
\begin{equation}
V = \begin{pmatrix}
i e^{i\sigma} & \mathcal{R}_{11}^0 e^{i\sigma} & \mathcal{R}_{21}^0 e^{i\sigma}  & \mathcal{R}_{31}^0 e^{i\sigma} & 0 & 0 \\
0 & \left( \mathcal{R}_{12}^0 + i \mathcal{R}_{13}^0 \right) & \left( \mathcal{R}_{22}^0 + i \mathcal{R}_{23}^0\right) & \left( \mathcal{R}_{32}^0 + i \mathcal{R}_{33}^0 \right) & 0 & 0 \\
0 & 0 & 0 & 0 &  e^{i\gamma} & i e^{i\gamma}
\end{pmatrix}, 
\end{equation}
and
\begin{equation}
\begin{pmatrix}
e^{i \sigma} h_1^+ \\
h_2^+ \\
e^{i \gamma} h_S^+
\end{pmatrix} = \begin{pmatrix}
G^+ \\
H^+ \\
h^+ \\
\end{pmatrix} = U \begin{pmatrix}
G^+ \\
H^+ \\
h^+ \\
\end{pmatrix}, \text{ with } U=\mathrm{diag}\left(e^{i\sigma},\,1,\,e^{i\gamma}\right).
\end{equation}
\end{subequations}
We want the scalars to be aligned in a way that the Goldstone bosons always occupy the first position. In the above example we assumed the C-III-a implementation so that it would be understandable what happens with complex vevs and inert doublets are considered. For simplicity, we are free to re-define the fields via $h_1 \to e^{i \sigma_1} \left[ e^{-i \sigma_1} h_1 \right]$, where $\sigma_1$ is the phase of vev $\left\langle h_1 \right\rangle = e^{i \sigma_1} \hat v_1$. This is done on purpose to simplify how the mass-squared matrices will look like. Apart from that, since $\hat v_S = 0$, we are free to rotate it by $\gamma$, which also diagonalises the neutral mass-squared matrix.

The $S$ and $T$ functions are given by:
\begin{subequations}
\begin{align}
\begin{split}
S  ={}& \frac{m_W^2 \sin {\theta_W}^2}{24 \alpha \pi^2 v^2} \Bigg[ \sum_{i=2}^n \left( 2 \sin {\theta_W}^2 - \left( U^\dagger U \right)_{ii} \right)^2 G \left( m_{H^\pm_i}^2, m_{H^\pm_i}^2, m_Z^2 \right)\\
& \hspace{60pt} + 2 \sum_{i=2}^{n-1} \sum_{j = i+1}^n\left| \left( U^\dagger U \right)_{ij} \right|^2 G \left( m_{H^\pm_i}^2, m_{H^\pm_{j}}^2, m_Z^2 \right)\\
& \hspace{60pt} + \sum_{i=2}^{n-1} \sum_{j = i+1}^n \left[ \mathbb{I}\mathrm{m} \left( V^\dagger V \right)_{ij} \right]^2 G \left( m_{H^0_{i}}^2, m_{H^0_{j}}^2, m_Z^2 \right)\\
& \hspace{60pt} - 2 \sum_{i=2}^n \left( U^\dagger U \right)_{ii} \ln{m_{H^\pm_i}^2} + \sum_{i=2}^n \left( V^\dagger V \right)_{ii} \ln{m_{H^0_{i}}^2} - \ln{m_h^2}\\
& \hspace{60pt} + \sum_{i=2}^n \left[ \mathbb{I}\mathrm{m} \left( V^\dagger V \right)_{1i}  \right]^2 \hat G \left( m_{H^0_{i}}^2, m_Z^2 \right) - \hat G \left( m_h^2, m_Z^2 \right) \Bigg],
\end{split}\\
\begin{split}
T  ={}& \frac{1}{16 \alpha \pi^2 v^2} \Bigg[ \sum_{i=2}^n\, \sum_{j=2}^m\, \left| \left( U^\dagger V \right)_{ij} \right|^2 F \left( m_{H^\pm_i}^2, m_{H^0_{j}}^2 \right)\\
& \hspace{60pt}- \sum_{i=2}^{n-1}\, \sum_{j = i+1}^n\,\left[ \mathbb{I}\mathrm{m} \left( V^\dagger V \right)_{ij} \right]^2 F \left( m_{H^0_{i}}^2, m_{H^0_{j}}^2 \right)\\
& \hspace{60pt}- 2\, \sum_{i=2}^{n-1}\, \sum_{j = i+1}^n\,\left| \left( U^\dagger U \right)_{ii} \right|^2 F \left( m_{H^\pm_i}^2 , m_{H^\pm_{j}}^2 \right)\\ 
& \hspace{60pt}+ 3\, \sum_{i=2}^n\,\left[ \mathbb{I}\mathrm{m} \left( V^\dagger V \right)_{1i} \right]^2 \left[F \left( m_Z^2, m_{H^0_{i}}^2 \right) - F \left( m_W^2, m_{H^0_{i}}^2 \right)\right]\\ 
& \hspace{60pt}- 3 \left[F \left( m_Z^2, m_h^2 \right) - F \left( m_W^2, m_h^2 \right)\right]\Bigg],
\end{split}
\end{align}
\end{subequations}
where $\alpha$ is the fine-structure constant, and summation is over the physical scalars, excluding the Goldstone bosons, ${H^\pm_1} = {G^\pm}$ and ${H^0_1} = {G^0}$. The functions, which appear in the expressions above, are given by~\cite{Veltman:1977kh}:
\begin{equation}
F(I, J) \equiv
\begin{cases}
 \frac{I+J}{2}-\frac{I J}{I-J} \ln \frac{I}{J}, &  I \neq J, \\
 ~0, &  I = J, 
\end{cases}
\end{equation}
and~\cite{Grimus:2007if,Grimus:2008nb}:
\begin{subequations}
\begin{align}
G(I, J, Q) \equiv {} & -\frac{16}{3}+\frac{5(I+J)}{Q}-\frac{2(I-J)^{2}}{Q^{2}} \\ &+\frac{3}{Q}\left[\frac{I^{2}+J^{2}}{I-J}-\frac{I^{2}-J^{2}}{Q}+\frac{(I-J)^{3}}{3 Q^{2}}\right] \ln \frac{I}{J}+\frac{r}{Q^{3}} f(t, r),\\
\tilde{G}(I, J, Q) \equiv {} &-2+\left(\frac{I-J}{Q}-\frac{I+J}{I-J}\right) \ln \frac{I}{J}+\frac{f(t, r)}{Q},\\
\hat{G}(I, Q)  \equiv {} & G(I, Q, Q)+12 \tilde{G}(I, Q, Q),
\end{align}
\end{subequations}
where
\begin{equation}
f(t, r) \equiv 
\begin{cases}
\sqrt{r} \ln \left|\frac{t-\sqrt{r}}{t+\sqrt{r}}\right|, &  r>0, \\ 
0, &   r=0,\\ 
2 \sqrt{-r} \arctan \left( \frac{\sqrt{-r}}{t} \right), &   r<0,
\end{cases}
\end{equation}
and
\begin{subequations}
\begin{align}
t & \equiv I+J-Q,\\
r & \equiv Q^{2}-2 Q(I+J)+(I-J)^{2}.
\end{align}
\end{subequations}

\chapter{Dark Matter in the \texorpdfstring{\boldmath$S_3$}{S3}-symmetric three-Higgs-doublet model}\label{Ch:DM_3HDM}

The discovery of the Higgs boson could play a pivotal role in addressing the origin of DM, as models involving the Higgs portal~\cite{Eboli:2000ze,Godbole:2003it,Patt:2006fw,Kanemura:2010sh,Chu:2011be,Djouadi:2011aa,Djouadi:2012zc,Belanger:2013kya,Curtin:2013fra,Arcadi:2017kky} may offer insights into the DM puzzle. In such models, the Higgs boson can decay into invisible channels if kinematically allowed, where these decays may arise from a portal connecting the visible SM sector to the dark/inert sector. Extending the EW sector to include additional particles or interactions could provide a natural framework for understanding DM. These extensions could lead to new experimental signatures, such as missing transverse momentum from invisible Higgs decays, offering ways to probe the nature of DM through both direct detection and collider experiments. One of the most well-studied extended scalar sector models is the IDM, see Section~\ref{Sec:IDM}.

While the IDM provides an elegant solution, it does not tackle other shortcomings within the SM. Bearing in mind these limitations and the lack of experimental evidence for the IDM, one might consider extending the model to NHDMs that include more than two doublets. The most natural way to control free parameters in NHDMs is to impose symmetries~\cite{Ferreira:2008zy,Ivanov:2011ae,Ivanov:2012ry,Ivanov:2012fp,Keus:2013hya,Ivanov:2014doa,Pilaftsis:2016erj,deMedeirosVarzielas:2019rrp,Darvishi:2019dbh,Bree:2024edl,Kuncinas:2024zjq,Doring:2024kdg}. In three-Higgs-doublet models (3HDM), the potential discrete symmetries that can be implemented were classified in Ref.~\cite{Ivanov:2012fp}, with a more comprehensive classification of all possible Abelian and discrete non-Abelian symmetries provided in Ref.~\cite{Keus:2013hya}. In addition to reducing the number of free parameters, these symmetries can also play a crucial role in stabilising DM.

Models with multiple $SU(2)$ scalar doublets can exhibit vacua with vanishing vevs in the symmetry-imposed basis. If an underlying symmetry stabilises such vacua, these models may produce viable DM candidates, provided the symmetry forbids direct couplings between DM candidates and fermions. In the previous chapter we examined the 3HDM with the $S_3$ symmetry, which features diverse vacuum structures, including configurations with one or two vanishing vevs. These vacua can be stabilised by remnants of the $S_3$ symmetry after SSB. We shall study in some detail several implementations, providing the mass terms, couplings, and numerically analysing the parameter space of some cases, ensuring compatibility with theoretical and experimental constraints.

In Section~\ref{Sec:SectionDM} we discussed vacua of the $S_3$-symmetric 3HDM, which could, in principle, accommodate a DM candidate. Seven, see Table~\ref{Table:Diff_Cases}, different implementations were identified, however only three cases are ``good-to-go": no unwanted massless scalars and correct predictions from the Yukawa Lagrangian. The $S_3$ implementations of particular interest are: R-I-1, R-II-1a, C-III-a. The R-II-1a and C-III-a with $\lambda_4 \in \mathbb{R}$ were analysed in Refs.~\cite{Khater:2021wcx,Kuncinas:2022whn,Kuncinas:2023hpf}, on which this chapter is based. We focus on models which would result in a dominant DM contribution.

\section{Dark Matter in the three-Higgs-doublet models}\label{Sec:3HDM-review}

In the construction of scalar DM models, discrete symmetries, particularly $\mathbb{Z}_n$ symmetries~\cite{Ivanov:2012hc}, are frequently employed. Here are several 3HDMs that can incorporate DM:
\begin{itemize}

\item \textbf{IDM2 model with \cite{Grzadkowski:2009bt,Grzadkowski:2010au,Osland:2013sla} and without~\cite{Merchand:2019bod} CP violation}

The IDM2 model includes two active Higgs doublets, which generally introduce FCNCs. To control FCNCs, one approach is to impose a $\mathbb{Z}_2^\prime$ symmetry between the active doublets, where one of the doublets carries the same charge as the RH up-type quarks under $\mathbb{Z}_2^\prime$. This symmetry is softly broken, resulting in CP violation within the 2HDM. Additionally, an inert $SU(2)$ scalar doublet is introduced, stabilised by a separate $\mathbb{Z}_2$ symmetry. To reduce the number of free parameters, the ``dark democracy" limit is applied, which ensures that the inert doublet couples identically to both active doublets. Recent constraints from direct DM detection experiments, such as XENONnT~\cite{XENON:2023cxc} and LUX-ZEPLIN~\cite{LZ:2024zvo} (see Figure 16 of Ref.~\cite{Grzadkowski:2010au}), suggest that the viable DM mass range in the IDM2 is comparable to that of the IDM. A re-evaluation of the model, assuming real couplings,  was performed in Ref.~\cite{Merchand:2019bod}. There, two allowed DM mass regions were identified: 57--73 GeV and 500--1000 GeV.

\item \textbf{\boldmath$\mathbb{Z}_2$-3HDM with~\cite{Keus:2014jha,Keus:2014isa,Keus:2015xya,Cordero:2017owj,Dey:2023exa} and without~\cite{Cordero-Cid:2016krd,Cordero-Cid:2018man,Cordero-Cid:2020yba} CP violation}

The key distinction between these models and the IDM2 model is that they feature a single active Higgs doublet and two inert doublets. Since there is only one active doublet, FCNCs are naturally avoided, as the Yukawa Lagrangian remains SM-like. In the $\mathbb{Z}_2$-symmetric 3HDM, the scalar potential consists of a single bilinear term and eight quartic phase-dependent couplings. It has been argued that the phenomenology of the model remains unchanged if only the bilinear term and the three quartic couplings that respect the $\mathbb{Z}_2 \times \mathbb{Z}_2$ symmetry are considered. To simplify the analysis, given the complexity of the parameter space, further relations among couplings were imposed, along with some numerical simplifications. We refer to these models (due to the simplified parameter space) as ``truncated" $\mathbb{Z}_2$-3HDM. A thorough collider-focused study was conducted alongside several cosmological constraints. After imposing both collider and cosmological cuts, two allowed DM mass regions were identified. The lighter DM mass region falls within 53--75 GeV. Due to the presence of an additional inert doublet, the heavy DM mass region can be as low as 360 GeV, facilitated by co-annihilation channels involving additional inert scalars~\cite{Keus:2015xya}. Similar to the IDM, for DM masses exceeding 500 GeV, a precise tuning of the mass splittings between the inert sector scalars is required.

\item \textbf{\boldmath$\mathbb{Z}_2 \times \mathbb{Z}_2$-3HDM \cite{Hernandez-Sanchez:2020aop,Hernandez-Sanchez:2022dnn,Boto:2024tzp}}

The $\mathbb{Z}_2 \times \mathbb{Z}_2$-3HDM with real coefficients has been studied, where the symmetry remains unbroken if only a single non-vanishing vev is present. The scalar doublet acquiring this vev is a singlet under both $\mathbb{Z}_2$ symmetries. Due to the presence of two distinct $\mathbb{Z}_2$ symmetries, the lightest particles from the $\mathbb{Z}_2$-odd doublets serve as DM candidates. However, it is not possible to entirely prevent conversion processes between the two DM states. As a result, the relic density is predominantly determined by the lighter DM state, while the heavier DM candidate contributes only a few percent. A limited exploration of the parameter space was conducted in Refs.~\cite{Hernandez-Sanchez:2020aop,Hernandez-Sanchez:2022dnn}, identifying a lightest DM candidate within the 65--80 GeV mass range and a heavier state at the $\mathcal{O}(100)$ GeV scale. In light of direct DM detection constraints from XENONnT~\cite{XENON:2023cxc} and LUX-ZEPLIN~\cite{LZ:2022lsv}, the mass of the surviving lightest DM candidate is around 71--73 GeV. This region is ruled out in light of the recent UX-ZEPLIN~\cite{LZ:2024zvo} data. Furthermore, the upcoming fifteen-year observations of dwarf spheroidal galaxies by Fermi-LAT~\cite{Hess:2021cdp} could potentially rule out the currently viable parameter space. Recently, the model was revisited in Ref.~\cite{Boto:2024tzp} with an expanded parameter space, allowing scalars to be as heavy as 1 TeV. The allowed DM mass regions for each sector remain similar to those of the IDM.

\item \textbf{\boldmath$\mathbb{Z}_3$-3HDM \cite{Aranda:2019vda,Hernandez-Otero:2022dxd}}

In the $\mathbb{Z}_3$-3HDM~\cite{Aranda:2019vda}, the only vacuum that preserves the $\mathbb{Z}_3$ symmetry requires two vanishing vevs. This model features two mass-degenerate DM candidates originating from the same inert doublet, each carrying opposite CP numbers. The contributions of both DM candidates to the relic density are typically of the same order. A study of collider dynamics in Ref.~\cite{Hernandez-Otero:2022dxd} highlighted distinctive final-state spectra. Since both DM candidates belong to the same doublet, constraints from direct DM detection experiments necessitate careful control of trilinear gauge vertices involving the $Z$ boson. This requirement is naturally fulfilled if there is maximal mixing between the inert doublets. In Ref.~\cite{Hernandez-Otero:2022dxd}, the $\mathbb{Z}_3$ symmetry was assumed to be softly broken, which in turn made the interaction strength of the trilinear scalar-gauge vertex proportional to the soft-term coupling. Several benchmark scenarios were explored in Refs.~\cite{Aranda:2019vda,Hernandez-Otero:2022dxd}, with the allowed DM mass regions, consistent with collider and cosmological constraints, found to be 53--77 GeV or above approximately 420 GeV.

\item \textbf{\boldmath$S_3$-3HDM \cite{Khater:2021wcx,Kuncinas:2022whn,Kuncinas:2023hpf}}

In the $S_3$-symmetric 3HDM with real couplings, three possible vacuum configurations can lead to a viable DM candidate. If soft symmetry breaking is introduced, up to eight additional models emerge, see Table~\ref{Table:Diff_Cases_Sum}. One such model, referred to as R-II-1a, was covered in Ref.~\cite{Khater:2021wcx}. Heavy DM states are excluded since the portal coupling increases with DM mass, whereas it must remain close to zero to satisfy the relic density constraint.  Another scenario, C-III-a, permits spontaneous CP violation~\cite{Kuncinas:2022whn}. Constraints from indirect DM searches, depending on assumed DM halo distribution profiles, could entirely rule out this model~\cite{Kuncinas:2023hpf}. The third case, R-I-1, differs from the previous two by having only a single non-vanishing vev. When considering complex couplings, two additional implementations exist~\cite{Kuncinas:2023ycz}, with vev structures resembling those of the R-I-1 and C-III-a cases, both of which can accommodate DM. We cover the R-II-1a and C-III-a implementations in this chapter. The allowed regions are: $[52.7;\,82.9]~\text{GeV}$ for R-II-1a and $[28.9;\,41.9]~\text{GeV}$ for C-III-a.

\newpage
\item \textbf{\boldmath$S_3 \rtimes \mathbb{Z}_2 \sim O(2)$-3HDM \cite{Machado:2012gxi,Fortes:2014dca}}

A related case to the R-I-1 of the $S_3$-symmetric implementation was explored in Refs.~\cite{Machado:2012gxi,Fortes:2014dca}. There, the symmetry was extended to $S_3 \times \mathbb{Z}_2 $, more accurately denoted as $S_3 \rtimes \mathbb{Z}_2$, which is actually $O(2)$, by requiring $\lambda_4=0$. A soft term was introduced to lift mass degeneracies. Although a systematic numerical analysis was not performed, benchmark points in the DM mass range of 40--150 GeV were provided in Ref.~\cite{Fortes:2014dca}. We discuss the R-I-1 implementation in this chapter, but do not perform a numerical check, and note that evaluating this implementation at a higher order should result in a natural mass splitting, not requiring soft terms.

\item \textbf{\boldmath$U(1) \times U(1)$-3HDM \cite{Kuncinas:2024zjq}}

The $U(1) \times U(1)$-symmetric 3HDM is the most general real scalar 3HDM potential, as it does not permit complex couplings. In Ref.~\cite{Kuncinas:2024zjq} a numerical analysis was performed. Due to challenges imposed by elastic scattering via the $Z$ boson, this model should be viewed as a toy model. In this model, there are two distinct DM mass scales, each associated with a different inert doublet. While there appears to be a preference for the lighter doublet to contribute dominantly to the relic density, this is mainly due to the allowed decay channels of the heavier states into the lighter ones. These processes could, in principle, be suppressed by adjusting the parameters of the model. The analysis revealed that the region where both neutral dark scalars fall within the mass range of $\mathcal{O}(100)-\mathcal{O}(300)$~GeV is excluded. This study offers valuable insights into distinguishing models with continuous symmetries, specifically $U(1)$, from those with discrete symmetries, particularly within the context of 3HDMs. This model will be covered Section~\ref{Sec:U1U1_analysis}.

\item \textbf{CP4-3HDM \cite{Ivanov:2018srm}}

A DM candidate can also be stabilised by a CP symmetry combined with an internal symmetry. One approach involves utilising the CP4 symmetry~\cite{Ivanov:2015mwl}, which remains unbroken by the vacuum. This model features two mass-degenerate neutral DM candidates, protected by the underlying CP4 symmetry. In Ref.~\cite{Ivanov:2018srm}, it was assumed that the thermal evolution of these DM candidates occurs in the asymmetric regime~\cite{Petraki:2013wwa, Zurek:2013wia}; however, it is important to note that the two mass-degenerate DM states in the CP4-3HDM are not a particle–antiparticle pair. Compared to conventional asymmetric DM models, the CP4-3HDM introduces an additional conversion process between the DM candidates, providing slightly more freedom in its parameter space. However, this CP4 asymmetric DM scenario is only valid at the EW scale, as additional high-scale physics is required to generate the initial asymmetry between the DM states. A detailed numerical scan of the model space has not been performed.

\item \textbf{\boldmath$\Sigma(36)$-3HDM \cite{Deng:2025dcq}}

In this model, there are two DM candidates, associated with different $SU(2)$ doublets. The DM candidate is stabilised by the $S_3$ symmetry. Due to the highly constrained scalar potential of $\Sigma(36)$, one bilinear and three quartic couplings, the mass degeneracies between the scalar doublets is inevitable. As a result of the constrained space, some of the couplings, $e.g.$, the portal couplings, are expressed in terms of the mass-squared parameters. It is impossible to have a small portal coupling for the sub-62 GeV DM candidates, and as a result, the light DM mass region is ruled out by the constraints coming from the Higgs boson decay to the invisible channel. Apart from that the $\Sigma(36)$-symmetric 3HDM predicts suppressed Higgs di-photon decay rates. The direct DM detection rates are also in conflict with the model. One of the possibilities to comply with the above constraints is to introduce a soft symmetry-breaking parameter. Then, one has to consider a heavy inert sector with mass splittings of $\leq \mathcal{O}(10)$ GeV, eventually resulting in a clash with theoretical and experimental constraints. The model can account only for a sub-leading $\Omega h^2$ contribution.

\end{itemize}

The DM mass ranges of these models are schematically represented in Figure~\ref{Fig:DM_mass_ranges_different_models}.

\vspace{6pt}\begin{figure}[htb]
\begin{center}
\includegraphics[scale=0.68]{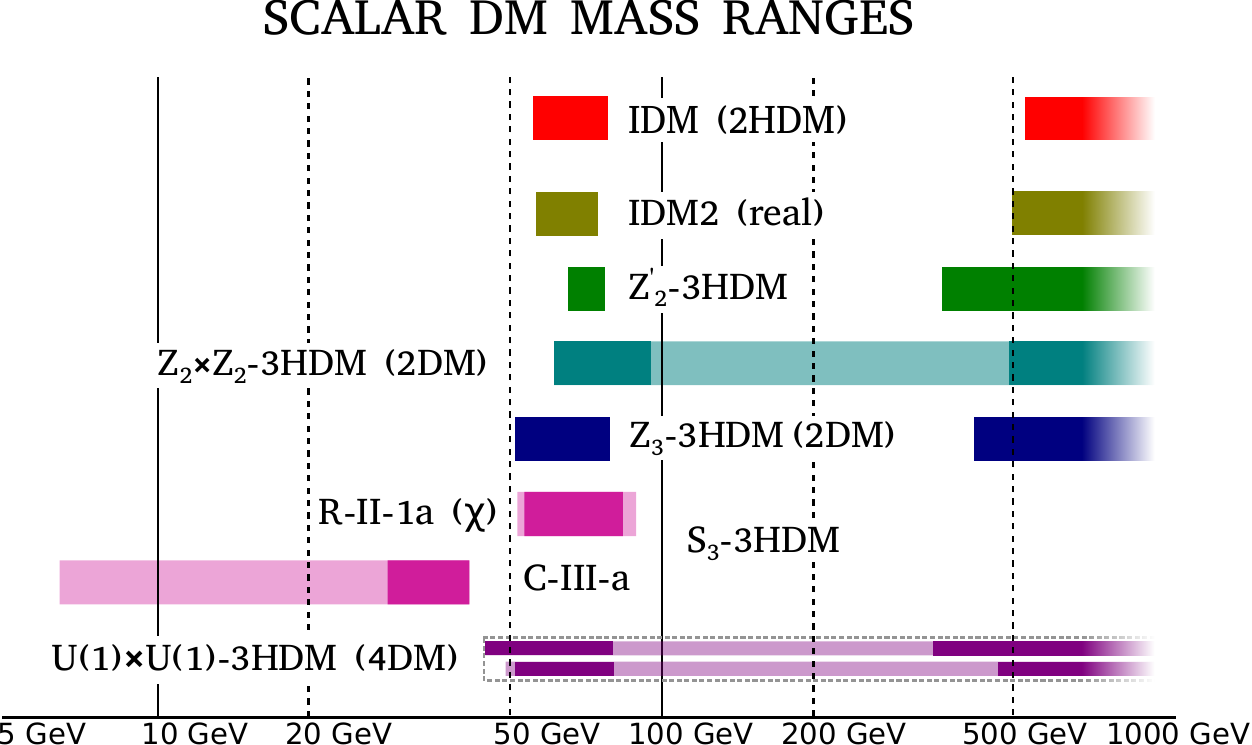}
\end{center}
\vspace*{-3mm}
\caption{Sketch of allowed DM mass ranges in the IDM and various 3HDMs up to 1 TeV. Red: IDM according to Refs.~\cite{Belyaev:2016lok,Kalinowski:2018ylg}. Olive: IDM2 with real coefficients~\cite{Merchand:2019bod}. Green: truncated $\mathbb{Z}_2$-3HDM without~\cite{Keus:2014jha,Keus:2014isa,Keus:2015xya,Cordero:2017owj,Dey:2023exa} and with~\cite{Cordero-Cid:2016krd,Cordero-Cid:2018man,Cordero-Cid:2020yba} CP violation. Teal: $\mathbb{Z}_2 \times \mathbb{Z}_2$ with two DM candidates~\cite{Boto:2024tzp}. The darker regions are consistent with solutions where a single doublet predominantly saturates the relic density. Navy: $\mathbb{Z}_3$ with two mass-degenerate DM candidates~\cite{Aranda:2019vda,Hernandez-Otero:2022dxd}. Pink: $S_3$-3HDM without CP violation, R-II-1a~\cite{Khater:2021wcx} and with CP violation, C-III-a~\cite{Kuncinas:2022whn}. The darker region is consistent with the re-evaluated constraints of the current work. Purple: the $U(1) \times U(1)$-symmetric model of Ref.~\cite{Kuncinas:2024zjq}. The two bands correspond to independent mass scales arising from different inert doublets. The darker regions represent the solutions where a single doublet primarily dominates the relic density, ensuring that its contribution to the overall DM relic density is significant. Combinations of mass scales in the intermediate mass region, of the lighter DM candidate 100--300 GeV and the heavier one in 100--450 GeV, are excluded.}
\label{Fig:DM_mass_ranges_different_models}
\end{figure}

\section{The R-I-1 implementation}

The R-I-1 vacuum is given by $(0,\,0,\,v)$. In the defining representation we have 
\begin{equation}
\begin{pmatrix}
\frac{1}{\sqrt{2}} & \frac{1}{\sqrt{6}} & \frac{1}{\sqrt{3}} \\
-\frac{1}{\sqrt{2}} & \frac{1}{\sqrt{6}} & \frac{1}{\sqrt{3}} \\
0 & -\frac{2}{\sqrt{3}} & \frac{1}{\sqrt{3}}
\end{pmatrix} \begin{pmatrix}
0 \\ 0 \\ v
\end{pmatrix} = \frac{1}{\sqrt{3}} \begin{pmatrix}
v \\
v \\
v
\end{pmatrix},
\end{equation}
which indicates that vacuum is invariant under the $S_3$ symmetry. We limit our discussion to the irreducible basis.

There is a single minimisation condition:
\begin{equation}
\mu_{SS}^2 = - \lambda_8 v^2.
\end{equation}

\subsection{The mass-squared matrices}

The charged mass-squared matrix is:
\begin{equation}
\mathcal{M}^2_\mathrm{Ch} = \mathrm{diag} \left(\mu_{11}^2 + \frac{v^2}{2} \lambda_5,\,  \mu_{11}^2 + \frac{v^2}{2} \lambda_5,\, 0  \right).
\end{equation}

There is no mixing in the neutral sector, and hence the mass-squared matrices can be split into two sectors:
\begin{equation}
\mathcal{M}_\eta^2 = \mathrm{diag} \left(\mu_{11}^2 + \frac{v^2}{2} (\lambda_5 + \lambda_6 + 2 \lambda_7),\,   \mu_{11}^2 + \frac{v^2}{2} (\lambda_5 + \lambda_6 + 2 \lambda_7),\, 2 v^2 \lambda_8  \right).
\end{equation}
and
\begin{equation}
\mathcal{M}_\chi^2 = \mathrm{diag} \left(\mu_{11}^2 + \frac{v^2}{2} (\lambda_5 + \lambda_6 - 2 \lambda_7),\,  \mu_{11}^2 + \frac{v^2}{2} (\lambda_5 + \lambda_6 - 2 \lambda_7),\, 0  \right).
\end{equation}

Due to no mixing the scalar states correspond to the mass eigenstates,
\begin{subequations}
\begin{align}
h_1 &= \begin{pmatrix}
h_1^\pm \\
\left( \eta_1 + i \chi_1 \right)/\sqrt{2}
\end{pmatrix}, \quad
h_2 = \begin{pmatrix}
h_2^\pm \\
\left( \eta_2 + i \chi_2 \right)/\sqrt{2}
\end{pmatrix},\\
h_S &= \begin{pmatrix}
G^\pm \\
\left( v + h + i G^0\right)/\sqrt{2}
\end{pmatrix}.
\end{align}
\end{subequations}
The $h_S$ doublet would then be the SM-like Higgs doublet. The mass-squared terms are given by:
\begin{subequations}\label{Eq:m2_terms}
\begin{align}
m_{h^\pm}^2 & = \mu_{11}^2 + \frac{v^2}{2} \lambda_5,\\
m_\eta^2 & = \mu_{11}^2 + \frac{v^2}{2} (\lambda_5 + \lambda_6 + 2 \lambda_7),\\
m_\chi^2 & = \mu_{11}^2 + \frac{v^2}{2} (\lambda_5 + \lambda_6 - 2 \lambda_7),\\
m_h^2 & = 2 v^2 \lambda_8 .
\end{align}
\end{subequations}

There are three pairs of mass-degenerate states between the $h_1$ and $h_2$ doublets. However, the scalar interactions shall differ for $h_1$ and $h_2$. The $h_1$ doublet exhibits a $\mathbb{Z}_2$ symmetry, $h_1 \to - h_1$, while $h_2$ does not. Such behaviour is caused by the $\lambda_4$ term. 

We can invert the mass-squared terms~\eqref{Eq:m2_terms} and solve for the couplings:
\begin{subequations}
\begin{align}
\lambda_5 & = \frac{2 \left( m_{h^\pm}^2 - \mu_{11}^2 \right)}{v^2},\\
\lambda_6 & = \frac{-2 m_{h^\pm}^2 + m_\eta^2 + m_\chi^2}{v^2},\\
\lambda_7 & = \frac{m_\eta^2 - m_\chi^2}{2v^2},\\
\lambda_8 & = \frac{m_h^2}{2 v^2},
\end{align}
\end{subequations}
while there are 5 free couplings $\{\mu_{11}^2,\, \lambda_1,\, \lambda_2,\, \lambda_3,\, \lambda_4 \}$.

\subsection{Interactions}

\vspace{10pt}
\textbf{Gauge couplings}
\vspace{5pt}

The kinetic Lagrangian is given by:
\begin{subequations}
\begin{align}
\begin{split}
\mathcal{L}_{VVH} =& \left[ \frac{g}{2 \cos \theta_W}m_ZZ_\mu Z^\mu + g m_W W_\mu^+ W^{\mu-} \right] h,
 \end{split}\\
\begin{split}
\mathcal{L}_{VHH} =& -\frac{ g}{2 \cos \theta_W}Z^\mu \eta_i \overset\leftrightarrow{\partial_\mu} \chi_i  - \frac{g}{2}\bigg\{ i W_\mu^+ \left[ ih_i^- \overset\leftrightarrow{\partial^\mu} \chi_i + h_i^- \overset\leftrightarrow{\partial^\mu} \eta_i  \right] + \mathrm{h.c.} \bigg\}\\
& + \left[ i e A^\mu + \frac{i g}{2} \frac{\cos (2\theta_W)}{\cos \theta_W} Z^\mu \right]  h_i^+ \overset\leftrightarrow{\partial_\mu} h_i^-  ,
\end{split}\\
\begin{split}
\mathcal{L}_{VVHH} =& \left[ \frac{g^2}{8 \cos^2\theta_W}Z_\mu Z^\mu + \frac{g^2}{4} W_\mu^+ W^{\mu-} \right] \left( \eta_i^2 + \chi_i^2 + h^2\right)\\
& + \bigg\{ \left[ \frac{e g}{2} A^\mu W_\mu^+ - \frac{g^2}{2} \frac{\sin^2\theta_W}{\cos \theta_W}Z^\mu W_\mu^+ \right] \left[ \eta_i h_i^- + i \chi_i h_i^- \right] + \mathrm{h.c.} \bigg\}\\
&+ \left[ e^2 A_\mu A^\mu + e g \frac{\cos (2\theta_W)}{\cos \theta_W}A_\mu Z^\mu + \frac{g^2}{4} \frac{\cos^2(2\theta_W)}{\cos^2\theta_W}Z_\mu Z^\mu + \frac{g^2}{2} W_\mu^- W^{\mu +} \right] \\
&\quad\times\left( h_i^-h_i^+ \right).
\end{split}
\end{align}
\end{subequations}

\vspace{10pt}
\textbf{Yukawa Lagrangian}
\vspace{5pt}

The only possibility is to assume that fermions transform as singlet under $S_3$.

\vspace{10pt}
\textbf{Scalar interactions}
\vspace{5pt}

For simplicity, the scalar couplings are presented with the symmetry factor, but without the overall coefficient ``$-i$". Trilinear couplings involving only neutral fields are:
\begin{subequations}
\begin{align}
g(\eta_1^2 \eta_2) &= 3 v \lambda_4,\\
g(\eta_2^3) &= - 3 v \lambda_4,\\
g(\chi_1^2 \eta_2) &= g(\eta_1 \chi_1  \chi_2) = v \lambda_4,\\
g(\eta_2 \chi_2^2) &= - v \lambda_4,\\
g(\eta_1^2 h) &= g(\eta_2^2 h) = v (\lambda_5 + \lambda_6 + 2 \lambda_7),\\
g(\chi_1^2 h) &= g(\chi_2^2 h) = v (\lambda_5 + \lambda_6 - 2 \lambda_7),\\
g(h^3) &= 6 v \lambda_8.
\end{align}
\end{subequations}

Trilinear couplings involving charged fields are:
\begin{subequations}
\begin{align}
g(\eta_1 h^+_1 h^-_2) &= g(\eta_2 h_1^+ h_1^-) = v \lambda_4,\\
g(\eta_2 h_2^+ h_2^-) &= - v \lambda_4,\\
g(h h_1^+ h_1^-) &= g(h h_2^+ h_2^-) = v \lambda_5.
\end{align}
\end{subequations}

Quartic couplings involving only neutral fields are:
\begin{subequations}
\begin{align}
g(\eta_1^4) &= g(\chi_1^4) =  g(\eta_2^4) =  g(\chi_2^4) = 6 \left( \lambda_1 + \lambda_3 \right),\\
g(\eta_1^2 \chi_1^2) &= g(\eta_2^2 \chi_2^2)  = g(\eta_1^2 \eta_2^2) = g(\chi_1^2 \chi_2^2) =  2 \left( \lambda_1 + \lambda_3 \right),\\
g(\eta_1^2 \chi_2^2)  &= g(\eta_2^2 \chi_1^2) =  2 \lambda_1 - 4 \lambda_2 - 2 \lambda_3,\\
g(\eta_1 \chi_1 \chi_2  h) &=  g(\chi_1^2 \eta_2 h) =  \lambda_4,\\
g(\chi_2^2 \eta_2 h) &= -\lambda_4,\\
g(\eta_1^2 \eta_2 h) &= 3 \lambda_4,\\
g(\eta_2^3 h) &= - 3 \lambda_4,\\
g(\chi_1^2 h^2) &= g(\chi_2^2 h^2) = (\lambda_5 + \lambda_6 - 2 \lambda_7),\\
g(\eta_1^2 h^2) &= g(\eta_2^2 h^2) = (\lambda_5 + \lambda_6 + 2 \lambda_7),\\
g(\eta_1 \chi_1 \eta_2  \chi_2)  &= 2 \left( \lambda_2 + \lambda_3 \right),\\
g(h^4) &= 6 \lambda_8.
\end{align}
\end{subequations}

Quartic couplings involving neutral and charged fields are:
\begin{subequations}
\begin{align}
g(\eta_1^2 h_1^+ h_1^-) &= g(\chi_1^2 h_1^+ h_1^-) = g(\eta_2^2  h_2^+ h_2^-) = g(\chi_2^2 h_2^+ h_2^-) =  2 \left( \lambda_1 + \lambda_3 \right),\\
g(\eta_1^2 h_2^+ h_2^-) &=  g(\chi_1^2 h_2^+ h_2^-) = g(\eta_2^2 h_1^+ h_1^-) = g(\chi_2^2 h_1^+ h_1^-) =   2 \left( \lambda_1 - \lambda_3 \right),\\
g(\eta_1 \eta_2 h_1^+ h_2^-) &= g(\chi_1 \chi_2 h_1^+ h_2^-) =  2 \lambda_3,\\
g(\eta_1 \chi_2  h_2^+ h_1^-) &= g(\chi_1 \eta_2 h_1^+ h_2^-) =  2 i \lambda_2,\\
g(\eta_1 h h_1^+ h_2^-) &= g(\eta_2 h h_1^+ h_1^-) = \lambda_4,\\
g(\eta_2 h h_2^+ h_2^-) &= - \lambda_4,\\
g(h^2 h_1^+ h_1^-) &= g(h^2 h_2^+ h_2^-) =\lambda_5.
\end{align}
\end{subequations}

Quartic couplings involving only charged fields are:
\begin{subequations}
\begin{align}
g(h_1^+ h_1^+ h_1^- h_1^-) &= g(h_2^+ h_2^+ h_2^- h_2^-) = 4 \left( \lambda_1 + \lambda_3 \right),\\
g(h_1^+ h_1^+ h_2^- h_2^-) &= 4 \left( \lambda_2 + \lambda_3 \right),\\
g(h_1^+ h_2^+ h_1^- h_2^-) &= 2 \left( \lambda_1 - \lambda_2 \right).
\end{align}
\end{subequations}

We recall that \mbox{$S_3 \cong \mathbb{Z}_3 \rtimes \mathbb{Z}_2$}. From the interactions above, one can identify that $h_1$ respects $\mathbb{Z}_2$---there are no interactions involving an odd number of physical states associated with the $h_1$ doublet. On the other hand, the scalars associated with the $h_2$ doublet are not restricted to appear in pairs, \textit{e.g.} there are interactions like $\eta_2 \chi_2^2$ or $\chi^2 \eta h$. Due to additional interactions involving the $\lambda_4$ term, the mass degeneracy between the scalars associated with the $h_1$ and $h_2$ doublets will be lifted. This splitting will be proportional to $\delta m_i \sim (\lambda_4)^2/(16 \pi^2)$. Therefore, we can assume that the mass-squared parameters of $h_2$ can be expressed as, $m_i^2 \to m_i^2 + \delta m_i^2 (\lambda_4)$.

\subsection[The R-I-1 implementation with complex \texorpdfstring{$\lambda_4$}{l4}]{The R-I-1 implementation with complex \boldmath$\lambda_4$}

Now, assume that the $\lambda_4$ coupling becomes complex. In such case it is impossible to rotate away the argument of $\lambda_4$. The CP-odd invariants do not vanish, see eqs.~\eqref{Eq:I6_inv} and \eqref{Eq:I2Y3Z_inv}. In the case of $\lambda_4 \in \mathbb{C}$ we get identical minimisation conditions and masses to those of $\lambda_4 \in \mathbb{R}$. However, the scalar couplings shall differ.

\vspace{10pt}
\textbf{Scalar interactions}
\vspace{5pt}

Apart from the substitution of $\lambda_4 \to \lambda_4^\mathrm{R}$, there are several new trilinear couplings:
\begin{subequations}
\begin{align}
g(\eta_1^2 \chi_2) & = v \lambda_4^\mathrm{I},\\
g(\eta_2^2 \chi_2) & = g(\eta_1 \eta_2 \chi_1) = -v \lambda_4^\mathrm{I},\\
g(\chi_2^3) & = 3 v \lambda_4^\mathrm{I},\\
g(\chi_1^2 \chi_2) & = - 3 v \lambda_4^\mathrm{I},\\
g(\chi_1 h_1^+ h_2^-) &= g(\chi_2 h_1^+ h_1^-) = -v \lambda_4^\mathrm{I},\\
g(\chi_2 h_2^+ h_2^-) & = v \lambda_4^\mathrm{I},
\end{align}
\end{subequations}
while the new quartic couplings involving $\lambda_4^\mathrm{I}$ are:
\begin{subequations}
\begin{align}
g(\eta_1^2 \chi_2 h) & = -\frac{1}{2}\lambda_4^\mathrm{I},\\
g(\eta_2^2 \chi_2 h) & = g(\chi_2^3 h) = \frac{1}{2}\lambda_4^\mathrm{I},\\
g(\chi_1^2 \chi_2 h) & = -\frac{3}{2}\lambda_4^\mathrm{I},\\
g(\eta_1 \eta_2 \chi_1 h) & = -\lambda_4^\mathrm{I},\\
g(\chi_1 h h_1^+ h_2^-) &= g(\chi_2 h h_1^+ h_1^-) = -\lambda_4^\mathrm{I},\\
g(\chi_2 h h_2^+ h_2^-) &= \lambda_4^\mathrm{I}.
\end{align}
\end{subequations}

\section{The R-II-1a implementation}

The R-II-1a vacuum is given by $(0,\, v_2,\, v_S)$. The minimisation conditions are:
\begin{subequations}\label{Eq:RII1a_Min_Cond}
\begin{align}
\mu_{SS}^2&= \frac{1}{2}\lambda _4\frac{ v_2^3}{v_S}-\frac{1}{2} (\lambda_5 + \lambda_6 + 2 \lambda_7) v_2^2-\lambda _8 v_S^2, \\
\mu_{11}^2&= -\left( \lambda _1+ \lambda _3\right) v_2^2+\frac{3}{2} \lambda _4 v_2 v_S-\frac{1}{2} (\lambda_5 + \lambda_6 + 2 \lambda_7) v_S^2.
\end{align}
\end{subequations}

The $\mathbb{Z}_2$ symmetry is preserved for $h_1 \to - h_1$, and hence the inert doublet will be associated with $h_1$. This implementation was covered in Ref.~\cite{Khater:2021wcx}.

The R-II-1a implementation does not allow for $\lambda_4 \in \mathbb{C}$.

\subsection{The mass-squared parameters}

The $h_S$ doublet is the only one that couples to fermions. Establishing $v_S$ as a reference point, we define the Higgs-basis rotation angle as:
\begin{equation}
\tan\beta = \frac{v_2}{v_S} = \frac{v \sin \beta}{v \cos \beta}.
\end{equation}
After an appropriate rephasing of the scalar doublets, we choose $ v_S > 0 $. Since $ v_2 $ can be negative, the Higgs-basis rotation angle is constrained to the range $ \beta \in \left[-\frac{\pi}{2}, \frac{\pi}{2}\right] $.
Then,
\begin{equation}\label{eq.Rbeta}
\begin{aligned}
\mathcal{R}_\beta = \frac{1}{v} \begin{pmatrix}
v & 0 & 0 \\
0 & v_2 & v_S \\
0 & -v_S & v_2
\end{pmatrix} = \begin{pmatrix}
1 & 0 & 0\\
0 & \sin \beta & \cos \beta \\
0 & -\cos \beta & \sin \beta
\end{pmatrix},
\end{aligned}
\end{equation}
and the vevs in the Higgs basis are
\begin{equation}
\mathcal{R}_\beta \begin{pmatrix}
0 \\
v_2  \\
v_S 
\end{pmatrix} = \begin{pmatrix}
0 \\ v \\ 0
\end{pmatrix}.
\end{equation}

\vspace{10pt}
\textbf{Charged sector}
\vspace{5pt}

In the basis of $\{h_1^+,\,h_2^+,\,h_S^+ \}$, the charged mass-squared matrix is:
\begin{equation}
\mathcal{M}^2_\mathrm{Ch}
=\begin{pmatrix}
\left(\mathcal{M}_\mathrm{Ch}^2\right)_{11} & 0 & 0 \\
0 & \left(\mathcal{M}_\mathrm{Ch}^2\right)_{22} & \left(\mathcal{M}_\mathrm{Ch}^2\right)_{23} \\
0 & \left(\mathcal{M}_\mathrm{Ch}^2\right)_{23} & \left(\mathcal{M}_\mathrm{Ch}^2\right)_{33}
\end{pmatrix},
\end{equation}
where
\begin{subequations}
\begin{align}
\left(\mathcal{M}_\mathrm{Ch}^2\right)_{11} &= -2\lambda_3v_2^2+\frac{5}{2}\lambda_4v_2v_S-\frac{1}{2}(\lambda_6+2\lambda_7)v_S^2, \\
\left(\mathcal{M}_\mathrm{Ch}^2\right)_{22} &= \frac{1}{2} v_S \left[ \lambda_4v_2-(\lambda_6+2\lambda_7)v_S \right], \\
\left(\mathcal{M}_\mathrm{Ch}^2\right)_{23} &= -\frac{1}{2} v_2 \left[\lambda_4v_2-(\lambda_6+2\lambda_7)v_S \right], \\
\left(\mathcal{M}_\mathrm{Ch}^2\right)_{33} &= \frac{1}{2} \frac{v_2^2}{v_S} \left[ \lambda_4v_2-(\lambda_6+2\lambda_7)v_S \right].
\end{align}
\end{subequations}
It is diagonalisable by going into the Higgs basis, see eq.~\eqref{eq.Rbeta}. We define the scalars:
\begin{subequations}
\begin{align}
h^+ & = h_1^+,\\
G^+ & = \sin\beta\,h_2^+ + \cos\beta\,h_S^+,\\ 
H^+ & = -\cos\beta\,h_2^+ + \sin\beta\,h_S^+,
\end{align}
\end{subequations}
with masses:
\begin{subequations}
\begin{align}
m^2_{h^+} &= -2\lambda_3v_2^2+\frac{5}{2}\lambda_4v_2v_S-\frac{1}{2}(\lambda_6+2\lambda_7)v_S^2, \\
m^2_{H^+} &=  \frac{v^2}{2 v_S} \left[ \lambda_4 v_2 - \left( \lambda_6 + 2\lambda_7 \right) v_S \right].
\end{align}
\end{subequations}

\vspace{10pt}
\textbf{Neutral inert sector}
\vspace{5pt}

The masses of the neutral states associated with $h_1$ are given by:
\begin{subequations}\label{Eq:RII1a_hS_Masses}
\begin{align}
m^2_{\eta} &= \frac{9}{2} \lambda_4 v_2 v_S,\\
m^2_{\chi} &= -2(\lambda_2+\lambda_3)v_2^2+\frac{5}{2}\lambda_4v_2v_S -2\lambda_7v_S^2.
\end{align}
\end{subequations}

\vspace{10pt}
\textbf{Neutral active sector}
\vspace{5pt}

In the basis of $\{\eta_2, \,\eta_S,\, \chi_2, \,\chi_S\}$, the neutral mass-squared matrix is block-diagonal:
\begin{equation}
\mathcal{M}_\mathrm{N}^2 = \mathrm{diag}\left(\mathcal{M}_{\eta_{2S}}^2 ,\,\mathcal{M}_{\chi_{2S}}^2\right),
\end{equation}
For the CP-odd sector we have:
\begin{equation}
\mathcal{M}^2_{\chi_{2S}}
=\begin{pmatrix}
\left(\mathcal{M}^2_{\chi_{2S}}\right)_{11} & \left(\mathcal{M}^2_{\chi_{2S}}\right)_{12}\vspace{2pt} \\ 
\left(\mathcal{M}^2_{\chi_{2S}}\right)_{12} & \left(\mathcal{M}^2_{\chi_{2S}}\right)_{22}
\end{pmatrix},
\end{equation}
where
\begin{subequations}
\begin{align}
\left(\mathcal{M}^2_{\chi_{2S}}\right)_{11} &= \frac{1}{2} v_S \left( \lambda_4v_2-4\lambda_7v_S \right), \\
\left(\mathcal{M}^2_{\chi_{2S}}\right)_{12} &= -\frac{1}{2} v_2 \left(\lambda_4v_2-4\lambda_7v_S \right), \\
\left(\mathcal{M}^2_{\chi_{2S}}\right)_{22} &= \frac{v_2^2}{2v_S} \left( \lambda_4v_2-4\lambda_7v_S \right).
\end{align}
\end{subequations}
The $\mathcal{M}^2_{\chi_{2S}}$ matrix can be diagonalised via $\mathcal{R}_\beta$ of eq.~\eqref{eq.Rbeta}. The two CP-odd states are:
\begin{subequations}
\begin{align}
G^0 & = \sin\beta\,\chi_2 + \cos\beta\,\chi_S,\\ 
A & = -\cos\beta\,\chi_2 + \sin\beta\,\chi_S,
\end{align}
\end{subequations}
where the mass-squared parameter of $A$ is
\begin{equation}
m^2_A = \frac{v^2}{2v_S} \left( \lambda_4v_2-4\lambda_7v_S \right).
\end{equation}

For the CP-even sector we get:
\begin{equation}\label{Eq:RII1aNeutralM2eta}
\mathcal{M}^2_{\eta_{2S}}
=\begin{pmatrix}
\left(\mathcal{M}^2_{\eta_{2S}}\right)_{11} & \left(\mathcal{M}^2_{\eta_{2S}}\right)_{12}\vspace{2pt} \\
\left(\mathcal{M}^2_{\eta_{2S}}\right)_{12} & \left(\mathcal{M}^2_{\eta_{2S}}\right)_{22}
\end{pmatrix},
\end{equation}
where
\begin{subequations}
\begin{align}
\left(\mathcal{M}^2_{\eta_{2S}}\right)_{11} &= \frac{1}{2} v_2 \left[ 4 \left( \lambda_1 + \lambda_3 \right) v_2 - 3\lambda_4 v_S \right], \\
\left(\mathcal{M}^2_{\eta_{2S}}\right)_{12} &= -\frac{1}{2} v_2 \left[ 3\lambda_4 v_2 - 2(\lambda_5 + \lambda_6 + 2 \lambda_7) v_S   \right], \\
\left(\mathcal{M}^2_{\eta_{2S}}\right)_{22} &= \frac{1}{2v_S} \left( \lambda_4 v_2^3 + 4\lambda_8 v_S^3 \right).
\end{align}
\end{subequations}
There are two possibilities to diagonalise it. If we do not go into the Higgs basis, the diagonalisation matrix is given by:
\begin{equation}
\mathcal{R}_\alpha = \begin{pmatrix}
\cos \alpha & \sin \alpha \\
-\sin \alpha & \cos \alpha
\end{pmatrix},
\end{equation}
with
\begin{equation}\label{Eq:RII1a_t2a}
\tan(2{\alpha}) =  \frac{2v_2v_S(-3\lambda_4v_2+2(\lambda_5 + \lambda_6 + 2 \lambda_7) v_S)}{4(\lambda_1+\lambda_3)v_2^2v_S-\lambda_4 \left( v_2^3+3v_2v_S^2 \right)-4\lambda_8v_S^3}.
\end{equation}
The physical states are:
\begin{subequations}
\begin{align}
h & = \cos{\alpha}\,\eta_2 + \sin{\alpha}\,\eta_S,\\ 
H & = -\sin{\alpha}\,\eta_2 + \cos{\alpha}\,\eta_S,
\end{align}
\end{subequations}
with masses: 
\begin{subequations}
\begin{align}
m_h^2 &= \frac{1}{4 v_S^2} \left[ 4 \left( \lambda_1 + \lambda_3 \right) v_2^2 v_S^2 + \lambda_4 v_2 v_S\left( v_2^2 - 3 v_S^2 \right) + 4\lambda_8 v_S^4  - v_S \Delta \right],\\
m_H^2 &= \frac{1}{4 v_S^2} \left[ 4 \left( \lambda_1 + \lambda_3 \right) v_2^2 v_S^2 + \lambda_4 v_2 v_S\left( v_2^2 - 3 v_S^2 \right) + 4\lambda_8 v_S^4  + v_S \Delta \right],
\end{align}
\end{subequations}
where
\begin{equation}\label{Eq:RII1a_Delta_mhmH}
\begin{aligned}
\Delta^2 =&\, 16 \left( \lambda_1 + \lambda_3\right)^2 v_2^4 v_S^2-8 \left( \lambda_1 + \lambda_3\right) v_2^2 v_S \left[ \lambda_4 \left( v_2^3 + 3 v_2 v_S^2 \right) + 4 \lambda_8 v_S^3 \right]\\
& + 16  (\lambda_5 + \lambda_6 + 2 \lambda_7)^2 v_2^2 v_S^4 - 48 \lambda_4  (\lambda_5 + \lambda_6 + 2 \lambda_7) v_2^3 v_S^3\\
& + \lambda_4^2 \left( v_2^6 + 42 v_2^4 v_S^2 + 9 v_2^2 v_S^4 \right)+ 8 \lambda_4 \lambda_8 v_2 v_S^3\left( v_2^2+3 v_S^2 \right) + 16 \lambda_8^2 v_S^6.
\end{aligned}
\end{equation}
We shall require that the lighter state $h$ is the SM-like Higgs boson.

Provided that the other active sectors are diagonalisable by going into the Higgs basis, we can define the diagonalisation matrix to be
\begin{equation}
\mathcal{R}_{\alpha^\prime} = \begin{pmatrix}
\cos \alpha^\prime & \sin \alpha^\prime \\
-\sin \alpha^\prime & \cos \alpha^\prime
\end{pmatrix},
\end{equation}
where the rotation angle $\alpha^\prime$ is given in terms of the Higgs basis transformation:
\begin{equation}\label{Eq:RII1a_AlphaPrime}
\alpha^\prime = \alpha + \beta - \frac{\pi}{2}.
\end{equation}

\vspace{10pt}
\textbf{Physical states}
\vspace{5pt}

After identifying the physical states, we can express the $ SU(2) $ scalar doublets in terms of these states as follows:
\begin{subequations}\label{RII1aHMEGen}
\begin{align}
h_1 &= \begin{pmatrix}
h^+ \\
\frac{1}{\sqrt{2}} \left( \eta + i \chi \right)
\end{pmatrix},\\
h_2 &= \begin{pmatrix}
 \sin \beta\,G^+ - \cos \beta\,H^+\\
\frac{1}{\sqrt{2}} \left( \sin \beta\,v + \cos{\alpha}\,h - \sin{\alpha}\,H  + i\left( \sin\beta\,G^0 - \cos\beta\,A\right)\right)
\end{pmatrix},\\
h_S &= \begin{pmatrix}
 \cos \beta\,G^+ + \sin \beta\,H^+\\
\frac{1}{\sqrt{2}} \left( \cos\beta\,v + \sin{\alpha}\,h + \cos{\alpha}\,H  + i\left( \cos\beta\,G^0 + \sin\beta\,A\right)\right)
\end{pmatrix},
\end{align}
\end{subequations}
or equivalently
\begin{subequations}\label{RII1aHB}
\begin{align}
h_1^\mathrm{HB} &= \begin{pmatrix}
h^+ \\
\frac{1}{\sqrt{2}} \left( \eta + i \chi \right)
\end{pmatrix},\\
h_2^\mathrm{HB} &= \begin{pmatrix}
 G^+ \\
\frac{1}{\sqrt{2}} \left( v +  \sin(\alpha+\beta)\,h + \cos(\alpha+\beta)\,H  + i G^0 \right)
\end{pmatrix},\\
h_3^\mathrm{HB} &= \begin{pmatrix}
 H^+\\
\frac{1}{\sqrt{2}} \left( -\cos(\alpha+\beta)\,h + \sin(\alpha+\beta)\,H  + i A \right)
\end{pmatrix}.
\end{align}
\end{subequations}

By inverting the mass-squared parameters, we can express the quartic couplings in terms of the vevs and masses:
\begin{subequations}\label{eq.R-II-1aInvertedCouplings}
\begin{align}
\lambda_1&=\frac{v^2 \left[ 9 \left( m_{h^+}^2+\cos^2 {\alpha}\,  m_h^2+\sin^2{\alpha}\,  m_H^2 \right)-m_{\eta}^2 \right]-9 m_{H^+}^2 v_S^2}{18 v^2 v_2^2}, \\
\lambda_2&=\frac{ \left(m_{h^+}^2 - m_{\chi}^2 \right)v^2 + \left(m_A^2-m_{H^+}^2 \right)v_S^2}{2 v^2 v_2^2}, \\
\lambda_3&=\frac{ \left(4 m_ \eta ^2-9 m_{h^+}^2 \right)v^2+9 m_{H^+}^2 v_S^2}{18 v^2 v_2^2}, \\
\lambda_4&=\frac{2 m_ \eta ^2}{9 v_2 v_S}, \label{Eq:R-II-1aInvertedCouplings-lam4}\\
\lambda_5 &=\frac{2 m_{H^+}^2}{v^2}+\frac{v_2 m_ \eta ^2-\frac{9}{2} \sin (2\alpha) v_S (m_H^2-m_h^2)}{9 v_2 v_S^2}, \\
\lambda_6&= \frac{m_A^2-2 m_{H^+}^2}{v^2} + \frac{m_ \eta ^2}{9 v_S^2}, \\
\lambda_7&=\frac{1}{18} \left(\frac{m_ \eta ^2}{v_S^2}-\frac{9 m_A^2}{v^2}\right), \\
\lambda_8&=\frac{9 v_S^2 \left(\sin^2 {\alpha}\,  m_h^2+\cos^2 {\alpha}\,  m_H^2\right)-v_2^2 m_ \eta ^2}{18 v_S^4}.
\end{align}
\end{subequations}

The R-II-1a implementation is invariant under a simultaneous transformation of
\begin{equation}\label{Eq:Param_sym}
\begin{aligned}
&\beta \to - \beta, ~ \alpha \to \pi - \alpha,\\
&\lambda_4 \to - \lambda_4,\\
&\{H^\pm,\,H,\,A\} \to -\{H^\pm,\,H,\,A\},
\end{aligned}
\end{equation}
which can be adopted to reduce ranges of the parameter space.

\subsection{Interactions}

\vspace{10pt}
\textbf{Gauge couplings}
\vspace{5pt}

Substituting the $SU(2)$ doublets in terms of the mass eigenstates of eq.~\eqref{RII1aHMEGen} into the kinetic Lagrangian yields:
\begin{subequations}
\begin{align}\label{LVVH-RII1a}
\begin{split}
\mathcal{L}_{VVH} =& \left[ \frac{g}{2 \cos \theta_W}m_ZZ_\mu Z^\mu + g m_W W_\mu^+ W^{\mu-} \right] \left[ \sin (\alpha + \beta)h + \cos (\alpha + \beta)H \right],
 \end{split}\\
\begin{split}\label{LVHH-RII1a}
\mathcal{L}_{VHH} =& -\frac{ g}{2 \cos \theta_W}Z^\mu \left[ \eta \overset\leftrightarrow{\partial_\mu} \chi  - \cos (\alpha+\beta)h \overset\leftrightarrow{\partial_\mu} A + \sin (\alpha+\beta) H \overset\leftrightarrow{\partial_\mu} A \right]\\
& - \frac{g}{2}\bigg\{ i W_\mu^+ \left[ ih^- \overset\leftrightarrow{\partial^\mu} \chi + h^- \overset\leftrightarrow{\partial^\mu} \eta  - \cos (\alpha+\beta)H^- \overset\leftrightarrow{\partial^\mu} h \right. \\ & \hspace{60pt} \left.  
+ \sin (\alpha+\beta)H^-\overset\leftrightarrow{\partial^\mu}H + i H^- \overset\leftrightarrow{\partial^\mu} A \right] + \mathrm{h.c.} \bigg\}\\
& + \left[ i e A^\mu + \frac{i g}{2} \frac{\cos (2\theta_W)}{\cos \theta_W} Z^\mu \right] \left( h^+ \overset\leftrightarrow{\partial_\mu} h^- + H^+ \overset\leftrightarrow{\partial_\mu} H^-  \right),
\end{split}\\
\begin{split}\label{LVVHH-RII1a}
\mathcal{L}_{VVHH} =& \left[ \frac{g^2}{8 \cos^2\theta_W}Z_\mu Z^\mu + \frac{g^2}{4} W_\mu^+ W^{\mu-} \right] \left( \eta^2 + \chi^2 + h^2 + H^2 +A^2\right)\\
& + \bigg\{ \left[ \frac{e g}{2} A^\mu W_\mu^+ - \frac{g^2}{2} \frac{\sin^2\theta_W}{\cos \theta_W}Z^\mu W_\mu^+ \right] \left[ \eta h^- + i \chi h^- - \cos (\alpha+\beta)hH^- \right. \\ 
& \hspace{145pt} \left. + \sin (\alpha+\beta)HH^- + iAH^- \right] + \mathrm{h.c.} \bigg\}\\
&+ \left[ e^2 A_\mu A^\mu + e g \frac{\cos (2\theta_W)}{\cos \theta_W}A_\mu Z^\mu + \frac{g^2}{4} \frac{\cos^2(2\theta_W)}{\cos^2\theta_W}Z_\mu Z^\mu + \frac{g^2}{2} W_\mu^- W^{\mu +} \right] \\
&\times\left( h^-h^+ + H^-H^+ \right).
\end{split}
\end{align}
\end{subequations}

Given that the $ h $ scalar is associated with the SM-like Higgs boson, the SM-like limit is attained when the interactions $ hVV $ satisfy the condition:
\begin{equation}\label{Eq:RII1a_SV_SM}
\sin(\alpha+\beta)=1, \quad \cos(\alpha+\beta)=0.
\end{equation}

\vspace{10pt}
\textbf{Yukawa Lagrangian}
\vspace{5pt}

Although there are two non-zero vevs, we still have to assume that all fermions transform as singlets under $S_3$. Otherwise, unrealistic structure of the CKM matrix is predicted, see eq.~\eqref{Eq:DifferentMDoublets} for the structure.  Since all fermions couple exclusively to $h_S$, tree-level FCNCs are absent. The structure of the Yukawa couplings resembles that of the Type-I 2HDM, except that our definition of $ \tan\beta $ is inverted:
\begin{equation}\label{Eq:Tan_beta_relation_S3}
\left(\tan\beta\right)_\text{R-II-1a}=\left(\frac{1}{\tan\beta}\right)_\text{Type-I 2HDM}.
\end{equation}

The scalar-fermion couplings of R-II-1a are:
\begin{subequations}\label{Eq:R-II-1a-Yukawa_Neutral}
\begin{align}
g \left( h \bar{f}f \right) & = -i \frac{m_f}{v}\frac{\sin\alpha}{\cos\beta},\\
g \left( H \bar{f}f \right) & = -i \frac{m_f}{v}\frac{\cos\alpha}{\cos\beta},\\ 
g \left( A \bar{u}u \right) & = - \gamma_5 \frac{m_{u}}{v}\tan \beta, \quad
g \left( A \bar{d}d \right) = \gamma_5 \frac{m_{d}}{v}\tan \beta.
\end{align}
\end{subequations}
For the leptonic sector, the Dirac mass terms would result in similar relations. The SM-like limit for $h$ is restored at 
\begin{equation}\label{Eq:RII1a_FS_SM}
\frac{\sin\alpha}{\cos\beta}= 1.
\end{equation}

The charged scalar-fermion couplings are given by:
\begin{subequations}\label{Eq:R-II-1a-Yukawa_Charged}
\begin{align}
g \left( H^+ \bar{u}_i d_j \right) & = i \frac{\sqrt{2}}{v} \tan \beta \left[ P_L m_u - P_R m_d  \right] \left( V_\mathrm{CKM} \right)_{ij},\\
g \left( H^- \bar{d}_i u_j \right) & = i \frac{\sqrt{2}}{v} \tan \beta \left[ P_R m_u - P_L m_d  \right] \left( V_\mathrm{CKM}^\dagger \right)_{ji},\\
g \left( H^+ \bar{\nu} l  \right) & = - i \frac{\sqrt{2} m_l}{v} \tan \beta P_R,\\
g \left( H^- \bar{l} \nu  \right) & = - i \frac{\sqrt{2} m_l}{v} \tan \beta P_L.
\end{align}
\end{subequations}

\vspace{10pt}
\textbf{Scalar interactions}
\vspace{5pt}

The scalar couplings are presented with the symmetry factor, but without the overall coefficient ``$-i$". Trilinear scalar couplings involving the same species are:
\begin{subequations}
\begin{align}
\begin{split}
g\left( h h h \right) & = 3 v \bigg[\cos^3 \alpha \left(2 \left(\lambda _1+\lambda _3\right) \sin \beta-\lambda _4 \cos \beta\right)\\
&\hspace{35pt} + \left((\lambda_5 + \lambda_6  + 2 \lambda_7) \cos \beta-3 \lambda _4 \sin \beta\right)\cos^2 \alpha \sin \alpha\\
&\hspace{35pt} + (\lambda_5 + \lambda_6  + 2 \lambda_7) \cos \alpha\sin^2 \alpha \sin \beta +2 \lambda _8 \sin^3 \alpha \cos \beta\bigg]\label{Eq.R_II_1a_hhh},
\end{split}\\
\begin{split}
g\left( H H H \right) & =-3 v \bigg[\sin^3 \alpha \left(2 \left(\lambda _1+\lambda _3\right) \sin \beta-\lambda _4 \cos \beta\right)\\
&\hspace{42pt} + \left(3 \lambda _4 \sin \beta-(\lambda_5 + \lambda_6  + 2 \lambda_7) \cos \beta\right)\cos \alpha \sin^2 \alpha \\
&\hspace{42pt} + (\lambda_5 + \lambda_6  + 2 \lambda_7) \cos^2 \alpha\sin \alpha \sin \beta -2 \lambda _8 \cos^3 \alpha \cos \beta\bigg].
\end{split}
\end{align} 
\end{subequations}

Other trilinear couplings involving the neutral fields are:
\begin{subequations}
\begin{align}
\begin{split}g\left( \eta  \eta  h \right) & = v \bigg[\sin \alpha \left(3 \lambda _4 \sin \beta + (\lambda_5 + \lambda_6  + 2 \lambda_7) \cos \beta\right) \\
&  \hspace{30pt}+2 \left(\lambda _1+\lambda _3\right)\cos \alpha \sin \beta   + 3\lambda _4  \cos \alpha \cos \beta  \bigg]\label{Eq.R_II_1a_hetaeta},\end{split}\\
\begin{split}g\left( \eta  \eta  H \right) & = v \bigg[ \cos \alpha \left(3 \lambda _4 \sin \beta + (\lambda_5 + \lambda_6  + 2 \lambda_7) \cos \beta\right) \\
&  \hspace{30pt}- 2 \left(\lambda _1+\lambda _3\right)\sin \alpha \sin \beta   - 3  \lambda _4  \sin \alpha  \cos \beta \bigg],\end{split}\\
\begin{split}g\left( \chi  \chi  h \right) & = v \bigg[\sin \alpha \left(\lambda _4 \sin \beta + (\lambda_5 + \lambda_6 - 2 \lambda_7) \cos \beta\right) \\
&  \hspace{30pt} + 2 \left(\lambda _1-2 \lambda _2-\lambda _3\right)\cos \alpha \sin \beta   + \lambda _4 \cos \alpha \cos \beta  \bigg]\label{Eq.R_II_1a_hchichi},\end{split}\\
\begin{split}g\left( \chi  \chi  H \right) & = v \bigg[\cos \alpha \left(\lambda _4 \sin \beta + (\lambda_5 + \lambda_6 - 2 \lambda_7) \cos \beta\right) \\
&  \hspace{30pt} -2 \left(\lambda _1-2 \lambda _2-\lambda _3\right)\sin \alpha  \sin \beta - \lambda _4 \sin \alpha  \cos \beta \bigg],\end{split}\\
g\left( \eta  \chi  A \right) & = -v \left[ \lambda _4 \cos (2 \beta)+\left(\lambda _2+\lambda _3-\lambda _7\right) \sin (2 \beta) \right],\\
\begin{split}
g\left( h h H \right) & = -v \bigg[\cos^3 \alpha \left(3 \lambda _4 \sin \beta-(\lambda_5 + \lambda_6  + 2 \lambda_7) \cos \beta\right) \\
& \hspace{35pt} +\cos^2 \alpha \sin \alpha \left(2\left(3 \lambda _1 + 3\lambda _3-\lambda_5 - \lambda_6  - 2 \lambda_7\right) \sin \beta-3 \lambda _4 \cos \beta\right)\\
& \hspace{35pt}  +(\lambda_5 + \lambda_6  + 2 \lambda_7) \sin^3 \alpha \sin \beta \\
& \hspace{35pt}  - 2 \cos \alpha \sin^2 \alpha \left(3 \lambda _4 \sin \beta+\left(3 \lambda _8-\lambda_5 - \lambda_6  - 2 \lambda_7\right) \cos \beta\right)\bigg],
\end{split}\\
\begin{split}
g\left( h H H \right) & = v \bigg[-\sin^3 \alpha \left(3 \lambda _4 \sin \beta-(\lambda_5 + \lambda_6  + 2 \lambda_7) \cos \beta \right)\\
& \hspace{35pt}  +\cos \alpha \sin^2 \alpha \left(2\left(3 \lambda _1 + 3\lambda _3-\lambda_5 - \lambda_6  - 2 \lambda_7\right) \sin \beta-3 \lambda _4 \cos \beta\right)\\
& \hspace{35pt}  +(\lambda_5 + \lambda_6  + 2 \lambda_7) \cos^3 \alpha \sin \beta \\
& \hspace{35pt} + 2 \cos^2 \alpha \sin \alpha \left(3 \lambda _4 \sin \beta+\left(3 \lambda _8-\lambda_5 - \lambda_6  - 2 \lambda_7\right) \cos \beta\right)\bigg],
\end{split}\\
\begin{split}
g\left( A A h \right) & = v \bigg[ \left( \lambda_4 \left( 2 \cos \beta \sin ^2\beta - \cos^3\beta \right) + (\lambda_5 + \lambda_6 - 2 \lambda_7) \sin^3\beta \right)\cos \alpha \\
&\hspace{35pt} +2 \cos \alpha \cos ^2\beta \sin \beta \left(\lambda _1+\lambda _3-2 \lambda _7\right)\\ 
&\hspace{35pt}  -\frac{1}{2}\left( \lambda _4 \sin (2 \beta) +4 \left(2 \lambda _7 - \lambda _8\right) \sin ^2\beta \right)\sin \alpha \cos \beta \\ 
&\hspace{35pt}  + (\lambda_5 + \lambda_6 - 2 \lambda_7) \cos ^3\beta \sin \alpha  \bigg],
\end{split}\\
\begin{split}
g\left( A A H \right) & = v \bigg[ -\left( \lambda_4 \left( 2 \cos \beta \sin ^2\beta - \cos^3\beta \right) + (\lambda_5 + \lambda_6 - 2 \lambda_7) \sin^3\beta \right)\sin \alpha \\
&\hspace{35pt} - 2 \sin \alpha \cos ^2\beta \sin \beta \left(\lambda _1+\lambda _3-2 \lambda _7\right)\\ 
&\hspace{35pt}  -\frac{1}{2}\left( \lambda _4 \sin (2 \beta) +4 \left(2 \lambda _7 - \lambda _8\right) \sin ^2\beta \right)\cos \alpha \cos \beta \\ 
&\hspace{35pt}  + (\lambda_5 + \lambda_6 - 2 \lambda_7) \cos ^3\beta \cos \alpha  \bigg].
\end{split}
\end{align} 
\end{subequations}

Trilinear couplings involving the charged fields are:
\begin{subequations}
\begin{align}
\begin{split}g\left( \eta  h^\pm H^\mp \right) & = -\frac{1}{4} v \bigg[4 \lambda _4 \cos (2 \beta)+\left(4 \lambda _3-\lambda _6-2 \lambda _7\right) \sin (2 \beta)\bigg],\end{split}\\
g\left( \chi  h^\pm H^\mp \right) & = \mp\frac{1}{4} i v \left(4 \lambda _2+\lambda _6-2 \lambda _7\right) \sin (2 \beta),\\
\begin{split}
g\left( h H^\pm H^\mp \right) & = -v \bigg[\left( \lambda_4 \left( \cos^3\beta -2 \cos \beta \sin ^2\beta\right)-\lambda _5 \sin^3\beta\right)\cos \alpha  \\
&\hspace{40pt} -\cos \alpha \cos ^2\beta \sin \beta \left(2 \lambda_1+2 \lambda_3-\lambda _6-2 \lambda _7\right) \\
&\hspace{40pt}+\left( \lambda _4 \cos ^2\beta\sin \beta-\lambda _5 \cos^3\beta + \lambda_7 \sin \beta \sin (2 \beta)\right)\sin \alpha\\
&\hspace{40pt}  + \sin \alpha \cos \beta \sin^2 \beta  \left( \lambda_6 - 2 \lambda_8 \right)  \bigg],
\end{split}\\
\begin{split}
g\left( H H^\pm H^\mp \right) & = v \bigg[\left( \lambda_4 \left( \cos^3\beta -2 \cos \beta \sin ^2\beta\right)-\lambda _5 \sin^3\beta\right)\sin \alpha  \\
&\hspace{40pt} -\sin \alpha \cos ^2\beta \sin \beta \left(2 \lambda_1+2 \lambda_3-\lambda _6-2 \lambda _7\right) \\
&\hspace{40pt}+\left( \lambda _4 \cos ^2\beta\sin \beta-\lambda _5 \cos^3\beta + \lambda_7 \sin \beta \sin (2 \beta)\right)\cos \alpha\\
&\hspace{40pt}  + \cos \alpha \cos \beta \sin^2 \beta  \left( \lambda_6 - 2 \lambda_8 \right)  \bigg],
\end{split}\\
\begin{split}g\left( h h^\pm h^\mp \right) & = v \bigg[\cos \alpha \left(2 \left(\lambda _1-\lambda _3\right) \sin \beta+\lambda _4 \cos \beta\right)\\
&\hspace{30pt} +\sin \alpha \left(\lambda _4 \sin \beta+\lambda _5 \cos \beta\right)\bigg],\end{split}\\
\begin{split}g\left( H h^\pm h^\mp \right) & = v \bigg[ - \sin \alpha \left(2 \left(\lambda _1-\lambda _3\right) \sin \beta+\lambda _4 \cos \beta\right)\\
&\hspace{30pt} + \cos \alpha \left(\lambda _4 \sin \beta+\lambda _5 \cos \beta\right)\bigg].\end{split}
\end{align} 
\end{subequations}

Quartic couplings involving the same species are:
{\setlength{\belowdisplayskip}{0pt} \setlength{\belowdisplayshortskip}{0pt}
\setlength{\abovedisplayskip}{0pt} \setlength{\abovedisplayshortskip}{0pt}
\begin{subequations}
\begin{align}
g\left( \eta  \eta  \eta  \eta  \right) & = g\left( \chi  \chi  \chi  \chi  \right) = 6 \left(\lambda _1+\lambda _3\right)\label{Eq.R_II_1a_QCLim_l1l3},\\
\begin{split} g\left( h h h h \right) & = 6 \bigg[\left(\lambda _1+\lambda _3\right) \cos^4 \alpha-2 \lambda _4 \cos^3 \alpha \sin \alpha\\
&\hspace{30pt} + \frac{1}{4}(\lambda_5 + \lambda_6  + 2 \lambda_7) \sin^2(2 \alpha) +\lambda _8 \sin^4 \alpha  \bigg]\label{Eq.R_II_1a_hhhh},\end{split}\\
\begin{split} g\left( H H H H \right) & = 6 \bigg[\left(\lambda _1+\lambda _3\right) \sin^4 \alpha + 2 \lambda _4 \cos \alpha \sin^3\alpha\\
&\hspace{30pt} + \frac{1}{4}(\lambda_5 + \lambda_6  + 2 \lambda_7) \sin^2(2 \alpha)  + \lambda _8 \cos^4 \alpha \bigg],\end{split}\\
\begin{split} g\left( A A A A \right) &= 6 \bigg[ \left(\lambda _1+\lambda _3\right) \cos^4\beta  + 2 \lambda _4 \cos^3\beta  \sin \beta \\
&\hspace{30pt} + \frac{1}{4}(\lambda_5 + \lambda_6  + 2 \lambda_7) \sin^2 (2 \beta) + \lambda _8 \sin^4\beta \bigg].\end{split}
\end{align} 
\end{subequations}}

Quartic couplings involving only the neutral fields are:
{
\setlength{\belowdisplayskip}{0pt} \setlength{\belowdisplayshortskip}{0pt}
\setlength{\abovedisplayskip}{0pt} \setlength{\abovedisplayshortskip}{0pt}
\begin{subequations}
\begin{align}
g\left( \eta  \eta  \chi  \chi  \right) & = 2 \left(\lambda _1+\lambda _3\right),\\
g\left( \eta  \eta  A A \right) & = 2 \left(\lambda _1-2 \lambda _2-\lambda _3\right) \cos ^2\beta-\lambda _4 \sin (2 \beta)+(\lambda_5 + \lambda_6 - 2 \lambda_7) \sin^2\beta,\\
g\left( \chi  \chi  A A \right) & = 2 \left(\lambda _1+\lambda _3\right) \cos^2\beta-3 \lambda _4 \sin (2 \beta)+(\lambda_5 + \lambda_6  + 2 \lambda_7) \sin^2\beta,\\
g\left( \eta  \eta  h h \right) & = 2 \left(\lambda _1+\lambda _3\right) \cos^2 \alpha+3 \lambda _4 \sin (2 \alpha)+(\lambda_5 + \lambda_6  + 2 \lambda_7) \sin ^2\alpha,\\
g\left( \eta  \eta  h H \right) & = -\frac{1}{2}\left(2 \lambda _1+2 \lambda _3 - \lambda_5 - \lambda_6  - 2 \lambda_7 \right) \sin (2 \alpha)+3 \lambda _4 \cos(2\alpha)  ,\\
g\left( \eta  \eta  H H \right) & = 2 \left(\lambda _1+\lambda _3\right) \sin^2\alpha-3 \lambda _4 \sin (2 \alpha)+(\lambda_5 + \lambda_6  + 2 \lambda_7) \cos^2\alpha,\\
g\left( \chi  \chi  h h \right) & = 2 \left(\lambda _1-2 \lambda _2-\lambda _3\right) \cos^2\alpha+ \lambda _4 \sin (2 \alpha)+ \lambda_5 + \lambda_6 - 2 \lambda_7 \sin^2\alpha,\\
g\left( \chi  \chi  h H \right) & = \lambda _4 \cos(2\alpha) -\frac{1}{2}\left(2 \lambda _1-4 \lambda _2-2 \lambda _3-\lambda_5 - \lambda_6 + 2 \lambda_7\right) \sin (2 \alpha),\\
g\left( \chi  \chi  H H \right) & = 2 \left(\lambda _1-2 \lambda _2-\lambda _3\right) \sin^2\alpha-\lambda _4 \sin (2 \alpha)+(\lambda_5 + \lambda_6 - 2 \lambda_7) \cos^2\alpha,\\
g\left( \eta  \chi  h A \right) & = -\cos \alpha \left(2 \left(\lambda _2+\lambda _3\right) \cos \beta-\lambda _4 \sin \beta\right)-\left(\lambda _4 \cos \beta-2 \lambda _7 \sin \beta\right)\sin \alpha,\\
g\left( \eta  \chi  H A \right) & = \sin \alpha \left(2 \left(\lambda _2+\lambda _3\right) \cos \beta-\lambda _4 \sin \beta\right)- \left(\lambda _4 \cos \beta-2 \lambda _7 \sin \beta\right)\cos \alpha,\\
\begin{split}
g\left( h h h H \right) & = -3 \cos \alpha \Big[\lambda _4 \cos (3 \alpha) + \left(\lambda _1+\lambda _3 -\lambda _8\right)\sin \alpha \\
&\hspace{60pt}+  \left(\lambda _1+\lambda _3-\lambda_5 - \lambda_6 - 2 \lambda_7+\lambda _8\right) \cos(2\alpha) \sin \alpha\Big],
\end{split}\\
g\left( h h H H \right) & = \frac{1}{4} \Big[3 \lambda _1+3 \lambda _3+6 \lambda _4 \sin (4 \alpha)+ (\lambda_5 + \lambda_6  + 2 \lambda_7)+3 \lambda _8 \\
&\hspace{30pt} -3 \left(\lambda _1+\lambda _3-\lambda_5 - \lambda_6 - 2 \lambda_7+\lambda _8\right) \cos (4 \alpha)\Big] ,\\
g\left( h H H H \right) & = -\frac{3}{2} \sin \alpha \Big[2 \lambda _4 \sin (3 \alpha)+\left(\lambda _1+\lambda _3+ \lambda_5 + \lambda_6  + 2 \lambda_7-3 \lambda _8\right) \cos \alpha\\
&\hspace{65pt} -\left(\lambda _1+\lambda _3-\lambda_5 - \lambda_6 - 2 \lambda_7+\lambda _8\right) \cos (3 \alpha)\Big] ,\\
\begin{split}
g\left( A A h h \right) & = - \sin (2 \alpha) \cos \beta \left(\lambda _4 \cos \beta+4 \lambda _7 \sin \beta\right)\\
&\hspace{13pt} + \sin^2 \alpha \left[(\lambda_5 + \lambda_6 - 2 \lambda_7) \cos^2\beta+2 \lambda _8 \sin^2\beta\right]\\
&\hspace{13pt} + \cos^2\alpha\left[ 2 \left(\lambda _1+\lambda _3\right) \cos^2\beta +\lambda _4 \sin (2 \beta) + (\lambda_5 + \lambda_6 - 2 \lambda_7) \sin^2\beta\right],
\end{split}\raisetag{2\baselineskip}\\
\begin{split}
g\left( A A h H \right) & = -\frac{1}{2}\sin (2 \alpha) \left(\lambda _1+\lambda _3+\lambda _4 \sin (2 \beta)-\lambda _8 \right)\\
&\hspace{13pt} -\frac{1}{2}\sin (2 \alpha) \cos (2 \beta) \left(\lambda _1+\lambda _3-\lambda_5 - \lambda_6 + 2 \lambda_7+\lambda _8\right) \\
&\hspace{13pt}-\cos (2 \alpha) \cos \beta \left(\lambda _4 \cos \beta + 4 \lambda _7 \sin \beta\right),
\end{split}\\
\begin{split}
g\left( A A H H \right) & = \sin (2 \alpha) \cos \beta \left(\lambda _4 \cos \beta+4 \lambda _7 \sin \beta\right)\\
 &\hspace{13pt} + \cos^2 \alpha \left[(\lambda_5 + \lambda_6 - 2 \lambda_7) \cos^2\beta+2 \lambda _8 \sin^2\beta\right]\\
 &\hspace{13pt} + \sin^2\alpha\left[ 2 \left(\lambda _1+\lambda _3\right) \cos^2\beta +\lambda _4 \sin (2 \beta) + (\lambda_5 + \lambda_6 - 2 \lambda_7) \sin^2\beta\right],
\end{split}\raisetag{2\baselineskip}
\end{align} 
\end{subequations}}

Quartic couplings involving both neutral and charged fields are:
{
\setlength{\belowdisplayskip}{0pt} \setlength{\belowdisplayshortskip}{0pt}
\setlength{\abovedisplayskip}{0pt} \setlength{\abovedisplayshortskip}{0pt}
\begin{subequations}
\begin{align}
g\left( \eta  \eta  h^\pm h^\mp \right) & =  g\left( \chi  \chi  h^\pm h^\mp \right) = 2 \left(\lambda_1+\lambda_3\right),\\
g\left( \eta  \eta  H^\pm H^\mp \right) & = g\left( \chi  \chi  H^\mp H^\pm \right) =2 \left(\lambda_1-\lambda_3\right) \cos^2\beta-\lambda _4 \sin (2 \beta)+ \lambda _5 \sin^2\beta ,\\
\begin{split}g\left( \eta  h h^\pm H^\mp \right) & = - \left(2 \lambda _3 \cos \beta-\lambda _4 \sin \beta\right)\cos \alpha-\lambda _4\sin \alpha \cos \beta\\
& \hspace{20pt}+\frac{1}{2} \left(\lambda _6+2 \lambda _7\right) \sin \alpha\sin \beta , \end{split}\\
\begin{split}g\left( \eta  H h^\pm H^\mp \right) & = \left(2 \lambda _3 \cos \beta-\lambda _4 \sin \beta\right)\sin \alpha-\lambda _4 \cos \alpha \cos \beta\\
& \hspace{20pt}+\frac{1}{2} \left(\lambda _6+2 \lambda _7\right) \cos \alpha \sin \beta,\end{split}\\
g\left( \chi  h h^\pm H^\mp \right) & = \mp i \left[2 \lambda _2 \cos \alpha \cos \beta+\frac{1}{2} \left(\lambda _6-2 \lambda _7\right) \sin \alpha\sin \beta \right],\\
g\left( \chi  H h^\pm H^\mp \right) & = \mp i \left[-2 \lambda _2 \sin \alpha \cos \beta+\frac{1}{2} \left(\lambda _6-2 \lambda _7\right) \cos \alpha \sin \beta\right],\\
g\left( h h h^\pm h^\mp \right) & =2 \left(\lambda _1-\lambda _3\right) \cos^2\alpha+\lambda _4 \sin (2 \alpha)+\lambda _5 \sin^2\alpha,\\
g\left( h H h^\pm h^\mp \right) & =-\frac{1}{2}\left(2\lambda _1-2\lambda _3-\lambda _5\right) \sin (2 \alpha)+\lambda _4 \cos(2\alpha) ,\\
g\left( H H h^\pm h^\mp \right) & =2 \left(\lambda _1-\lambda _3\right) \sin^2\alpha-\lambda _4 \sin (2 \alpha)+\lambda _5 \cos^2\alpha,\\
g\left( A A h^\pm h^\mp \right) & = 2 \left(\lambda _1-\lambda _3\right) \cos^2\beta-\lambda _4 \sin (2 \beta)+\lambda _5 \sin^2\beta,\\
g\left( \eta   A h^\pm H^\mp \right) & = \pm i \left[-2 \lambda _2 \cos^2\beta+\frac{1}{2} \left(\lambda _6-2 \lambda _7\right) \sin^2\beta\right],\\
g\left( \chi   A h^\pm H^\mp \right) & =2 \lambda _3 \cos^2\beta-\lambda _4 \sin (2 \beta)+ \frac{1}{2} \left(\lambda _6+2 \lambda _7\right) \sin^2\beta,\\
\begin{split}
g\left( h h H^\pm H^\mp \right) & = \cos^2\alpha \left[2 \left(\lambda _1+\lambda _3\right) \cos^2\beta+\lambda _4 \sin (2 \beta)+\lambda _5 \sin^2\beta\right]\\
&\hspace{15pt} -\lambda _4 \sin (2 \alpha) \cos^2\beta + \lambda _5 \sin^2\alpha \cos^2\beta\\
&\hspace{15pt} -\frac{1}{2}\left(\lambda_6 + 2\lambda_7 \right)\sin (2 \alpha) \sin (2 \beta)+2 \lambda _8 \sin^2\alpha \sin^2\beta ,
\end{split}\\
\begin{split}
g\left( h H H^\pm H^\mp \right) & =-\frac{1}{2} \left(\lambda _1+\lambda _3+\lambda _4 \sin (2 \beta)-\lambda _8\right)\sin (2 \alpha) \\
&\hspace{15pt}-\frac{1}{2} \left(\lambda _1+\lambda _3-\lambda _5+\lambda _8\right) \cos (2 \beta) \sin (2 \alpha) \\
&\hspace{15pt} - \frac{1}{2}\left[\lambda _4 \cos \beta+\left(\lambda _6+2 \lambda _7\right)\right] \cos (2 \alpha) \sin (2\beta),
\end{split}\\
\begin{split}
g\left( H H H^\pm H^\mp \right) & = \sin^2\alpha \left[2 \left(\lambda _1+\lambda _3\right) \cos^2\beta+\lambda _4 \sin (2 \beta)+\lambda _5 \sin^2\beta\right]\\
&\hspace{15pt} +\lambda _4 \sin (2 \alpha) \cos^2\beta + \lambda _5 \cos^2\alpha \cos^2\beta\\
&\hspace{15pt} +\frac{1}{2}\left(\lambda_6 + 2\lambda_7 \right)\sin (2 \alpha) \sin (2 \beta)+2 \lambda _8 \cos^2\alpha \sin^2\beta ,
\end{split}\\
\begin{split}
g\left( A A H^\pm H^\mp \right) & = 2 \Big[\left(\lambda _1+\lambda _3\right) \cos^4\beta+2 \lambda _4 \cos^3\beta \sin \beta\\
&\hspace{30pt}+\frac{1}{4}(\lambda_5 + \lambda_6  + 2 \lambda_7) \sin^2 (2 \beta)+\lambda _8 \sin^4\beta\Big].\end{split}
\end{align} 
\end{subequations}}

Quartic couplings involving only the charged fields are:
\begin{subequations}
\begin{align}
g\left( h^\pm h^\pm h^\mp h^\mp \right) & = 4 \left(\lambda _1+\lambda _3\right),\\
\begin{split}g\left( H^\pm H^\pm H^\mp H^\mp \right) & = 4 \Big[\left(\lambda _1+\lambda _3\right) \cos^4\beta+2 \lambda _4 \cos^3\beta \sin \beta\\
& \hspace{30pt} +\frac{1}{4}(\lambda_5 + \lambda_6  + 2 \lambda_7) \sin^2 (2 \beta)+\lambda _8 \sin^4\beta\Big],\end{split}\\
g\left( h^\pm h^\pm H^\mp H^\mp \right) & = 4 \left[\left(\lambda _2+\lambda _3\right) \cos^2\beta-\frac{1}{2} \lambda _4 \sin (2 \beta)+\lambda _7 \sin^2\beta\right],\\
g\left( h^\pm h^\mp H^\pm H^\mp \right) & = 2 \left(\lambda _1-\lambda _2\right) \cos^2\beta-2 \lambda _4 \sin (2 \beta)+\left(\lambda _5+\lambda _6\right) \sin^2\beta.
\end{align} 
\end{subequations}

\section{The C-III-a implementation}

The C-III-a vacuum is given by $(0,\,\hat{v}_2 e^{i \sigma},\,\hat{v}_S)$. This vacuum is reminiscent of the R-II-1a vacuum, $\{0,\,v_2,\,v_S\} \in \mathbb{R}$. Now, $v_2$ is allowed to be complex. Due to the complex phase, entries of the mass-squared matrices will become complex. It is convenient to re-define the $h_2$ doublet by extracting the overall phase,
\begin{equation} \label{eq:phase}
h_2= e^{i\sigma} e^{-i\sigma} \left(
\begin{array}{c}h_2^{+}\\  \frac{1}{\sqrt{2}}(\hat v_2 e^{i \sigma}+\eta_2 +i \chi_2)
\end{array}\right) = e^{i\sigma}\left(
\begin{array}{c}h_2^{\prime+}\\ \frac{1}{\sqrt{2}}(\hat v_2+\eta_2^\prime+i \chi_2^\prime)
\end{array}\right).
\end{equation}
We shall omit the primes on the fields.

The minimisation conditions are:
\begin{subequations}\label{Eq: C-III-a-Min_Con}
\begin{align}
\mu_{SS}^2 &= -\frac{1}{2}(\lambda_5 + \lambda_6 - 2 \lambda_7)v_2^2 - \lambda_8 v_S^2,\\
\mu_{11}^2 &= - \left( \lambda_1 + \lambda_3\right)v_2^2 - \frac{1}{2}\left( \lambda_5 + \lambda_6 - 2 \lambda_7 - 8 \cos^2 \sigma \lambda_7 \right)v_S^2,\\
\lambda_4 &= \frac{4 \cos \sigma v_S}{v_2}\lambda_7.
\label{eq:c-iii-a-lam4-lam7}
\end{align}
\end{subequations}

Compared to the R-II-1a implementation, this vacuum introduces one additional parameter: a non-vanishing relative phase $\sigma$. This results in an additional constraint between two quartic terms, as given in eq.~(\ref{eq:c-iii-a-lam4-lam7}). In the limit of $\cos \sigma = 1$, the expressions for $ \mu_{SS}^2 $ and $ \mu_{11}^2 $ become identical to the R-II-1a ones, see eq.~\eqref{Eq:RII1a_Min_Cond}.

Only $h_S$ couples to fermions. As a consequence, we choose the Higgs basis rotation to be that of eq.~\eqref{eq.Rbeta}. The DM candidate resides in the $h_1$ doublet and is stabilised by the $\mathbb{Z}_2: h_1 \to - h_1$ symmetry. The C-III-a case was covered in Ref.~\cite{Kuncinas:2022whn}.

\subsection{The mass-squared matrices}\label{Sec:CIIIa_M2_deriv}

\vspace{10pt}
\textbf{Charged sector}
\vspace{5pt}

We start by considering the charged mass-squared matrix in the $\{h_1^+,\,h_2^+,\,h_S^+ \}$ basis:
\begin{equation}
\mathcal{M}^2_\mathrm{Ch}
=\begin{pmatrix}
\left(\mathcal{M}_\mathrm{Ch}^2\right)_{11} & 0 & 0 \\
0 & \left(\mathcal{M}_\mathrm{Ch}^2\right)_{22} & \left(\mathcal{M}_\mathrm{Ch}^2\right)_{23} \\
0 & \left(\mathcal{M}_\mathrm{Ch}^2\right)_{23} & \left(\mathcal{M}_\mathrm{Ch}^2\right)_{33}
\end{pmatrix},
\end{equation}
where
\begin{subequations}
\begin{align}
\left(\mathcal{M}_\mathrm{Ch}^2\right)_{11} &= -2\lambda_3\hat v_2^2 -\frac{1}{2}\left[\lambda_6 - 10\lambda_7 - 8 \lambda_7 \cos(2\sigma)\right]\hat v_S^2, \\
\left(\mathcal{M}_\mathrm{Ch}^2\right)_{22} &= -\frac{1}{2}(\lambda_6-2\lambda_7)\hat v_S^2 , \\
\left(\mathcal{M}_\mathrm{Ch}^2\right)_{23} &= \frac{1}{2}(\lambda_6-2\lambda_7)\hat v_2 \hat v_S, \\
\left(\mathcal{M}_\mathrm{Ch}^2\right)_{33} &= -\frac{1}{2}(\lambda_6-2\lambda_7)\hat v_2^2.
\end{align}
\end{subequations}
The charged mass-squared matrix is diagonalisable by going into the Higgs basis, $\mathcal{R}_\beta$. In terms of the mass eigenstates, the scalar fields are:
\begin{subequations}
\begin{align}
h^+ & = h_1^+,\\
G^+ & = \sin\beta\,h_2^+ + \cos\beta\,h_S^+,\\ 
H^+ & = -\cos\beta\,h_2^+ + \sin\beta\,h_S^+,
\end{align}
\end{subequations}
with masses:
\begin{subequations} \label{Eq:C-III-c-masses-ch}
\begin{align}
m^2_{h^+} &= -2\lambda_3\hat v_2^2 -\frac{1}{2}\left[\lambda_6 - 10\lambda_7 - 8 \lambda_7 \cos(2\sigma)\right]\hat v_S^2, \\
m^2_{H^+} &= -\frac{1}{2}(\lambda_6-2\lambda_7)v^2.
\end{align}
\end{subequations}

\vspace{10pt}
\textbf{Neutral inert sector}
\vspace{5pt}

In the $\{\eta_1,\,\chi_1\}$ basis, the neutral mass-squared matrix is:
\begin{equation}
\mathcal{M}^2_\mathrm{N1}
=\begin{pmatrix}
\left(\mathcal{M}_\mathrm{N1}^2\right)_{11} & \left(\mathcal{M}_\mathrm{N1}^2\right)_{12} \\
\left(\mathcal{M}_\mathrm{N1}^2\right)_{12} & \left(\mathcal{M}_\mathrm{N1}^2\right)_{22}
\end{pmatrix},
\end{equation}
where
\begin{subequations}
\begin{align}
\left(\mathcal{M}_\mathrm{N1}^2\right)_{11} &= -2 \left( \lambda_2 + \lambda_3 \right)\sin^2\sigma \hat v_2^2 + 2 \lambda_7 \left[ 5 + 4 \cos(2\sigma) \right] \hat v_S^2, \\
\left(\mathcal{M}_\mathrm{N1}^2\right)_{12} &= \left[ \left( \lambda_2 + \lambda_3 \right)\hat v_2^2 + 2\lambda_7 \hat v_S^2 \right] \sin(2\sigma), \\
\left(\mathcal{M}_\mathrm{N1}^2\right)_{22} &= -2\left[\left( \lambda_2 + \lambda_3 \right)\hat v_2^2 - 4\lambda_7 \hat v_S^2 \right] \cos^2\sigma.
\end{align}
\end{subequations}
It is diagonalisable by
\begin{equation}
\mathcal{R}_\gamma = \begin{pmatrix}
\cos \gamma & \sin \gamma \\
-\sin \gamma & \cos \gamma
\end{pmatrix},
\end{equation}
where
\begin{equation} \label{Eq:tan_2gamma}
\tan (2 \gamma) = \frac{\left[ \left( \lambda_2 + \lambda_3 \right)\hat v_2^2 + 2 \lambda_7 \hat v_S^2 \right]\sin(2\sigma)}{ \left( \lambda_2 + \lambda_3 \right)\cos(2\sigma) \hat v_2^2 + \lambda_7 \left[ 3 + 2 \cos(2\sigma) \right] \hat v_S^2}.
\end{equation}

The physical neutral states associated with $h_1$  are as follows:
\begin{subequations}
\begin{align}
\varphi_1 & = \cos\gamma\,\eta_1 + \sin\gamma\,\chi_1,\\ 
\varphi_2 & = -\sin\gamma\,\eta_1+ \cos\gamma\,\chi_1,
\end{align}
\end{subequations}
with the mass-squared parameters
\begin{equation}
m_{\varphi_i}^2  = -\left( \lambda_2 + \lambda_3 \right)\hat v_2^2  
+ \lambda_7 \left[ 7 + 6 \cos(2\sigma) \right]\hat v_S^2 \mp \Delta,
\end{equation}
where
\begin{equation} \label{Eq:C-III-a-inert-splitting}
\Delta^2 = \left[ \left( \lambda_2 + \lambda_3\right) \hat v_2^2 
+ \lambda_7 \left( 2 + 3 \cos(2\sigma) \right) \hat v_S^2\right]^2 + 9 \lambda_7 ^2 \sin^2 (2\sigma) \hat v_S^4. 
\end{equation}
Equivalently, these can be presented as
\begin{subequations} \label{Eq:inert-masses}
\begin{align}
m^2_{\varphi_1}=&-2(\lambda_2+\lambda_3)\hat v_2^2\sin^2(\gamma-\sigma) \nonumber \\
&+\lambda_7\hat v_S^2[7+6\cos(2\sigma)+3\cos(2\gamma)+2\cos(2\gamma-2\sigma)], \\
m^2_{\varphi_2}=&-2(\lambda_2+\lambda_3)\hat v_2^2\cos^2(\gamma-\sigma) \nonumber \\
&+\lambda_7\hat v_S^2[7+6\cos(2\sigma)-3\cos(2\gamma)-2\cos(2\gamma-2\sigma)].
\label{Eq:m_varphi2}
\end{align}
\end{subequations}

By considering $\Delta^2=0$, we can check the mass-degenerate limit, in which
\begin{subequations}
\begin{equation}
 \left( \lambda_2 + \lambda_3\right) \tan^2 \beta + \lambda_7 \left[ 2 + 3 \cos(2\sigma) \right] \to 0,
\end{equation}
if simultaneously
\begin{equation}
\lambda_7 \sin(2\sigma)\to0.
\end{equation}
\end{subequations}
Solving the above yields that the mass-degenerate limit is reached for $\lambda_2+\lambda_3=\lambda_7=0$, corresponding to massless states. 

The mass-squared parameters of the $\{h^\pm,\,H^\pm,\,\varphi_1,\,\varphi_2\}$ states are defined in terms of the $\{\lambda_2,\,\lambda_3,\,\lambda_6,\,\lambda_7\}$ couplings. Accounting for vevs, one can determine the rotation angle $\gamma$ of eq.~(\ref{Eq:tan_2gamma}). This procedure results in a quadratic equation, yielding two sets of couplings for the same set of masses and $ \sigma $. To gain more control over the input, one can sacrifice one of the mass-squared parameters, replacing it by the rotation angle $\gamma$. This allows for the input of the masses while resulting in linear equations for the $\lambda$'s, thereby providing unambiguous couplings. Solving eqs.~\eqref{Eq:C-III-c-masses-ch} and \eqref{Eq:inert-masses} in terms of the quartic couplings yields:
\begin{subequations} \label{Eq:C-III-a-lambdas-a}
\begin{align}
\begin{split}
\lambda_2 &= \frac{m_{h^+}^2}{2 v_2^2} - \frac{m_{H^+}^2 v_S^2}{2v^2 v_2^2}\\ &~~~ - \frac{ \left( m_{\varphi_1}^2 + m_{\varphi_2}^2\right)\left[ 2 + \cos (2\sigma) \right] + \left( m_{\varphi_2}^2 - m_{\varphi_1}^2 \right)\left[ 2 \cos (2\gamma) + \cos (2\gamma - 2\sigma) \right]}{12 \cos^2 \sigma v_2^2},
\end{split}\\
\begin{split}
\lambda_3 &= -\frac{m_{h^+}^2}{2 v_2^2} +  \frac{m_{H^+}^2 v_S^2}{2v^2 v_2^2} + \frac{\left( m_{\varphi_1}^2 + m_{\varphi_2}^2 \right)\sin \sigma - \left( m_{\varphi_2}^2 - m_{\varphi_1}^2 \right)\sin(2\gamma-\sigma)}{6 v_2^2 \sin \sigma},
\end{split}\\
\lambda_6 & = -\frac{2m_{H^+}^2 }{v^2} + \frac{\left( m_{\varphi_1}^2 + m_{\varphi_2}^2 \right)\sin \sigma - \left( m_{\varphi_2}^2 - m_{\varphi_1}^2 \right)\sin(2\gamma-\sigma)}{12 v_S^2 \cos^2 \sigma \sin \sigma},\\
\lambda_7 & = \frac{\left( m_{\varphi_1}^2 + m_{\varphi_2}^2 \right)\sin \sigma - \left( m_{\varphi_2}^2 - m_{\varphi_1}^2 \right)\sin(2\gamma-\sigma)}{24 v_S^2 \cos^2\sigma \sin \sigma}. \label{Eq:C-III-a-lambda7}
\end{align}
\end{subequations}

Note that expressions like $\alpha(\gamma,\sigma) m^2_{\varphi_1}+\beta(\gamma,\sigma) m^2_{\varphi_2}$ can be expressed in terms of $A(\gamma,\sigma) m^2_{\varphi_1}+B(\gamma,\sigma) m^2_{\varphi_2}$, as long as $\alpha+\beta g = A+B g$, where $g$ is the ratio of the two coefficients:
\begin{subequations}\label{eq:m_phi2_vs_m_phi1}
\begin{equation}
f_+(\sigma,\gamma) m^2_{\varphi_1} = f_-(\sigma,\gamma) m^2_{\varphi_2},
\end{equation}
with
\begin{equation}
f_\pm(\sigma, \gamma) = [3+2\cos(2\sigma)]\sin(2\gamma-2\sigma)+\sin(2\gamma)\pm\sin(2\sigma).
\end{equation}
\end{subequations}
As a matter of fact, we can express the contributions to the $ \lambda $s that involve $ m^2_{\varphi_1} $ and $ m^2_{\varphi_2} $ in various ways.

In the left panel of Figure~\ref{Fig:m2_vs_m1} we highlight in color the regions where $ m^2_{\varphi_2} > m^2_{\varphi_1} $. The red edge corresponds to the limit where $ m^2_{\varphi_1} / m^2_{\varphi_2} \to 0 $. In the white and grey regions, the ratio is either negative (white) or below unity (grey). In fact, the grey region is identical to the colored region after a solid rotation by $\pi$, i.e., $ \{\sigma,\,\gamma\} \to \{\pi - \sigma, \, \pi/2 - \gamma\} $, which is equivalent to interchanging the two coefficients in eq.~(\ref{eq:m_phi2_vs_m_phi1}).

\begin{figure}[htb]
\begin{center}
\includegraphics[scale=0.35]{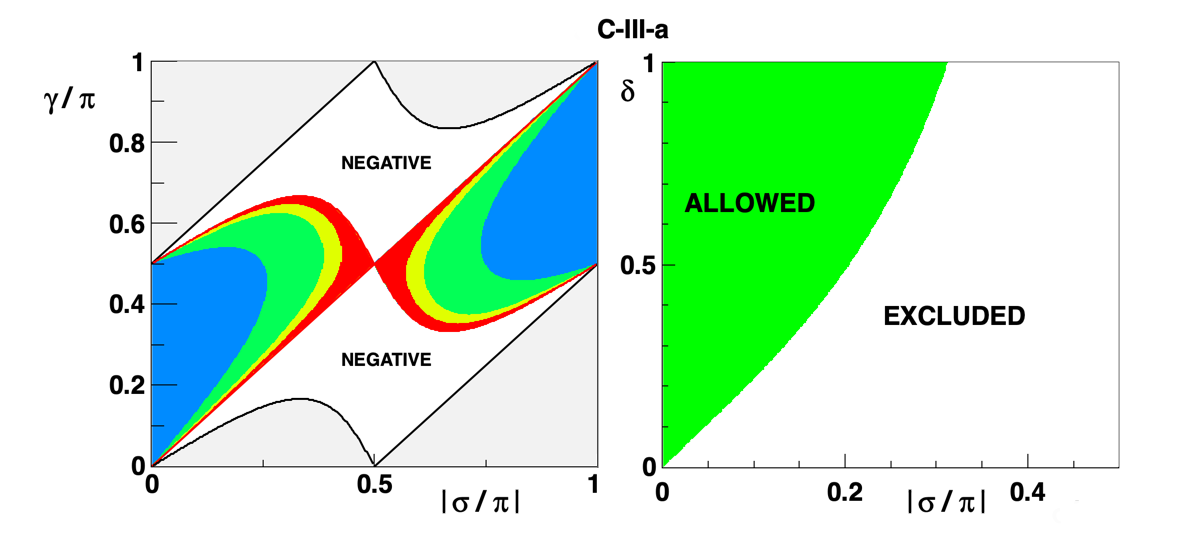}
\end{center}
\vspace*{-8mm}
\caption{Determining the parameter space of the C-III-a case based on the input of $m^2_{\varphi_i}$. Left panel: the ratio $ g(\gamma,\sigma) = m^2_{\varphi_2} / m^2_{\varphi_1} = f_+(\sigma,\gamma) / f_-(\sigma,\gamma) $ is shown for $ g(\gamma,\sigma) > 1 $. Contours are displayed at 2 (transition from blue to green), 5 (transition from green to yellow), and 10 (transition from yellow to red). According to eq.~(\ref{eq:m_phi2_vs_m_phi1}), the ratio depends on $ \sigma $ only through $ \sin(2\sigma) $ and $ \cos(2\sigma) $, meaning it is identical for $ \sigma $ and $ \sigma + \pi $, as illustrated. Right panel: the mass gap $\delta$ of eq.~\eqref{Eq:CIIIa_mass_gap_delta} as a function of the $\sigma$ phase. Figures taken from Ref.~\cite{Kuncinas:2022whn}.}
\label{Fig:m2_vs_m1}
\end{figure}

Let us consider the mass gap between $\varphi_i$ states. We can eliminate $\gamma$ from eqs.~\eqref{Eq:C-III-a-lambdas-a}. Then we should guarantee that the argument should be positive,
\begin{equation}
9(m_{\varphi_2}^2-m_{\varphi_1}^2)^2-4m_{\varphi_1}^2m_{\varphi_2}^2\tan^2\sigma>0.
\end{equation}
For finite values of $ \sigma $, this condition can be reformulated as a constraint on the mass gap:
\begin{equation}\label{Eq:CIIIa_mass_gap_delta}
\delta=\frac{m_{\varphi_2}^2-m_{\varphi_1}^2}{\sqrt{m_{\varphi_1}^2m_{\varphi_2}^2}}>\frac{2}{3}|\tan\sigma|.
\end{equation}
The $\delta$ parameter is plotted in the right panel of Figure~\ref{Fig:m2_vs_m1}.

For a fixed value of $ \sigma $, the absolute mass gap scales proportionally to the absolute mass scale. This presents a challenge for the high-mass DM region, where a small mass splitting is necessary.

\newpage
\vspace{10pt}
\textbf{Neutral active sector}
\vspace{5pt}

The neutral mass-squared matrix of $h_2$ and $h_S$  in the basis of $\{\eta_2, \,\eta_S,\, \chi_2, \,\chi_S\}$ is: 
\begin{equation}
\mathcal{M}^2_\mathrm{N2S}
=\begin{pmatrix}
\left(\mathcal{M}^2_\mathrm{N2S}\right)_{11} & \left(\mathcal{M}^2_\mathrm{N2S}\right)_{12} & \left(\mathcal{M}^2_\mathrm{N2S}\right)_{13} & \left(\mathcal{M}^2_\mathrm{N2S}\right)_{14}\vspace{2pt} \\ 
\left(\mathcal{M}^2_\mathrm{N2S}\right)_{12} & \left(\mathcal{M}^2_\mathrm{N2S}\right)_{22} & \left(\mathcal{M}^2_\mathrm{N2S}\right)_{14} & \left(\mathcal{M}^2_\mathrm{N2S}\right)_{24}\vspace{2pt} \\ 
\left(\mathcal{M}^2_\mathrm{N2S}\right)_{13} & \left(\mathcal{M}^2_\mathrm{N2S}\right)_{14} & \left(\mathcal{M}^2_\mathrm{N2S}\right)_{33} & \left(\mathcal{M}^2_\mathrm{N2S}\right)_{34}\vspace{2pt} \\ 
\left(\mathcal{M}^2_\mathrm{N2S}\right)_{14} & \left(\mathcal{M}^2_\mathrm{N2S}\right)_{24} & \left(\mathcal{M}^2_\mathrm{N2S}\right)_{34} & \left(\mathcal{M}^2_\mathrm{N2S}\right)_{44}\vspace{2pt} \\ 
\end{pmatrix},
\end{equation}
where
\begin{subequations}
\begin{align}
\left(\mathcal{M}^2_\mathrm{N2S}\right)_{11} &= 2 \left( \lambda_1 + \lambda_3 \right)\hat v_2^2 - 6 \lambda_7 \cos^2 \sigma \hat v_S^2,\\
\left(\mathcal{M}^2_\mathrm{N2S}\right)_{12} &= \left( \lambda_5 + \lambda_6 - 2 \lambda_7 - 2 \lambda_7 \cos^2 \sigma \right) \hat v_2 \hat v_S,\\
\left(\mathcal{M}^2_\mathrm{N2S}\right)_{13} &= \lambda_7 \sin (2\sigma) \hat v_S^2,\\
\left(\mathcal{M}^2_\mathrm{N2S}\right)_{14} &= -\lambda_7 \sin (2\sigma) \hat v_2 \hat v_S,\\
\left(\mathcal{M}^2_\mathrm{N2S}\right)_{22} &= 2 \left( \lambda_7 \cos^2 \sigma \hat v_2^2 + \lambda_8 \hat v_S^2 \right),\\
\left(\mathcal{M}^2_\mathrm{N2S}\right)_{24} &= \lambda_7 \sin (2\sigma) \hat v_2^2,\\
\left(\mathcal{M}^2_\mathrm{N2S}\right)_{33} &= 2\lambda_7 \sin^2 \sigma \hat v_S^2,\\
\left(\mathcal{M}^2_\mathrm{N2S}\right)_{34} &= -2 \lambda_7 \sin^2 \sigma \hat v_2 \hat v_S,\\
\left(\mathcal{M}^2_\mathrm{N2S}\right)_{44} &= 2 \lambda_7 \sin^2 \sigma \hat v_2^2.
\end{align}
\end{subequations}
The physical states will be expressed in terms of combinations of all fields $\{\eta_2, \,\eta_S,\, \chi_2, \,\chi_S\}$. To identify the physical states, we begin by transforming the scalars into the Higgs basis:
\begin{equation}\label{Eq:C-III-a_NeutralActiveHBRot}
\begin{pmatrix}
\phi_1 \\
\phi_2 \\
G^0 \\
\phi_3
\end{pmatrix} = \mathcal{I}_2 \otimes \begin{pmatrix}
\sin \beta & \cos \beta \\
-\cos \beta & \sin \beta
\end{pmatrix} \begin{pmatrix}
\eta_2 \\
\eta_S \\
\chi_2 \\
\chi_S
\end{pmatrix}.
\end{equation}
In the new basis the Goldstone boson, $G^0$, can be identified. The mass-squared matrix of the remaining three $\phi_i$ fields is:
\begin{equation} \label{Eq:C-III-a_M_sq_3by3}
\mathcal{M}^{2}_\mathrm{\phi}
=\begin{pmatrix}
\left( \mathcal{M}^{2}_\mathrm{\phi} \right)_{11} & \left( \mathcal{M}^{2}_\mathrm{\phi} \right)_{12} & 0 \\
\left( \mathcal{M}^{2}_\mathrm{\phi} \right)_{12} & \left( \mathcal{M}^{2}_\mathrm{\phi} \right)_{22} & \left( \mathcal{M}^{2}_\mathrm{\phi} \right)_{23} \\
0 & \left( \mathcal{M}^{2}_\mathrm{\phi} \right)_{23} & \left( \mathcal{M}^{2}_\mathrm{\phi} \right)_{33}
\end{pmatrix},
\end{equation}
where 
\begin{subequations}
\begin{align}
(\mathcal{M}^2_\mathrm{\phi})_{11} &= \frac{2}{v^2} \left[ \left( \lambda_1 + \lambda_3 \right)v_2^4 + \left( \lambda_5 + \lambda_6 - 2 \lambda_7 - 4 \lambda_7 \cos^2 \sigma \right)v_2^2v_S^2 + \lambda_8 v_S^4 \right] , \label{Eq:RII1a_Mphi11}\\
\begin{split}(\mathcal{M}^2_\mathrm{\phi})_{12} &= -\frac{1}{v^2} \Big[ \left(2\lambda_1 + 2\lambda_3 - \lambda_5 - \lambda_6 + 2 \lambda_7 \right)v_2^3 v_S\\
&\hspace{40pt} + \left( \lambda_5 + \lambda_6 - 2 \lambda_7  - 8 \lambda_7 \cos^2 \sigma - 2 \lambda_8 \right)v_2 v_S^3 \Big],\label{Eq:RII1a_Mphi12}\end{split}\\
\begin{split}(\mathcal{M}^2_\mathrm{\phi})_{22} &= \frac{2}{v^2} \Big[ \left( \lambda_1 + \lambda_3 - \lambda_5 - \lambda_6 + 2 \lambda_7 + 2 \lambda_7 \cos^2\sigma+ \lambda_8 \right)v_2^2 v_S^2\\
&\hspace{40pt}  + \lambda_7 \cos^2 \sigma \left( v_2^4-3v_S^4 \right) \Big],\end{split}\\
(\mathcal{M}^2_\mathrm{\phi})_{23} &= v^2 \lambda_7 \sin(2\sigma),\\
(\mathcal{M}^2_\mathrm{\phi})_{33} &= 2 v^2 \lambda_7 \sin^2 \sigma.
\end{align}
\end{subequations}
Then, we diagonalise $\mathcal{M}_\phi^2$ by
\begin{equation}\label{Eq:C-III-a_NeutralActiveHBRotR0Diag}
\begin{pmatrix}
H_1 \\
H_2 \\
H_3
\end{pmatrix} = \mathcal{R}^0 \begin{pmatrix}
\phi_1 \\
\phi_2 \\
\phi_3
\end{pmatrix},
\end{equation}
where the rotation transformation is
\begin{equation}\label{Eq:RII1a_RotR0}
\mathcal{R}^0 = \begin{pmatrix}
1 & 0 & 0 \\
0 & \cos {\theta_3} & \sin {\theta_3} \\
0 & -\sin {\theta_3} & \cos {\theta_3}
\end{pmatrix} \begin{pmatrix}
\cos {\theta_2} & 0 & \sin {\theta_2} \\
0 & 1 & 0 \\
-\sin {\theta_2} & 0 & \cos {\theta_2}
\end{pmatrix} \begin{pmatrix}
\cos {\theta_1} & \sin {\theta_1} & 0 \\
-\sin {\theta_1} & \cos {\theta_1} & 0 \\
0 & 0 & 1\\
\end{pmatrix}.
\end{equation}
We assume that the neutral scalars $H_i$ are ordered as $m_{H_i} \leq m_{H_{i+1}}$. 

When $\lambda$'s are used as input, one could proceed to diagonalise $\mathcal{M}_\phi^2$. However, to gain more control over the physical aspects of the implementation, it is preferable to start with one or two masses as input, along with several mixing angles, and then determine $\lambda$'s afterwards. Equations \eqref{Eq:C-III-a_NeutralActiveHBRot} and \eqref{Eq:C-III-a_NeutralActiveHBRotR0Diag} connect the remaining $\lambda$'s with te mass-squared parameters:
\begin{subequations}\label{Eq:C-III-a_CouplingsInverted}
\begin{align}
\begin{split}
\lambda_1 & = \frac{m_{h^+}^2}{2 v_2^2} - \frac{m_{H^+}^2 v_S^2}{2v^2 v_2^2} - \frac{\left( m_{\varphi_1}^2 + m_{\varphi_2}^2 \right)\sin \sigma - \left( m_{\varphi_2}^2 - m_{\varphi_1}^2 \right)\sin(2\gamma - \sigma)}{24 v_2^2 \sin \sigma}\\
& ~~~ + \frac{1}{2 v^2 v_2^2}\sum_{i=1}^3  \left( \mathcal{R}^0_{i1} v_2 - \mathcal{R}^0_{i2} v_S \right)^2 m_{H_i}^2,
\end{split}\raisetag{70pt}\\
\begin{split}
\lambda_5 & = \frac{2m_{H^+}^2 }{v^2} + \frac{\left( m_{\varphi_1}^2 + m_{\varphi_2}^2 \right)\sin \sigma - \left( m_{\varphi_2}^2 - m_{\varphi_1}^2 \right)\sin(2\gamma - \sigma)}{12 v_S^2 \sin \sigma}\\
& ~~~  + \frac{1}{v^2 v_2 v_S} \sum_{i=1}^3  \left( \mathcal{R}^0_{i2} v_2 + \mathcal{R}^0_{i1} v_S \right)\left( \mathcal{R}^0_{i1} v_2 - \mathcal{R}^0_{i2} v_S \right) m_{H_i}^2,
\end{split}\\
\begin{split}
\lambda_8 & = -\frac{\left[\left( m_{\varphi_1}^2 + m_{\varphi_2}^2 \right)\sin \sigma - \left( m_{\varphi_2}^2 - m_{\varphi_1}^2 \right)\sin(2\gamma - \sigma)\right] v_2^2}{24 v_S^4 \sin \sigma}\\
& ~~~ +\frac{1}{2 v^2\hat v_S^2}\sum_i  (\mathcal{R}^0_{i2}\hat v_2+\mathcal{R}^0_{i1}\hat v_S)^2 m_{H_i}^2 .
\end{split}
\end{align}
\end{subequations}
In addition, the $\mathcal{R}^0$ diagonalisation matrix should satisfy the following criteria:
\begin{subequations}\label{Eq:C-III-a_ActiveSectorAnglesConstraints}
\begin{align}
\sum_i \mathcal{R}^0_{i1}\mathcal{R}^0_{i3} m_{H_i}^2&=0, \label{Eq:non-inert-constraint1}\\
\sum_i \mathcal{R}^0_{i2}\mathcal{R}^0_{i3} m_{H_i}^2&=v^2\sin(2\sigma)\lambda_7, \label{Eq:non-inert-constraint2}\\\
\sum_i \left(\mathcal{R}^0_{i3}\right)^2m_{H_i}^2&=2v^2\sin^2\sigma \lambda_7. \label{Eq:non-inert-constraint3}
\end{align}
\end{subequations}
The latter two conditions establish a relationship between the mass scales of the $m_{H_i}^2$ masses and the $m_{\varphi_i}$ masses through the $\lambda_7$ coupling. This way, the $m_{H_i}^2$ parameters can be expressed in terms of:
\begin{subequations}
\begin{align}
m_{H_1}^2 & = \Phi \left( \sin \theta_1 \cot \theta_2 + \tan \sigma\right), \label{Eq:CIIIa_mH12_theta}\\
m_{H_2}^2 & = \Phi \left( - \sin \theta_1 \tan \theta_2 + \frac{\cos \theta_1 \cot \theta_3}{\cos \theta_2}   + \tan \sigma\right),\\
m_{H_3}^2 & = \Phi \left( - \sin \theta_1 \tan \theta_2 - \frac{\cos \theta_1 \tan \theta_3}{\cos \theta_2} + \tan \sigma\right),
\end{align}
\end{subequations}
where 
\begin{equation}
\Phi = \frac{m_{\varphi_1}^2 v^2 \sin \left( 2 \gamma - 2 \sigma \right) \sin \sigma }{ 3 v_S^2 \left[ \sin \left( 2 \gamma - 3 \sigma \right) + 2 \sin\left( 2 \gamma - \sigma \right) - \sin \sigma \right]}.
\end{equation}
The parameter $\Phi$ can take both negative and positive values. Some of the conditions of eq.~\eqref{Eq:C-III-a_ActiveSectorAnglesConstraints} can be re-written:
\begin{subequations}
\begin{align}
\eqref{Eq:non-inert-constraint2}:&\quad \Phi = v^2\sin(2\sigma)\lambda_7,\\
\eqref{Eq:non-inert-constraint3}:&\quad \Phi \tan \sigma = 2v^2\sin^2\sigma \lambda_7.
\end{align}
\end{subequations}

In a parameter scan, it is preferable to keep $ m_{H_1} $ fixed at its experimental value, 125~GeV. The constraints in eq.~\eqref{Eq:C-III-a_ActiveSectorAnglesConstraints} allow for different choices of input parameters: one mass and two angles, two masses and one angle, or three masses. From eq.~\eqref{Eq:CIIIa_mH12_theta}, which expresses $ m_{H_1}^2 $ as $ f(\theta_1, \theta_2) $, it follows that not all combinations of masses and angles can be chosen arbitrarily as input. In Ref.~\cite{Kuncinas:2022whn} we used $ \theta_2 $ and $ \theta_3 $ as input parameters, along with $ m_{H_1} $, associated with the SM-like Higgs boson.

\newpage
\vspace{10pt}
\textbf{Physical states}
\vspace{5pt}

The $SU(2)$ scalar doublets in terms of the mass eigenstates are given by:
\begin{subequations}\label{Eq:C-III-a_Expanded_Mass_Eigenstates}
\begin{align}
h_1 &= e^{i \gamma}\begin{pmatrix}
h^+ \\
\left(\varphi_1 + i \varphi_2\right)/\sqrt{2}
\end{pmatrix},\\
\begin{split}
h_2 & = e^{i \sigma}\begin{pmatrix}
\sin \beta\,G^+ - \cos \beta\,H^+\\
\left( \sin \beta\, v  + i \sin \beta\, G^0 + \sum_{i=1}^3 \left[ \sin \beta\,\mathcal{R}^0_{i1} - \cos \beta\, \left( \mathcal{R}^0_{i2} + i \mathcal{R}^0_{i3} \right)\right] H_i  \right) /\sqrt{2}
\end{pmatrix}\\
& = e^{i \sigma}\begin{pmatrix}
\sin \beta\,G^+ - \cos \beta\,H^+\\
\left( \sin \beta\, v  + i \sin \beta\, G^0 + \sum_{i=1}^3 A_{2i} H_i \right)/\sqrt{2}
\end{pmatrix},
\end{split}\\
\begin{split}
h_S & = \begin{pmatrix}
\cos \beta\, G^+ + \sin \beta\, H^+\\
 \left( \cos \beta\, v + i \cos \beta\,G^0 + \sum_{i=1}^3 \left[ \cos \beta\,\mathcal{R}^0_{i1} + \sin \beta\,\left( \mathcal{R}^0_{i2} + i \mathcal{R}^0_{i3} \right) \right] H_i \right)/\sqrt{2}
\end{pmatrix}\\
& = \begin{pmatrix}
\cos \beta\, G^+ + \sin \beta\, H^+\\
\left( \cos \beta\, v + i \cos \beta\,G^0 + \sum_{i=1}^3 A_{Si} H_i \right)/\sqrt{2}
\end{pmatrix},
\end{split}\label{Eq:h_S-expansion}
\end{align}
\end{subequations} 
where $A_{ij}$ is a complex quantity, defined in terms of the Higgs basis transformation $\beta$ angle and the $\mathcal{R}^0$ diagonalisation matrix. Since $v_1 = 0$, we can extract the $\gamma$ phase from the $h_1$ doublet. Such parameterisation lets the $\varphi_i$ states to be interpreted as mass eigenstates; it does not change the definition of the $\gamma$ angle of eq.~\eqref{Eq:tan_2gamma}.

\subsection{Interactions}

\vspace{10pt}
\textbf{Gauge couplings}
\vspace{5pt}

The gauge-scalar interactions are given by the following kinetic Lagrangian:
\begin{subequations}
\begin{align}
\begin{split}
\mathcal{L}_{VVH} =& \left[ \frac{g}{2 \cos \theta_W }m_ZZ_\mu Z^\mu + g m_W W_\mu^+ W^{\mu-} \right] \sum_{i=1}^3 \mathcal{R}_{i1}^0 H_i ,\label{Eq:CIIIa_VVH}
\end{split}\\
\begin{split}
\mathcal{L}_{VHH} =&-\frac{ g}{2 \cos \theta_W }Z^\mu \left( \sum_{i<j=2}^3 \left( \mathcal{R}_{i2}^0\mathcal{R}_{j3}^0-\mathcal{R}_{i3}^0\mathcal{R}_{j2}^0 \right) H_i  \overset\leftrightarrow{\partial_\mu} H_j 
+ \varphi_1  \overset\leftrightarrow{\partial_\mu} \varphi_2  \right)\\
& - \frac{g}{2}\bigg\{ i W_\mu^+ \left(\sum_{i=1}^3  \left( \mathcal{R}_{i2}^0 + i \mathcal{R}_{i3}^0 \right) H^- \overset\leftrightarrow{\partial^\mu} H_i 
+ h^- \overset\leftrightarrow{\partial^\mu} 
(\varphi_1 + i \varphi_2) \right) + \mathrm{h.c.} \bigg\}\\
& + \left[ i e A^\mu + \frac{i g}{2} \frac{\cos (2\theta_W) }{\cos \theta_W } Z^\mu \right] \left( H^+ \overset\leftrightarrow{\partial_\mu} H^- + h^+ \overset\leftrightarrow{\partial_\mu} h^- \right),\raisetag{101pt}
\end{split}\\
\begin{split}
\mathcal{L}_{VVHH} =& \left[ \frac{g^2}{8 \cos^2 \theta_W }Z_\mu Z^\mu + \frac{g^2}{4} W_\mu^+ W^{\mu-} \right] \left( H_1^2 + H_2^2 + H_3^2 + \varphi_1^2 + \varphi_2^2\right)\\
& + \bigg\{ \left[ \frac{e g}{2} A^\mu W_\mu^+ - \frac{g^2}{2} \frac{\sin^2 {\theta_W}}{\cos \theta_W }Z^\mu W_\mu^+ \right] \bigg( \sum_{i=1}^3  \left( \mathcal{R}_{i2}^0 + i \mathcal{R}_{i3}^0 \right) H_i H^- \\ 
& \hspace{175pt} + (\varphi_1 + i \varphi_2) h^-\bigg) + \mathrm{h.c.} \bigg\}\\
&+ \left[ e^2 A_\mu A^\mu + e g \frac{\cos (2\theta_W) }{\cos \theta_W }A_\mu Z^\mu + \frac{g^2}{4} \frac{\cos^2 (2\theta_W)}{\cos \theta_W ^2}Z_\mu Z^\mu + \frac{g^2}{2} W_\mu^- W^{\mu +} \right]\\
&\hspace{15pt} \times\left(H^-H^+ + h^-h^+ \right).\raisetag{110pt}\label{Eq:CIIIa_SSVV}
\end{split}
\end{align}
\end{subequations}

By observing eq.~\eqref{Eq:CIIIa_VVH}, we can identify the SM-like Higgs boson to be the one which couples to the gauge bosons with $\mathcal{R}_{i1}^0  \approx 1$. We recall that the rotation matrix  $\mathcal{R}^0$, see eq.~\eqref{Eq:RII1a_RotR0}, is orthogonal, and hence $(\mathcal{R}^0_{i1})^2+(\mathcal{R}^0_{i2})^2+(\mathcal{R}^0_{i3})^2=1$. This condition implies that all other entries of the column $1$ and row $i$ have to vanish. In eq.~\eqref{Eq:C-III-a_NeutralActiveHBRot} we may identify $\phi_1$ as the SM-like Higgs boson (since it would transform together with the Goldstone boson $G^0$), provided it is already a mass eigenstate. To guarantee this, $\left( \mathcal{M}_{\phi}^2 \right)_{12}$ of eq.~\eqref{Eq:RII1a_Mphi12} would have to be zero.

\vspace{10pt}
\textbf{Yukawa Lagrangian}
\vspace{5pt}

There are several possibilities to construct the Yukawa Lagrangian based on the $S_3$ representations of the fermions. However, only the trivial representation results in a realistic prediction.  NFC is ensured since the $S_3$ symmetry restricts fermion couplings to a single scalar doublet. The scalar-fermion couplings, accounting for $``-i"$, are:
\begin{subequations}\label{Eq:C-III-a-Yukawa}
\begin{align}
g \left( H_i \bar{u} u \right) &= \frac{m_u}{v} \left[ -i\left( \mathcal{R}_{i1}^0 + \mathcal{R}_{i2}^0 \tan \beta \right) - \gamma_5 \mathcal{R}_{i3}^0 \tan \beta \right],\\ 
g \left( H_i \bar{d} d \right) &= \frac{m_d}{v} \left[ -i\left( \mathcal{R}_{i1}^0 + \mathcal{R}_{i2}^0 \tan \beta \right) + \gamma_5 \mathcal{R}_{i3}^0 \tan \beta \right].
\end{align}
\end{subequations}
The leptonic Dirac mass terms result in similar relations. Notice that in the above equations $H_i$ couples to fermions as a CP-indefinite state, \textit{i.e.}, the first term transforms as a CP-even quantity while the second one, due to the presence of $\gamma_5$, as a CP-odd quantity. As a result, the scalar-fermion decay rate is given by
\begin{equation}\label{Eq:CIIIa_Yukawa_Hiff}
\begin{split}
\Gamma \left( H_i \to  \bar{f} f \right) = \frac{N_c m_{H_i} m_f^2}{8 \pi v^2} &\bigg[ \left( 1- 4 \frac{m_f^2}{m_{H_i}^2} \right)^{3/2} | \mathcal{R}_{i1}^0 + \mathcal{R}_{i2}^0 \tan \beta |^2\\ &~\,~+ \left( 1- 4 \frac{m_f^2}{m_{H_i}^2} \right)^{1/2} | \mathcal{R}_{i3}^0 \tan \beta |^2 \bigg],
\end{split}
\end{equation}
Given that $H_1$ is assumed to be the SM-like Higgs boson, we can express the decay rate as a ratio normalised to the SM value:
\begin{equation}\label{Eq:CIIIa_Gamma_SFF_Ratio}
\kappa_{ff}^2 \approx | \mathcal{R}_{11}^0 + \mathcal{R}_{12}^0 \tan \beta |^2 + \left( 1- 4 \frac{m_f^2}{m_{h_\mathrm{SM}}^2} \right)^{-1} | \mathcal{R}_{13}^0 \tan \beta |^2.
\end{equation}
This quantity will serve as a measure of the SM-like limit for the fermion couplings.

The charged scalar-fermion couplings are identical to those of the R-II-1a implementation, see eq.~\eqref{Eq:R-II-1a-Yukawa_Charged}.

\vspace{10pt}
\textbf{Scalar interactions}
\vspace{5pt}

The scalar couplings are presented with the symmetry factor, but without the overall coefficient ``$-i$". To simplify notation we introduce a permutation function
\begin{equation}
P_{\overline{m} n o}(i,j,k) = \sum\limits_{\substack{t_\alpha \in \{i,j,k\}}} A^\ast_{mt_1} A_{nt_2} A_{ot_3}, 
\end{equation}
where the $A$'s are coefficients of the field expansions, see eq.~\eqref{Eq:C-III-a_Expanded_Mass_Eigenstates} and the $\{i,j,k\}$ indices are carried by the $H_i H_j H_k$ fields. The barred index $\overline{m}$ indicates a conjugated $A$ parameter. For example, the $P_{\overline{2} 2 S}(i,j,k) $ function is given by
\begin{equation}
\begin{aligned}
P_{\overline{2} 2 S}(i,j,k) ={}& A^\ast_{2i} (A_{2j} A_{Sk}+A_{2k}A_{Sj}) +A^\ast_{2j}(A_{2i}A_{Sk}+A_{2k}A_{Si})\\
&+A^\ast_{2k}(A_{2i}A_{Sj}+A_{2j}A_{Si}).
\end{aligned}
\end{equation}
The order of the $\{m,\, n,\, p\}$ indices is arbitrary, 
\begin{equation}
P_{\overline{m} n o}(i,j,k) = P_{n \overline{m} o}(i,j,k) = P_{n o \overline{m}}(i,j,k), \text{ and interchange of }n\leftrightarrow o.
\end{equation}
Apart from that,
\begin{equation}
\left(P_{\overline{m} n o}(i,j,k)\right)^\ast = P_{m \overline{n} \overline{o}}(i,j,k).
\end{equation}

Since the number of the involved $H_i$ scalars in a vertex varies from two to four, there are also permutation functions $P_{\overline{m} n}(i,j)$ and $P_{\overline{m} \overline{n} o p}(i,j,k,\ell)$. For example,
\begin{equation}
P_{2S}(i,j)=A_{2i}A_{Sj}+A_{Si}A_{2j}.
\end{equation}

Trilinear couplings involving the neutral fields are:
\begin{subequations}
\begin{align}
\begin{split}
g\left( H_i H_j H_k \right) &= v \Big[\frac{1}{2}\left( \lambda_1 + \lambda_3 \right) P_{\overline{2}22}(i,j,k)\sin \beta\\
&\hspace{30pt} - \frac{1}{4}\lambda_4 e^{i \sigma} (  P_{\overline{2} 22}(i,j,k) \cos \beta
+ \sin \beta\left[ P_{\overline{S} 22}(i,j,k) + 2 P_{\overline{S2}2}(i,j,k) \right] ) \\
&\hspace{30pt} + \frac{1}{4} \left( \lambda_5 + \lambda_6 \right) \left[  P_{\overline{2} 2S}(i,j,k)  \cos \beta
+  P_{2S\overline{S}}(i,j,k) \sin \beta \right]\\
&\hspace{30pt} + \frac{1}{2}\lambda_7 e^{2 i \sigma} \left[  P_{\overline{S} 22}(i,j,k)  \cos \beta
+  P_{2\overline{SS}}(i,j,k)\sin \beta\right]\\
&\hspace{30pt} + \frac{1}{2}\lambda_8  P_{S\overline{SS}}(i,j,k)\cos \beta + \mathrm{h.c.} \Big]\label{Eq.C_III_a_HiHjHk},
\end{split}\\
\begin{split}
g\left( \varphi_1 \varphi_1 H_i \right) &= v \Big[ \lambda_1  A_{2i} \sin \beta   - \lambda_2  \left( 1-e^{-2i \left( \gamma - \sigma \right)} \right) A_{2i} \sin \beta   
+ \lambda_3  e^{2i\left(\sigma-\gamma\right)} A_{2i} \sin \beta  \\
&\hspace{30pt}+\frac{1}{2}\lambda_4 \left[ e^{i\sigma}\left( 2+e^{-2i\gamma}\right) A_{2i} \cos \beta   + e^{-i \sigma}\left( 2 + e^{-2i\left(\gamma-\sigma\right)}\right) A_{Si} \sin \beta    \right]\\
&\hspace{30pt}+ \frac{1}{2}\left( \lambda_5 + \lambda_6 \right)  A_{Si} \cos \beta   + \lambda_7 e^{-2i \gamma} A_{Si} \cos \beta   + \mathrm{h.c.}  \Big],
\end{split}\label{Eq.C_III_a_varphi1varphi1Hi}\\
\begin{split}
g\left( \varphi_1 \varphi_2 H_i \right) 
&= -v \Big[ i e^{-2i\gamma}(\left( \lambda_2 + \lambda_3 \right)e^{2i\sigma} A_{2i} \sin \beta   
+ \frac{1}{2}  \lambda_4 e^{i\sigma}\left(  A_{2i} \cos \beta   
+  A_{Si} \sin \beta   \right)\\
&\hspace{75pt} + \lambda_7  A_{Si} \cos \beta  ) + \mathrm{h.c.}\Big],
\end{split}\label{Eq.C_III_a_varphi1varphi2Hi}\\
\begin{split}
g\left( \varphi_2 \varphi_2 H_i \right) &= v \Big[ \lambda_1  A_{2i} \sin \beta   - \lambda_2  \left( 1+e^{-2i \left( \gamma - \sigma \right)} \right) A_{2i} \sin \beta   - \lambda_3  e^{2i\left(\sigma-\gamma\right)} A_{2i} \sin \beta  \\
&\hspace{30pt}+\frac{1}{2}\lambda_4 \left[ e^{i\sigma}\left( 2-e^{-2i\gamma}\right) A_{2i} \cos \beta   + e^{-i \sigma}\left( 2 - e^{-2i\left(\gamma-\sigma\right)}\right) A_{Si} \sin \beta    \right]\\
&\hspace{30pt}+ \frac{1}{2}\left( \lambda_5 + \lambda_6 \right)  A_{Si} \cos \beta   - \lambda_7 e^{-2i \gamma} A_{Si} \cos \beta   + \mathrm{h.c.}  \Big]\label{Eq.C_III_a_varphi2varphi2Hi}.
\end{split}
\end{align} 
\end{subequations}

Trilinear couplings involving the charged fields are:
\begin{subequations}
\begin{align}
\begin{split}
g\left( \varphi_1 h^+ H^- \right) &= v \Big[ \frac{1}{2}\lambda_2 \left( 1 - e^{2i\left(\gamma - \sigma \right)} \right)\sin (2\beta) - \frac{1}{2}\lambda_3 \left( 1 + e^{2i\left(\gamma - \sigma \right)} \right)\sin (2\beta)\\
&\hspace{30pt} + \lambda_4 \left[ -\frac{1}{2}e^{-i\sigma}\left( 1 + e^{2i\gamma}\right)\cos^2\beta + e^{i\gamma} \cos (\gamma - \sigma) \sin^2 \beta \right]\\
&\hspace{30pt} + \frac{1}{4}\lambda_6 \sin (2\beta) + \frac{1}{2}\lambda_7 e^{2i\gamma}\sin (2\beta)  \Big],
\end{split}\\
\begin{split}
g\left( \varphi_2 h^+ H^- \right) &= v \Big[ -i\frac{1}{2}\lambda_2 \left( 1 + e^{2i\left(\gamma - \sigma \right)} \right)\sin (2\beta) + i\frac{1}{2}\lambda_3 \left( 1 - e^{2i\left(\gamma - \sigma \right)} \right)\sin (2\beta)\\
&\hspace{30pt} + \lambda_4 \left( e^{i\left( \gamma - \sigma\right)}\sin \gamma \cos^2\beta - e^{i\gamma} \sin (\gamma - \sigma) \sin^2 \beta \right)\\
&\hspace{30pt} -i\frac{1}{4}\lambda_6 \sin (2\beta) + i\frac{1}{2}\lambda_7 e^{2i\gamma}\sin (2\beta)  \Big],
\end{split}\\
\begin{split}g \left( H_i h^\pm h^\mp \right) &= v \Big[ \left( \lambda_1 - \lambda_3 \right) A_{2i} \sin \beta+ \frac{1}{2} \lambda_4 e^{i \sigma} \left(  A_{2i} \cos \beta   +  A_{Si}^\ast \sin \beta   \right)\\
&\hspace{30pt}    + \frac{1}{2} \lambda_5  A_{Si} \cos \beta   + \mathrm{h.c.} \Big],\end{split}\\
\begin{split}
g\left( H_i H^\pm H^\mp  \right) &= v \Big[ \left( \lambda_1 + \lambda_3 \right)\cos^2\beta  A_{2i} \sin \beta\\
&\hspace{30pt}   - \frac{1}{2}\lambda_4 e^{i \sigma} \cos \beta \left[ \cos (2 \beta)A_{2i} + \frac{1}{2}\sin (2\beta)A_{Si}^\ast -  A_{2i}^\ast \sin^2 \beta \right]\\
&\hspace{30pt} +\frac{1}{2}\lambda_5 \left(  A_{Si} \cos^3\beta +  A_{2i} \sin^3 \beta  \right)\\
&\hspace{30pt} - \frac{1}{4}\lambda_6 \sin (2\beta) \left(  A_{2i} \cos \beta   +  A_{Si} \sin \beta   \right) \\
&\hspace{30pt} - \frac{1}{2}\lambda_7 e^{2i\sigma}\sin (2\beta)\left(  A_{2i} \cos \beta   +  A_{Si} ^\ast \sin \beta  \right)\\
&\hspace{30pt} + \lambda_8 A_{Si} \cos \beta \sin^2 \beta  + \mathrm{h.c.} \Big].
\end{split}
\end{align} 
\end{subequations}

The quartic couplings involving only the neutral fields are:
\begin{subequations}
\begin{align}
g \left( \varphi_1 \varphi_1 \varphi_1 \varphi_1 \right) &= g \left( \varphi_2 \varphi_2 \varphi_2 \varphi_2 \right) = 6 \left( \lambda_1 + \lambda_3 \right)\label{Eq.C_III_a_4varphi},\\
g \left( \varphi_1 \varphi_1 \varphi_2 \varphi_2 \right) &= 2 \left( \lambda_1 + \lambda_3 \right),\\
\begin{split}
g \left( \varphi_1 \varphi_1 H_i H_j \right) 
&= \frac{1}{2}\lambda_1 P_{2\overline{2}}(i,j) + \frac{1}{2} \lambda_2 \left[ e^{2i\left(\sigma - \gamma \right)}P_{22}(i,j)  
- P_{2\overline{2}}(i,j)\right]
\\
&\hspace{15pt}+ \frac{1}{2}\lambda_3 e^{2i\left( \sigma - \gamma \right)} P_{22}(i,j)\\
&\hspace{15pt}  + \lambda_4 \left( \frac{1}{2}e^{i\left( \sigma - 2 \gamma \right)}P_{2S}(i,j)  
+ e^{i \sigma} P_{2\overline{S}}(i,j)  \right)\\
&\hspace{15pt} + \frac{1}{4} \left( \lambda_5 + \lambda_6 \right) P_{S\overline{S}}(i,j) + \frac{1}{2} \lambda_7 e^{2i \gamma }P_{\overline{SS}}(i,j)  + \mathrm{h.c.},
\end{split}\\
\begin{split}
g \left( \varphi_1 \varphi_2 H_i H_j \right) 
&= -\frac{i}{2}\left(\lambda_2 + \lambda_3 \right)e^{2i \left(\sigma - \gamma\right)}P_{22}(i,j) 
- \frac{i}{2}\lambda_4 e^{i \left( \sigma - 2\gamma \right)}P_{2S}(i,j)\\
&\hspace{15pt} + \frac{i}{2}\lambda_7 e^{2i\gamma}P_{\overline{SS}}(i,j) +  \mathrm{h.c.},
\end{split}\\
\begin{split}
g \left( \varphi_2 \varphi_2 H_i H_j \right) 
&= \frac{1}{2}\lambda_1 P_{2\overline{2}}(i,j) - \frac{1}{2} \lambda_2 \left[ e^{2i\left(\sigma - \gamma \right)}P_{22}(i,j)  
+ P_{2\overline{2}}(i,j)\right]\\
&\hspace{15pt}- \frac{1}{2}\lambda_3 e^{2i\left( \sigma - \gamma \right)} P_{22}(i,j)\\
&\hspace{15pt}  - \lambda_4 \left( \frac{1}{2}e^{i\left( \sigma - 2 \gamma \right)}P_{2S}(i,j)  
- e^{i \sigma} P_{2\overline{S}}(i,j)  \right)\\
&\hspace{15pt} + \frac{1}{4} \left( \lambda_5 + \lambda_6 \right) P_{S\overline{S}}(i,j) - \frac{1}{2} \lambda_7 e^{2i \gamma }P_{\overline{SS}}(i,j) + \mathrm{h.c.},
\end{split}\\
\begin{split}
g \left( H_i H_j H_k H_l \right) 
&= \frac{1}{8}\left( \lambda_1 + \lambda_3 \right) P_{22\overline{22}}(i,j,k,l) 
- \frac{1}{4} \lambda_4 e^{i \sigma} P_{22\overline{2S}}(i,j,k,l) \\
&\hspace{15pt} + \frac{1}{8}\left( \lambda_5 + \lambda_6 \right) P_{2\overline{2}S\overline{S}}(i,j,k,l)  
+ \frac{1}{4} \lambda_7 e^{2i\sigma} P_{22\overline{SS}}(i,j,k,l) \\
&\hspace{15pt} + \frac{1}{8} \lambda_8 P_{SS\overline{SS}}(i,j,k,l)  + \mathrm{h.c.}\label{Eq.C_III_a_HiHjHkHl}
\end{split}
\end{align} 
\end{subequations}

The quartic couplings involving both neutral and charged fields are:
\begin{subequations}
\begin{align}
g \left( \varphi_i \varphi_i h^\pm h^\mp \right) &= 2 \left( \lambda_1 + \lambda_3 \right),\\
g \left( \varphi_i \varphi_i H^\pm H^\mp \right) &= 2 \left( \lambda_1 - \lambda_3 \right)\cos^2\beta - \lambda_4 \cos \sigma\sin (2\beta) + \lambda_5 \sin^2 \beta,\\
\begin{split}
g \left( \varphi_1 H_i h^+ H^- \right) 
&= \lambda_2 \cos \beta \left( A_{2i} - e^{2i\left( \gamma - \sigma \right)}A_{2i}^\ast \right) 
-\lambda_3 \cos \beta \left( A_{2i} + e^{2i\left( \gamma - \sigma \right)}A_{2i}^\ast \right)\\
&\hspace{15pt} - \frac{1}{2}\lambda_4 \left[ \cos \beta \left( e^{-i \sigma}A_{Si} + e^{i\left( 2\gamma - \sigma \right)}A_{Si}^\ast \right) - \sin \beta\left( e^{i \sigma}A_{2i} + e^{i\left( 2\gamma - \sigma \right)}A_{2i}^\ast \right) \right]\\
&\hspace{15pt} + \frac{1}{2} \lambda_6  A_{Si} \sin \beta   
+ \lambda_7 e^{2i \gamma} A_{Si}^\ast \sin \beta  ,
\end{split}\\
\begin{split}
g \left( \varphi_2 H_i h^+ H^- \right) 
&= -i\lambda_2 \cos \beta \left( A_{2i} + e^{2i\left( \gamma - \sigma \right)}A_{2i}^\ast \right) 
+ i\lambda_3 \cos \beta \left( A_{2i} - e^{2i\left( \gamma - \sigma \right)}A_{2i}^\ast \right)\\
&\hspace{15pt} + \frac{i}{2}\lambda_4 \left[ \cos \beta \left( e^{-i \sigma}A_{Si} - e^{i\left(2\gamma - \sigma \right)}A_{Si}^\ast \right) - \sin \beta\left( e^{i \sigma}A_{2i} - e^{i\left(2\gamma - \sigma \right)}A_{2i}^\ast \right) \right]\\
&\hspace{15pt} - \frac{i}{2} \lambda_6  A_{Si} \sin \beta   
+ i\lambda_7 e^{2i \gamma} A_{Si}^\ast \sin \beta  ,
\end{split}\\
g \left( H_i H_j h^\pm h^\mp \right) 
&= \frac{1}{2}\left( \lambda_1 - \lambda_3 \right)P_{2\overline{2}}(i,j) 
+ \frac{1}{2} \lambda_4 e^{i \sigma} P_{2\overline{S}}(i,j)
+ \frac{1}{4} \lambda_5 P_{S\overline{S}}(i,j) + \mathrm{h.c.},\\
\begin{split}
g \left( H_i H_j H^\pm H^\mp \right) 
&= \frac{1}{2}\left( \lambda_1 + \lambda_3 \right)  P_{2\overline{2}}(i,j) \cos^2\beta \\
&\hspace{15pt}- \frac{1}{2}\lambda_4 e^{i\sigma}\cos \beta 
\left[  P_{2\overline{S}}(i,j) \cos \beta -  P_{2\overline{2}}(i,j) \sin \beta\right]\\
&\hspace{15pt} + \frac{1}{4} \lambda_5 \left[  P_{S\overline{S}}(i,j) \cos^2\beta
+  P_{2\overline{2}}(i,j) \sin^2 \beta\right] 
- \frac{1}{4}\lambda_6 \sin (2\beta)P_{\overline{2}S}(i,j) \\
&\hspace{15pt} - \frac{1}{2}\lambda_7 e^{2i\sigma} \sin (2\beta) P_{2\overline{S}}(i,j) 
+\frac{1}{2}\lambda_8  P_{S\overline{S}}(i,j) \sin^2 \beta  + \mathrm{h.c.}
\end{split}
\end{align} 
\end{subequations}
The quartic couplings involving only the charged fields are:
\begin{subequations}
\begin{align}
g\left( h^\pm h^\pm h^\mp h^\mp \right) & = 4 \left( \lambda_1 + \lambda_3 \right),\label{Eq:hphphmhm}\\
\begin{split} g\left( H^\pm H^\pm H^\mp H^\mp \right) & = 4 \Big[ \left( \lambda_1 + \lambda_3\right)\cos^4\beta  + 2 \lambda_4 \cos \sigma\cos^3\beta \sin \beta\\
& \hspace{25pt}+ \frac{1}{4} \left( \lambda_5 + \lambda_6 + 2 \lambda_7 -4 \lambda_7 \sin^2 \sigma \right)\sin^2 (2\beta) + \lambda_8 \sin^4 \beta \Big],\end{split}\\
g\left( h^+ h^+ H^- H^- \right) & = 4 e^{2i(\gamma-\sigma)}\left[ \left( \lambda_2 + \lambda_3 \right)\cos^2\beta - \frac{1}{2}e^{i \sigma} \lambda_4 \sin (2\beta) + e^{2 i \sigma} \lambda_7 \sin^2 \beta \right],\\
g\left( h^\pm h^\mp H^\pm H^\mp \right) & = \left[ 2 \left( \lambda_1 - \lambda_2 \right)\cos^2\beta - 2\lambda_4 \cos \sigma\sin (2\beta) + \left( \lambda_5 + \lambda_6 \right) \sin^2 \beta \right].
\end{align} 
\end{subequations}

\subsection[The C-III-a implementation with complex \texorpdfstring{$\lambda_4$}{l4}]{The C-III-a implementation with complex \boldmath$\lambda_4$}

It is possible to have the C-III-a vacuum along with a $\lambda_4 \in \mathbb{C}$. In this case the minimisation conditions take a different form:
\begin{subequations}
\begin{align}
\mu_{SS}^2 ={}& - \frac{1}{2} (\lambda_5 + \lambda_6 + 2 \lambda_7) v_2^2 + \lambda_4^\mathrm{R} \frac{v_2^3}{\cos \sigma v_S} - 2 \lambda_8 v_S^2,\\
\mu_{11}^2 ={}& - \left( \lambda_1 + \lambda_3 \right) v_2^2 + \lambda_4^\mathrm{R} \frac{3 v_2 v_S}{2 \cos \sigma} - \frac{1}{2}\left( \lambda_5 + \lambda_6 + 2 \lambda_7 + 8 \lambda_7 \sin^2 \sigma \right) v_S^2,\\
\lambda_4^\mathrm{I} ={}& - \lambda_4^\mathrm{R} \tan \sigma + \lambda_7 \frac{4 \sin \sigma v_S}{v_2},
\end{align}
\end{subequations}
The discussion is identical to the C-III-a case.

\vspace{10pt}
\textbf{The mass-squared matrices}
\vspace{5pt}
    
The charged mass-squared matrix in the $\{h_1^+,\,h_2^+,\,h_S^+ \}$ basis is given by:
\begin{equation}
\mathcal{M}^2_\mathrm{Ch}
=\begin{pmatrix}
\left(\mathcal{M}_\mathrm{Ch}^2\right)_{11} & 0 & 0 \\
0 & \left(\mathcal{M}_\mathrm{Ch}^2\right)_{22} & \left(\mathcal{M}_\mathrm{Ch}^2\right)_{23} \\
0 & \left(\mathcal{M}_\mathrm{Ch}^2\right)_{23} & \left(\mathcal{M}_\mathrm{Ch}^2\right)_{33}
\end{pmatrix},
\end{equation}
where
\begin{subequations}
\begin{align}
\left(\mathcal{M}_\mathrm{Ch}^2\right)_{11} &= -2\lambda_3\hat v_2^2 + \frac{5}{2}\lambda_4^\mathrm{R} \frac{v_2 v_S}{\cos \sigma} -\frac{1}{2}\left[\lambda_6 + 10\lambda_7 - 8 \lambda_7 \cos(2\sigma)\right]\hat v_S^2, \\
\left(\mathcal{M}_\mathrm{Ch}^2\right)_{22} &= \frac{1}{2} \lambda_4^\mathrm{R} \frac{v_2 v_S}{\cos \sigma}-\frac{1}{2}(\lambda_6+2\lambda_7)\hat v_S^2 , \\
\left(\mathcal{M}_\mathrm{Ch}^2\right)_{23} &= -\frac{1}{2} \lambda_4^\mathrm{R} \frac{v_2^2}{\cos \sigma} +  \frac{1}{2}(\lambda_6+2\lambda_7)v_2 v_S, \\
\left(\mathcal{M}_\mathrm{Ch}^2\right)_{33} &= \frac{1}{2} \lambda_4^\mathrm{R} \frac{v_2^3 }{\cos \sigma v_S}-\frac{1}{2}(\lambda_6+2\lambda_7) v_2^2.
\end{align}
\end{subequations}

The charged mass eigenstates are:
\begin{subequations}
\begin{align}
h^+ & = h_1^+,\\
G^+ & = \sin\beta\,h_2^+ + \cos\beta\,h_S^+,\\ 
H^+ & = -\cos\beta\,h_2^+ + \sin\beta\,h_S^+,
\end{align}
\end{subequations}
with masses:
\begin{subequations} \label{Eq:C-III-c-C-masses-ch}
\begin{align}
m^2_{h^+} &= -2\lambda_3\hat v_2^2 + \frac{5}{2}\lambda_4^\mathrm{R} \frac{v_2 v_S}{\cos \sigma} -\frac{1}{2}\left[\lambda_6 + 10\lambda_7 - 8 \lambda_7 \cos(2\sigma)\right]\hat v_S^2, \\
m^2_{H^+} &= \frac{1}{2}(\lambda_4^\mathrm{R} \frac{\tan \beta}{\cos \sigma} - \lambda_6 - 2\lambda_7)v^2.
\end{align}
\end{subequations}

The inert sector  neutral mass-squared matrix in the $\{\eta_1,\,\chi_1\}$ basis is:
\begin{equation}
\mathcal{M}^2_\mathrm{N1}
=\begin{pmatrix}
\left(\mathcal{M}_\mathrm{N1}^2\right)_{11} & \left(\mathcal{M}_\mathrm{N1}^2\right)_{12} \\
\left(\mathcal{M}_\mathrm{N1}^2\right)_{12} & \left(\mathcal{M}_\mathrm{N1}^2\right)_{22}
\end{pmatrix},
\end{equation}
where
\begin{subequations}
\begin{align}
\left(\mathcal{M}_\mathrm{N1}^2\right)_{11} &= -2 \sin^2\sigma \left[ \left( \lambda_2 + \lambda_3 \right) \hat v_2^2 + 4 \lambda_7 \hat v_S^2\right] + \frac{1}{2} \lambda_4^\mathrm{R} \frac{7 + 2 \cos(2 \sigma)}{\cos \sigma}\hat v_2 \hat v_S, \\
\left(\mathcal{M}_\mathrm{N1}^2\right)_{12} &= \sin (2 \sigma)\left[ \left( \lambda_2 + \lambda_3 \right) \hat v_2^2 - 2\lambda_7 \hat v_S^2\right] + 2 \sin \sigma \lambda_4^\mathrm{R} \hat v_2 \hat v_S, \\
\left(\mathcal{M}_\mathrm{N1}^2\right)_{22} &=  -2 \left( \lambda_2 + \lambda_3 \right) \cos^2 \sigma \hat v_2^2 + \frac{1}{2} \lambda_4^\mathrm{R} \frac{ 7 - 2 \cos(2 \sigma)}{\cos \sigma} \hat v_2 \hat v_S - 2\lambda_7 \left( 5 - 4\cos(2\sigma) \right) \hat v_S^2.
\end{align}
\end{subequations}
The $\gamma$ diagonalisation angle is given by
{\setlength{\belowdisplayskip}{0pt} \setlength{\belowdisplayshortskip}{0pt}
\setlength{\abovedisplayskip}{0pt} \setlength{\abovedisplayshortskip}{0pt}
\begin{equation}
\begin{aligned}
\tan (2 \gamma) ={}& \bigg\{ 2 \sin(2\sigma)\left[ \cos \sigma \left(  \lambda_2 + \lambda_3 \right) \hat v_2^2 - 2 \cos \sigma  \lambda_7 v_S^2 + \lambda_4^\mathrm{R} \hat v_2 \hat v_S \right] \bigg\} / \\
&\bigg\{ \left( \lambda_2 + \lambda_3 \right) \left[ \cos \sigma + \cos(3 \sigma) \right] \hat v_2^2 + 2 \lambda_4^\mathrm{R} \cos(2 \sigma) \hat v_2 \hat v_S\\
&\hspace{140pt} + 2 \lambda_7 \left[ 2 \cos \sigma - \cos(3 \sigma) \right] \hat v_S^2 \bigg\}.
\end{aligned}
\end{equation}}

The physical neutral states are: 
\begin{subequations}
\begin{align}
\varphi_1 & = \cos\gamma\,\eta_1 + \sin\gamma\,\chi_1,\\ 
\varphi_2 & = -\sin\gamma\,\eta_1+ \cos\gamma\,\chi_1,
\end{align}
\end{subequations}
with masses
\begin{subequations}
\begin{align}
m^2_{\varphi_1}={}& -\left( \lambda_2 + \lambda_3 \right) \hat v_2 + \frac{7}{2} \lambda_4^\mathrm{R} \frac{1}{\cos \sigma} \hat v_2 \hat v_S - \lambda_7 \left( 7 -6 \cos (2\sigma) \right) \hat v_S^2 - \frac{1}{\sqrt{2} \cos^4 \sigma} \Delta,\\
m^2_{\varphi_2}={}& -\left( \lambda_2 + \lambda_3 \right) \hat v_2 + \frac{7}{2} \lambda_4^\mathrm{R} \frac{1}{\cos \sigma} \hat v_2 \hat v_S - \lambda_7 \left( 7 -6 \cos (2\sigma) \right) \hat v_S^2 + \frac{1}{\sqrt{2} \cos^4 \sigma} \Delta,\\
\end{align}
\end{subequations}
where
\begin{equation}
\begin{aligned}
\Delta^2 = \cos ^6\sigma \bigg[&
  \left(2 (\lambda_4^\mathrm{R})^2 - (\lambda_2+\lambda_3)\lambda_7 +  (\lambda_2+\lambda_3) \lambda_7 \left[ 2 \cos (2 \sigma )+3 \cos (4 \sigma )\right]\right)\hat v_2^2 \hat v_S^2\\
&+4 \lambda_4^\mathrm{R} (\lambda_2+\lambda_3) \cos \sigma\hat v_2^3 \hat v_S-2 \lambda_4^\mathrm{R} \lambda_7   \left[\cos \sigma-3 \cos (3 \sigma )\right]\hat v_2\hat v_S^3\\
&+2  (\lambda_2+\lambda_3)^2 \cos ^2 \sigma \hat v_2^4+\lambda_7^2 \left[ 7+\cos (2 \sigma )-6 \cos (4 \sigma ) \right]\hat v_S^4 \bigg].
\end{aligned}
\end{equation}

Finally, the neutral mass-squared matrix in the basis of $\{\eta_2, \,\eta_S,\, \chi_2, \,\chi_S\}$ is given by: 
\begin{equation}
\mathcal{M}^2_\mathrm{N2S}
=\begin{pmatrix}
\left(\mathcal{M}^2_\mathrm{N2S}\right)_{11} & \left(\mathcal{M}^2_\mathrm{N2S}\right)_{12} & \left(\mathcal{M}^2_\mathrm{N2S}\right)_{13} & \left(\mathcal{M}^2_\mathrm{N2S}\right)_{14}\vspace{2pt} \\ 
\left(\mathcal{M}^2_\mathrm{N2S}\right)_{12} & \left(\mathcal{M}^2_\mathrm{N2S}\right)_{22} & \left(\mathcal{M}^2_\mathrm{N2S}\right)_{14} & \left(\mathcal{M}^2_\mathrm{N2S}\right)_{24}\vspace{2pt} \\ 
\left(\mathcal{M}^2_\mathrm{N2S}\right)_{13} & \left(\mathcal{M}^2_\mathrm{N2S}\right)_{14} & \left(\mathcal{M}^2_\mathrm{N2S}\right)_{33} & \left(\mathcal{M}^2_\mathrm{N2S}\right)_{34}\vspace{2pt} \\ 
\left(\mathcal{M}^2_\mathrm{N2S}\right)_{14} & \left(\mathcal{M}^2_\mathrm{N2S}\right)_{24} & \left(\mathcal{M}^2_\mathrm{N2S}\right)_{34} & \left(\mathcal{M}^2_\mathrm{N2S}\right)_{44}\vspace{2pt} \\ 
\end{pmatrix},
\end{equation}
where
\begin{subequations}
\begin{align}
\left(\mathcal{M}^2_\mathrm{N2S}\right)_{11} &= 2 \left( \lambda_1 + \lambda_3 \right)\hat v_2^2 - \frac{3}{2} \lambda_4^\mathrm{R} \frac{\hat v_2 \hat v_S}{\cos \sigma} + 6 \lambda_7 \sin^2 \sigma \hat v_S^2,\\
\left(\mathcal{M}^2_\mathrm{N2S}\right)_{12} &= \frac{3}{2}\lambda_4^\mathrm{R} \frac{\hat v_2^2}{\cos \sigma} +  \left( \lambda_5 + \lambda_6 + 2 \lambda_7 + 2 \lambda_7 \sin^2 \sigma \right) \hat v_2 \hat v_S,\\
\left(\mathcal{M}^2_\mathrm{N2S}\right)_{13} &= \lambda_7 \sin (2\sigma) \hat v_S^2,\\
\left(\mathcal{M}^2_\mathrm{N2S}\right)_{14} &= -\lambda_7 \sin (2\sigma) \hat v_2 \hat v_S,\\
\left(\mathcal{M}^2_\mathrm{N2S}\right)_{22} &= \lambda_4^\mathrm{R} \frac{\hat v_2^3}{2 \cos \sigma \hat v_S}  - 2\lambda_7 \sin^2 \sigma \hat v_2^2 + 2 \lambda_8 \hat v_S^2,\\
\left(\mathcal{M}^2_\mathrm{N2S}\right)_{24} &= \lambda_7 \sin (2\sigma) \hat v_2^2,\\
\left(\mathcal{M}^2_\mathrm{N2S}\right)_{33} &= \frac{1}{2} \lambda_4^\mathrm{R} \frac{\hat v_2 \hat v_S}{\sin \sigma} - 2 \lambda_7 \cos^2 \sigma \hat v_S^2,\\
\left(\mathcal{M}^2_\mathrm{N2S}\right)_{34} &= - \frac{1}{2}\left( \lambda_4^\mathrm{R} \frac{\hat v_2}{\cos \sigma} + 4 \lambda_7 \cos^2 \sigma \hat v_S  \right) \hat v_2,\\
\left(\mathcal{M}^2_\mathrm{N2S}\right)_{44} &= \frac{1}{2} \lambda_4^\mathrm{R} \frac{\hat v_2^3 }{\cos \sigma \hat v_S} - 2 \lambda_7 \cos^2 \sigma \hat v_2^2.
\end{align}
\end{subequations}
In the Higgs basis, see eqs.~\eqref{Eq:C-III-a_NeutralActiveHBRot} and~\eqref{Eq:C-III-a_M_sq_3by3}, we have: 
\begin{subequations}
\begin{align}
(\mathcal{M}^2_\mathrm{\phi})_{11} &= \frac{2}{v^2} \left[ \left( \lambda_1 + \lambda_3 \right)v_2^4 - \lambda_4^\mathrm{R} \frac{\hat v_2^3 \hat v_S}{\cos \sigma} + \left( \lambda_5 + \lambda_6 + 2 \lambda_7 + 4 \lambda_7 \sin^2 \sigma \right)v_2^2v_S^2 + \lambda_8 v_S^4 \right],\\
\begin{split}
(\mathcal{M}^2_\mathrm{\phi})_{12} &= -\frac{\hat v_2}{v^2} \Big[ \left(2\lambda_1 + 2\lambda_3 - \lambda_5 - \lambda_6 - 2 \lambda_7 \right)v_2^2 v_S +\lambda_4^\mathrm{R} \frac{\hat v_2 \left(  \hat v_2^2 - 3 \hat v_S^2\right)}{\cos \sigma}\\
& \hspace{45pt}+ \left( \lambda_5 + \lambda_6 + 2 \lambda_7  + 8 \lambda_7 \sin^2 \sigma - 2 \lambda_8 \right) v_S^3 \Big],
\end{split}\\
\begin{split}
(\mathcal{M}^2_\mathrm{\phi})_{22} &= \frac{1}{2 v^2 \hat v_S} \Big[ \left( \lambda_1 + \lambda_3 - \lambda_5 - \lambda_6 - 2 \lambda_7 - 2 \lambda_7 \sin^2\sigma + \lambda_8 \right)v_2^2 v_S^3\\
& \hspace{50pt} + \lambda_4^\mathrm{R} \frac{\hat v_2 \left( \hat v_2^4 + 6 \hat v_2^2 \hat v_S^2 - 3 \hat v_S^4 \right)}{\cos \sigma}\\
& \hspace{50pt} -4\lambda_7 \sin^2 \sigma  \left( v_2^4-3v_S^4 \right)v_S \Big],
\end{split}\\
(\mathcal{M}^2_\mathrm{\phi})_{23} &= v^2 \lambda_7 \sin(2\sigma),\\
(\mathcal{M}^2_\mathrm{\phi})_{33} &= \frac{v^2}{2 \cos \sigma} \left( \lambda_4^\mathrm{R} \frac{\hat v_2}{\hat v_S} - 4 \lambda_7 \cos^3 \sigma \right).
\end{align}
\end{subequations}

\subsection{Relation of the C-III-a implementation to other cases}

The C-III-a model can be connected to various other $S_3$-based implementations by considering specific limits that result in vacua where $ v_1 $ is not proportional to $ v_2 $, and where the vevs $ v_2 $ or $ v_S $ do not vanish. However, these relationships cannot always be established. In the limit of $\cos\sigma=0$, the mass splitting between the neutral inert states, see eq.~(\ref{Eq:C-III-a-inert-splitting}), becomes
\begin{equation}
\Delta = |\left( \lambda_2 + \lambda_3 \right) v_2^2 - \lambda_7 v_S^2|.
\end{equation}
Apart from that, $\lambda_4$ should vanish, see eq.~\eqref{eq:c-iii-a-lam4-lam7}, resulting in the $O(2)$ symmetry. Then, C-III-a with $\cos\sigma=0$ becomes equivalent to C-III-f, $(\pm i v_1,\, i v_2,\, v_S)$, or C-III-g, $(\pm i v_1,\, -i v_2,\, v_S)$, depending on in which quadrant the $\sigma$ phase is, with $v_1 \ll v$. 

For $\Delta=0$, the mass degeneracy limit, we get $\lambda_7 = \left( \lambda_2 + \lambda_3 \right) \tan^2 \beta$. In this case, irrespective of the $\lambda_7$ coupling we get two massless states. This case is then equivalent to C-IV-b, $( v_1,\, \pm i v_2,\, v_S)$, with $v_1 \ll v$. However, C-IV-b increases the underlying symmetry to that of $O(2)$. So, where does the other massless state originate from? It should be noted that one of the mass eigenvalues of C-IV-b explicitly depends on $ v_1^2 $.

Other vacua of the form $(0,\,v_2,\,v_S)$ can also be reached. The R-II-1a is a special case and will be discussed separately. The only other real implementation is R-III, $ (v_1,\, v_2,\, v_S) $. However, it is impossible to reach this case starting from C-III-a, as R-III would require both $ \sigma = 0 $ and $ \lambda_4 = 0 $. Then, the only option would be to force $ \lambda_7 = 0 $, which is not a requirement for R-III. Moving to the complex vacua, the C-III-d, $(\pm i v_1,\, v_2,\, v_S)$ and C-III-e, $(\pm i v_1,\, -v_2,\, v_S)$, cases are not reachable because the minimisation conditions of these depend on the $\lambda_2+\lambda_3$ term, whereas those of C-III-a do not. Another implementation is C-IV-d, $( v_1 e^{i \sigma_1},\, \pm v_2 e^{i \sigma_1},\, v_S)$, which is, in fact, CP conserving due to the underlying $O(2) \times U(1)$ symmetry. Finally, when $\lambda_2 + \lambda_3=0$ along with $\lambda_7=0$ is satisfied, C-III-a can be viewed as a special case of C-V, $(v_1 e^{i \sigma_1},\, v_2 e^{i \sigma_2},\, v_S)$, which does not result in CP violation. 

An overview of the above relations is summarised in Table~\ref{Table:CIIIaReducedComparison}.

{{\renewcommand{\arraystretch}{1.05}
\begin{table}[b!]
\caption{Relations of C-III-a to other $S_3$-based implementations. Different vacua, in most cases, are examined in the limit of $\hat v_1 \to 0$ along with general minimisation conditions. Massless states, either in terms of a single scalar field, $ m_{X_i} $, or in terms of the mixing of fields, $ m_{X_i-X_j} $, are presented.}
\label{Table:CIIIaReducedComparison}
\begin{center}
\begin{tabular}{|c|c|l|}
\hline\hline
Model & Conditions & Comments \\ \hline \hline
\begin{tabular}[c]{@{}c@{}}R-II-1a\\ \footnotesize $(0,\, v_2,\, v_S )$\end{tabular} & \begin{tabular}[c]{@{}c@{}} $\sigma=0$ \end{tabular} & Special point in R-II-1a, $\lambda_4 = 4 \lambda_7 v_S/v_2$. \\ \hline
\begin{tabular}[c]{@{}c@{}}R-III\\ \footnotesize$(v_1,\, v_2,\, v_S )$ \end{tabular} & \begin{tabular}[c]{@{}c@{}}   \end{tabular}& \begin{tabular}[l]{@{}l@{}}Not reachable, $\lambda_7 \neq 0$ in R-III.\end{tabular} \\ \hline \hline
\begin{tabular}[c]{@{}c@{}}C-III-d,e\\ \footnotesize$(\pm i v_1,\, v_2,\, v_S)$,\\ \footnotesize$(\pm i v_1,\, -v_2,\, v_S)$ \end{tabular} & \begin{tabular}[c]{@{}c@{}} \end{tabular} & \begin{tabular}[l]{@{}l@{}}Not reachable. There are no\\ vanishing couplings in C-III-d,e.\end{tabular}\\ \hline
\begin{tabular}[c]{@{}c@{}}C-III-f,g\\ \footnotesize$(\pm i v_1,\, i v_2,\, v_S)$,\\ \footnotesize$(\pm i v_1,\, -i v_2,\, v_S)$ \end{tabular} & \begin{tabular}[c]{@{}c@{}} $\sigma = \pm \pi/2$,\\  $\lambda_4=0$\end{tabular} & \begin{tabular}[c]{@{}l@{}}Additional $O(2)$ symmetry; $m_{\chi_{1}}=0$. \end{tabular} \\ \hline
\begin{tabular}[c]{@{}c@{}}C-IV-b\\ \footnotesize$( v_1,\, \pm i v_2,\, v_S)$ \end{tabular} &\begin{tabular}[c]{@{}c@{}}$\sigma = \pm \pi/2,$\\  $\lambda_4=0$,\\ $\lambda_7 = \left( \lambda_2 + \lambda_3 \right) v_2^2 / v_S^2$ \end{tabular} & \begin{tabular}[l]{@{}l@{}} Exact C-IV-b: additional $O(2)$ symmetry.\\ 
C-III-a limit: another massless state;\\ \hspace{68pt}  $m_{\eta_1}=m_{\chi_1}=0$. \end{tabular}  \\ \hline
\begin{tabular}[c]{@{}c@{}}C-IV-d\\ \footnotesize $( v_1 e^{i \sigma_1},\, \pm v_2 e^{i \sigma_1},\, v_S)$ \end{tabular} & $\lambda_4=\lambda_7=0$ & \begin{tabular}[l]{@{}l@{}}  Additional $O(2) \times U(1)$ symmetry;\\ \qquad $m_{\eta_1-\chi_1}=m_{\chi_2-\chi_S}=0.$ \end{tabular} \\ \hline
\begin{tabular}[c]{@{}c@{}}C-V\\ \footnotesize$(v_1 e^{i \sigma_1},\, v_2 e^{i \sigma_2},\, v_S)$ \end{tabular} & \begin{tabular}[c]{@{}c@{}} $\lambda_2+\lambda_3=\lambda_4=\lambda_7 = 0$ \end{tabular} & \begin{tabular}[l]{@{}l@{}}  Additional $SU(2)$ symmetry;\\ $m_{\eta_1}=m_{\chi_1}=m_{\chi_2-\chi_S}=0$. \end{tabular} \\ \hline \hline 
\end{tabular}
\end{center}
\end{table}}

Both R-II-1a and C-III-a share a similar structure of the vevs, $(0,\,v_2,\,v_S)$:
\begin{equation*}
\text{R-II-1a: }(0,\, v_2,\, v_S ),\quad \text{C-III-a: }(0,\, v_2 e^{i \sigma},\, v_S ).
\end{equation*}

In R-II-1a, there is no mixing between the neutral states associated with $h_1$, while in C-III-a these states mix. In addition, in R-II-1a the neutral states are of definite CP parities, while in the C-III-a implementation the three neutral physical states mix together to produce CP-indefinite states. One might expect that all the masses and mixings of R-II-1a could be recovered from those of C-III-a by simply taking the limit of $\sigma \to 0$. However, this is not the case. This raises the question of why the R-II-1a case cannot be trivially obtained from C-III-a by setting the $\sigma$ phase to zero.

The explanation is straightforward: one only needs to examine the minimisation condition that arises from the variation of $\sigma$,
\begin{equation}
\sin \sigma \left( \lambda_4 \sin \beta - 4 \lambda_7 \cos \beta \cos \sigma \right)  = 0,
\label{minsig}
\end{equation}
assuming that $\beta \neq \{0,\,\pi/2\}$. The minimisation conditions are satisfied for either $\sigma=0$, leading to the real solution R-II-1a, or for $\lambda_4$ related to $\lambda_7$ by  eq.~(\ref{eq:c-iii-a-lam4-lam7}). Note that R-II-Ia does not require a condition relating $\lambda_4$ to $\lambda_7$. Imposing both results in $S_3$ being increased to $O(2)$.

\section[Imposing constraints in \texorpdfstring{$S_3$}{S3} implementations]{Imposing constraints in \boldmath$S_3$ implementations}\label{Sec:S3_DM_evaluation_checks}

In the numerical parameter scan, theoretical and experimental constraints are enforced, and accordingly, specific selection criteria (Cuts) are defined and implemented:
\begin{itemize}
\setlength\itemsep{0.2em}				  
\item Cut~1: perturbativity, stability, unitarity checks, LEP constraints;
\item Cut~2: SM-like gauge and Yukawa sector, $S$ and $T$ parameters, $B$ physics;
\item Cut~3: other relevant SM-like Higgs boson properties, DM relic density, direct and indirect searches;
\end{itemize}
with each of the subsequent constraint being superimposed over the previous ones.

The mass of the SM-like Higgs particle is fixed at $125$ GeV~\cite{ParticleDataGroup:2024cfk}. In the extended Higgs sector studies it is usually assumed that the charged scalars should not be lighter than 80~GeV~\cite{Pierce:2007ut,Arbey:2017gmh}. We consider a more generous lower bound of 70~GeV. Measurements of the gauge bosons widths at LEP~\cite{Schael:2013ita} prevent the decays of $W^\pm$ and $Z$ into scalars. Therefore, we constrain such decays by putting limits on the masses of the 3HDM scalars.

The general algorithm is depicted in Figure~\ref{Fig:Algorithm}.

We will now discuss the numerical scans presented in Refs.~\cite{Khater:2021wcx,Kuncinas:2022whn,Kuncinas:2023hpf}.

\vspace{10pt}
\textbf{R-II-1a}
\vspace{5pt}

The R-II-1a implementation is analysed using the following input:
\begin{itemize}
\setlength\itemsep{0.5em}
\item Mass of the SM-like Higgs is fixed at $m_h \approx 125$ GeV;
\item The charged scalar masses $\{m_{h^\pm},\, m_{H^\pm}\} \in [0.07,~1]$ TeV;
\item The inert sector masses $\{m_{\eta},\, m_{\chi}\} \in [0,~1]$ TeV. Either $\eta$ or $\chi$, whichever is the lightest state, could be a DM candidate;
\item The active sector masses $\{m_H,~m_A\} \in [m_h,~1~\mathrm{TeV}]$;
\item The angles are evaluated in the following ranges: $\beta \in [-\frac{\pi}{2},~\frac{\pi}{2}]$ and $\alpha \in [0,~\pi]$.
\end{itemize}

\begin{figure}[htb]
\begin{center}
\includegraphics[scale=0.8]{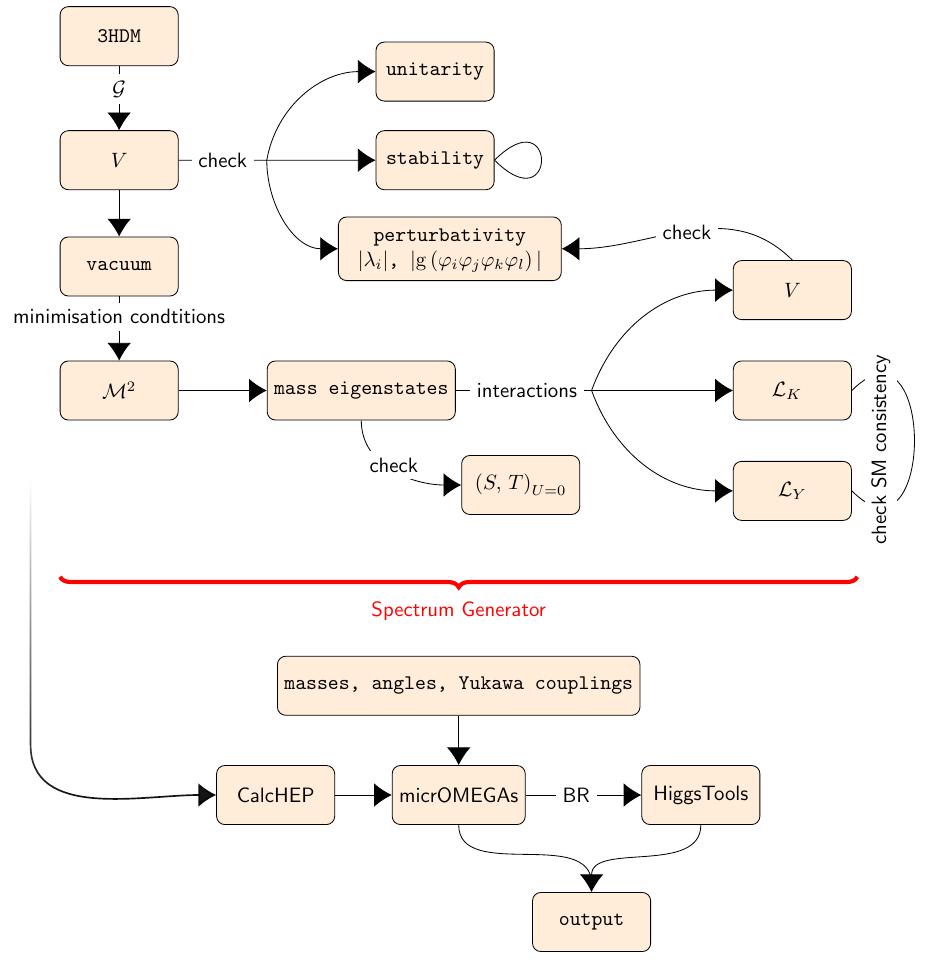}
\end{center}
\vspace*{-4mm}
\caption{ A cartoon showing a general approach to how different implementations are evaluated based on several constraints.}
\label{Fig:Algorithm}
\end{figure}

\vspace{10pt}
\textbf{C-III-a}
\vspace{5pt}

For the C-III-a implementation we use the following input:
\begin{itemize}
\item The lightest $H_i$ state is the SM-like Higgs, $m_{H_1} \approx 125$ GeV;
\item The charged scalar masses $\{ m_{h^+},~m_{H^+}\} \in [0.07,~1]$ TeV;
\item The mass of the DM candidate $m_{\varphi_1} \in [0,~1]$ TeV;
\item The angles are evaluated in the following ranges: $\beta \in [0,~\pi/2]$, $\sigma \in [-\pi,~\pi]$, \mbox{$\gamma \in [0,~\pi]$}, $\theta_2 \in [-\pi/2,~\pi/2]$, $\theta_3 \in [-\pi/2,~\pi/2]$.
\end{itemize}
As can be noted, not all masses are used as input. The values of $\{m_{H_2},\,m_{H_3},\,m_{\varphi_2},\,\theta_1\}$ are calculated based on the input.

\subsection{Cut 1 constraints}

The Cut~1 theory constraints were discussed in Section~\ref{Sec:Constraints_3HDM_gen}. In most cases, quartic couplings involving inert scalars can be examined to establish limits on the couplings, defining the perturbativity limit. For example, for R-II-1a we have $g\left( \eta  \eta  \eta  \eta  \right) = 6 \left(\lambda _1+\lambda _3\right)$, see eq.~\eqref{Eq.R_II_1a_QCLim_l1l3}. In light of the perturbativity limit, $|g| \leq 4\pi$, we obtain \mbox{$|\lambda_1 + \lambda_3| \leq 2/3 \pi$}. Evaluating the perturbativity limit for other couplings is more complicated.

\begin{figure}[htb]
\begin{center}
\includegraphics[scale=0.255]{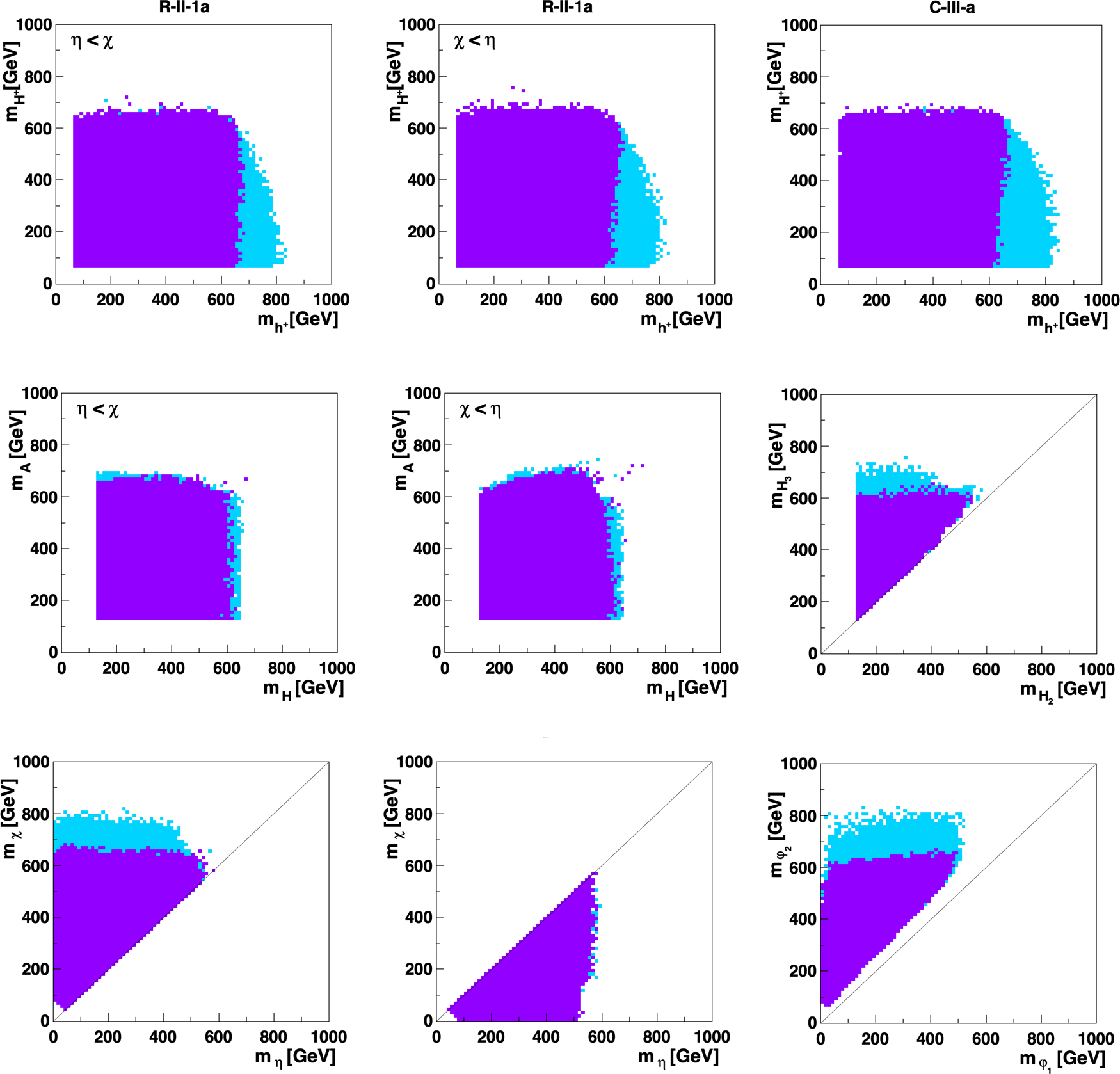}
\end{center}
\vspace*{-8mm}
\caption{Scatter plots of masses of the R-II-1a (first two columns) and C-III-a (last column) implementations that satisfy the theory constraints of Cut~1. Top row: the charged sector. Middle row: the active neutral sector. Bottom row: the inert neutral sector. The sky-blue region accommodates the $16\pi$ unitarity constraint, whereas the dark-indigo region satisfies the $8\pi$ constraint. It turns out that the points passing all of the checks satisfy the unitarity constraint at a value of $8\pi$. Figures taken from Refs.~\cite{Khater:2021wcx,Kuncinas:2022whn}.}
\label{Fig:C-III-a-masses-th_constraints}
\end{figure}

As illustrated in Figure~\ref{Fig:C-III-a-masses-th_constraints}, imposing theoretical constraints eliminates certain regions of the parameter space. If no free bilinear parameters are present, the unitarity, perturbativity, and stability constraints are useful for determining the upper bounds of the scalar masses, since squared masses are proportional to the quartic couplings multiplied by the vevs. However, when free bilinear terms are present, the quartic couplings become sub-leading at higher masses. Experimental constraints will further restrict the allowed regions of the parameter space.

In our model, fermions couple exclusively to the $h_S$ doublet, which allows us to place a constraint on the value of $\tan \beta$. From the definition of the Yukawa couplings, \mbox{$Y_f = \frac{\sqrt{2} m_f}{v \cos \beta}$} and the perturbativity condition $|Y_f| \leq 4\pi$, the most stringent bound arises from the heaviest fermion, which would be the top quark. Based on these values, we obtain $|\tan \beta| \leq 12.6$. While Landau poles would appear near this non-perturbative region, other constraints prevent the parameters from approaching this limit.

\subsection{Cut 2 constraints}

The Cut~2 constraints are applied to the parameter points that pass the Cut~1 constraints:
\begin{itemize}

\item \textbf{SM-like limit}

The SM-like limit refers to the scenario where the CP-even boson, which resides in the only doublet that acquires a vev in the Higgs basis, is already a physical mass eigenstate. As a result, it does not mix with the other neutral scalars. This doublet also contains the would-be Goldstone bosons $G^{\pm}$ and $G^0$. This limit is considered special and is termed the SM-like limit because the CP-even neutral boson behaves similarly to the SM Higgs boson in many aspects. 

\begin{figure}[htb]
\begin{center}
\includegraphics[scale=0.35]{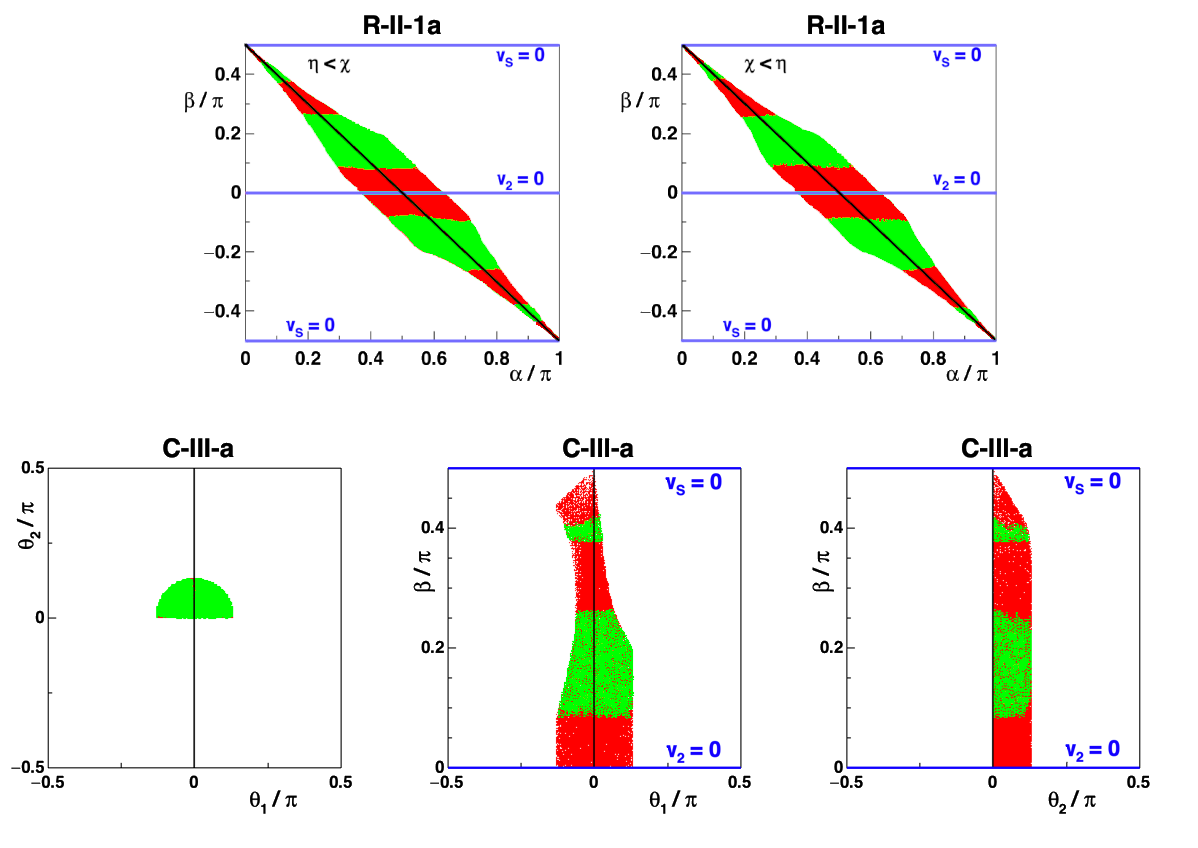}
\end{center}
\vspace*{-4mm}
\caption{Constraints on the angles of R-II-1a (top) and C-III-a (bottom) from the gauge and Yukawa couplings. The white region is excluded. Lime: values satisfying Cut~2, plotted over the red-coloured background. Figures taken from Refs.~\cite{Khater:2021wcx,Kuncinas:2022whn}.}
\label{Fig:SM-Like_Limit}
\end{figure}

Apart from identifying the 125 GeV state in the studied implementations, one needs to verify that it couples to the SM sector in consistency with the experimental observations. We shall adopt the following 3-$\sigma$ bounds from the PDG~\cite{ParticleDataGroup:2024cfk}: $\kappa^2_{VV} \in\{1.00 \pm 0.08 \}$, which comes from $h_\mathrm{SM} W^+ W^-$ and $\kappa^2_{ff} \in\{0.99 \pm 0.12 \}$, which comes from $h_\mathrm{SM} b\bar b$. Note that PDG lists the average for the fermion coupling as $0.94 \pm 0.05$ and for the gauge boson coupling as $1.023 \pm 0.026$. The 3-$\sigma$ allowed regions are presented in Figure~\ref{Fig:SM-Like_Limit}.

For R-II-1a, the SM-like limit is reached for  $\sin(\alpha + \beta) = 1$, where $\alpha = -\beta + \frac{\pi}{2}$. In this case, as seen from eq.~\eqref{LVVH-RII1a}, only the scalar $h$ has the $VVH$ couplings, and their strengths match those of the SM Higgs. Additionally, from eq.~\eqref{Eq:R-II-1a-Yukawa_Neutral}, in the limit of $\sin \alpha = \cos \beta$ it should be evident that the $h$ state would couple to fermions with the same strength as the SM Higgs boson. 

For C-III-a, we have $\kappa^2_{VV} = \cos\theta_1\cos\theta_2$. However, even if this constraint is satisfied, it does not guarantee that the non-SM trilinear $Z H_1 H_i$ and $W^\pm H^\mp H_1$, and quartic $A W^\pm H^\mp H_1$, $Z W^\pm H^\mp H_1$ couplings vanish. We recall that $H_1$ is the SM-like Higgs boson. The $\kappa_{ff}^2$ is given by the $\{\theta_1,\, \theta_2, \, \beta\}$ set of angles.

\item \textbf{Electroweak oblique parameters}

 The electroweak oblique parameters ($S$ and $T$) are evaluated at the limit of $U=0$. Sufficient mass splittings of the extended scalar sector can provide a significant contribution. The constraints for the C-III-a implementation were presented in Section~\ref{Sec:SectionSTU}.

\item \textbf{\boldmath$B$ physics constraints}

Rare $B$-meson decays, which receive their leading contributions in the SM from one-loop diagrams, serve as crucial constraints on many popular BSM scenarios. Among these, the inclusive radiative decay of $\bar B\to X(s)\gamma$ is particularly noteworthy. The significance of charged scalar exchange in influencing the $\bar B\to X(s)\gamma$ decay rate has been recognised since the late 1980s~\cite{Grinstein:1987pu,Hou:1988gv,Grinstein:1990tj}. The rate is calculated by expanding the relevant Wilson coefficients in powers of $ \alpha_s / (4\pi) $, following these steps: (1) matching the coefficients to the full theory at a high scale ($\mu_0 \sim m_W$ or $ m_t$), (2) evolving them down to the low scale $\mu_b \sim m_b$, during which operator mixing occurs, (3) determining the matrix elements at the low scale~\cite{Buras:1993xp,Ciafaloni:1997un,Ciuchini:1997xe,Borzumati:1998tg,Bobeth:1999ww,Bobeth:1999mk,Gambino:2001ew,Cheung:2003pw,Misiak:2004ew,Czakon:2006ss,Hermann:2012fc,Misiak:2015xwa,Misiak:2017bgg,Misiak:2020vlo,Czaja:2023ren}.

Although 3HDMs contain two physical charged Higgs bosons, in the implementations under consideration, only one of them couples to fermions, while the other resides in the inert sector. This suggests that we can adopt the approach of Misiak and Steinhauser~\cite{Misiak:2006ab,Misiak:2017bgg} as used for the 2HDM, where the active charged scalar has relative Yukawa couplings, in the notation of Ref.~\cite{Misiak:2006ab},
\begin{equation} \label{Eq:Au-Ad}
A_u=A_d=\tan\beta.
\end{equation}
We recall that the $\tan\beta$ here is the inverse of the ``usual" $\tan\beta$, see eq.~\eqref{Eq:Tan_beta_relation_S3}. According to eq.~(\ref{Eq:Au-Ad}), the relevant couplings match those of the 2HDM \mbox{Type-I} model, except that we focus on small values of $\tan\beta$. The $\bar{B} \to X(s)\gamma$ constraint excludes values of $|\tan\beta|\geq 4$. Once Cut 3 is applied, the allowed range will be further constrained.

The current world average extrapolated to the photon energy $E_\gamma  = 1.6~\text{GeV}$ is $\mathrm{Br} \left( \bar B\to X(s)\gamma  \right) \times 10^4 = 3.49 \pm 0.19$~\cite{HFLAV:2022esi,ParticleDataGroup:2024cfk}. We note that the most recent results from Belle~II~\cite{Belle-II:2022hys} are not included in the world average, but are consistent with the SM and world averages. We apply an $(n=3)$-$\sigma$ tolerance along with an additional ten percent computational uncertainty:
\begin{equation}
\begin{aligned}
\mathrm{Br} \left( \bar B\to X(s)\gamma  \right) \times 10^4 &=  3.49 \pm \sqrt{(3.49 \times 0.19)^2 + (0.19\,n)^2}\,.
\end{aligned}
\end{equation}
The acceptable region, is then $[2.62;\,4.36 ]$. We show in Figure~\ref{Fig:bsgamma-impact} the regions in the $\tan\beta{-}m_{H^+}$ parameter plane that are not excluded by this constraint.

Due to $A_u A_d = \tan^2 \beta > 0$, there exists a region where negative interference occurs between the SM-type contributions and the charged Higgs loop. As the value of $\tan\beta$ increases for a fixed value of the charged scalar mass, the branching ratio initially decreases before rising again, as shown in the lower right-hand corner of the left panel of Figure~\ref{Fig:bsgamma-impact}. However, this interference region is excluded by Cut~3. Conversely, for any fixed $\tan\beta$, the rate approaches the SM value as $m_{H^+}$ becomes sufficiently large.

\begin{figure}[h]
\begin{center}
\includegraphics[scale=0.3]{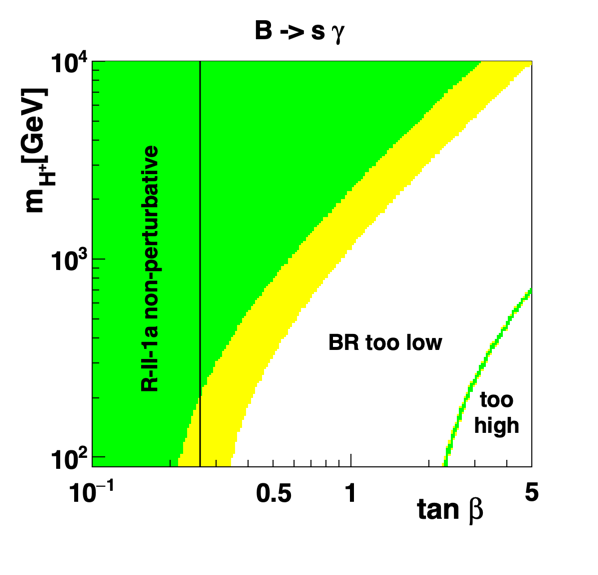}
\includegraphics[scale=0.3]{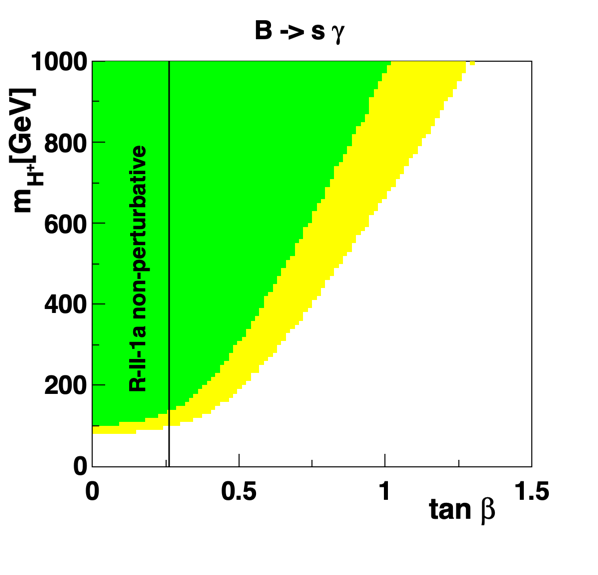}
\end{center}
\vspace*{-4mm}
\caption{Regions in the $\tan\beta{-}m_{H^+}$ plane that satisfy the $ \bar{B} \to X(s)\gamma $ constraint. Left: A logarithmic scale representation extending to larger values of $\tan\beta$ and $m_{H^+}$. Right: A linear scale representation focusing on the small $\tan\beta$ region. The yellow region represents a 3-$\sigma$ tolerance with respect to the experimental rate, while the lime regions correspond to a 2-$\sigma$ bound. The vertical line at $\tan\beta = 0.26$ indicates the lower bound on $|\tan\beta|$ that is consistent with $|\lambda_4| < 4\pi$ for the R-II-1a implementation. Figures taken from Ref.~\cite{Khater:2021wcx}.}
\label{Fig:bsgamma-impact}
\end{figure}

\end{itemize}

Let us consider the C-III-a implementation. After applying Cut~2, the allowed mass ranges shown in Figure~\ref{Fig:C-III-a-masses-th_constraints} are further reduced. The mass scatter plots that satisfy both Cut~1 and Cut~2 are presented in Figure~\ref{Fig:C-III-Cut2}. The most significant reduction in the allowed parameter space occurs in the charged sector. The $\bar{B} \to X(s)\gamma$ constraint introduces cuts in two regions of the charged Higgs masses: to the left and right of the allowed 3-$\sigma$ yellow region. For relatively light charged scalars $m_{H^\pm} \lesssim 300~\text{GeV}$, heavier states with $m_{h^\pm} > 600~\text{GeV}$ are allowed by the $\bar{B} \to X(s)\gamma$ constraint, but are excluded by the SM-like constraints and electroweak precision observables. The upper-right corner of the inert neutral sector is excluded due to the $ \bar{B} \to X(s)\gamma $ constraint. This is somewhat unexpected, as the constraint on the charged scalars would typically not limit the parameter space of the neutral scalar sector in the IDM. However, in the \mbox{C-III-a} implementation the parameters are highly constrained. For the heavy active neutral sector, we observe that Cut~2 results in an upper bound for $m_{H_3}$.

\begin{figure}[htb]
\begin{center}
\includegraphics[scale=0.25]{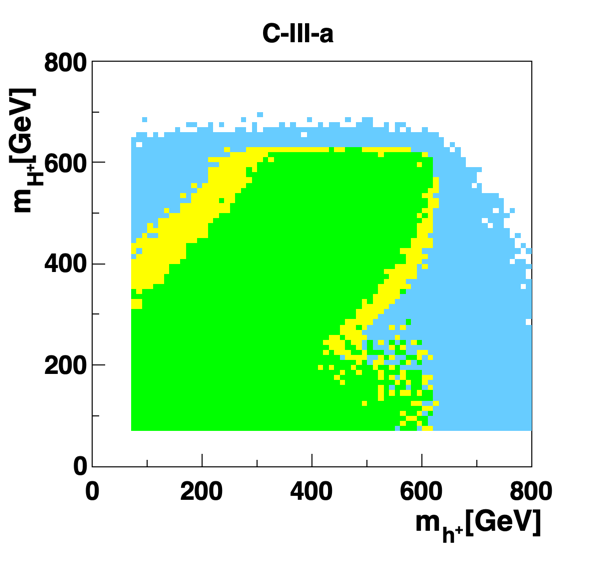}
\includegraphics[scale=0.25]{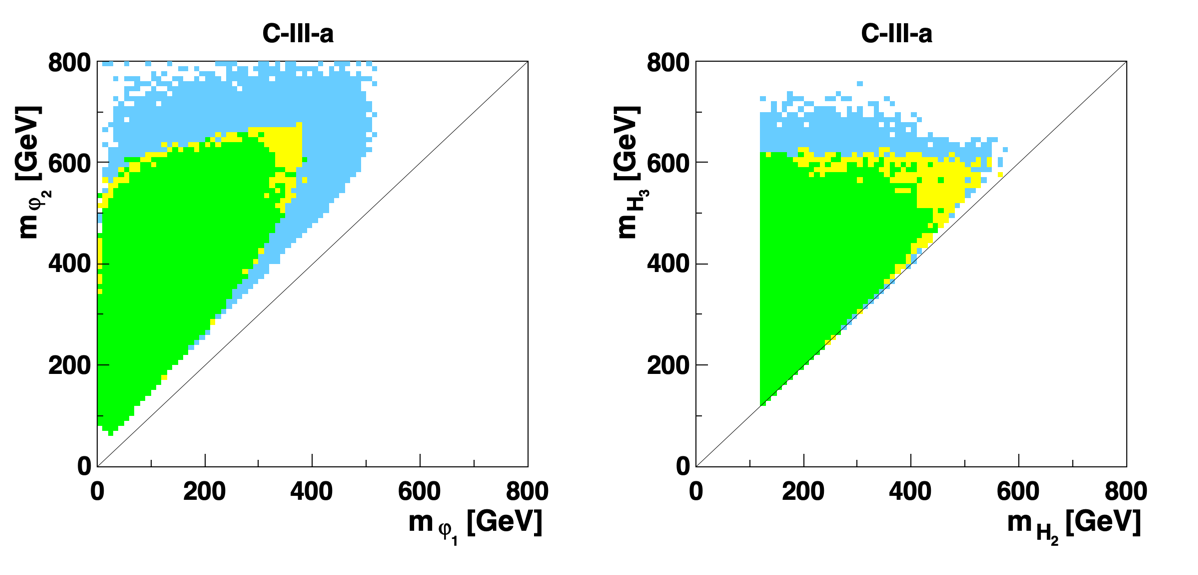}
\end{center}
\vspace*{-8mm}
\caption{Scatter plots of masses that satisfy Cut~1 and Cut~2 constraints. Left: the charged sector. Middle: the inert neutral sector. Right: the active heavy neutral sector. The sky-blue region satisfies Cut~1 constraint. The yellow region represents a 3-$\sigma$ tolerance with respect to Cut~2, while the lime region indicates where the model lies within the 2-\(\sigma\) bound of these values. Figures taken from Ref.~\cite{Kuncinas:2022whn}.}
\label{Fig:C-III-Cut2}
\end{figure}

\subsection{Cut 3 constraints}

We now move on to the Cut 3 constraints, which are based on both LHC data and astroparticle observations. In order to check the LHC search results, we implement some na\"ive checks within the spectrum generator and furthermore rely on $\mathsf{HiggsTools}$~\cite{Bahl:2022igd}, which utilises $\mathsf{HiggsBounds}$~\cite{Bechtle:2008jh,Bechtle:2011sb,Bechtle:2012lvg,Bechtle:2013wla,Bechtle:2015pma,Bechtle:2020pkv,Bahl:2021yhk} and $\mathsf{HiggsSignals}$~\cite{Bechtle:2013xfa,Stal:2013hwa,Bechtle:2014ewa,Bechtle:2020uwn}. For the evaluation of the DM related observables we utilise $\mathtt{micrOMEGAs~6.1.15}$~\cite{Belanger:2004yn,Belanger:2006is,Belanger:2008sj,Belanger:2010pz,Belanger:2013oya,Belanger:2014vza,Barducci:2016pcb,Belanger:2018ccd,Alguero:2023zol}.

Furthermore, we examine the Higgs self-interactions, which may become a key test in the future.

\vspace{10pt}
\textbf{LHC Higgs constraints}
\vspace{5pt}

First of all, we impose the condition that the full width of the SM-like Higgs particle, $h$ or $H_1$, must be within $\Gamma_{h_\mathrm{SM}}^\mathrm{exp} = 3.7^{+1.9}_{-1.4}$~MeV~\cite{ParticleDataGroup:2024cfk}. In the SM, the total width of the Higgs boson is approximately 4 MeV. The upper value, $\Gamma_{h_\mathrm{SM}}^\mathrm{exp} = 5.6~\text{MeV}$, is used in several initial checks, implemented within the spectrum generator. 

We shall focus on the $H_1$ SM-like Higgs boson of the C-III-a implementation since it acts as a CP-indefinite state. In addition to the SM-like limit involving fermions and gauge bosons, several channels are compared with the experimental results:
\begin{itemize}

\item \textbf{Decay \boldmath$\{h,\,H_1\} \to g g $ }

In the SM case, gluon fusion is the primary mechanism for the Higgs boson production. Due to (lack of) experimental constraints, we do not explicitly take this channel into account. At the leading order the rate is given by Refs.~\cite{Wilczek:1977zn,Georgi:1977gs,Ellis:1979jy,Rizzo:1979mf}:
\begin{equation}\label{Eq:H1_to_gg}
\Gamma\left(H_1 \rightarrow g g \right) =\frac{\alpha_S^2 m_{H_1}^3}{128 \pi^3 v^2}
\left( \left| \sum_f C_{\bar{f}fH_1}^S \mathcal{F}_{1/2}^S(\tau_f) \right|^2 + \left| \sum_f C_{\bar{f}fH_1}^P \mathcal{F}_{1/2}^P(\tau_f) \right|^2 \right),
\end{equation}
where $\mathcal{F}_{1/2}$ is the spin-dependent function:
\begin{subequations}
\begin{align}
\mathcal{F}_{1} &=2+3 \tau+3 \tau(2-\tau) f(\tau), \\
\mathcal{F}_{1 / 2}^i &= \left\{ \begin{aligned}
& -2 \tau[1+(1-\tau) f(\tau)], & i = S,\\
& -2 \tau f(\tau), & i = P,
\end{aligned}\right. \\
\mathcal{F}_{0} &=\tau[1-\tau f(\tau)],
\end{align}
\end{subequations}
where
\begin{equation}
\tau_i = \frac{4 m_i^2}{m_{H_1}^2},
\end{equation}
and 
\begin{equation}
f(\tau)=\left\{
\begin{aligned}
&\arcsin ^{2}\left(\frac{1}{\sqrt{\tau}}\right), & \tau \geq 1, \\
&-\frac{1}{4}\left[\ln \left(\frac{1+\sqrt{1-\tau}}{1-\sqrt{1-\tau}}\right)-i \pi\right]^{2}, & \tau<1.
\end{aligned}\right.
\end{equation}
The $C$'s of eq.~\eqref{Eq:H1_to_gg} represent couplings normalised to their respective SM values,
\begin{equation}
\begin{aligned}
\mathcal{L}_\text{int}^\prime  =\,&-\frac{ m_f}{v}  C_{\bar{f}fH_1}^S \bar{f}fH_1 -i\frac{ m_d}{v}  C_{\bar{d}d H_1}^P \bar{d} \gamma_5 dH_1 + i\frac{ m_u}{v}  C_{\bar{u}u H_1}^P \bar{u} \gamma_5 u H_1\\
&+  g m_W C_{W^+W^-H_1} W_\mu^+ W^{\mu -} H_1  - \frac{2   m_{\varphi^\pm_i}^2}{v} C_{ \varphi^+_i \varphi^-_i h}  \varphi^+_i \varphi^-_i H_1.
\end{aligned}
\end{equation}
In the case of R-II-1a, the above equations are simplified due to the fact that there is no CP violation. 

The decay width of this process can either increase or decrease compared to the SM case. This behavior is caused by an extra factor in the amplitude, $g_{\bar{f}fh}^\text{R-II-1a} = g_{\bar{f}fh}^\text{SM}\sin\alpha/\cos\beta$. In the case of C-III-a, the factor is more involved, see eq.~\eqref{Eq:CIIIa_Gamma_SFF_Ratio}. The two-gluon constraint does not play a crucial role in the performed scans. After applying all of the scans, the surviving data points are within the range of $[0.9,\,1.1]$, when normalised to the SM rates.

\item \textbf{Decay \boldmath$\{h,\,H_1\} \to \gamma \gamma$}

The di-photon partial decay width is altered by the contributions of charged-scalar loops, which are absent in the SM. The one-loop width has been previously determined in Refs.~\cite{Ellis:1975ap, Shifman:1979eb, Gunion:1989we}:
\begin{equation}\label{Eq:h_gammagamma}
\begin{aligned}
\Gamma\left(H_1 \rightarrow \gamma \gamma\right)&=\frac{\alpha^2 m_{H_1}^3}{256 \pi^3 v^2}\Bigg(\bigg|\sum_f  Q_f^2 N_c C_{\bar{f}fH_1}^S \mathcal{F}_{1 / 2}^S\left(\tau_{f}\right) + C_{W^+W^-H_1} \mathcal{F}_1\left(\tau_{W^\pm}\right) \\ 
&\hspace{35pt} + \sum_{\varphi^\pm_i}C_{ \varphi^+_i \varphi^-_i H_1} \mathcal{F}_0\left(\tau_{\varphi^\pm}\right) \bigg|^{2}  + \bigg|\sum_f Q_f^2 N_c C_{\bar{f}fH_1}^P \mathcal{F}_{1 / 2}^P\left(\tau_{f}\right)\bigg|^{2} \Bigg),
\end{aligned}
\end{equation}
where $Q_f$ is the electric charge of the fermion. The normalised coefficients $C$'s and the spin-dependent loop functions $\mathcal{F}$ were previously defined in the discussion on di-gluon decays.

Given the previous discussion on gluons, we do not attempt to precisely determine the two-gluon production factor; instead, we approximate the strength of the di-photon channel by:
\begin{equation}\label{Eq:Diphoton_strength}
\mu_{\gamma\gamma} \approx \frac{\Gamma\left( \{h,\,H_1\} \to \gamma \gamma \right)}{\Gamma^\text{exp}\left( h_{SM} \to \gamma \gamma \right)} \frac{\Gamma^\text{exp} \left( h_{SM} \right)}{\Gamma\left( \{h,\,H_1\} \right)},
\end{equation}
where $\mu_{\gamma \gamma} = 1.10 \pm 0.06$~\cite{ParticleDataGroup:2024cfk}. We evaluate this constraint while accounting for an additional ten percent computational uncertainty and enforcing a $(n=3)\text{-}\sigma$ tolerance,
\begin{equation}\label{Eq:Diphoton_Bounds}
\mu_{\gamma\gamma} = 1.11 \pm \sqrt{(1.11\times0.1)^2 + (0.1n)^2},
\end{equation}
which corresponds to the 3-$\sigma$ range of $[0.78;\,1.42]$. 

In both implementations, the di-photon branching ratio is enhanced for light charged scalars associated with the inert sector. As the mass of the inert charged inert scalar grows, the branching ratio decreases accordingly. Conversely, for heavier charged active sector scalars, the di-photon branching ratio increases.

\item \textbf{Decay \boldmath $\{h,\,H_1\} \to\mathrm{invisible}$}

The final colliders-related constraint we want to cover are the invisible decays~\cite{CMS:2022dwd,ATLAS:2022vkf,ATLAS:2023tkt} of the SM-like Higgs bosons $\{h,\,H_1\}$.  These can decay to lighter scalars, \textit{e.g.}. in C-III-a: $H_1 \to \varphi_i \varphi_j$, if such decays are kinematically allowed, $m_{\varphi_i} + m_{\varphi_j} \lesssim 125\text{ GeV}$,
\begin{equation}\label{Eq.DecayWidth_SSS}
\begin{aligned}
\Gamma\left( H_1 \to \varphi_i \varphi_j \right) ={}& \frac{2-\delta_{ij}}{32 \pi m_{H_1}^3} \left| g_{H_1 \varphi_i \varphi_j} \right|^2 \\
& \times \sqrt{\left[ m_{H_1}^2 - \left( m_{\varphi_i} + m_{\varphi_j} \right)^2 \right]\left[ m_{H_1}^2 - \left( m_{\varphi_i} - m_{\varphi_j} \right)^2 \right]},
\end{aligned}
\end{equation}
with a symmetry factor $(2-\delta_{ij})$.

These processes can significantly increase the total width of the SM-like Higgs state. Once again, consider the C-III-a implementation: if kinematically accessible, there are three possible decay channels $H_1 \to \{\varphi_1 \varphi_1,\,\varphi_2 \varphi_2,\,\varphi_1 \varphi_2 \}$. In the R-II-1a model, due to CP conservation, there are only two invisible decay channels. Knowing how light the inert neutral states can be simplifies the analysis considerably. In our calculations we adopt the PDG~\cite{ParticleDataGroup:2024cfk} suggested value of \mbox{$\text{Br}^\mathrm{exp}\left( h_{SM} \to \text{inv.} \right)<0.107$}. We note that depending on which processes one considers, there are more severe constraints on the invisible Higgs boson channel than those appearing in the PDG, set by ATLAS~\cite{ATLAS:2022yvh} and CMS~\cite{CMS:2022qva}. The experimental constraint applies directly if only a single (invisible) decay channel is kinematically available.

In a simplified approximation, a particle will escape a 30-meter detector, not accounting for the time dilation factor, if its decay width is $\Gamma^\text{tot} \lesssim 6.6 \times 10^{-18}$ GeV. Depending on the total width of the inert states two distinct scenarios emerge:
\begin{itemize}
\item If $\Gamma^\text{tot}(\varphi_j) \gtrsim 6.6 \times 10^{\,\text{-}18}$~GeV, particle will decay within the detector into two DM candidates and two (off-shell) gauge bosons, subsequently also decaying. In this case only the Higgs boson decay to two DM candidates channel will contribute to the invisible decay channel.
\item For $\Gamma^\text{tot}(\varphi_j) \lesssim 6.6 \times 10^{\,\text{-}18}$~GeV, the invisible branching ratio will be given by a sum of different components. For example, in the case of R-II-1a, the SM-like state $h$ can decay to both $\eta$ and $\chi$, yielding
\begin{equation}
\text{Br}\left( h \to \text{inv.} \right) = \frac{ \Gamma\left(h \to \eta \eta \right) +  \Gamma\left(h \to \chi \chi \right) }{\Gamma\left( h \right)}.
\end{equation}
\end{itemize}

\item \textbf{CP properties of C-III-a}

In the case of the C-III-a implementation, the SM-like Higgs boson interacts with fermions as a CP-indefinite state, see eqs.~\eqref{Eq:C-III-a-Yukawa}. We can compare the CP-indefinite couplings against those of the LHC bounds~\cite{ATLAS:2020ior,CMS:2021sdq}. In Figure~\ref{Fig:CP_Hff}, the CP-even and CP-odd couplings for C-III-a are presented. Current LHC data indicates a preference for positive CP-even Yukawa couplings ($\kappa_f$). The parameter space of C-III-a, after applying all constraints, fall well within the 2-$\sigma$ measurements.

\begin{figure}[htb]
\begin{center}
\includegraphics[scale=0.4]{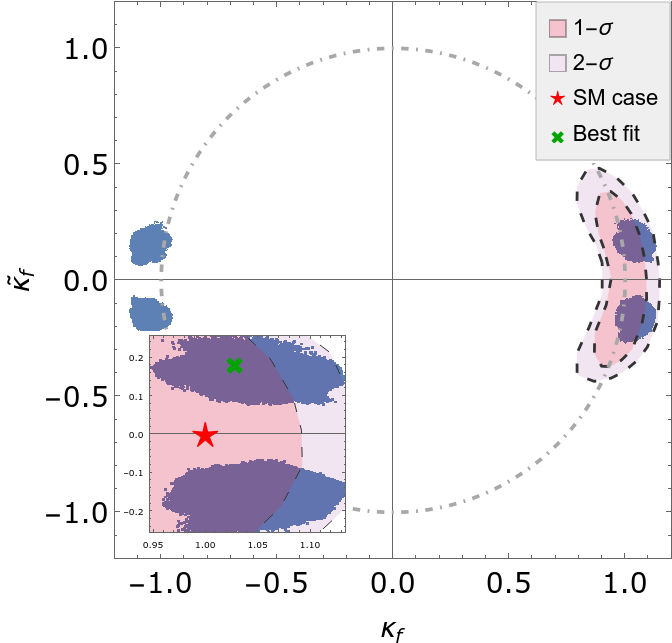}							  
\end{center}
\vspace*{-4mm}
\caption{Exploring the CP properties of the SM-like Higgs boson-fermion couplings in the C-III-a implementation. The neutral scalar-fermion interactions are parameterised in terms of $\mathcal{L} \supset-\frac{m_f}{v} \overline{\psi_f} \phi^0 \left(\kappa_f+i \gamma_5 \tilde{\kappa}_f\right) \psi_f$. The best fit of the LHC data~\cite{ATLAS:2020ior,CMS:2021sdq} is shown in the lower-left quadrant. Figure taken from Ref.~\cite{Kuncinas:2023hpf}.}
\label{Fig:CP_Hff}
\end{figure}

\end{itemize}

\vspace{10pt}
\textbf{Self interactions}
\vspace{5pt}

In the SM, the trilinear and quadrilinear self-interactions of the Higgs boson are~\cite{Boudjema:1995cb}
\begin{equation}
g(h^3_\mathrm{SM})=3m_{h_\mathrm{SM}} ^2/v,\quad
g(h_\mathrm{SM}^4)
= g(h_\mathrm{SM}^3)/v.
\end{equation}

First consider the R-II-1a case. The relevant couplings were provided in eqs.~\eqref{Eq.R_II_1a_hhh} and \eqref{Eq.R_II_1a_hhhh}. We expanded them in terms of the $\{m_h^2,\, m_H^2,\, m_\eta^2\}$ masses:
\begin{subequations}
\begin{align}
g(h^3)&= \frac{3 m_h^2}{v} \left[ \sin\left(\alpha+\beta\right) + \frac{2 \cos^2\left( \alpha + \beta \right)\cos\left(\alpha-\beta\right)}{\sin\left(2\beta\right)} \right] + \frac{2 m_\eta^2 \cos^3(\alpha + \beta)}{3 v \sin(2\beta)\cos^2(\beta)},\\
\begin{split}g(h^4)&= \frac{3 m_h^2}{v^2} \left[ \sin\left(\alpha+\beta\right) + \frac{2 \cos^2\left( \alpha + \beta \right)\cos\left(\alpha-\beta\right)}{\sin\left(2\beta\right)} \right]^2\\
&\hspace{15pt}+ \frac{3 m_H^2 \cos^2(\alpha + \beta) \sin^2(2\alpha)}{v^2 \sin^2(2\beta)}+ \frac{2 m_\eta^2 \cos^3(\alpha + \beta) \left[ 3 \cos(\alpha) \cot(\beta) + \sin(\alpha) \right]}{3 v^2 \sin(2\beta)\cos^3(\beta)}.\raisetag{4\normalbaselineskip}\end{split}
\end{align}
\end{subequations}
Imposing $\alpha+\beta=\pi/2$  reproduces the SM-like Higgs self-couplings.

For C-III-a these couplings were given by eqs.~\eqref{Eq.C_III_a_HiHjHk} and \eqref{Eq.C_III_a_HiHjHkHl}. Invoking the expressions for the $\lambda$'s and having expressed $m_{\varphi_2}^2$ according to eq.~\eqref{eq:m_phi2_vs_m_phi1} we find that:
\begin{equation}
g(H_1^3) = \frac{1}{v} \left[ m_{H_1}^2 B_{H_1} + m_{H_2}^2 B_{H_2} + m_{H_3}^2 B_{H_3} + m_{\varphi_1}^2 B_{\varphi_1}  \right],
\end{equation}
where $B_i$ represent coefficients that are expressed in terms of angles. For instance,
\begin{equation} \label{Eq:trilin-A_H1}
\begin{aligned}
B_{H_1} =& \frac{3}{8} \cos^3 \theta_2 \bigg\lbrace 2 \cos \theta_1 \left[  \cos(2 \theta_2) + 5\right] \\  & \hspace{50pt}- \frac{2 \cos^2 \theta_2 \sin(2\beta - 3 \theta_1 ) - \cos(2\beta) \sin \theta_1 \left[ \cos(2 \theta_2) - 7 \right] }{\sin \beta \cos \beta}\bigg\rbrace.
\end{aligned}
\end{equation}
The quartic self-interactions have a similar structure to the trilinear ones.

In the future, trilinear Higgs self-interactions could serve as a key test for BSM physics. For this purpose, we present in Figure~\ref{Fig:g_hhh_norm} the trilinear couplings normalised to the SM value.  Notice that in C-III-a these can be significantly larger, compared to the SM ones.

\begin{figure}[htb]
\begin{center}
\includegraphics[scale=0.3]{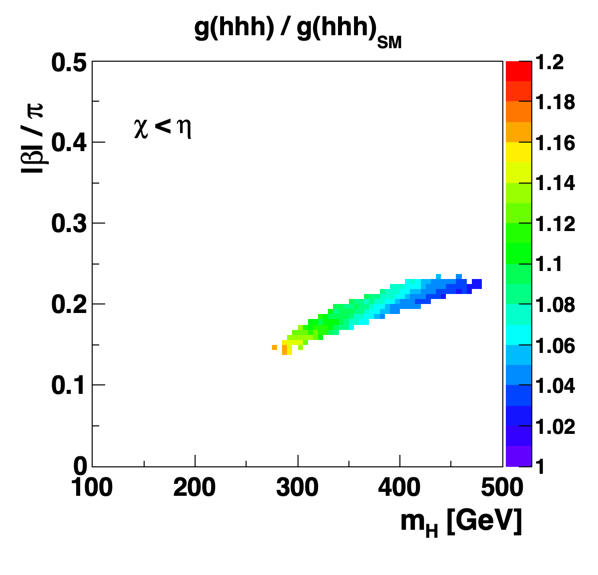}	
\includegraphics[scale=0.3]{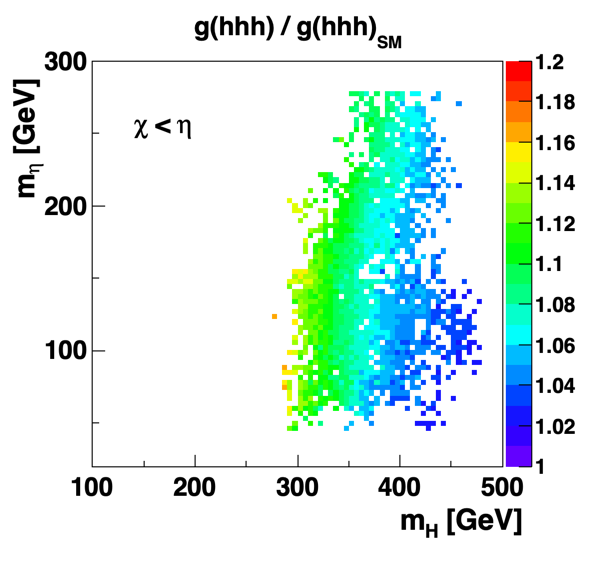}	
\includegraphics[scale=0.3]{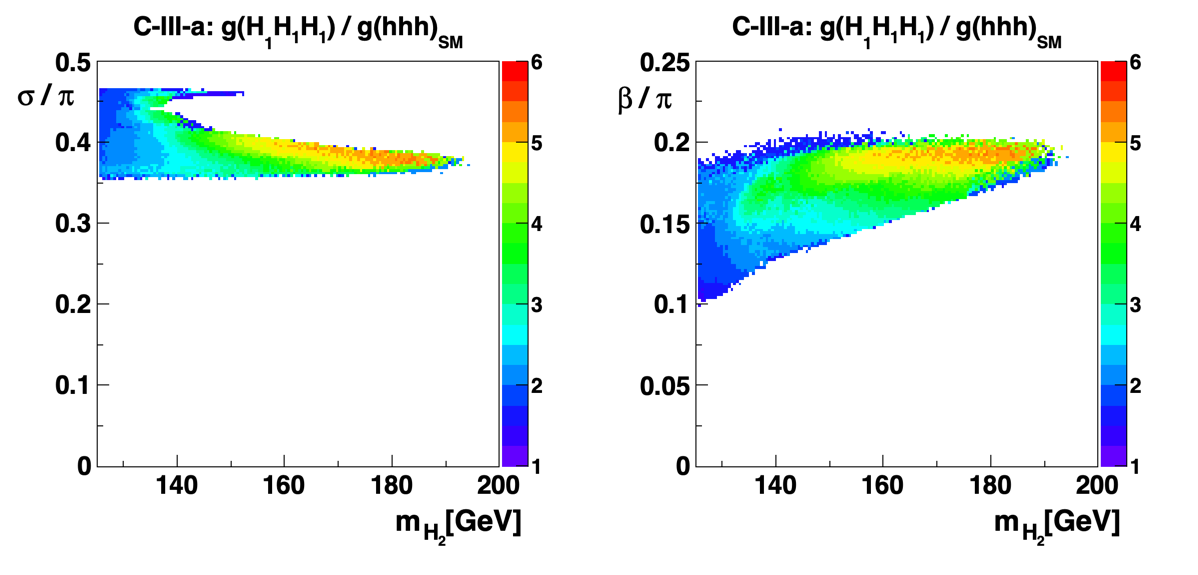}								  
\end{center}
\vspace*{-8mm}
\caption{Trilinear self-interactions of the SM-like Higgs boson normalised to the SM value, after applying Cut~3, for the (top) R-II-1a implementation and (bottom) C-III-a implementation. Maximum values of the normalised trilinear couplings are represented by the colour bar. Figures taken from Refs.~\cite{Khater:2021wcx,Kuncinas:2022whn}.}
\label{Fig:g_hhh_norm}
\end{figure}

\vspace{10pt}
\textbf{\textbf{Astroparticle checks}}
\vspace{5pt}

Finally, we discuss DM-related checks. We consider the $\Lambda_\mathrm{CDM}$ model with a freeze-out scenario. The CDM relic density along with the decay widths discussed above and other astroparticle observables are evaluated using $\mathsf{micrOMEGAs~6.1.15}$~\cite{Belanger:2004yn,Belanger:2006is,Belanger:2008sj,Belanger:2010pz,Belanger:2013oya,Belanger:2014vza,Barducci:2016pcb,Belanger:2018ccd,Alguero:2023zol}. The 't~Hooft-Feynman gauge is adopted, and switches are set to the $\mathsf{VZdecay=VWdecay=2}$ position, identifying that the three-body final states will be computed for annihilation an coannihilation processes. Apart from that, we employ the $\mathsf{fast=-1}$ switch in the boundary regions, indicating that a highly accurate calculation is performed. The steering $\mathsf{CalcHEP}$~\cite{Belyaev:2012qa} model files were produced manually.

The performed DM related checks are:
\begin{itemize}

\item \textbf{Relic density}

We adopt the CDM relic density value of $0.1200 \pm 0.0012$~\cite{Planck:2018vyg,ParticleDataGroup:2024cfk}. We evaluate the relic density parameter while accounting for an additional ten percent computational uncertainty and enforcing a $(n=3)\text{-}\sigma$ tolerance,
\begin{equation} \label{Eq:Omega-value}
\begin{aligned}
\Omega h^2 &= 0.1200 \pm \sqrt{ \left( 0.1200 \times 0.1\right)^2  + (0.0012\,n)^2}.
\end{aligned}
\end{equation}
This corresponds to the $[0.1075;\,0.1325]$ region.

Let us revisit the results of Figure~\ref{Fig:mass-ranges}. In the IDM, two regions that are consistent with the CDM relic density were found: the intermediate-mass region and the high-mass region. The high-mass region results in correct predictions for the relic density due to two main reasons:
\begin{itemize}
\item Near-mass degeneracy between the scalars of the inert sector. Small mass differences result in weak (not as in EW) couplings, causing the contributions from different inert scalars to the annihilation processes to be suppressed, thereby yielding an acceptable relic density.
\item The ability to select the Higgs boson portal coupling $\lambda_L$ freely. This parameter governs the trilinear $XXh$ and quartic $XXhh$ couplings and must remain sufficiently small for masses of order 500 GeV, but the absolute value should increase with the inert sector masses. In this context, $X$ represents the scalars from the inert sector.
\end{itemize}
In this region, the main DM annihilation mechanism is into the charged gauge bosons, which is controlled by the gauge couplings. The relic density can be kept within acceptable limits by suppressing annihilations through an intermediate $h$ boson or into a pair of $h$ bosons, as shown in Figure~\ref{Fig:PortalHMR}. The necessary relic abundance is obtained by carefully adjusting the mass splittings.

\begin{figure}[h]
\begin{center}
\includegraphics[scale=0.6]{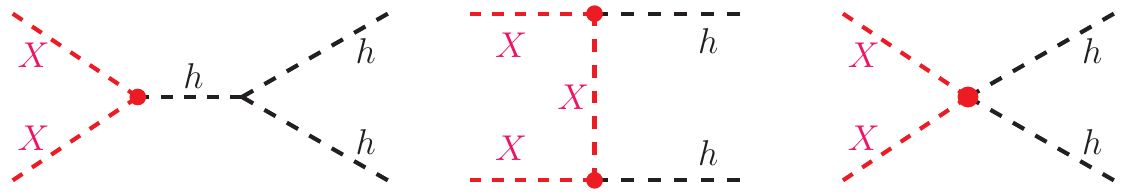}
\end{center}
\vspace*{-4mm}
\caption{Contribution to $X$ annihilation channels at high DM masses.}
\label{Fig:PortalHMR}
\end{figure}

Whereas the IDM and the 3HDMs share many similarities, it is not always possible to simultaneously tune the mass-splitting between the inert sector and the portal coupling in the 3HDMs. The portal couplings are crucial for the phenomenology of the Early Universe. In the 3HDMs, in general, there is not a single portal coupling, but rather two independent ones, $XXh$ and $XXhh$, and additional ones involving other physical scalars. Based on the symmetry and vacuum configuration, the portal couplings might be related among themselves and/or given by the scalar masses. This is exactly what happens in both R-II-1a and C-III-a implementations.

Let us examine a simplified scenario involving heavy inert scalars in the R-II-1a case. Suppose the SM-like scenario with a small perturbation to the first order in $\delta$, $\alpha=\pi/2-\beta+\delta$. In the limit of $\delta\to0$, the portal couplings become:
\begin{equation}  \label{Eq:portal-vs-mX_sq}
g(X X h)/v=g(X X h h) =\frac{1}{v^2}\left[ m_h^2 + 2 m_X^2 \right].
\end{equation}
A striking feature is that the value of the portal couplings increases as the DM candidate $X$ becomes heavy. Here, we assume that $X=\{\eta,\, \chi\}$. The values of the portal couplings are displayed in Figure~\ref{Fig:quartic-coupling}. As illustrated in this figure, the correlation with DM mass, see eq.~(\ref{Eq:portal-vs-mX_sq}), is qualitatively supported by the parameter points that pass Cut~2. In the full R-II-1a implementation, additional annihilation channels involving active scalars become available. For instance, the heavy active scalar $H$ scalar contributes to the $X,X \to h,h$ process via the $s$-channel. Additionally, there are other accessible channels involving both active and inert scalars. These large portal couplings lead to a rapid annihilation of DM in the Early Universe, effectively excluding the possibility of achieving the experimentally observed DM relic density for heavy DM candidates.

\begin{figure}[htb]
\begin{center}
\includegraphics[scale=0.30]{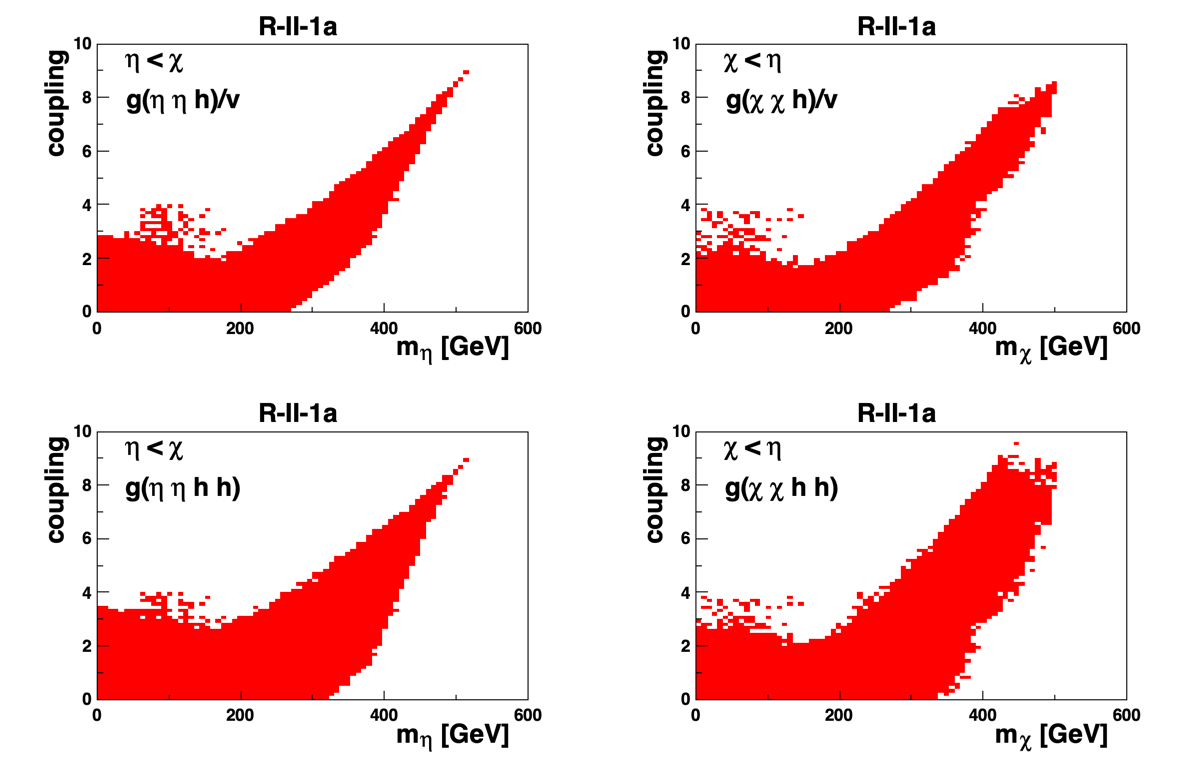}
\end{center}
\vspace*{-4mm}
\caption{The case of R-II-1a. The absolute value of the trilinear portal coupling (shown at the top) and the quartic portal coupling (shown at the bottom) as functions of the mass of the DM candidate. Figure taken from Ref.~\cite{Khater:2021wcx}.}
\label{Fig:quartic-coupling}
\end{figure}

What about the C-III-a implementation? The previous discussion still applies. Apart from that, in Section~\ref{Sec:CIIIa_M2_deriv} we observed that achieving mass-degeneracy is not possible. A mass gap is always present. For heavy states with $m_{\varphi_1} \geq 300~\text{GeV}$, the application of the Cut~1 constraint results in the formation of a mass gap, where $m_{\varphi_2} \approx m_{\varphi_1} + 70~\text{GeV}$.

The relic density as a function of the DM mass can be inspected in Figure~\ref{Fig:Omega}.

\begin{figure}[h]
\begin{center}
\includegraphics[scale=0.25]{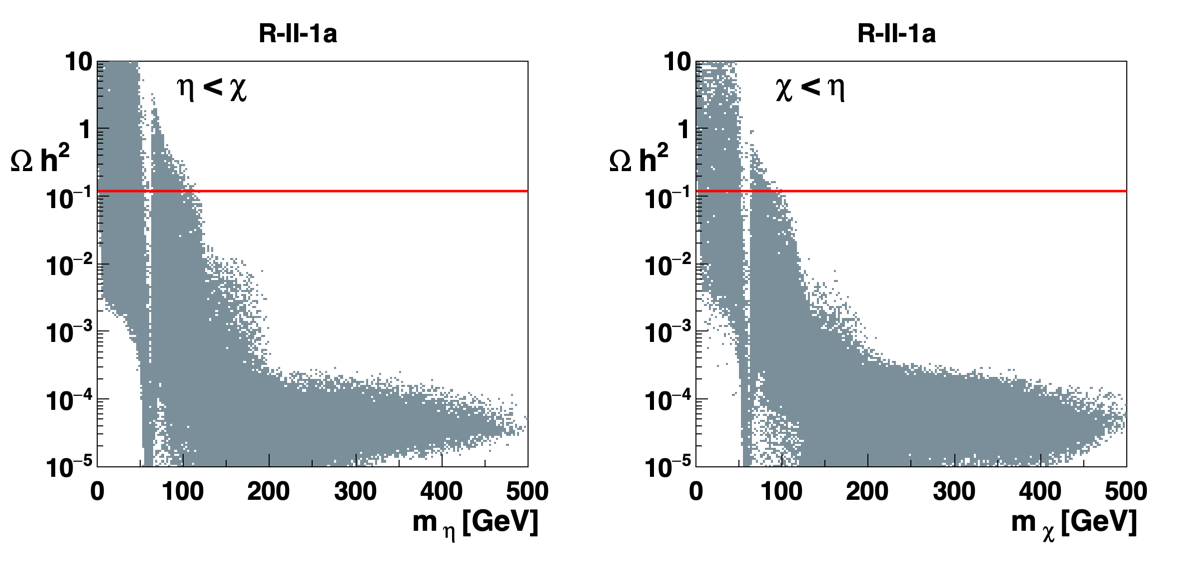}
\includegraphics[scale=0.25]{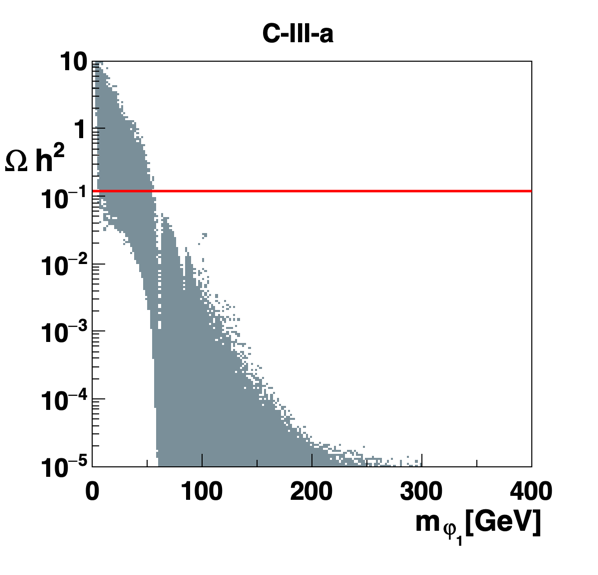}
\end{center}
\vspace*{-8mm}
\caption{Dark matter relic density for the R-II-1a model (first two panels) and the C-III-a model (last panel) after satisfying Cut~2 constraints. In the high-mass region the DM relic density is shown to be way too low. The red line indicates the observed relic density according to the Planck collaboration~\cite{Planck:2018vyg}. Figures taken from Refs.~\cite{Khater:2021wcx,Kuncinas:2022whn}.}
\label{Fig:Omega}
\end{figure}

\item \textbf{Direct DM detection}

For the direct DM searches we rely on the results of XENONnT~\cite{XENON:2020kmp,XENON:2022ltv,XENON:2023cxc} and LUX-ZEPLIN~\cite{LZ:2021xov,LZ:2022lsv,LZ:2024zvo}. It both implementations it is possible to fine-tune parameters so that the direct DM detection would be low. For the results consider Figure~\ref{Fig:S3_DD} adopted from Ref.~\cite{Kuncinas:2023hpf} and re-evaluated in light of the new report from LUX-ZEPELIN~\cite{LZ:2024zvo}. Both implementations allow for extremely small values of the DM-nucleon cross-section.

\begin{figure}[h]
\begin{center}
\includegraphics[scale=0.29]{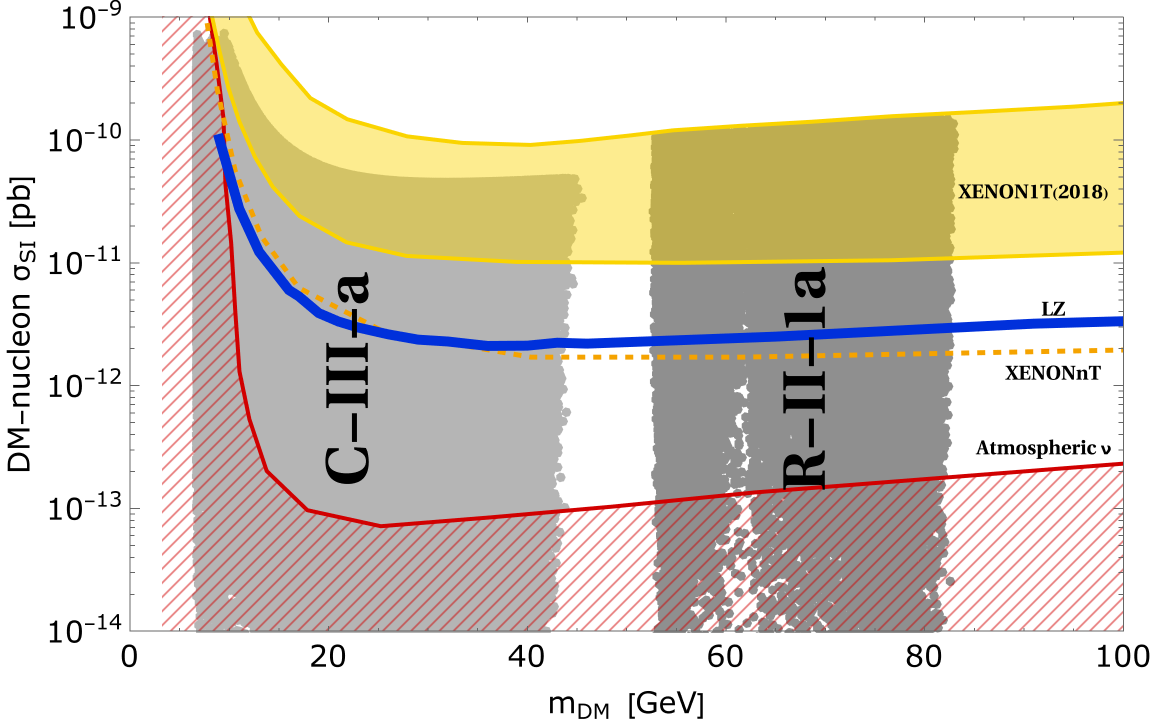}
\end{center}
\vspace*{-8mm}
\caption{The spin-independent DM-nucleon cross-section compatible with XENON1T data at 90\% C.L. (yellow band). The blue line represents the LUX-ZEPLIN~\cite{LZ:2024zvo} non-observation of the DM signal. The dashed line represents the future sensitivity of XENONnT~\cite{XENON:2020kmp}. The points (light grey for C-III-a and grey for R-II-1a) represent cases that satisfy Cut~3. The red line indicates the neutrino floor. }
\label{Fig:S3_DD}
\end{figure}

An interesting feature of C-III-a is that it satisfies the direct DM detection criteria in the regions \mbox{$m_{\varphi_1} \in [6;\, 360]~\text{GeV}$} and at $m_{\varphi_1} \approx 1~\text{GeV}$. This can be attributed to:
\begin{itemize}
\item Interference between different portal $\varphi_1 \varphi_1 H_i$ couplings;
\item $h_i$ couples to fermion as CP-indefinite state.
\end{itemize}
The relative importance of these effects is influenced by the input parameters.

\item \textbf{Indirect DM detection}

The light DM mass region is strongly constrained by the indirect DM detection experiments~\cite{Hess:2021cdp}, ruling out the canonical cross-section of $10^{-26} \text{cm$^3$/s}$~\cite{Fermi-LAT:2015att}. It is not straightforward to apply the experimental DM annihilation bounds to the studied implementations. We note that the two implementations are analysed at tree level.

In the R-II-1a implementation, for light DM candidates, two dominant DM annihilation channels are into a $b \bar{b}$ pair or a $W^+ W^-$ pair. In both these channels, the dominant branching ratio can be as low as 0.38. In the C-III-a implementation, for light DM candidates, the dominant DM annihilation channel is into a $b \bar{b}$ pair, with a branching ratio $\text{Br} \geq 0.8$.  The C-III-a model would be excluded if the Navarro–Frenk–White (NFW) DM distribution profile~\cite{Navarro:1995iw,Navarro:1996gj} of $\rho = 0.3 \, \text{GeV/cm}^3$ was considered; though these constraints are not definitively confirmed. Due to different limitations, we allow the DM self-annihilation cross-section to vary within a magnitude of order one around the experimental bounds. The results are presented in Figure~\ref{Fig:InDD}.

\begin{figure}[h]
\begin{center}
\includegraphics[scale=0.34]{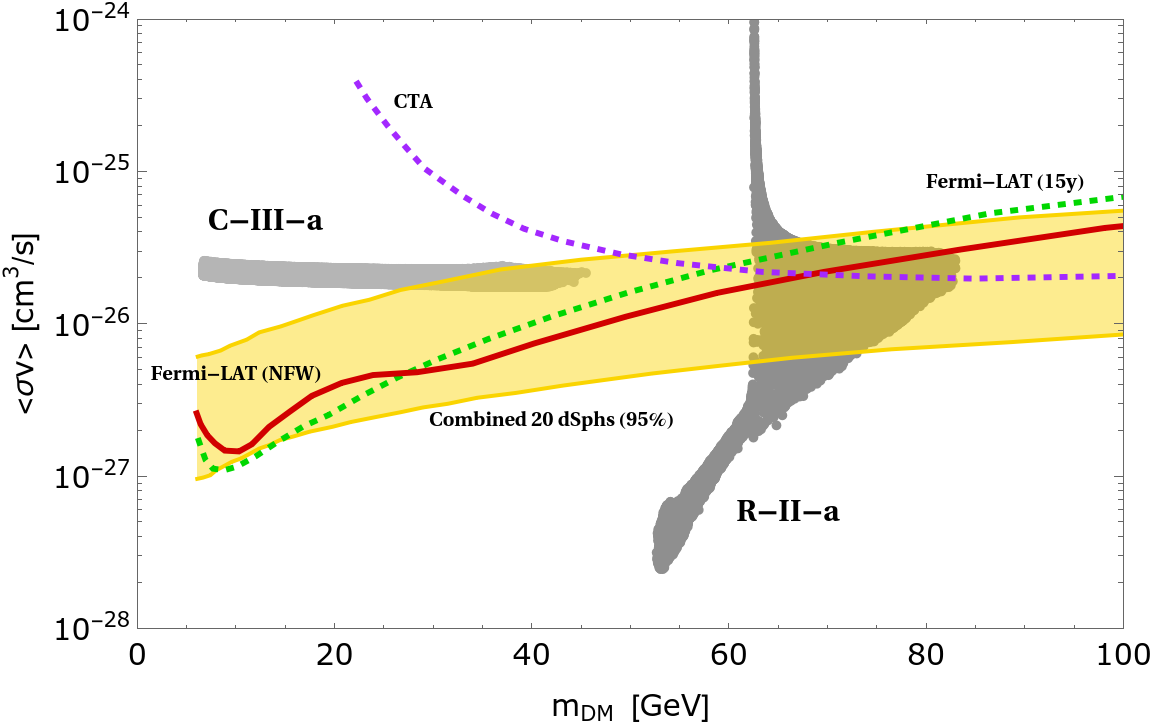}
\end{center}
\vspace*{-4mm}
\caption{
The DM self-annihilation cross-section is plotted as a function of the DM mass. The yellow band indicates the bounds at 90\% C.L. that are consistent with the observation of 20 dwarf spheroidal galaxies (dSphs). The red line represents the Fermi-LAT limits, assuming the NFW profile with a density of $\rho = 0.3\text{ GeV/cm}^3$. The dashed lines show the expected sensitivities for the future, with Fermi-LAT (green) and the Cherenkov Telescope Array (CTA) (purple). The light grey points correspond to C-III-a, while the grey points correspond to R-II-1a, both of which pass all other checks. Figure taken from Ref.~\cite{Kuncinas:2023hpf}.}
\label{Fig:InDD}
\end{figure}

\end{itemize}

\section{Discussing two implementations}

A sample size of $10^6$ points satisfying all constraints was considered sufficient in Refs.~\cite{Khater:2021wcx,Kuncinas:2022whn,Kuncinas:2023hpf}. We re-evaluate data from Ref.~\cite{Kuncinas:2023hpf} by incorporating the latest constraints; we do not attempt to regenerate the parameter space to reach the $10^6$ points threshold.

\vspace{10pt}
\textbf{R-II-1a}
\vspace{5pt}

Within the model, two potential DM candidates exist: $\eta$ and $\chi$. 

For the scenario where the $\eta$ scalar is the lightest, see Figure~\ref{Fig:Omega}, it looks as if there could be viable solutions in the mass range $m_\eta \in [2;\,120]~\text{GeV}$. However, in the lower part of this interval, constraints from the Higgs invisible branching ratio and the relic DM density render this region incompatible with the experimental data.  

Different mass domains are permitted when considering the three key constraints individually: the relic density, direct DM detection, LHC data. The viable region would be where all these domains intersect, which corresponds to satisfying Cut~3. Figure~\ref{Fig:Cut3_constraints} illustrates the allowed mass regions when imposing two constraints at a time. However, the overlap of these regions does not necessarily indicate viable solutions, as there are seven additional parameters influencing the analysis: masses, diagonalisation angles. In fact, after applying all Cut~3 constraints simultaneously, no viable parameter space remains.

\begin{figure}[h]
\begin{center}
\hspace*{-10pt}\includegraphics[scale=0.37]{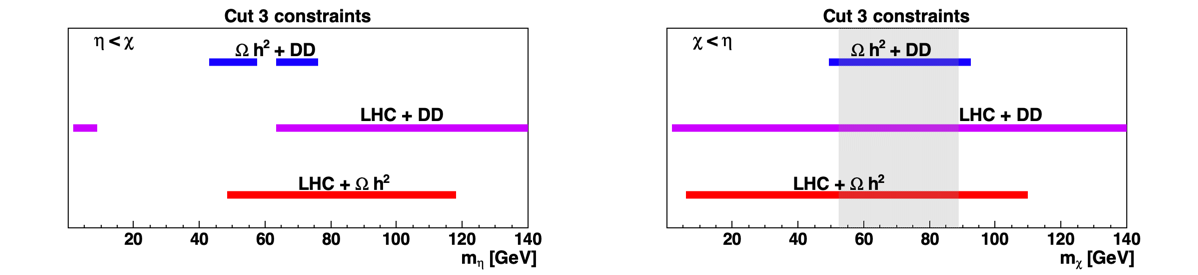}																
\end{center}
\vspace*{-4mm}
\caption{Allowed mass regions of the R-II-1a DM candidate. Blue: relic density and direct DM detection. Purple: LHC Higgs constraints and  direct DM detection. Red: relic density and LHC Higgs constraints. Grey: all of the Cut~3 constraints are met. It is important to note that additional input parameters, which further refine the viable parameter space, are not displayed.  Figures taken from Ref.~\cite{Khater:2021wcx}.}
\label{Fig:Cut3_constraints}
\end{figure}

For the scenario where the $\chi$ scalar is the lightest, the $\Omega h^2$ distribution shifts slightly towards lower relic density values, as illustrated in Figure~\ref{Fig:Omega}. Consequently, the mass range compatible with the relic density constraint is $m_\chi \in [2,\,105]~\text{GeV}$.  In contrast, for the $\eta$ case with masses below 40 GeV, imposing LHC-related constraints results in all $\Omega h^2$ values exceeding 0.22. This limitation does not affect the $\chi$ case, where $\Omega h^2$ can be as low as \(\approx 0.07\). However, despite this advantage, the sub-40 GeV region remains incompatible with Cut~3.  In the high-mass DM region the R-II-1a model does not permit for solutions compatible with the observed relic density, since it saturates at $\Omega h^2 \sim \mathcal{O}(10^{-4})$.

In Ref.~\cite{Kuncinas:2023hpf} it was observed that after applying Cut~3 the allowed DM mass region of the R-II-1a implementation remained nearly unchanged compared to  Ref.~\cite{Khater:2021wcx}, narrowing slightly from $m_\mathrm{DM} \in [52.5;\,89.2]~\text{GeV}$ to $m_\mathrm{DM} \in [53.1;\,83.8]~\text{GeV}$. However, upon re-evaluating the data with updated constraints, we find a significantly more constrained range of $m_\mathrm{DM} \in [52.7;\,82.9]~\text{GeV}$.

The parameter space surviving all of the constraints is displayed in Figure~\ref{Fig:Cut4}.

\begin{figure}[htb]
\begin{center}
\includegraphics[scale=0.23]{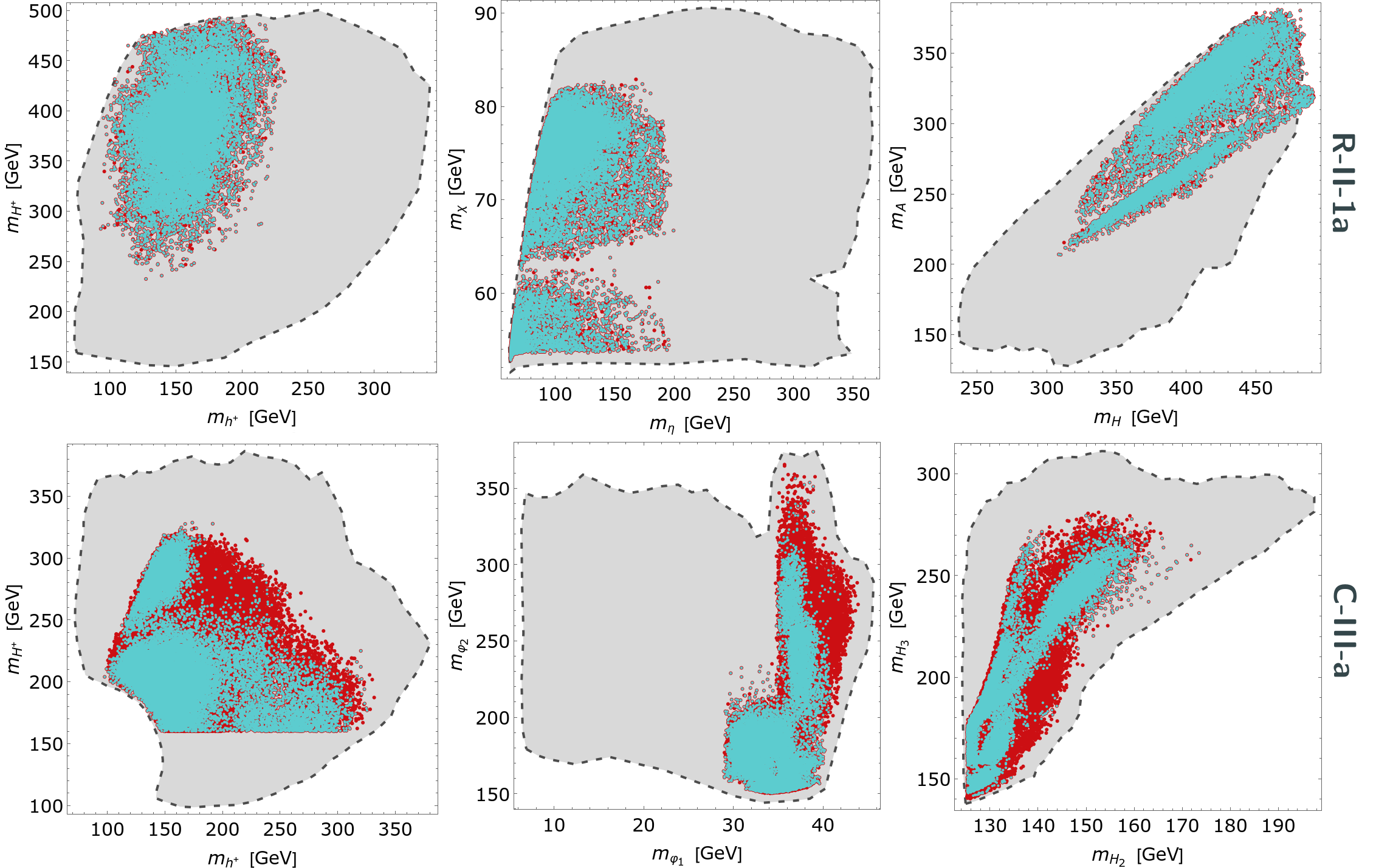}
\end{center}
\vspace*{-8mm}
\caption{Scatter plots showing mass distributions that comply with constraints. Left column: charged sector. Middle column: inert neutral sector. Right column: active neutral sector. The grey region satisfies Cut~3 of Refs.~\cite{Khater:2021wcx,Kuncinas:2022whn}. The red region relies on $\mathsf{HiggsTools}$ and indirect DM constraints of Ref.~\cite{Kuncinas:2023hpf}. The cyan region is the  re-evaluated region based on the PDG~\cite{ParticleDataGroup:2024cfk} values along with constraints from LUX-ZEPELIN~\cite{LZ:2024zvo}.}
\label{Fig:Cut4}
\end{figure}

\vspace{10pt}
\textbf{C-III-a}
\vspace{5pt}

The case of the C-III-a presents a more intriguing scenario as it permits relatively light DM candidates. When the DM candidate is sufficiently light, specifically $m_{\varphi_1} \leq m_{H_1}/2$, the decay of the SM-like Higgs boson into the inert sector, through the channel $H_1 \to \varphi_1 \varphi_1 $ becomes significant. The substantial branching ratio of this decay mode has a significant effect on the total width of the SM-like Higgs boson, which is further constrained by Cut~3.  One might initially expect that in the sub-$(m_{H_1}/2)$ region, the most stringent constraint would arise from the invisible decay width of the SM-like Higgs boson, necessitating fine-tuning of the portal couplings. Actually, both the relic density and direct DM detection constraints impose even stricter limitations in this mass range.

The charged scalar masses are not significantly restricted by the Cuts. However, there are notable restrictions on the masses of the neutral inert sector. The $\Omega h^2$ constraint imposes an upper limit of $m_{\varphi_1} < 55~\text{GeV}$ and a lower limit of $m_{\varphi_1} > 6~\text{GeV}$. Additionally, $\Omega h^2$ checks also permit $m_{\varphi_1} \approx 1~\text{GeV}$.  From the LHC checks, states lighter than \mbox{$m_{\varphi_2} \approx 110~\text{GeV}$} are restricted. These constraints are sensitive to the total width of the Higgs boson. A mass gap of $m_{\varphi_2} \approx m_{\varphi_1} + 110~\text{GeV}$ develops for $m_{\varphi_1} < m_{H_1}/2$. Furthermore, the allowed masses of the neutral active sector are pushed away from the mass degenerate limit by both the LHC and $\Omega h^2$ constraints. This results in the condition \( m_{H_3} > m_{H_2} + 20~\text{GeV} \).

The other scalars residing in the inert doublet predominantly decay into $\varphi_1$) and on-shell gauge bosons. However, due to constraints imposed by Cut~3, lower bounds are introduced on the masses of both these scalars, effectively ruling out co-annihilations into gauge bosons or the production of off-shell gauge bosons. 

The heavier non-inert neutral states primarily decay into the DM candidate. This decay mode is especially dominant for the $H_2$ scalar, where $\mathrm{Br}(H_2 \to \varphi_1 \varphi_1)$ may exceed 0.99. Additionally, the $H^+$ state predominantly decays into the inert scalars, specifically $H^+ \to h^+ \varphi_1$, alongside the more familiar decays from the 2HDM, such as $H^+ \to \{ t\bar{b}, \nu\bar{\ell} \}$.

To summarise, the dominant decay channel for all of the scalars, except for those of the SM-like Higgs boson, involve states that include at least one DM candidate. These decays would be accompanied by large missing transverse momentum in the detector. This feature is generally true for the inert scalars and the neutral components, but is only partially true for the $H^\pm$ decays, depending on the specific parameters. It would be valuable to further refine the available parameter space of the charged state, particularly focusing on the $m_{H^\pm}{-}\beta$ plane. Various experimental studies~\cite{Arbey:2017gmh, ATLAS:2018gfm, CMS:2019bfg, CMS:2020imj, ATLAS:2021upq} could help provide more restrictive bounds. 

If the C-III-a implementation was realised in Nature, detecting it would be quite challenging, since most of the scalars decay to states with at least a single DM candidate, leading to decays accompanied by large missing transverse momentum in the detector. Moreover, there seems to be limited potential for observing a signal through direct DM detection, making it exceedingly difficult to detect any direct interactions between DM and nucleons under the current experimental capabilities. Thus, new experimental strategies or more sensitive detectors would be required to probe this model more effectively.

In Ref.~\cite{Kuncinas:2022whn}, the viable DM mass range was identified as $m_\mathrm{DM} \in [6.5;\,44.5]~\text{GeV}$. As discussed in Ref.~\cite{Kuncinas:2023hpf}, collider searches had a limited impact on constraining the parameter space. Yet, the experimental bounds from the indirect DM detection could completely rule out the C-III-a case entirely, depending on the assumed DM halo distribution profiles. Applying more conservative constraints reduces the allowed DM mass range to  \mbox{$m_\mathrm{DM} \in [28.9;\,44.3]~\text{GeV}$}. Upon re-evaluating the data from Ref.~\cite{Kuncinas:2022whn} we find a slightly reduced range, \mbox{$m_\mathrm{DM} \in [28.9;\,41.9]~\text{GeV}$}. Notably, only about 30 per cent of the dataset survives, predominantly due to the updated constraints from LUX-ZEPELIN~\cite{LZ:2024zvo}. The parameter space surviving all of the constraints is displayed in Figure~\ref{Fig:Cut4}.

\chapter{Continuous symmetries of the three-Higgs-doublet models}\label{Ch:U1_3HDM}

There are many possibilities to implement DM within 3HDMs. Over the past few years, substantial progress has been made in understanding the symmetries that can be realised in the 3HDMs~\cite{Ferreira:2008zy,Ivanov:2011ae,Ivanov:2012ry,Ivanov:2012fp,Keus:2013hya,Ivanov:2014doa,Pilaftsis:2016erj,deMedeirosVarzielas:2019rrp,Darvishi:2019dbh,Bree:2024edl,Kuncinas:2024zjq,Doring:2024kdg}, some of which could accommodate DM. The NHDMs with continuous symmetries have not been explored extensively. In Ref.~\cite{Kuncinas:2024zjq} our goal was to classify all possible embeddings of a $U(1)$ symmetry in 3HDMs, presenting and comparing distinct possibilities, in contrast to the common stabilisation of DM in terms of discrete symmetries~\cite{Ivanov:2012hc,Jurciukonis:2022oru}. We examined these symmetries and investigated whether they could potentially accommodate DM candidates, which shall be discussed in this chapter. We will demonstrate that in the neutral sector, mass degeneracies always occur when a $U(1)$ symmetry remains unbroken, similar to what happens in the 2HDM, see Section~\ref{Sec:IDM}. Depending on the specific $U(1)$ symmetry that remains unbroken, there may be one or even two pairs of mass-degenerate neutral states. When CP symmetry is conserved, these degenerate pairs will have opposite CP parities---one even and the other odd. We recall that the mass-degenerate states are problematic for the direct DM searches~\cite{Barbieri:2006dq,Hambye:2009pw,Arina:2009um,Escudero:2016gzx}.

Some of the 3HDMs under consideration lead to multi-component DM models, which have been widely explored in various physical scenarios~\cite{Boehm:2003ha,Ma:2006uv,Hur:2007ur,Cao:2007fy,Zurek:2008qg,Profumo:2009tb,Batell:2010bp,Liu:2011aa,Belanger:2011ww,Belanger:2012vp, Medvedev:2013vsa,Esch:2014jpa,Biswas:2015sva,Cai:2015zza,Arcadi:2016kmk,Ahmed:2017dbb,Bernal:2018aon,Poulin:2018kap,YaserAyazi:2018lrv,Bhattacharya:2018cgx,Elahi:2019jeo,Borah:2019aeq,Nanda:2019nqy,Hall:2019rld,Betancur:2020fdl,DuttaBanik:2020jrj,Chakrabarty:2021kmr,Choi:2021yps,DiazSaez:2021pmg,Hall:2021zsk,Mohamadnejad:2021tke,Yaguna:2021rds,Ho:2022erb,Das:2022oyx,BasiBeneito:2022qxd}, offering a more complex and nuanced approach to understanding DM. Models with additional $SU(2)$ scalar singlets~\cite{Silveira:1985rk,McDonald:1993ex,Burgess:2000yq,Barger:2007im,Andreas:2008xy,GAMBIT:2017gge} have also shown great potential in addressing the DM puzzle and share many similarities with our framework. The singlet extended models can accommodate two- and three-component scalar singlet DM candidates~\cite{Barger:2008jx,Drozd:2011aa,Modak:2013jya,Belanger:2014bga,Bhattacharya:2016ysw,Bhattacharya:2017fid,Pandey:2017quk,Belanger:2020hyh,Coito:2021fgo,Yaguna:2021vhb,Belanger:2021lwd,Belanger:2022esk}, offering a broader range of possibilities for explaining DM interactions and properties.

Finite subgroups of $SU(3)$ are widely used in particle physics, including, but not limited to, model building, study of flavour symmetries~\cite{Luhn:2007yr,Zwicky:2009vt,Altarelli:2010gt,Albright:2010ap,Parattu:2010cy,Grimus:2011fk,deAdelhartToorop:2011re,Ferreira:2012ri,Feruglio:2012cw,King:2013eh,King:2014nza,Ludl:2014axa,King:2017guk,Kobayashi:2018vbk,Novichkov:2018ovf,Novichkov:2018nkm,Feruglio:2019ybq,Xing:2020ijf}, stabilisation of DM. The classification of these finite subgroups began some decades ago~\cite{Miller:1916,Fairbairn:1964sga,Bovier:1980ga,Bovier:1980gc,Fairbairn:1982jx}, and more recently continued in Refs.~\cite{Luhn:2007uq,Escobar:2008vc,Ishimori:2010au}. Generators of finite groups up to order 2000  are known~\cite{Grimus:2010ak,Ludl:2010bj,Ludl:2011gn,Jurciukonis:2017mjp}, where the $\mathsf{SmallGrp}$ library~\cite{SmallGrp} of $\mathsf{GAP}$ was utilised. While being not the most elegant approach, one could try to brute-force different generators, and their products to identify symmetries of 3HDMs.

In this chapter we outline continuous symmetries and their implementations. We do not consider custodial symmetries. For a more detailed review one can consider Ref.~\cite{Kuncinas:2024zjq}.

\section{Philosophy}\label{Sec:Philosophy}

Our primary objective is to identify potential DM candidates by examining different vacua, ensuring that at least one vev vanishes. For a comprehensive coverage, we consider all possible implementations, including scenarios that allow for SSB of continuous symmetries, though, our main focus will be on vacua that preserve the underlying symmetry.

The symmetry-breaking patterns for the discussed $U(1)$-3HDMs are illustrated in Figure~\ref{Fig:Symmetry_breaking} in terms of maximal symmetries. All $U(1)$-symmetric potentials can be derived by imposing additional constraints on a $\mathbb{Z}_2$-symmetric or a $\mathbb{Z}_3$-symmetric 3HDM. Both of these symmetries can lead to a $U(1)_1$-symmetric potential, while only $\mathbb{Z}_2$ and not $\mathbb{Z}_3$ can also give rise to a $U(1)_2$-symmetric potential. By imposing further constraints, the $U(1)_2$-symmetric potential can become symmetric under a bigger $U(1) \times \mathbb{Z}_2$ group. Both branches ultimately lead to a $U(1) \times U(1)$-symmetric potential, which represents the most general real 3HDM. While discrete symmetries can appear in the symmetry breaking patterns of Figure~\ref{Fig:Symmetry_breaking}, we focus on the relevant branches that only include continuous symmetries. For instance, both $\mathbb{Z}_2 \times \mathbb{Z}_2$ and $\mathbb{Z}_4$ symmetries can be extended to $U(1) \times \mathbb{Z}_2$.

\vspace{3pt}\begin{figure}[htb]
\begin{center}
\includegraphics[scale=1.16]{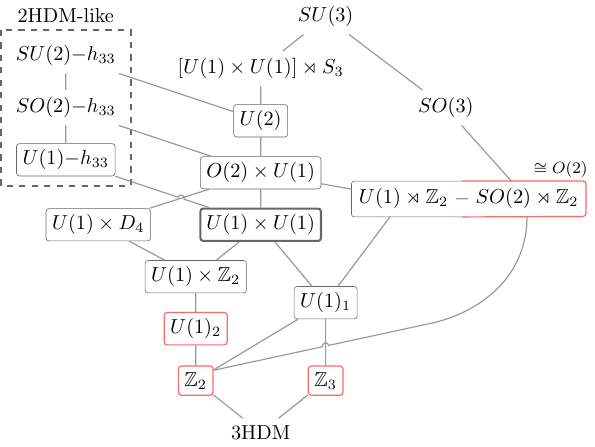}
\end{center}
\vspace*{-6mm}
\caption{The considered 3HDMs follow a hierarchy of symmetry-breaking patterns, where symmetries increase from bottom to top by imposing specific relations among the scalar potential coefficients. Lines between cases indicate connections through symmetries, meaning that the scalar potential at an upper level is contained within that of a lower level, provided the correct charge assignments are chosen at each step. For the $O(2)$, $\{U(1) \rtimes \mathbb{Z}_2,\, SO(2) \rtimes \mathbb{Z}_2\}$ cases, the horizontal line within the box signifies a change of basis. Models enclosed in rectangles indicate scenarios where the vacuum preserves the full underlying symmetry. Red rectangles highlight cases where CP violation is possible, assuming at least one of the vevs vanishes. The upper-left panel refers to hypothesised 2HDM-like cases. It was suggested that $U(1) \times D_4$ due to a non-trivial intersection of the groups may be more appropriately expressed using a quotient group structure~\cite{Ivanonv_pr}. Figure taken from Ref.~\cite{Kuncinas:2024zjq}.}
\label{Fig:Symmetry_breaking}
\end{figure}

One could also consider an alternative construction involving a 2HDM sector with a $U(1)$, $SO(2)$ or $SU(2)$-symmetric potential, alongside a separate SM-like potential. Technically, this would not constitute a 3HDM, as the two sectors remain distinct. Interaction between these sectors could occur via a messenger field, such as gauge bosons or potentially through fermions. 
In this scenario, the underlying symmetry must prevent bilinear terms from coupling between the two sectors and must also forbid quartic terms that would mix these sectors. However, we are not aware of any symmetry or mechanism that could naturally enforce this structure. As a result, we will not explore this possibility further.

Following Ref.~\cite{Ivanov:2011ae}, there are multiple ways to assign $U(1)$ charges to three doublets:
\vspace*{-10pt}\begin{subequations}\label{Eq:U11_U12_charges}
\begin{align}
\begin{split}
U(1)_1 (\alpha) \equiv {}& \mathrm{diag} \left(e^{-i \alpha},\, e^{i \alpha},\, 1 \right) \\
 ={}& e^{i \alpha} \mathrm{diag} \left(e^{-2i \alpha},\,1,\, e^{-i\alpha} \right) = e^{-i \alpha} \mathrm{diag} \left(1,\, e^{2i \alpha},\, e^{i\alpha} \right),\label{Eq:U11_U12_charges_U1a}
\end{split}\\
\begin{split}
U(1)_2 (\beta) \equiv {}& \mathrm{diag} \left(e^{i \beta/3},\, e^{i \beta/3},\, e^{-2 i \beta/3} \right)\\
={}& e^{i \beta/3}\mathrm{diag} \left(1,\, 1,\, e^{-i \beta} \right) = e^{- 2i \beta/3}\mathrm{diag} \left(e^{i \beta},\, e^{i \beta},\, 1 \right).\label{Eq:U11_U12_charges_U1b}
\end{split}
\end{align}
\end{subequations}
In the above, the overall phase factors can be absorbed by the $U(1)_Y$ group. An alternative approach would be to select any pair of the following $U(1)$ charge assignments, 
\begin{subequations}\label{Eq:U1i_charges}
\begin{align}
U(1)_{h_1} ={}&  \mathrm{diag} \left(e^{i \alpha},\,1,\,1\right),\\
U(1)_{h_2} ={}&  \mathrm{diag} \left(1,\,e^{i \alpha},\,1\right),\\
U(1)_{h_3} ={}&  \mathrm{diag} \left(1,\,1,\,e^{i \alpha}\right),
\end{align}
\end{subequations}
ensuring that the chosen charges are consistent with the desired $U(1)$ symmetry structure,
\vspace*{-10pt}\begin{subequations}
\begin{align}
U(1)_{h_1} ={}& e^{i \alpha/3} ~ U(1)_1 ( -{\alpha}/{2}) ~  U(1)_2 ({\alpha}/{2}), \\
U(1)_{h_2} ={}&  e^{i \alpha/3} ~ U(1)_1 ( {\alpha}/{2}) ~  U(1)_2 ({\alpha}/{2}), \\
U(1)_{h_3} ={}&  e^{i \frac{1}{3} (2 \pi +\alpha)} ~ U(1)_1 ( \pi ) ~  U(1)_2 (\pi - \alpha).
\end{align}
\end{subequations}
 
Some of the continuous symmetries appear as a direct product with discrete symmetries, leading to hybrid symmetry structures that constrain the scalar potential in specific ways. For example, cases like $U(1) \times \mathbb{Z}_2$ or $U(1) \rtimes \mathbb{Z}_2$ introduce additional restrictions on the potential while still allowing for viable DM scenarios. Indeed, by imposing additional symmetries on top of the $U(1)$ symmetries, the underlying symmetry group is enlarged. This process further constrains the scalar potential, reducing the number of free parameters and allowed interactions. Such extensions can lead to more predictive models, influencing the stability of DM candidates and the phenomenology of the model.

As additional symmetries are imposed on top of the lower continuous symmetries, the potential becomes more constrained. While initially this can lead to more predictive models, there is a limit to how much symmetry can be imposed before the potential is too restrictive to satisfy all the necessary constraints, \textit{e.g.}, the correct relic density and being unable to tune the portal coupling, which would result  in a high DM-nucleon cross-section and a strongly coupled DM to the SM-like Higgs boson. An example of expanding symmetries is given in Figure~\ref{Fig:Symmetries_Couplings}. There, the minimisation conditions are taken into consideration. A full picture of the allowed terms will be presented in Table~\ref{Table:contU1_Diff_Cases}.

\begin{figure}[htb]
\begin{center}
\includegraphics[scale=1]{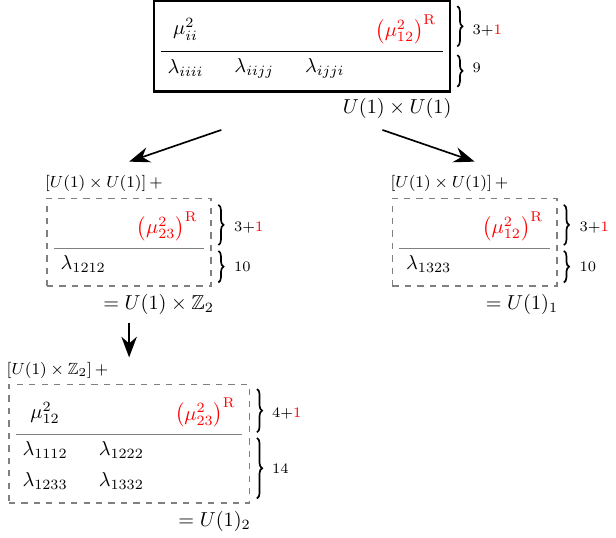}
\end{center}
\vspace*{-3mm}
\caption{Allowed couplings for some $U(1)$-based 3HDMs. The red terms indicate potential soft symmetry-breaking terms. The underlying symmetry for each case is listed beneath the corresponding block, and the total numbers of bilinear and quartic terms are provided to the right of the blocks. Figure taken from Ref.~\cite{Kuncinas:2024zjq}.}
\label{Fig:Symmetries_Couplings}
\end{figure}

We assume the following field content of the $SU(2)$ scalar doublets:
\begin{equation}\label{Eq:Extract_phase}
h_i = e^{i \sigma_i} \begin{pmatrix}
h_i^+ \\
\frac{1}{\sqrt{2}}\left( \hat{v}_i  + \eta_i + i \chi_i \right)
\end{pmatrix},
\end{equation}
\textit{i.e.}, we extract the vacuum phase and re-define the scalar fields. As before, due to the overall $U(1)_Y$ symmetry a single $\sigma_i$ can be rotated away. Since our focus is on DM, we will assume that at least one of the vevs vanishes. Furthermore, we focus on implementations where the neutral states of the potential inert scalar doublet do not mix with the non-inert ones. We do not consider such scenarios since those require fine-tuning of DM stability, such as those where the DM candidate lifetime needs to exceed the age of the Universe. As a result, we are interested in vacua  $(\hat v_1 e^{i \sigma},\, \hat v_2,\, 0)$ and $(v,\, 0,\, 0)$, and in possible permutations of entries of these. 

Permitting both complex vevs and complex couplings,
\begin{equation}
\mu_{ij}^2 \equiv (\mu_{ij}^2)^\mathrm{R} + i (\mu_{ij}^2)^\mathrm{I} \text{ and } \lambda_{ijkl} \equiv \lambda_{ijkl}^\mathrm{R} + i \lambda_{ijkl}^\mathrm{I},
\end{equation}
could lead to the presence of at least one unphysical phase.

In cases of spontaneous CP violation, all phases can be absorbed into the vevs. A basis transformation can render the vevs real. These re-phasings of the doublets do not correspond to symmetries; they are merely a choice of basis and have no physical consequences. However, these transformations could modify the form of the scalar potential, \textit{e.g.}, the underlying symmetry might permit phase-sensitive couplings that, due to the symmetry, must be real in a specific basis. In such cases, some couplings may split into multiple terms because of a phase originating from the vacuum. Whenever this occurs, we prefer to keep the vacuum complex, allowing for complex coefficients alongside complex vevs. This applies to the following symmetries: $O(2) \times U(1)$, $O(2)$, $U(1) \times D_4$, and $SO(3)$.

In Table~\ref{Table:contU1_Diff_Cases} we summarise symmetries to be discussed.

{{\renewcommand{\arraystretch}{1.23}
\setlength\LTcapwidth{\linewidth}
\begin{center}
\begin{longtable}[htb]{|c|c|c|l|} 
\caption{ Considered 3HDMs. In the third column the numbers of independent bilinear plus quartic couplings are provided. The notation of $V_0$, see eq.~\eqref{V_U1xU1}, stands for allowed couplings $\{\mu_{ii}^2,~\lambda_{iiii},~ \lambda_{iijj},~ \lambda_{ijji}\}$, and $V_{O(2) \times U(1)}$, see eq.~\eqref{Eq:V_U1_U1_S2}, requiring $\{\mu_{11}^2 = \mu_{22}^2,~ \mu_{33}^2,~ \lambda_{1111} = \lambda_{2222},~\lambda_{3333},~\lambda_{1122},~\lambda_{1133} = \lambda_{2233},~\lambda_{1221},~\lambda_{1331} = \lambda_{2332}\}$, and also $\Lambda=\frac{1}{2}\left( 2 \lambda_{1111} - \lambda_{1221} - \lambda_{1122} \right)$. }
\label{Table:contU1_Diff_Cases} \\ \hline\hline
\begin{tabular}[l]{@{}c@{}} Underlying \\ symmetry \end{tabular} & Reference & \begin{tabular}[l]{@{}c@{}} Indep. \\ couplings \end{tabular} & \begin{tabular}[l]{@{}c@{}} Allowed couplings and \\ necessary relations \end{tabular} \\ \hline\hline
$\mathbb{Z}_2$ & Table~\ref{Table:Z2_cases}  & 4 + 17 & \begin{tabular}[l]{@{}l@{}} $V_0,~\mu_{12}^2,~\lambda_{ijij},~\lambda_{1112},~\lambda_{1222},~\lambda_{1233},~$ \\$\lambda_{1323},~\lambda_{1332}$ \end{tabular} \\ \hline 
$U(1)_2$ & Table~\ref{Table:U12_Cases}  &  4 + 14 & \begin{tabular}[l]{@{}l@{}} $V_0,~\mu_{12}^2,~\lambda_{1212},~\lambda_{1112},~\lambda_{1222},$\\$\lambda_{1233},~\lambda_{1332}$\end{tabular}  \\ \hline
$\mathbb{Z}_3$ & Table~\ref{Table:Z3_cases}  &  3 + 12 & \begin{tabular}[l]{@{}l@{}} $V_0,~ \lambda_{1323},~\lambda_{1213},~\lambda_{1232}$ \end{tabular} \\ \hline
$U(1)_1$ & Table~\ref{Table:U11_Cases}  &  3 + 10 & $V_0,~\lambda_{1323}$\\ \hline
$U(1) \times \mathbb{Z}_2$ & Table~\ref{Table:U1Z2_Cases}  &  3 + 10 & $V_0,~\lambda_{1212}$\\ \hline
$U(1) \times U(1)$ & Table~\ref{Table:U1_U1_Cases}  &  3 + 9 & $V_0$ \\ \hline
$O(2)_{\scriptscriptstyle[SO(2) \rtimes \mathbb{Z}_2]}$ &   Table~\ref{Table:O2_Cases_1}  &  2 + 7 & \begin{tabular}[l]{@{}l@{}} $V_{O(2) \times U(1)},~\lambda_{1212}= \Lambda,~\lambda_{1313}=\lambda_{2323} $ \end{tabular}\\ \hline
$O(2)_{\scriptscriptstyle[U(1) \rtimes \mathbb{Z}_2]}$ & Table~\ref{Table:O2_Cases_2}  &  2 + 7 & \begin{tabular}[l]{@{}l@{}} $V_{O(2) \times U(1)},~\lambda_{1323} $ \end{tabular}\\ \hline 
$U(1) \times D_4$ & Table~\ref{Table:U1Z2_S2_Cases}  &  2 + 7 & \begin{tabular}[l]{@{}l@{}} $V_{O(2) \times U(1)},~ \mathbb{R}\mathrm{e}(\lambda_{1212})$ \end{tabular}\\ \hline
$O(2)_{\scriptscriptstyle [SO(2) \rtimes \mathbb{Z}_2]} \times U(1)$\hspace*{-0.5pt} & Table~\ref{Table:U1U1S2_Cases_1}  &  2 + 6 & \begin{tabular}[l]{@{}l@{}} $V_{O(2) \times U(1)},~\lambda_{1212} = \Lambda$ \end{tabular} \\  \hline 
$O(2)_{\scriptscriptstyle [U(1) \rtimes \mathbb{Z}_2]} \times U(1)$\hspace*{-0.5pt} & Table~\ref{Table:U1U1S2_Cases_2}  &  2 + 6 & $V_{O(2) \times U(1)}$ \\  \hline 
$U(2)$ & Table~\ref{Table:U2_Cases}  &  2 + 5 & \begin{tabular}[l]{@{}l@{}} $\mu_{11}^2 = \mu_{22}^2,~ \mu_{33}^2,~ \lambda_{1111} = \lambda_{2222},~\lambda_{3333},$ \\ $\Lambda=0,~\lambda_{1133} = \lambda_{2233},~\lambda_{1221}$\\$\lambda_{1331} = \lambda_{2332}$ \end{tabular}\\ \hline
$\left[ U(1) \times U(1) \right] \rtimes S_3$ & Table~\ref{Table:U11S3_Cases}  &  1 + 3 & \begin{tabular}[l]{@{}l@{}}$\mu_{11}^2 = \mu_{22}^2 = \mu_{33}^2,~\lambda_{1111} = \lambda_{2222} = \lambda_{3333},$ \\ $\lambda_{1122} = \lambda_{1133} = \lambda_{2233},$ \\$\lambda_{1221} = \lambda_{1331} = \lambda_{2332}$ \end{tabular}\\ \hline
$SO(3)$ & Table~\ref{Table:SO3_Cases}  &  1 + 3 & \begin{tabular}[l]{@{}l@{}}$\mu_{11}^2 = \mu_{22}^2 = \mu_{33}^2,~\lambda_{1111} = \lambda_{2222} = \lambda_{3333},$ \\ $\lambda_{1122} = \lambda_{1133} = \lambda_{2233},$\\$\lambda_{1221} = \lambda_{1331} = \lambda_{2332}, ~\lambda_{ijij} = \Lambda$\end{tabular}\\ \hline
$SU(3)$ & Table~\ref{Table:SU3_Cases}  &  1 + 2 & \begin{tabular}[l]{@{}l@{}}$\mu_{11}^2 = \mu_{22}^2 = \mu_{33}^2,~\lambda_{1111} = \lambda_{2222} = \lambda_{3333},$ \\ $\lambda_{1122} = \lambda_{1133} = \lambda_{2233},$ \\ $\lambda_{ijji} = 2 \lambda_{1111} - \lambda_{1122}$\end{tabular}\\ \hline \hline
\end{longtable}
\end{center}}

Symmetries typically reduce the number of independent parameters, causing some couplings or masses to vanish or become related. Requiring invariance under different symmetries may lead to the same scalar potential, while symmetries with the same number of independent couplings may differ in their $SU(2)$ doublet structures. Experimentally, they could be distinguished through distinct coupling patterns or mass relations among physical states, the latter being a common feature of the models discussed here.

\section[Comparison with the \texorpdfstring{$\mathbb{Z}_2\text{-3HDM and}\; \mathbb{Z}_3\text{-3HDM}$}{Z2-3HDM and Z3-3HDM}]{Comparison with the \boldmath$\mathbb{Z}_2\text{-3HDM and}\; \mathbb{Z}_3\text{-3HDM}$}\label{Sec:Pot_Zn}

By examining Figure~\ref{Fig:Symmetry_breaking}, or equivalently Table~\ref{Table:contU1_Diff_Cases}, we can identify that the origin of all continuous symmetries can be traced to either $\mathbb{Z}_2$ or $\mathbb{Z}_3$, which are the least symmetric groups. For completeness, we consider both these discrete symmetries, as they can serve as a foundation for discussing the more symmetric models by imposing conditions on the couplings, \textit{i.e.}, all the models with continuous symmetries discussed in the following sections can be derived from the $\mathbb{Z}_2$-symmetric or $\mathbb{Z}_3$-symmetric 3HDM by requiring certain couplings to vanish or enforcing specific relations among them. 

To illustrate this, for instance, consider the $U(1)_1$-symmetric 3HDM (to be covered in Section~\ref{Sec:Pot_U11}), see Table~\ref{Table:contU1_Diff_Cases}. Starting with the $\mathbb{Z}_2$-symmetric model, we need to impose the conditions
\begin{equation*}
\mu_{12}^2 = 0 \text{ and }\lambda_{ijij} = \lambda_{1112} = \lambda_{1222} = \lambda_{1233} = \lambda_{1332} = 0.
\end{equation*}
Alternatively, taking the $\mathbb{Z}_3$-symmetric model as the reference point, the conditions are:
\begin{equation*}
\lambda_{1213} = \lambda_{1232} = 0.
\end{equation*}
These conditions can then be substituted into the mass-squared matrices of either the $\mathbb{Z}_2$-symmetric or the $\mathbb{Z}_3$-symmetric 3HDM. We had explicitly verified that this procedure holds for all symmetry-breaking patterns presented in Figure~\ref{Fig:Symmetry_breaking}.

We are interested in cases involving CP violation, and thus, we will examine two possibilities: explicit CP violation and spontaneous CP violation.

\vspace{10pt}
\textbf{\boldmath$\mathbb{Z}_2$-symmetric 3HDM}
\vspace{5pt}

Without loss of generality, one can assign the following $\mathbb{Z}_2$ charges to the $SU(2)$ scalar doublets:
\begin{equation}\label{Eq:Z2_charges}
\mathbb{Z}_2: ~ h_1\to h_1,~~ h_2\to h_2,~~ h_3\to -h_3.
\end{equation}
The phase-sensitive part of the scalar potential is given by:
\begin{equation}\label{Eq:Z2_Vph}
\begin{split}
V_{\mathbb{Z}_2}^\text{ph} ={}& \mu_{12}^2 h_{12} + \sum_{i<j} \lambda_{ijij} h_{ij}^2 + \lambda_{1112} h_{11}h_{12} + \lambda_{1222} h_{12}h_{22}\\
&  + \lambda_{1233} h_{12}h_{33} + \lambda_{1323} h_{13}h_{23} + \lambda_{1332} h_{13}h_{32} + \mathrm{h.c.}
\end{split}
\end{equation} 
We note that in Ref.~\cite{Darvishi:2019dbh} there is a coupling missing in the case of the $\mathbb{Z}_2$-symmetric 3HDM. For $\mathbb{Z}_2$ the missing terms is $\lambda_{1223}$, while for the $\mathbb{Z}_2^\prime$ entry the missing coupling is $\lambda_{1321}$. This was clarified with the authors.

Due to the scalar potential being symmetric under the interchange of $h_1 \leftrightarrow h_2$, some permutations of vevs lead to identical physical models. A vanishing vev does not automatically guarantee no mixing between the inert and active states, unless it is $\left\langle h_3 \right\rangle = 0$, or equivalently $\left\langle h_1 \right\rangle = \left\langle h_2 \right\rangle = 0$, which is then stabilised by $\mathbb{Z}_2$.

Different implementations are summarised in Table~\ref{Table:Z2_cases}. The mass-squared matrices can be found in Appendix A of Ref.~\cite{Kuncinas:2024zjq}.

{\renewcommand{\arraystretch}{1.3}
\begin{table}[htb]
\caption{ Different implementations within the $\mathbb{Z}_2$-symmetric 3HDM. In the second column the ``$\checkmark$" indicates that the symmetry is not spontaneously broken. The scalar potential is provided in the third column with soft terms if there are massless states. In the fourth column mixing patterns are presented; fields within braces mix to form block-diagonal structures. Additional information is presented in the last column. In all cases complex parameters are present, they may result in CP violation.}
\label{Table:Z2_cases}
\begin{center}
\begin{tabular}{|c|c|c|c|c|} \hline\hline
Vacuum & SYM & $V$ & \begin{tabular}[l]{@{}c@{}} Mixing of the \\ neutral states\end{tabular} & Comments \\ \hline
$(\hat v_1 e^{i \sigma},\, \hat v_2,\, 0)$ & $\checkmark$ & $V_{\mathbb{Z}_2}$ & \footnotesize\begin{tabular}[l]{@{}c@{}} $\{\eta_1,\, \eta_2,\, \chi_1,\, \chi_2\}$ \\ ${-}\{\eta_3,\,\chi_3\}$ \end{tabular} & -\\ \hline
$(0,\, \hat v_2,\, \hat v_3)$ & $-$ & $V_{\mathbb{Z}_2}$ &  total mixing & No obvious DM \\ \hline
$(0,\, \hat v_2 e^{i \sigma},\, \hat v_3)$ & $-$ & $V_{\mathbb{Z}_2}$  & \footnotesize\begin{tabular}[l]{@{}c@{}} $\{\eta_1,\,\eta_2,\, \eta_3,\, \chi_1\}$ \\ ${-}\{\chi_2\}{-}\{\chi_3\}$ \end{tabular} & \footnotesize\begin{tabular}[l]{@{}c@{}} $m_{\chi_2} = m_{\chi_3}=0$ \\ No obvious DM  \end{tabular} \\ \hline 
$(0,\, \hat v_2 e^{i \sigma},\, \hat v_3)$ & $-$ & $V_{\mathbb{Z}_2} + (\mu_{23}^2)^\mathrm{R}$  & total mixing & No obvious DM \\ \hline 
$(v,\, 0,\, 0)$ & $\checkmark$ & $V_{\mathbb{Z}_2}$ & \footnotesize\begin{tabular}[l]{@{}c@{}} $\{\eta_1,\, \eta_2,\, \chi_2\}{-}\{\chi_1\}$ \\ ${-}\{\eta_3,\,\chi_3\}$ \end{tabular} & - \\ \hline
$(0,\, 0,\, v)$ & $\checkmark$ & $V_{\mathbb{Z}_2}$ & \footnotesize\begin{tabular}[l]{@{}c@{}} $\{\eta_1,\, \eta_2,\, \chi_1,\, \chi_2\}$ \\ ${-}\{\eta_3\}{-}\{\chi_3\}$ \end{tabular} & - \\ \hline \hline
\end{tabular} 
\end{center}
\end{table}}

\vspace{10pt}
\textbf{\boldmath$\mathbb{Z}_3$-symmetric 3HDM}
\vspace{5pt}

By assigning the following $\mathbb{Z}_3$ charges to the scalar doublets,
\begin{equation}\label{Eq:Z3_gen_ph_tr}
h_1 \to h_1, ~ h_2 \to e^{i \frac{2\pi}{3}} h_2, ~ e^{i \frac{4\pi}{3}} h_3 \to h_3,
\end{equation}
the phase-sensitive part of the potential yields:
\begin{equation}
V_{\mathbb{Z}_3}^\text{ph} = \lambda_{1323} h_{13} h_{23} + \lambda_{1213} h_{12} h_{13} + \lambda_{1232} h_{12} h_{32} + \mathrm{h.c.}
\end{equation}

By comparing $V_{\mathbb{Z}_3}^\text{ph}$ to $V_{\mathbb{Z}_2}^\text{ph}$ we can identify two new couplings $\{\lambda_{1213},\, \lambda_{1232}\}$, while several of the quartic couplings of the $\mathbb{Z}_2$-symmetric scalar potential are not allowed by the $\mathbb{Z}_3$ symmetry. Actually, the presence of these two couplings, and, as a consequence, a non-existence of a unitary transformation to the basis of $V_{\mathbb{Z}_2}^\text{ph}$, indicates that $\mathbb{Z}_2$ and $\mathbb{Z}_3$ are different symmetries and do not share a common origin apart from the most general (without symmetry constraints) 3HDM.

One can identify a permutation symmetry applicable to the indices of the doublets in eq.~\eqref{Eq:Z3_gen_ph_tr}. This indicates that it is enough to consider only two distinct vacuum configurations.

Different implementations are summarised in Table~\ref{Table:Z3_cases}. The mass-squared matrices can be found in Appendix B of Ref.~\cite{Kuncinas:2024zjq}.

{\renewcommand{\arraystretch}{1.3}
\begin{table}[htb]
\caption{ Different implementations within the $\mathbb{Z}_3$-symmetric 3HDM. In the second column the ``$\checkmark$" indicates that the symmetry is not spontaneously broken. The scalar potential is provided in the third column with soft terms if there are massless states. In the fourth column mixing patterns are presented; fields within braces mix to form block-diagonal structures. Additional information is presented in the last column. It is not possible to have simultaneously CP violation and a DM candidate.}
\label{Table:Z3_cases}
\begin{center}
\begin{tabular}{|c|c|c|c|c|} \hline\hline
Vacuum & SYM &$V$ & \begin{tabular}[l]{@{}c@{}} Mixing of the \\ neutral states\end{tabular} & Comments \\ \hline
$(\hat{v}_1,\, \hat{v}_2,\, 0)$ & $-$ &  $V_{\mathbb{Z}_3}$ & total mixing & No obvious DM \\ \hline
$(\hat{v}_1 e^{i \sigma},\, \hat{v}_2,\, 0)$ & $-$ &  $V_{U(1)_1}$ & \footnotesize\begin{tabular}[l]{@{}c@{}} $\{\eta_1,\, \eta_2\}{-}\{\eta_3\}$\\${-}\{\chi_1\}{-}\{\chi_2\}{-}\{\chi_3\}$ \end{tabular} & $m_{\chi_1} = m_{\chi_2} = 0$ \\ \hline
$(\hat{v}_1 e^{i \pi/3},\, \hat{v}_2,\, 0)$  & $-$ &  $V_{\mathbb{Z}_3}$ & total mixing & No obvious DM \\ \hline
$(\hat v_1 e^{i \sigma},\, \hat v_2,\, 0)$ & $-$ & $V_{U(1)_1} + (\mu_{12}^2)^\mathrm{R}$ & \footnotesize\begin{tabular}[l]{@{}c@{}} $\{\eta_1,\, \eta_2\}{-}\{\eta_3\}$\\ ${-}\{\chi_1,\, \chi_2\}{-}\{\chi_3\}$ \end{tabular} & - \\ \hline
$(v,\,0,\,0)$ & $\checkmark$ & $V_{\mathbb{Z}_3}$ & \footnotesize\begin{tabular}[l]{@{}c@{}} $\{\eta_1\}{-}\{\chi_1\}$ \\ ${-}\{\eta_2,\ \eta_3,\, \chi_2,\, \chi_3\}$ \end{tabular} & \footnotesize\begin{tabular}[l]{@{}c@{}} Two pairs of \\ mass-degenerate states \end{tabular} \\ \hline \hline
\end{tabular} 
\end{center}
\end{table}}

\section[\texorpdfstring{$U(1)\times U(1)$}{U(1) x U(1)}-symmetric 3HDM]{\boldmath$U(1)\times U(1)$-symmetric 3HDM}\label{Sec:Pot_U1U1}

For example, consider that the $SU(2)$ scalar doublets acquire charges
\begin{equation*}
\begin{aligned}
U(1)_1 \times U(1)_2 :{}&  \mathrm{diag} \left( e^{-i \alpha},\, e^{i \alpha},\, e^{-i \beta} \right) = e^{-i \alpha} \mathrm{diag} \left( 1,\, e^{2 i \alpha},\, e^{i (\alpha-\beta)} \right)\\
& = {} e^{i \alpha} \mathrm{diag} \left( e^{-2 i \alpha},\, 1,\, e^{-i (\alpha+\beta)} \right) = e^{-i \beta} \mathrm{diag} \left( e^{-i (\alpha-\beta)},\, e^{i (\alpha+\beta)},\, 1 \right).
\end{aligned}
\end{equation*}
For values of $\alpha = \pi$ and $\beta = 3\pi$ there is a non-trivial intersection, \textit{i.e.}, $(-1,\,-1,\,-1)$. Such intersection indicates that one should be cautious when considering a direct product of the two $U(1)$ groups. Strictly speaking, it can only be defined when the only common element between the two $U(1)$ groups is the identity element,  \textit{i.e.}, $(1,\,1,\,1)$.

The most general $U(1) \times U(1)$-symmetric 3HDM scalar potential is:
\begin{equation}\label{V_U1xU1}
V_0 = \sum_i \mu_{ii}^2 h_{ii} + \sum_i \lambda_{iiii} h_{ii}^2 + \sum_{i<j} \lambda_{iijj} h_{ii} h_{jj} + \sum_{i<j} \lambda_{ijji} h_{ij} h_{ji}.
\end{equation}

In addition to defining the scalar potential, we must consider different implementations arising from various vacuum configurations. The form of the scalar potential is highly symmetric---
it remains unchanged under any of the $S_3$ re-labelings of the scalar doublets. This indicates that there is no need to examine different permutations of vevs. For instance, the scalar content and interactions of implementations $(v,\,0,\,0)$, $(0,\,v,\,0)$ and $(0,\,0,\,v)$ are identical. Furthermore, since the form of the potential does not allow for complex couplings, any complex phase coming from vevs is thus unphysical. This property allows us to restrict the discussion to real vacua. In table~\ref{Table:U1_U1_Cases} possible cases are summarised.

{{\renewcommand{\arraystretch}{1.3}
\begin{table}[htb]
\caption{Different implementations within the $U(1)\times U(1)$-symmetric 3HDM. In the second column the ``$\checkmark$" indicates that the symmetry is not spontaneously broken. The scalar potential is provided in the third column with soft terms if there are massless states. In the fourth column mixing patterns are presented; fields within braces mix to form a block-diagonal structure. Additional information is presented in the last column. There is no CP violation in any of the $U(1)\times U(1)$-symmetric 3HDM implementations.}
\label{Table:U1_U1_Cases}
\begin{center}
\begin{tabular}{|c|c|c|c|c|} \hline\hline
Vacuum & SYM & $V$ & \begin{tabular}[l]{@{}c@{}} Mixing of the \\ neutral states\end{tabular} & Comments \\ \hline
$(\hat{v}_1,\, \hat{v}_2,\, 0)$ & - & $V_0$ & \footnotesize\begin{tabular}[l]{@{}c@{}} $\{\eta_1,\, \eta_2\}{-}\{\eta_3\}$\\${-}\{\chi_1\}{-}\{\chi_2\}{-}\{\chi_3\}$ \end{tabular} & \footnotesize\begin{tabular}[l]{@{}c@{}} $m_{\eta_3} = m_{\chi_3}$ \\ $m_{\chi_1} = m_{\chi_2} = 0$ \end{tabular} \\ \hline
$(\hat{v}_1,\, \hat{v}_2,\, 0)$ & - & $V_0 + (\mu_{12}^2)^\mathrm{R}$ & \footnotesize\begin{tabular}[l]{@{}c@{}} $\{\eta_1,\, \eta_2\}{-}\{\eta_3\}$ \\ ${-}\{\chi_1,\, \chi_2\}{-}\{\chi_3\}$ \end{tabular} & $m_{\eta_3} = m_{\chi_3}$ \\ \hline
$(v,\, 0,\, 0)$  & $\checkmark$ & $V_0$ & diagonal & \footnotesize\begin{tabular}[l]{@{}c@{}} $m_{\eta_2} = m_{\chi_2},~m_{\eta_3} = m_{\chi_3}$ \end{tabular} \\ \hline \hline
\end{tabular}\vspace*{-9pt}
\end{center}
\end{table}}

\newpage
\bigskip\textbf{Case of \boldmath$(v,\,0,\,0)$} \labeltext{Case of $(v,\,0,\,0)$}{Sec:U1U1_v00}

There is a single minimisation condition:
\begin{equation} \label{Eq:U1U1_MinCond3}
\mu_{11}^2 = - \lambda_{1111} v^2.
\end{equation}

The charged mass-squared matrix in the basis $\{h_1^\pm,\, h_2^\pm,\, h_3^\pm \}$ is:
\begin{equation}
\mathcal{M}_\mathrm{Ch}^2 = \mathrm{diag}\left(0,\, \mu_{22}^2 + \frac{1}{2} \lambda_{1122} v^2,\, \mu_{33}^2 + \frac{1}{2} \lambda_{1133} v^2   \right).
\end{equation}

The neutral mass-squared matrix in the basis $\{\eta_1,\, \eta_2,\, \eta_3,\, \chi_1,\, \chi_2,\, \chi_3 \}$ is:
\begin{equation}
\mathcal{M}_\mathrm{N}^2 = \mathrm{diag} \bigg(m_h^2,\, m_{H_2}^2,\, m_{H_3}^2,\, 0,\,  m_{H_2}^2,\, m_{H_3}^2   \bigg),
\end{equation}
where
\begin{subequations}\label{Eq:MN2_U1_U1}
\begin{align}
m_h^2 ={}& 2 \lambda_{1111} v^2,\\
 m_{H_i}^2 ={}& \mu_{ii}^2 + \frac{1}{2} \left( \lambda_{11ii} + \lambda_{1ii1} \right) v^2, \text{ for } i=\{2,\,3\}.
\end{align}
\end{subequations}

In this implementation:
\begin{itemize}
\item $\{ \mu_{22}^2,\, \mu_{33}^2,\, \lambda_{1111},\, \lambda_{1122},\, \lambda_{1133},\, \lambda_{1221},\, \lambda_{1331}\}$ contribute to five scalar masses;
\item $\{\lambda_{2222},\, \lambda_{2233},\, \lambda_{2332},\,$ $\lambda_{3333} \}$ appear only in scalar interactions.
\end{itemize}

\section[\texorpdfstring{$U(1)_1$}{U(1)1}-symmetric 3HDM]{\boldmath$U(1)_1$-symmetric 3HDM}\label{Sec:Pot_U11}

The $U(1)_1$-symmetric scalar potential is given by
\begin{equation}
\begin{aligned}
V_{U(1)_1} ={}& V_0 + V^\mathrm{ph}_{U(1)_1}\\
={}& \sum_i \mu_{ii}^2 h_{ii} + \sum_i \lambda_{iiii} h_{ii}^2 + \sum_{i<j} \lambda_{iijj} h_{ii} h_{jj} + \sum_{i<j} \lambda_{ijji} h_{ij} h_{ji}\\
& + \left\lbrace \lambda_{1323} h_{13} h_{23} + \mathrm{h.c.} \right\rbrace.
\end{aligned}
\end{equation}
As depicted in Figure~\ref{Fig:Symmetry_breaking}, the potential can be derived by imposing a $\mathbb{Z}_2 \times \mathbb{Z}_3$ symmetry.

The scalar potential is not symmetric under all permutations of vevs, and hence we shall consider different options. However, the form of the potential is left intact after permutations between $h_1$ and $h_2$. Since there is a single complex coupling $\lambda_{1323}$ sensitive to the phases of all three doublets, it is always possible to absorb all phases by a doublet with a vanishing vev. Even if one were to assume an implementation without a vanishing vev, the minimisation conditions would force the complex quartic coupling to become real. Therefore, we shall assume that $\lambda_{1323} \in \mathbb{R}$, except when soft symmetry-breaking terms are introduced.

The discussion of the $U(1)_1$-symmetric 3HDM is summarised in Table~\ref{Table:U11_Cases}. The only interesting cases are those with two vanishing vevs since all other cases result in SSB of $U(1)_1$. All discussed cases are CP conserving.

{\renewcommand{\arraystretch}{1.3}
\begin{table}[htb]
\caption{Similar to Table~\ref{Table:U1_U1_Cases}, but now for $U(1)_1$. In one case the minimisation conditions can lead to a higher symmetry. None of these cases violates CP.}
\label{Table:U11_Cases}
\begin{center}
\begin{tabular}{|c|c|c|c|c|} \hline\hline
Vacuum & SYM &$V$ & \begin{tabular}[l]{@{}c@{}} Mixing of the \\ neutral states\end{tabular} & Comments \\ \hline
$(\hat{v}_1,\, \hat{v}_2,\, 0)$ & $-$ &  $V_{U(1)_1}$ & \footnotesize\begin{tabular}[l]{@{}c@{}} $\{\eta_1,\, \eta_2\}{-}\{\eta_3\}$\\${-}\{\chi_1\}{-}\{\chi_2\}{-}\{\chi_3\}$ \end{tabular} & $m_{\chi_1} = m_{\chi_2} = 0$ \\ \hline
$(\hat{v}_1,\, \hat{v}_2,\, 0)$ & $-$ & $V_{U(1)_1} + (\mu_{12}^2)^\mathrm{R}$ & \footnotesize\begin{tabular}[l]{@{}c@{}} $\{\eta_1,\, \eta_2\}{-}\{\eta_3\}$\\ ${-}\{\chi_1,\, \chi_2\}{-}\{\chi_3\}$ \end{tabular} & - \\ \hline
$(0,\, \hat v_2,\, \hat v_3)$  & $-$ & $V_0$ & \footnotesize\begin{tabular}[l]{@{}c@{}} $\{\eta_1\}{-}\{\eta_2,\, \eta_3\}$ \\ ${-}\{\chi_1\}{-}\{\chi_2\}{-}\{\chi_3\}$ \end{tabular} & \footnotesize\begin{tabular}[l]{@{}c@{}} $m_{\eta_1} = m_{\chi_1}$ \\ $m_{\chi_2} = m_{\chi_3} = 0$ \end{tabular} \\ \hline
$(v,\,0,\,0)$ & $\checkmark$ & $V_{U(1)_1}$  & diagonal & \footnotesize\begin{tabular}[l]{@{}c@{}} $m_{\eta_2} = m_{\chi_2},~m_{\eta_3} = m_{\chi_3}$ \end{tabular} \\ \hline
$(0,\,0,\,v)$ & $\checkmark$ & $V_{U(1)_1}$ & \footnotesize\begin{tabular}[l]{@{}c@{}} $\{\eta_1,\, \eta_2\}{-}\{\eta_3\}$ \\ ${-}\{\chi_1,\, \chi_2\}{-}\{\chi_3\}$ \end{tabular} & \footnotesize\begin{tabular}[l]{@{}c@{}} Two pairs of \\ mass-degenerate states \end{tabular} \\ \hline \hline
\end{tabular} 
\end{center}
\end{table}

\bigskip\textbf{Case of \boldmath$(v,\,0,\,0)$}

The mass-squared matrices are identical to the ones of $(v,\,0,\,0)$ of $U(1) \times U(1)$, see \ref{Sec:U1U1_v00} in Section~\ref{Sec:Pot_U1U1}. Due to the presence of the phase-sensitive quartic coupling, it is physically different from $U(1) \times U(1)$, \textit{e.g.}, there is a scalar interaction $\eta_2 \eta_3^2 \sim  \lambda_{1323}$ present.

In this implementation:
\begin{itemize}
\item $\{ \mu_{22}^2,\, \mu_{33}^2,\, \lambda_{1111},\, \lambda_{1122},\, \lambda_{1133},\, \lambda_{1221},\, \lambda_{1331}\}$ contribute to five scalar masses;
\item $\{\lambda_{2222},\, \lambda_{2233},\, \lambda_{2332},\,\lambda_{3333},\, \lambda_{1323} \}$ appear only in scalar interactions.
\end{itemize}

\bigskip\textbf{Case of \boldmath$(0,\,0,\,v)$}

The minimisation condition is:
\begin{equation}
\mu_{33}^2 = - \lambda_{3333} v^2,
\end{equation}

The charged mass-squared matrix in the basis $\{h_1^\pm,\, h_2^\pm,\, h_3^\pm \}$ is diagonal:
\begin{equation}
\mathcal{M}_\mathrm{Ch}^2 = \mathrm{diag}\left( \mu_{11}^2 + \frac{1}{2} \lambda_{1133} v^2 ,\, \mu_{22}^2 + \frac{1}{2} \lambda_{2233} v^2 ,\, 0   \right).
\end{equation}

The neutral mass-squared matrix in the basis $\{\eta_1,\, \eta_2,\, \eta_3,\, \chi_1,\, \chi_2,\, \chi_3 \}$ is:
\begin{equation}
\mathcal{M}_\mathrm{N}^2 = \mathrm{diag}\Bigg(  \begin{pmatrix}
(\mathcal{M}_\mathrm{N}^2)_{11} & (\mathcal{M}_\mathrm{N}^2)_{12} \\
(\mathcal{M}_\mathrm{N}^2)_{12}& (\mathcal{M}_\mathrm{N}^2)_{22} \end{pmatrix},\, (\mathcal{M}_\mathrm{N}^2)_{33},\, \begin{pmatrix}
(\mathcal{M}_\mathrm{N}^2)_{11} & -(\mathcal{M}_\mathrm{N}^2)_{12}\\
-(\mathcal{M}_\mathrm{N}^2)_{12} & (\mathcal{M}_\mathrm{N}^2)_{22}
\end{pmatrix},\, 0 \Bigg),
\end{equation}
where
\begin{subequations}
\begin{align}
(\mathcal{M}_\mathrm{N}^2)_{ii} ={}& \mu_{ii}^2 + \frac{1}{2} \left( \lambda_{ii33} + \lambda_{i33i} \right) v^2, \text{ for } i=\{1,\,2\},\\
(\mathcal{M}_\mathrm{N}^2)_{12} ={}& \frac{1}{2} \lambda_{1323} v^2,\\
(\mathcal{M}_\mathrm{N}^2)_{33} ={}& 2 \lambda_{3333} v^2.
\end{align}
\end{subequations}

The two-by-two matrices have identical eigenvalues:
\begin{equation}
m_{H_i}^2 =  \frac{1}{4} \left[ 2\left( \mu_{11}^2 + \mu_{22}^2 \right) +  v^2 \left(\lambda_{1133} + \lambda_{2233} + \lambda_{1331} + \lambda_{2332}\right) \pm \Delta \right],
\end{equation}
where
\begin{equation}
\begin{aligned}
\Delta^2 ={}& v^4 \left[  4 \left(\lambda_{1323}^\mathrm{R}\right)^2 + \left( \lambda_{1133} - \lambda_{2233} + \lambda_{1331} - \lambda_{2332} \right)^2 \right]\\
&  + 4 v^2 \left( \lambda_{1133} - \lambda_{2233} + \lambda_{1331} - \lambda_{2332} \right)\left( \mu_{11}^2 - \mu_{22}^2 \right) + 4 \left( \mu_{11}^2 - \mu_{22}^2 \right)^2.
\end{aligned}
\end{equation}

In this implementation:
\begin{itemize}
\item  $\{ \mu_{11}^2,\, \mu_{22}^2,\, \lambda_{1133},\, \lambda_{1331},\, \lambda_{2233},\, \lambda_{2332},\, \lambda_{1323}\}$ contribute to five scalar masses;
\item  $\{\lambda_{1111},\, \lambda_{1122},\, \lambda_{1221},\,$ $ \lambda_{2222}\}$ appear only in scalar interactions.
\end{itemize}

\section[\texorpdfstring{$U(1)_2$}{U(1)2}-symmetric 3HDM]{\boldmath$U(1)_2$-symmetric 3HDM}\label{Sec:Pot_U12}

The $U(1)_2$-symmetric 3HDM permits six phase-dependent couplings instead of just one of the $U(1)_1$-symmetric 3HDM:
\begin{equation}\label{Eq:Vph_U12}
\begin{aligned}
V_{U(1)_2} ={}& V_0 + V_{U(1)_2}^\text{ph}\\
={}& \sum_i \mu_{ii}^2 h_{ii} + \sum_i \lambda_{iiii} h_{ii}^2 + \sum_{i<j} \lambda_{iijj} h_{ii} h_{jj} + \sum_{i<j} \lambda_{ijji} h_{ij} h_{ji}\\
&  +  \Big\{\mu_{12}^2 h_{12} + \lambda_{1212} h_{12}^2 + \lambda_{1112} h_{11}h_{12}+ \lambda_{1222} h_{12}h_{22}\\
& \hspace{20pt} + \lambda_{1233} h_{12}h_{33} + \lambda_{1332} h_{13}h_{32} + \mathrm{h.c.} \Big\}.
\end{aligned}
\end{equation}
In contrast to the models considered above, explicit CP violation generally occurs. Due to $\left\langle h_i\right\rangle = 0$, it may be possible to rotate away one complex phase. Due to CP violation we need to consider two cases: real vevs and complex couplings, complex vevs and real couplings. The interchange of indices $1 \leftrightarrow 2$ of the scalar doublets leads to identical solutions, simplifying the number of different vacua we need to consider. The different implementations are summarised in Table~\ref{Table:U12_Cases}.

{{\renewcommand{\arraystretch}{1.3}
\begin{table}[htb]
\caption{ Similar to Table~\ref{Table:U1_U1_Cases}, but now for $U(1)_2$. In all cases complex parameters are present, and may result in CP violation. In cases of two non-vanishing vevs without hats, $v_i$, the vevs can be real or complex.}
\label{Table:U12_Cases}
\begin{center}
\begin{tabular}{|c|c|c|c|c|} \hline\hline
Vacuum & SYM & $V$ & \begin{tabular}[l]{@{}c@{}} Mixing of the \\ neutral states\end{tabular} & Comments \\ \hline
$(\hat v_1 e^{i \sigma},\, \hat v_2,\, 0)$ & $\checkmark$ & $V_{U(1)_2}$ & \footnotesize\begin{tabular}[l]{@{}c@{}} $\{\eta_1,\, \eta_2,\, \chi_1,\, \chi_2\}$ \\ ${-}\{\eta_3\}{-}\{\chi_3\}$ \end{tabular} & $m_{\eta_3} = m_{\chi_3}$ \\ \hline
$(0,\, \hat v_2 e^{i\sigma},\, \hat v_3)$ & $-$ & $V_{U(1)_2}$ & \footnotesize\begin{tabular}[l]{@{}c@{}} $\{\eta_1,\,\eta_2,\, \eta_3, \chi_1\}$ \\ ${-}\{\chi_2\}{-}\{\chi_3\}$ \end{tabular} & \footnotesize\begin{tabular}[l]{@{}c@{}} $m_{\chi_2} = m_{\chi_3} = 0 $ \\ No obvious DM \end{tabular} \\ \hline
$(0,\, \hat{v}_2,\, \hat{v}_3)$ & $-$ & $V_{U(1)_2} + (\mu_{23}^2)^\mathrm{R}$ & \footnotesize\begin{tabular}[l]{@{}c@{}} $\{\eta_1,\,\eta_2,\,\eta_3,\,\chi_1\}$\\${-}\{\chi_2,\, \chi_3\}$ \end{tabular} & \footnotesize No obvious DM  \\ \hline 
$(v,\, 0,\, 0)$ & $\checkmark$ & $V_{U(1)_2}$ & \footnotesize\begin{tabular}[l]{@{}c@{}} $\{\eta_1,\, \eta_2,\, \chi_2\}$ \\ ${-}\{\eta_3\}{-}\{\chi_1\}{-}\{\chi_3\}$ \end{tabular} & $m_{\eta_3} = m_{\chi_3}$ \\ \hline
$(0,\, 0,\, v)$ & $\checkmark$ & $V_{U(1)_2}$ & \footnotesize\begin{tabular}[l]{@{}c@{}} $\{\eta_1,\, \eta_2,\, \chi_1,\, \chi_2\}$ \\ ${-}\{\eta_3\}{-}\{\chi_3\}$ \end{tabular} & \footnotesize\begin{tabular}[l]{@{}c@{}} Two pairs of \\ mass-degenerate \\ neutral states\end{tabular} \\ \hline \hline
\end{tabular} 
\end{center}
\end{table}}

\newpage

For the vacuum $(v_1,\, v_2,\, 0)$, a single complex parameter, either $(\mu_{12}^2)^\mathrm{I}$ or $\sigma$, is sufficient. As a result we have to consider: real vevs and complex couplings, complex vevs and real couplings. In both cases, the neutral sector of the $h_3$ doublet contains a mass-degenerate pair.

\bigskip\textbf{Case of \boldmath$(\hat{v}_1,\, \hat{v}_2,\, 0)$  and complex couplings}

In order to simplify the off-diagonal elements of the mass-squared matrices, we define the ratio of two vevs as, \textit{cf.} $\tan \beta$ in the 2HDM~\eqref{Eq:Tan_b}:
\begin{equation}
\hat v_{i/j} \equiv \hat v_i / \hat v_j.
\end{equation}

The minimisation conditions are:
\begin{subequations}\label{Eq:U12_Min_Con_1}
\begin{align}
(\mu_{12}^2)^\mathrm{I} ={}& -\frac{1}{2} \left[ \lambda_{1112}^\mathrm{I} \hat{v}_1^2 + \left( 2 \lambda_{1212}^\mathrm{I} \hat{v}_1 + \lambda_{1222}^\mathrm{I} \hat{v}_2 \right)\hat{v}_2  \right],\\
\begin{split}
(\mu_{12}^2)^\mathrm{R} ={}& -\frac{1}{2}\Bigg[ 3 \lambda_{1112}^\mathrm{R} \hat v_1^2 + 2\left( \mu_{11}^2 + \lambda_{1111} \hat v_1^2 \right) \hat v_{1/2} \\
& \hspace{30pt} + \left( \lambda_{1122} + 2 \lambda_{1212}^\mathrm{R} + \lambda_{1221} \right)  \hat v_1 \hat v_2 +  \lambda_{1222}^\mathrm{R} \hat v_2^2 \Bigg],
\end{split}\\
\mu_{22}^2 ={}&   \left( \mu_{11}^2 \hat{v}_1^2 + \lambda_{1111} \hat{v}_1^4 + \lambda_{1112}^\mathrm{R} \hat v_1^3 \hat v_2 - \lambda_{1222}^\mathrm{R} \hat v_1 \hat v_2^3 - \lambda_{2222} \hat{v}_2^4 \right)\frac{1}{\hat{v}_2^2}.
\end{align}
\end{subequations}

The charged mass-squared matrix in the basis $\{h_1^\pm,\, h_2^\pm,\, h_3^\pm \}$ is:
\begin{equation}\label{Eq:MCh2_v1_v2_0}
\mathcal{M}_\mathrm{Ch}^2 = \begin{pmatrix}
\left( \mathcal{M}_\textrm{Ch}^2 \right)_{11} & -\left( \mathcal{M}_\textrm{Ch}^2 \right)_{11} \hat v_{1/2} & 0 \\
-\left( \mathcal{M}_\textrm{Ch}^2 \right)_{11} \hat v_{1/2} & \left( \mathcal{M}_\textrm{Ch}^2 \right)_{11} \hat v_{1/2}^2 & 0 \\
 0 & 0 & \left( \mathcal{M}_\textrm{Ch}^2 \right)_{33}
\end{pmatrix},
\end{equation}
where 
\begin{subequations}
\begin{align}
\left( \mathcal{M}_\textrm{Ch}^2 \right)_{11} ={}& \mu_{11}^2 + \lambda_{1111} \hat v_1^2 + \lambda_{1112}^\mathrm{R}\hat v_1 \hat v_2 + \frac{1}{2} \lambda_{1122} \hat v_2^2,\\
\left( \mathcal{M}_\textrm{Ch}^2 \right)_{33} ={}& \mu_{33}^2 + \frac{1}{2}\lambda_{1133} \hat v_1^2 + \lambda_{1233}^\mathrm{R} \hat v_1 \hat v_2 + \frac{1}{2} \lambda_{2233} \hat v_2^2.
\end{align}
\end{subequations}

The neutral mass-squared matrix in the basis $\{\eta_1,\, \eta_2,\, \chi_1,\, \chi_2\}{-}\{\eta_3,\,\chi_3\}$ is:
\begin{equation}\label{Eq:MN2_v1_v2_0}
\mathcal{M}_\mathrm{N}^2 = \scriptstyle \begin{pmatrix}
(\mathcal{M}_\mathrm{N}^2)_{11} & (\mathcal{M}_\mathrm{N}^2)_{12}  & (\mathcal{M}_\mathrm{N}^2)_{13} & -(\mathcal{M}_\mathrm{N}^2)_{13} \hat v_{1/2} & 0 & 0\\
(\mathcal{M}_\mathrm{N}^2)_{12} & (\mathcal{M}_\mathrm{N}^2)_{22}  & (\mathcal{M}_\mathrm{N}^2)_{23} & -(\mathcal{M}_\mathrm{N}^2)_{23} \hat v_{1/2} & 0 & 0\\
(\mathcal{M}_\mathrm{N}^2)_{13} & (\mathcal{M}_\mathrm{N}^2)_{23}  & (\mathcal{M}_\mathrm{N}^2)_{33} & -(\mathcal{M}_\mathrm{N}^2)_{33} \hat v_{1/2} & 0 & 0\\
-(\mathcal{M}_\mathrm{N}^2)_{13} \hat v_{1/2} & -(\mathcal{M}_\mathrm{N}^2)_{23} \hat v_{1/2}  & -(\mathcal{M}_\mathrm{N}^2)_{33} \hat v_{1/2} & (\mathcal{M}_\mathrm{N}^2)_{33} \hat v_{1/2}^2 & 0 & 0\\
0 & 0 & 0 &  0 & (\mathcal{M}_\mathrm{N}^2)_{44} & 0 \\
0 & 0 & 0 & 0 & 0 & (\mathcal{M}_\mathrm{N}^2)_{44}  \\
\end{pmatrix},
\end{equation}
where
\begin{subequations}
\begin{align}
\begin{split}
(\mathcal{M}_\mathrm{N}^2)_{11} ={}& \mu_{11}^2 + 3 \lambda_{1111} \hat v_1^2 + 3 \lambda_{1112}^\mathrm{R} \hat v_1 \hat v_2 + \frac{1}{2}\left( \lambda_{1122} + 2 \lambda_{1212}^\mathrm{R} + \lambda_{1221} \right) \hat v_2^2,
\end{split}\\ 
\begin{split}
(\mathcal{M}_\mathrm{N}^2)_{12} ={}& - \left( \mu_{11}^2 + \lambda_{1111} \hat v_1^2 \right)\hat v_{1/2}^2 + \lambda_{1222}^\mathrm{R} \hat v_2^2+ \frac{1}{2} \left(  \lambda_{1122} + 2\lambda_{1212}^\mathrm{R} + \lambda_{1221}\right) \hat v_1 \hat v_2,
\end{split}\\
\begin{split}
(\mathcal{M}_\mathrm{N}^2)_{13} ={}& \lambda_{1112}^\mathrm{I} \hat v_1 \hat v_2 + \lambda_{1212}^\mathrm{I}\hat v_2^2,
\end{split}\\
\begin{split}
(\mathcal{M}_\mathrm{N}^2)_{22} ={}& \left( \mu_{11}^2 + \lambda_{1111} \hat v_1^2 \right)\hat v_{1/2}^2  + 2 \lambda_{2222} \hat v_2^2 + \frac{1}{2} \left( \lambda_{1122} + \lambda_{1221} + 2 \lambda_{1212}^\mathrm{R} \right) \hat v_1^2 \\
& +  \lambda_{1112}^\mathrm{R} v_1^2 \hat v_{1/2} + 2 \lambda_{1222}^\mathrm{R} \hat v_1 \hat v_2,
\end{split}\\
\begin{split}
(\mathcal{M}_\mathrm{N}^2)_{23} ={}& \lambda_{1212}^\mathrm{I} \hat v_1 \hat v_2 + \lambda_{1222}^\mathrm{I} \hat v_2^2,
\end{split}\\
\begin{split}
(\mathcal{M}_\mathrm{N}^2)_{33} ={}& \mu_{11}^2 + \lambda_{1111} \hat v_1^2 + \lambda_{1112}^\mathrm{R} \hat v_1 \hat v_2 + \frac{1}{2}\left( \lambda_{1122} - 2 \lambda_{1212}^\mathrm{R} + \lambda_{1221} \right) \hat v_2^2,
\end{split}\\
\begin{split}
(\mathcal{M}_\mathrm{N}^2)_{44} ={}& \mu_{33}^2 + \frac{1}{2} \left( \lambda_{1133} + \lambda_{1331} \right) \hat v_1^2 + \frac{1}{2} \left( \lambda_{2233} + \lambda_{2332} \right) \hat v_2^2\\
& + \left( \lambda_{1233}^\mathrm{R} + \lambda_{1332}^\mathrm{R} \right) \hat v_1 \hat v_2.
\end{split}
\end{align}
\end{subequations}

Although $\left\langle h_3 \right\rangle = 0$, it is not possible to rotate away one of the imaginary parts of the quartic couplings, as those fields always appear in pairs proportional to $h_{33}$. Due to the complexity of the analytical expressions for the mass-squared parameters, we do not provide the mass eigenvalues explicitly.

In this implementation:
\begin{itemize}
\item  $\{ \mu_{11}^2,\, \mu_{33}^2,\, \lambda_{1111},\, \lambda_{1122},\, \lambda_{1133},\, \lambda_{1221},\, \lambda_{1331},\, \lambda_{2222},\, \lambda_{2233},\,$  $ \lambda_{2332},\,\lambda_{1112},\, \lambda_{1212},\, \lambda_{1222},$ \\ \hspace*{5pt}$\lambda_{1233}^\mathrm{R},\, \lambda_{1332}^\mathrm{R}\}$ contribute to six scalar masses;
\item  $\{\lambda_{3333},\,$ $ \lambda_{1233}^\mathrm{I},\, \lambda_{1332}^\mathrm{I} \}$ appear only in scalar interactions.
\end{itemize}

It is important to note that this case includes an implementation where all coefficients are real. The two scenarios---one with complex couplings and real vevs, and the other with only real coefficients---are physically distinct. The former allows for CP violation in the scalar sector, while the latter does not.

\newpage
\bigskip\textbf{Case of \boldmath$(\hat{v}_1 e^{i \sigma},\, \hat{v}_2,\, 0)$ and real couplings}

Now, consider the case with complex vevs and real couplings, \textit{i.e.}, $(\mu_{12}^2)^\mathrm{I}=\lambda_{1212}^\mathrm{I} = \lambda_{1112}^\mathrm{I} = \lambda_{1222}^\mathrm{I}= \lambda_{1233}^\mathrm{I}= \lambda_{1332}^\mathrm{I}=0$. The minimisation conditions are given by:
\begin{subequations}\label{Eq:U12_Min_Con_2}
\begin{align}
\mu_{11}^2 ={}& -\lambda_{1111} \hat v_1^2 - \cos \sigma \lambda_{1112}^\mathrm{R} \hat v_1 \hat v_2 - \frac{1}{2}\left( \lambda_{1122} - 2 \lambda_{1212}^\mathrm{R} + \lambda_{1221} \right)\hat v_2^2,\\
(\mu_{12}^2)^\mathrm{R} ={}& -\frac{1}{2} \left[  \lambda_{1112}^\mathrm{R} \hat v_1^2 + \left( 4 \cos \sigma \lambda_{1212}^\mathrm{R} \hat v_1 + \lambda_{1222}^\mathrm{R} \hat v_2\right) \hat v_2 \right],\\
\mu_{22}^2 ={}& - \left( \cos \sigma \lambda_{1222}^\mathrm{R} \hat v_1 + \lambda_{2222} \hat v_2 \right) \hat v_2 - \frac{1}{2}\left( \lambda_{1122} - 2 \lambda_{1212}^\mathrm{R} + \lambda_{1221} \right)\hat v_1^2.
\end{align}
\end{subequations}

The form of the charged mass-squared matrix is the same as in eq.~\eqref{Eq:MCh2_v1_v2_0}. This indicates that it was beneficial to extract the phases as in eq.~\eqref{Eq:Extract_phase}. Now, the elements of the charged mass-squared matrix are: 
\begin{subequations}
\begin{align}
\left( \mathcal{M}_\textrm{Ch}^2 \right)_{11} ={}& \frac{1}{2} \left( 2 \lambda_{1212}^\mathrm{R} - \lambda_{1221} \right) \hat v_2^2,\\
\left( \mathcal{M}_\textrm{Ch}^2 \right)_{33} ={}& \mu_{33}^2 + \frac{1}{2}\lambda_{1133} \hat v_1^2 + \cos\sigma\lambda_{1233}^\mathrm{R} \hat v_1 \hat v_2 + \frac{1}{2} \lambda_{2233} \hat v_2^2.
\end{align}
\end{subequations}

The structure of the neutral mass-squared matrix is identical to eq.~\eqref{Eq:MN2_v1_v2_0}. The elements of the mass-squared matrix are given by:
\begin{subequations}
\begin{align}
(\mathcal{M}_\mathrm{N}^2)_{11} ={}& 2 \left[ \lambda_{1111} \hat v_1^2 + \cos \sigma \left( \lambda_{1112}^\mathrm{R} \hat v_1 + \cos \sigma \lambda_{1212}^\mathrm{R} \hat v_2 \right) \hat v_2^2 \right],\\ 
\begin{split}
(\mathcal{M}_\mathrm{N}^2)_{12} ={}& \left(  \lambda_{1122} - 2\sin^2 \sigma\lambda_{1212}^\mathrm{R}  + \lambda_{1221}\right) \hat v_1 \hat v_2 + \cos \sigma \left( \lambda_{1112}^\mathrm{R} \hat v_1^2 + \lambda_{1222}^\mathrm{R} \hat v_2^2 \right),
\end{split}\\
\begin{split}
(\mathcal{M}_
\mathrm{N}^2)_{13} ={}& -\sin \sigma \left( \lambda_{1112}^\mathrm{R} \hat v_1 + 2 \cos \sigma \lambda_{1212}^\mathrm{R} \hat v_2 \right) \hat v_2,
\end{split}\\
(\mathcal{M}_\mathrm{N}^2)_{22} ={}& 2 \left( \cos^2 \sigma \lambda_{1212}^\mathrm{R} \hat v_1^2 + \cos \sigma \lambda_{1222}^\mathrm{R} \hat v_1 \hat v_2 + \lambda_{2222} \hat v_2^2  \right),\\
\begin{split}
(\mathcal{M}_\mathrm{N}^2)_{23} ={}& -\sin \sigma \left( \lambda_{1222}^\mathrm{R} \hat v_2 + 2 \cos \sigma \lambda_{1212}^\mathrm{R} \hat v_1 \right) \hat v_2,
\end{split}\\
(\mathcal{M}_\mathrm{N}^2)_{33} ={}& 2 \sin^2 \sigma \lambda_{1212}^\mathrm{R} \hat v_2^2,\\
\begin{split}
(\mathcal{M}_\mathrm{N}^2)_{44} ={}& \mu_{33}^2 + \frac{1}{2} \left( \lambda_{1133} + \lambda_{1331} \right) \hat v_1^2 + \frac{1}{2} \left( \lambda_{2233} + \lambda_{2332} \right) \hat v_2^2\\
& + \cos \sigma \left( \lambda_{1233}^\mathrm{R} + \lambda_{1332}^\mathrm{R} \right) \hat v_1 \hat v_2.
\end{split}
\end{align}
\end{subequations}

Except that now $\lambda_{ijkl}^\mathrm{I}=0$, the counting of couplings is identical to the  $(\hat{v}_1,\, \hat{v}_2,\, 0)$ implementation. Due to the presence of a complex phase, CP can be spontaneously violated in the scalar sector.

\bigskip\textbf{Case of \boldmath$(v,\,0,\,0)$}

The vacuum resembles that of the Higgs basis. However, due to mixing, the would-be SM-like Higgs field, $\eta_1$, is not physical since it mixes with the $\{\eta_2,\, \chi_2\}$ fields. This feature is caused by the presence of the $\lambda_{1112}$ term. In other words, $\lambda_{1112}=0$ results in the $\eta_1$ state being associated with the SM-like Higgs boson.

The minimisation conditions are:
\begin{subequations}
\begin{align}
\mu_{11}^2 ={}& - \lambda_{1111} v^2,\\
\mu_{12}^2 ={}& - \frac{1}{2} \lambda_{1112} v^2,
\end{align}
\end{subequations}
where $\mu_{12}^2$ is complex.

The charged mass-squared matrix in the basis $\{h_1^\pm,\, h_2^\pm,\, h_3^\pm \}$ is:
\begin{equation}
\mathcal{M}_\mathrm{Ch}^2 = \mathrm{diag}\left(0,\, \mu_{22}^2 + \frac{1}{2} \lambda_{1122} v^2 ,\, \mu_{33}^2 + \frac{1}{2} \lambda_{1133} v^2   \right).
\end{equation}

The neutral mass-squared matrix in the basis $\{\eta_1,\, \eta_2,\, \chi_1,\, \chi_2 \}{-}\{\eta_3,\, \chi_3\}$ is:
\begin{equation}
\mathcal{M}_\mathrm{N}^2 = \begin{pmatrix}
(\mathcal{M}_\mathrm{N}^2)_{11} & \mathbb{R}\mathrm{e}\big((\mathcal{M}_\mathrm{N}^2)_{12} \big) & 0 & -\mathbb{I}\mathrm{m}\big((\mathcal{M}_\mathrm{N}^2)_{12} \big) & 0  & 0\\
\mathbb{R}\mathrm{e}\big((\mathcal{M}_\mathrm{N}^2)_{12} \big) & (\mathcal{M}_\mathrm{N}^2)_{22}  & 0 & (\mathcal{M}_\mathrm{N}^2)_{24} & 0  & 0\\
0 & 0 & 0 & 0 & 0 & 0 \\
-\mathbb{I}\mathrm{m}\big((\mathcal{M}_\mathrm{N}^2)_{12} \big) & (\mathcal{M}_\mathrm{N}^2)_{24}  & 0 & (\mathcal{M}_\mathrm{N}^2)_{44} & 0 & 0\\
0 & 0 & 0 & 0 & (\mathcal{M}_\mathrm{N}^2)_{55} & 0 \\
0 & 0 & 0 & 0 & 0 & (\mathcal{M}_\mathrm{N}^2)_{55} 
\end{pmatrix},
\end{equation}
where
\begin{subequations}
\begin{align}
(\mathcal{M}_\mathrm{N}^2)_{11} ={}& 2 \lambda_{1111} v^2,\\
(\mathcal{M}_\mathrm{N}^2)_{12} ={}& \lambda_{1112} v^2,\\
(\mathcal{M}_\mathrm{N}^2)_{22} ={}& \mu_{22}^2 + \frac{1}{2} \left( \lambda_{1122} + 2 \lambda_{1212}^\mathrm{R} + \lambda_{1221} \right)v^2,\\
(\mathcal{M}_\mathrm{N}^2)_{24} ={}& -\lambda_{1212}^\mathrm{I} v^2,\\
(\mathcal{M}_\mathrm{N}^2)_{44} ={}& \mu_{22}^2 + \frac{1}{2} \left( \lambda_{1122} - 2 \lambda_{1212}^\mathrm{R} + \lambda_{1221} \right)v^2,\\
(\mathcal{M}_\mathrm{N}^2)_{55} ={}& \mu_{33}^2 + \frac{1}{2} \left( \lambda_{1133} + \lambda_{1331} \right)v^2.
\end{align}
\end{subequations}

Mixing between the $\{\eta_1,\, \eta_2, \chi_2\}$ scalars is caused by the $\{\lambda_{1112}^\mathrm{I},\, \lambda_{1212}^\mathrm{I}\}$ couplings. Due to the fact that $\left\langle h_2 \right\rangle = 0$, without loss of generality, it is possible to re-phase the $h_2$ doublet in order to absorb one of the arguments of the quartic couplings.

In this implementation:
\begin{itemize}
\item  $\{ \mu_{22}^2,\, \mu_{33}^2,\, \lambda_{1111},\, \lambda_{1122},\, \lambda_{1133},\, \lambda_{1221},\, \lambda_{1331},\, \lambda_{1112},\, \lambda_{1212} \}$ give rise to six scalar masses;
\item  $\{\lambda_{2222},\, \lambda_{2233},\, \lambda_{2332},\,$ $\lambda_{3333},\, \lambda_{1222},\, \lambda_{1233},\, \lambda_{1332}\}$ appear only in scalar interactions.
\end{itemize}

\bigskip\textbf{Case of \boldmath$(0,\,0,\,v)$}

There is a single minimisation condition given by:
\begin{equation}
\mu_{33}^2 = - \lambda_{3333} v^2.
\end{equation}

The charged mass-squared matrix in the basis $\{h_1^\pm,\, h_2^\pm,\, h_3^\pm \}$ is:
\begin{equation}
\mathcal{M}_\mathrm{Ch}^2 = \begin{pmatrix}
\left( \mathcal{M}_\textrm{Ch}^2 \right)_{11}  & \left( \mathcal{M}_\textrm{Ch}^2 \right)_{12}^\ast  & 0 \\
\left( \mathcal{M}_\textrm{Ch}^2 \right)_{12}  & \left( \mathcal{M}_\textrm{Ch}^2 \right)_{22} & 0 \\
 0 & 0 & 0
\end{pmatrix},
\end{equation}
where 
\begin{equation}
\left( \mathcal{M}_\textrm{Ch}^2 \right)_{ij} = \mu_{ii}^2 + \frac{1}{2} \lambda_{ij33} v^2.
\end{equation}

The neutral mass-squared matrix in the basis $\{\eta_1,\, \eta_2,\, \chi_1,\, \chi_2\}{-}\{\eta_3\}{-}\{\chi_3\}$ is:
\begin{equation}
\mathcal{M}_\mathrm{N}^2 = \begin{pmatrix}
(\mathcal{M}_\mathrm{N}^2)_{11} & \mathbb{R}\mathrm{e}\left((\mathcal{M}_\mathrm{N}^2)_{12} \right) & 0 & -\mathbb{I}\mathrm{m}\left((\mathcal{M}_\mathrm{N}^2)_{12} \right) & 0  & 0\\
\mathbb{R}\mathrm{e}\left((\mathcal{M}_\mathrm{N}^2)_{12} \right) & (\mathcal{M}_\mathrm{N}^2)_{22} & \mathbb{I}\mathrm{m}\left((\mathcal{M}_\mathrm{N}^2)_{12} \right) & 0 & 0  & 0 \\
0 & \mathbb{I}\mathrm{m}\left((\mathcal{M}_\mathrm{N}^2)_{12} \right) & (\mathcal{M}_\mathrm{N}^2)_{11} & \mathbb{R}\mathrm{e}\left((\mathcal{M}_\mathrm{N}^2)_{12} \right) & 0  & 0\\
-\mathbb{I}\mathrm{m}\left((\mathcal{M}_\mathrm{N}^2)_{12} \right) & 0 & \mathbb{R}\mathrm{e}\left((\mathcal{M}_\mathrm{N}^2)_{12} \right) & (\mathcal{M}_\mathrm{N}^2)_{22} & 0  & 0\\
0 & 0 & 0 & 0 & (\mathcal{M}_\mathrm{N}^2)_{33} & 0 \\
0 & 0 & 0 & 0 & 0 & 0  \\
\end{pmatrix},
\end{equation}
where
\begin{subequations}
\begin{align}
(\mathcal{M}_\mathrm{N}^2)_{ij} ={}& \mu_{ij}^2 + \frac{1}{2} \left( \lambda_{ij33} + \lambda_{i33j} \right) v^2,\text{ for } \{i,\, j\}=\{1,\,2\},\\
(\mathcal{M}_\mathrm{N}^2)_{33} ={}& 2 \lambda_{3333} v^2.
\end{align}
\end{subequations}

Due to the non-vanishing $\{\left(\mu_{12}^2 \right)^\mathrm{I},\, \lambda_{1233}^\mathrm{I},\, \lambda_{1332}^\mathrm{I}\}$ couplings, there is mixing between the $\{\eta_1,\, \eta_2,\, \chi_1,\, \chi_2\}$ states. Two out of the three imaginary couplings causing mixing between the pairs of $\eta$ and $\chi$ fields can be absorbed. The physical states are pairwise degenerate:
\begin{equation}
m_{H_i}^2 =  \frac{1}{4} \left[ 2\left( \mu_{11}^2 + \mu_{22}^2 \right) +  v^2 \left(\lambda_{1133} + \lambda_{2233} + \lambda_{1331} + \lambda_{2332}\right) \pm \Delta \right],~\text{for }i=\{1,\dots,4\},
\end{equation}
where
\begin{equation}
\begin{aligned}
\Delta^2 ={}&  \left[ v^2 \left( \lambda_{1133} - \lambda_{2233} + \lambda_{1331} - \lambda_{2332} \right) + 2 \left( \mu_{11}^2 - \mu_{22}^2\right)\right]^2\\
&  + 4  \left| v^2 \left( \lambda_{1233} + \lambda_{1332}\right) + 2 \mu_{12}^2\right|^2.
\end{aligned}
\end{equation}

In this implementation:
\begin{itemize}
\item  $\{ \mu_{11}^2,\, \mu_{12}^2,\, \mu_{22}^2,\, \lambda_{1133},\, \lambda_{1331},\, \lambda_{2233},\, \lambda_{2332},\, \lambda_{3333},\, \lambda_{1233},\, \lambda_{1332} \}$ contribute to six scalar masses;
\item  $\{\lambda_{1111},\,  \lambda_{1122},\, $ $\lambda_{1221},\, \lambda_{2222},\, \lambda_{1112},\, \lambda_{1212},\, \lambda_{1222}\}$ appear only in scalar interactions.
\end{itemize}

\section[\texorpdfstring{$U(1) \times \mathbb{Z}_2$}{U(1) x Z2}-symmetric 3HDM]{\boldmath$U(1) \times \mathbb{Z}_2$-symmetric 3HDM}\label{Sec:Pot_U1Z2}

The previously discussed $U(1)_2$ symmetry can be extended by a $\mathbb{Z}_2$ symmetry,
\begin{equation}
h_1 \to -h_1, \quad h_2 \to h_2, \quad h_3 \to h_3.
\end{equation}
An equivalent implementation could have been chosen with $h_2$ being odd under $\mathbb{Z}_2$, instead of $h_1$. Regardless of the $\mathbb{Z}_2$ charge, the $U(1) \times \mathbb{Z}_2$-symmetric scalar potential is
\begin{equation}\label{Eq:V_U1Z2}
\begin{aligned}
V_{U(1) \times \mathbb{Z}_2} ={}& V_0 + V^\mathrm{ph}_{U(1) \times \mathbb{Z}_2}\\
={}& \sum_i \mu_{ii}^2 h_{ii} + \sum_i \lambda_{iiii} h_{ii}^2 + \sum_{i<j} \lambda_{iijj} h_{ii} h_{jj} + \sum_{i<j} \lambda_{ijji} h_{ij} h_{ji}\\
& + \left\lbrace \lambda_{1212} h_{12}^2 + \mathrm{h.c.} \right\rbrace.
\end{aligned}
\end{equation}

The form of above potential does not change under the exchange of indices, $1 \leftrightarrow 2$. This observation reduces the number of distinct vacua we have to consider. Only one term appears in the phase-sensitive part, similar (not the term, but case) to the $U(1)_1$-symmetric model. While one might think that a basis transformation could relate the two models, this is not the case. Additionally, the phase from the single phase-dependent coupling $\lambda_{1212}$ is unphysical and can be rotated away. We assume that $\lambda_{1212} \in \mathbb{R}$, unless soft symmetry-breaking terms are introduced. Different implementations are summarised in Table~\ref{Table:U1Z2_Cases}.

{{\renewcommand{\arraystretch}{1.3}
\begin{table}[htb]
\caption{Similar to Table~\ref{Table:U1_U1_Cases}, but now for $U(1) \times \mathbb{Z}_2$. None of these cases violates CP.}
\label{Table:U1Z2_Cases}
\begin{center}
\begin{tabular}{|c|c|c|c|c|} \hline\hline
Vacuum & SYM & $V$ & \begin{tabular}[l]{@{}c@{}} Mixing of the \\ neutral states\end{tabular} & Comments \\ \hline
$(\hat{v}_1,\, \hat{v}_2,\, 0)$ & $-$ & $V_{U(1) \times \mathbb{Z}_2}$ & \footnotesize\begin{tabular}[l]{@{}c@{}} $\{\eta_1,\, \eta_2\}{-}\{\eta_3\}$ \\ ${-}\{\chi_1,\, \chi_2\}{-}\{\chi_3\}$ \end{tabular} & $m_{\eta_3} = m_{\chi_3}$ \\ \hline
$(0,\, \hat{v}_2,\, \hat{v}_3)$ & $-$ & $V_{U(1) \times \mathbb{Z}_2}$ & \footnotesize\begin{tabular}[l]{@{}c@{}} $\{\eta_1\}{-}\{\eta_2,\, \eta_3\}$ \\ ${-}\{\chi_1\}{-}\{\chi_2\}{-}\{\chi_3\}$ \end{tabular} & $m_{\chi_2} = m_{\chi_3} = 0$ \\ \hline
$(0,\, \hat{v}_2,\, \hat{v}_3)$ & $-$ & $V_{U(1) \times \mathbb{Z}_2} + (\mu_{23}^2)^\mathrm{R}$ & \footnotesize\begin{tabular}[l]{@{}c@{}} $\{\eta_1\}{-}\{\eta_2,\, \eta_3\}$ \\ ${-}\{\chi_1\}{-}\{\chi_2, \,\chi_3\}$ \end{tabular} & - \\ \hline
$(v, \,0, \, 0)$ & $\checkmark$ & $V_{U(1) \times \mathbb{Z}_2}$ & diagonal & $m_{\eta_3} = m_{\chi_3}$ \\ \hline
$(0, \,0, \, v)$  & $\checkmark$ & $V_{U(1) \times \mathbb{Z}_2}$ & diagonal & \footnotesize\begin{tabular}[l]{@{}c@{}} $m_{\eta_1} = m_{\chi_1},~m_{\eta_2} = m_{\chi_2}$ \end{tabular} \\ \hline \hline
\end{tabular} 
\end{center}
\end{table}}

\bigskip\textbf{Case of \boldmath$(v,\,0,\,0)$}

The minimsiation condition is:
\begin{equation}
\mu_{11}^2 =- \lambda_{1111} v^2.
\end{equation}

The charged mass-squared matrix in the basis $\{h_1^\pm,\, h_2^\pm,\, h_3^\pm \}$ is:
\begin{equation}
\mathcal{M}_\mathrm{Ch}^2 = \mathrm{diag}\left(0,\, \mu_{22}^2 + \frac{1}{2} \lambda_{1122} v^2 ,\, \mu_{33}^2 + \frac{1}{2} \lambda_{1133} v^2   \right).
\end{equation}

The neutral mass-squared matrix in the basis $\{\eta_1,\, \eta_2,\, \eta_3,\, \chi_1,\, \chi_2,\, \chi_3 \}$ is:
\begin{equation}
\mathcal{M}_\mathrm{N}^2 = \mathrm{diag} \bigg( m_h^2 ,\, m_{\eta_2}^2,\, m_{H_3}^2,\, 0 ,\, m_{\chi_2}^2,\, m_{H_3}^2  \bigg),
\end{equation}
where
\begin{subequations}
\begin{align}
m_h^2 ={}& 2 \lambda_{1111} v^2,\\
m_{\eta_2}^2 ={}& \mu_{22}^2 + \frac{1}{2} \left( \lambda_{1122} + 2 \lambda_{1212} + \lambda_{1221} \right)v^2,\\
m_{H_3}^2 ={}& \mu_{33}^2 + \frac{1}{2} \left( \lambda_{1133} + \lambda_{1331} \right)v^2,\\
m_{\chi_2}^2 ={}& \mu_{22}^2 + \frac{1}{2} \left( \lambda_{1122} - 2 \lambda_{1212} + \lambda_{1221} \right)v^2.
\end{align}
\end{subequations}

In this implementation:
\begin{itemize}
\item $\{ \mu_{22}^2,\, \mu_{33}^2,\, \lambda_{1111},\, \lambda_{1122},\, \lambda_{1133},\, \lambda_{1221},\, \lambda_{1331},\,\lambda_{1212}\}$  contribute to six scalar masses;
\item $\{\lambda_{2222},\, \lambda_{2233},\,$ $ \lambda_{2332},\, \lambda_{3333}\}$  appear only in scalar interactions.
\end{itemize}

\bigskip\textbf{Case of \boldmath$(0,\,0,\,v)$}

The minimsiation condition is:
\begin{equation}
\mu_{33}^2 =- \lambda_{3333} v^2.
\end{equation}

After re-labeling the indices of the scalar doublets, this case becomes identical, both in terms of the charged and neutral mass-squared matrices, to \ref{Sec:U1U1_v00} of $U(1) \times U(1)$ in Section~\ref{Sec:Pot_U1U1}. However it is physically different from the $U(1) \times U(1)$ model. For example, consider the following derivative, assuming $(0,\,0,\,v)$:
\begin{equation}
\dfrac{\partial V}{\partial \chi_1 \,\partial \chi_2 \,\partial \eta_2} = \lambda_{1122} (0) + \lambda_{1221}\,\eta_2 + \lambda_{1212} (\eta_1 - i \chi_1),
\end{equation}
which indicates that the $\lambda_{1212}$ terms is responsible for additional scalar interactions.

In this implementation:
\begin{itemize}
\item  $\{ \mu_{11}^2,\, \mu_{22}^2,\, \lambda_{1133},\, \lambda_{1331},\, \lambda_{2233},\, \lambda_{2332},\, \lambda_{3333}\}$ contribute to five scalar masses;
\item  $\{\lambda_{1111}$ $\lambda_{1122},\, \lambda_{1221},\,\lambda_{2222},$ $\lambda_{1212}\}$ appear only in scalar interactions.
\end{itemize}

\section[\texorpdfstring{$O(2)$}{O(2)}-symmetric 3HDM]{\boldmath$O(2)$-symmetric 3HDM}\label{Sec:Pot_O2}

The $O(2)$-symmetric 3HDM can be written down in two different forms. We follow the notation of Ref.~\cite{deMedeirosVarzielas:2019rrp}. In Ref.~\cite{Darvishi:2019dbh}, the group was presented as $SO(2)$, where the authors assumed all couplings to be real, $\mathrm{CP1} \times SO(2)$, arguing that the symmetry finds its origin in the 2HDM.  Following the work of Ref.~\cite{deMedeirosVarzielas:2019rrp}, we consider two bases:
\begin{itemize}
\item $O(2)$ can be given in terms of the orthogonal rotations,
\begin{equation}\label{Eq:Rep_O2_SO2_Z2}
\begin{pmatrix}
h_1 \\
h_2 \\
h_3
\end{pmatrix} = \begin{pmatrix}
\cos \theta & \sin \theta & 0\\
\sin \theta & -\cos \theta & 0\\
0 & 0 & 1
\end{pmatrix} \begin{pmatrix}
h_1 ^\prime \\
h_2 ^\prime \\
h_3 ^\prime
\end{pmatrix}.
\end{equation}
Invariance under this transformation results in the scalar potential of:
\begin{equation}\label{Eq:V_O2_SO2_Z2_II_cons}
\begin{aligned}
V_{SO(2) \rtimes \mathbb{Z}_2} ={}& \mu_{11}^2 (h_{11} + h_{22}) + \mu_{33}^2 h_{33} + \lambda_{1111} (h_{11} + h_{22})^2 + \lambda_{3333} h_{33}^2\\
& + \lambda_{1133} (h_{11} h_{33} + h_{22} h_{33}) + \lambda_{1221} (h_{12} h_{21} - h_{11}h_{22})\\
& + \lambda_{1331} (h_{13} h_{31} + h_{23} h_{32}) +  \lambda_{1212} \left( h_{12}^2 + h_{21}^2  - 2 h_{11} h_{22}\right)\\
& + \{\lambda_{1313} \left( h_{13}^2 + h_{23}^2 \right) + \mathrm{h.c.} \},
\end{aligned}
\end{equation}
which is the form equivalent to the one of Ref.~\cite{deMedeirosVarzielas:2019rrp}. Alternatively, we can rename $\lambda_{1212}$ as $\Lambda$,
\begin{equation}\label{Eq:Incr_sym}
\Lambda \equiv  \frac{1}{2}\left( 2\lambda_{1111} - \lambda_{1122}  - \lambda_{1221}\right),
\end{equation}
which yields $\lambda_{1122} = 2 (\lambda_{1111} - \frac{1}{2} \lambda_{1221} - \lambda_{1212})$. Then, the scalar potential becomes:
\begin{equation}\label{Eq:V_O2_SO2_Z2}
\begin{aligned}
V_{SO(2) \rtimes \mathbb{Z}_2} ={}& \mu_{11}^2 (h_{11} + h_{22}) + \mu_{33}^2 h_{33} + \lambda_{1111} (h_{11}^2 + h_{22}^2) + \lambda_{3333} h_{33}^2 + \lambda_{1122} h_{11} h_{22}\\
& + \lambda_{1133} (h_{11} h_{33} + h_{22} h_{33}) + \lambda_{1221} h_{12} h_{21} + \lambda_{1331} (h_{13} h_{31} + h_{23} h_{32})\\
& + \Lambda \left( h_{12}^2 + h_{21}^2 \right) + \{\lambda_{1313} \left( h_{13}^2 + h_{23}^2 \right) + \mathrm{h.c.} \}.
\end{aligned}
\end{equation}

\item  $O(2)$ can be given in terms of the $U(1)_1$ re-phasing transformation,
\begin{equation}\label{Eq:Rep_O2_U1_Z2}
\begin{pmatrix}
h_1 \\
h_2 \\
h_3
\end{pmatrix} = \begin{pmatrix}
0 & e^{- i \theta} & 0 \\
e^{i \theta} & 0 & 0 \\
0 & 0 & 1
\end{pmatrix} \begin{pmatrix}
h_1 ^\prime \\
h_2 ^\prime \\
h_3 ^\prime
\end{pmatrix},
\end{equation}
which combines $U(1)_1$ and $\mathbb{Z}_2$ transformations. To be more precise, it is actually $S_2$. We recall that $\mathbb{Z}_2 \cong S_2$. The scalar potential is given by:
\begin{equation}\label{Eq:V_O2_U1_Z2}
\begin{aligned}
V_{U(1)_1 \rtimes \mathbb{Z}_2} ={}& \mu_{11}^2 (h_{11} + h_{22}) + \mu_{33}^2 h_{33} + \lambda_{1111} (h_{11}^2 + h_{22}^2) + \lambda_{3333} h_{33}^2 + \lambda_{1122} h_{11} h_{22}\\
& + \lambda_{1133} (h_{11} h_{33} + h_{22} h_{33}) + \lambda_{1221} h_{12} h_{21} + \lambda_{1331} (h_{13} h_{31} + h_{23} h_{32})\\
& + \left\lbrace \lambda_{1323} h_{13} h_{23} + \mathrm{h.c.} \right\rbrace.
\end{aligned}
\end{equation}
\end{itemize}

The above $O(2)$-symmetric 3HDM potentials differ by the $h_{ij}^2 \leftrightarrow h_{13}h_{23}$ terms. Nevertheless, these two representations,  eqs.~\eqref{Eq:Rep_O2_SO2_Z2} and \eqref{Eq:Rep_O2_U1_Z2}, are connected via
\begin{equation}\label{Eq:Rot_O2_SO2_U1}
\begin{pmatrix}
\cos \theta & \sin \theta & 0 \\
\sin \theta & -\cos \theta & 0 \\
0 & 0 & 1
\end{pmatrix} = \frac{1}{\sqrt{2}} \begin{pmatrix}
1 & 1 & 0 \\
i & -i & 0 \\
0 & 0 & \sqrt{2}
\end{pmatrix} \begin{pmatrix}
0 & e^{- i \theta} & 0 \\
e^{i \theta} & 0 & 0 \\
0 & 0 & 1
\end{pmatrix}
\frac{1}{\sqrt{2}} \begin{pmatrix}
1 & -i & 0 \\
1 & i & 0 \\
0 & 0 & \sqrt{2}
\end{pmatrix}.
\end{equation}
which indicates that both $V_{SO(2) \rtimes \mathbb{Z}_2}$ and $V_{U(1)_1 \rtimes \mathbb{Z}_2}$ have an identical structure in a common basis.

Different vacuum configurations are summarised in Tables~\ref{Table:O2_Cases_1} and~\ref{Table:O2_Cases_2}.

{\renewcommand{\arraystretch}{1.3}
\begin{table}[htb]
\caption{Similar to Table~\ref{Table:U1_U1_Cases}, but now for $SO(2) \rtimes \mathbb{Z}_2$, with the scalar potential given by eq.~\eqref{Eq:V_O2_SO2_Z2}. There may be CP violation in the $(0,\, \hat{v}_2,\, \hat{v}_3)$ implementation with $\mu_{23}^2$.}
\label{Table:O2_Cases_1}
\begin{center}
\begin{tabular}{|c|c|c|c|c|} \hline\hline
Vacuum & SYM &$V$ & \begin{tabular}[l]{@{}c@{}} Mixing of the \\ neutral states\end{tabular} & Comments \\ \hline
$(\hat{v}_1 e^{i \sigma},\, \hat{v}_2,\, 0)$ & $-$ &  $V_{SO(2) \rtimes \mathbb{Z}_2},\,\Lambda=0$ & \footnotesize\begin{tabular}[l]{@{}c@{}} $\{\eta_1,\, \eta_2\}{-}\{\eta_3,\, \chi_3\}$\\${-}\{\chi_1\}{-}\{\chi_2\}$ \end{tabular} & \footnotesize$m_{\eta_{(1,2)}} = m_{\chi_1} = m_{\chi_2} = 0$ \\ \hline
$(\frac{v}{\sqrt{2}} e^{i \sigma},\, \pm \frac{v}{\sqrt{2}},\, 0)$ & $-$ &  $V_{SO(2) \rtimes \mathbb{Z}_2} + \mu_{12}^2$ & \footnotesize\begin{tabular}[l]{@{}c@{}} $\{\eta_1,\, \eta_2,\, \chi_1,\, \chi_2\}$\\${-}\{\eta_3,\, \chi_3\}$ \end{tabular} & - \\ \hline
$(i\frac{v}{\sqrt{2}},\, \pm \frac{v}{\sqrt{2}},\, 0)$ & $-$ &  $V_{SO(2) \rtimes \mathbb{Z}_2}$ & \footnotesize\begin{tabular}[l]{@{}c@{}} $\{\eta_1,\, \eta_2\}{-}\{\eta_3\}$\\${-}\{\chi_1,\,\chi_2\}{-}\{\chi_3\}$ \end{tabular} & $m_{\eta_{(1,2)}} = m_{\chi_{(1,2)}}$  \\ \hline
$(\hat{v}_1,\, \hat{v}_2,\, 0)$ & $-$ &  $V_{SO(2) \rtimes \mathbb{Z}_2}$ & \footnotesize\begin{tabular}[l]{@{}c@{}} $\{\eta_1,\, \eta_2\}{-}\{\eta_3\}$\\${-}\{\chi_1,\,\chi_2\}{-}\{\chi_3\}$ \end{tabular} & $m_{\eta_{(1,2)}} = m_{\chi_{(1,2)}} = 0$ \\ \hline
$(0,\, \hat{v}_2,\, \hat{v}_3)$ & $-$& $V_{SO(2) \rtimes \mathbb{Z}_2}$ & \footnotesize\begin{tabular}[l]{@{}c@{}} $\{\eta_1\}{-}\{\eta_2,\, \eta_3\}$\\${-}\{\chi_1\}{-}\{\chi_2,\,\chi_3\}$ \end{tabular} & $ m_{\eta_1} = m_{\chi_{(2,3)}} = 0$ \\ \hline
$(0,\, \hat{v}_2,\, \hat{v}_3)$ & $-$& $V_{SO(2) \rtimes \mathbb{Z}_2} + \mu_{23}^2$ & \footnotesize\begin{tabular}[l]{@{}c@{}} $\{\eta_1,\, \chi_1\}$\\${-}\{\eta_2,\, \eta_3,\,\chi_2,\,\chi_3\}$ \end{tabular} & - \\ \hline
$(v,\,0,\,0)$ & $-$ & $V_{SO(2) \rtimes \mathbb{Z}_2}$  & diagonal & $m_{\eta_2} = m_{\chi_1} = 0$\\ \hline
$(v,\,0,\,0)$ & $-$ & $V_{SO(2) \rtimes \mathbb{Z}_2} + (\mu_{23}^2)^\mathrm{R}$  & \footnotesize\begin{tabular}[l]{@{}c@{}} $\{\eta_1\}{-}\{\chi_1\}$ \\  ${-}\{\eta_2,\, \eta_3,\, \chi_2,\, \chi_3\}$ \end{tabular} & -\\ \hline
$(0,\,0,\,v)$ & $\checkmark$ & $V_{SO(2) \rtimes \mathbb{Z}_2}$ & diagonal & \footnotesize\begin{tabular}[l]{@{}c@{}} $m_{h_1^+} = m_{h_2^+}$, \\ $m_{\eta_1} = m_{\eta_2},~m_{\chi_1} = m_{\chi_2}$ \end{tabular} \\ \hline \hline
\end{tabular} 
\end{center}
\end{table}}

{\renewcommand{\arraystretch}{1.3}
\begin{table}[htb]
\caption{Similar to Table~\ref{Table:U1_U1_Cases}, but now for $U(1)_1 \rtimes \mathbb{Z}_2$, with the scalar potential given by eq.~\eqref{Eq:V_O2_U1_Z2}. In one case the minimisation conditions can lead to a higher $O(2) \times U(1)$ symmetry. None of these cases violates CP.}
\label{Table:O2_Cases_2}
\begin{center}
\begin{tabular}{|c|c|c|c|c|} \hline\hline
Vacuum & SYM &$V$ & \begin{tabular}[l]{@{}c@{}} Mixing of the \\ neutral states\end{tabular} & Comments \\ \hline
$(\hat{v}_1 ,\, \hat{v}_2,\, 0)$ & $-$ &  $V_{U(1)_1 \rtimes \mathbb{Z}_2},\,\Lambda=0$ & \footnotesize\begin{tabular}[l]{@{}c@{}} $\{\eta_1,\, \eta_2\}{-}\{\eta_3,\,\chi_3\}$\\${-}\{\chi_1\}{-}\{\chi_2\}$ \end{tabular} & \footnotesize\begin{tabular}[l]{@{}c@{}} $m_{\eta_{(1,2)}} = m_{\chi_1}= m_{\chi_2} = 0$ \end{tabular} \\ \hline
$(\hat{v}_1 ,\, \hat{v}_2,\, 0)$ & $-$ &  $V_{U(1)_1 \rtimes \mathbb{Z}_2} + (\mu_{12}^2)^\mathrm{R}$ & \footnotesize\begin{tabular}[l]{@{}c@{}} $\{\eta_1,\, \eta_2\}{-}\{\eta_3,\,\chi_3\}$\\${-}\{\chi_1,\,\chi_2\}$ \end{tabular} & - \\ \hline
$(\frac{v}{\sqrt{2}} ,\, \pm \frac{v}{\sqrt{2}},\, 0)$ & $-$ &  $V_{U(1)_1 \rtimes \mathbb{Z}_2}$ & \footnotesize\begin{tabular}[l]{@{}c@{}} $\{\eta_1,\, \eta_2\}{-}\{\eta_3,\,\chi_3\}$\\${-}\{\chi_1\}{-}\{\chi_2\}$ \end{tabular} & \footnotesize $m_{\chi_1} = m_{\chi_2} = 0$ \\ \hline
$(\frac{v}{\sqrt{2}} ,\, \pm \frac{v}{\sqrt{2}},\, 0)$ & $-$ &  $V_{U(1)_1 \rtimes \mathbb{Z}_2} + (\mu_{12}^2)^\mathrm{R}$ & \footnotesize\begin{tabular}[l]{@{}c@{}} $\{\eta_1,\, \eta_2\}{-}\{\eta_3,\,\chi_3\}$\\${-}\{\chi_1,\,\chi_2\}$ \end{tabular} & - \\ \hline
$(0,\, \hat v_2,\, \hat v_3)$ & $-$& $V_{O(2) \times U(1)}$ & \footnotesize\begin{tabular}[l]{@{}c@{}} $\{\eta_1\}{-}\{\eta_2,\, \eta_3\}$\\${-}\{\chi_1\}{-}\{\chi_2\}{-}\{\chi_3\}$ \end{tabular} & - \\ \hline
$(v,\,0,\,0)$ & $-$ & $V_{U(1)_1 \rtimes \mathbb{Z}_2}$  & diagonal & \footnotesize\begin{tabular}[l]{@{}c@{}} $m_{\eta_2} = m_{\chi_2},\,m_{\eta_3} = m_{\chi_3}$ \end{tabular} \\ \hline
$(0,\,0,\,v)$ & $\checkmark$ & $V_{U(1)_1 \rtimes \mathbb{Z}_2}$ & \footnotesize\begin{tabular}[l]{@{}c@{}} $\{\eta_1,\, \eta_2\}{-}\{\eta_3\}$\\${-}\{\chi_1,\,\chi_2\}{-}\{\chi_3\}$ \end{tabular} & \footnotesize\begin{tabular}[l]{@{}c@{}} $m_{h_1^+} = m_{h_2^+}$ \\ Two pairs of neutral \\ mass-degenerate states \end{tabular} \\ \hline \hline
\end{tabular} 
\end{center}
\end{table}}

\bigskip\textbf{\boldmath$ SO(2) \rtimes \mathbb{Z}_2$: case of \boldmath$(0,\,0,\,v)$}

The minimisation condition is given by:
\begin{equation}
\mu_{33}^2 = - \lambda_{3333} v^2,
\end{equation}

The charged mass-squared matrix in the basis $\{h_1^\pm,\, h_2^\pm,\, h_3^\pm \}$ is:
\begin{equation}
\mathcal{M}_\mathrm{Ch}^2 = \mathrm{diag} \left( \mu_{11}^2 + \frac{1}{2} \lambda_{1133}v^2 ,\, \mu_{11}^2 + \frac{1}{2} \lambda_{1133}v^2 ,\, 0 \right).
\end{equation}

The neutral mass-squared matrix in the basis $\{\eta_1,\, \eta_2,\, \eta_3,\, \chi_1,\, \chi_2,\, \chi_3 \}$ is:
\begin{equation}
\mathcal{M}_\mathrm{N}^2 = \mathrm{diag} \left( m_\eta^2,\, m_\eta^2,\, m_h^2,\, m_\chi^2,\, m_\chi^2,\, 0 \right),
\end{equation}
where
\begin{subequations}
\begin{align}
m_\eta^2 ={}& \mu_{11}^2 + \frac{1}{2} \left( \lambda_{1133} + 2 \lambda_{1313}^\mathrm{R} + \lambda_{1331} \right) v^2,\\
m_h^2 ={}& 2 \lambda_{3333} v^2,\\
m_\chi^2 ={}& \mu_{11}^2 + \frac{1}{2} \left( \lambda_{1133} - 2 \lambda_{1313}^\mathrm{R} + \lambda_{1331} \right) v^2,
\end{align}
\end{subequations}
and the $h$ state is associated with the SM-like Higgs boson. 

In this implementation:
\begin{itemize}
\item  $\{ \mu_{11}^2,\, \lambda_{1133},\, \lambda_{1331},\, \lambda_{3333}, \lambda_{1313}^\mathrm{R} \}$ contribute to four scalar masses;
\item  $\{\lambda_{1111},\, \lambda_{1122},\, \lambda_{1221}\}$ appear only in scalar interactions.
\end{itemize}

\bigskip\textbf{\boldmath$ U(1)_1 \rtimes \mathbb{Z}_2$: case of \boldmath$(0,\,0,\,v)$}

The minimisation condition is:
\begin{equation}
\mu_{33}^2 = - \lambda_{3333} v^2.
\end{equation}

The charged mass-squared matrix in the basis $\{h_1^\pm,\, h_2^\pm,\, h_3^\pm \}$ is:
\begin{equation}
\mathcal{M}_\mathrm{Ch}^2 = \mathrm{diag} \left( \mu_{11}^2 + \frac{1}{2} \lambda_{1133}v^2 ,\, \mu_{11}^2 + \frac{1}{2} \lambda_{1133}v^2 ,\, 0 \right).
\end{equation}

The neutral mass-squared matrix in the basis $\{\eta_1,\, \eta_2,\, \eta_3,\, \chi_1,\, \chi_2,\, \chi_3 \}$ is:
\begin{equation}
\mathcal{M}_\mathrm{N}^2 = \mathrm{diag}\Bigg( \begin{pmatrix}
(\mathcal{M}_\mathrm{N}^2)_{11} & (\mathcal{M}_\mathrm{N}^2)_{12}\\
(\mathcal{M}_\mathrm{N}^2)_{12} & (\mathcal{M}_\mathrm{N}^2)_{11}
\end{pmatrix},\, (\mathcal{M}_\mathrm{N}^2)_{33},\,
\begin{pmatrix}
(\mathcal{M}_\mathrm{N}^2)_{11} & -(\mathcal{M}_\mathrm{N}^2)_{12}\\
-(\mathcal{M}_\mathrm{N}^2)_{12}  & (\mathcal{M}_\mathrm{N}^2)_{11}
\end{pmatrix},\, 0 \Bigg),
\end{equation}
where
\begin{subequations}
\begin{align}
(\mathcal{M}_\mathrm{N}^2)_{11} ={}& \mu_{11}^2 + \frac{1}{2} \left( \lambda_{1133} + \lambda_{1331} \right) v^2,\\
(\mathcal{M}_\mathrm{N}^2)_{12} ={}& \frac{1}{2} \lambda_{1323} v^2,\\
(\mathcal{M}_\mathrm{N}^2)_{33} ={}& 2 \lambda_{3333} v^2.
\end{align}
\end{subequations}
The mass-squared parameters of the mixed states are:
\begin{equation}
m_{H_i}^2 = \mu_{11}^2 + \frac{1}{2} v^2 \left( \lambda_{1133} + \lambda_{1331} \pm \lambda_{1323}^\mathrm{R} \right).
\end{equation}

In this implementation:
\begin{itemize}
\item  $\{ \mu_{11}^2,\, \lambda_{1133},\, \lambda_{1331},\, \lambda_{3333},\, \lambda_{1323} \}$ contribute to four scalar masses;
\item  $\{\lambda_{1111},\, \lambda_{1122},\, \lambda_{1221}\}$ appear only in scalar interactions.
\end{itemize}

\section[\texorpdfstring{$ O(2) \times U(1)$}{ O(2) x U(1)}-symmetric 3HDM]{\boldmath$ O(2) \times U(1)$-symmetric 3HDM}\label{Sec:Ident_O2_U1}

We have already stumbled across the $O(2) \times U(1)$ symmetry in the context of additional continuous symmetries of the $S_3$-symmetric scalar potential, when the minimisation conditions require $\lambda_4 = \lambda_7 = 0$, see Section~\ref{Sec:S3_br_cont_symm}. Actually this case was mentioned (in terms of operators but the symmetry was not identified) previously in Ref.~\cite{Ivanov:2012ry,Ivanov:2012fp}, but was rejected since it resulted in a continuous symmetry and the authors were interested in discrete symmetries. The symmetry was later identified in Ref.~\cite{KunMT} as $SO(2) \times U(1)$ and the notation was corrected in Ref.~\cite{Kuncinas:2020wrn} to that of $O(2) \times U(1)$. Later, it was once again re-discovered in Ref.~\cite{Kuncinas:2024zjq}.

\subsection{Discussion}

There are several possible paths to identify $O(2) \times U(1)$. Consider that we start with the $U(1) \times U(1)$-symmetric scalar potential. Then, we can enforce an additional $S_2$ symmetry acting on a pair of doublets $\{h_1,\, h_2\}$, which we shall denote $S_2(h_1,h_2)$. As a result,  $\left[ U(1) \times U(1) \right] \rtimes S_2$ can be expressed in terms of the transformation
\begin{equation}\label{Eq:U1_U1_S2_Rgen}
\begin{pmatrix}
h_1 \\
h_2 \\
h_3
\end{pmatrix} = \begin{pmatrix}
0 & e^{i \theta_1} & 0 \\
e^{i \theta_2} & 0 & 0 \\
0 & 0 & 1
\end{pmatrix} \begin{pmatrix}
h_1 ^\prime \\
h_2 ^\prime \\
h_3 ^\prime
\end{pmatrix},
\end{equation}
with $V(h_i^\prime) = V(h_i)$. The scalar potential, invariant under this transformation, is given by:
\begin{equation}\label{Eq:V_U1_U1_S2}
\begin{aligned}
V ={}& \mu_{11}^2 (h_{11} + h_{22}) + \mu_{33}^2 h_{33} + \lambda_{1111} (h_{11}^2 + h_{22}^2) + \lambda_{3333} h_{33}^2 + \lambda_{1122} h_{11} h_{22}\\
& + \lambda_{1133} (h_{11} h_{33} + h_{22} h_{33}) + \lambda_{1221} h_{12} h_{21} + \lambda_{1331} (h_{13} h_{31} + h_{23} h_{32}).
\end{aligned}
\end{equation}
Actually, $U(1)_1 \rtimes S_2(h_1,h_3)$ (or equivalently $U(1)_1 \rtimes S_2(h_2,h_3)$) or else $U(1)_2 \rtimes S_2(h_1,h_3)$ (or equivalently $U(1)_2 \rtimes S_2(h_2,h_3)$) result in an identical structure of the scalar potential.

The generators of $\left[ U(1) \times U(1) \right] \rtimes S_2$ are given by:
\begin{subequations}
\begin{align}
g_1 ={}& \mathrm{diag}(e^{i \theta_1},\, 1,\, 1), \\
g_2 ={}& \mathrm{diag}(1,\, e^{i \theta_2},\, 1), \\
g_3 ={}& \begin{pmatrix}
0 & 1 & 0\\
1 & 0 & 0\\
0 & 0 & 1
\end{pmatrix},
\end{align}
\end{subequations}
where the phase generators are not unique due to the overall $U(1)_Y$ symmetry of the potential. Since $S_2$ is in the reducible basis and $S_2 \cong \mathbb{Z}_2$, we can write the scalar potential in yet another basis,
\begin{equation}\label{Eq:Rot_tr_S2_to_Z2_3D}
\begin{pmatrix}
h_1 \\
h_2 \\
h_3
\end{pmatrix} = \begin{pmatrix}
\cos \theta & \sin \theta & 0\\
-\sin \theta & \cos \theta & 0\\
0 & 0 & 1
\end{pmatrix} \begin{pmatrix}
h_1^\prime \\
h_2^\prime \\
h_3^\prime
\end{pmatrix},
\end{equation}
with $\theta= \pm \pi/4$. Regardless of the sign of the $\theta$ rotation, the scalar potential of eq.~\eqref{Eq:V_U1_U1_S2} can be presented as
\begin{equation}\label{Eq:V_U1_U1_Z2}
\begin{aligned}
V ={}& \mu_{11}^2 (h_{11} + h_{22}) + \mu_{33}^2 h_{33} + \lambda_{a} (h_{11}^2 + h_{22}^2) + \lambda_b h_{12}h_{21}  + \lambda_c h_{11} h_{22} + \lambda_{3333} h_{33}^2\\
&  +\lambda_{1133} (h_{11} h_{33} + h_{22} h_{33})   + \lambda_{1331} (h_{13} h_{31} + h_{23} h_{32})  + \frac{1}{2}\Lambda\left( h_{12}^2 + h_{21}^2 \right),
\end{aligned}
\end{equation}
where
\begin{subequations}\label{Eq:la_lb_lc}
\begin{align}
\lambda_a ={}& \frac{1}{2} \lambda_{1111} + \frac{1}{4} \lambda_{1122} + \frac{1}{4} \lambda_{1221},\\
\lambda_b ={}& \lambda_{1111} - \frac{1}{2} \lambda_{1122} + \frac{1}{2} \lambda_{1221},\\
\lambda_c ={}& \lambda_{1111} + \frac{1}{2} \lambda_{1122} - \frac{1}{2} \lambda_{1221},\\
\Lambda ={}& \frac{1}{2}\left( 2\lambda_{1111} - \lambda_{1122}  - \lambda_{1221}\right) = -2 \lambda_a + \lambda_b + \lambda_c,
\end{align}
\end{subequations}
where $\Lambda$ was defined in eq.~\eqref{Eq:Incr_sym}. In this basis, the transformation of eq.~\eqref{Eq:U1_U1_S2_Rgen} becomes
\begin{equation}
\begin{aligned}
&\begin{pmatrix}
\cos ( \pm\frac{\pi}{4}) & \sin (\pm \frac{\pi}{4}) & 0\\
-\sin ( \pm\frac{\pi}{4}) & \cos ( \pm\frac{\pi}{4}) & 0\\
0 & 0 & 1
\end{pmatrix}
\begin{pmatrix}
0 & e^{i \theta_1} & 0\\
e^{i \theta_2} & 0 & 0\\
0 & 0 & 1
\end{pmatrix}
\begin{pmatrix}
\cos (\pm \frac{\pi}{4}) & \sin (\pm \frac{\pi}{4}) & 0\\
-\sin (\pm \frac{\pi}{4}) & \cos ( \pm\frac{\pi}{4}) & 0\\
0 & 0 & 1
\end{pmatrix}^\dagger= \\
&\hspace{120pt}= \frac{1}{2}\begin{pmatrix}
\pm\left( e^{i \theta_1} + e^{i \theta_2} \right) & \left( e^{i \theta_1} - e^{i \theta_2} \right) & 0 \\
-\left( e^{i \theta_1} - e^{i \theta_2} \right) & \mp\left( e^{i \theta_1} + e^{i \theta_2} \right) & 0 \\
0 & 0 & 2
\end{pmatrix} \subset U(2).
\end{aligned} 
\end{equation}
Note that the most general $U(2)$ transformation can be parameterised in terms of four independent angles.

Another possibility to get the $O(2) \times U(1)$-symmetric potential has its origin in the $S_3$-symmetric 3HDM. As was mentioned in Section~\ref{Sec:S3_br_cont_symm}, the $S_3$-symemtric scalar potential with $\lambda_4 = \lambda_7 = 0$ results in a larger $O(2) \times U(1)$ symmetry. One could impose the $\mathbb{Z}_2$ or the $U(1)$ symmetry on $S_3$ to eliminate the phase-sensitive part. Notably, $S_3 \times \mathbb{Z}_2 \cong D_6$, which generators can be presented in terms of the phase transformations and a permutation symmetry acting on two doublets, with some flexibility in signs. The largest realisable cyclic group for 3HDMs is $\mathbb{Z}_4$, while the higher ones result a $U(1)_1$-symmetric potential. Consequently, $D_n$'s for $n \geq 5$ result in eq.~\eqref{Eq:V_U1_U1_S2}.

Finally, since $O(2) \times U(1)$ contains the $O(2)$ group, it would be instructive to discuss what happens when the $U(1)$ group is applied, which could be viewed as a way to remove the phase-sensitive couplings from the $O(2)$-symmetric 3HDM. We emphasise that there are two distinct representations of $O(2)$, these were given by eqs.~\eqref{Eq:Rep_O2_SO2_Z2} and \eqref{Eq:Rep_O2_U1_Z2}. By applying $U(1)_2$ or $U(1)_{h_3}$ to the $SO(2) \rtimes \mathbb{Z}_2$ scalar potential of eq.~\eqref{Eq:V_O2_SO2_Z2}, the $\lambda_{1313}$ coupling is forced to vanish:
\begin{equation}\label{Eq:V_O2_U1_expl_O2}
\begin{aligned}
V_{\left[ SO(2) \rtimes \mathbb{Z}_2 \right] \times U(1) } ={}& \mu_{11}^2 (h_{11} + h_{22}) + \mu_{33}^2 h_{33} + \lambda_{1111} (h_{11}^2 + h_{22}^2) + \lambda_{3333} h_{33}^2 + \lambda_{1122} h_{11} h_{22}\\
& + \lambda_{1133} (h_{11} h_{33} + h_{22} h_{33}) + \lambda_{1221} h_{12} h_{21} + \lambda_{1331} (h_{13} h_{31} + h_{23} h_{32})\\
& + \Lambda \left( h_{12}^2 + h_{21}^2 \right).
\end{aligned}
\end{equation}
The scalar potential is then symmetric under 
\begin{equation}
\begin{pmatrix}
h_1 \\
h_2 \\
h_3
\end{pmatrix} = \begin{pmatrix}
e^{i \beta} & 0 & 0 \\
0 & e^{i \beta} & 0 \\
0 & 0 & 1
\end{pmatrix} \begin{pmatrix}
 \cos \alpha &  \sin \alpha & 0\\
 \sin \alpha &  -\cos \alpha & 0 \\
 0 & 0 & 1
\end{pmatrix} \begin{pmatrix}
h_1 ^\prime \\
h_2 ^\prime \\
h_3 ^\prime
\end{pmatrix}.
\end{equation}

Alternatively, if one starts with the $U(1)_1 \rtimes \mathbb{Z}_2$-symmetric 3HDM from eq.~\eqref{Eq:V_O2_U1_Z2}, the $h_{13}h_{23}$ term must be removed. Given the underlying $U(1)_1$ symmetry in the scalar potential, this term can be eliminated by applying $U(1)_2$ or any of the $U(1)_{h_i}$ symmetries, resulting in the scalar potential of:
\begin{equation}\label{Eq:V_O2_U1_expl_U1}
\begin{aligned}
V_{\left[ U(1)_1 \rtimes \mathbb{Z}_2 \right] \times U(1) } ={}& \mu_{11}^2 (h_{11} + h_{22}) + \mu_{33}^2 h_{33} + \lambda_{1111} (h_{11}^2 + h_{22}^2) + \lambda_{3333} h_{33}^2 + \lambda_{1122} h_{11} h_{22}\\
& + \lambda_{1133} (h_{11} h_{33} + h_{22} h_{33}) + \lambda_{1221} h_{12} h_{21} + \lambda_{1331} (h_{13} h_{31} + h_{23} h_{32}),
\end{aligned}
\end{equation}
which is exactly the one presented in eq.~\eqref{Eq:V_U1_U1_S2}. These two are connected via a basis transformation
\begin{equation}\label{Eq:h_Rot1_h}
\begin{array}{@{}c@{}c@{}c@{}c@{}}
\begin{pmatrix}
h_1 \\
h_2 \\
h_3
\end{pmatrix}
&{}={}&
\frac{1}{\sqrt{2}}\begin{pmatrix}
-e^{-i \frac{\pi}{4}} & -e^{i \frac{\pi}{4}} & 0 \\
e^{-i \frac{\pi}{4}} & -e^{i \frac{\pi}{4}} & 0 \\
0 & 0 & \sqrt{2}
\end{pmatrix}
&
\begin{pmatrix}
h_1 ^\prime \\
h_2 ^\prime \\
h_3 ^\prime
\end{pmatrix}.
\\[1ex]
\mathclap{\scriptstyle \left[ U(1)_1 \rtimes \mathbb{Z}_2 \right] \times U(1) } &&
\mathclap{ } &
\mathclap{\scriptstyle \left[ SO(2) \rtimes \mathbb{Z}_2 \right] \times U(1)}
\end{array}
\end{equation}

Let us consider yet another approach of deriving the $O(2) \times U(1)$-symmetric scalar potential. Notably, the scalar potential of eq.~\eqref{Eq:V_U1_U1_S2} can be derived from the most general 3HDM by enforcing invariance under the transformation given by:
\begin{equation}\label{Eq:O2_U1_ex_pr1}
g= \begin{pmatrix}
0 & e^{i \alpha} & 0 \\
-e^{-i \alpha} & 0 & 0\\
0 & 0 & 1
\end{pmatrix}.
\end{equation}
Applying the generator four times we get $g^4=e$. This generator defines a $\mathbb{Z}_4$ symmetry, while the presence of a free phase $\alpha$ suggests a family of groups parameterised by $\alpha$. In Refs.~\cite{Ivanov:2012ry,Ivanov:2012fp} there was an attempt to apply the $Q_8$ symmetry,
\begin{equation}\label{Eq:O2_U1_ex_pr2}
Q_8 = \left\langle a=\mathrm{diag}(i,\,-i,\,1),~b= \begin{pmatrix}
0 & e^{i \alpha} & 0 \\
-e^{-i \alpha} & 0 & 0\\
0 & 0 & 1
\end{pmatrix} \Bigg| ~ a^4=\mathcal{I}_4,~a^2=b^2,~aba=b\right\rangle,
\end{equation}
to the 3HDM. It was concluded that $Q_8$ could not be realised, as the potential gained a higher continuous symmetry. Notice that $b$ of eq.~\eqref{Eq:O2_U1_ex_pr2} is identical to $g$ of eq.~\eqref{Eq:O2_U1_ex_pr1}. Apart from that, the scalar potential invariant under the $b$ generator is automatically invariant under the $a$ generator. Therefore, the scalar potential under a specific representation of $Q_8$ is shown to be symmetric under $O(2) \times U(1)$.

In the bilinear space formalism, the quartic part of the scalar potential is
\begin{equation}
V = \Lambda_0 r_0^2 + L_i r_0 r_i + \Lambda_{ij} r_i r_j,
\end{equation}
where $r_i$ are the gauge-invariant bilinear combinations in the Gell-Mann basis. We shall assume that
\begin{equation}
\Lambda = \begin{pmatrix}
\Lambda_0 & L_i \\
L_i^\mathrm{T} & \Lambda_{ij}
\end{pmatrix}.
\end{equation}
We adopt the notation from Ref.~\cite{deMedeirosVarzielas:2019rrp}, as it provides essential tools for identifying the underlying symmetries in a basis-independent manner. 

For $O(2) \times U(1)$ we have:
\begin{equation}
\Lambda=
\begin{pmatrix}
 \lambda_a & 0 & 0 & 0 & \multirow{4}{*}{\vdots}  & \lambda_b \\
 0 & \lambda_{1221} & 0 & 0 &  & 0 \\
 0 & 0 & \lambda_{1221} & 0 &  & 0 \\
 0 & 0 & 0 & 2 \lambda_{1111}-\lambda_{1122} &  & 0\\
   \multicolumn{4}{c}{\dots}& \mathcal{I}_4 \otimes \lambda_{1331} & \dots \\[10pt]
 \lambda_b & 0 & 0 & 0 & \vdots & \lambda_c
\end{pmatrix},
\end{equation} 
where
\begin{subequations}\label{Eq:lambda_abc}
\begin{align}
\lambda_a ={}& \frac{1}{3} \left( 2 \lambda_{1111} + \lambda_{1122} + 2 \lambda_{1133} + \lambda_{3333} \right),\\
\lambda_b ={}& \frac{1}{3} \left( 2 \lambda_{1111} + \lambda_{1122} - \lambda_{1133} - 2\lambda_{3333} \right),\\
\lambda_c ={}& \frac{1}{3} \left( 2 \lambda_{1111} + \lambda_{1122} - 4 \lambda_{1133} + 4\lambda_{3333} \right).
\end{align}
\end{subequations}
The eigenvalue degeneracy pattern of $\Lambda_{ij}$ is $1+1+2+4$.

\subsection{Different implementations}\label{Sec:Pot_O2_U1}

Due to the fact that two representations of the $O(2)$ group were discussed, we shall specify which of the $O(2)$ representations is considered. For $\left[ SO(2) \rtimes \mathbb{Z}_2 \right] \times U(1)$ the scalar potential was provided in eq.~\eqref{Eq:V_O2_U1_expl_O2}, while for $\left[ U(1)_1 \rtimes \mathbb{Z}_2 \right] \times U(1)$ the scalar potential was given by eq.~\eqref{Eq:V_U1_U1_S2}. These are connected via the basis transformation of eq.~\eqref{Eq:h_Rot1_h}.

While analysing different vacua it is important to not misjudge which vacuum configurations one has to take into account. In the case of the $U(1) \times U(1)$-symmetric potential we covered two vacua, $(\hat v_1,\, \hat v_2,\, 0)$ and $(v,\,0,\,0)$. Since $O(2) \times U(1)$ is a more symmetric potential, one might be tempted to conclude that the discussion should be generalised to these two vacua. In reality, one has to account for permutations of different configurations. This could be understood since in $U(1) \times U(1)$ there are two dimension-one groups involved, while in the case of the $O(2) \times U(1)$ group, $O(2)$ is of dimension two. A group of dimension two treats the two doublets, under which they transform, differently from the other doublet. If we analyse a group of dimension three, then the findings from the $U(1) \times U(1)$-symmetric case can be generalised to this scenario.

Different implementations of $O(2) \times U(1)$ are presented in Table~\ref{Table:U1U1S2_Cases_1} and Table~\ref{Table:U1U1S2_Cases_1}.

{{\renewcommand{\arraystretch}{1.3}
\begin{table}[htb]
\caption{Similar to Table~\ref{Table:U1_U1_Cases}, but now for $\mathcal{G}_1 \equiv \left[ SO(2) \rtimes \mathbb{Z}_2 \right] \times U(1)$, with the scalar potential given by eq.~\eqref{Eq:V_O2_U1_expl_O2}. None of these cases violates CP.}
\label{Table:U1U1S2_Cases_1}
\begin{center}
\begin{tabular}{|c|c|c|c|c|} \hline\hline
Vacuum & SYM & $V$ & \begin{tabular}[l]{@{}c@{}} Mixing of the \\ neutral states\end{tabular} & Comments \\ \hline
$(\frac{v}{\sqrt{2}} e^{i \sigma},\, \pm \frac{v}{\sqrt{2}},\, 0)$ & $-$& $V_{\mathcal{G}_1}  + \mu_{12}^2$ & \footnotesize\begin{tabular}[l]{@{}c@{}} $\{\eta_1,\, \eta_2,\, \chi_1,\, \chi_2\}$ \\ ${-}\{\eta_3\}{-}\{\chi_3\}$ \end{tabular} & \begin{tabular}[l]{@{}c@{}} $m_{\eta_3} = m_{\chi_3}$\end{tabular} \\ \hline
$(i\frac{v}{\sqrt{2}},\, \pm \frac{v}{\sqrt{2}},\, 0)$ & $-$& $V_{\mathcal{G}_1}$ & \footnotesize\begin{tabular}[l]{@{}c@{}} $\{\eta_1,\, \eta_2\}{-}\{\eta_3\}$\\${-}\{\chi_1,\, \chi_2\}{-}\{\chi_3\}$ \end{tabular} & \footnotesize\begin{tabular}[l]{@{}c@{}} $m_{\eta_{(1,2)}}$ = $m_{\chi_{(1,2)}},~ m_{\eta_3} = m_{\chi_3}$ \end{tabular} \\ \hline
$(\hat{v}_1,\, \hat{v}_2,\, 0)$ & $-$& $V_{\mathcal{G}_1}$ & \footnotesize\begin{tabular}[l]{@{}c@{}} $\{\eta_1,\, \eta_2\}{-}\{\eta_3\}$\\${-}\{\chi_1,\, \chi_2\}{-}\{\chi_3\}$ \end{tabular} & \footnotesize\begin{tabular}[l]{@{}c@{}} $m_{\eta_3} = m_{\chi_3}$ \\ $m_{\eta_{(1,2)}}= m_{\chi_{(1,2)}} = 0$ \end{tabular} \\ \hline
$(0,\, \hat{v}_2,\, \hat{v}_3)$ & $-$& $V_{\mathcal{G}_1}$ & \footnotesize\begin{tabular}[l]{@{}c@{}} $\{\eta_1\}{-}\{\eta_2,\, \eta_3\}$\\${-}\{\chi_1\}{-}\{\chi_2\}{-}\{\chi_3\}$ \end{tabular} & \begin{tabular}[l]{@{}c@{}} $ m_{\eta_1} = m_{\chi_2} = m_{\chi_3} = 0$ \end{tabular} \\ \hline
$(0,\, \hat{v}_2,\, \hat{v}_3)$ & $-$& $V_{\mathcal{G}_1} + (\mu_{23}^2)^\mathrm{R} $ & \footnotesize\begin{tabular}[l]{@{}c@{}} $\{\eta_1\}{-}\{\eta_2,\, \eta_3\}$ \\ ${-}\{\chi_1\}{-}\{\chi_2,\, \chi_3\}$ \end{tabular} & - \\ \hline
$(v,\, 0,\, 0)$  & $-$ & $V_{\mathcal{G}_1}$ & diagonal & \footnotesize\begin{tabular}[l]{@{}c@{}} $m_{\eta_3} = m_{\chi_3}$ \\ $m_{\eta_2}= m_{\chi_1} = 0$ \end{tabular} \\ \hline
$(v,\, 0,\, 0)$  & $-$ & $V_{\mathcal{G}_1} + (\mu_{23}^2)^\mathrm{R} $ & \footnotesize\begin{tabular}[l]{@{}c@{}} $\{\eta_1\}{-}\{\eta_2,\, \eta_3\}$ \\ ${-}\{\chi_1\}{-}\{\chi_2,\, \chi_3\}$ \end{tabular} &  - \\ \hline
$(0,\, 0,\, v)$  & $\checkmark$ & $V_{\mathcal{G}_1}$ & diagonal & \footnotesize\begin{tabular}[l]{@{}c@{}} $ m_{h^+_1} = m_{h^+_2} $, \\ $m_{\eta_1} = m_{\eta_2} = m_{\chi_1} = m_{\chi_2}$ \end{tabular} \\ \hline \hline
\end{tabular}
\end{center}
\end{table}}

{{\renewcommand{\arraystretch}{1.3}
\begin{table}[h]
\caption{Similar to Table~\ref{Table:U1_U1_Cases}, but now for $\mathcal{G}_2 \equiv \left[ U(1)_1 \rtimes \mathbb{Z}_2 \right] \times U(1)$, with the scalar potential given by eq.~\eqref{Eq:V_U1_U1_S2}. In one case the minimisation conditions can lead to a higher symmetry. None of these cases violates CP.}
\label{Table:U1U1S2_Cases_2}
\begin{center}
\begin{tabular}{|c|c|c|c|c|} \hline\hline
Vacuum & SYM & $V$ & \begin{tabular}[l]{@{}c@{}} Mixing of the \\ neutral states\end{tabular} & Comments \\ \hline
$(\hat{v}_1,\, \hat{v}_2,\, 0)$ & $-$& $V_{U(2)}$ & \footnotesize\begin{tabular}[l]{@{}c@{}} $\{\eta_1,\, \eta_2\}{-}\{\eta_3\}$\\${-}\{\chi_1\}{-}\{\chi_2\}{-}\{\chi_3\}$ \end{tabular} & \footnotesize\begin{tabular}[l]{@{}c@{}} $m_{\eta_3} = m_{\chi_3}$ \\ $m_{\eta_{(1,2)}}= m_{\chi_1} = m_{\chi_2} = 0$ \end{tabular} \\ \hline
$(\hat{v}_1,\, \hat{v}_2,\, 0)$ & $-$& $V_{\mathcal{G}_2}  + (\mu_{12}^2)^\mathrm{R}$ & \footnotesize\begin{tabular}[l]{@{}c@{}} $\{\eta_1,\, \eta_2\}{-}\{\eta_3\}$ \\ ${-}\{\chi_1,\, \chi_2\}{-}\{\chi_3\}$ \end{tabular} & \begin{tabular}[l]{@{}c@{}} $m_{\eta_3} = m_{\chi_3}$ \end{tabular} \\ \hline
$(\frac{v}{\sqrt{2}},\, \pm \frac{v}{\sqrt{2}},\, 0)$ & $-$ & $V_{\mathcal{G}_2}$ & \footnotesize\begin{tabular}[l]{@{}c@{}} $\{\eta_1,\, \eta_2\}{-}\{\eta_3\}$\\${-}\{\chi_1\}{-}\{\chi_2\}{-}\{\chi_3\}$ \end{tabular} & \footnotesize\begin{tabular}[l]{@{}c@{}} $m_{\eta_3} = m_{\chi_3}$ \\ $ m_{\chi_1} = m_{\chi_2} = 0$ \end{tabular} \\ \hline
$(\frac{v}{\sqrt{2}},\, \pm \frac{v}{\sqrt{2}},\, 0)$ & $-$ & $V_{\mathcal{G}_2} + (\mu_{12}^2)^\mathrm{R}$ & \footnotesize\begin{tabular}[l]{@{}c@{}} $\{\eta_1,\, \eta_2\}{-}\{\eta_3\}$ \\ ${-}\{\chi_1,\, \chi_2\}{-}\{\chi_3\}$ \end{tabular} & \begin{tabular}[l]{@{}c@{}} $m_{\eta_3} = m_{\chi_3}$ \end{tabular} \\ \hline
$(0,\, \hat{v}_2,\, \hat{v}_3)$ & $-$& $V_{\mathcal{G}_2}$ & \footnotesize\begin{tabular}[l]{@{}c@{}} $\{\eta_1\}{-}\{\eta_2,\, \eta_3\}$\\${-}\{\chi_1\}{-}\{\chi_2\}{-}\{\chi_3\}$ \end{tabular} & \footnotesize\begin{tabular}[l]{@{}c@{}} $m_{\eta_1} = m_{\chi_1}$ \\ $ m_{\chi_2} = m_{\chi_3} = 0$ \end{tabular} \\ \hline
$(0,\, \hat{v}_2,\, \hat{v}_3)$ & $-$& $V_{\mathcal{G}_2} + (\mu_{23}^2)^\mathrm{R} $ & \footnotesize\begin{tabular}[l]{@{}c@{}} $\{\eta_1\}{-}\{\eta_2,\, \eta_3\}$ \\ ${-}\{\chi_1\}{-}\{\chi_2,\, \chi_3\}$ \end{tabular} & \begin{tabular}[l]{@{}c@{}} $m_{\eta_1} = m_{\chi_1}$ \end{tabular} \\ \hline
$(v,\, 0,\, 0)$  & $-$ & $V_{\mathcal{G}_2}$ & diagonal & \footnotesize\begin{tabular}[l]{@{}c@{}} $m_{\eta_2} = m_{\chi_2},~m_{\eta_3} = m_{\chi_3}$ \end{tabular} \\ \hline
$(0,\, 0,\, v)$  & $\checkmark$ & $V_{\mathcal{G}_2}$ & diagonal & \footnotesize\begin{tabular}[l]{@{}c@{}} $ m_{h^+_1} = m_{h^+_2} $, \\ $m_{\eta_1} = m_{\eta_2} = m_{\chi_1} = m_{\chi_2}$ \end{tabular} \\ \hline \hline
\end{tabular}
\end{center}
\end{table}}

\newpage

\bigskip\textbf{\boldmath$\left[ SO(2) \rtimes \mathbb{Z}_2 \right] \times U(1)$: case of \boldmath$(0,\,0,\,v)$}

The minimisation condition is given by
\begin{equation}
\mu_{33}^2 = - \lambda_{3333} v^2.
\end{equation}

The charged mass-squared matrix in the basis $\{h_1^\pm,\, h_2^\pm,\, h_3^\pm \}$ is:
\begin{equation}
\mathcal{M}_\mathrm{Ch}^2 = \mathrm{diag} \left( \mu_{11}^2 + \frac{1}{2} \lambda_{1133}v^2,\, \mu_{11}^2 + \frac{1}{2} \lambda_{1133}v^2,\, 0 \right).
\end{equation}

The neutral mass-squared matrix in the basis $\{\eta_1,\, \eta_2,\, \eta_3,\, \chi_1,\, \chi_2,\, \chi_3 \}$ is:
\begin{equation} \label{Eq:masses}
\mathcal{M}_\mathrm{N}^2 = \mathrm{diag} \left( m_H^2,\, m_H^2,\, m_h^2,\, m_H^2,\, m_H^2,\, 0 \right),
\end{equation}
where
\begin{subequations}\label{Eq:MN2_U1_U1_S2}
\begin{align}
m_H^2 ={}& \mu_{11}^2 + \frac{1}{2} \left( \lambda_{1133} + \lambda_{1331} \right) v^2,\\
m_h^2 ={}& 2 \lambda_{3333} v^2,
\end{align}
\end{subequations}
where $h$ is the SM-like Higgs boson.

In this implementation:
\begin{itemize}
\item $\{ \mu_{11}^2,\, \lambda_{1133},\, \lambda_{1331},\, \lambda_{3333} \}$  contribute to five scalar masses;
\item $\{\lambda_{1111},\, \lambda_{1122},\, \lambda_{1221}\}$  appear only in scalar interactions.
\end{itemize}

The above discussion also holds for the case of $\left[ U(1)_1 \rtimes \mathbb{Z}_2 \right] \times U(1)$.

\section[\texorpdfstring{$U(1) \times D_4$}{U(1) x D4}-symmetric 3HDM]{\boldmath$U(1) \times D_4$-symmetric 3HDM }\label{Sec:Pot_U1_Z2_S2}

The $U(1) \times D_4$-symmetric (this could be more appropriately described as a quotient group~\cite{Ivanonv_pr}) scalar potential can be derived from the $U(1) \times \mathbb{Z}_2$-symmetric potential by applying an $S_2$ symmetry. We believe that this case was considered for the first time in Ref.~\cite{Kuncinas:2024zjq}. 

\subsection{Discussion}

Consider the phase-sensitive part of the $U(1) \times \mathbb{Z}_2$-symmetric 3HDM,
\begin{equation*}
V^\mathrm{ph}_{U(1) \times \mathbb{Z}_2} = \lambda_{1212} h_{12}^2 + \mathrm{h.c.}
\end{equation*}
Invariance under $S_2(h_1, h_2)$ forces the coupling to become real, but not to vanish. The $\left[ U(1) \times \mathbb{Z}_2 \right] \rtimes S_2$-symmetric potential is then given by
\begin{equation}\label{Eq:V_U1_Z2_S2}
\begin{aligned}
V ={}& \mu_{11}^2 (h_{11} + h_{22}) + \mu_{33}^2 h_{33} + \lambda_{1111} (h_{11}^2 + h_{22}^2) + \lambda_{3333} h_{33}^2 + \lambda_{1122} h_{11} h_{22}\\
& + \lambda_{1133} (h_{11} h_{33} + h_{22} h_{33}) + \lambda_{1221} h_{12} h_{21} + \lambda_{1331} (h_{13} h_{31} + h_{23} h_{32})\\
& + \lambda_{1212} (h_{12}^2 + h_{21}^2).
\end{aligned}
\end{equation}
This scalar potential resembles the one of $O(2) \times U(1)$, see eq.~\eqref{Eq:V_U1_U1_Z2}, with
\begin{equation}
\lambda_a = \lambda_{1111},~ \lambda_{b} = \lambda_{1221},~ \lambda_{c} = \lambda_{1122},~\Lambda = \lambda_{1212},
\end{equation}
where $\Lambda$ is promoted to an independent coupling. This already indicates that the potential in question can be increased to the $O(2) \times U(1)$-symmetric 3HDM.

The generating set of $\left[ U(1) \times \mathbb{Z}_2 \right] \rtimes S_2$ can be identified to be:
\begin{subequations}\label{Eq:Gen_U1_Z2_S2}
\begin{align}
g_1 ={}& \mathrm{diag}(1,\, 1,\, e^{i \alpha}), \\
g_2 ={}& \mathrm{diag}(-1,\, 1,\, 1), \\
g_3 ={}& \begin{pmatrix}
0 & 1 & 0\\
1 & 0 & 0\\
0 & 0 & 1
\end{pmatrix}.
\end{align}
\end{subequations}
However, invariance of the scalar potential under
\begin{equation}
\begin{pmatrix}
h_1 \\
h_2 \\
h_3
\end{pmatrix} = \begin{pmatrix}
0 & -e^{i \alpha} & 0 \\
e^{i \alpha} & 0 & 0 \\
0 & 0 & 1
\end{pmatrix} \begin{pmatrix}
h_1 ^\prime \\
h_2 ^\prime \\
h_3 ^\prime
\end{pmatrix}
\end{equation}
suggests that the $\left\lbrace \lambda_{1112} \left( h_{11}h_{12} - h_{21}h_{22}\right) + \mathrm{h.c.} \right\rbrace$ term should be present. Nevertheless, this term can be shown to be redundant by going into a new basis, given by
\begin{equation}
\begin{pmatrix}
h_1 \\
h_2 \\
h_3
\end{pmatrix} = \frac{1}{\sqrt{2}}\begin{pmatrix}
e^{\frac{i \pi}{4}} & e^{\frac{i \pi}{4}} & 0 \\
-e^{-\frac{i \pi}{4}} & e^{-\frac{i \pi}{4}} & 0 \\
0 & 0 & \sqrt{2}
\end{pmatrix} \begin{pmatrix}
h_1 ^\prime \\
h_2 ^\prime \\
h_3 ^\prime
\end{pmatrix}.
\end{equation}
In this basis, $\lambda_{1212}\in \mathbb{C}$, but an additional  re-phasing can render it real, consistent with the scalar potential of eq.~\eqref{Eq:V_U1_Z2_S2}.

As an exercise, let us identify the group generated by $\{g_2,\, g_3\}$,
\begin{equation}
\{g_{2}^2,\, g_{2},\, g_{3},\, g_{2} g_{3},\, g_{3} g_{2},\, g_{3} g_{2} g_{3},\, g_{2} g_{3} g_{2},\, (g_{2} g_{3})^2\},
\end{equation}
which is of order eight. There are five distinct groups of order eight, up to isomorphism. Three of these are Abelian, $\{\mathbb{Z}_8,\, \mathbb{Z}_4 \times \mathbb{Z}_2,\, \mathbb{Z}_2^3\}$,  and two of them are non-Abelian, $\{Q_8,\, D_4\}$. Abelian groups do not lead to structures similar to the scalar potential in eq.~\eqref{Eq:V_U1_Z2_S2}. The case of the non-Abelian groups is more interesting. Actually, the above set of generators corresponds to those of $D_4 \cong \mathbb{Z}_4 \rtimes \mathbb{Z}_2$,
\begin{align}\label{Eq:Gen_U1_Z2_S2_2}
a ={}& g_2 g_3 = \begin{pmatrix}
0 & -1 & 0 \\
1 & 0 & 0\\
0 & 0 & 1
\end{pmatrix}, \quad
b = g_3 = \begin{pmatrix}
0 & 1 & 0\\
1 & 0 & 0\\
0 & 0 & 1
\end{pmatrix}.
\end{align}
The final group we want to consider is the quaternion group $Q_8$. The quaternion group $Q_8$ can be realised in a two-dimensional matrix representation over the complex numbers, $Q_8 \to \mathrm{SL}(2,\mathbb{C})$,
\begin{equation}
Q_8 = \left\langle \mathcal{I}_3,\, \begin{pmatrix}
 i &  0 &  0  \\
 0 & -i &  0 \\
 0 &  0 &  1
\end{pmatrix},\, \begin{pmatrix}
 0 &  1 &  0  \\
-1 &  0 &  0 \\
 0 &  0 &  1
\end{pmatrix},\, \begin{pmatrix}
 0 &  i &  0  \\
 i &  0 &  0 \\
 0 &  0 &  1
\end{pmatrix} \right\rangle.
\end{equation}
In fact, an even larger group of order sixteen, the Pauli group $\mathcal{P}_1$ is also realised. The scalar potentials of these match the potential of eq.~\eqref{Eq:V_U1_Z2_S2}.

If we identify $D_4$ with $\{g_2,\, g_3\}$, then the center,
\begin{equation}
Z(\mathcal{G}) = \left\lbrace z \in \mathcal{G} | \forall g \in \mathcal{G}, zg = gz \right\rbrace,
\end{equation}
of $D_4$ is
\begin{equation}
Z(D_4) = \{e,\,g_2g_3g_2g_3\} = \{ \mathcal{I}_3,\, \mathrm{diag}(-1,\,-1,\,1)\}.
\end{equation}
Notice that we can re-phase $g_1$ to match $g_2g_3g_2g_3$, therefore verifying that there is a trivial intersection between $U(1)$ and $D_4$. By factoring out the common element we get $D_4/Z(D(4)) \cong V_4 \cong \mathbb{Z}_2 \times \mathbb{Z}_2$. Due to lack of experience we shall stick to naming the group $U(1) \times D_4$.

In the bilinear formalism we have:
\begin{equation}
\Lambda=
\scriptstyle\begin{pmatrix}
 \lambda_a & 0 & 0 & 0 & 0 & 0 & 0 & 0 & \lambda_b \\
 0 & \lambda_{1221}+2\lambda_{1212} & 0 & 0 & 0 & 0 & 0 & 0 & 0 \\
 0 & 0 & \lambda_{1221}-2\lambda_{1212} & 0 & 0 & 0 & 0 & 0 & 0 \\
 0 & 0 & 0 & 2 \lambda_{1111}-\lambda_{1122} & 0 & 0 & 0 & 0 & 0 \\
 0 & 0 & 0 & 0 & \lambda_{1331} & 0 & 0 & 0 & 0 \\
 0 & 0 & 0 & 0 & 0 & \lambda_{1331} & 0 & 0 & 0 \\
 0 & 0 & 0 & 0 & 0 & 0 & \lambda_{1331} & 0 & 0 \\
 0 & 0 & 0 & 0 & 0 & 0 & 0 & \lambda_{1331} & 0 \\
 \lambda_b & 0 & 0 & 0 & 0 & 0 & 0 & 0 & \lambda_c
\end{pmatrix},
\end{equation}
where,
\begin{subequations}
\begin{align}
\lambda_a ={}& \frac{1}{3} \left( 2 \lambda_{1111} + \lambda_{1122} + 2 \lambda_{1133} + \lambda_{3333} \right),\\
\lambda_b ={}& \frac{1}{3} \left( 2 \lambda_{1111} + \lambda_{1122} - \lambda_{1133} - 2\lambda_{3333} \right),\\
\lambda_c ={}& \frac{1}{3} \left( 2 \lambda_{1111} + \lambda_{1122} - 4 \lambda_{1133} + 4\lambda_{3333} \right).
\end{align}
\end{subequations}
The degeneracy pattern of eigenvalues of $\Lambda_{ij}$ is $1+1+1+1+4$.

\subsection{Different implementations}\label{Sec:Pot_U1_D4}

Different vacua are summarised in Table~\ref{Table:U1Z2_S2_Cases}.

\bigskip\textbf{Case of \boldmath$(0,\,0,\,v)$}\labeltext{Case of $(0,\,0,\,v)$}{Sec:U1Z2S2_00v}

The minimisation condition is
\begin{equation}
\mu_{33}^2 = - \lambda_{3333} v^2.
\end{equation}

The charged mass-squared matrix in the basis $\{h_1^\pm,\, h_2^\pm,\, h_3^\pm \}$ is:
\begin{equation}
\mathcal{M}_\mathrm{Ch}^2 = \mathrm{diag} \left( \mu_{11}^2 + \frac{1}{2} \lambda_{1133}v^2,\, \mu_{11}^2 + \frac{1}{2} \lambda_{1133}v^2,\, 0 \right).
\end{equation}

The neutral mass-squared matrix in the basis $\{\eta_1,\, \eta_2,\, \eta_3,\, \chi_1,\, \chi_2,\, \chi_3 \}$ is:
\begin{equation}
\mathcal{M}_\mathrm{N}^2 = \mathrm{diag} \left( m_H^2,\, m_H^2,\, m_h^2,\, m_H^2,\, m_H^2,\, 0 \right),
\end{equation}
where
\begin{subequations}
\begin{align}
m_H^2 ={}& \mu_{11}^2 + \frac{1}{2} \left( \lambda_{1133} + \lambda_{1331} \right) v^2,\\
m_h^2 ={}& 2 \lambda_{3333} v^2,
\end{align}
\end{subequations}
and the $h$ state is associated with the SM-like Higgs boson.

In this implementation:
\begin{itemize}
\item  $\{ \mu_{11}^2,\, \lambda_{1133},\, \lambda_{1331},\, \lambda_{3333} \}$ contribute to three scalar masses;
\item  $\{\lambda_{1111},\, \lambda_{1122},\, \lambda_{1221},\, \lambda_{1212}\}$  appear only in scalar interactions.
\end{itemize}

{{\renewcommand{\arraystretch}{1.3}
\begin{table}[htb]
\caption{Similar to Table~\ref{Table:U1_U1_Cases}, but now for $U(1) \times D_4$. In one case the minimisation conditions can lead to a higher symmetry. None of these cases violates CP.}
\label{Table:U1Z2_S2_Cases}
\begin{center}
\begin{tabular}{|c|c|c|c|c|} \hline\hline
Vacuum & SYM & $V$ & \begin{tabular}[l]{@{}c@{}} Mixing of the \\ neutral states\end{tabular} & Comments \\ \hline
$(\hat{v}_1,\, \hat{v}_2,\, 0)$ & $-$ & $V_{O(2)\times U(1)}$ & \footnotesize\begin{tabular}[l]{@{}c@{}} $\{\eta_1,\, \eta_2\}{-}\{\eta_3\}$  \\ ${-}\{\chi_1,\, \chi_2\}{-}\{\chi_3\}$ \end{tabular} & \footnotesize\begin{tabular}[l]{@{}c@{}}  $m_{\eta_3} = m_{\chi_3}$ \\ $m_{\eta_{(1,2)}} = m_{\chi_{(1,2)}}=0$ \\\end{tabular} \\ \hline
$(\hat{v}_1,\, \hat{v}_2,\, 0)$ & $-$ & $V_{U(1) \times D_4} + (\mu_{12}^2)^\mathrm{R}$ & \footnotesize\begin{tabular}[l]{@{}c@{}} $\{\eta_1,\, \eta_2\}{-}\{\eta_3\}$ \\ ${-}\{\chi_1,\, \chi_2\}{-}\{\chi_3\}$ \end{tabular} & \begin{tabular}[l]{@{}c@{}} $m_{\eta_3} = m_{\chi_3}$ \end{tabular} \\ \hline
$ (\frac{v}{\sqrt{2}},\, \pm \frac{v}{\sqrt{2}} ,\, 0)$ & $-$ & $V_{U(1) \times D_4}$ & \footnotesize\begin{tabular}[l]{@{}c@{}} $\{\eta_1,\, \eta_2\}{-}\{\eta_3\}$ \\ ${-}\{\chi_1,\, \chi_2\}{-}\{\chi_3\}$ \end{tabular} & $m_{\eta_3} = m_{\chi_3}$  \\ \hline
$(0,\, \hat{v}_2,\, \hat{v}_3)$ & $-$ & $V_{U(1) \times D_4}$ & \footnotesize\begin{tabular}[l]{@{}c@{}} $\{\eta_1\}{-}\{\eta_2,\, \eta_3\}$ \\ ${-}\{\chi_1\}{-}\{\chi_2\}{-}\{\chi_3\}$ \end{tabular} & $m_{\chi_2} = m_{\chi_3}=0$ \\ \hline
$(0,\, \hat{v}_2,\, \hat{v}_3)$ & $-$ & $V_{U(1) \times D_4} + (\mu_{23}^2)^\mathrm{R}$ & \footnotesize\begin{tabular}[l]{@{}c@{}} $\{\eta_1\}{-}\{\eta_2,\, \eta_3\}$ \\ ${-}\{\chi_1\}{-}\{\chi_2, \,\chi_3\}$ \end{tabular} & - \\ \hline
$(v, \,0, \, 0)$ & $-$ & $V_{U(1) \times D_4}$ & diagonal & $m_{\eta_3} = m_{\chi_3}$ \\ \hline
$(0, \,0, \, v)$  & $\checkmark$ & $V_{U(1) \times D_4}$ & diagonal & \footnotesize\begin{tabular}[l]{@{}c@{}}  $m_{h_1^+} = m_{h_2^+}$, \\ $m_{\eta_1} = m_{\eta_2}=$ \\ $= m_{\chi_1}= m_{\chi_2}$ \end{tabular} \\ \hline \hline
\end{tabular} 
\end{center}
\end{table}}								   
\section[\texorpdfstring{$U(2)$}{U(2)}-symmetric 3HDM]{\boldmath$U(2)$-symmetric 3HDM}\label{Sec:Pot_U2}

Another possible symmetry of dimension two is $U(2)$, with the scalar potential given by
\begin{equation}
\begin{aligned}
V_{U(2)} ={}& \mu_{11}^2 (h_{11} + h_{22}) + \mu_{33}^2 h_{33} + \lambda_{1111} (h_{11}^2 + h_{22}^2) + \lambda_{3333} h_{33}^2 + \left( 2 \lambda_{1111} - \lambda_{1221} \right) h_{11} h_{22}\\
&  + \lambda_{1133} (h_{11} h_{33} + h_{22} h_{33}) + \lambda_{1221} h_{12} h_{21}+ \lambda_{1331} (h_{13} h_{31} + h_{23} h_{32}).
\end{aligned}
\end{equation}

\newpage 	 
All the cases with different vacuum configurations are summarised in Table~\ref{Table:U2_Cases}. 

{{\renewcommand{\arraystretch}{1.3}
\begin{table}[htb]
\caption{Similar to Table~\ref{Table:U1_U1_Cases}, but now for $U(2)$. None of these cases violates CP.}
\label{Table:U2_Cases}
\begin{center}
\begin{tabular}{|c|c|c|c|c|} \hline\hline
Vacuum & SYM & $V$ & \begin{tabular}[l]{@{}c@{}} Mixing of the \\ neutral states\end{tabular} & Comments \\ \hline
$(\hat{v}_1,\, \hat{v}_2,\, 0)$ & $-$ & $V_{U(2)}$ & \footnotesize\begin{tabular}[l]{@{}c@{}} $\{\eta_1,\, \eta_2\}{-}\{\eta_3\}$ \\ ${-}\{\chi_1\}{-}\{\chi_2\}{-}\{\chi_3\}$ \end{tabular} & \footnotesize\begin{tabular}[l]{@{}c@{}} $m_{\eta_3} = m_{\chi_3}$ \\ $m_{\eta_{(1,2)}}=m_{\chi_1} = m_{\chi_2} = 0$ \\  \end{tabular} \\ \hline
$(\frac{v}{\sqrt{2}},\, \pm \frac{v}{\sqrt{2}} ,\, 0)$ & $-$ & $V_{U(2)} + \left( \mu_{12}^2 \right)^\mathrm{R}$ & \footnotesize\begin{tabular}[l]{@{}c@{}} $\{\eta_1,\, \eta_2\}{-}\{\eta_3\}$ \\ ${-}\{\chi_1,\, \chi_2\}{-}\{\chi_3\}$ \end{tabular} & \footnotesize\begin{tabular}[l]{@{}c@{}} $m_{\eta_{(1,2)}}=m_{\chi_{(1,2)}},~m_{\eta_3} = m_{\chi_3}$ \end{tabular}\\ \hline
$(0,\, \hat{v}_2,\, \hat{v}_3)$ & $-$ & $V_{U(2)}$ & \footnotesize\begin{tabular}[l]{@{}c@{}} $\{\eta_1\}{-}\{\eta_2,\,\eta_3\}$ \\ ${-}\{\chi_1\}{-}\{\chi_2\}{-}\{\chi_3\}$ \end{tabular} & \footnotesize\begin{tabular}[l]{@{}c@{}} $m_{\eta_1} = m_{\chi_i}=0$ \end{tabular} \\ \hline
$(0,\, \hat{v}_2,\, \hat{v}_3)$ & $-$ & $V_{U(2)} + \left( \mu_{23}^2 \right)^\mathrm{R}$ & \footnotesize\begin{tabular}[l]{@{}c@{}} $\{\eta_1\}{-}\{\eta_2,\,\eta_3\}$ \\ ${-}\{\chi_1\}{-}\{\chi_2,\,\chi_3\}$ \end{tabular} &  $m_{\eta_1} = m_{\chi_1}$  \\ \hline
$(v,\, 0,\, 0)$ & $-$ & $V_{U(2)}$ & diagonal &  \footnotesize\begin{tabular}[l]{@{}c@{}} $m_{\eta_3} = m_{\chi_3}$ \\ $m_{\eta_2}=m_{\chi_1} = m_{\chi_2} = 0$ \end{tabular} \\ \hline
$(v,\, 0,\, 0)$ & $-$ & $V_{U(2)} + \left( \mu_{23}^2 \right)^\mathrm{R}$ & \footnotesize\begin{tabular}[l]{@{}c@{}} $\{\eta_1\}{-}\{\eta_2,\, \eta_3\}$ \\ ${-}\{\chi_1\}{-}\{\chi_2,\,\chi_3\}$ \end{tabular} &  \footnotesize\begin{tabular}[l]{@{}c@{}} Two pairs of neutral \\ mass-degenerate states \end{tabular}\\ \hline
$(0,\, 0,\, v)$ & $\checkmark$ & $V_{U(2)}$ & diagonal & \footnotesize\begin{tabular}[l]{@{}c@{}} $m_{h_1^+} = m_{h_2^+}$, \\ $m_{\eta_1}  = m_{\eta_2} = m_{\chi_1} = m_{\chi_2}$ \end{tabular} \\ \hline
\end{tabular}  
\end{center}
\end{table}}

\bigskip\textbf{Case of \boldmath$(0,\,0,\,v)$}

The minimisation condition is:
\begin{equation}
\mu_{11}^2 = - \lambda_{1111} v^2.
\end{equation}

This implementation is equivalent in both the charged and neutral mass-squared matrices to \ref{Sec:U1Z2S2_00v} of $U(1) \times D_4$ in Section~\ref{Sec:Pot_U1_Z2_S2}. Although they share the same mass eigenstates, the physical content is different. For example, in the $U(1) \times D_4$ model there are non-vanishing vertices $\eta_1 \eta_2 \chi_1 \chi_2 $ and $h_1^\pm h_1^\pm h_2^\mp h_2^\mp $, which are proportional to the $\lambda_{1212}$ coupling. However, in the $U(2)$-symmetric model these physical interactions are not present due to $\lambda_{1212}=0$.

In this implementation:
\begin{itemize}
\item $\{ \mu_{11}^2,\, \lambda_{1133},\, \lambda_{1331},\, \lambda_{3333} \}$  contribute to three scalar masses;
\item $\{\lambda_{1111},\, \lambda_{1122}\}$ appear only in scalar interactions.
\end{itemize}

\section[\texorpdfstring{$SO(3)$}{O(3)}-symmetric 3HDM]{\boldmath$SO(3)$-symmetric 3HDM}\label{Sec:Pot_SO3}

Next, we examine models stabilised by continuous symmetries of dimension three. Since we now focus on symmetries that act on all doublets, vevs of these will always result in SSB of the underlying continuous symmetry, which is not our focus. Here, a single case with different vanishing vevs suffices, without considering different permutations of vevs. We begin with the $SO(3)$-symmetric 3HDM, the scalar potential of which is:
\begin{equation}\label{Eq:V_SO3}
\begin{aligned}
V_{SO(3)} ={}& \mu_{11}^2 (h_{11} + h_{22} + h_{33}) + \lambda_{1111} (h_{11}^2 + h_{22}^2 + h_{33}^2)\\
& + \lambda_{1122} (h_{11} h_{22} + h_{22} h_{33} + h_{33} h_{11}) + \lambda_{1221} (h_{12} h_{21} + h_{23} h_{32} + h_{31} h_{13})\\
& + \Lambda (h_{12}^2 + h_{21}^2 + h_{23}^2 + h_{32}^2 + h_{31}^2 + h_{13}^2)\\
={}& \mu_{11}^2 \sum_i h_{ii} + \lambda_{1111} \sum_i h_{ii}^2  + \lambda_{1122} \sum_{i<j} h_{ii} h_{jj} + \lambda_{1221} \sum_{i<j} h_{ij} h_{ji}\\
& + \Lambda \sum_{i<j} (h_{ij}^2 + h_{ji}^2),
\end{aligned}
\end{equation}
where $\Lambda$ was defined earlier in eq.~\eqref{Eq:Incr_sym}.

Different implementations are summarised in Table~\ref{Table:SO3_Cases}. 

{{\renewcommand{\arraystretch}{1.25}
\begin{table}[htb]
\caption{Similar to Table~\ref{Table:U1_U1_Cases}, but now for $SO(3)$. None of these cases violates CP.}
\label{Table:SO3_Cases}
\begin{center}
\begin{tabular}{|c|c|c|c|c|} \hline\hline
Vacuum & SYM & $V$ & \begin{tabular}[l]{@{}c@{}} Mixing of the \\ neutral states\end{tabular} & Comments \\ \hline
$(\frac{v}{\sqrt{2}} e^{i \sigma},\, \pm \frac{v}{\sqrt{2}} ,\, 0)$ & $-$ & $V_{SO(3)} + \mu_{12}^2 $ & \footnotesize\begin{tabular}[l]{@{}c@{}} $\{\eta_1,\, \eta_2,\,\chi_1,\, \chi_2\}$ \\ ${-}\{\eta_3,\, \chi_3\}$ \end{tabular} & - \\ \hline
$(i \frac{v}{\sqrt{2}},\, \pm \frac{v}{\sqrt{2}} ,\, 0)$ & $-$ &  $V_{SO(3)} + \left( \mu_{12}^2 \right)^\mathrm{I}$  & \footnotesize\begin{tabular}[l]{@{}c@{}} $\{\eta_1,\, \eta_2\}{-}\{\eta_3\}$ \\ ${-}\{\chi_1,\,\chi_2\}{-}\{\chi_3\}$ \end{tabular} & \footnotesize\begin{tabular}[l]{@{}c@{}} $m_{\eta_{(1,2)}} = m_{\chi_{(1,2)}},~m_{\eta_3} = m_{\chi_3}$ \end{tabular} \\ \hline
$(\hat{v}_1,\, \hat{v}_2,\, 0)$ & $-$ & $V_{SO(3)}$ & \footnotesize\begin{tabular}[l]{@{}c@{}} $\{\eta_1,\, \eta_2\}{-}\{\eta_3\}$ \\ ${-}\{\chi_1,\,\chi_2\}{-}\{\chi_3\}$ \end{tabular} & \footnotesize\begin{tabular}[l]{@{}c@{}} $m_{h^+_{(1,2)}} = m_{h_3^+},~m_{\chi_{(1,2)}}^\prime = m_{\chi_3}$ \\ $m_{\eta_{(1,2)}} = m_{\eta_3} = m_{\chi_{(1,2)}}=0$ \end{tabular} \\ \hline
$(v,\, 0,\, 0)$ & $-$ & $V_{SO(3)}$ & diagonal &  \footnotesize\begin{tabular}[l]{@{}c@{}} $m_{h_2^+} = m_{h_3^+},~m_{\chi_2} = m_{\chi_3}$ \\ $m_{\eta_2} = m_{\eta_3} = m_{\chi_1}=0$ \end{tabular}\\ \hline
$(v,\, 0,\, 0)$ & $-$ & $V_{SO(3)} + \left( \mu_{23}^2 \right)^\mathrm{R}$ & \footnotesize\begin{tabular}[l]{@{}c@{}} $\{\eta_1\}{-}\{\eta_2,\, \eta_3\}$ \\ ${-}\{\chi_1\}{-}\{\chi_2,\, \chi_3\}$ \end{tabular} &  \footnotesize\begin{tabular}[l]{@{}c@{}} Two pairs of neutral \\ mass-degenerate states \end{tabular}\\ \hline
\end{tabular}  
\end{center}
\end{table}}

\section[\texorpdfstring{$[ U(1) \times U(1)] \rtimes S_3$}{U(1) x U(1) x S3}-symmetric 3HDM]{\boldmath$[ U(1) \times U(1)] \rtimes S_3$-symmetric 3HDM}\label{Sec:Pot_U1_U1_S3}

Beyond the $U(1) \times U(1)$ group, we can also consider permutations of all three doublets, given by the $S_3$ group. In this case the scalar potential becomes:
\begin{equation}\label{Eq:V_U1_U1_S3}
\begin{aligned}
V_{[ U(1) \times U(1)] \rtimes S_3} ={}& \mu_{11}^2 (h_{11} + h_{22} + h_{33}) + \lambda_{1111} (h_{11}^2 + h_{22}^2 + h_{33}^2)\\
& + \lambda_{1122} (h_{11} h_{22} + h_{22} h_{33} + h_{33} h_{11}) + \lambda_{1221} (h_{12} h_{21} + h_{23} h_{32} + h_{31} h_{13})\\
={}& \mu_{11}^2 \sum_i h_{ii} + \lambda_{1111} \sum_i h_{ii}^2  + \lambda_{1122} \sum_{i<j} h_{ii} h_{jj} + \lambda_{1221} \sum_{i<j} h_{ij} h_{ji}.
\end{aligned}
\end{equation}
Applying an additional permutation, either $S_2(h_1, h_3)$ or $S_2(h_2, h_3)$, to the potential in eq.~\eqref{Eq:V_U1_U1_S2} results in $\left[ O(2) \times U(1) \right] \rtimes S_2$, which has an identical scalar potential to eq.~\eqref{Eq:V_U1_U1_S3}.

The total number of free couplings of eq.~\eqref{Eq:V_U1_U1_S3} matches that of the $SO(3)$-symmetric scalar potential, as seen in eq.~\eqref{Eq:V_SO3}: one bilinear and three independent quartic couplings. However, the invariance under the $SO(3)$ symmetry strictly necessitates the inclusion of the $\Lambda(h_{ij}^2 + h_{ji}^2)$ term.

For both the $O(2) \times U(1)$-symmetric and the $U(1) \times D_4$-symmetric scalar potentials we considered which discrete symmetries could yield an identical scalar potential. For the case of $[ U(1) \times U(1)] \rtimes S_3$, consider the following generator:
\begin{equation}
g= \begin{pmatrix}
0 & e^{i \theta_1} & 0 \\
0 & 0 & e^{i \theta_2}\\
e^{-i (\theta_1 + \theta_2)} & 0 & 0
\end{pmatrix},
\end{equation}
which results in $g^3 = e$.  This corresponds to the family of $\mathbb{Z}_3$ representations parameterised by $\theta_i$. The above generator was examined in Refs.~\cite{Ivanov:2012ry,Ivanov:2012fp}, as part of a larger extension $\left[ \mathbb{Z}_2 \times \mathbb{Z}_2\right] \rtimes \mathbb{Z}_3$, identified as tetrahedral symmetry. Fixing the $\theta_i$ phases led the authors of Ref.~\cite{Ivanov:2012fp} to a discrete $A_4$ symmetry, which would allow for additional terms like $h_{ij}^2$. Without fixing the phases, and hence restricting the structure of the potential, results in the potential of eq.~\eqref{Eq:V_U1_U1_S3}, which is symmetric under $\left[ U(1) \times U(1)\right] \rtimes S_3$.

Since this case was not presented in the basis-independent classification of Ref.~\cite{deMedeirosVarzielas:2019rrp}, we provide the quartic part in the bilinear formalism
\begin{equation}
\Lambda = \mathrm{diag} \left( \lambda_{1111} + \lambda_{1122},\, \mathcal{I}_2 \otimes \lambda_{1221},\, 2 \lambda_{1111} - \lambda_{1122},\, \mathcal{I}_4 \otimes \lambda_{1221},\, 2 \lambda_{1111} - \lambda_{1122}  \right).
\end{equation}
The eigenvalue degeneracy pattern of $\Lambda_{ij}$ is $2+6$.

Since $[ U(1) \times U(1)] \rtimes S_3$ is of dimension three, there are no vacuum configurations which would preserve the underlying symmetry. With less freedom, the scalar potential might start to render some unphysical results. For example, consider that there is a single vanishing vev present. Then, there are two possibilities to implement such vacuum: $(\hat v_1,\, \hat v_2,\, 0)$ and $ (\frac{v}{\sqrt{2}},\, \pm \frac{v}{\sqrt{2}} ,\, 0)$. In the case of the $(\hat v_1,\, \hat v_2,\, 0)$ vacuum, the minimisation conditions require vanishing of the $\Lambda$ coupling, which would result in the $SU(3)$-symmetric potential, which we shall cover in the next section. For the $(\hat v_1,\, \hat v_2,\, 0)$ vacuum, the minimisation conditions are not uniquely defined; instead of fixing $\Lambda=0$, we can relate vevs, forcing the vacuum to become $ (\frac{v}{\sqrt{2}},\, \pm \frac{v}{\sqrt{2}} ,\, 0)$. In this case, the mass-squared parameters are given by:
\begin{subequations}
\begin{align}
m_{H_1}^2 ={}& \frac{1}{2} \left( 2 \lambda_{1111} - \lambda_{1122} - \lambda_{1221} \right) v^2,\\
m_{H_2}^2 ={}& \frac{1}{2} \left( 2 \lambda_{1111} + \lambda_{1122} + \lambda_{1221} \right) v^2,\\
m_{\eta_3}^2 = m_{\chi_3}^2 ={}& -\frac{1}{4} \left( 2 \lambda_{1111} - \lambda_{1122} - \lambda_{1221} \right) v^2.
\end{align}
\end{subequations}
It is not possible for all mass-squared parameters to be positive definite at the same time. The presence of mass-squared parameters with opposite signs, $m_{H_1}^2 = -2 m_{\eta_3}^2$, indicates a saddle point, making this case unphysical. One way to resolve this is by introducing soft symmetry-breaking terms.

In Table~\ref{Table:U11S3_Cases} different implementations are summarised. 

{{\renewcommand{\arraystretch}{1.22}
\begin{table}[htb]
\caption{Similar to Table~\ref{Table:U1_U1_Cases}, but now for $\mathcal{G}_3 \equiv \left[ U(1) \times U(1) \right] \rtimes S_3 $. In one case the minimisation conditions lead to a higher symmetry. None of these cases violates~CP.}
\label{Table:U11S3_Cases}
\begin{center}
\begin{tabular}{|c|c|c|c|c|} \hline\hline
Vacuum & SYM & $V$ & \begin{tabular}[l]{@{}c@{}} Mixing of the \\ neutral states\end{tabular} & Comments \\ \hline
$(\hat{v}_1,\, \hat{v}_2,\, 0)$ & $-$& $V_{SU(3)}$ & \footnotesize\begin{tabular}[l]{@{}c@{}} $\{\eta_1,\, \eta_2\}{-}\{\eta_3\}$\\${-}\{\chi_1\}{-}\{\chi_2\}{-}\{\chi_3\}$ \end{tabular} & \footnotesize\begin{tabular}[l]{@{}c@{}} $m_{h^+_{(1,2)}} = m_{h^+_3}$ \\ Five neutral massless states \end{tabular} \\ \hline
$(\hat{v}_1,\, \hat{v}_2,\, 0)$ & $-$& $V_{\mathcal{G}_3}  + (\mu_{12}^2)^\mathrm{R}$ & \footnotesize\begin{tabular}[l]{@{}c@{}} $\{\eta_1,\, \eta_2\} {-}\{\eta_3\}$ \\ ${-}\{\chi_1,\, \chi_2\}{-}\{\chi_3\}$ \end{tabular} & \begin{tabular}[l]{@{}c@{}} $m_{h^+_{(1,2)}} = m_{h^+_3},$ \\ $ m_{\eta_3} = m_{\chi_3} = m_{\chi_{(2,3)}}$ \end{tabular} \\ \hline
$(\frac{v}{\sqrt{2}},\, \pm \frac{v}{\sqrt{2}},\, 0)$ & $-$ & $V_{\mathcal{G}_3}$ & \footnotesize\begin{tabular}[l]{@{}c@{}} $\{\eta_1,\, \eta_2\}{-}\{\eta_3\}$\\${-}\{\chi_1\}{-}\{\chi_2\}{-}\{\chi_3\}$ \end{tabular} & \footnotesize\begin{tabular}[l]{@{}c@{}} $m_{\eta_3} = m_{\chi_3}$ \\ $ m_{\chi_1} = m_{\chi_2} = 0$ \\ A negative $m_\mathrm{neutral}^2$  \end{tabular} \\ \hline
$(\frac{v}{\sqrt{2}},\, \pm \frac{v}{\sqrt{2}},\, 0)$ & $-$ & $V_{\mathcal{G}_3} + (\mu_{12}^2)^\mathrm{R}$ & \footnotesize\begin{tabular}[l]{@{}c@{}} $\{\eta_1,\, \eta_2\} {-}\{\eta_3\}$ \\ ${-}\{\chi_1,\, \chi_2\}{-}\{\chi_3\}$ \end{tabular} & \begin{tabular}[l]{@{}c@{}} $m_{\eta_3} = m_{\chi_3}$ \end{tabular} \\ \hline
$(v,\, 0,\, 0)$  & $-$ & $V_{\mathcal{G}_3}$ & diagonal & \footnotesize\begin{tabular}[l]{@{}c@{}} $ m_{h^+_2} = m_{h^+_3}, $ \\ $m_{\eta_2} = m_{\eta_3} = m_{\chi_2} = m_{\chi_3}$ \end{tabular} \\ \hline\hline
\end{tabular}
\end{center}
\end{table}}

\section[\texorpdfstring{$SU(3)$}{SU(3)}-symmetric 3HDM]{\boldmath$SU(3)$-symmetric 3HDM}\label{Sec:Pot_SU3}

By imposing $\Lambda=0$ to $V_{[ U(1) \times U(1)] \rtimes S_3}$ of eq.~\eqref{Eq:V_U1_U1_S3}, we get the $SU(3)$-symmetric 3HDM:
\begin{equation}\label{Eq:V_SU3}
\begin{aligned}
V_{SU(3)} ={}& \mu_{11}^2 (h_{11} + h_{22} + h_{33}) + \lambda_{1111} (h_{11} + h_{22} + h_{33})^2\\
& +\lambda_{1221} (h_{12} h_{21} + h_{23} h_{32} + h_{31} h_{13} - h_{11} h_{22} - h_{22} h_{33} - h_{33} h_{11})\\
={}& \mu_{11}^2 \sum_i h_{ii} + \lambda_{1111} \left( \sum_i h_{ii} \right) ^2 + \lambda_{1221} \sum_{i<j} (h_{ij} h_{ji} -h_{ii} h_{jj} ).
\end{aligned}
\end{equation}
Alternatively, this is the $SU(3) \times U(1)$-symmetric 3HDM, in the notation of Ref.~\cite{Darvishi:2019dbh}.

Vacuum configurations with at least one vanishing vev are summarised in Table~\ref{Table:SU3_Cases}.

{{\renewcommand{\arraystretch}{1.22}
\begin{table}[htb]
\caption{Similar to Table~\ref{Table:U1_U1_Cases}, but now for $SU(3)$. None of these cases violates CP.}
\label{Table:SU3_Cases}
\begin{center}
\begin{tabular}{|c|c|c|c|c|} \hline\hline
Vacuum & SYM & $V$ & \begin{tabular}[l]{@{}c@{}} Mixing of the \\ neutral states\end{tabular} & Comments \\ \hline
$(\hat{v}_1,\, \hat{v}_2,\, 0)$ & $-$ & $V_{SU(3)}$ & \footnotesize\begin{tabular}[l]{@{}c@{}} $\{\eta_1,\, \eta_2\}{-}\{\eta_3\}$ \\ ${-}\{\chi_1\}{-}\{\chi_2\}{-}\{\chi_3\}$ \end{tabular} & \footnotesize\begin{tabular}[l]{@{}c@{}} $m_{h^+_{(1,2)}} = m_{h_3^+}$ \\ $m_{\eta_{(1,2)}} = m_{\eta_3} = m_{\chi_i}=0$ \end{tabular} \\ \hline
$(\frac{v}{\sqrt{2}},\, \pm \frac{v}{\sqrt{2}} ,\, 0)$ & $-$ & $V_{SU(3)} + \left( \mu_{12}^2 \right)^\mathrm{R}$ & \footnotesize\begin{tabular}[l]{@{}c@{}} $\{\eta_1,\, \eta_2\}{-}\{\eta_3\}$ \\ ${-}\{\chi_1,\, \chi_2\}{-}\{\chi_3\}$ \end{tabular} & \footnotesize\begin{tabular}[l]{@{}c@{}} $m_{\eta_{(1,2)}} = m_{\chi_{(1,2)}},~m_{\eta_3} = m_{\chi_3}$\end{tabular}\\ \hline
$(v,\, 0,\, 0)$ & $-$ & $V_{SU(3)}$ & diagonal &  \footnotesize\begin{tabular}[l]{@{}c@{}} $m_{h_2^+} = m_{h_3^+}$ \\ $m_{\eta_2} = m_{\eta_3} = m_{\chi_i}=0$ \end{tabular}\\ \hline
$(v,\, 0,\, 0)$ & $-$ & $V_{SU(3)} + \left( \mu_{23}^2 \right)^\mathrm{R}$ & \footnotesize\begin{tabular}[l]{@{}c@{}} $\{\eta_1\}{-}\{\eta_2,\, \eta_3\}$ \\ ${-}\{\chi_1\}{-}\{\chi_2,\, \chi_3\}$ \end{tabular} &  \footnotesize\begin{tabular}[l]{@{}c@{}} $m_{H_1}^2 = - m_{H_2}^2$ \end{tabular}\\ \hline
\end{tabular}  
\end{center}\vspace*{-10pt}
\end{table}}

\newpage

\section{The Dark Matter candidates}\label{Sec:Different_DM_cases}

To sum up, we identified the following cases, {\it without} SSB:
{\quad\begin{longtable}{llll}
\hyperref[Table:U1_U1_Cases]{$\bullet \quad U(1) \times U(1)$}\hspace{20pt} & $(v,\,0,\,0)$ &  & \\
\hyperref[Table:U11_Cases]{$\bullet \quad U(1)_1$} & $(v,\,0,\,0)$ & $(0,\,0,\,v)$ & \\
\hyperref[Table:U1Z2_Cases]{$\bullet \quad U(1) \times \mathbb{Z}_2$} & $(v,\,0,\,0)$ & $(0,\,0,\,v)$ & \\
\hyperref[Table:U12_Cases]{$\bullet \quad U(1)_2$} & $(v,\,0,\,0)$ & $(0,\,0,\,v)$ &  $(v_1,\,v_2,\,0)$ \\
\hyperref[Table:U1U1S2_Cases_1]{$\bullet \quad O(2) \times U(1)$} & & $(0,\,0,\,v)$ & \\
\hyperref[Table:O2_Cases_1]{$\bullet \quad O(2)$} & & $(0,\,0,\,v)$ & \\
\hyperref[Table:U1Z2_S2_Cases]{$\bullet \quad U(1) \times D_4$} & & $(0,\,0,\,v)$ & \\
\hyperref[Table:U2_Cases]{$\bullet \quad U(2)$} & & $(0,\,0,\,v)$ & \\
\addtocounter{table}{-1}
\end{longtable}}
\vspace{-32pt}\begin{flushleft}While  \hyperref[Table:U11S3_Cases]{$\left[ U(1) \times U(1) \right] \rtimes S_3$}, \hyperref[Table:SO3_Cases]{$SO(3)$}, \hyperref[Table:SU3_Cases]{$SU(3)$} always result in SSB.\end{flushleft}

A key feature of $U(1)$-stabilised DM candidates is a characteristic mass-degeneracy pattern in the neutral sector. This arises from the structure of the allowed $SU(2)$ singlets, $h_{ij}$. Consider a general scenario where the vacuum of the $h_3$ doublet is stabilised, $\langle h_3 \rangle = 0$, by an underlying symmetry. The mass-squared parameters for $h_3$ receive contributions only from the quartic terms $ h_{ij} h_{33}$ and $h_{i3} h_{3j}$, as well as the bilinear term $h_{33}$. However, the quartic $ \lambda_{3333}$ term does not contribute. Consequently, the mass-squared matrix entries associated with the $h_3$ doublet remain invariant under the interchange of the two neutral fields, $\eta_3 \leftrightarrow \chi_3$, leading to $h_3^\dagger h_3 \supset \eta_3^2 + \chi_3^2$. As a result, in NHDMs, where DM stability is ensured by continuous symmetries and it is essential that the underlying symmetry remains unbroken, there is necessarily at least a single pair of mass-degenerate DM candidates, $m_{\eta_3}^2 = m_{\chi_3}^2$, present.

Now, if there are two vev-less doublets, which are stabilised by a $U(1)$ symmetry, two independent degeneracy patterns emerge in the neutral sector. In some scenarios, this results in two inert sectors with no mixing in the mass matrix, leading to two pairs of degenerate scalar fields. However, this does not necessarily indicate that there are no interactions between these sectors.

As we have seen in the previous sections, some of the implementations might share identical mass-squared matrices, which indicates that the mass eigenstates are also identical, leading, in most cases, to identical physical couplings with gauge bosons and fermion. So, the question of whether two implementations are identical, gets simplified to trying to understand if the scalar interactions are different. Although a rigorous check might be necessary to determine different physical implementations, in Ref.~\cite{Kuncinas:2024zjq} we relied on two criteria---first, one can compare the mass spectrum, identifying mass degeneracy patters, and next, if two implementations share an identical form, scalar couplings can be compared numerically. Actually, in Ref.~\cite{Kuncinas:2024zjq} we considered a more generalised case, by inspecting couplings of \textit{all} implementations independently.

Different cases exhibiting mass-degenerate pairs are summarised in Table~\ref{Table:Mass_degen_patterns}. 

For instance, consider the $U(1)_1$-symmetric 3HDM. There are two implementations, see Table~\ref{Table:Mass_degen_patterns}, with two vanishing vevs, $(v,\, 0,\, 0)$ and $(0,\, 0,\, v)$, which preserve the underlying symmetry. For the $(v,\, 0,\, 0)$ case, the neutral mass-squared matrix remains diagonal, whereas in the $(0,\, 0,\, v)$ case, mixing occurs between the $h_1$ and $h_2$ doublets due to the $\lambda_{1323}$ term. Consequently, the latter scenario features a trilinear coupling between the SM-like Higgs boson, $h_3$, and the states associated with the other two doublets. This distinction verifies that the two cases are physically different.

{{\renewcommand{\arraystretch}{1.2}
\setlength\LTcapwidth{\linewidth}
\begin{table}[h]
\caption{Mass degeneracy patterns of implementations without SSB. Some shared features of different blocks are indicated above each block. In the last column, a combination of fields, or a single field, responsible for the mass eigenstate of the SM-like Higgs boson is presented.}
\begin{center}
\begin{threeparttable}
\begin{tabular}{|c|c|c|c|c|}\hline\hline
Model & Vacuum & Symmetry & Reference & $h_\text{SM}$ \\ \hline\hline
\multicolumn{5}{c}{$m_{\eta_3}=m_{\chi_3}$} {\rule{0pt}{15pt}} \\ \hline
I-a & $(v,\, 0,\, 0)$ & $U(1) \times \mathbb{Z}_2$ & Table~\ref{Table:U1Z2_Cases} & $\eta_1$ \\ \hline
I-b & $(v,\, 0,\, 0)$ & $U(1)_2$ & Table~\ref{Table:U12_Cases} & $\{\eta_1,\, \eta_2,\, \chi_2\}$ \\ \hline
I-c & $(\hat{v}_1,\, \hat{v}_2,\, 0)$ & $U(1)_2$ & Table~\ref{Table:U12_Cases} & $\{\eta_1,\, \eta_2 ,\, \chi_1,\, \chi_2\}$\\ \hline
I-d & $(\hat{v}_1 e^{i\sigma},\, \hat{v}_2,\, 0)$ & $U(1)_2$ & Table~\ref{Table:U12_Cases} &  $\{\eta_1,\, \eta_2 ,\, \chi_1,\, \chi_2\}$ \\ \hline
\multicolumn{5}{c}{$m_{\eta_i}=m_{\chi_i}, \text{ for } \left\langle h_i\right\rangle = 0$} {\rule{0pt}{15pt}} \\ \hline
II-a &$(v,\, 0,\, 0)$ & $U(1)\times U(1)$ & Table~\ref{Table:U1_U1_Cases} &  $\eta_1$\\ \hline
II-b & $(v,\, 0,\, 0)$ & $U(1)_1$ & Table~\ref{Table:U11_Cases} &  $\eta_1$\\ \hline
II-c & $(0,\, 0,\, v)$ & $U(1) \times \mathbb{Z}_2$ & Table~\ref{Table:U1Z2_Cases} & $\eta_3$ \\ \hline \hline
\multicolumn{5}{c}{$m_{\{\eta_1,\eta_2\}} = m_{\{\chi_1,\chi_2\}}$, $\eta_i$ and $\chi_i$ mix separately} {\rule{0pt}{15pt}} \\ \hline
II-d & $(0,\, 0,\,v)$ & $U(1)_1$ & Table~\ref{Table:U11_Cases} & $\eta_3$\\ \hline
II-e & $(0,\, 0,\, v)$ & $O(2)$ & Tables~\ref{Table:O2_Cases_1} and~\ref{Table:O2_Cases_2} & $\eta_3$ \\ \hline
\multicolumn{5}{c}{$m_{\{\eta_1,\eta_2,\chi_1,\chi_2\}}$, all states mix together} {\rule{0pt}{15pt}} \\ \hline
II-f & $(0,\, 0,\, v)$ & $U(1)_2$ & Table~\ref{Table:U12_Cases} & $\eta_3$ \\ \hline\hline
\multicolumn{5}{c}{$m_{\eta_1}=m_{\eta_2}=m_{\chi_1}=m_{\chi_2}$ $^\alpha$} {\rule{0pt}{15pt}} \\ \hline
II-g & $(0,\, 0,\, v)$ & $O(2) \times U(1)$ & Tables~\ref{Table:U1U1S2_Cases_1} and~\ref{Table:U1U1S2_Cases_2} & $\eta_3$ \\ \hline
II-h & $(0,\, 0,\, v)$ & $U(1) \times D_4$ & Table~\ref{Table:U1Z2_S2_Cases} & $\eta_3$ \\ \hline \hline
II-i & $(0,\, 0,\, v)$ & $U(2)$ & Table~\ref{Table:U2_Cases} & $\eta_3$ \\ \hline
\end{tabular}
\begin{tablenotes}
\item [$\alpha$] In these cases the potential is completely symmetric under interchange $h_1\leftrightarrow h_2$. This symmetry remains unbroken in these implementations. There is also an unbroken symmetry for $\eta_i\leftrightarrow\chi_i$, CP is conserved. We have two pairs of identical fields with different names.\end{tablenotes}
\end{threeparttable}
\end{center}
\label{Table:Mass_degen_patterns}
\end{table}}

There are four distinct implementations of $U(1)_2$, which preserve the underlying symmetry. By analysing the mass-squared parameters of the $(v,\, 0,\, 0)$ and $(0,\, 0,\, v)$ cases, one finds that these cases exhibit different numbers of mass-degenerate states, as shown in Table~\ref{Table:Mass_degen_patterns}. Furthermore, the case $(v_1,\, v_2,\, 0)$ is related to $(v,\, 0,\, 0)$ via a Higgs basis transformation. Due to the CP indefinite states, implementations with and without CP violation yield different physical cases. To confirm whether $(\hat{v}_1,\, \hat{v}_2,\, 0)$ and $ (\hat{v}_1 e^{i \sigma},\, \hat{v}_2,\, 0)$ represent different scenarios, the latter case should be examined to determine if it imposes additional constraints. It is important to emphasise that spontaneous CP violation can be represented  by a complex potential and real vevs. However, in the considered cases, the phases of the potential terms are related. 

Although different implementations stabilised by the same underlying symmetry correspond to a different physical parameter space, it is not guaranteed that this argument holds among different symmetries. In conclusion, to distinguish different implementations one can not rely solely on the mass-degeneracy patterns. 

In Table~\ref{Table:Couplings_tri_quart}, we summarise the total number of trilinear and quartic couplings, excluding vertices involving Goldstone bosons. The goal is to determine whether certain implementations of different models may lead to physically identical scenarios. A recurring pattern of ``7 + 32" couplings emerges in several implementations: $U(1) \times U(1)$, $U(1) \times \mathbb{Z}_2$, $O(2) \times U(1)$, and $U(2)$. While the mass-degeneracy patterns, listed in Table~\ref{Table:Mass_degen_patterns}, generally differ among these implementations, two cases, $O(2) \times U(1)$ and $U(2)$, share the same degeneracy pattern. To assess their physical distinctness, one could investigate whether some physical couplings are related by a constant factor across these two cases. Upon a more thorough verification, we confirm that the coupling patterns are distinct for all four models. Thus, we conclude that they, and all the analysed models and their implementations, as listed at the beginning of this section, represent physically distinct scenarios.  We identified thirteen distinct cases that could accommodate DM. A comprehensive study of various DM models would necessitate a detailed numerical analysis.

\begin{table}[htb]
\caption{
The table presents the physical scalar couplings (Goldstone bosons are not accounted) for different implementations that preserve the underlying symmetry. Each entry corresponds to the total number of non-vanishing trilinear and quartic couplings. Some of the couplings might be related by simple constants, \textit{e.g.}, $g(X) = C g(Y)$. The number of independent couplings are given in parentheses. We do not double-count Hermitian conjugate couplings, meaning that terms like \( g(H_i H_j^+ H_k^-) \) and \( g(H_i H_j^- H_k^+)^* \) are treated as a single entry in the parentheses. The two \( U(1) \times U(1) \) implementations exhibit identical structures upon re-labeling the fields, while the $U(1)_2$ model is the only one that features CP violation.}
\label{Table:Couplings_tri_quart}
\begin{center}
\begin{tabular}{|c|c|c|c|c|} \hline\hline
Symmetry & $(v,\, 0,\, 0)$ & $(0,\, 0,\, v)$ & $(v_1,\, v_2,\, 0)$ \\ \hline
$ \quad U(1) \times U(1)$&	\multicolumn{2}{c|}{7 + 32 (5 + 10)} &\\ \hline 
$ \quad U(1)_1$ &	15 + 40 (8 + 13) &	9 + 53 (6 + 24) &	\\ \hline  
$ \quad U(1) \times \mathbb{Z}_2$ &	7 + 32 (6 + 11)&	 7 + 38 (5 + 16) &	\\ \hline 
$ \quad U(1)_2$ &  \begin{tabular}[l]{@{}c@{}} $\mathbb{R}:16 + 41\,(13 + 24)$ \\ $\mathbb{C}:24 + 55\,(20 + 40)$  \end{tabular} &	\begin{tabular}[l]{@{}c@{}} $\mathbb{R}:10 + 55\,(7 + 30)$ \\ $\mathbb{C}:12 + 73\,(8 + 36)$  \end{tabular} &	\begin{tabular}[l]{@{}c@{}} $\mathbb{R}:16 + 41\,(13 + 24)$ \\ $\mathbb{C}:24 + 55\,(20 + 40)$  \end{tabular}\\ \hline
$ \quad O(2) \times U(1)$ &	& 7 + 32 (3 + 7)	&	\\ \hline   
$ \quad O(2)$ &	& 7 + 39 (4 + 12)	&	\\ \hline   
$ \quad U(1) \times D_4$ &	&	7 + 34 (3 + 11) &	\\ \hline   
$ \quad U(2)$ &	&	7 + 32 (3 + 6) &	\\ \hline   
\end{tabular} 
\end{center}
\end{table}}

\section[Dark Matter in the \texorpdfstring{$U(1)\times U(1)$}{U(1) x U(1)} 3HDM]{Dark Matter in the \boldmath$U(1)\times U(1)$ 3HDM}\label{Sec:U1U1_analysis}

After identifying and classifying various models and their different implementations, we turn our focus to a numerical scan of the $U(1) \times U(1)$-symmetric 3HDM, because it represents the most general real scalar potential. As outlined in Section~\ref{Sec:Different_DM_cases}, we know that different models display distinct physical properties. It is important to emphasize that our study of the $U(1) \times U(1)$-symmetric 3HDM is intended primarily for illustrative purposes. We consider it a toy model that serves to highlight the key features and physical distinctions among different implementations, rather than as a definitive description of the underlying physics. We assume a conventional freeze-out mechanism similar to that used in the IDM-like scenarios, except that we now have four DM candidates grouped into two pairs of mass-degenerate states. Alternatively, one could consider these states as complex scalars, though in our framework both approaches lead to equivalent outcomes.

Since WIMPs do not carry an electric charge, they generally do not interact with atomic electrons. If two DM candidates are mass-degenerate, they can undergo elastic scattering (scatter without any change in states) with atomic nuclei~\cite{Goodman:1984dc}. In models with mass-degenerate DM candidates, based on implementations, there is an unsuppressed cross-section for spin-independent elastic scattering of nuclei, imposing tight constraint from experiments. It should be noted that the cross-section depends on both the local WIMP density and its speed distribution in the detector's frame.

For example, the elastic scattering of nuclei in the \mbox{$U(1) \times U(1)$}-symmetric model is given by the $\eta_i q \xrightarrow[]{Z} \chi_i q$ processes, where $m_{\eta_i} = m_{\chi_i}$. The key challenge in direct DM detection arises from the spin-independent elastic scattering cross section mediated by the $Z$ boson. In the NHDMs the interaction strength is fixed by the gauge coupling, which is roughly, $g_{Z\eta\chi} \sim 0.37$. The current direct DM detection experiments effectively rule out couplings even several orders of magnitude below the provided value. To reconcile theory with experimental observations, a common solution is to introduce kinematic suppression by ensuring that the mass splitting between the two nearly degenerate particles exceeds a few hundred keV. This way the elastic scattering process, otherwise induced by the $Z$ boson exchange, becomes kinematically forbidden. This constraint is crucial regardless of the DM candidate mass.

We stress that the $U(1) \times U(1)$-symmetric model possesses several challenges and thus should be viewed as a toy model.

\subsection{Mass-squared parameters}

We assume that the vacuum is given by $(v,\,0,\,0)$, which does not spontaneously break the underlying $U(1) \times U(1)$ symmetry. Since the mass-squared matrix is block-diagonal, the scalar states coincide with the mass eigenstates. The mass of the SM-like Higgs boson is:
\begin{equation}
m_h^2 = 2 \lambda_{1111} v^2.
\end{equation}
The mass-squared parameters of the charged states are:
\begin{equation}
m_{h_i^+}^2 = \mu_{ii}^2 + \frac{1}{2} \lambda_{11ii}v^2,\text{ for }i=\{2,3\}.
\end{equation}
The neutral states associated with $h_2$ and $h_3$ are mass-degenerate:
\begin{equation}
m_{H_i}^2 \equiv m_{\eta_i}^2 = m_{\chi_i}^2  = \mu_{ii}^2 + \frac{1}{2} \left( \lambda_{11ii} + \lambda_{1ii1} \right) v^2, \text{ for } i=\{2,\,3\}.
\end{equation}
The masses of the charged and neutral states are connected:
\begin{equation}\label{Eq:U1U1_mHNi}
m_{H_i}^2 = m_{h_i^+}^2 + \frac{v^2}{2} \lambda_{1ii1}, \text{ for } i=\{2,\,3\}.
\end{equation}
We require $\lambda_{1ii1}<0$, as otherwise the charged states would be lighter than the neutral ones. Also, without loss of generality, we may assume that states respect the following mass-hierarchy: $m_{H_2}^2 \leq m_{H_3}^2$. However, this hierarchy does not guarantee that the charged states should also be ordered.

\subsection{The kinetic Lagrangian}

The kinetic Lagrangian for the $(U(1) \times U(1)$-symmetric 3HDM is analogous to that of the IDM, with the only difference being the presence of an extra copy of the doublet with a vanishing vev. Excluding the Goldstone bosons, the kinetic Lagrangian is given by:
\begin{subequations}
\begin{align}
\begin{split}
\mathcal{L}_{VVH} =& \left[ \frac{g}{2 \cos \theta_W}m_ZZ_\mu Z^\mu + g m_W W_\mu^+ W^{\mu-} \right] h,
 \end{split}\\
\begin{split}
\mathcal{L}_{VHH} =& -\frac{ g}{2 \cos \theta_W}Z^\mu \eta_i \overset\leftrightarrow{\partial_\mu} \chi_i  - \frac{g}{2}\bigg\{ i W_\mu^+ \left[ ih_i^- \overset\leftrightarrow{\partial^\mu} \chi_i + h_i^- \overset\leftrightarrow{\partial^\mu} \eta_i  \right] + \mathrm{h.c.} \bigg\}\\
& + \left[ i e A^\mu + \frac{i g}{2} \frac{\cos (2\theta_W)}{\cos \theta_W} Z^\mu \right]  h_i^+ \overset\leftrightarrow{\partial_\mu} h_i^-  ,
\end{split}\\
\begin{split}
\mathcal{L}_{VVHH} =& \left[ \frac{g^2}{8 \cos^2\theta_W}Z_\mu Z^\mu + \frac{g^2}{4} W_\mu^+ W^{\mu-} \right] \left( \eta_i^2 + \chi_i^2 + h^2\right)\\
& + \bigg\{ \left[ \frac{e g}{2} A^\mu W_\mu^+ - \frac{g^2}{2} \frac{\sin^2\theta_W}{\cos \theta_W}Z^\mu W_\mu^+ \right] \left[ \eta_i h_i^- + i \chi_i h_i^- \right] + \mathrm{h.c.} \bigg\}\\
&+ \left[ e^2 A_\mu A^\mu + e g \frac{\cos (2\theta_W)}{\cos \theta_W}A_\mu Z^\mu + \frac{g^2}{4} \frac{\cos^2(2\theta_W)}{\cos^2\theta_W}Z_\mu Z^\mu + \frac{g^2}{2} W_\mu^- W^{\mu +} \right] \\
&\quad~\times\left( h_i^-h_i^+ \right),
\end{split}
\end{align}
\end{subequations}
where the index $``i"$  is summed over the doublets $h_2$ and $h_3$.

In the $U(1) \times U(1)$-symmetric 3HDM, there are two pairs of neutral mass-degenerate states. Consequently, in the early Universe, processes like $\eta_i \rightarrow Z \chi_i$ or $\chi_i \rightarrow Z \eta_i$ did not occur. Moreover, if the masses satisfy $m_{H_i} < m_Z/2$, then the decay rate for $Z \rightarrow \eta_i \chi_i$ is subject to constraints imposed by the invisible width of the $Z$ boson.

\subsection{The scalar interactions}\label{Sec:U1xU1_SSS_SSSS}

The scalar couplings are presented with the symmetry factor, but without the overall coefficient ``$-i$". Trilinear scalar couplings are:
\begin{subequations}
\begin{align}
g (\eta_i^2 h) = g(\chi_i^2h) ={}& v \left( \lambda_{11ii} + \lambda_{1ii1} \right),\label{Eq:U1xU1_portal_3}\\
g(h^3) ={}& 6 v \lambda_{1111},\\
g(h_i^+ h_i^- h) ={}& v \lambda_{11ii}.
\end{align}
\end{subequations}

Quartic interactions, with $i \neq j$, are:
\begin{subequations}
\begin{align}
g(\eta_i^4) = g(\chi_i^4) ={}& 6 \lambda_{iiii},\\
g(h^4) ={}& 6 \lambda_{1111},\\
g(\eta_i^2 \chi_i^2) = g(\eta_i^2 h^+_i h^-_i) = g(\chi_i^2 h^+_i h_i^-) ={}& 2 \lambda_{iiii},\\
g(\eta_i^2 \eta_j^2) = g(\chi_i^2 \eta_j^2) = g(\chi_i^2 \chi_j^2) = g(h_i^+ h_i^- h_j^+ h_j^-) ={}& \lambda_{iijj} + \lambda_{ijji},\\
g(\eta_i^2 h^2) = g(\chi_i^2 h^2) ={}& \lambda_{11ii} + \lambda_{1ii1}, \label{Eq:U1xU1_portal_4}\\
g(\eta^2_i h_j^+ h_j^-) = g(\chi^2_i h_j^+ h_j^-) ={}& \lambda_{iijj},\\
g(h_i^+ h_i^- h_i^+ h_i^-) ={}& 4 \lambda_{iiii},\\
g(\eta_i \eta_j h_i^+ h_j^-) = g(\chi_i \chi_j h_i^+ h_j^-) ={}& \frac{1}{2} \lambda_{ijji},\\
g(h^2 h_i^+ h_i^-) ={}& \lambda_{11ii},\label{Eq:U1xU1_portal_Ch}\\
g(\chi_i \eta_j h^+_j h^-_i) ={}& \frac{i}{2} \lambda_{ijji},\\
g(\eta_i \chi_j h^+_j h^-_i) ={}& -\frac{i}{2} \lambda_{ijji}.   
\end{align}
\end{subequations}
All these couplings, apart from the last two, $\chi_i \eta_j h^+_j h^-_i$ and $\eta_i \chi_j h^+_j h^-_i$, are symmetric under the interchange $\eta_i \leftrightarrow \chi_i$, while the latter two are related by complex conjugation. Due to the underlying symmetry, there are no trilinear interactions between $h_2$ and $h_3$.

\subsection{The Yukawa Lagrangian}

We emphasise that the unbroken $U(1) \times U(1)$ symmetry acts solely on the $\{h_2,\, h_3\}$ doublets. Suppose that the extension of the model is non-minimal, \textit{e.g.}, there are non-SM fermion. Then, one of the $\{h_2,\, h_3\}$ doublets could couple to such fermions. We do not consider such cases. Instead, we restrict ourselves to the SM-like fermionic sector that couples exclusively to the $h_1$ doublet. In doing so, the generic SM Yukawa Lagrangian is reproduced.

\subsection{Comparison with the Inert Doublet Model}

In the IDM, with $h_1$ and $h_2$ transforming under $\mathbb{Z}_2$, the scalar potential is given by:
\begin{equation}
\begin{aligned}
V_\mathrm{IDM} ={}& \mu_{11}^2 h_{11} + \mu_{22}^2 h_{22} + \lambda_{1111} h_{11}^2 + \lambda_{2222} h_{22}^2 + \lambda_{1122} h_{11} h_{22} + \lambda_{1221} h_{12} h_{21}\\
&  + \lambda_{1212} \left( h_{12}^2 + h_{21}^2 \right),
\end{aligned}
\end{equation}
where all parameters are taken to be real.

For simplicity, let us split the $U(1) \times U(1)$-symmetric scalar potential into several parts, which would resemble the IDM-like structure:
\begin{equation}\label{Eq:V_U1xU1_expl}
\begin{aligned}
V_{U(1) \times U(1)} =\hspace{10pt}& \frac{1}{2}\mu_{11}^2 h_{11} + \mu_{22}^2 h_{22} + \frac{1}{2}\lambda_{1111} h_{11}^2 + \lambda_{2222} h_{22}^2 + \lambda_{1122} h_{11} h_{22} + \lambda_{1221} h_{12} h_{21}\\
 +& \frac{1}{2}\mu_{11}^2 h_{11} + \mu_{33}^2 h_{33} + \frac{1}{2}\lambda_{1111} h_{11}^2 + \lambda_{3333} h_{33}^2 + \lambda_{1133} h_{11} h_{33} + \lambda_{1331} h_{13} h_{31}\\
 +& \lambda_{2233} h_{22} h_{33} + \lambda_{2332} h_{23} h_{32}.
\end{aligned}
\end{equation}
This way it is easy to compare the two potentials. Each of the first two lines corresponds to an IDM-like potential, with $\lambda_{1212}=0$. Actually, in this case, as should be expected, the underlying symmetry is increased to $U(1)$. Although there are less free parameters, it is still possible to tune several couplings, like the portal coupling.

Unlike the IDM, the $(U(1) \times U(1)$-symmetric model features several new interactions between the inert doublets, see the last line of eq.~\eqref{Eq:V_U1xU1_expl}. These interactions play a crucial role in governing the annihilation of $h_3$ into $h_2$. Moreover, since $\{\lambda_{2233},\, \lambda_{2332}\}$ are not multiplied by $h_1$, these terms contribute solely to the quartic interactions between $h_2$ and $h_3$.

Apart from the $U(1)$-symmetric 2HDM, the Replicated Inert Doublet Model (RIDM) also closely resembles the $U(1) \times U(1)$-symmetric 3HDM. The RIDM was presented in Ref.~\cite{Haber:2018iwr} with the scalar potential given by
\begin{equation}
\begin{aligned}
V_\mathrm{RIDM} ={}& \mu_{11}^2 h_{11} + \mu_{22}^2 \left( h_{22} + h_{33} \right) + \lambda_{1111} h_{11}^2 + \lambda_{2222} \left( h_{22} + h_{33} \right)^2\\
& + \lambda_{1122} \left( h_{11} h_{22} + h_{11} h_{33} \right) + \lambda_{1221} \left( h_{12} h_{21} + h_{13} h_{31} \right)\\
& + \lambda_{1212} \left( h_{12}^2 + h_{21}^2 + h_{13}^2 + h_{31}^2 \right).
\end{aligned}
\end{equation}
In the RIDM, for the vacuum $(v,\,0,\,0)$, the mass-squared parameters are given by:
\begin{subequations}
\begin{align}
m_{h_2^\pm}^2 = m_{h_3^\pm}^2 ={}& \mu_{22}^2 + \frac{1}{2} \lambda_{1122} v^2,\\
m_{\eta_2}^2 = m_{\eta_3}^2 ={}& \mu_{22}^2 + \frac{1}{2} \left( \lambda_{1122} + \lambda_{1221} + 2 \lambda_{1212} \right)v^2,\\
m_{\chi_2}^2 = m_{\chi_3}^2 ={}& \mu_{22}^2 + \frac{1}{2} \left( \lambda_{1122} + \lambda_{1221} - 2 \lambda_{1212} \right)v^2.
\end{align}
\end{subequations}
When $\lambda_{1212} = 0$, all the neutral states arising from the inert doublets become mass-degenerate. This degeneracy pattern can be attributed to the global underlying $O(4)$ symmetry.

\subsection{Numerical studies}

The parameter space of the $U(1) \times U(1)$-symmetric model was scanned, taking into account:
\begin{itemize}
\item Scalars can be as heavy as 2 TeV;
\item The DM candidates correspond to the neutral states of the $h_2$ and $h_3$ doublets, with the mass hierarchy $m_{H_2} \leq m_{H_3}$;
\item We consider that: $m_{H_i^\pm} \geq 70\text{ GeV}$ and  $m_{H_i} > \frac{1}{2}m_Z$;
\item Unitarity, perturbativity, stability. We verified the stability conditions of Ref.~\cite{Faro:2019vcd}, which were later re-evaluated in Ref.~\cite{Boto:2022uwv}. We found the unitarity condition to be consistent with Ref.~\cite{Bento:2022vsb};
\item The electroweak oblique parameters $S$ and $T$ are evaluated following the technique of Refs.~\cite{Grimus:2007if,Grimus:2008nb} at $U=0$;
\item The Higgs boson invisible decay is $\text{Br} \left(h \to \mathrm{invisible} \right)\leq 0.107$~\cite{CMS:2022dwd,ATLAS:2022vkf,ATLAS:2023tkt,ParticleDataGroup:2024cfk};
\item LHC search results implemented in $\mathsf{HiggsTools}$~\cite{Bahl:2022igd}, utilising $\mathsf{HiggsBounds}$~\cite{Bechtle:2008jh,Bechtle:2011sb,Bechtle:2012lvg,Bechtle:2013wla,Bechtle:2015pma,Bechtle:2020pkv,Bahl:2021yhk} and $\mathsf{HiggsSignals}$~\cite{Bechtle:2013xfa,Stal:2013hwa,Bechtle:2014ewa,Bechtle:2020uwn}; 
\item DM relic density is assumed to be consistent with the Planck measurements~\cite{Planck:2018vyg}. We evaluate the total relic density as $\Omega_\mathrm{DM}h^2 = \Omega_2 h^2 + \Omega_3 h^2$, where $\Omega_i$ are contributions of the $h_2$ and $h_3$ doublets;
\item Direct DM searches based on XENONnT~\cite{XENON:2020kmp,XENON:2022ltv,XENON:2023cxc} and LUX-ZEPLIN~\cite{LZ:2021xov,LZ:2022lsv,LZ:2024zvo};
\item Indirect DM searches based on Ref.~\cite{Hess:2021cdp}.
\end{itemize}
We allow for a 3-$\sigma$ tolerance with the numerical values taken from the PDG~\cite{ParticleDataGroup:2024cfk}. Apart from that, we take into account a 10\% tolerance, due to theoretical and computational uncertainties. For the evaluation of the DM observables we utilise $\mathtt{micrOMEGAs~6.1.15}$~\cite{Belanger:2004yn,Belanger:2006is,Belanger:2008sj,Belanger:2010pz,Belanger:2013oya,Belanger:2014vza,Barducci:2016pcb,Belanger:2018ccd,Alguero:2023zol}. We accumulated a total of $10^6$ parameter points simultaneously satisfying all constraints. For additional discussion see Section~\ref{Sec:S3_DM_evaluation_checks}.

Contributions to the relic density from the two sectors are illustrated in Figure~\ref{Fig:Omega_mHi} as functions of the DM mass. In a scenario where only the lighter inert doublet, $h_2 $, contributes to the DM relic density, the resulting relic density plot resembles that of the IDM. However, with the inclusion of the second inert doublet, $h_3$, additional quartic interactions between the two inert sectors become possible, given by interactions $g(h_2 h_2 h_3 h_3) = f(\lambda_{2233},\, \lambda_{2332})$. These quartic interactions facilitate the annihilation of the heavier doublet into the lighter one, which is evident from the observed region of low relic density in panel (b) compared to panel (a) of Figure~\ref{Fig:Omega_mHi}. The viable mass ranges are displayed in Figure~\ref{Fig:U1xU1_masses}.

\vspace*{30pt}		  
\begin{figure}[htb]
\begin{center}
\includegraphics[scale=0.255]{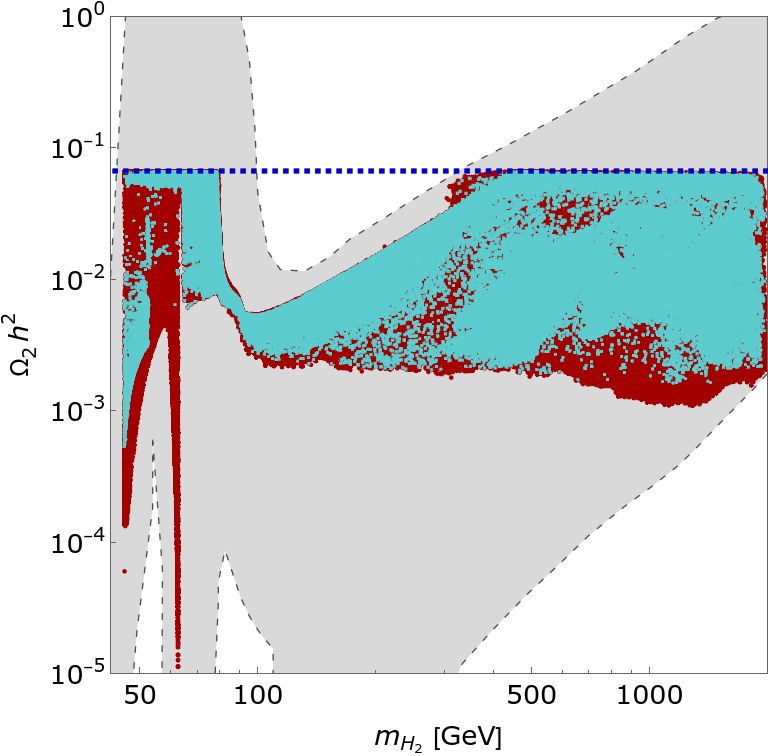}
\includegraphics[scale=0.255]{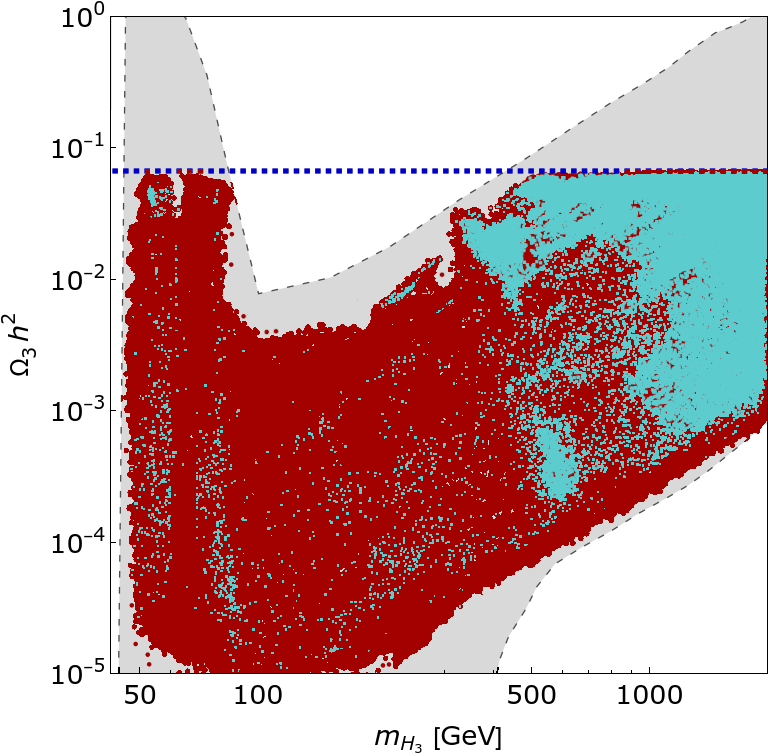}
\includegraphics[scale=0.255]{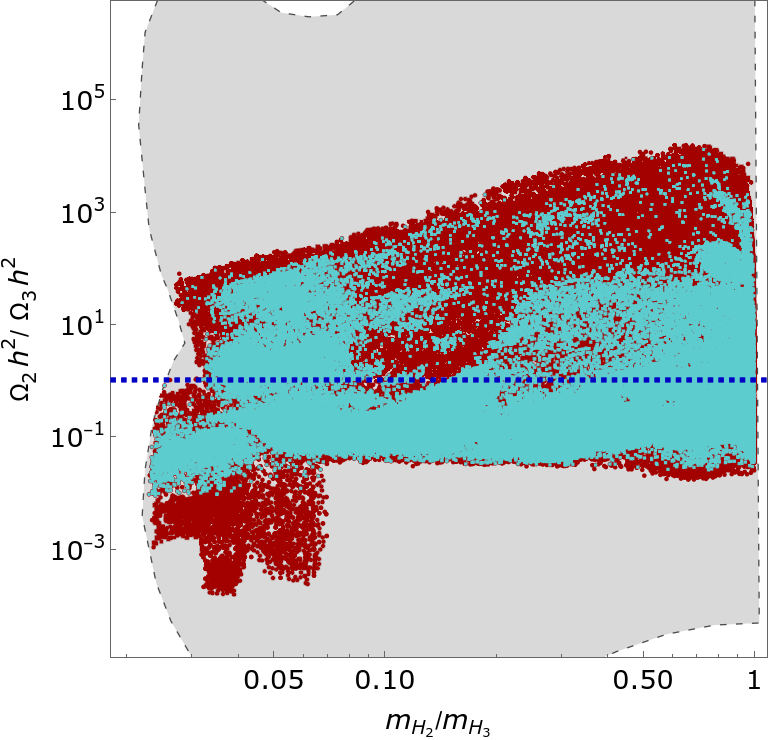}
\put (-370,132) {\small (a)}
\put (-220,132) {\small (b)}
\put (- 65,132) {\small (c)}
\end{center}
\vspace*{-3mm}
\caption{ Panels (a), (b): relic density as a function of mass for the two DM candidates. The  horizontal blue line indicates the maximal relic density, $\Omega h^2 \approx 2 \Omega_i h^2$. Panel (c): ratio of the relic density components, $\Omega_2 h^2$ and $\Omega_3 h^2$, as a function of the ratio of the DM masses. Now, the blue line indicates equal contribution the relic density, $\Omega_2 h^2 = \Omega_3 h^2$. The grey blob represents parameter points satisfying only theoretical constraints. Red points correspond to parameters also satisfying experimental constraints, as presented in Ref.~\cite{Kuncinas:2024zjq}. The cyan points correspond to updated scans, mainly including the updated LUX-ZEPLIN~\cite{LZ:2024zvo} results. }
\label{Fig:Omega_mHi}
\end{figure}

\begin{figure}[htb]
\begin{center}
\includegraphics[scale=0.3]{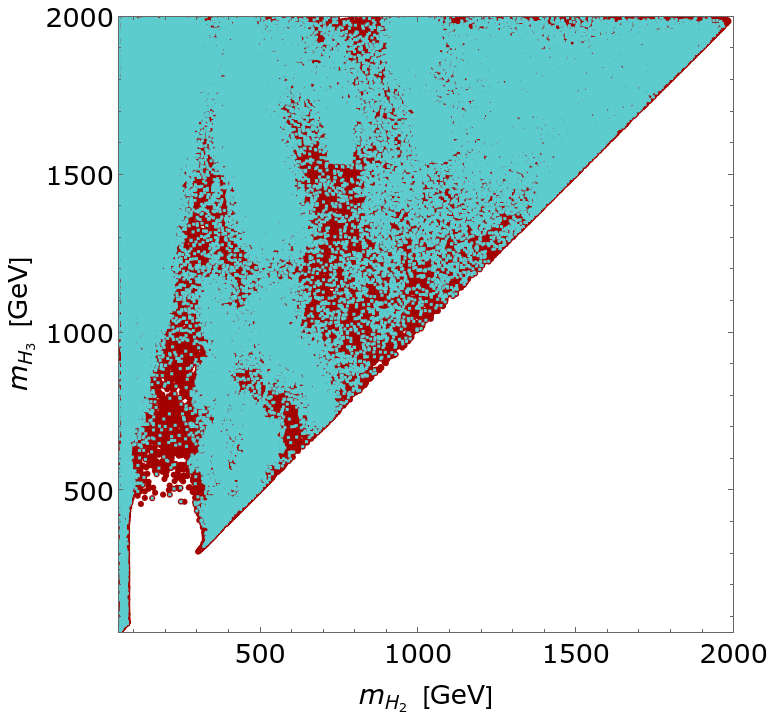}
\includegraphics[scale=0.3]{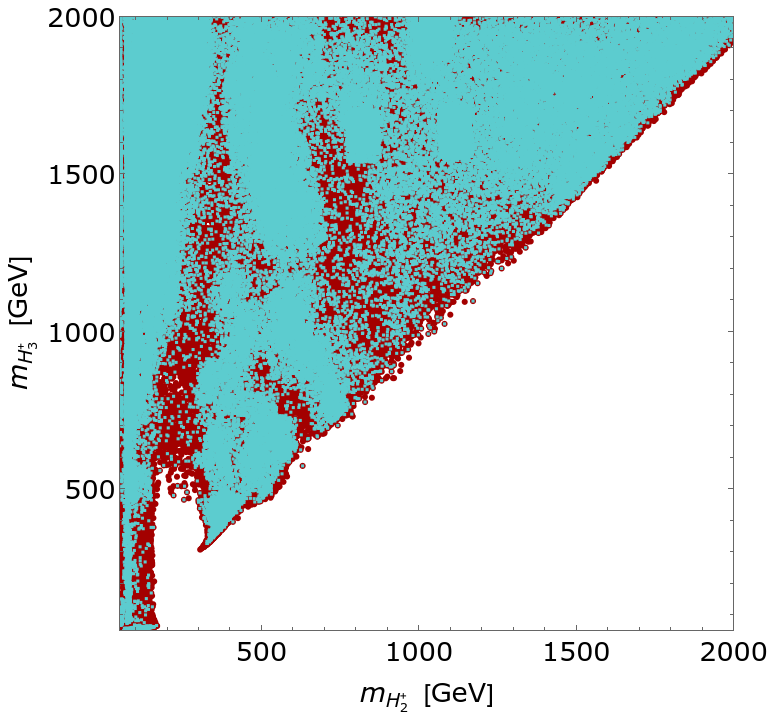}
\end{center}
\vspace*{-3mm}
\caption{ Mass scatter plots of parameters satisfying all constraints. Left: masses of the neutral inert sector. The labeling is $m_{H_2} \leq m_{H_3}$.  Right: masses of the charged inert sectors. Red points correspond to parameters evaluated in Ref.~\cite{Kuncinas:2024zjq}. The cyan points correspond to updated scans. No parameter space in the planes of masses was ruled out.}
\label{Fig:U1xU1_masses}
\end{figure}

\newpage

Since two scalar sectors contribute to $\Omega h^2$, it is useful to compare their individual contributions. This comparison is presented in panel (c) of Figure~\ref{Fig:Omega_mHi}. Additionally, we examined scenarios where both mass scales and their respective contributions to $\Omega h^2$ are nearly equal, within 5\%. Two distinct regions satisfy this criterion: $ [50;\, 80]$ GeV and $[300;\, 2000]$ GeV. We recall that in the IDM there is a region starting at $ m_\mathrm{DM} \sim 500$ GeV, which is compatible with all constraints. In the case of the  $U(1) \times U(1)$-symmetric 3HDM, it starts at $m_\mathrm{DM} \sim 300$ GeV. This bound aligns with that of the truncated $\mathbb{Z}_2$-symmetric 3HDM, where $m_\mathrm{DM} \sim 360$ GeV, as illustrated in Figure~\ref{Fig:DM_mass_ranges_different_models}.

In Ref.~\cite{Kuncinas:2024zjq}, the dominant part of the parameter space was found to be incompatible with the direct DM detection constraints.  In the low-mass region, the sum of the couplings \mbox{$(\lambda_{11ii} + \lambda_{1ii1})$}, the Higgs portal coupling, must be kept small, as indicated in eqs.~\eqref{Eq:U1xU1_portal_3} and \eqref{Eq:U1xU1_portal_4}. Meanwhile, the coupling \( \lambda_{1ii1} \) plays a crucial role in determining the mass splitting between the neutral and charged states, as shown in eq.~\eqref{Eq:U1U1_mHNi}. The updated constraints, coming from LUX-ZEPLIN~\cite{LZ:2024zvo} further rule out some parts of the parameter space, \textit{e.g.}, notice that in Figure~\ref{Fig:Omega_mHi} there are no cyan-coloured points in panel (a) extending to the bottom for $m_{H_2} \approx m_{h}/2$. In contrast, indirect DM detection constraints are sub-leading.

Let us summarise several aspects of the $U(1)\times U(1)$-symmetric 3HDM:
\begin{itemize}  \setlength\itemsep{2.3pt}
\item The key distinction from the conventional $\mathbb{Z}_2$ stabilisation of DM is the existence of pairs of mass-degenerate neutral states with opposite CP parities, which can also be interpreted as complex neutral scalars;
\item When the DM masses simultaneously fall within the intervals $m_{H_2} \in [100;\,300]$ GeV and $m_{H_3} \in [100;\,450]$ GeV, the annihilation into gauge bosons in the Early Universe becomes too efficient, leading to a conflict with observations;
\item Assuming that $m_{H_2} \sim m_{H_3}$ and $\Omega_2 h^2 \sim \Omega_3 h^2 $, there are two regions: $[50;\, 80]$ GeV and $[300;\, 2000]$ GeV;
\item Regions were $\Omega h^2 $ is dominated by a single state, see panel (c) in Figure~\ref{Fig:Omega_mHi} are: $[45;\,80]$ GeV and $[330;\,2000]$ GeV for $h_2$, $[52;\,80]$ GeV and $[470;\,2000]$~GeV for $h_3$.
\end{itemize}

As stated earlier, our goal was not to systematically scrutinize the model, but rather to treat it as a toy model; we recall that there are some challenges present from elastic scattering with nuclei. Given the presence of two dark doublets, exploring different production channels at colliders could be of interest. While the model itself is simple, different implementations might lead to distinct observable signals. A key difference between DM candidates stabilised by continuous symmetries and those stabilised by $\mathbb{Z}_2$ is mass degeneracy. Generally, both the CP-even and CP-odd components of the same sector contribute equally to astrophysical and cosmological constraints. This does not necessarily indicate that both components are equally abundant. Mechanisms could exist that lead to local variations in relative abundance, and it is also possible that the mass-degenerate states decouple at different temperatures.

\chapter{Conclusions}
\label{chapter:conclusions}

In summary, the SM’s inability to address several key issues, most notably, the existence and nature of DM and additional sources of CP violation, has driven the exploration of NHDMs. These models introduce a multitude of degrees of freedom, prompting researchers to search for underlying patterns that can simplify their complexity.

In particular, 3HDM with an underlying $S_3$ symmetry offers a fertile ground for addressing DM and CP violation challenges, alas not both simultaneously. While the IDM suffers from the fact that its underlying $\mathbb{Z}_2$ symmetry prohibits complex phases, in the $S_3$-symmetric 3HDM the implementations capable of accommodating a DM candidate force fermions to transform as singlets under $S_3$, which is the SM-like case. Through a detailed analysis of various vacuum configurations, both with real and complex parameters, and with and without soft symmetry breaking, several promising scenarios have been identified. Some of these scenarios provide viable DM candidates, stabilised by remnant symmetries of the $S_3$ group, while others yield interesting patterns of CP violation, either explicit or spontaneous, that have implications for the flavour structure and the generation of the CKM matrix. Furthermore, the presence of light scalars in the C-V implementation with $\lambda_4 \in \mathbb{C}$ is quite intriguing, although such states might be ruled out by some unconsidered (in the work) experimental channels. 

We performed an updated numerical analysis of two cases of the $S_3$-symmetric 3HDM. The R-II-1a implementation was shown to be capable of accommodating a DM candidate within the 52.7--82.9 GeV range, while for C-III-a, the corresponding updated range is 28.9--41.9 GeV, which may altogether be ruled out, depending on the DM halo distribution profile. In addition, we briefly examined two other implementations---R-I-1, with and without complex $\lambda_4$, and  C-III-a with $\lambda_4 \in \mathbb{C}$. It would be interesting to cover these cases numerically in the near future, and we already got some hints how the parameter space looks like for these two cases.

From another perspective, it is also interesting to consider DM stabilisation mechanisms other than by the usual discrete symmetries, specifically stabilisation by continuous symmetries. We identified all possible ways of embedding a $U(1)$ symmetry in 3HDMs that can accommodate DM. All these models contain mass-degenerate pairs of DM candidates due to an unbroken continuous symmetry. We recall that broken continuous symmetries result in Goldstone bosons. An intriguing outcome was the rediscovery and identification of some ``forgotten”  (and also some new) models in terms of the $U(1)$-based symmetries, which indicates that many open questions in 3HDMs remain. It would be interesting to perform an analysis similar to the one we did for continuous symmetries, but this time focusing on the discrete symmetries of 3HDMs, to investigate what additional stabilisation mechanisms exist beyond $\{\mathbb{Z}_2,\, \mathbb{Z}_3,\, S_3 \}$.

We also performed a numerical exploration of the $U(1) \times U(1)$-symmetric 3HDM. Although this choice might not have been optimal, the analysis provides useful insights into how to discriminate models  with $U(1)$ symmetries from those with discrete symmetries in  the context of 3HDMs. The model features a multi-component DM sector, with two independent mass scales. After imposing the relevant experimental constraints we found viable solutions throughout a broad DM mass range, 45--2000 GeV, the latter being an upper limit of our scan. We found that only the region where \textit{both} neutral dark scalars fall within the range  $\mathcal{O}(100)-\mathcal{O}(300)$~GeV is excluded. After revising the dataset, mainly incorporating the latest LUX-ZEPLIN constraints, we concluded that some parameter space was ruled out. Usually, both the CP-even and the CP-odd components of the same sector contribute equally to the astrophysical and cosmological constraints. However, this does not necessarily imply that both components are equally abundant. There could be mechanisms leading to local variations in their relative abundances, or the components of the mass-degenerate states might have decoupled in the Early Universe at different temperatures.

In summary, the rich phenomenology of the studied models, including the light neutral scalars, mass-degenerate DM candidates, and distinct patterns of scalar couplings, makes them particularly attractive for future studies. Detailed numerical scans and further experimental investigations will be crucial to fully understand the implications of these models. Ultimately, the work reviewed here not only enhances our understanding of how extended scalar sectors can accommodate DM and CP violation but also provides a robust framework and many concrete results that can be used by model builders in the ongoing quest for physics beyond the SM.

Our world would have been more interesting if the Higgs boson had not turned out to be around 125 GeV; such a discovery might have opened up a broader landscape for phenomenological work. Regardless, current experimental evidence supports a single, well-defined model that ``accurately” describes the observed properties of the Higgs boson.  After all, there is only one model that Nature prefers! This reality reaffirms the robustness of the SM while simultaneously challenges us to refine our approaches to uncover any subtle signs of BSM.

\newpage
\phantomsection
\bibliographystyle{JHEP}
\addcontentsline{toc}{chapter}{Bibliography}
\bibliography{ref}

\providecommand{\href}[2]{#2}\begingroup\raggedright\begin{thebibliography}{100}

\bibitem{Emmanuel-Costa:2016vej}
D.~Emmanuel-Costa, O.~M. Ogreid, P.~Osland and M.~N. Rebelo, \emph{{Spontaneous
  symmetry breaking in the $S_3$-symmetric scalar sector}},
  \href{http://dx.doi.org/10.1007/JHEP08(2016)169,
  10.1007/JHEP02(2016)154}{\emph{JHEP} {\bf 02} (2016) 154},
  [\href{http://arxiv.org/abs/1601.04654}{{\tt 1601.04654}}].

\bibitem{Das:2014fea}
D.~Das and U.~K. Dey, \emph{{Analysis of an extended scalar sector with $S_3$
  symmetry}}, \href{http://dx.doi.org/10.1103/PhysRevD.91.039905,
  10.1103/PhysRevD.89.095025}{\emph{Phys. Rev.} {\bf D89} (2014) 095025},
  [\href{http://arxiv.org/abs/1404.2491}{{\tt 1404.2491}}].

\bibitem{ParticleDataGroup:2024cfk}
S.~Navas et~al., \emph{{Review of particle physics}},
  \href{http://dx.doi.org/10.1103/PhysRevD.110.030001}{\emph{Phys. Rev. D} {\bf
  110} (2024) 030001}.

\bibitem{LHCHiggsCrossSectionWorkingGroup:2013rie}
J.~R. Andersen et~al., \emph{{Handbook of LHC Higgs Cross Sections: 3. Higgs
  Properties}},  \href{http://arxiv.org/abs/1307.1347}{{\tt 1307.1347}}.

\bibitem{ATLAS:2022vkf}
\emph{{A detailed map of Higgs boson interactions by the ATLAS experiment ten
  years after the discovery}},
  \href{http://dx.doi.org/10.1038/s41586-022-04893-w}{\emph{Nature} {\bf 607}
  (2022) 52--59}, [\href{http://arxiv.org/abs/2207.00092}{{\tt 2207.00092}}].

\bibitem{CMS:2022dwd}
\emph{{A portrait of the Higgs boson by the CMS experiment ten years after the
  discovery}},
  \href{http://dx.doi.org/10.1038/s41586-022-04892-x}{\emph{Nature} {\bf 607}
  (2022) 60--68}, [\href{http://arxiv.org/abs/2207.00043}{{\tt 2207.00043}}].

\bibitem{Planck_CMB_link}
P.~Collaboration, ``Planck image gallery.''
  \url{https://www.cosmos.esa.int/web/planck/picture-gallery}, July 2018.

\bibitem{Fields:2019pfx}
B.~D. Fields, K.~A. Olive, T.-H. Yeh and C.~Young, \emph{{Big-Bang
  Nucleosynthesis after Planck}},
  \href{http://dx.doi.org/10.1088/1475-7516/2020/03/010}{\emph{JCAP} {\bf 03}
  (2020) 010}, [\href{http://arxiv.org/abs/1912.01132}{{\tt 1912.01132}}].

\bibitem{Kuncinas:2023ycz}
A.~Kun\v{c}inas, O.~M. Ogreid, P.~Osland and M.~N. Rebelo, \emph{{Complex
  S$_{3}$-symmetric 3HDM}},
  \href{http://dx.doi.org/10.1007/JHEP07(2023)013}{\emph{JHEP} {\bf 07} (2023)
  013}, [\href{http://arxiv.org/abs/2302.07210}{{\tt 2302.07210}}].

\bibitem{Belyaev:2016lok}
A.~Belyaev, G.~Cacciapaglia, I.~P. Ivanov, F.~Rojas-Abatte and M.~Thomas,
  \emph{{Anatomy of the Inert Two Higgs Doublet Model in the light of the LHC
  and non-LHC Dark Matter Searches}},
  \href{http://dx.doi.org/10.1103/PhysRevD.97.035011}{\emph{Phys. Rev.} {\bf
  D97} (2018) 035011}, [\href{http://arxiv.org/abs/1612.00511}{{\tt
  1612.00511}}].

\bibitem{Kalinowski:2018ylg}
J.~Kalinowski, W.~Kotlarski, T.~Robens, D.~Sokolowska and A.~F. Zarnecki,
  \emph{{Benchmarking the Inert Doublet Model for $e^+ e^-$ colliders}},
  \href{http://dx.doi.org/10.1007/JHEP12(2018)081}{\emph{JHEP} {\bf 12} (2018)
  081}, [\href{http://arxiv.org/abs/1809.07712}{{\tt 1809.07712}}].

\bibitem{Merchand:2019bod}
M.~Merchand and M.~Sher, \emph{{Constraints on the Parameter Space in an Inert
  Doublet Model with two Active Doublets}},
  \href{http://dx.doi.org/10.1007/JHEP03(2020)108}{\emph{JHEP} {\bf 03} (2020)
  108}, [\href{http://arxiv.org/abs/1911.06477}{{\tt 1911.06477}}].

\bibitem{Keus:2014jha}
V.~Keus, S.~F. King, S.~Moretti and D.~Sokolowska, \emph{{Dark Matter with Two
  Inert Doublets plus One Higgs Doublet}},
  \href{http://dx.doi.org/10.1007/JHEP11(2014)016}{\emph{JHEP} {\bf 11} (2014)
  016}, [\href{http://arxiv.org/abs/1407.7859}{{\tt 1407.7859}}].

\bibitem{Keus:2014isa}
V.~Keus, S.~F. King and S.~Moretti, \emph{{Phenomenology of the inert (2+1) and
  (4+2) Higgs doublet models}},
  \href{http://dx.doi.org/10.1103/PhysRevD.90.075015}{\emph{Phys. Rev. D} {\bf
  90} (2014) 075015}, [\href{http://arxiv.org/abs/1408.0796}{{\tt 1408.0796}}].

\bibitem{Keus:2015xya}
V.~Keus, S.~F. King, S.~Moretti and D.~Sokolowska, \emph{{Observable Heavy
  Higgs Dark Matter}},
  \href{http://dx.doi.org/10.1007/JHEP11(2015)003}{\emph{JHEP} {\bf 11} (2015)
  003}, [\href{http://arxiv.org/abs/1507.08433}{{\tt 1507.08433}}].

\bibitem{Cordero:2017owj}
A.~Cordero, J.~Hernandez-Sanchez, V.~Keus, S.~F. King, S.~Moretti, D.~Rojas
  et~al., \emph{{Dark Matter Signals at the LHC from a 3HDM}},
  \href{http://dx.doi.org/10.1007/JHEP05(2018)030}{\emph{JHEP} {\bf 05} (2018)
  030}, [\href{http://arxiv.org/abs/1712.09598}{{\tt 1712.09598}}].

\bibitem{Dey:2023exa}
A.~Dey, V.~Keus, S.~Moretti and C.~Shepherd-Themistocleous, \emph{{A smoking
  gun signature of the 3HDM}},
  \href{http://dx.doi.org/10.1007/JHEP07(2024)038}{\emph{JHEP} {\bf 07} (2024)
  038}, [\href{http://arxiv.org/abs/2310.06593}{{\tt 2310.06593}}].

\bibitem{Cordero-Cid:2016krd}
A.~Cordero-Cid, J.~Hernández-Sánchez, V.~Keus, S.~F. King, S.~Moretti,
  D.~Rojas et~al., \emph{{CP violating scalar Dark Matter}},
  \href{http://dx.doi.org/10.1007/JHEP12(2016)014}{\emph{JHEP} {\bf 12} (2016)
  014}, [\href{http://arxiv.org/abs/1608.01673}{{\tt 1608.01673}}].

\bibitem{Cordero-Cid:2018man}
A.~Cordero-Cid, J.~Hern\'andez-S\'anchez, V.~Keus, S.~Moretti, D.~Rojas and
  D.~Soko\l{}owska, \emph{{Lepton collider indirect signatures of dark
  CP-violation}},
  \href{http://dx.doi.org/10.1140/epjc/s10052-020-7689-0}{\emph{Eur. Phys. J.
  C} {\bf 80} (2020) 135}, [\href{http://arxiv.org/abs/1812.00820}{{\tt
  1812.00820}}].

\bibitem{Cordero-Cid:2020yba}
A.~Cordero-Cid, J.~Hern\'andez-S\'anchez, V.~Keus, S.~Moretti, D.~Rojas-Ciofalo
  and D.~Soko\l{}owska, \emph{{Collider signatures of dark $CP$-violation}},
  \href{http://dx.doi.org/10.1103/PhysRevD.101.095023}{\emph{Phys. Rev. D} {\bf
  101} (2020) 095023}, [\href{http://arxiv.org/abs/2002.04616}{{\tt
  2002.04616}}].

\bibitem{Boto:2024tzp}
R.~Boto, P.~N. Figueiredo, J.~C. Romao and J.~P. Silva, \emph{{Novel two
  component dark matter features in the Z$_{2}$ \texttimes{} Z$_{2}$ 3HDM}},
  \href{http://dx.doi.org/10.1007/JHEP11(2024)108}{\emph{JHEP} {\bf 11} (2024)
  108}, [\href{http://arxiv.org/abs/2407.15933}{{\tt 2407.15933}}].

\bibitem{Aranda:2019vda}
A.~Aranda, D.~Hern\'andez-Otero, J.~Hern\'andez-Sanchez, V.~Keus, S.~Moretti,
  D.~Rojas-Ciofalo et~al., \emph{{Z$_3$ symmetric inert ( 2+1 )-Higgs-doublet
  model}}, \href{http://dx.doi.org/10.1103/PhysRevD.103.015023}{\emph{Phys.
  Rev. D} {\bf 103} (2021) 015023},
  [\href{http://arxiv.org/abs/1907.12470}{{\tt 1907.12470}}].

\bibitem{Hernandez-Otero:2022dxd}
D.~Hern\'andez-Otero, J.~Hern\'andez-S\'anchez, S.~Moretti and T.~Shindou,
  \emph{{The $Z_3$ soft breaking in the I(2+1)HDM and its probes at present and
  future colliders}},  \href{http://arxiv.org/abs/2203.06323}{{\tt
  2203.06323}}.

\bibitem{Khater:2021wcx}
W.~Khater, A.~Kun\v{c}inas, O.~M. Ogreid, P.~Osland and M.~N. Rebelo,
  \emph{{Dark matter in three-Higgs-doublet models with S$_{3}$ symmetry}},
  \href{http://dx.doi.org/10.1007/JHEP01(2022)120}{\emph{JHEP} {\bf 01} (2022)
  120}, [\href{http://arxiv.org/abs/2108.07026}{{\tt 2108.07026}}].

\bibitem{Kuncinas:2022whn}
A.~Kun\v{c}inas, O.~M. Ogreid, P.~Osland and M.~N. Rebelo, \emph{{Dark matter
  in a CP-violating three-Higgs-doublet model with S3 symmetry}},
  \href{http://dx.doi.org/10.1103/PhysRevD.106.075002}{\emph{Phys. Rev. D} {\bf
  106} (2022) 075002}, [\href{http://arxiv.org/abs/2204.05684}{{\tt
  2204.05684}}].

\bibitem{Kuncinas:2024zjq}
A.~Kun\v{c}inas, P.~Osland and M.~N. Rebelo, \emph{{U(1)-charged Dark Matter in
  three-Higgs-doublet models}},
  \href{http://dx.doi.org/10.1007/JHEP11(2024)086}{\emph{JHEP} {\bf 11} (2024)
  086}, [\href{http://arxiv.org/abs/2408.02728}{{\tt 2408.02728}}].

\bibitem{ATLAS:2020ior}
G.~Aad et~al., \emph{{$CP$ Properties of Higgs Boson Interactions with Top
  Quarks in the $t\bar{t}H$ and $tH$ Processes Using $H \rightarrow
  \gamma\gamma$ with the ATLAS Detector}},
  \href{http://dx.doi.org/10.1103/PhysRevLett.125.061802}{\emph{Phys. Rev.
  Lett.} {\bf 125} (2020) 061802}, [\href{http://arxiv.org/abs/2004.04545}{{\tt
  2004.04545}}].

\bibitem{CMS:2021sdq}
A.~Tumasyan et~al., \emph{{Analysis of the $CP$ structure of the Yukawa
  coupling between the Higgs boson and $\tau$ leptons in proton-proton
  collisions at $ \sqrt{s} $ = 13 TeV}},
  \href{http://dx.doi.org/10.1007/JHEP06(2022)012}{\emph{JHEP} {\bf 06} (2022)
  012}, [\href{http://arxiv.org/abs/2110.04836}{{\tt 2110.04836}}].

\bibitem{Kuncinas:2023hpf}
A.~Kun\v{c}inas, O.~M. Ogreid, P.~Osland and M.~N.~Rebelo, \emph{{Revisiting
  two dark matter candidates in $S_3$-symmetric three-Higgs-doublet models}},
  \href{http://dx.doi.org/10.22323/1.431.0031}{\emph{PoS} {\bf DISCRETE2022}
  (2024) 031}, [\href{http://arxiv.org/abs/2301.12194}{{\tt 2301.12194}}].

\bibitem{Planck:2018vyg}
N.~Aghanim et~al., \emph{{Planck 2018 results. VI. Cosmological parameters}},
  \href{http://dx.doi.org/10.1051/0004-6361/201833910}{\emph{Astron.
  Astrophys.} {\bf 641} (2020) A6},
  [\href{http://arxiv.org/abs/1807.06209}{{\tt 1807.06209}}].

\bibitem{LZ:2024zvo}
J.~Aalbers et~al., \emph{{Dark Matter Search Results from 4.2 Tonne-Years of
  Exposure of the LUX-ZEPLIN (LZ) Experiment}},
  \href{http://arxiv.org/abs/2410.17036}{{\tt 2410.17036}}.

\bibitem{XENON:2020kmp}
E.~Aprile et~al., \emph{{Projected WIMP sensitivity of the XENONnT dark matter
  experiment}},
  \href{http://dx.doi.org/10.1088/1475-7516/2020/11/031}{\emph{JCAP} {\bf 11}
  (2020) 031}, [\href{http://arxiv.org/abs/2007.08796}{{\tt 2007.08796}}].

\bibitem{Ivanonv_pr}
{I. P. Ivanov, private communication}.

\bibitem{ATLAS:2012yve}
G.~Aad et~al., \emph{{Observation of a new particle in the search for the
  Standard Model Higgs boson with the ATLAS detector at the LHC}},
  \href{http://dx.doi.org/10.1016/j.physletb.2012.08.020}{\emph{Phys. Lett. B}
  {\bf 716} (2012) 1--29}, [\href{http://arxiv.org/abs/1207.7214}{{\tt
  1207.7214}}].

\bibitem{CMS:2012qbp}
S.~Chatrchyan et~al., \emph{{Observation of a New Boson at a Mass of 125 GeV
  with the CMS Experiment at the LHC}},
  \href{http://dx.doi.org/10.1016/j.physletb.2012.08.021}{\emph{Phys. Lett. B}
  {\bf 716} (2012) 30--61}, [\href{http://arxiv.org/abs/1207.7235}{{\tt
  1207.7235}}].

\bibitem{Dirac:1927dy}
P.~A.~M. Dirac, \emph{{Quantum theory of emission and absorption of
  radiation}}, \href{http://dx.doi.org/10.1098/rspa.1927.0039}{\emph{Proc. Roy.
  Soc. Lond. A} {\bf 114} (1927) 243}.

\bibitem{Dirac:1928hu}
P.~A.~M. Dirac, \emph{{The quantum theory of the electron}},
  \href{http://dx.doi.org/10.1098/rspa.1928.0023}{\emph{Proc. Roy. Soc. Lond.
  A} {\bf 117} (1928) 610--624}.

\bibitem{Dirac:1930ek}
P.~A.~M. Dirac, \emph{{A Theory of Electrons and Protons}},
  \href{http://dx.doi.org/10.1098/rspa.1930.0013}{\emph{Proc. Roy. Soc. Lond.
  A} {\bf 126} (1930) 360--365}.

\bibitem{Dirac:1931kp}
P.~A.~M. Dirac, \emph{{Quantised singularities in the electromagnetic field,}},
  \href{http://dx.doi.org/10.1098/rspa.1931.0130}{\emph{Proc. Roy. Soc. Lond.
  A} {\bf 133} (1931) 60--72}.

\bibitem{Weyl:1929}
H.~Weyl, \emph{Gravitation and the electron},
  \href{http://dx.doi.org/10.1073/pnas.15.4.323}{\emph{Proceedings of the
  National Academy of Sciences} {\bf 15} (1929) 323--334}.

\bibitem{Anderson:1933mb}
C.~D. Anderson, \emph{{The Positive Electron}},
  \href{http://dx.doi.org/10.1103/PhysRev.43.491}{\emph{Phys. Rev.} {\bf 43}
  (1933) 491--494}.

\bibitem{Yang:1954ek}
C.-N. Yang and R.~L. Mills, \emph{{Conservation of Isotopic Spin and Isotopic
  Gauge Invariance}},
  \href{http://dx.doi.org/10.1103/PhysRev.96.191}{\emph{Phys. Rev.} {\bf 96}
  (1954) 191--195}.

\bibitem{Glashow:1959wxa}
S.~L. Glashow, \emph{{The renormalizability of vector meson interactions}},
  \href{http://dx.doi.org/10.1016/0029-5582(59)90196-8}{\emph{Nucl. Phys.} {\bf
  10} (1959) 107--117}.

\bibitem{Salam:1959zz}
A.~Salam and J.~C. Ward, \emph{{Weak and electromagnetic interactions}},
  \href{http://dx.doi.org/10.1007/BF02726525}{\emph{Nuovo Cim.} {\bf 11} (1959)
  568--577}.

\bibitem{Glashow:1961tr}
S.~L. Glashow, \emph{{Partial Symmetries of Weak Interactions}},
  \href{http://dx.doi.org/10.1016/0029-5582(61)90469-2}{\emph{Nucl. Phys.} {\bf
  22} (1961) 579--588}.

\bibitem{Englert:1964et}
F.~Englert and R.~Brout, \emph{{Broken Symmetry and the Mass of Gauge Vector
  Mesons}}, \href{http://dx.doi.org/10.1103/PhysRevLett.13.321}{\emph{Phys.
  Rev. Lett.} {\bf 13} (1964) 321--323}.

\bibitem{Higgs:1964pj}
P.~W. Higgs, \emph{{Broken Symmetries and the Masses of Gauge Bosons}},
  \href{http://dx.doi.org/10.1103/PhysRevLett.13.508}{\emph{Phys. Rev. Lett.}
  {\bf 13} (1964) 508--509}.

\bibitem{Guralnik:1964eu}
G.~S. Guralnik, C.~R. Hagen and T.~W.~B. Kibble, \emph{{Global Conservation
  Laws and Massless Particles}},
  \href{http://dx.doi.org/10.1103/PhysRevLett.13.585}{\emph{Phys. Rev. Lett.}
  {\bf 13} (1964) 585--587}.

\bibitem{Weinberg:1967tq}
S.~Weinberg, \emph{{A Model of Leptons}},
  \href{http://dx.doi.org/10.1103/PhysRevLett.19.1264}{\emph{Phys. Rev. Lett.}
  {\bf 19} (1967) 1264--1266}.

\bibitem{Salam:1968rm}
A.~Salam, \emph{{Weak and Electromagnetic Interactions}},
  \href{http://dx.doi.org/10.1142/9789812795915_0034}{\emph{Conf. Proc. C} {\bf
  680519} (1968) 367--377}.

\bibitem{Yukawa:1935xg}
H.~Yukawa, \emph{{On the Interaction of Elementary Particles I}},
  \href{http://dx.doi.org/10.1143/PTPS.1.1}{\emph{Proc. Phys. Math. Soc. Jap.}
  {\bf 17} (1935) 48--57}.

\bibitem{Lattes:1947mw}
C.~M.~G. Lattes, H.~Muirhead, G.~P.~S. Occhialini and C.~F. Powell,
  \emph{{Processes Involving Charged Mesons}},
  \href{http://dx.doi.org/10.1038/159694a0}{\emph{Nature} {\bf 159} (1947)
  694--697}.

\bibitem{Gell-Mann:1964ewy}
M.~Gell-Mann, \emph{{A Schematic Model of Baryons and Mesons}},
  \href{http://dx.doi.org/10.1016/S0031-9163(64)92001-3}{\emph{Phys. Lett.}
  {\bf 8} (1964) 214--215}.

\bibitem{Zweig:1964ruk}
G.~Zweig, \emph{{An SU(3) model for strong interaction symmetry and its
  breaking. Version 1}}.
\newblock 1, 1964.
\newblock 10.17181/CERN-TH-401.

\bibitem{Zweig:1964jf}
G.~Zweig, \emph{{An SU(3) model for strong interaction symmetry and its
  breaking. Version 2}}.
\newblock 2, 1964.
\newblock 10.17181/CERN-TH-412.

\bibitem{Fritzsch:1973pi}
H.~Fritzsch, M.~Gell-Mann and H.~Leutwyler, \emph{{Advantages of the Color
  Octet Gluon Picture}},
  \href{http://dx.doi.org/10.1016/0370-2693(73)90625-4}{\emph{Phys. Lett. B}
  {\bf 47} (1973) 365--368}.

\bibitem{Gross:1973id}
D.~J. Gross and F.~Wilczek, \emph{{Ultraviolet Behavior of Nonabelian Gauge
  Theories}}, \href{http://dx.doi.org/10.1103/PhysRevLett.30.1343}{\emph{Phys.
  Rev. Lett.} {\bf 30} (1973) 1343--1346}.

\bibitem{Politzer:1973fx}
H.~D. Politzer, \emph{{Reliable Perturbative Results for Strong
  Interactions?}},
  \href{http://dx.doi.org/10.1103/PhysRevLett.30.1346}{\emph{Phys. Rev. Lett.}
  {\bf 30} (1973) 1346--1349}.

\bibitem{Iliopoulos:2025fhr}
J.~Iliopoulos, \emph{{From Many Models to ONE THEORY}},
  \href{http://arxiv.org/abs/2501.10233}{{\tt 2501.10233}}.

\bibitem{Cheng:1984vwu}
T.-P. Cheng and L.-F. Li, \emph{{Gauge Theory of Elementary Particle Physics}}.
\newblock Oxford University Press, Oxford, UK, 1984.

\bibitem{Langacker:2010zza}
P.~Langacker, \emph{{The standard model and beyond}}.
\newblock Boca Raton, USA: CRC Pr. (2010) 663 p, 2010.

\bibitem{Pal:2014xrq}
P.~B. Pal, \emph{{An Introductory Course of Particle Physics}}.
\newblock CRC Press, 7, 2014,
  \href{http://dx.doi.org/10.1201/b17199}{10.1201/b17199}.

\bibitem{Pais:1952zz}
A.~Pais, \emph{{Some Remarks on the V-Particles}},
  \href{http://dx.doi.org/10.1103/PhysRev.86.663}{\emph{Phys. Rev.} {\bf 86}
  (1952) 663--672}.

\bibitem{Gell-Mann:1953hzm}
M.~Gell-Mann, \emph{{Isotopic Spin and New Unstable Particles}},
  \href{http://dx.doi.org/10.1103/PhysRev.92.833}{\emph{Phys. Rev.} {\bf 92}
  (1953) 833--834}.

\bibitem{Nakano:1953zz}
T.~Nakano and K.~Nishijima, \emph{{Charge Independence for V-particles}},
  \href{http://dx.doi.org/10.1143/PTP.10.581}{\emph{Prog. Theor. Phys.} {\bf
  10} (1953) 581--582}.

\bibitem{Wu:1957my}
C.~S. Wu, E.~Ambler, R.~W. Hayward, D.~D. Hoppes and R.~P. Hudson,
  \emph{{Experimental Test of Parity Conservation in $\beta$ Decay}},
  \href{http://dx.doi.org/10.1103/PhysRev.105.1413}{\emph{Phys. Rev.} {\bf 105}
  (1957) 1413--1414}.

\bibitem{Garwin:1957hc}
R.~L. Garwin, L.~M. Lederman and M.~Weinrich, \emph{{Observations of the
  Failure of Conservation of Parity and Charge Conjugation in Meson Decays: The
  Magnetic Moment of the Free Muon}},
  \href{http://dx.doi.org/10.1103/PhysRev.105.1415}{\emph{Phys. Rev.} {\bf 105}
  (1957) 1415--1417}.

\bibitem{Friedman:1957mz}
J.~I. Friedman and V.~L. Telegdi, \emph{{Nuclear Emulsion Evidence for Parity
  Nonconservation in the Decay Chain $\pi^+ \to \mu^+ \to e^+$}},
  \href{http://dx.doi.org/10.1103/PhysRev.106.1290}{\emph{Phys. Rev.} {\bf 106}
  (1957) 1290--1293}.

\bibitem{Feynman:1958ty}
R.~P. Feynman and M.~Gell-Mann, \emph{{Theory of Fermi interaction}},
  \href{http://dx.doi.org/10.1103/PhysRev.109.193}{\emph{Phys. Rev.} {\bf 109}
  (1958) 193--198}.

\bibitem{Sudarshan:1958vf}
E.~C.~G. Sudarshan and R.~e. Marshak, \emph{{Chirality invariance and the
  universal Fermi interaction}},
  \href{http://dx.doi.org/10.1103/PhysRev.109.1860.2}{\emph{Phys. Rev.} {\bf
  109} (1958) 1860--1860}.

\bibitem{Ginzburg:1950sr}
V.~L. Ginzburg and L.~D. Landau, \emph{{On the Theory of superconductivity}},
  \href{http://dx.doi.org/10.1016/b978-0-08-010586-4.50078-x}{\emph{Zh. Eksp.
  Teor. Fiz.} {\bf 20} (1950) 1064--1082}.

\bibitem{Meissner:1933ela}
W.~Meissner and R.~Ochsenfeld, \emph{{Ein neuer Effekt bei Eintritt der
  Supraleitf\"ahigkeit}},
  \href{http://dx.doi.org/10.1007/BF01504252}{\emph{Naturwiss.} {\bf 21} (1933)
  787--788}.

\bibitem{Nambu:1960tm}
Y.~Nambu, \emph{{Quasiparticles and Gauge Invariance in the Theory of
  Superconductivity}},
  \href{http://dx.doi.org/10.1103/PhysRev.117.648}{\emph{Phys. Rev.} {\bf 117}
  (1960) 648--663}.

\bibitem{Goldstone:1961eq}
J.~Goldstone, \emph{{Field Theories with Superconductor Solutions}},
  \href{http://dx.doi.org/10.1007/BF02812722}{\emph{Nuovo Cim.} {\bf 19} (1961)
  154--164}.

\bibitem{tHooft:1971qjg}
G.~'t~Hooft, \emph{{Renormalizable Lagrangians for Massive Yang-Mills Fields}},
  \href{http://dx.doi.org/10.1016/0550-3213(71)90139-8}{\emph{Nucl. Phys.} {\bf
  B35} (1971) 167--188}.

\bibitem{Goldstone:1962es}
J.~Goldstone, A.~Salam and S.~Weinberg, \emph{{Broken Symmetries}},
  \href{http://dx.doi.org/10.1103/PhysRev.127.965}{\emph{Phys. Rev.} {\bf 127}
  (1962) 965--970}.

\bibitem{Coleman:1973jx}
S.~R. Coleman and E.~J. Weinberg, \emph{{Radiative Corrections as the Origin of
  Spontaneous Symmetry Breaking}},
  \href{http://dx.doi.org/10.1103/PhysRevD.7.1888}{\emph{Phys. Rev. D} {\bf 7}
  (1973) 1888--1910}.

\bibitem{Glashow:1976nt}
S.~L. Glashow and S.~Weinberg, \emph{{Natural Conservation Laws for Neutral
  Currents}}, \href{http://dx.doi.org/10.1103/PhysRevD.15.1958}{\emph{Phys.
  Rev. D} {\bf 15} (1977) 1958}.

\bibitem{Paschos:1976ay}
E.~A. Paschos, \emph{{Diagonal Neutral Currents}},
  \href{http://dx.doi.org/10.1103/PhysRevD.15.1966}{\emph{Phys. Rev. D} {\bf
  15} (1977) 1966}.

\bibitem{Cabibbo:1963yz}
N.~Cabibbo, \emph{{Unitary Symmetry and Leptonic Decays}},
  \href{http://dx.doi.org/10.1103/PhysRevLett.10.531}{\emph{Phys. Rev. Lett.}
  {\bf 10} (1963) 531--533}.

\bibitem{Kobayashi:1973fv}
M.~Kobayashi and T.~Maskawa, \emph{{CP Violation in the Renormalizable Theory
  of Weak Interaction}},
  \href{http://dx.doi.org/10.1143/PTP.49.652}{\emph{Prog. Theor. Phys.} {\bf
  49} (1973) 652--657}.

\bibitem{Chau:1984fp}
L.-L. Chau and W.-Y. Keung, \emph{{Comments on the Parametrization of the
  Kobayashi-Maskawa Matrix}},
  \href{http://dx.doi.org/10.1103/PhysRevLett.53.1802}{\emph{Phys. Rev. Lett.}
  {\bf 53} (1984) 1802}.

\bibitem{Belfatto:2019swo}
B.~Belfatto, R.~Beradze and Z.~Berezhiani, \emph{{The CKM unitarity problem: A
  trace of new physics at the TeV scale?}},
  \href{http://dx.doi.org/10.1140/epjc/s10052-020-7691-6}{\emph{Eur. Phys. J.
  C} {\bf 80} (2020) 149}, [\href{http://arxiv.org/abs/1906.02714}{{\tt
  1906.02714}}].

\bibitem{Coutinho:2019aiy}
A.~M. Coutinho, A.~Crivellin and C.~A. Manzari, \emph{{Global Fit to Modified
  Neutrino Couplings and the Cabibbo-Angle Anomaly}},
  \href{http://dx.doi.org/10.1103/PhysRevLett.125.071802}{\emph{Phys. Rev.
  Lett.} {\bf 125} (2020) 071802}, [\href{http://arxiv.org/abs/1912.08823}{{\tt
  1912.08823}}].

\bibitem{Crivellin:2020lzu}
A.~Crivellin and M.~Hoferichter, \emph{{\ensuremath{\beta} Decays as Sensitive
  Probes of Lepton Flavor Universality}},
  \href{http://dx.doi.org/10.1103/PhysRevLett.125.111801}{\emph{Phys. Rev.
  Lett.} {\bf 125} (2020) 111801}, [\href{http://arxiv.org/abs/2002.07184}{{\tt
  2002.07184}}].

\bibitem{Glashow:1970gm}
S.~L. Glashow, J.~Iliopoulos and L.~Maiani, \emph{{Weak Interactions with
  Lepton-Hadron Symmetry}},
  \href{http://dx.doi.org/10.1103/PhysRevD.2.1285}{\emph{Phys. Rev. D} {\bf 2}
  (1970) 1285--1292}.

\bibitem{Pontecorvo:1967fh}
B.~Pontecorvo, \emph{{Neutrino Experiments and the Problem of Conservation of
  Leptonic Charge}}, {\emph{Sov. Phys. JETP} {\bf 26} (1968) 984--988}.

\bibitem{Maki:1962mu}
Z.~Maki, M.~Nakagawa and S.~Sakata, \emph{{Remarks on the unified model of
  elementary particles}},
  \href{http://dx.doi.org/10.1143/PTP.28.870}{\emph{Prog. Theor. Phys.} {\bf
  28} (1962) 870--880}.

\bibitem{Esteban:2024eli}
I.~Esteban, M.~C. Gonzalez-Garcia, M.~Maltoni, I.~Martinez-Soler, J.~P.
  Pinheiro and T.~Schwetz, \emph{{NuFit-6.0: Updated global analysis of
  three-flavor neutrino oscillations}},
  \href{http://arxiv.org/abs/2410.05380}{{\tt 2410.05380}}.

\bibitem{Gunion:1989we}
J.~F. Gunion, H.~E. Haber, G.~L. Kane and S.~Dawson, \emph{{The Higgs Hunter's
  Guide}}, vol.~80.
\newblock 2000,
  \href{http://dx.doi.org/10.1201/9780429496448}{10.1201/9780429496448}.

\bibitem{Carena:2002es}
M.~Carena and H.~E. Haber, \emph{{Higgs Boson Theory and Phenomenology}},
  \href{http://dx.doi.org/10.1016/S0146-6410(02)00177-1}{\emph{Prog. Part.
  Nucl. Phys.} {\bf 50} (2003) 63--152},
  [\href{http://arxiv.org/abs/hep-ph/0208209}{{\tt hep-ph/0208209}}].

\bibitem{Djouadi:2005gi}
A.~Djouadi, \emph{{The Anatomy of electro-weak symmetry breaking. I: The Higgs
  boson in the standard model}},
  \href{http://dx.doi.org/10.1016/j.physrep.2007.10.004}{\emph{Phys. Rept.}
  {\bf 457} (2008) 1--216}, [\href{http://arxiv.org/abs/hep-ph/0503172}{{\tt
  hep-ph/0503172}}].

\bibitem{Denner:2011mq}
A.~Denner, S.~Heinemeyer, I.~Puljak, D.~Rebuzzi and M.~Spira, \emph{{Standard
  Model Higgs-Boson Branching Ratios with Uncertainties}},
  \href{http://dx.doi.org/10.1140/epjc/s10052-011-1753-8}{\emph{Eur. Phys. J.
  C} {\bf 71} (2011) 1753}, [\href{http://arxiv.org/abs/1107.5909}{{\tt
  1107.5909}}].

\bibitem{LHCHiggsCrossSectionWorkingGroup:2016ypw}
D.~de~Florian et~al., \emph{{Handbook of LHC Higgs Cross Sections: 4.
  Deciphering the Nature of the Higgs Sector}},
  \href{http://arxiv.org/abs/1610.07922}{{\tt 1610.07922}}.

\bibitem{ATLAS:2023yqk}
G.~Aad et~al., \emph{{Evidence for the Higgs Boson Decay to a Z Boson and a
  Photon at the LHC}},
  \href{http://dx.doi.org/10.1103/PhysRevLett.132.021803}{\emph{Phys. Rev.
  Lett.} {\bf 132} (2024) 021803}, [\href{http://arxiv.org/abs/2309.03501}{{\tt
  2309.03501}}].

\bibitem{Spira:1995rr}
M.~Spira, A.~Djouadi, D.~Graudenz and P.~M. Zerwas, \emph{{Higgs boson
  production at the LHC}},
  \href{http://dx.doi.org/10.1016/0550-3213(95)00379-7}{\emph{Nucl. Phys. B}
  {\bf 453} (1995) 17--82}, [\href{http://arxiv.org/abs/hep-ph/9504378}{{\tt
  hep-ph/9504378}}].

\bibitem{Jacob:1959at}
M.~Jacob and G.~C. Wick, \emph{{On the general theory of collisions for
  particles with spin}},
  \href{http://dx.doi.org/10.1016/0003-4916(59)90051-X}{\emph{Annals Phys.}
  {\bf 7} (1959) 404--428}.

\bibitem{Balachandran:1968zza}
A.~P. Balachandran and J.~Nuyts, \emph{{Simultaneous partial-wave expansion in
  the Mandelstamm variables: Crossing symmetry for partial waves}},
  \href{http://dx.doi.org/10.1103/PhysRev.172.1821}{\emph{Phys. Rev.} {\bf 172}
  (1968) 1821--1827}.

\bibitem{Luscher:1988gc}
M.~Luscher and P.~Weisz, \emph{{Is There a Strong Interaction Sector in the
  Standard Lattice Higgs Model?}},
  \href{http://dx.doi.org/10.1016/0370-2693(88)91799-6}{\emph{Phys. Lett.} {\bf
  B212} (1988) 472--478}.

\bibitem{Cornwall:1974km}
J.~M. Cornwall, D.~N. Levin and G.~Tiktopoulos, \emph{{Derivation of Gauge
  Invariance from High-Energy Unitarity Bounds on the s Matrix}},
  \href{http://dx.doi.org/10.1103/PhysRevD.10.1145,
  10.1103/PhysRevD.11.972}{\emph{Phys. Rev.} {\bf D10} (1974) 1145}.

\bibitem{Vayonakis:1976vz}
C.~E. Vayonakis, \emph{{Born Helicity Amplitudes and Cross-Sections in
  Nonabelian Gauge Theories}},
  \href{http://dx.doi.org/10.1007/BF02746538}{\emph{Lett. Nuovo Cim.} {\bf 17}
  (1976) 383}.

\bibitem{Lee:1977eg}
B.~W. Lee, C.~Quigg and H.~Thacker, \emph{{Weak Interactions at Very
  High-Energies: The Role of the Higgs Boson Mass}},
  \href{http://dx.doi.org/10.1103/PhysRevD.16.1519}{\emph{Phys. Rev. D} {\bf
  16} (1977) 1519}.

\bibitem{Chanowitz:1985hj}
M.~S. Chanowitz and M.~K. Gaillard, \emph{{The TeV Physics of Strongly
  Interacting W's and Z's}},
  \href{http://dx.doi.org/10.1016/0550-3213(85)90580-2}{\emph{Nucl. Phys. B}
  {\bf 261} (1985) 379--431}.

\bibitem{Lee:1977yc}
B.~W. Lee, C.~Quigg and H.~B. Thacker, \emph{{The Strength of Weak Interactions
  at Very High-Energies and the Higgs Boson Mass}},
  \href{http://dx.doi.org/10.1103/PhysRevLett.38.883}{\emph{Phys. Rev. Lett.}
  {\bf 38} (1977) 883--885}.

\bibitem{Ross:1975fq}
D.~A. Ross and M.~J.~G. Veltman, \emph{{Neutral Currents in Neutrino
  Experiments}},
  \href{http://dx.doi.org/10.1016/0550-3213(75)90485-X}{\emph{Nucl. Phys.} {\bf
  B95} (1975) 135--147}.

\bibitem{Sikivie:1980hm}
P.~Sikivie, L.~Susskind, M.~B. Voloshin and V.~I. Zakharov, \emph{{Isospin
  Breaking in Technicolor Models}},
  \href{http://dx.doi.org/10.1016/0550-3213(80)90214-X}{\emph{Nucl. Phys. B}
  {\bf 173} (1980) 189--207}.

\bibitem{Langacker:1980js}
P.~Langacker, \emph{{Grand Unified Theories and Proton Decay}},
  \href{http://dx.doi.org/10.1016/0370-1573(81)90059-4}{\emph{Phys. Rept.} {\bf
  72} (1981) 185}.

\bibitem{Peskin:1990zt}
M.~E. Peskin and T.~Takeuchi, \emph{{A New constraint on a strongly interacting
  Higgs sector}},
  \href{http://dx.doi.org/10.1103/PhysRevLett.65.964}{\emph{Phys. Rev. Lett.}
  {\bf 65} (1990) 964--967}.

\bibitem{Peskin:1991sw}
M.~E. Peskin and T.~Takeuchi, \emph{{Estimation of oblique electroweak
  corrections}}, \href{http://dx.doi.org/10.1103/PhysRevD.46.381}{\emph{Phys.
  Rev. D} {\bf 46} (1992) 381--409}.

\bibitem{tHooft:1973mfk}
G.~'t~Hooft, \emph{{Dimensional regularization and the renormalization group}},
  \href{http://dx.doi.org/10.1016/0550-3213(73)90376-3}{\emph{Nucl. Phys. B}
  {\bf 61} (1973) 455--468}.

\bibitem{Weinberg:1973xwm}
S.~Weinberg, \emph{{New approach to the renormalization group}},
  \href{http://dx.doi.org/10.1103/PhysRevD.8.3497}{\emph{Phys. Rev. D} {\bf 8}
  (1973) 3497--3509}.

\bibitem{Plehn:2017fdg}
M.~Bauer and T.~Plehn, \emph{{Yet Another Introduction to Dark Matter}: {The
  Particle Physics Approach}}, vol.~959 of \emph{Lecture Notes in Physics}.
\newblock Springer, 2019,
  \href{http://dx.doi.org/10.1007/978-3-030-16234-4}{10.1007/978-3-030-16234-4}.

\bibitem{Martin:2013tda}
J.~Martin, C.~Ringeval and V.~Vennin, \emph{{Encyclop\ae{}dia Inflationaris}:
  {Opiparous Edition}},
  \href{http://dx.doi.org/10.1016/j.dark.2024.101653}{\emph{Phys. Dark Univ.}
  {\bf 5-6} (2014) 75--235}, [\href{http://arxiv.org/abs/1303.3787}{{\tt
  1303.3787}}].

\bibitem{Lee:1977ua}
B.~W. Lee and S.~Weinberg, \emph{{Cosmological Lower Bound on Heavy Neutrino
  Masses}}, \href{http://dx.doi.org/10.1103/PhysRevLett.39.165}{\emph{Phys.
  Rev. Lett.} {\bf 39} (1977) 165--168}.

\bibitem{Sasaki:2018dmp}
M.~Sasaki, T.~Suyama, T.~Tanaka and S.~Yokoyama, \emph{{Primordial black
  holes\textemdash{}perspectives in gravitational wave astronomy}},
  \href{http://dx.doi.org/10.1088/1361-6382/aaa7b4}{\emph{Class. Quant. Grav.}
  {\bf 35} (2018) 063001}, [\href{http://arxiv.org/abs/1801.05235}{{\tt
  1801.05235}}].

\bibitem{Carr:2020xqk}
B.~Carr and F.~Kuhnel, \emph{{Primordial Black Holes as Dark Matter: Recent
  Developments}},
  \href{http://dx.doi.org/10.1146/annurev-nucl-050520-125911}{\emph{Ann. Rev.
  Nucl. Part. Sci.} {\bf 70} (2020) 355--394},
  [\href{http://arxiv.org/abs/2006.02838}{{\tt 2006.02838}}].

\bibitem{Green:2020jor}
A.~M. Green and B.~J. Kavanagh, \emph{{Primordial Black Holes as a dark matter
  candidate}}, \href{http://dx.doi.org/10.1088/1361-6471/abc534}{\emph{J. Phys.
  G} {\bf 48} (2021) 043001}, [\href{http://arxiv.org/abs/2007.10722}{{\tt
  2007.10722}}].

\bibitem{Moroi:1993mb}
T.~Moroi, H.~Murayama and M.~Yamaguchi, \emph{{Cosmological constraints on the
  light stable gravitino}},
  \href{http://dx.doi.org/10.1016/0370-2693(93)91434-O}{\emph{Phys. Lett. B}
  {\bf 303} (1993) 289--294}.

\bibitem{Feng:2004mt}
J.~L. Feng, S.~Su and F.~Takayama, \emph{{Supergravity with a gravitino LSP}},
  \href{http://dx.doi.org/10.1103/PhysRevD.70.075019}{\emph{Phys. Rev. D} {\bf
  70} (2004) 075019}, [\href{http://arxiv.org/abs/hep-ph/0404231}{{\tt
  hep-ph/0404231}}].

\bibitem{Rychkov:2007uq}
V.~S. Rychkov and A.~Strumia, \emph{{Thermal production of gravitinos}},
  \href{http://dx.doi.org/10.1103/PhysRevD.75.075011}{\emph{Phys. Rev. D} {\bf
  75} (2007) 075011}, [\href{http://arxiv.org/abs/hep-ph/0701104}{{\tt
  hep-ph/0701104}}].

\bibitem{Ellwanger:2009dp}
U.~Ellwanger, C.~Hugonie and A.~M. Teixeira, \emph{{The Next-to-Minimal
  Supersymmetric Standard Model}},
  \href{http://dx.doi.org/10.1016/j.physrep.2010.07.001}{\emph{Phys. Rept.}
  {\bf 496} (2010) 1--77}, [\href{http://arxiv.org/abs/0910.1785}{{\tt
  0910.1785}}].

\bibitem{Lesgourgues:2006nd}
J.~Lesgourgues and S.~Pastor, \emph{{Massive neutrinos and cosmology}},
  \href{http://dx.doi.org/10.1016/j.physrep.2006.04.001}{\emph{Phys. Rept.}
  {\bf 429} (2006) 307--379},
  [\href{http://arxiv.org/abs/astro-ph/0603494}{{\tt astro-ph/0603494}}].

\bibitem{Boyarsky:2009ix}
A.~Boyarsky, O.~Ruchayskiy and M.~Shaposhnikov, \emph{{The Role of sterile
  neutrinos in cosmology and astrophysics}},
  \href{http://dx.doi.org/10.1146/annurev.nucl.010909.083654}{\emph{Ann. Rev.
  Nucl. Part. Sci.} {\bf 59} (2009) 191--214},
  [\href{http://arxiv.org/abs/0901.0011}{{\tt 0901.0011}}].

\bibitem{Boyarsky:2018tvu}
A.~Boyarsky, M.~Drewes, T.~Lasserre, S.~Mertens and O.~Ruchayskiy,
  \emph{{Sterile neutrino Dark Matter}},
  \href{http://dx.doi.org/10.1016/j.ppnp.2018.07.004}{\emph{Prog. Part. Nucl.
  Phys.} {\bf 104} (2019) 1--45}, [\href{http://arxiv.org/abs/1807.07938}{{\tt
  1807.07938}}].

\bibitem{Bringmann:2022aim}
T.~Bringmann, P.~F. Depta, M.~Hufnagel, J.~Kersten, J.~T. Ruderman and
  K.~Schmidt-Hoberg, \emph{{Minimal sterile neutrino dark matter}},
  \href{http://dx.doi.org/10.1103/PhysRevD.107.L071702}{\emph{Phys. Rev. D}
  {\bf 107} (2023) L071702}, [\href{http://arxiv.org/abs/2206.10630}{{\tt
  2206.10630}}].

\bibitem{Kim:2008hd}
J.~E. Kim and G.~Carosi, \emph{{Axions and the Strong CP Problem}},
  \href{http://dx.doi.org/10.1103/RevModPhys.82.557}{\emph{Rev. Mod. Phys.}
  {\bf 82} (2010) 557--602}, [\href{http://arxiv.org/abs/0807.3125}{{\tt
  0807.3125}}].

\bibitem{Jaeckel:2010ni}
J.~Jaeckel and A.~Ringwald, \emph{{The Low-Energy Frontier of Particle
  Physics}},
  \href{http://dx.doi.org/10.1146/annurev.nucl.012809.104433}{\emph{Ann. Rev.
  Nucl. Part. Sci.} {\bf 60} (2010) 405--437},
  [\href{http://arxiv.org/abs/1002.0329}{{\tt 1002.0329}}].

\bibitem{Arias:2012az}
P.~Arias, D.~Cadamuro, M.~Goodsell, J.~Jaeckel, J.~Redondo and A.~Ringwald,
  \emph{{WISPy Cold Dark Matter}},
  \href{http://dx.doi.org/10.1088/1475-7516/2012/06/013}{\emph{JCAP} {\bf 06}
  (2012) 013}, [\href{http://arxiv.org/abs/1201.5902}{{\tt 1201.5902}}].

\bibitem{Marsh:2015xka}
D.~J.~E. Marsh, \emph{{Axion Cosmology}},
  \href{http://dx.doi.org/10.1016/j.physrep.2016.06.005}{\emph{Phys. Rept.}
  {\bf 643} (2016) 1--79}, [\href{http://arxiv.org/abs/1510.07633}{{\tt
  1510.07633}}].

\bibitem{Irastorza:2018dyq}
I.~G. Irastorza and J.~Redondo, \emph{{New experimental approaches in the
  search for axion-like particles}},
  \href{http://dx.doi.org/10.1016/j.ppnp.2018.05.003}{\emph{Prog. Part. Nucl.
  Phys.} {\bf 102} (2018) 89--159},
  [\href{http://arxiv.org/abs/1801.08127}{{\tt 1801.08127}}].

\bibitem{Ferreira:2020fam}
E.~G.~M. Ferreira, \emph{{Ultra-light dark matter}},
  \href{http://dx.doi.org/10.1007/s00159-021-00135-6}{\emph{Astron. Astrophys.
  Rev.} {\bf 29} (2021) 7}, [\href{http://arxiv.org/abs/2005.03254}{{\tt
  2005.03254}}].

\bibitem{Burnell:2005hm}
F.~Burnell and G.~D. Kribs, \emph{{The Abundance of Kaluza-Klein dark matter
  with coannihilation}},
  \href{http://dx.doi.org/10.1103/PhysRevD.73.015001}{\emph{Phys. Rev. D} {\bf
  73} (2006) 015001}, [\href{http://arxiv.org/abs/hep-ph/0509118}{{\tt
  hep-ph/0509118}}].

\bibitem{Kong:2005hn}
K.~Kong and K.~T. Matchev, \emph{{Precise calculation of the relic density of
  Kaluza-Klein dark matter in universal extra dimensions}},
  \href{http://dx.doi.org/10.1088/1126-6708/2006/01/038}{\emph{JHEP} {\bf 01}
  (2006) 038}, [\href{http://arxiv.org/abs/hep-ph/0509119}{{\tt
  hep-ph/0509119}}].

\bibitem{Hooper:2007qk}
D.~Hooper and S.~Profumo, \emph{{Dark Matter and Collider Phenomenology of
  Universal Extra Dimensions}},
  \href{http://dx.doi.org/10.1016/j.physrep.2007.09.003}{\emph{Phys. Rept.}
  {\bf 453} (2007) 29--115}, [\href{http://arxiv.org/abs/hep-ph/0701197}{{\tt
  hep-ph/0701197}}].

\bibitem{Belanger:2010yx}
G.~Belanger, M.~Kakizaki and A.~Pukhov, \emph{{Dark matter in UED: The Role of
  the second KK level}},
  \href{http://dx.doi.org/10.1088/1475-7516/2011/02/009}{\emph{JCAP} {\bf 02}
  (2011) 009}, [\href{http://arxiv.org/abs/1012.2577}{{\tt 1012.2577}}].

\bibitem{Sanders:2002pf}
R.~H. Sanders and S.~S. McGaugh, \emph{{Modified Newtonian dynamics as an
  alternative to dark matter}},
  \href{http://dx.doi.org/10.1146/annurev.astro.40.060401.093923}{\emph{Ann.
  Rev. Astron. Astrophys.} {\bf 40} (2002) 263--317},
  [\href{http://arxiv.org/abs/astro-ph/0204521}{{\tt astro-ph/0204521}}].

\bibitem{Bekenstein:2004ne}
J.~D. Bekenstein, \emph{{Relativistic gravitation theory for the MOND
  paradigm}}, \href{http://dx.doi.org/10.1103/PhysRevD.70.083509}{\emph{Phys.
  Rev. D} {\bf 70} (2004) 083509},
  [\href{http://arxiv.org/abs/astro-ph/0403694}{{\tt astro-ph/0403694}}].

\bibitem{Clifton:2011jh}
T.~Clifton, P.~G. Ferreira, A.~Padilla and C.~Skordis, \emph{{Modified Gravity
  and Cosmology}},
  \href{http://dx.doi.org/10.1016/j.physrep.2012.01.001}{\emph{Phys. Rept.}
  {\bf 513} (2012) 1--189}, [\href{http://arxiv.org/abs/1106.2476}{{\tt
  1106.2476}}].

\bibitem{Bertone:2016nfn}
G.~Bertone and D.~Hooper, \emph{{History of dark matter}},
  \href{http://dx.doi.org/10.1103/RevModPhys.90.045002}{\emph{Rev. Mod. Phys.}
  {\bf 90} (2018) 045002}, [\href{http://arxiv.org/abs/1605.04909}{{\tt
  1605.04909}}].

\bibitem{Arcadi:2017kky}
G.~Arcadi, M.~Dutra, P.~Ghosh, M.~Lindner, Y.~Mambrini, M.~Pierre et~al.,
  \emph{{The waning of the WIMP? A review of models, searches, and
  constraints}},
  \href{http://dx.doi.org/10.1140/epjc/s10052-018-5662-y}{\emph{Eur. Phys. J.
  C} {\bf 78} (2018) 203}, [\href{http://arxiv.org/abs/1703.07364}{{\tt
  1703.07364}}].

\bibitem{Roszkowski:2017nbc}
L.~Roszkowski, E.~M. Sessolo and S.~Trojanowski, \emph{{WIMP dark matter
  candidates and searches\textemdash{}current status and future prospects}},
  \href{http://dx.doi.org/10.1088/1361-6633/aab913}{\emph{Rept. Prog. Phys.}
  {\bf 81} (2018) 066201}, [\href{http://arxiv.org/abs/1707.06277}{{\tt
  1707.06277}}].

\bibitem{Cirelli:2024ssz}
M.~Cirelli, A.~Strumia and J.~Zupan, \emph{{Dark Matter}},
  \href{http://arxiv.org/abs/2406.01705}{{\tt 2406.01705}}.

\bibitem{Carlson:1992fn}
E.~D. Carlson, M.~E. Machacek and L.~J. Hall, \emph{{Self-interacting dark
  matter}}, \href{http://dx.doi.org/10.1086/171833}{\emph{Astrophys. J.} {\bf
  398} (1992) 43--52}.

\bibitem{Moroi:1999zb}
T.~Moroi and L.~Randall, \emph{{Wino cold dark matter from anomaly mediated
  SUSY breaking}},
  \href{http://dx.doi.org/10.1016/S0550-3213(99)00748-8}{\emph{Nucl. Phys. B}
  {\bf 570} (2000) 455--472}, [\href{http://arxiv.org/abs/hep-ph/9906527}{{\tt
  hep-ph/9906527}}].

\bibitem{Lin:2000qq}
W.~B. Lin, D.~H. Huang, X.~Zhang and R.~H. Brandenberger, \emph{{Nonthermal
  production of WIMPs and the subgalactic structure of the universe}},
  \href{http://dx.doi.org/10.1103/PhysRevLett.86.954}{\emph{Phys. Rev. Lett.}
  {\bf 86} (2001) 954}, [\href{http://arxiv.org/abs/astro-ph/0009003}{{\tt
  astro-ph/0009003}}].

\bibitem{McDonald:2001vt}
J.~McDonald, \emph{{Thermally generated gauge singlet scalars as
  selfinteracting dark matter}},
  \href{http://dx.doi.org/10.1103/PhysRevLett.88.091304}{\emph{Phys. Rev.
  Lett.} {\bf 88} (2002) 091304},
  [\href{http://arxiv.org/abs/hep-ph/0106249}{{\tt hep-ph/0106249}}].

\bibitem{Hall:2009bx}
L.~J. Hall, K.~Jedamzik, J.~March-Russell and S.~M. West, \emph{{Freeze-In
  Production of FIMP Dark Matter}},
  \href{http://dx.doi.org/10.1007/JHEP03(2010)080}{\emph{JHEP} {\bf 03} (2010)
  080}, [\href{http://arxiv.org/abs/0911.1120}{{\tt 0911.1120}}].

\bibitem{Petraki:2013wwa}
K.~Petraki and R.~R. Volkas, \emph{{Review of asymmetric dark matter}},
  \href{http://dx.doi.org/10.1142/S0217751X13300287}{\emph{Int. J. Mod. Phys.
  A} {\bf 28} (2013) 1330028}, [\href{http://arxiv.org/abs/1305.4939}{{\tt
  1305.4939}}].

\bibitem{Zurek:2013wia}
K.~M. Zurek, \emph{{Asymmetric Dark Matter: Theories, Signatures, and
  Constraints}},
  \href{http://dx.doi.org/10.1016/j.physrep.2013.12.001}{\emph{Phys. Rept.}
  {\bf 537} (2014) 91--121}, [\href{http://arxiv.org/abs/1308.0338}{{\tt
  1308.0338}}].

\bibitem{Li:2013nal}
B.~Li, T.~Rindler-Daller and P.~R. Shapiro, \emph{{Cosmological Constraints on
  Bose-Einstein-Condensed Scalar Field Dark Matter}},
  \href{http://dx.doi.org/10.1103/PhysRevD.89.083536}{\emph{Phys. Rev. D} {\bf
  89} (2014) 083536}, [\href{http://arxiv.org/abs/1310.6061}{{\tt 1310.6061}}].

\bibitem{Hochberg:2014dra}
Y.~Hochberg, E.~Kuflik, T.~Volansky and J.~G. Wacker, \emph{{Mechanism for
  Thermal Relic Dark Matter of Strongly Interacting Massive Particles}},
  \href{http://dx.doi.org/10.1103/PhysRevLett.113.171301}{\emph{Phys. Rev.
  Lett.} {\bf 113} (2014) 171301}, [\href{http://arxiv.org/abs/1402.5143}{{\tt
  1402.5143}}].

\bibitem{Co:2015pka}
R.~T. Co, F.~D'Eramo, L.~J. Hall and D.~Pappadopulo, \emph{{Freeze-In Dark
  Matter with Displaced Signatures at Colliders}},
  \href{http://dx.doi.org/10.1088/1475-7516/2015/12/024}{\emph{JCAP} {\bf 12}
  (2015) 024}, [\href{http://arxiv.org/abs/1506.07532}{{\tt 1506.07532}}].

\bibitem{Kuflik:2015isi}
E.~Kuflik, M.~Perelstein, N.~R.-L. Lorier and Y.-D. Tsai, \emph{{Elastically
  Decoupling Dark Matter}},
  \href{http://dx.doi.org/10.1103/PhysRevLett.116.221302}{\emph{Phys. Rev.
  Lett.} {\bf 116} (2016) 221302}, [\href{http://arxiv.org/abs/1512.04545}{{\tt
  1512.04545}}].

\bibitem{Pappadopulo:2016pkp}
D.~Pappadopulo, J.~T. Ruderman and G.~Trevisan, \emph{{Dark matter freeze-out
  in a nonrelativistic sector}},
  \href{http://dx.doi.org/10.1103/PhysRevD.94.035005}{\emph{Phys. Rev. D} {\bf
  94} (2016) 035005}, [\href{http://arxiv.org/abs/1602.04219}{{\tt
  1602.04219}}].

\bibitem{Tulin:2017ara}
S.~Tulin and H.-B. Yu, \emph{{Dark Matter Self-interactions and Small Scale
  Structure}},
  \href{http://dx.doi.org/10.1016/j.physrep.2017.11.004}{\emph{Phys. Rept.}
  {\bf 730} (2018) 1--57}, [\href{http://arxiv.org/abs/1705.02358}{{\tt
  1705.02358}}].

\bibitem{Kramer:2020sbb}
E.~D. Kramer, E.~Kuflik, N.~Levi, N.~J. Outmezguine and J.~T. Ruderman,
  \emph{{Heavy Thermal Dark Matter from a New Collision Mechanism}},
  \href{http://dx.doi.org/10.1103/PhysRevLett.126.081802}{\emph{Phys. Rev.
  Lett.} {\bf 126} (2021) 081802}, [\href{http://arxiv.org/abs/2003.04900}{{\tt
  2003.04900}}].

\bibitem{DAgnolo:2020mpt}
R.~T. D'Agnolo, D.~Liu, J.~T. Ruderman and P.-J. Wang, \emph{{Forbidden dark
  matter annihilations into Standard Model particles}},
  \href{http://dx.doi.org/10.1007/JHEP06(2021)103}{\emph{JHEP} {\bf 06} (2021)
  103}, [\href{http://arxiv.org/abs/2012.11766}{{\tt 2012.11766}}].

\bibitem{Bringmann:2021tjr}
T.~Bringmann, P.~F. Depta, M.~Hufnagel, J.~T. Ruderman and K.~Schmidt-Hoberg,
  \emph{{Dark Matter from Exponential Growth}},
  \href{http://dx.doi.org/10.1103/PhysRevLett.127.191802}{\emph{Phys. Rev.
  Lett.} {\bf 127} (2021) 191802}, [\href{http://arxiv.org/abs/2103.16572}{{\tt
  2103.16572}}].

\bibitem{Puetter:2022ucx}
L.~Puetter, J.~T. Ruderman, E.~Salvioni and B.~Shakya, \emph{{Bouncing dark
  matter}}, \href{http://dx.doi.org/10.1103/PhysRevD.109.023032}{\emph{Phys.
  Rev. D} {\bf 109} (2024) 023032},
  [\href{http://arxiv.org/abs/2208.08453}{{\tt 2208.08453}}].

\bibitem{deBernardis:2000sbo}
P.~de~Bernardis et~al., \emph{{A Flat universe from high resolution maps of the
  cosmic microwave background radiation}},
  \href{http://dx.doi.org/10.1038/35010035}{\emph{Nature} {\bf 404} (2000)
  955--959}, [\href{http://arxiv.org/abs/astro-ph/0004404}{{\tt
  astro-ph/0004404}}].

\bibitem{Hinshaw:2012aka}
G.~Hinshaw et~al., \emph{{Nine-Year Wilkinson Microwave Anisotropy Probe (WMAP)
  Observations: Cosmological Parameter Results}},
  \href{http://dx.doi.org/10.1088/0067-0049/208/2/19}{\emph{Astrophys. J.
  Suppl.} {\bf 208} (2013) 19}, [\href{http://arxiv.org/abs/1212.5226}{{\tt
  1212.5226}}].

\bibitem{Aghanim:2018eyx}
N.~Aghanim et~al., \emph{{Planck 2018 results. VI. Cosmological parameters}},
  \href{http://dx.doi.org/10.1051/0004-6361/201833910}{\emph{Astron.
  Astrophys.} {\bf 641} (2020) A6},
  [\href{http://arxiv.org/abs/1807.06209}{{\tt 1807.06209}}].

\bibitem{Bahcall:1995cy}
N.~A. Bahcall, L.~M. Lubin and V.~Dorman, \emph{{Where is the dark matter?}},
  \href{http://dx.doi.org/10.1086/309577}{\emph{Astrophys. J. Lett.} {\bf 447}
  (1995) L81}, [\href{http://arxiv.org/abs/astro-ph/9506041}{{\tt
  astro-ph/9506041}}].

\bibitem{Verde:2001sf}
L.~Verde et~al., \emph{{The 2dF Galaxy Redshift Survey: The Bias of galaxies
  and the density of the Universe}},
  \href{http://dx.doi.org/10.1046/j.1365-8711.2002.05620.x}{\emph{Mon. Not.
  Roy. Astron. Soc.} {\bf 335} (2002) 432},
  [\href{http://arxiv.org/abs/astro-ph/0112161}{{\tt astro-ph/0112161}}].

\bibitem{Nesseris:2007pa}
S.~Nesseris and L.~Perivolaropoulos, \emph{{Testing Lambda CDM with the Growth
  Function delta(a): Current Constraints}},
  \href{http://dx.doi.org/10.1103/PhysRevD.77.023504}{\emph{Phys. Rev. D} {\bf
  77} (2008) 023504}, [\href{http://arxiv.org/abs/0710.1092}{{\tt 0710.1092}}].

\bibitem{Addison:2013haa}
G.~E. Addison, G.~Hinshaw and M.~Halpern, \emph{{Cosmological constraints from
  baryon acoustic oscillations and clustering of large-scale structure}},
  \href{http://dx.doi.org/10.1093/mnras/stt1687}{\emph{Mon. Not. Roy. Astron.
  Soc.} {\bf 436} (2013) 1674--1683},
  [\href{http://arxiv.org/abs/1304.6984}{{\tt 1304.6984}}].

\bibitem{More:2014uva}
S.~More, H.~Miyatake, R.~Mandelbaum, M.~Takada, D.~Spergel, J.~Brownstein
  et~al., \emph{{The Weak Lensing Signal and the Clustering of BOSS Galaxies
  II: Astrophysical and Cosmological Constraints}},
  \href{http://dx.doi.org/10.1088/0004-637X/806/1/2}{\emph{Astrophys. J.} {\bf
  806} (2015) 2}, [\href{http://arxiv.org/abs/1407.1856}{{\tt 1407.1856}}].

\bibitem{Riess:1998cb}
A.~G. Riess et~al., \emph{{Observational evidence from supernovae for an
  accelerating universe and a cosmological constant}},
  \href{http://dx.doi.org/10.1086/300499}{\emph{Astron. J.} {\bf 116} (1998)
  1009--1038}, [\href{http://arxiv.org/abs/astro-ph/9805201}{{\tt
  astro-ph/9805201}}].

\bibitem{Perlmutter:1998np}
S.~Perlmutter et~al., \emph{{Measurements of $\Omega$ and $\Lambda$ from 42
  high redshift supernovae}},
  \href{http://dx.doi.org/10.1086/307221}{\emph{Astrophys. J.} {\bf 517} (1999)
  565--586}, [\href{http://arxiv.org/abs/astro-ph/9812133}{{\tt
  astro-ph/9812133}}].

\bibitem{Zwicky:1933gu}
F.~Zwicky, \emph{{Die Rotverschiebung von extragalaktischen Nebeln}},
  \href{http://dx.doi.org/10.1007/s10714-008-0707-4}{\emph{Helv. Phys. Acta}
  {\bf 6} (1933) 110--127}.

\bibitem{Trimble:1987ee}
V.~Trimble, \emph{{Existence and Nature of Dark Matter in the Universe}},
  \href{http://dx.doi.org/10.1146/annurev.aa.25.090187.002233}{\emph{Ann. Rev.
  Astron. Astrophys.} {\bf 25} (1987) 425--472}.

\bibitem{Sofue:2000jx}
Y.~Sofue and V.~Rubin, \emph{{Rotation curves of spiral galaxies}},
  \href{http://dx.doi.org/10.1146/annurev.astro.39.1.137}{\emph{Ann. Rev.
  Astron. Astrophys.} {\bf 39} (2001) 137--174},
  [\href{http://arxiv.org/abs/astro-ph/0010594}{{\tt astro-ph/0010594}}].

\bibitem{Clowe:2006eq}
D.~Clowe, M.~Bradac, A.~H. Gonzalez, M.~Markevitch, S.~W. Randall, C.~Jones
  et~al., \emph{{A direct empirical proof of the existence of dark matter}},
  \href{http://dx.doi.org/10.1086/508162}{\emph{Astrophys. J. Lett.} {\bf 648}
  (2006) L109--L113}, [\href{http://arxiv.org/abs/astro-ph/0608407}{{\tt
  astro-ph/0608407}}].

\bibitem{Allen:2011zs}
S.~W. Allen, A.~E. Evrard and A.~B. Mantz, \emph{{Cosmological Parameters from
  Observations of Galaxy Clusters}},
  \href{http://dx.doi.org/10.1146/annurev-astro-081710-102514}{\emph{Ann. Rev.
  Astron. Astrophys.} {\bf 49} (2011) 409--470},
  [\href{http://arxiv.org/abs/1103.4829}{{\tt 1103.4829}}].

\bibitem{Salucci:2018hqu}
P.~Salucci, \emph{{The distribution of dark matter in galaxies}},
  \href{http://dx.doi.org/10.1007/s00159-018-0113-1}{\emph{Astron. Astrophys.
  Rev.} {\bf 27} (2019) 2}, [\href{http://arxiv.org/abs/1811.08843}{{\tt
  1811.08843}}].

\bibitem{Simon:2019nxf}
J.~D. Simon, \emph{{The Faintest Dwarf Galaxies}},
  \href{http://dx.doi.org/10.1146/annurev-astro-091918-104453}{\emph{Ann. Rev.
  Astron. Astrophys.} {\bf 57} (2019) 375--415},
  [\href{http://arxiv.org/abs/1901.05465}{{\tt 1901.05465}}].

\bibitem{McDermott:2010pa}
S.~D. McDermott, H.-B. Yu and K.~M. Zurek, \emph{{Turning off the Lights: How
  Dark is Dark Matter?}},
  \href{http://dx.doi.org/10.1103/PhysRevD.83.063509}{\emph{Phys. Rev. D} {\bf
  83} (2011) 063509}, [\href{http://arxiv.org/abs/1011.2907}{{\tt 1011.2907}}].

\bibitem{Dolgov:2013una}
A.~Dolgov, S.~Dubovsky, G.~Rubtsov and I.~Tkachev, \emph{{Constraints on
  millicharged particles from Planck data}},
  \href{http://dx.doi.org/10.1103/PhysRevD.88.117701}{\emph{Phys. Rev. D} {\bf
  88} (2013) 117701}, [\href{http://arxiv.org/abs/1310.2376}{{\tt 1310.2376}}].

\bibitem{Randall:2007ph}
S.~W. Randall, M.~Markevitch, D.~Clowe, A.~H. Gonzalez and M.~Bradac,
  \emph{{Constraints on the Self-Interaction Cross-Section of Dark Matter from
  Numerical Simulations of the Merging Galaxy Cluster 1E 0657-56}},
  \href{http://dx.doi.org/10.1086/587859}{\emph{Astrophys. J.} {\bf 679} (2008)
  1173--1180}, [\href{http://arxiv.org/abs/0704.0261}{{\tt 0704.0261}}].

\bibitem{Harvey:2015hha}
D.~Harvey, R.~Massey, T.~Kitching, A.~Taylor and E.~Tittley, \emph{{The
  non-gravitational interactions of dark matter in colliding galaxy clusters}},
  \href{http://dx.doi.org/10.1126/science.1261381}{\emph{Science} {\bf 347}
  (2015) 1462--1465}, [\href{http://arxiv.org/abs/1503.07675}{{\tt
  1503.07675}}].

\bibitem{WebCMBsim}
\emph{{CMB Simulator}}.
\newblock
  \href{https://plancksatellite.org.uk/cmb-sim/}{[https://plancksatellite.org.uk/cmb-sim/]}.

\bibitem{WebCMBsim2}
\emph{{CMB Analyzer}}.
\newblock
  \href{https://lambda.gsfc.nasa.gov/education/bau\_documentation.html}{[https://lambda.gsfc.nasa.gov/education/bau\_documentation.html]}.

\bibitem{Tremaine:1979we}
S.~Tremaine and J.~Gunn, \emph{{Dynamical Role of Light Neutral Leptons in
  Cosmology}}, \href{http://dx.doi.org/10.1103/PhysRevLett.42.407}{\emph{Phys.
  Rev. Lett.} {\bf 42} (1979) 407--410}.

\bibitem{White:1984yj}
S.~D. White, C.~Frenk and M.~Davis, \emph{{Clustering in a Neutrino Dominated
  Universe}}, \href{http://dx.doi.org/10.1086/161425}{\emph{Astrophys. J.
  Lett.} {\bf 274} (1983) L1--L5}.

\bibitem{Abazajian:2005xn}
K.~Abazajian, \emph{{Linear cosmological structure limits on warm dark
  matter}}, \href{http://dx.doi.org/10.1103/PhysRevD.73.063513}{\emph{Phys.
  Rev. D} {\bf 73} (2006) 063513},
  [\href{http://arxiv.org/abs/astro-ph/0512631}{{\tt astro-ph/0512631}}].

\bibitem{Viel:2013fqw}
M.~Viel, G.~D. Becker, J.~S. Bolton and M.~G. Haehnelt, \emph{{Warm dark matter
  as a solution to the small scale crisis: New constraints from high redshift
  Lyman-\ensuremath{\alpha} forest data}},
  \href{http://dx.doi.org/10.1103/PhysRevD.88.043502}{\emph{Phys. Rev. D} {\bf
  88} (2013) 043502}, [\href{http://arxiv.org/abs/1306.2314}{{\tt 1306.2314}}].

\bibitem{Schneider:2013wwa}
A.~Schneider, D.~Anderhalden, A.~Maccio and J.~Diemand, \emph{{Warm dark matter
  does not do better than cold dark matter in solving small-scale
  inconsistencies}}, \href{http://dx.doi.org/10.1093/mnrasl/slu034}{\emph{Mon.
  Not. Roy. Astron. Soc.} {\bf 441} (2014) 6},
  [\href{http://arxiv.org/abs/1309.5960}{{\tt 1309.5960}}].

\bibitem{Lovell:2011rd}
M.~R. Lovell, V.~Eke, C.~S. Frenk, L.~Gao, A.~Jenkins, T.~Theuns et~al.,
  \emph{{The Haloes of Bright Satellite Galaxies in a Warm Dark Matter
  Universe}},
  \href{http://dx.doi.org/10.1111/j.1365-2966.2011.20200.x}{\emph{Mon. Not.
  Roy. Astron. Soc.} {\bf 420} (2012) 2318--2324},
  [\href{http://arxiv.org/abs/1104.2929}{{\tt 1104.2929}}].

\bibitem{Ghigna:1998vn}
S.~Ghigna, B.~Moore, F.~Governato, G.~Lake, T.~R. Quinn and J.~Stadel,
  \emph{{Dark matter halos within clusters}},
  \href{http://dx.doi.org/10.1046/j.1365-8711.1998.01918.x}{\emph{Mon. Not.
  Roy. Astron. Soc.} {\bf 300} (1998) 146--162},
  [\href{http://arxiv.org/abs/astro-ph/9801192}{{\tt astro-ph/9801192}}].

\bibitem{Moore:1999nt}
B.~Moore, S.~Ghigna, F.~Governato, G.~Lake, T.~R. Quinn, J.~Stadel et~al.,
  \emph{{Dark matter substructure within galactic halos}},
  \href{http://dx.doi.org/10.1086/312287}{\emph{Astrophys. J. Lett.} {\bf 524}
  (1999) L19--L22}, [\href{http://arxiv.org/abs/astro-ph/9907411}{{\tt
  astro-ph/9907411}}].

\bibitem{Klypin:1999uc}
A.~A. Klypin, A.~V. Kravtsov, O.~Valenzuela and F.~Prada, \emph{{Where are the
  missing Galactic satellites?}},
  \href{http://dx.doi.org/10.1086/307643}{\emph{Astrophys. J.} {\bf 522} (1999)
  82--92}, [\href{http://arxiv.org/abs/astro-ph/9901240}{{\tt
  astro-ph/9901240}}].

\bibitem{Baer:2014eja}
H.~Baer, K.-Y. Choi, J.~E. Kim and L.~Roszkowski, \emph{{Dark matter production
  in the early Universe: beyond the thermal WIMP paradigm}},
  \href{http://dx.doi.org/10.1016/j.physrep.2014.10.002}{\emph{Phys. Rept.}
  {\bf 555} (2015) 1--60}, [\href{http://arxiv.org/abs/1407.0017}{{\tt
  1407.0017}}].

\bibitem{Workgroup:2017lvb}
T.~Bringmann et~al., \emph{{DarkBit: A GAMBIT module for computing dark matter
  observables and likelihoods}},
  \href{http://dx.doi.org/10.1140/epjc/s10052-017-5155-4}{\emph{Eur. Phys. J.
  C} {\bf 77} (2017) 831}, [\href{http://arxiv.org/abs/1705.07920}{{\tt
  1705.07920}}].

\bibitem{Gondolo:2004sc}
P.~Gondolo, J.~Edsjo, P.~Ullio, L.~Bergstrom, M.~Schelke and E.~A. Baltz,
  \emph{{DarkSUSY: Computing supersymmetric dark matter properties
  numerically}},
  \href{http://dx.doi.org/10.1088/1475-7516/2004/07/008}{\emph{JCAP} {\bf 07}
  (2004) 008}, [\href{http://arxiv.org/abs/astro-ph/0406204}{{\tt
  astro-ph/0406204}}].

\bibitem{Belanger:2004yn}
G.~B\'elanger, F.~Boudjema, A.~Pukhov and A.~Semenov, \emph{{micrOMEGAs:
  Version 1.3}},
  \href{http://dx.doi.org/10.1016/j.cpc.2005.12.005}{\emph{Comput. Phys.
  Commun.} {\bf 174} (2006) 577--604},
  [\href{http://arxiv.org/abs/hep-ph/0405253}{{\tt hep-ph/0405253}}].

\bibitem{Kawasaki:2004yh}
M.~Kawasaki, K.~Kohri and T.~Moroi, \emph{{Hadronic decay of late - decaying
  particles and Big-Bang Nucleosynthesis}},
  \href{http://dx.doi.org/10.1016/j.physletb.2005.08.045}{\emph{Phys. Lett. B}
  {\bf 625} (2005) 7--12}, [\href{http://arxiv.org/abs/astro-ph/0402490}{{\tt
  astro-ph/0402490}}].

\bibitem{Kawasaki:2004qu}
M.~Kawasaki, K.~Kohri and T.~Moroi, \emph{{Big-Bang nucleosynthesis and
  hadronic decay of long-lived massive particles}},
  \href{http://dx.doi.org/10.1103/PhysRevD.71.083502}{\emph{Phys. Rev. D} {\bf
  71} (2005) 083502}, [\href{http://arxiv.org/abs/astro-ph/0408426}{{\tt
  astro-ph/0408426}}].

\bibitem{Jedamzik:2006xz}
K.~Jedamzik, \emph{{Big bang nucleosynthesis constraints on hadronically and
  electromagnetically decaying relic neutral particles}},
  \href{http://dx.doi.org/10.1103/PhysRevD.74.103509}{\emph{Phys. Rev. D} {\bf
  74} (2006) 103509}, [\href{http://arxiv.org/abs/hep-ph/0604251}{{\tt
  hep-ph/0604251}}].

\bibitem{Hoyle:1954zz}
F.~Hoyle, \emph{{On Nuclear Reactions Occuring in Very Hot Stars. 1. The
  Synthesis of Elements from Carbon to Nickel}},
  \href{http://dx.doi.org/10.1086/190005}{\emph{Astrophys. J. Suppl.} {\bf 1}
  (1954) 121--146}.

\bibitem{Burbidge:1957vc}
M.~E. Burbidge, G.~Burbidge, W.~A. Fowler and F.~Hoyle, \emph{{Synthesis of the
  elements in stars}},
  \href{http://dx.doi.org/10.1103/RevModPhys.29.547}{\emph{Rev. Mod. Phys.}
  {\bf 29} (1957) 547--650}.

\bibitem{Coc:2003ce}
A.~Coc, E.~Vangioni-Flam, P.~Descouvemont, A.~Adahchour and C.~Angulo,
  \emph{{Updated Big Bang nucleosynthesis confronted to WMAP observations and
  to the abundance of light elements}},
  \href{http://dx.doi.org/10.1086/380121}{\emph{Astrophys. J.} {\bf 600} (2004)
  544--552}, [\href{http://arxiv.org/abs/astro-ph/0309480}{{\tt
  astro-ph/0309480}}].

\bibitem{Angulo:2005mi}
C.~Angulo et~al., \emph{{The $Be^{7}(d,p)2\alpha$ cross section at big bang
  energies and the primordial $Li^{7}$ abundances}},
  \href{http://dx.doi.org/10.1086/491732}{\emph{Astrophys. J. Lett.} {\bf 630}
  (2005) L105--L108}, [\href{http://arxiv.org/abs/astro-ph/0508454}{{\tt
  astro-ph/0508454}}].

\bibitem{Cyburt:2008kw}
R.~H. Cyburt, B.~D. Fields and K.~A. Olive, \emph{{An Update on the big bang
  nucleosynthesis prediction for Li-7: The problem worsens}},
  \href{http://dx.doi.org/10.1088/1475-7516/2008/11/012}{\emph{JCAP} {\bf 11}
  (2008) 012}, [\href{http://arxiv.org/abs/0808.2818}{{\tt 0808.2818}}].

\bibitem{Fields:2011zzb}
B.~D. Fields, \emph{{The primordial lithium problem}},
  \href{http://dx.doi.org/10.1146/annurev-nucl-102010-130445}{\emph{Ann. Rev.
  Nucl. Part. Sci.} {\bf 61} (2011) 47--68},
  [\href{http://arxiv.org/abs/1203.3551}{{\tt 1203.3551}}].

\bibitem{joshi1997elements}
A.~Joshi, \emph{Elements of Group Theory for Physicists}.
\newblock New Age International, 1997.

\bibitem{jones2020groups}
H.~Jones, \emph{Groups, Representations and Physics}.
\newblock CRC Press, 2020.

\bibitem{James_Liebeck_2001}
G.~James and M.~Liebeck, \emph{Representations and Characters of Groups}.
\newblock Cambridge University Press, 2~ed., 2001.

\bibitem{sapir2007grouptheoryproblems}
M.~V. Sapir, \emph{Some group theory problems},
  \href{http://arxiv.org/abs/0704.2899}{{\tt 0704.2899}}.

\bibitem{cocke2019databasegroupsequivalentcharacter}
W.~Cocke, S.~Goldstein and M.~Stemper, \emph{{A Database of Groups with
  Equivalent Character Tables}},  \href{http://arxiv.org/abs/1907.07633}{{\tt
  1907.07633}}.

\bibitem{GAP4}
\emph{{GAP -- Groups, Algorithms, and Programming, Version 4.13.1},
  \url{https://www.gap-system.org}},  2024.

\bibitem{Ishimori:2010au}
H.~Ishimori, T.~Kobayashi, H.~Ohki, Y.~Shimizu, H.~Okada and M.~Tanimoto,
  \emph{{Non-Abelian Discrete Symmetries in Particle Physics}},
  \href{http://dx.doi.org/10.1143/PTPS.183.1}{\emph{Prog. Theor. Phys. Suppl.}
  {\bf 183} (2010) 1--163}, [\href{http://arxiv.org/abs/1003.3552}{{\tt
  1003.3552}}].

\bibitem{Espinoza:2018itz}
C.~Espinoza, E.~A. Garc\'es, M.~Mondrag\'on and H.~Reyes-Gonz\'alez, \emph{{The
  $S3$ Symmetric Model with a Dark Scalar}},
  \href{http://dx.doi.org/10.1016/j.physletb.2018.11.028}{\emph{Phys. Lett. B}
  {\bf 788} (2019) 185--191}, [\href{http://arxiv.org/abs/1804.01879}{{\tt
  1804.01879}}].

\bibitem{Espinoza:2020qyf}
C.~Espinoza and M.~Mondrag\'on, \emph{{Prospects of Indirect Detection for the
  Heavy S3 Dark Doublet}},  \href{http://arxiv.org/abs/2008.11792}{{\tt
  2008.11792}}.

\bibitem{Shao:2023oxt}
J.~Shao and I.~P. Ivanov, \emph{{Symmetries for the 4HDM: extensions of cyclic
  groups}}, \href{http://dx.doi.org/10.1007/JHEP10(2023)070}{\emph{JHEP} {\bf
  10} (2023) 070}, [\href{http://arxiv.org/abs/2305.05207}{{\tt 2305.05207}}].

\bibitem{Shao:2024ibu}
J.~Shao, I.~P. Ivanov and M.~Korhonen, \emph{{Symmetries for the 4HDM: II.
  Extensions by rephasing groups}},
  \href{http://dx.doi.org/10.1088/1751-8121/ad7340}{\emph{J. Phys. A} {\bf 57}
  (2024) 385401}, [\href{http://arxiv.org/abs/2404.10349}{{\tt 2404.10349}}].

\bibitem{Sloane:77}
N.~J.~A. Sloane, \emph{Error-correcting codes and invariant theory: New
  applications of a nineteenth-century technique}, {\emph{The American
  Mathematical Monthly} {\bf 84} (1977) 82--107}.

\bibitem{Patera:78}
J.~Patera, R.~T. Sharp and P.~Winternitz, \emph{Polynomial irreducible tensors
  for point groups}, \href{http://dx.doi.org/10.1063/1.523595}{\emph{Journal of
  Mathematical Physics} {\bf 19} (11, 1978) 2362--2376}.

\bibitem{Derman:1978rx}
E.~Derman, \emph{{Flavor Unification, $\tau$ Decay and $b$ Decay Within the Six
  Quark Six Lepton {Weinberg-Salam} Model}},
  \href{http://dx.doi.org/10.1103/PhysRevD.19.317}{\emph{Phys. Rev. D} {\bf 19}
  (1979) 317--329}.

\bibitem{Derman:1979nf}
E.~Derman and H.-S. Tsao, \emph{{SU(2) X U(1) X S($n$) Flavor Dynamics and a
  Bound on the Number of Flavors}},
  \href{http://dx.doi.org/10.1103/PhysRevD.20.1207}{\emph{Phys. Rev. D} {\bf
  20} (1979) 1207}.

\bibitem{Lee:1973iz}
T.~D. Lee, \emph{{A Theory of Spontaneous T Violation}},
  \href{http://dx.doi.org/10.1103/PhysRevD.8.1226}{\emph{Phys. Rev. D} {\bf 8}
  (1973) 1226--1239}.

\bibitem{Branco:2011iw}
G.~C. Branco, P.~M. Ferreira, L.~Lavoura, M.~N. Rebelo, M.~Sher and J.~P.
  Silva, \emph{{Theory and phenomenology of two-Higgs-doublet models}},
  \href{http://dx.doi.org/10.1016/j.physrep.2012.02.002}{\emph{Phys. Rept.}
  {\bf 516} (2012) 1--102}, [\href{http://arxiv.org/abs/1106.0034}{{\tt
  1106.0034}}].

\bibitem{Martin:1997ns}
S.~P. Martin, \emph{{A Supersymmetry primer}},
  \href{http://dx.doi.org/10.1142/9789812839657_0001}{\emph{Adv. Ser. Direct.
  High Energy Phys.} {\bf 18} (1998) 1--98},
  [\href{http://arxiv.org/abs/hep-ph/9709356}{{\tt hep-ph/9709356}}].

\bibitem{Djouadi:2005gj}
A.~Djouadi, \emph{{The Anatomy of electro-weak symmetry breaking. II. The Higgs
  bosons in the minimal supersymmetric model}},
  \href{http://dx.doi.org/10.1016/j.physrep.2007.10.005}{\emph{Phys. Rept.}
  {\bf 459} (2008) 1--241}, [\href{http://arxiv.org/abs/hep-ph/0503173}{{\tt
  hep-ph/0503173}}].

\bibitem{Deshpande:1977rw}
N.~G. Deshpande and E.~Ma, \emph{{Pattern of Symmetry Breaking with Two Higgs
  Doublets}}, \href{http://dx.doi.org/10.1103/PhysRevD.18.2574}{\emph{Phys.
  Rev.} {\bf D18} (1978) 2574}.

\bibitem{Silveira:1985rk}
V.~Silveira and A.~Zee, \emph{{Scalar Phantoms}},
  \href{http://dx.doi.org/10.1016/0370-2693(85)90624-0}{\emph{Phys. Lett.} {\bf
  161B} (1985) 136--140}.

\bibitem{Barbieri:2006dq}
R.~Barbieri, L.~J. Hall and V.~S. Rychkov, \emph{{Improved naturalness with a
  heavy Higgs: An Alternative road to LHC physics}},
  \href{http://dx.doi.org/10.1103/PhysRevD.74.015007}{\emph{Phys. Rev.} {\bf
  D74} (2006) 015007}, [\href{http://arxiv.org/abs/hep-ph/0603188}{{\tt
  hep-ph/0603188}}].

\bibitem{LopezHonorez:2006gr}
L.~Lopez~Honorez, E.~Nezri, J.~F. Oliver and M.~H.~G. Tytgat, \emph{{The Inert
  Doublet Model: An Archetype for Dark Matter}},
  \href{http://dx.doi.org/10.1088/1475-7516/2007/02/028}{\emph{JCAP} {\bf 0702}
  (2007) 028}, [\href{http://arxiv.org/abs/hep-ph/0612275}{{\tt
  hep-ph/0612275}}].

\bibitem{Branco:1980sz}
G.~C. Branco, \emph{{Spontaneous {CP} Nonconservation and Natural Flavor
  Conservation: A Minimal Model}},
  \href{http://dx.doi.org/10.1103/PhysRevD.22.2901}{\emph{Phys. Rev. D} {\bf
  22} (1980) 2901}.

\bibitem{Branco:1985aq}
G.~C. Branco and M.~N. Rebelo, \emph{{The Higgs Mass in a Model With Two Scalar
  Doublets and Spontaneous {CP} Violation}},
  \href{http://dx.doi.org/10.1016/0370-2693(85)91476-5}{\emph{Phys. Lett. B}
  {\bf 160} (1985) 117--120}.

\bibitem{Weinberg:1990me}
S.~Weinberg, \emph{{Unitarity Constraints on {CP} Nonconservation in Higgs
  Exchange}}, \href{http://dx.doi.org/10.1103/PhysRevD.42.860}{\emph{Phys. Rev.
  D} {\bf 42} (1990) 860--866}.

\bibitem{Pilaftsis:1999qt}
A.~Pilaftsis and C.~E.~M. Wagner, \emph{{Higgs bosons in the minimal
  supersymmetric standard model with explicit CP violation}},
  \href{http://dx.doi.org/10.1016/S0550-3213(99)00261-8}{\emph{Nucl. Phys. B}
  {\bf 553} (1999) 3--42}, [\href{http://arxiv.org/abs/hep-ph/9902371}{{\tt
  hep-ph/9902371}}].

\bibitem{Gavela:1993ts}
M.~B. Gavela, P.~Hernandez, J.~Orloff and O.~Pene, \emph{{Standard model CP
  violation and baryon asymmetry}},
  \href{http://dx.doi.org/10.1142/S0217732394000629}{\emph{Mod. Phys. Lett. A}
  {\bf 9} (1994) 795--810}, [\href{http://arxiv.org/abs/hep-ph/9312215}{{\tt
  hep-ph/9312215}}].

\bibitem{Farakos:1994kx}
K.~Farakos, K.~Kajantie, K.~Rummukainen and M.~E. Shaposhnikov, \emph{{3-D
  physics and the electroweak phase transition: Perturbation theory}},
  \href{http://dx.doi.org/10.1016/0550-3213(94)90173-2}{\emph{Nucl. Phys. B}
  {\bf 425} (1994) 67--109}, [\href{http://arxiv.org/abs/hep-ph/9404201}{{\tt
  hep-ph/9404201}}].

\bibitem{Gavela:1994dt}
M.~B. Gavela, P.~Hernandez, J.~Orloff, O.~Pene and C.~Quimbay, \emph{{Standard
  model CP violation and baryon asymmetry. Part 2: Finite temperature}},
  \href{http://dx.doi.org/10.1016/0550-3213(94)00410-2}{\emph{Nucl. Phys. B}
  {\bf 430} (1994) 382--426}, [\href{http://arxiv.org/abs/hep-ph/9406289}{{\tt
  hep-ph/9406289}}].

\bibitem{Turok:1990zg}
N.~Turok and J.~Zadrozny, \emph{{Electroweak baryogenesis in the two doublet
  model}}, \href{http://dx.doi.org/10.1016/0550-3213(91)90356-3}{\emph{Nucl.
  Phys. B} {\bf 358} (1991) 471--493}.

\bibitem{Funakubo:1993jg}
K.~Funakubo, A.~Kakuto and K.~Takenaga, \emph{{The Effective potential of
  electroweak theory with two massless Higgs doublets at finite temperature}},
  \href{http://dx.doi.org/10.1143/PTP.91.341}{\emph{Prog. Theor. Phys.} {\bf
  91} (1994) 341--352}, [\href{http://arxiv.org/abs/hep-ph/9310267}{{\tt
  hep-ph/9310267}}].

\bibitem{Joyce:1994zt}
M.~Joyce, T.~Prokopec and N.~Turok, \emph{{Nonlocal electroweak baryogenesis.
  Part 2: The Classical regime}},
  \href{http://dx.doi.org/10.1103/PhysRevD.53.2958}{\emph{Phys. Rev. D} {\bf
  53} (1996) 2958--2980}, [\href{http://arxiv.org/abs/hep-ph/9410282}{{\tt
  hep-ph/9410282}}].

\bibitem{Cline:1995dg}
J.~M. Cline, K.~Kainulainen and A.~P. Vischer, \emph{{Dynamics of two Higgs
  doublet CP violation and baryogenesis at the electroweak phase transition}},
  \href{http://dx.doi.org/10.1103/PhysRevD.54.2451}{\emph{Phys. Rev. D} {\bf
  54} (1996) 2451--2472}, [\href{http://arxiv.org/abs/hep-ph/9506284}{{\tt
  hep-ph/9506284}}].

\bibitem{Cline:1996mga}
J.~M. Cline and P.-A. Lemieux, \emph{{Electroweak phase transition in two Higgs
  doublet models}},
  \href{http://dx.doi.org/10.1103/PhysRevD.55.3873}{\emph{Phys. Rev. D} {\bf
  55} (1997) 3873--3881}, [\href{http://arxiv.org/abs/hep-ph/9609240}{{\tt
  hep-ph/9609240}}].

\bibitem{Fromme:2006cm}
L.~Fromme, S.~J. Huber and M.~Seniuch, \emph{{Baryogenesis in the two-Higgs
  doublet model}},
  \href{http://dx.doi.org/10.1088/1126-6708/2006/11/038}{\emph{JHEP} {\bf 11}
  (2006) 038}, [\href{http://arxiv.org/abs/hep-ph/0605242}{{\tt
  hep-ph/0605242}}].

\bibitem{Peccei:1977hh}
R.~D. Peccei and H.~R. Quinn, \emph{{CP Conservation in the Presence of
  Instantons}},
  \href{http://dx.doi.org/10.1103/PhysRevLett.38.1440}{\emph{Phys. Rev. Lett.}
  {\bf 38} (1977) 1440--1443}.

\bibitem{Peccei:1977ur}
R.~D. Peccei and H.~R. Quinn, \emph{{Constraints Imposed by CP Conservation in
  the Presence of Instantons}},
  \href{http://dx.doi.org/10.1103/PhysRevD.16.1791}{\emph{Phys. Rev. D} {\bf
  16} (1977) 1791--1797}.

\bibitem{Weinberg:1977ma}
S.~Weinberg, \emph{{A New Light Boson?}},
  \href{http://dx.doi.org/10.1103/PhysRevLett.40.223}{\emph{Phys. Rev. Lett.}
  {\bf 40} (1978) 223--226}.

\bibitem{Wilczek:1977pj}
F.~Wilczek, \emph{{Problem of Strong $P$ and $T$ Invariance in the Presence of
  Instantons}}, \href{http://dx.doi.org/10.1103/PhysRevLett.40.279}{\emph{Phys.
  Rev. Lett.} {\bf 40} (1978) 279--282}.

\bibitem{Tao:1996vb}
Z.-j. Tao, \emph{{Radiative seesaw mechanism at weak scale}},
  \href{http://dx.doi.org/10.1103/PhysRevD.54.5693}{\emph{Phys. Rev. D} {\bf
  54} (1996) 5693--5697}, [\href{http://arxiv.org/abs/hep-ph/9603309}{{\tt
  hep-ph/9603309}}].

\bibitem{Barbieri:2000gf}
R.~Barbieri and A.~Strumia, \emph{{The 'LEP paradox'}},  in \emph{{4th
  Rencontres du Vietnam}: {Physics at Extreme Energies (Particle Physics and
  Astrophysics)}}, 7, 2000.
\newblock \href{http://arxiv.org/abs/hep-ph/0007265}{{\tt hep-ph/0007265}}.

\bibitem{Ma:2006km}
E.~Ma, \emph{{Verifiable radiative seesaw mechanism of neutrino mass and dark
  matter}}, \href{http://dx.doi.org/10.1103/PhysRevD.73.077301}{\emph{Phys.
  Rev.} {\bf D73} (2006) 077301},
  [\href{http://arxiv.org/abs/hep-ph/0601225}{{\tt hep-ph/0601225}}].

\bibitem{Ma:2006fn}
E.~Ma, \emph{{Common origin of neutrino mass, dark matter, and baryogenesis}},
  \href{http://dx.doi.org/10.1142/S0217732306021141}{\emph{Mod. Phys. Lett. A}
  {\bf 21} (2006) 1777--1782}, [\href{http://arxiv.org/abs/hep-ph/0605180}{{\tt
  hep-ph/0605180}}].

\bibitem{Casas:2006bd}
J.~A. Casas, J.~R. Espinosa and I.~Hidalgo, \emph{{Expectations for LHC from
  naturalness: modified versus SM Higgs sector}},
  \href{http://dx.doi.org/10.1016/j.nuclphysb.2007.04.024}{\emph{Nucl. Phys. B}
  {\bf 777} (2007) 226--252}, [\href{http://arxiv.org/abs/hep-ph/0607279}{{\tt
  hep-ph/0607279}}].

\bibitem{Hambye:2007vf}
T.~Hambye and M.~H.~G. Tytgat, \emph{{Electroweak symmetry breaking induced by
  dark matter}},
  \href{http://dx.doi.org/10.1016/j.physletb.2007.11.069}{\emph{Phys. Lett.}
  {\bf B659} (2008) 651--655}, [\href{http://arxiv.org/abs/0707.0633}{{\tt
  0707.0633}}].

\bibitem{Pomarol:1993mu}
A.~Pomarol and R.~Vega, \emph{{Constraints on CP violation in the Higgs sector
  from the rho parameter}},
  \href{http://dx.doi.org/10.1016/0550-3213(94)90611-4}{\emph{Nucl. Phys. B}
  {\bf 413} (1994) 3--15}, [\href{http://arxiv.org/abs/hep-ph/9305272}{{\tt
  hep-ph/9305272}}].

\bibitem{Ivanov:2005hg}
I.~P. Ivanov, \emph{{Two-Higgs-doublet model from the group-theoretic
  perspective}},
  \href{http://dx.doi.org/10.1016/j.physletb.2005.10.015}{\emph{Phys. Lett. B}
  {\bf 632} (2006) 360--365}, [\href{http://arxiv.org/abs/hep-ph/0507132}{{\tt
  hep-ph/0507132}}].

\bibitem{Ivanov:2006yq}
I.~P. Ivanov, \emph{{Minkowski space structure of the Higgs potential in
  2HDM}}, \href{http://dx.doi.org/10.1103/PhysRevD.75.035001}{\emph{Phys. Rev.
  D} {\bf 75} (2007) 035001}, [\href{http://arxiv.org/abs/hep-ph/0609018}{{\tt
  hep-ph/0609018}}].

\bibitem{Gerard:2007kn}
J.~M. Gerard and M.~Herquet, \emph{{A Twisted custodial symmetry in the
  two-Higgs-doublet model}},
  \href{http://dx.doi.org/10.1103/PhysRevLett.98.251802}{\emph{Phys. Rev.
  Lett.} {\bf 98} (2007) 251802},
  [\href{http://arxiv.org/abs/hep-ph/0703051}{{\tt hep-ph/0703051}}].

\bibitem{Ivanov:2007de}
I.~P. Ivanov, \emph{{Minkowski space structure of the Higgs potential in 2HDM.
  II. Minima, symmetries, and topology}},
  \href{http://dx.doi.org/10.1103/PhysRevD.77.015017}{\emph{Phys. Rev. D} {\bf
  77} (2008) 015017}, [\href{http://arxiv.org/abs/0710.3490}{{\tt 0710.3490}}].

\bibitem{Ferreira:2009wh}
P.~M. Ferreira, H.~E. Haber and J.~P. Silva, \emph{{Generalized CP symmetries
  and special regions of parameter space in the two-Higgs-doublet model}},
  \href{http://dx.doi.org/10.1103/PhysRevD.79.116004}{\emph{Phys. Rev. D} {\bf
  79} (2009) 116004}, [\href{http://arxiv.org/abs/0902.1537}{{\tt 0902.1537}}].

\bibitem{Ferreira:2010yh}
P.~M. Ferreira, H.~E. Haber, M.~Maniatis, O.~Nachtmann and J.~P. Silva,
  \emph{{Geometric picture of generalized-CP and Higgs-family transformations
  in the two-Higgs-doublet model}},
  \href{http://dx.doi.org/10.1142/S0217751X11051494}{\emph{Int. J. Mod. Phys.
  A} {\bf 26} (2011) 769--808}, [\href{http://arxiv.org/abs/1010.0935}{{\tt
  1010.0935}}].

\bibitem{Battye:2011jj}
R.~A. Battye, G.~D. Brawn and A.~Pilaftsis, \emph{{Vacuum Topology of the Two
  Higgs Doublet Model}},
  \href{http://dx.doi.org/10.1007/JHEP08(2011)020}{\emph{JHEP} {\bf 08} (2011)
  020}, [\href{http://arxiv.org/abs/1106.3482}{{\tt 1106.3482}}].

\bibitem{Pilaftsis:2011ed}
A.~Pilaftsis, \emph{{On the Classification of Accidental Symmetries of the Two
  Higgs Doublet Model Potential}},
  \href{http://dx.doi.org/10.1016/j.physletb.2011.11.047}{\emph{Phys. Lett. B}
  {\bf 706} (2012) 465--469}, [\href{http://arxiv.org/abs/1109.3787}{{\tt
  1109.3787}}].

\bibitem{BhupalDev:2014bir}
P.~S. Bhupal~Dev and A.~Pilaftsis, \emph{{Maximally Symmetric Two Higgs Doublet
  Model with Natural Standard Model Alignment}},
  \href{http://dx.doi.org/10.1007/JHEP12(2014)024}{\emph{JHEP} {\bf 12} (2014)
  024}, [\href{http://arxiv.org/abs/1408.3405}{{\tt 1408.3405}}].

\bibitem{Pilaftsis:2016erj}
A.~Pilaftsis, \emph{{Symmetries for standard model alignment in multi-Higgs
  doublet models}},
  \href{http://dx.doi.org/10.1103/PhysRevD.93.075012}{\emph{Phys. Rev. D} {\bf
  93} (2016) 075012}, [\href{http://arxiv.org/abs/1602.02017}{{\tt
  1602.02017}}].

\bibitem{Haber:2018iwr}
H.~E. Haber, O.~M. Ogreid, P.~Osland and M.~N. Rebelo, \emph{{Symmetries and
  Mass Degeneracies in the Scalar Sector}},
  \href{http://dx.doi.org/10.1007/JHEP01(2019)042}{\emph{JHEP} {\bf 01} (2019)
  042}, [\href{http://arxiv.org/abs/1808.08629}{{\tt 1808.08629}}].

\bibitem{Darvishi:2019dbh}
N.~Darvishi and A.~Pilaftsis, \emph{{Classifying Accidental Symmetries in
  Multi-Higgs Doublet Models}},
  \href{http://dx.doi.org/10.1103/PhysRevD.101.095008}{\emph{Phys. Rev. D} {\bf
  101} (2020) 095008}, [\href{http://arxiv.org/abs/1912.00887}{{\tt
  1912.00887}}].

\bibitem{Bento:2020jei}
M.~P. Bento, R.~Boto, J.~P. Silva and A.~Trautner, \emph{{A fully basis
  invariant Symmetry Map of the 2HDM}},
  \href{http://dx.doi.org/10.1007/JHEP02(2021)220}{\emph{JHEP} {\bf 21} (2020)
  229}, [\href{http://arxiv.org/abs/2009.01264}{{\tt 2009.01264}}].

\bibitem{Ferreira:2020ana}
P.~M. Ferreira, B.~Grzadkowski, O.~M. Ogreid and P.~Osland, \emph{{Symmetries
  of the 2HDM: an invariant formulation and consequences}},
  \href{http://dx.doi.org/10.1007/JHEP02(2021)196}{\emph{JHEP} {\bf 02} (2021)
  196}, [\href{http://arxiv.org/abs/2010.13698}{{\tt 2010.13698}}].

\bibitem{Ferreira:2022gjh}
P.~M. Ferreira, B.~Grzadkowski, O.~M. Ogreid and P.~Osland, \emph{{Softly
  broken symmetries in the 2HDM -- an invariant formulation}},
  \href{http://dx.doi.org/10.1007/JHEP01(2023)143}{\emph{JHEP} {\bf 01} (2023)
  143}, [\href{http://arxiv.org/abs/2209.00152}{{\tt 2209.00152}}].

\bibitem{Ferreira:2023dke}
P.~M. Ferreira, B.~Grzadkowski, O.~M. Ogreid and P.~Osland, \emph{{New
  symmetries of the two-Higgs-doublet model}},
  \href{http://dx.doi.org/10.1140/epjc/s10052-024-12561-8}{\emph{Eur. Phys. J.
  C} {\bf 84} (2024) 234}, [\href{http://arxiv.org/abs/2306.02410}{{\tt
  2306.02410}}].

\bibitem{Doring:2024kdg}
C.~D\"oring and A.~Trautner, \emph{{Symmetries from outer automorphisms and
  unorthodox group extensions}},  \href{http://arxiv.org/abs/2410.11052}{{\tt
  2410.11052}}.

\bibitem{Trautner:2025yxz}
A.~Trautner, \emph{{Goofy is the new Normal}},
  \href{http://arxiv.org/abs/2505.00099}{{\tt 2505.00099}}.

\bibitem{Kallosh:1995hi}
R.~Kallosh, A.~D. Linde, D.~A. Linde and L.~Susskind, \emph{{Gravity and global
  symmetries}}, \href{http://dx.doi.org/10.1103/PhysRevD.52.912}{\emph{Phys.
  Rev. D} {\bf 52} (1995) 912--935},
  [\href{http://arxiv.org/abs/hep-th/9502069}{{\tt hep-th/9502069}}].

\bibitem{Banks:2010zn}
T.~Banks and N.~Seiberg, \emph{{Symmetries and Strings in Field Theory and
  Gravity}}, \href{http://dx.doi.org/10.1103/PhysRevD.83.084019}{\emph{Phys.
  Rev. D} {\bf 83} (2011) 084019}, [\href{http://arxiv.org/abs/1011.5120}{{\tt
  1011.5120}}].

\bibitem{Harlow:2018jwu}
D.~Harlow and H.~Ooguri, \emph{{Constraints on Symmetries from Holography}},
  \href{http://dx.doi.org/10.1103/PhysRevLett.122.191601}{\emph{Phys. Rev.
  Lett.} {\bf 122} (2019) 191601}, [\href{http://arxiv.org/abs/1810.05337}{{\tt
  1810.05337}}].

\bibitem{Harlow:2018tng}
D.~Harlow and H.~Ooguri, \emph{{Symmetries in quantum field theory and quantum
  gravity}}, \href{http://dx.doi.org/10.1007/s00220-021-04040-y}{\emph{Commun.
  Math. Phys.} {\bf 383} (2021) 1669--1804},
  [\href{http://arxiv.org/abs/1810.05338}{{\tt 1810.05338}}].

\bibitem{Haber:1993an}
H.~E. Haber and R.~Hempfling, \emph{{The Renormalization group improved Higgs
  sector of the minimal supersymmetric model}},
  \href{http://dx.doi.org/10.1103/PhysRevD.48.4280}{\emph{Phys. Rev. D} {\bf
  48} (1993) 4280--4309}, [\href{http://arxiv.org/abs/hep-ph/9307201}{{\tt
  hep-ph/9307201}}].

\bibitem{Wu:1994ja}
Y.~L. Wu and L.~Wolfenstein, \emph{{Sources of CP violation in the two Higgs
  doublet model}},
  \href{http://dx.doi.org/10.1103/PhysRevLett.73.1762}{\emph{Phys. Rev. Lett.}
  {\bf 73} (1994) 1762--1764}, [\href{http://arxiv.org/abs/hep-ph/9409421}{{\tt
  hep-ph/9409421}}].

\bibitem{Davidson:2005cw}
S.~Davidson and H.~E. Haber, \emph{{Basis-independent methods for the
  two-Higgs-doublet model}},
  \href{http://dx.doi.org/10.1103/PhysRevD.72.099902}{\emph{Phys. Rev. D} {\bf
  72} (2005) 035004}, [\href{http://arxiv.org/abs/hep-ph/0504050}{{\tt
  hep-ph/0504050}}].

\bibitem{Olaussen:2010aq}
K.~Olaussen, P.~Osland and M.~A. Solberg, \emph{{Symmetry and Mass Degeneration
  in Multi-Higgs-Doublet Models}},
  \href{http://dx.doi.org/10.1007/JHEP07(2011)020}{\emph{JHEP} {\bf 07} (2011)
  020}, [\href{http://arxiv.org/abs/1007.1424}{{\tt 1007.1424}}].

\bibitem{Botella:1994cs}
F.~J. Botella and J.~P. Silva, \emph{{Jarlskog - like invariants for theories
  with scalars and fermions}},  \href{http://arxiv.org/abs/hep-ph/9411288}{{\tt
  hep-ph/9411288}}.

\bibitem{Branco:1999fs}
G.~C. Branco, L.~Lavoura and J.~P. Silva, \emph{{CP Violation}}, vol.~103.
\newblock 1999,
  \href{http://dx.doi.org/10.1093/oso/9780198503996.001.0001}{10.1093/oso/9780198503996.001.0001}.

\bibitem{Nishi:2006tg}
C.~C. Nishi, \emph{{CP violation conditions in N-Higgs-doublet potentials}},
  \href{http://dx.doi.org/10.1103/PhysRevD.76.119901}{\emph{Phys. Rev. D} {\bf
  74} (2006) 036003}, [\href{http://arxiv.org/abs/hep-ph/0605153}{{\tt
  hep-ph/0605153}}].

\bibitem{Nishi:2007nh}
C.~C. Nishi, \emph{{The Structure of potentials with N Higgs doublets}},
  \href{http://dx.doi.org/10.1103/PhysRevD.76.055013}{\emph{Phys. Rev. D} {\bf
  76} (2007) 055013}, [\href{http://arxiv.org/abs/0706.2685}{{\tt 0706.2685}}].

\bibitem{Ivanov:2010ww}
I.~P. Ivanov and C.~C. Nishi, \emph{{Properties of the general NHDM. I. The
  Orbit space}},
  \href{http://dx.doi.org/10.1103/PhysRevD.82.015014}{\emph{Phys. Rev. D} {\bf
  82} (2010) 015014}, [\href{http://arxiv.org/abs/1004.1799}{{\tt 1004.1799}}].

\bibitem{Maniatis:2014oza}
M.~Maniatis and O.~Nachtmann, \emph{{Stability and symmetry breaking in the
  general three-Higgs-doublet model}},
  \href{http://dx.doi.org/10.1007/JHEP10(2015)149}{\emph{JHEP} {\bf 02} (2015)
  058}, [\href{http://arxiv.org/abs/1408.6833}{{\tt 1408.6833}}].

\bibitem{Velhinho:1994np}
J.~Velhinho, R.~Santos and A.~Barroso, \emph{{Tree level vacuum stability in
  two Higgs doublet models}},
  \href{http://dx.doi.org/10.1016/0370-2693(94)91109-6}{\emph{Phys. Lett. B}
  {\bf 322} (1994) 213--218}.

\bibitem{Nagel:2004sw}
F.~Nagel, \emph{{New aspects of gauge-boson couplings and the Higgs sector}}.
\newblock PhD thesis, University of Heidelberg, 2004.
\newblock
  \href{http://www.ub.uni-heidelberg.de/archiv/4803}{[10.11588/heidok.00004803]}.

\bibitem{Maniatis:2006fs}
M.~Maniatis, A.~von Manteuffel, O.~Nachtmann and F.~Nagel, \emph{{Stability and
  symmetry breaking in the general two-Higgs-doublet model}},
  \href{http://dx.doi.org/10.1140/epjc/s10052-006-0016-6}{\emph{Eur. Phys. J.
  C} {\bf 48} (2006) 805--823},
  [\href{http://arxiv.org/abs/hep-ph/0605184}{{\tt hep-ph/0605184}}].

\bibitem{Maniatis:2006jd}
M.~Maniatis, A.~von Manteuffel and O.~Nachtmann, \emph{{Determining the global
  minimum of Higgs potentials via Groebner bases: Applied to the NMSSM}},
  \href{http://dx.doi.org/10.1140/epjc/s10052-006-0186-2}{\emph{Eur. Phys. J.
  C} {\bf 49} (2007) 1067--1076},
  [\href{http://arxiv.org/abs/hep-ph/0608314}{{\tt hep-ph/0608314}}].

\bibitem{Maniatis:2007vn}
M.~Maniatis, A.~von Manteuffel and O.~Nachtmann, \emph{{CP violation in the
  general two-Higgs-doublet model: A Geometric view}},
  \href{http://dx.doi.org/10.1140/epjc/s10052-008-0712-5}{\emph{Eur. Phys. J.
  C} {\bf 57} (2008) 719--738}, [\href{http://arxiv.org/abs/0707.3344}{{\tt
  0707.3344}}].

\bibitem{Nishi:2007dv}
C.~C. Nishi, \emph{{Physical parameters and basis transformations in the
  Two-Higgs-Doublet model}},
  \href{http://dx.doi.org/10.1103/PhysRevD.77.055009}{\emph{Phys. Rev. D} {\bf
  77} (2008) 055009}, [\href{http://arxiv.org/abs/0712.4260}{{\tt 0712.4260}}].

\bibitem{Dixit:1987at}
V.~V. Dixit and M.~Sher, \emph{{Variation of the Fermi Constant and Primordial
  Nucleosynthesis}},
  \href{http://dx.doi.org/10.1103/PhysRevD.37.1097}{\emph{Phys. Rev. D} {\bf
  37} (1988) 1097}.

\bibitem{Kujat:1999rk}
J.~Kujat and R.~J. Scherrer, \emph{{The Effect of time variation in the Higgs
  vacuum expectation value on the cosmic microwave background}},
  \href{http://dx.doi.org/10.1103/PhysRevD.62.023510}{\emph{Phys. Rev. D} {\bf
  62} (2000) 023510}, [\href{http://arxiv.org/abs/astro-ph/9912174}{{\tt
  astro-ph/9912174}}].

\bibitem{Uzan:2002vq}
J.-P. Uzan, \emph{{The Fundamental Constants and Their Variation: Observational
  Status and Theoretical Motivations}},
  \href{http://dx.doi.org/10.1103/RevModPhys.75.403}{\emph{Rev. Mod. Phys.}
  {\bf 75} (2003) 403}, [\href{http://arxiv.org/abs/hep-ph/0205340}{{\tt
  hep-ph/0205340}}].

\bibitem{Yoo:2002vw}
J.~Yoo and R.~J. Scherrer, \emph{{Big bang nucleosynthesis and cosmic microwave
  background constraints on the time variation of the Higgs vacuum expectation
  value}}, \href{http://dx.doi.org/10.1103/PhysRevD.67.043517}{\emph{Phys. Rev.
  D} {\bf 67} (2003) 043517},
  [\href{http://arxiv.org/abs/astro-ph/0211545}{{\tt astro-ph/0211545}}].

\bibitem{Casadio:2007ip}
R.~Casadio, P.~L. Iafelice and G.~P. Vacca, \emph{{Non-adiabatic quantum
  effects from a Standard Model time-dependent Higgs vev}},
  \href{http://dx.doi.org/10.1016/j.nuclphysb.2007.05.015}{\emph{Nucl. Phys. B}
  {\bf 783} (2007) 1--30}, [\href{http://arxiv.org/abs/hep-th/0702175}{{\tt
  hep-th/0702175}}].

\bibitem{Fung:2021wbz}
L.~W.~H. Fung, L.~Li, T.~Liu, H.~N. Luu, Y.-C. Qiu and S.~H.~H. Tye,
  \emph{{Axi-Higgs cosmology}},
  \href{http://dx.doi.org/10.1088/1475-7516/2021/08/057}{\emph{JCAP} {\bf 08}
  (2021) 057}, [\href{http://arxiv.org/abs/2102.11257}{{\tt 2102.11257}}].

\bibitem{Aoki:2009ha}
M.~Aoki, S.~Kanemura, K.~Tsumura and K.~Yagyu, \emph{{Models of Yukawa
  interaction in the two Higgs doublet model, and their collider
  phenomenology}},
  \href{http://dx.doi.org/10.1103/PhysRevD.80.015017}{\emph{Phys. Rev. D} {\bf
  80} (2009) 015017}, [\href{http://arxiv.org/abs/0902.4665}{{\tt 0902.4665}}].

\bibitem{Eriksson:2009ws}
D.~Eriksson, J.~Rathsman and O.~Stal, \emph{{2HDMC: Two-Higgs-Doublet Model
  Calculator Physics and Manual}},
  \href{http://dx.doi.org/10.1016/j.cpc.2009.09.011}{\emph{Comput. Phys.
  Commun.} {\bf 181} (2010) 189--205},
  [\href{http://arxiv.org/abs/0902.0851}{{\tt 0902.0851}}].

\bibitem{Haber:2006ue}
H.~E. Haber and D.~O'Neil, \emph{{Basis-independent methods for the
  two-Higgs-doublet model. II. The Significance of tan$\beta$}},
  \href{http://dx.doi.org/10.1103/PhysRevD.74.015018}{\emph{Phys. Rev. D} {\bf
  74} (2006) 015018}, [\href{http://arxiv.org/abs/hep-ph/0602242}{{\tt
  hep-ph/0602242}}].

\bibitem{Pich:2009sp}
A.~Pich and P.~Tuzon, \emph{{Yukawa Alignment in the Two-Higgs-Doublet Model}},
  \href{http://dx.doi.org/10.1103/PhysRevD.80.091702}{\emph{Phys. Rev. D} {\bf
  80} (2009) 091702}, [\href{http://arxiv.org/abs/0908.1554}{{\tt 0908.1554}}].

\bibitem{Ferreira:2010xe}
P.~M. Ferreira, L.~Lavoura and J.~P. Silva, \emph{{Renormalization-group
  constraints on Yukawa alignment in multi-Higgs-doublet models}},
  \href{http://dx.doi.org/10.1016/j.physletb.2010.04.033}{\emph{Phys. Lett. B}
  {\bf 688} (2010) 341--344}, [\href{http://arxiv.org/abs/1001.2561}{{\tt
  1001.2561}}].

\bibitem{Bijnens:2011gd}
J.~Bijnens, J.~Lu and J.~Rathsman, \emph{{Constraining General Two Higgs
  Doublet Models by the Evolution of Yukawa Couplings}},
  \href{http://dx.doi.org/10.1007/JHEP05(2012)118}{\emph{JHEP} {\bf 05} (2012)
  118}, [\href{http://arxiv.org/abs/1111.5760}{{\tt 1111.5760}}].

\bibitem{Celis:2013rcs}
A.~Celis, V.~Ilisie and A.~Pich, \emph{{LHC constraints on two-Higgs doublet
  models}}, \href{http://dx.doi.org/10.1007/JHEP07(2013)053}{\emph{JHEP} {\bf
  07} (2013) 053}, [\href{http://arxiv.org/abs/1302.4022}{{\tt 1302.4022}}].

\bibitem{Lee:1966ik}
T.~D. Lee and G.~C. Wick, \emph{{Space Inversion, Time Reversal, and Other
  Discrete Symmetries in Local Field Theories}},
  \href{http://dx.doi.org/10.1103/PhysRev.148.1385}{\emph{Phys. Rev.} {\bf 148}
  (1966) 1385--1404}.

\bibitem{Ecker:1981wv}
G.~Ecker, W.~Grimus and W.~Konetschny, \emph{{Quark Mass Matrices in Left-right
  Symmetric Gauge Theories}},
  \href{http://dx.doi.org/10.1016/0550-3213(81)90309-6}{\emph{Nucl. Phys. B}
  {\bf 191} (1981) 465--492}.

\bibitem{Ecker:1987qp}
G.~Ecker, W.~Grimus and H.~Neufeld, \emph{{A Standard Form for Generalized {CP}
  Transformations}},
  \href{http://dx.doi.org/10.1088/0305-4470/20/12/010}{\emph{J. Phys. A} {\bf
  20} (1987) L807}.

\bibitem{Neufeld:1987wa}
H.~Neufeld, W.~Grimus and G.~Ecker, \emph{{Generalized {CP} Invariance, Neutral
  Flavor Conservation and the Structure of the Mixing Matrix}},
  \href{http://dx.doi.org/10.1142/S0217751X88000254}{\emph{Int. J. Mod. Phys.
  A} {\bf 3} (1988) 603--616}.

\bibitem{Ferreira:2025ymc}
P.~M. Ferreira and T.~F. Pinto, \emph{{One-loop pseudoscalar mass in a 2HDM
  with a $Z_3$ symmetry}},  \href{http://arxiv.org/abs/2504.11602}{{\tt
  2504.11602}}.

\bibitem{Ivanov:2012fp}
I.~P. Ivanov and E.~Vdovin, \emph{{Classification of finite reparametrization
  symmetry groups in the three-Higgs-doublet model}},
  \href{http://dx.doi.org/10.1140/epjc/s10052-013-2309-x}{\emph{Eur. Phys. J.
  C} {\bf 73} (2013) 2309}, [\href{http://arxiv.org/abs/1210.6553}{{\tt
  1210.6553}}].

\bibitem{Pierce:2007ut}
A.~Pierce and J.~Thaler, \emph{{Natural Dark Matter from an Unnatural Higgs
  Boson and New Colored Particles at the TeV Scale}},
  \href{http://dx.doi.org/10.1088/1126-6708/2007/08/026}{\emph{JHEP} {\bf 08}
  (2007) 026}, [\href{http://arxiv.org/abs/hep-ph/0703056}{{\tt
  hep-ph/0703056}}].

\bibitem{Arbey:2017gmh}
A.~Arbey, F.~Mahmoudi, O.~St{\aa}l and T.~Stefaniak, \emph{{Status of the
  Charged Higgs Boson in Two Higgs Doublet Models}},
  \href{http://dx.doi.org/10.1140/epjc/s10052-018-5651-1}{\emph{Eur. Phys. J.
  C} {\bf 78} (2018) 182}, [\href{http://arxiv.org/abs/1706.07414}{{\tt
  1706.07414}}].

\bibitem{Ginzburg:2002wt}
I.~F. Ginzburg, M.~Krawczyk and P.~Osland, \emph{{Two Higgs doublet models with
  CP violation}},  in \emph{{International Workshop on Linear Colliders (LCWS
  2002)}}, pp.~703--706, 11, 2002.
\newblock \href{http://arxiv.org/abs/hep-ph/0211371}{{\tt hep-ph/0211371}}.

\bibitem{Khater:2003wq}
W.~Khater and P.~Osland, \emph{{CP violation in top quark production at the LHC
  and two Higgs doublet models}},
  \href{http://dx.doi.org/10.1016/S0550-3213(03)00300-6}{\emph{Nucl. Phys. B}
  {\bf 661} (2003) 209--234}, [\href{http://arxiv.org/abs/hep-ph/0302004}{{\tt
  hep-ph/0302004}}].

\bibitem{ElKaffas:2006gdt}
A.~W. El~Kaffas, W.~Khater, O.~M. Ogreid and P.~Osland, \emph{{Consistency of
  the two Higgs doublet model and CP violation in top production at the LHC}},
  \href{http://dx.doi.org/10.1016/j.nuclphysb.2007.03.041}{\emph{Nucl. Phys. B}
  {\bf 775} (2007) 45--77}, [\href{http://arxiv.org/abs/hep-ph/0605142}{{\tt
  hep-ph/0605142}}].

\bibitem{Grzadkowski:2009iz}
B.~Grzadkowski and P.~Osland, \emph{{Tempered Two-Higgs-Doublet Model}},
  \href{http://dx.doi.org/10.1103/PhysRevD.82.125026}{\emph{Phys. Rev. D} {\bf
  82} (2010) 125026}, [\href{http://arxiv.org/abs/0910.4068}{{\tt 0910.4068}}].

\bibitem{Arhrib:2010ju}
A.~Arhrib, E.~Christova, H.~Eberl and E.~Ginina, \emph{{CP violation in charged
  Higgs production and decays in the Complex Two Higgs Doublet Model}},
  \href{http://dx.doi.org/10.1007/JHEP04(2011)089}{\emph{JHEP} {\bf 04} (2011)
  089}, [\href{http://arxiv.org/abs/1011.6560}{{\tt 1011.6560}}].

\bibitem{Cirelli:2005uq}
M.~Cirelli, N.~Fornengo and A.~Strumia, \emph{{Minimal dark matter}},
  \href{http://dx.doi.org/10.1016/j.nuclphysb.2006.07.012}{\emph{Nucl. Phys.}
  {\bf B753} (2006) 178--194}, [\href{http://arxiv.org/abs/hep-ph/0512090}{{\tt
  hep-ph/0512090}}].

\bibitem{Cirelli:2007xd}
M.~Cirelli, A.~Strumia and M.~Tamburini, \emph{{Cosmology and Astrophysics of
  Minimal Dark Matter}},
  \href{http://dx.doi.org/10.1016/j.nuclphysb.2007.07.023}{\emph{Nucl. Phys.}
  {\bf B787} (2007) 152--175}, [\href{http://arxiv.org/abs/0706.4071}{{\tt
  0706.4071}}].

\bibitem{Cirelli:2008id}
M.~Cirelli, R.~Franceschini and A.~Strumia, \emph{{Minimal Dark Matter
  predictions for galactic positrons, anti-protons, photons}},
  \href{http://dx.doi.org/10.1016/j.nuclphysb.2008.03.013}{\emph{Nucl. Phys.}
  {\bf B800} (2008) 204--220}, [\href{http://arxiv.org/abs/0802.3378}{{\tt
  0802.3378}}].

\bibitem{Cirelli:2009uv}
M.~Cirelli and A.~Strumia, \emph{{Minimal Dark Matter: Model and results}},
  \href{http://dx.doi.org/10.1088/1367-2630/11/10/105005}{\emph{New J. Phys.}
  {\bf 11} (2009) 105005}, [\href{http://arxiv.org/abs/0903.3381}{{\tt
  0903.3381}}].

\bibitem{Hambye:2009pw}
T.~Hambye, F.~S. Ling, L.~Lopez~Honorez and J.~Rocher, \emph{{Scalar Multiplet
  Dark Matter}}, \href{http://dx.doi.org/10.1007/JHEP05(2010)066,
  10.1088/1126-6708/2009/07/090}{\emph{JHEP} {\bf 07} (2009) 090},
  [\href{http://arxiv.org/abs/0903.4010}{{\tt 0903.4010}}].

\bibitem{Cai:2012kt}
Y.~Cai, W.~Chao and S.~Yang, \emph{{Scalar Septuplet Dark Matter and Enhanced
  $h\rightarrow \gamma\gamma$ Decay Rate}},
  \href{http://dx.doi.org/10.1007/JHEP12(2012)043}{\emph{JHEP} {\bf 12} (2012)
  043}, [\href{http://arxiv.org/abs/1208.3949}{{\tt 1208.3949}}].

\bibitem{Earl:2013jsa}
K.~Earl, K.~Hartling, H.~E. Logan and T.~Pilkington, \emph{{Constraining models
  with a large scalar multiplet}},
  \href{http://dx.doi.org/10.1103/PhysRevD.88.015002}{\emph{Phys. Rev.} {\bf
  D88} (2013) 015002}, [\href{http://arxiv.org/abs/1303.1244}{{\tt
  1303.1244}}].

\bibitem{Garcia-Cely:2015dda}
C.~Garcia-Cely, A.~Ibarra, A.~S. Lamperstorfer and M.~H.~G. Tytgat,
  \emph{{Gamma-rays from Heavy Minimal Dark Matter}},
  \href{http://dx.doi.org/10.1088/1475-7516/2015/10/058}{\emph{JCAP} {\bf 1510}
  (2015) 058}, [\href{http://arxiv.org/abs/1507.05536}{{\tt 1507.05536}}].

\bibitem{DelNobile:2015bqo}
E.~Del~Nobile, M.~Nardecchia and P.~Panci, \emph{{Millicharge or Decay: A
  Critical Take on Minimal Dark Matter}},
  \href{http://dx.doi.org/10.1088/1475-7516/2016/04/048}{\emph{JCAP} {\bf 1604}
  (2016) 048}, [\href{http://arxiv.org/abs/1512.05353}{{\tt 1512.05353}}].

\bibitem{McDonald:1993ex}
J.~McDonald, \emph{{Gauge singlet scalars as cold dark matter}},
  \href{http://dx.doi.org/10.1103/PhysRevD.50.3637}{\emph{Phys. Rev.} {\bf D50}
  (1994) 3637--3649}, [\href{http://arxiv.org/abs/hep-ph/0702143}{{\tt
  hep-ph/0702143}}].

\bibitem{Burgess:2000yq}
C.~P. Burgess, M.~Pospelov and T.~ter Veldhuis, \emph{{The Minimal model of
  nonbaryonic dark matter: A Singlet scalar}},
  \href{http://dx.doi.org/10.1016/S0550-3213(01)00513-2}{\emph{Nucl. Phys.}
  {\bf B619} (2001) 709--728}, [\href{http://arxiv.org/abs/hep-ph/0011335}{{\tt
  hep-ph/0011335}}].

\bibitem{Patt:2006fw}
B.~Patt and F.~Wilczek, \emph{{Higgs-field portal into hidden sectors}},
  \href{http://arxiv.org/abs/hep-ph/0605188}{{\tt hep-ph/0605188}}.

\bibitem{Barger:2007im}
V.~Barger, P.~Langacker, M.~McCaskey, M.~J. Ramsey-Musolf and G.~Shaughnessy,
  \emph{{LHC Phenomenology of an Extended Standard Model with a Real Scalar
  Singlet}}, \href{http://dx.doi.org/10.1103/PhysRevD.77.035005}{\emph{Phys.
  Rev.} {\bf D77} (2008) 035005}, [\href{http://arxiv.org/abs/0706.4311}{{\tt
  0706.4311}}].

\bibitem{Andreas:2008xy}
S.~Andreas, T.~Hambye and M.~H.~G. Tytgat, \emph{{WIMP dark matter, Higgs
  exchange and DAMA}},
  \href{http://dx.doi.org/10.1088/1475-7516/2008/10/034}{\emph{JCAP} {\bf 0810}
  (2008) 034}, [\href{http://arxiv.org/abs/0808.0255}{{\tt 0808.0255}}].

\bibitem{Cline:2013gha}
J.~M. Cline, K.~Kainulainen, P.~Scott and C.~Weniger, \emph{{Update on scalar
  singlet dark matter}}, \href{http://dx.doi.org/10.1103/PhysRevD.92.039906,
  10.1103/PhysRevD.88.055025}{\emph{Phys. Rev.} {\bf D88} (2013) 055025},
  [\href{http://arxiv.org/abs/1306.4710}{{\tt 1306.4710}}].

\bibitem{Feng:2014vea}
L.~Feng, S.~Profumo and L.~Ubaldi, \emph{{Closing in on singlet scalar dark
  matter: LUX, invisible Higgs decays and gamma-ray lines}},
  \href{http://dx.doi.org/10.1007/JHEP03(2015)045}{\emph{JHEP} {\bf 03} (2015)
  045}, [\href{http://arxiv.org/abs/1412.1105}{{\tt 1412.1105}}].

\bibitem{Beniwal:2015sdl}
A.~Beniwal, F.~Rajec, C.~Savage, P.~Scott, C.~Weniger, M.~White et~al.,
  \emph{{Combined analysis of effective Higgs portal dark matter models}},
  \href{http://dx.doi.org/10.1103/PhysRevD.93.115016}{\emph{Phys. Rev.} {\bf
  D93} (2016) 115016}, [\href{http://arxiv.org/abs/1512.06458}{{\tt
  1512.06458}}].

\bibitem{Cuoco:2016jqt}
A.~Cuoco, B.~Eiteneuer, J.~Heisig and M.~Krämer, \emph{{A global fit of the
  $\gamma$-ray galactic center excess within the scalar singlet Higgs portal
  model}}, \href{http://dx.doi.org/10.1088/1475-7516/2016/06/050}{\emph{JCAP}
  {\bf 1606} (2016) 050}, [\href{http://arxiv.org/abs/1603.08228}{{\tt
  1603.08228}}].

\bibitem{He:2016mls}
X.-G. He and J.~Tandean, \emph{{New LUX and PandaX-II Results Illuminating the
  Simplest Higgs-Portal Dark Matter Models}},
  \href{http://dx.doi.org/10.1007/JHEP12(2016)074}{\emph{JHEP} {\bf 12} (2016)
  074}, [\href{http://arxiv.org/abs/1609.03551}{{\tt 1609.03551}}].

\bibitem{Casas:2017jjg}
J.~A. Casas, D.~G. Cerdeño, J.~M. Moreno and J.~Quilis, \emph{{Reopening the
  Higgs portal for single scalar dark matter}},
  \href{http://dx.doi.org/10.1007/JHEP05(2017)036}{\emph{JHEP} {\bf 05} (2017)
  036}, [\href{http://arxiv.org/abs/1701.08134}{{\tt 1701.08134}}].

\bibitem{Athron:2017kgt}
P.~Athron et~al., \emph{{Status of the scalar singlet dark matter model}},
  \href{http://dx.doi.org/10.1140/epjc/s10052-017-5113-1}{\emph{Eur. Phys. J.}
  {\bf C77} (2017) 568}, [\href{http://arxiv.org/abs/1705.07931}{{\tt
  1705.07931}}].

\bibitem{Belanger:2012zr}
G.~B\'elanger, K.~Kannike, A.~Pukhov and M.~Raidal, \emph{{$Z_3$ Scalar Singlet
  Dark Matter}},
  \href{http://dx.doi.org/10.1088/1475-7516/2013/01/022}{\emph{JCAP} {\bf 1301}
  (2013) 022}, [\href{http://arxiv.org/abs/1211.1014}{{\tt 1211.1014}}].

\bibitem{Ko:2014nha}
P.~Ko and Y.~Tang, \emph{{Self-interacting scalar dark matter with local $Z_3$
  symmetry}},
  \href{http://dx.doi.org/10.1088/1475-7516/2014/05/047}{\emph{JCAP} {\bf 1405}
  (2014) 047}, [\href{http://arxiv.org/abs/1402.6449}{{\tt 1402.6449}}].

\bibitem{Majumdar:2006nt}
D.~Majumdar and A.~Ghosal, \emph{{Dark Matter candidate in a Heavy Higgs Model
  - Direct Detection Rates}},
  \href{http://dx.doi.org/10.1142/S0217732308025954}{\emph{Mod. Phys. Lett.}
  {\bf A23} (2008) 2011--2022},
  [\href{http://arxiv.org/abs/hep-ph/0607067}{{\tt hep-ph/0607067}}].

\bibitem{Gustafsson:2007pc}
M.~Gustafsson, E.~Lundstrom, L.~Bergstrom and J.~Edsjo, \emph{{Significant
  Gamma Lines from Inert Higgs Dark Matter}},
  \href{http://dx.doi.org/10.1103/PhysRevLett.99.041301}{\emph{Phys. Rev.
  Lett.} {\bf 99} (2007) 041301},
  [\href{http://arxiv.org/abs/astro-ph/0703512}{{\tt astro-ph/0703512}}].

\bibitem{Cao:2007rm}
Q.-H. Cao, E.~Ma and G.~Rajasekaran, \emph{{Observing the Dark Scalar Doublet
  and its Impact on the Standard-Model Higgs Boson at Colliders}},
  \href{http://dx.doi.org/10.1103/PhysRevD.76.095011}{\emph{Phys. Rev.} {\bf
  D76} (2007) 095011}, [\href{http://arxiv.org/abs/0708.2939}{{\tt
  0708.2939}}].

\bibitem{Lundstrom:2008ai}
E.~Lundstrom, M.~Gustafsson and J.~Edsjo, \emph{{The Inert Doublet Model and
  LEP II Limits}},
  \href{http://dx.doi.org/10.1103/PhysRevD.79.035013}{\emph{Phys. Rev.} {\bf
  D79} (2009) 035013}, [\href{http://arxiv.org/abs/0810.3924}{{\tt
  0810.3924}}].

\bibitem{Agrawal:2008xz}
P.~Agrawal, E.~M. Dolle and C.~A. Krenke, \emph{{Signals of Inert Doublet Dark
  Matter in Neutrino Telescopes}},
  \href{http://dx.doi.org/10.1103/PhysRevD.79.015015}{\emph{Phys. Rev.} {\bf
  D79} (2009) 015015}, [\href{http://arxiv.org/abs/0811.1798}{{\tt
  0811.1798}}].

\bibitem{Nezri:2009jd}
E.~Nezri, M.~H.~G. Tytgat and G.~Vertongen, \emph{{e+ and anti-p from inert
  doublet model dark matter}},
  \href{http://dx.doi.org/10.1088/1475-7516/2009/04/014}{\emph{JCAP} {\bf 0904}
  (2009) 014}, [\href{http://arxiv.org/abs/0901.2556}{{\tt 0901.2556}}].

\bibitem{Dolle:2009fn}
E.~M. Dolle and S.~Su, \emph{{The Inert Dark Matter}},
  \href{http://dx.doi.org/10.1103/PhysRevD.80.055012}{\emph{Phys. Rev.} {\bf
  D80} (2009) 055012}, [\href{http://arxiv.org/abs/0906.1609}{{\tt
  0906.1609}}].

\bibitem{Arina:2009um}
C.~Arina, F.-S. Ling and M.~H.~G. Tytgat, \emph{{IDM and iDM or The Inert
  Doublet Model and Inelastic Dark Matter}},
  \href{http://dx.doi.org/10.1088/1475-7516/2009/10/018}{\emph{JCAP} {\bf 10}
  (2009) 018}, [\href{http://arxiv.org/abs/0907.0430}{{\tt 0907.0430}}].

\bibitem{Dolle:2009ft}
E.~Dolle, X.~Miao, S.~Su and B.~Thomas, \emph{{Dilepton Signals in the Inert
  Doublet Model}},
  \href{http://dx.doi.org/10.1103/PhysRevD.81.035003}{\emph{Phys. Rev.} {\bf
  D81} (2010) 035003}, [\href{http://arxiv.org/abs/0909.3094}{{\tt
  0909.3094}}].

\bibitem{Honorez:2010re}
L.~Lopez~Honorez and C.~E. Yaguna, \emph{{The inert doublet model of dark
  matter revisited}},
  \href{http://dx.doi.org/10.1007/JHEP09(2010)046}{\emph{JHEP} {\bf 09} (2010)
  046}, [\href{http://arxiv.org/abs/1003.3125}{{\tt 1003.3125}}].

\bibitem{Miao:2010rg}
X.~Miao, S.~Su and B.~Thomas, \emph{{Trilepton Signals in the Inert Doublet
  Model}}, \href{http://dx.doi.org/10.1103/PhysRevD.82.035009}{\emph{Phys.
  Rev.} {\bf D82} (2010) 035009}, [\href{http://arxiv.org/abs/1005.0090}{{\tt
  1005.0090}}].

\bibitem{LopezHonorez:2010tb}
L.~Lopez~Honorez and C.~E. Yaguna, \emph{{A new viable region of the inert
  doublet model}},
  \href{http://dx.doi.org/10.1088/1475-7516/2011/01/002}{\emph{JCAP} {\bf 1101}
  (2011) 002}, [\href{http://arxiv.org/abs/1011.1411}{{\tt 1011.1411}}].

\bibitem{Sokolowska:2011sb}
D.~Sokolowska, \emph{{Dark Matter Data and Constraints on Quartic Couplings in
  IDM}},  \href{http://arxiv.org/abs/1107.1991}{{\tt 1107.1991}}.

\bibitem{Swiezewska:2012eh}
B.~Swiezewska and M.~Krawczyk, \emph{{Diphoton rate in the inert doublet model
  with a 125 GeV Higgs boson}},
  \href{http://dx.doi.org/10.1103/PhysRevD.88.035019}{\emph{Phys. Rev.} {\bf
  D88} (2013) 035019}, [\href{http://arxiv.org/abs/1212.4100}{{\tt
  1212.4100}}].

\bibitem{Goudelis:2013uca}
A.~Goudelis, B.~Herrmann and O.~St{\aa}l, \emph{{Dark matter in the Inert
  Doublet Model after the discovery of a Higgs-like boson at the LHC}},
  \href{http://dx.doi.org/10.1007/JHEP09(2013)106}{\emph{JHEP} {\bf 09} (2013)
  106}, [\href{http://arxiv.org/abs/1303.3010}{{\tt 1303.3010}}].

\bibitem{Krawczyk:2013jta}
M.~Krawczyk, D.~Sokolowska, P.~Swaczyna and B.~Swiezewska, \emph{{Constraining
  Inert Dark Matter by $R_{\gamma\gamma}$ and WMAP data}},
  \href{http://dx.doi.org/10.1007/JHEP09(2013)055}{\emph{JHEP} {\bf 09} (2013)
  055}, [\href{http://arxiv.org/abs/1305.6266}{{\tt 1305.6266}}].

\bibitem{Arhrib:2013ela}
A.~Arhrib, Y.-L.~S. Tsai, Q.~Yuan and T.-C. Yuan, \emph{{An Updated Analysis of
  Inert Higgs Doublet Model in light of the Recent Results from LUX, PLANCK,
  AMS-02 and LHC}},
  \href{http://dx.doi.org/10.1088/1475-7516/2014/06/030}{\emph{JCAP} {\bf 1406}
  (2014) 030}, [\href{http://arxiv.org/abs/1310.0358}{{\tt 1310.0358}}].

\bibitem{Ilnicka:2015jba}
A.~Ilnicka, M.~Krawczyk and T.~Robens, \emph{{Inert Doublet Model in light of
  LHC Run I and astrophysical data}},
  \href{http://dx.doi.org/10.1103/PhysRevD.93.055026}{\emph{Phys. Rev.} {\bf
  D93} (2016) 055026}, [\href{http://arxiv.org/abs/1508.01671}{{\tt
  1508.01671}}].

\bibitem{Diaz:2015pyv}
M.~A. Díaz, B.~Koch and S.~Urrutia-Quiroga, \emph{{Constraints to Dark Matter
  from Inert Higgs Doublet Model}},
  \href{http://dx.doi.org/10.1155/2016/8278375}{\emph{Adv. High Energy Phys.}
  {\bf 2016} (2016) 8278375}, [\href{http://arxiv.org/abs/1511.04429}{{\tt
  1511.04429}}].

\bibitem{Aoki:2013lhm}
M.~Aoki, S.~Kanemura and H.~Yokoya, \emph{{Reconstruction of Inert Doublet
  Scalars at the International Linear Collider}},
  \href{http://dx.doi.org/10.1016/j.physletb.2013.07.011}{\emph{Phys. Lett.}
  {\bf B725} (2013) 302--309}, [\href{http://arxiv.org/abs/1303.6191}{{\tt
  1303.6191}}].

\bibitem{Hashemi:2015swh}
M.~Hashemi, M.~Krawczyk, S.~Najjari and A.~F. Żarnecki, \emph{{Production of
  Inert Scalars at the high energy $e^+ e^-$ colliders}},
  \href{http://arxiv.org/abs/1512.01175}{{\tt 1512.01175}}.

\bibitem{Kalinowski:2018kdn}
J.~Kalinowski, W.~Kotlarski, T.~Robens, D.~Sokolowska and A.~F. Zarnecki,
  \emph{{Exploring Inert Scalars at CLIC}},
  \href{http://dx.doi.org/10.1007/JHEP07(2019)053}{\emph{JHEP} {\bf 07} (2019)
  053}, [\href{http://arxiv.org/abs/1811.06952}{{\tt 1811.06952}}].

\bibitem{Dercks:2018wch}
D.~Dercks and T.~Robens, \emph{{Constraining the Inert Doublet Model using
  Vector Boson Fusion}},
  \href{http://dx.doi.org/10.1140/epjc/s10052-019-7436-6}{\emph{Eur. Phys. J.
  C} {\bf 79} (2019) 924}, [\href{http://arxiv.org/abs/1812.07913}{{\tt
  1812.07913}}].

\bibitem{Braathen:2024lyl}
J.~Braathen, M.~Gabelmann, T.~Robens and P.~Stylianou, \emph{{Probing the Inert
  Doublet Model via Vector-Boson Fusion at a Muon Collider}},
  \href{http://arxiv.org/abs/2411.13729}{{\tt 2411.13729}}.

\bibitem{XENON:2022ltv}
E.~Aprile et~al., \emph{{Search for New Physics in Electronic Recoil Data from
  XENONnT}},
  \href{http://dx.doi.org/10.1103/PhysRevLett.129.161805}{\emph{Phys. Rev.
  Lett.} {\bf 129} (2022) 161805}, [\href{http://arxiv.org/abs/2207.11330}{{\tt
  2207.11330}}].

\bibitem{XENON:2023cxc}
E.~Aprile et~al., \emph{{First Dark Matter Search with Nuclear Recoils from the
  XENONnT Experiment}},
  \href{http://dx.doi.org/10.1103/PhysRevLett.131.041003}{\emph{Phys. Rev.
  Lett.} {\bf 131} (2023) 041003}, [\href{http://arxiv.org/abs/2303.14729}{{\tt
  2303.14729}}].

\bibitem{LZ:2021xov}
D.~S. Akerib et~al., \emph{{Projected sensitivities of the LUX-ZEPLIN
  experiment to new physics via low-energy electron recoils}},
  \href{http://dx.doi.org/10.1103/PhysRevD.104.092009}{\emph{Phys. Rev. D} {\bf
  104} (2021) 092009}, [\href{http://arxiv.org/abs/2102.11740}{{\tt
  2102.11740}}].

\bibitem{LZ:2022lsv}
J.~Aalbers et~al., \emph{{First Dark Matter Search Results from the LUX-ZEPLIN
  (LZ) Experiment}},
  \href{http://dx.doi.org/10.1103/PhysRevLett.131.041002}{\emph{Phys. Rev.
  Lett.} {\bf 131} (2023) 041002}, [\href{http://arxiv.org/abs/2207.03764}{{\tt
  2207.03764}}].

\bibitem{Escudero:2016gzx}
M.~Escudero, A.~Berlin, D.~Hooper and M.-X. Lin, \emph{{Toward (Finally!)
  Ruling Out Z and Higgs Mediated Dark Matter Models}},
  \href{http://dx.doi.org/10.1088/1475-7516/2016/12/029}{\emph{JCAP} {\bf 12}
  (2016) 029}, [\href{http://arxiv.org/abs/1609.09079}{{\tt 1609.09079}}].

\bibitem{Alves:2016bib}
A.~Alves, D.~A. Camargo, A.~G. Dias, R.~Longas, C.~C. Nishi and F.~S. Queiroz,
  \emph{{Collider and Dark Matter Searches in the Inert Doublet Model from
  Peccei-Quinn Symmetry}},
  \href{http://dx.doi.org/10.1007/JHEP10(2016)015}{\emph{JHEP} {\bf 10} (2016)
  015}, [\href{http://arxiv.org/abs/1606.07086}{{\tt 1606.07086}}].

\bibitem{Jueid:2020rek}
A.~Jueid, J.~Kim, S.~Lee, S.~Y. Shim and J.~Song, \emph{{Phenomenology of the
  Inert Doublet Model with a global U(1) symmetry}},
  \href{http://dx.doi.org/10.1103/PhysRevD.102.075011}{\emph{Phys. Rev. D} {\bf
  102} (2020) 075011}, [\href{http://arxiv.org/abs/2006.10263}{{\tt
  2006.10263}}].

\bibitem{Koide:1982ax}
Y.~Koide, \emph{{A Fermion - Boson Composite Model of Quarks and Leptons}},
  \href{http://dx.doi.org/10.1016/0370-2693(83)90644-5}{\emph{Phys. Lett. B}
  {\bf 120} (1983) 161--165}.

\bibitem{Koide:1983qe}
Y.~Koide, \emph{{A New View of Quark and Lepton Mass Hierarchy}},
  \href{http://dx.doi.org/10.1103/PhysRevD.28.252}{\emph{Phys. Rev. D} {\bf 28}
  (1983) 252}.

\bibitem{Sakharov:1967dj}
A.~D. Sakharov, \emph{{Violation of CP Invariance, C asymmetry, and baryon
  asymmetry of the universe}},
  \href{http://dx.doi.org/10.1070/PU1991v034n05ABEH002497}{\emph{Pisma Zh.
  Eksp. Teor. Fiz.} {\bf 5} (1967) 32--35}.

\bibitem{Kuzmin:1985mm}
V.~A. Kuzmin, V.~A. Rubakov and M.~E. Shaposhnikov, \emph{{On the Anomalous
  Electroweak Baryon Number Nonconservation in the Early Universe}},
  \href{http://dx.doi.org/10.1016/0370-2693(85)91028-7}{\emph{Phys. Lett. B}
  {\bf 155} (1985) 36}.

\bibitem{Huet:1994jb}
P.~Huet and E.~Sather, \emph{{Electroweak baryogenesis and standard model CP
  violation}}, \href{http://dx.doi.org/10.1103/PhysRevD.51.379}{\emph{Phys.
  Rev. D} {\bf 51} (1995) 379--394},
  [\href{http://arxiv.org/abs/hep-ph/9404302}{{\tt hep-ph/9404302}}].

\bibitem{Graverini:2018riw}
E.~Graverini, \emph{{Flavour anomalies: a review}},
  \href{http://dx.doi.org/10.1088/1742-6596/1137/1/012025}{\emph{J. Phys. Conf.
  Ser.} {\bf 1137} (2019) 012025}, [\href{http://arxiv.org/abs/1807.11373}{{\tt
  1807.11373}}].

\bibitem{Belle-II:2018jsg}
W.~Altmannshofer et~al., \emph{{The Belle II Physics Book}},
  \href{http://arxiv.org/abs/1808.10567}{{\tt 1808.10567}}.

\bibitem{Grzadkowski:2009bt}
B.~Grzadkowski, O.~M. Ogreid and P.~Osland, \emph{{Natural Multi-Higgs Model
  with Dark Matter and CP Violation}},
  \href{http://dx.doi.org/10.1103/PhysRevD.80.055013}{\emph{Phys. Rev.} {\bf
  D80} (2009) 055013}, [\href{http://arxiv.org/abs/0904.2173}{{\tt
  0904.2173}}].

\bibitem{Grzadkowski:2010au}
B.~Grzadkowski, O.~M. Ogreid, P.~Osland, A.~Pukhov and M.~Purmohammadi,
  \emph{{Exploring the CP-Violating Inert-Doublet Model}},
  \href{http://dx.doi.org/10.1007/JHEP06(2011)003}{\emph{JHEP} {\bf 06} (2011)
  003}, [\href{http://arxiv.org/abs/1012.4680}{{\tt 1012.4680}}].

\bibitem{Azevedo:2018fmj}
D.~Azevedo, P.~M. Ferreira, M.~M. Muhlleitner, S.~Patel, R.~Santos and
  J.~Wittbrodt, \emph{{CP in the dark}},
  \href{http://dx.doi.org/10.1007/JHEP11(2018)091}{\emph{JHEP} {\bf 11} (2018)
  091}, [\href{http://arxiv.org/abs/1807.10322}{{\tt 1807.10322}}].

\bibitem{Ivanov:2018srm}
I.~P. Ivanov and M.~Laletin, \emph{{Dark matter from CP symmetry of order 4:
  evolution in the asymmetric regime}},
  \href{http://dx.doi.org/10.1088/1475-7516/2019/05/032}{\emph{JCAP} {\bf 05}
  (2019) 032}, [\href{http://arxiv.org/abs/1812.05525}{{\tt 1812.05525}}].

\bibitem{Biermann:2022meg}
L.~Biermann, M.~M\"uhlleitner and J.~M\"uller, \emph{{Electroweak Phase
  Transition in a Dark Sector with CP Violation}},
  \href{http://arxiv.org/abs/2204.13425}{{\tt 2204.13425}}.

\bibitem{Osland:2013sla}
P.~Osland, A.~Pukhov, G.~M. Pruna and M.~Purmohammadi, \emph{{Phenomenology of
  charged scalars in the CP-Violating Inert-Doublet Model}},
  \href{http://dx.doi.org/10.1007/JHEP04(2013)040}{\emph{JHEP} {\bf 04} (2013)
  040}, [\href{http://arxiv.org/abs/1302.3713}{{\tt 1302.3713}}].

\bibitem{Weinberg:1976hu}
S.~Weinberg, \emph{{Gauge Theory of CP Violation}},
  \href{http://dx.doi.org/10.1103/PhysRevLett.37.657}{\emph{Phys. Rev. Lett.}
  {\bf 37} (1976) 657}.

\bibitem{Deshpande:1976yp}
N.~G. Deshpande and E.~Ma, \emph{{Comment on Weinberg's Gauge Theory of CP
  Nonconservation}},
  \href{http://dx.doi.org/10.1103/PhysRevD.16.1583}{\emph{Phys. Rev. D} {\bf
  16} (1977) 1583}.

\bibitem{Gatto:1979mr}
R.~Gatto, G.~Morchio, G.~Sartori and F.~Strocchi, \emph{{Natural Flavor
  Conservation in Higgs Induced Neutral Currents and the Quark Mixing Angles}},
  \href{http://dx.doi.org/10.1016/0550-3213(80)90399-5}{\emph{Nucl. Phys. B}
  {\bf 163} (1980) 221--253}.

\bibitem{Branco:1979pv}
G.~C. Branco, \emph{{Spontaneous CP Violation in Theories with More Than Four
  Quarks}}, \href{http://dx.doi.org/10.1103/PhysRevLett.44.504}{\emph{Phys.
  Rev. Lett.} {\bf 44} (1980) 504}.

\bibitem{deAdelhartToorop:2010jxh}
R.~de~Adelhart~Toorop, F.~Bazzocchi, L.~Merlo and A.~Paris, \emph{{Constraining
  Flavour Symmetries At The EW Scale I: The A4 Higgs Potential}},
  \href{http://dx.doi.org/10.1007/JHEP03(2011)035}{\emph{JHEP} {\bf 03} (2011)
  035}, [\href{http://arxiv.org/abs/1012.1791}{{\tt 1012.1791}}].

\bibitem{Degee:2012sk}
A.~Degee, I.~P. Ivanov and V.~Keus, \emph{{Geometric minimization of highly
  symmetric potentials}},
  \href{http://dx.doi.org/10.1007/JHEP02(2013)125}{\emph{JHEP} {\bf 02} (2013)
  125}, [\href{http://arxiv.org/abs/1211.4989}{{\tt 1211.4989}}].

\bibitem{GonzalezFelipe:2013xok}
R.~Gonz\'alez~Felipe, H.~Ser\^odio and J.~P. Silva, \emph{{Models with three
  Higgs doublets in the triplet representations of $A_{4}$ or $S_{4}$}},
  \href{http://dx.doi.org/10.1103/PhysRevD.87.055010}{\emph{Phys. Rev. D} {\bf
  87} (2013) 055010}, [\href{http://arxiv.org/abs/1302.0861}{{\tt 1302.0861}}].

\bibitem{Branco:1983tn}
G.~C. Branco, J.~M. Gerard and W.~Grimus, \emph{{Geometrical T violation}},
  \href{http://dx.doi.org/10.1016/0370-2693(84)92024-0}{\emph{Phys. Lett. B}
  {\bf 136} (1984) 383--386}.

\bibitem{deMedeirosVarzielas:2011zw}
I.~de~Medeiros~Varzielas and D.~Emmanuel-Costa, \emph{{Geometrical CP
  Violation}}, \href{http://dx.doi.org/10.1103/PhysRevD.84.117901}{\emph{Phys.
  Rev. D} {\bf 84} (2011) 117901}, [\href{http://arxiv.org/abs/1106.5477}{{\tt
  1106.5477}}].

\bibitem{deMedeirosVarzielas:2012ylr}
I.~de~Medeiros~Varzielas, D.~Emmanuel-Costa and P.~Leser, \emph{{Geometrical CP
  Violation from Non-Renormalisable Scalar Potentials}},
  \href{http://dx.doi.org/10.1016/j.physletb.2012.08.008}{\emph{Phys. Lett. B}
  {\bf 716} (2012) 193--196}, [\href{http://arxiv.org/abs/1204.3633}{{\tt
  1204.3633}}].

\bibitem{Ma:2013xqa}
E.~Ma, \emph{{Neutrino Mixing and Geometric CP Violation with Delta(27)
  Symmetry}},
  \href{http://dx.doi.org/10.1016/j.physletb.2013.05.011}{\emph{Phys. Lett. B}
  {\bf 723} (2013) 161--163}, [\href{http://arxiv.org/abs/1304.1603}{{\tt
  1304.1603}}].

\bibitem{Nishi:2013jqa}
C.~C. Nishi, \emph{{Generalized $CP$ symmetries in $\Delta(27)$ flavor
  models}}, \href{http://dx.doi.org/10.1103/PhysRevD.88.033010}{\emph{Phys.
  Rev. D} {\bf 88} (2013) 033010}, [\href{http://arxiv.org/abs/1306.0877}{{\tt
  1306.0877}}].

\bibitem{Fallbacher:2015rea}
M.~Fallbacher and A.~Trautner, \emph{{Symmetries of symmetries and geometrical
  CP violation}},
  \href{http://dx.doi.org/10.1016/j.nuclphysb.2015.03.003}{\emph{Nucl. Phys. B}
  {\bf 894} (2015) 136--160}, [\href{http://arxiv.org/abs/1502.01829}{{\tt
  1502.01829}}].

\bibitem{Ivanov:2015mwl}
I.~P. Ivanov and J.~P. Silva, \emph{{$CP$-conserving multi-Higgs model with
  irremovable complex coefficients}},
  \href{http://dx.doi.org/10.1103/PhysRevD.93.095014}{\emph{Phys. Rev. D} {\bf
  93} (2016) 095014}, [\href{http://arxiv.org/abs/1512.09276}{{\tt
  1512.09276}}].

\bibitem{Ferreira:2017tvy}
P.~M. Ferreira, I.~P. Ivanov, E.~Jim\'enez, R.~Pasechnik and H.~Ser\^odio,
  \emph{{CP4 miracle: shaping Yukawa sector with CP symmetry of order four}},
  \href{http://dx.doi.org/10.1007/JHEP01(2018)065}{\emph{JHEP} {\bf 01} (2018)
  065}, [\href{http://arxiv.org/abs/1711.02042}{{\tt 1711.02042}}].

\bibitem{Ivanov:2018ime}
I.~P. Ivanov, C.~C. Nishi, J.~P. Silva and A.~Trautner, \emph{{Basis-invariant
  conditions for $CP$ symmetry of order four}},
  \href{http://dx.doi.org/10.1103/PhysRevD.99.015039}{\emph{Phys. Rev. D} {\bf
  99} (2019) 015039}, [\href{http://arxiv.org/abs/1810.13396}{{\tt
  1810.13396}}].

\bibitem{Ivanov:2021pnr}
I.~P. Ivanov and S.~A. Obodenko, \emph{{Constraining CP4 3HDM with Top Quark
  Decays}}, \href{http://dx.doi.org/10.3390/universe7060197}{\emph{Universe}
  {\bf 7} (2021) 197}, [\href{http://arxiv.org/abs/2104.11440}{{\tt
  2104.11440}}].

\bibitem{Ferreira:2008zy}
P.~M. Ferreira and J.~P. Silva, \emph{{Discrete and continuous symmetries in
  multi-Higgs-doublet models}},
  \href{http://dx.doi.org/10.1103/PhysRevD.78.116007}{\emph{Phys. Rev. D} {\bf
  78} (2008) 116007}, [\href{http://arxiv.org/abs/0809.2788}{{\tt 0809.2788}}].

\bibitem{Ivanov:2011ae}
I.~P. Ivanov, V.~Keus and E.~Vdovin, \emph{{Abelian symmetries in
  multi-Higgs-doublet models}},
  \href{http://dx.doi.org/10.1088/1751-8113/45/21/215201}{\emph{J. Phys. A}
  {\bf 45} (2012) 215201}, [\href{http://arxiv.org/abs/1112.1660}{{\tt
  1112.1660}}].

\bibitem{Ivanov:2012ry}
I.~P. Ivanov and E.~Vdovin, \emph{{Discrete symmetries in the
  three-Higgs-doublet model}},
  \href{http://dx.doi.org/10.1103/PhysRevD.86.095030}{\emph{Phys. Rev. D} {\bf
  86} (2012) 095030}, [\href{http://arxiv.org/abs/1206.7108}{{\tt 1206.7108}}].

\bibitem{Keus:2013hya}
V.~Keus, S.~F. King and S.~Moretti, \emph{{Three-Higgs-doublet models:
  symmetries, potentials and Higgs boson masses}},
  \href{http://dx.doi.org/10.1007/JHEP01(2014)052}{\emph{JHEP} {\bf 01} (2014)
  052}, [\href{http://arxiv.org/abs/1310.8253}{{\tt 1310.8253}}].

\bibitem{Ivanov:2014doa}
I.~P. Ivanov and C.~C. Nishi, \emph{{Symmetry breaking patterns in 3HDM}},
  \href{http://dx.doi.org/10.1007/JHEP01(2015)021}{\emph{JHEP} {\bf 01} (2015)
  021}, [\href{http://arxiv.org/abs/1410.6139}{{\tt 1410.6139}}].

\bibitem{deMedeirosVarzielas:2019rrp}
I.~de~Medeiros~Varzielas and I.~P. Ivanov, \emph{{Recognizing symmetries in a
  3HDM in a basis-independent way}},
  \href{http://dx.doi.org/10.1103/PhysRevD.100.015008}{\emph{Phys. Rev. D} {\bf
  100} (2019) 015008}, [\href{http://arxiv.org/abs/1903.11110}{{\tt
  1903.11110}}].

\bibitem{Bree:2024edl}
I.~Br\'ee, D.~D. Correia and J.~P. Silva, \emph{{Generalized CP symmetries in
  three-Higgs-doublet models}},
  \href{http://dx.doi.org/10.1103/PhysRevD.110.035028}{\emph{Phys. Rev. D} {\bf
  110} (2024) 035028}, [\href{http://arxiv.org/abs/2407.09615}{{\tt
  2407.09615}}].

\bibitem{Pakvasa:1977in}
S.~Pakvasa and H.~Sugawara, \emph{{Discrete Symmetry and Cabibbo Angle}},
  \href{http://dx.doi.org/10.1016/0370-2693(78)90172-7}{\emph{Phys. Lett. B}
  {\bf 73} (1978) 61--64}.

\bibitem{Kuncinas:2020wrn}
A.~Kun\v{c}inas, O.~Ogreid, P.~Osland and M.~Rebelo, \emph{{$S_3$-inspired
  three-Higgs-doublet models: A class with a complex vacuum}},
  \href{http://dx.doi.org/10.1103/PhysRevD.101.075052}{\emph{Phys. Rev. D} {\bf
  101} (2020) 075052}, [\href{http://arxiv.org/abs/2001.01994}{{\tt
  2001.01994}}].

\bibitem{Kubo:2004ps}
J.~Kubo, H.~Okada and F.~Sakamaki, \emph{{Higgs potential in minimal S(3)
  invariant extension of the standard model}},
  \href{http://dx.doi.org/10.1103/PhysRevD.70.036007}{\emph{Phys. Rev.} {\bf
  D70} (2004) 036007}, [\href{http://arxiv.org/abs/hep-ph/0402089}{{\tt
  hep-ph/0402089}}].

\bibitem{Teshima:2012cg}
T.~Teshima, \emph{{Higgs potential in $S_3$ invariant model for quark/lepton
  mass and mixing}},
  \href{http://dx.doi.org/10.1103/PhysRevD.85.105013}{\emph{Phys. Rev.} {\bf
  D85} (2012) 105013}, [\href{http://arxiv.org/abs/1202.4528}{{\tt
  1202.4528}}].

\bibitem{Zhao:2023hws}
D.~Zhao, I.~P. Ivanov, R.~Pasechnik and P.~Zhang, \emph{{Limiting FCNC induced
  by a CP symmetry of order 4}},
  \href{http://dx.doi.org/10.1007/JHEP04(2023)116}{\emph{JHEP} {\bf 04} (2023)
  116}, [\href{http://arxiv.org/abs/2302.03094}{{\tt 2302.03094}}].

\bibitem{Liu:2024aew}
B.~Liu, I.~P. Ivanov and J.~a. Gon\c{c}alves, \emph{{Tridiagonal scalar mass
  matrix in the CP4 3HDM and its implications}},
  \href{http://arxiv.org/abs/2409.05992}{{\tt 2409.05992}}.

\bibitem{Mendez:1991gp}
A.~Mendez and A.~Pomarol, \emph{{Signals of CP violation in the Higgs sector}},
  \href{http://dx.doi.org/10.1016/0370-2693(91)91836-K}{\emph{Phys. Lett. B}
  {\bf 272} (1991) 313--318}.

\bibitem{Lavoura:1994fv}
L.~Lavoura and J.~P. Silva, \emph{{Fundamental CP violating quantities in a
  SU(2) x U(1) model with many Higgs doublets}},
  \href{http://dx.doi.org/10.1103/PhysRevD.50.4619}{\emph{Phys. Rev. D} {\bf
  50} (1994) 4619--4624}, [\href{http://arxiv.org/abs/hep-ph/9404276}{{\tt
  hep-ph/9404276}}].

\bibitem{Branco:2005em}
G.~C. Branco, M.~N. Rebelo and J.~I. Silva-Marcos, \emph{{CP-odd invariants in
  models with several Higgs doublets}},
  \href{http://dx.doi.org/10.1016/j.physletb.2005.03.075}{\emph{Phys. Lett. B}
  {\bf 614} (2005) 187--194}, [\href{http://arxiv.org/abs/hep-ph/0502118}{{\tt
  hep-ph/0502118}}].

\bibitem{Gunion:2005ja}
J.~F. Gunion and H.~E. Haber, \emph{{Conditions for CP-violation in the general
  two-Higgs-doublet model}},
  \href{http://dx.doi.org/10.1103/PhysRevD.72.095002}{\emph{Phys. Rev. D} {\bf
  72} (2005) 095002}, [\href{http://arxiv.org/abs/hep-ph/0506227}{{\tt
  hep-ph/0506227}}].

\bibitem{deMedeirosVarzielas:2016rii}
I.~de~Medeiros~Varzielas, S.~F. King, C.~Luhn and T.~Neder, \emph{{CP-odd
  invariants for multi-Higgs models: applications with discrete symmetry}},
  \href{http://dx.doi.org/10.1103/PhysRevD.94.056007}{\emph{Phys. Rev. D} {\bf
  94} (2016) 056007}, [\href{http://arxiv.org/abs/1603.06942}{{\tt
  1603.06942}}].

\bibitem{Ogreid:2017alh}
O.~M. Ogreid, P.~Osland and M.~N. Rebelo, \emph{{A Simple Method to detect
  spontaneous CP Violation in multi-Higgs models}},
  \href{http://dx.doi.org/10.1007/JHEP08(2017)005}{\emph{JHEP} {\bf 08} (2017)
  005}, [\href{http://arxiv.org/abs/1701.04768}{{\tt 1701.04768}}].

\bibitem{Trautner:2018ipq}
A.~Trautner, \emph{{Systematic construction of basis invariants in the 2HDM}},
  \href{http://dx.doi.org/10.1007/JHEP05(2019)208}{\emph{JHEP} {\bf 05} (2019)
  208}, [\href{http://arxiv.org/abs/1812.02614}{{\tt 1812.02614}}].

\bibitem{Ivanov:2019kyh}
I.~P. Ivanov, C.~C. Nishi and A.~Trautner, \emph{{Beyond basis invariants}},
  \href{http://dx.doi.org/10.1140/epjc/s10052-019-6845-x}{\emph{Eur. Phys. J.
  C} {\bf 79} (2019) 315}, [\href{http://arxiv.org/abs/1901.11472}{{\tt
  1901.11472}}].

\bibitem{Kubo:2003iw}
J.~Kubo, A.~Mondragon, M.~Mondragon and E.~Rodriguez-Jauregui, \emph{{The
  Flavor symmetry}}, \href{http://dx.doi.org/10.1143/PTP.109.795}{\emph{Prog.
  Theor. Phys.} {\bf 109} (2003) 795--807},
  [\href{http://arxiv.org/abs/hep-ph/0302196}{{\tt hep-ph/0302196}}].

\bibitem{Teshima:2005bk}
T.~Teshima, \emph{{Flavor mass and mixing and S(3) symmetry: An S(3) invariant
  model reasonable to all}},
  \href{http://dx.doi.org/10.1103/PhysRevD.73.045019}{\emph{Phys. Rev. D} {\bf
  73} (2006) 045019}, [\href{http://arxiv.org/abs/hep-ph/0509094}{{\tt
  hep-ph/0509094}}].

\bibitem{Mondragon:2007nk}
A.~Mondragon, M.~Mondragon and E.~Peinado, \emph{{S(3)-flavour symmetry as
  realized in lepton flavour violating processes}},
  \href{http://dx.doi.org/10.1088/1751-8113/41/30/304035}{\emph{J. Phys. A}
  {\bf 41} (2008) 304035}, [\href{http://arxiv.org/abs/0712.1799}{{\tt
  0712.1799}}].

\bibitem{Mondragon:2007af}
A.~Mondragon, M.~Mondragon and E.~Peinado, \emph{{Lepton masses, mixings and
  FCNC in a minimal S(3)-invariant extension of the Standard Model}},
  \href{http://dx.doi.org/10.1103/PhysRevD.76.076003}{\emph{Phys. Rev. D} {\bf
  76} (2007) 076003}, [\href{http://arxiv.org/abs/0706.0354}{{\tt 0706.0354}}].

\bibitem{GonzalezCanales:2012blg}
F.~Gonzalez~Canales, A.~Mondragon and M.~Mondragon, \emph{{The $S_3$ Flavour
  Symmetry: Neutrino Masses and Mixings}},
  \href{http://dx.doi.org/10.1002/prop.201200121}{\emph{Fortsch. Phys.} {\bf
  61} (2013) 546--570}, [\href{http://arxiv.org/abs/1205.4755}{{\tt
  1205.4755}}].

\bibitem{Ma:2013zca}
E.~Ma and B.~Melic, \emph{{Updated $S_{3}$ model of quarks}},
  \href{http://dx.doi.org/10.1016/j.physletb.2013.07.015}{\emph{Phys. Lett. B}
  {\bf 725} (2013) 402--406}, [\href{http://arxiv.org/abs/1303.6928}{{\tt
  1303.6928}}].

\bibitem{GonzalezCanales:2013pdx}
F.~Gonz\'alez~Canales, A.~Mondrag\'on, M.~Mondrag\'on, U.~J. Salda\~na Salazar
  and L.~Velasco-Sevilla, \emph{{Quark sector of S3 models: classification and
  comparison with experimental data}},
  \href{http://dx.doi.org/10.1103/PhysRevD.88.096004}{\emph{Phys. Rev. D} {\bf
  88} (2013) 096004}, [\href{http://arxiv.org/abs/1304.6644}{{\tt 1304.6644}}].

\bibitem{Das:2015sca}
D.~Das, U.~K. Dey and P.~B. Pal, \emph{{$S_3$ symmetry and the quark mixing
  matrix}}, \href{http://dx.doi.org/10.1016/j.physletb.2015.12.038}{\emph{Phys.
  Lett. B} {\bf 753} (2016) 315--318},
  [\href{http://arxiv.org/abs/1507.06509}{{\tt 1507.06509}}].

\bibitem{Cruz:2017add}
A.~A. Cruz and M.~Mondrag\'on, \emph{{Neutrino masses, mixing, and leptogenesis
  in an S3 model}},  \href{http://arxiv.org/abs/1701.07929}{{\tt 1701.07929}}.

\bibitem{Jarlskog:1985ht}
C.~Jarlskog, \emph{{Commutator of the Quark Mass Matrices in the Standard
  Electroweak Model and a Measure of Maximal $CP$~Nonconservation}},
  \href{http://dx.doi.org/10.1103/PhysRevLett.55.1039}{\emph{Phys. Rev. Lett.}
  {\bf 55} (1985) 1039}.

\bibitem{Dunietz:1985uy}
I.~Dunietz, O.~W. Greenberg and D.-d. Wu, \emph{{A Priori Definition of Maximal
  CP Violation}},
  \href{http://dx.doi.org/10.1103/PhysRevLett.55.2935}{\emph{Phys. Rev. Lett.}
  {\bf 55} (1985) 2935}.

\bibitem{Bernabeu:1986fc}
J.~Bernabeu, G.~C. Branco and M.~Gronau, \emph{{CP Restrictions on Quark Mass
  Matrices}}, \href{http://dx.doi.org/10.1016/0370-2693(86)90659-3}{\emph{Phys.
  Lett. B} {\bf 169} (1986) 243--247}.

\bibitem{CMS:2020zge}
A.~M. Sirunyan et~al., \emph{{Inclusive search for highly boosted Higgs bosons
  decaying to bottom quark-antiquark pairs in proton-proton collisions at
  $\sqrt{s} =$ 13 TeV}},
  \href{http://dx.doi.org/10.1007/JHEP12(2020)085}{\emph{JHEP} {\bf 12} (2020)
  085}, [\href{http://arxiv.org/abs/2006.13251}{{\tt 2006.13251}}].

\bibitem{ATLAS:2020fcp}
G.~Aad et~al., \emph{{Measurements of $WH$ and $ZH$ production in the $H
  \rightarrow b\bar{b}$ decay channel in $pp$ collisions at 13 TeV with the
  ATLAS detector}},
  \href{http://dx.doi.org/10.1140/epjc/s10052-020-08677-2}{\emph{Eur. Phys. J.
  C} {\bf 81} (2021) 178}, [\href{http://arxiv.org/abs/2007.02873}{{\tt
  2007.02873}}].

\bibitem{ATLAS:2020jwz}
G.~Aad et~al., \emph{{Measurement of the associated production of a Higgs boson
  decaying into $b$-quarks with a vector boson at high transverse momentum in
  $pp$ collisions at $\sqrt{s} = 13$ TeV with the ATLAS detector}},
  \href{http://dx.doi.org/10.1016/j.physletb.2021.136204}{\emph{Phys. Lett. B}
  {\bf 816} (2021) 136204}, [\href{http://arxiv.org/abs/2008.02508}{{\tt
  2008.02508}}].

\bibitem{CMS:2021gfa}
A.~Tumasyan et~al., \emph{{Search for flavor-changing neutral current
  interactions of the top quark and the Higgs boson decaying to a bottom
  quark-antiquark pair at $ \sqrt{s} $ = 13 TeV}},
  \href{http://dx.doi.org/10.1007/JHEP02(2022)169}{\emph{JHEP} {\bf 02} (2022)
  169}, [\href{http://arxiv.org/abs/2112.09734}{{\tt 2112.09734}}].

\bibitem{ATLAS:2022gzn}
\emph{{Search for flavour-changing neutral current interactions of the top
  quark and the Higgs boson in events with a pair of $\tau$-leptons in pp
  collisions at $\sqrt{s}=13$ TeV with the ATLAS detector}},
  \href{http://arxiv.org/abs/2208.11415}{{\tt 2208.11415}}.

\bibitem{CMS:2022dbt}
\emph{{Search for $CP$ violation in ttH and tH production in multilepton
  channels in proton-proton collisions at $\sqrt{s}$ = 13 TeV}},
  \href{http://arxiv.org/abs/2208.02686}{{\tt 2208.02686}}.

\bibitem{CMS:2020osd}
A.~M. Sirunyan et~al., \emph{{Search for a light charged Higgs boson in the
  H$^\pm$ $\to $ cs channel in proton-proton collisions at $\sqrt{s} =$ 13
  TeV}}, \href{http://dx.doi.org/10.1103/PhysRevD.102.072001}{\emph{Phys. Rev.
  D} {\bf 102} (2020) 072001}, [\href{http://arxiv.org/abs/2005.08900}{{\tt
  2005.08900}}].

\bibitem{ATLAS:2021zyv}
\emph{{Search for a light charged Higgs boson in t$\rightarrow$H$^+$b decays,
  with H$^+$$\rightarrow$cb, in the lepton+jets final state in proton-proton
  collisions at $\sqrt{s}=13$ TeV with the ATLAS detector}}, .

\bibitem{Schael:2013ita}
S.~Schael et~al., \emph{{Electroweak Measurements in Electron-Positron
  Collisions at W-Boson-Pair Energies at LEP}},
  \href{http://dx.doi.org/10.1016/j.physrep.2013.07.004}{\emph{Phys. Rept.}
  {\bf 532} (2013) 119--244}, [\href{http://arxiv.org/abs/1302.3415}{{\tt
  1302.3415}}].

\bibitem{Fox:2017uwr}
P.~J. Fox and N.~Weiner, \emph{{Light Signals from a Lighter Higgs}},
  \href{http://dx.doi.org/10.1007/JHEP08(2018)025}{\emph{JHEP} {\bf 08} (2018)
  025}, [\href{http://arxiv.org/abs/1710.07649}{{\tt 1710.07649}}].

\bibitem{Cepeda:2021rql}
M.~Cepeda, S.~Gori, V.~M. Outschoorn and J.~Shelton, \emph{{Exotic Higgs
  Decays}},  \href{http://arxiv.org/abs/2111.12751}{{\tt 2111.12751}}.

\bibitem{Biekotter:2022jyr}
T.~Biek\"otter, S.~Heinemeyer and G.~Weiglein, \emph{{Mounting evidence for a
  95 GeV Higgs boson}},
  \href{http://dx.doi.org/10.1007/JHEP08(2022)201}{\emph{JHEP} {\bf 08} (2022)
  201}, [\href{http://arxiv.org/abs/2203.13180}{{\tt 2203.13180}}].

\bibitem{Robens:2022zgk}
T.~Robens, \emph{{A Short Overview on Low Mass Scalars at Future Lepton
  Colliders}}, \href{http://dx.doi.org/10.3390/universe8050286}{\emph{Universe}
  {\bf 8} (2022) 286}, [\href{http://arxiv.org/abs/2205.09687}{{\tt
  2205.09687}}].

\bibitem{Plantey:2022jdg}
R.~Plantey, O.~M. Ogreid, P.~Osland, M.~N. Rebelo and M.~A. Solberg,
  \emph{{Weinberg\textquoteright{}s 3HDM potential with spontaneous CP
  violation}}, \href{http://dx.doi.org/10.1103/PhysRevD.108.075029}{\emph{Phys.
  Rev. D} {\bf 108} (2023) 075029},
  [\href{http://arxiv.org/abs/2208.13594}{{\tt 2208.13594}}].

\bibitem{Branco:2015bfb}
G.~C. Branco and I.~P. Ivanov, \emph{{Group-theoretic restrictions on
  generation of CP-violation in multi-Higgs-doublet models}},
  \href{http://dx.doi.org/10.1007/JHEP01(2016)116}{\emph{JHEP} {\bf 01} (2016)
  116}, [\href{http://arxiv.org/abs/1511.02764}{{\tt 1511.02764}}].

\bibitem{Adulpravitchai:2011ei}
A.~Adulpravitchai, B.~Batell and J.~Pradler, \emph{{Non-Abelian Discrete Dark
  Matter}}, \href{http://dx.doi.org/10.1016/j.physletb.2011.04.015}{\emph{Phys.
  Lett. B} {\bf 700} (2011) 207--216},
  [\href{http://arxiv.org/abs/1103.3053}{{\tt 1103.3053}}].

\bibitem{Nierste:1995zx}
U.~Nierste and K.~Riesselmann, \emph{{Higgs sector renormalization group in the
  MS and OMS scheme: The Breakdown of perturbation theory for a heavy Higgs}},
  \href{http://dx.doi.org/10.1103/PhysRevD.53.6638}{\emph{Phys. Rev. D} {\bf
  53} (1996) 6638--6652}, [\href{http://arxiv.org/abs/hep-ph/9511407}{{\tt
  hep-ph/9511407}}].

\bibitem{Sher:1988mj}
M.~Sher, \emph{{Electroweak Higgs Potentials and Vacuum Stability}},
  \href{http://dx.doi.org/10.1016/0370-1573(89)90061-6}{\emph{Phys. Rept.} {\bf
  179} (1989) 273--418}.

\bibitem{Faro:2019vcd}
F.~S. Faro and I.~P. Ivanov, \emph{{Boundedness from below in the $U(1)\times
  U(1)$ three-Higgs-doublet model}},
  \href{http://dx.doi.org/10.1103/PhysRevD.100.035038}{\emph{Phys. Rev. D} {\bf
  100} (2019) 035038}, [\href{http://arxiv.org/abs/1907.01963}{{\tt
  1907.01963}}].

\bibitem{Marciano:1989ns}
W.~J. Marciano, G.~Valencia and S.~Willenbrock, \emph{{Renormalization Group
  Improved Unitarity Bounds on the Higgs Boson and Top Quark Masses}},
  \href{http://dx.doi.org/10.1103/PhysRevD.40.1725}{\emph{Phys. Rev.} {\bf D40}
  (1989) 1725}.

\bibitem{Grinstein:1991cd}
B.~Grinstein and M.~B. Wise, \emph{{Operator analysis for precision electroweak
  physics}}, \href{http://dx.doi.org/10.1016/0370-2693(91)90061-T}{\emph{Phys.
  Lett. B} {\bf 265} (1991) 326--334}.

\bibitem{Grimus:2007if}
W.~Grimus, L.~Lavoura, O.~M. Ogreid and P.~Osland, \emph{{A Precision
  constraint on multi-Higgs-doublet models}},
  \href{http://dx.doi.org/10.1088/0954-3899/35/7/075001}{\emph{J. Phys.} {\bf
  G35} (2008) 075001}, [\href{http://arxiv.org/abs/0711.4022}{{\tt
  0711.4022}}].

\bibitem{Grimus:2008nb}
W.~Grimus, L.~Lavoura, O.~M. Ogreid and P.~Osland, \emph{{The Oblique
  parameters in multi-Higgs-doublet models}},
  \href{http://dx.doi.org/10.1016/j.nuclphysb.2008.04.019}{\emph{Nucl. Phys.}
  {\bf B801} (2008) 81--96}, [\href{http://arxiv.org/abs/0802.4353}{{\tt
  0802.4353}}].

\bibitem{Veltman:1977kh}
M.~J.~G. Veltman, \emph{{Limit on Mass Differences in the Weinberg Model}},
  \href{http://dx.doi.org/10.1016/0550-3213(77)90342-X}{\emph{Nucl. Phys. B}
  {\bf 123} (1977) 89--99}.

\bibitem{Eboli:2000ze}
O.~J.~P. Eboli and D.~Zeppenfeld, \emph{{Observing an invisible Higgs boson}},
  \href{http://dx.doi.org/10.1016/S0370-2693(00)01213-2}{\emph{Phys. Lett. B}
  {\bf 495} (2000) 147--154}, [\href{http://arxiv.org/abs/hep-ph/0009158}{{\tt
  hep-ph/0009158}}].

\bibitem{Godbole:2003it}
R.~M. Godbole, M.~Guchait, K.~Mazumdar, S.~Moretti and D.~P. Roy, \emph{{Search
  for `invisible' Higgs signals at LHC via associated production with gauge
  bosons}}, \href{http://dx.doi.org/10.1016/j.physletb.2003.06.066}{\emph{Phys.
  Lett. B} {\bf 571} (2003) 184--192},
  [\href{http://arxiv.org/abs/hep-ph/0304137}{{\tt hep-ph/0304137}}].

\bibitem{Kanemura:2010sh}
S.~Kanemura, S.~Matsumoto, T.~Nabeshima and N.~Okada, \emph{{Can WIMP Dark
  Matter overcome the Nightmare Scenario?}},
  \href{http://dx.doi.org/10.1103/PhysRevD.82.055026}{\emph{Phys. Rev. D} {\bf
  82} (2010) 055026}, [\href{http://arxiv.org/abs/1005.5651}{{\tt 1005.5651}}].

\bibitem{Chu:2011be}
X.~Chu, T.~Hambye and M.~H.~G. Tytgat, \emph{{The Four Basic Ways of Creating
  Dark Matter Through a Portal}},
  \href{http://dx.doi.org/10.1088/1475-7516/2012/05/034}{\emph{JCAP} {\bf 05}
  (2012) 034}, [\href{http://arxiv.org/abs/1112.0493}{{\tt 1112.0493}}].

\bibitem{Djouadi:2011aa}
A.~Djouadi, O.~Lebedev, Y.~Mambrini and J.~Quevillon, \emph{{Implications of
  LHC searches for Higgs--portal dark matter}},
  \href{http://dx.doi.org/10.1016/j.physletb.2012.01.062}{\emph{Phys. Lett. B}
  {\bf 709} (2012) 65--69}, [\href{http://arxiv.org/abs/1112.3299}{{\tt
  1112.3299}}].

\bibitem{Djouadi:2012zc}
A.~Djouadi, A.~Falkowski, Y.~Mambrini and J.~Quevillon, \emph{{Direct Detection
  of Higgs-Portal Dark Matter at the LHC}},
  \href{http://dx.doi.org/10.1140/epjc/s10052-013-2455-1}{\emph{Eur. Phys. J.
  C} {\bf 73} (2013) 2455}, [\href{http://arxiv.org/abs/1205.3169}{{\tt
  1205.3169}}].

\bibitem{Belanger:2013kya}
G.~B\'elanger, B.~Dumont, U.~Ellwanger, J.~F. Gunion and S.~Kraml,
  \emph{{Status of invisible Higgs decays}},
  \href{http://dx.doi.org/10.1016/j.physletb.2013.05.024}{\emph{Phys. Lett. B}
  {\bf 723} (2013) 340--347}, [\href{http://arxiv.org/abs/1302.5694}{{\tt
  1302.5694}}].

\bibitem{Curtin:2013fra}
D.~Curtin et~al., \emph{{Exotic decays of the 125 GeV Higgs boson}},
  \href{http://dx.doi.org/10.1103/PhysRevD.90.075004}{\emph{Phys. Rev. D} {\bf
  90} (2014) 075004}, [\href{http://arxiv.org/abs/1312.4992}{{\tt 1312.4992}}].

\bibitem{Ivanov:2012hc}
I.~P. Ivanov and V.~Keus, \emph{{$Z_p$ scalar dark matter from
  multi-Higgs-doublet models}},
  \href{http://dx.doi.org/10.1103/PhysRevD.86.016004}{\emph{Phys. Rev. D} {\bf
  86} (2012) 016004}, [\href{http://arxiv.org/abs/1203.3426}{{\tt 1203.3426}}].

\bibitem{Hernandez-Sanchez:2020aop}
J.~Hernandez-Sanchez, V.~Keus, S.~Moretti, D.~Rojas-Ciofalo and D.~Sokolowska,
  \emph{{Complementary Probes of Two-component Dark Matter}},
  \href{http://arxiv.org/abs/2012.11621}{{\tt 2012.11621}}.

\bibitem{Hernandez-Sanchez:2022dnn}
J.~Hernandez-Sanchez, V.~Keus, S.~Moretti and D.~Sokolowska,
  \emph{{Complementary collider and astrophysical probes of multi-component
  Dark Matter}}, \href{http://dx.doi.org/10.1007/JHEP03(2023)045}{\emph{JHEP}
  {\bf 03} (2023) 045}, [\href{http://arxiv.org/abs/2202.10514}{{\tt
  2202.10514}}].

\bibitem{Hess:2021cdp}
H.~Abdalla et~al., \emph{{Combined dark matter searches towards dwarf
  spheroidal galaxies with Fermi-LAT, HAWC, H.E.S.S., MAGIC, and VERITAS}},
  \href{http://dx.doi.org/10.22323/1.395.0528}{\emph{PoS} {\bf ICRC2021} (2021)
  528}, [\href{http://arxiv.org/abs/2108.13646}{{\tt 2108.13646}}].

\bibitem{Machado:2012gxi}
A.~C.~B. Machado and V.~Pleitez, \emph{{A model with two inert scalar
  doublets}}, \href{http://dx.doi.org/10.1016/j.aop.2015.10.017}{\emph{Annals
  Phys.} {\bf 364} (2016) 53--67}, [\href{http://arxiv.org/abs/1205.0995}{{\tt
  1205.0995}}].

\bibitem{Fortes:2014dca}
E.~C. F.~S. Fortes, A.~C.~B. Machado, J.~Montaño and V.~Pleitez, \emph{{Scalar
  dark matter candidates in a two inert Higgs doublet model}},
  \href{http://dx.doi.org/10.1088/0954-3899/42/10/105003}{\emph{J. Phys.} {\bf
  G42} (2015) 105003}, [\href{http://arxiv.org/abs/1407.4749}{{\tt
  1407.4749}}].

\bibitem{Deng:2025dcq}
H.~Deng, R.~Boto, I.~P. Ivanov and J.~P. Silva, \emph{{Dark matter stabilized
  by a non-abelian group: lessons from the $\Sigma(36)$ 3HDM}},
  \href{http://arxiv.org/abs/2501.05929}{{\tt 2501.05929}}.

\bibitem{Grinstein:1987pu}
B.~Grinstein and M.~B. Wise, \emph{{Weak Radiative B Meson Decay as a Probe of
  the Higgs Sector}},
  \href{http://dx.doi.org/10.1016/0370-2693(88)90227-4}{\emph{Phys. Lett. B}
  {\bf 201} (1988) 274--278}.

\bibitem{Hou:1988gv}
W.-S. Hou and R.~Willey, \emph{{Effects of Extended Higgs Sector on Loop
  Induced $B$ Decays}},
  \href{http://dx.doi.org/10.1016/0550-3213(89)90434-3}{\emph{Nucl. Phys. B}
  {\bf 326} (1989) 54--72}.

\bibitem{Grinstein:1990tj}
B.~Grinstein, R.~P. Springer and M.~B. Wise, \emph{{Strong Interaction Effects
  in Weak Radiative $\bar{B}$ Meson Decay}},
  \href{http://dx.doi.org/10.1016/0550-3213(90)90350-M}{\emph{Nucl. Phys. B}
  {\bf 339} (1990) 269--309}.

\bibitem{Buras:1993xp}
A.~Buras, M.~Misiak, M.~Munz and S.~Pokorski, \emph{{Theoretical uncertainties
  and phenomenological aspects of B $\to$ X(s) gamma decay}},
  \href{http://dx.doi.org/10.1016/0550-3213(94)90299-2}{\emph{Nucl. Phys. B}
  {\bf 424} (1994) 374--398}, [\href{http://arxiv.org/abs/hep-ph/9311345}{{\tt
  hep-ph/9311345}}].

\bibitem{Ciafaloni:1997un}
P.~Ciafaloni, A.~Romanino and A.~Strumia, \emph{{Two loop QCD corrections to
  charged Higgs mediated b $\to$ s gamma decay}},
  \href{http://dx.doi.org/10.1016/S0550-3213(98)00190-4}{\emph{Nucl. Phys. B}
  {\bf 524} (1998) 361--376}, [\href{http://arxiv.org/abs/hep-ph/9710312}{{\tt
  hep-ph/9710312}}].

\bibitem{Ciuchini:1997xe}
M.~Ciuchini, G.~Degrassi, P.~Gambino and G.~Giudice, \emph{{Next-to-leading QCD
  corrections to $B \to X_s \gamma$: Standard model and two Higgs doublet
  model}}, \href{http://dx.doi.org/10.1016/S0550-3213(98)00244-2}{\emph{Nucl.
  Phys. B} {\bf 527} (1998) 21--43},
  [\href{http://arxiv.org/abs/hep-ph/9710335}{{\tt hep-ph/9710335}}].

\bibitem{Borzumati:1998tg}
F.~Borzumati and C.~Greub, \emph{{2HDMs predictions for anti-B $\to$ X(s) gamma
  in NLO QCD}}, \href{http://dx.doi.org/10.1103/PhysRevD.58.074004}{\emph{Phys.
  Rev. D} {\bf 58} (1998) 074004},
  [\href{http://arxiv.org/abs/hep-ph/9802391}{{\tt hep-ph/9802391}}].

\bibitem{Bobeth:1999ww}
C.~Bobeth, M.~Misiak and J.~Urban, \emph{{Matching conditions for $b \to s
  \gamma$ and $b \to s gluon$ in extensions of the standard model}},
  \href{http://dx.doi.org/10.1016/S0550-3213(99)00688-4}{\emph{Nucl. Phys. B}
  {\bf 567} (2000) 153--185}, [\href{http://arxiv.org/abs/hep-ph/9904413}{{\tt
  hep-ph/9904413}}].

\bibitem{Bobeth:1999mk}
C.~Bobeth, M.~Misiak and J.~Urban, \emph{{Photonic penguins at two loops and
  $m_t$ dependence of $BR[B \to X_s l^+ l^-]$}},
  \href{http://dx.doi.org/10.1016/S0550-3213(00)00007-9}{\emph{Nucl. Phys. B}
  {\bf 574} (2000) 291--330}, [\href{http://arxiv.org/abs/hep-ph/9910220}{{\tt
  hep-ph/9910220}}].

\bibitem{Gambino:2001ew}
P.~Gambino and M.~Misiak, \emph{{Quark mass effects in anti-B $\to$ X(s
  gamma)}}, \href{http://dx.doi.org/10.1016/S0550-3213(01)00347-9}{\emph{Nucl.
  Phys. B} {\bf 611} (2001) 338--366},
  [\href{http://arxiv.org/abs/hep-ph/0104034}{{\tt hep-ph/0104034}}].

\bibitem{Cheung:2003pw}
K.~Cheung and O.~C. Kong, \emph{{Can the two Higgs doublet model survive the
  constraint from the muon anomalous magnetic moment as suggested?}},
  \href{http://dx.doi.org/10.1103/PhysRevD.68.053003}{\emph{Phys. Rev. D} {\bf
  68} (2003) 053003}, [\href{http://arxiv.org/abs/hep-ph/0302111}{{\tt
  hep-ph/0302111}}].

\bibitem{Misiak:2004ew}
M.~Misiak and M.~Steinhauser, \emph{{Three loop matching of the dipole
  operators for $b \to s \gamma$ and $b \to s g$}},
  \href{http://dx.doi.org/10.1016/j.nuclphysb.2004.02.006}{\emph{Nucl. Phys. B}
  {\bf 683} (2004) 277--305}, [\href{http://arxiv.org/abs/hep-ph/0401041}{{\tt
  hep-ph/0401041}}].

\bibitem{Czakon:2006ss}
M.~Czakon, U.~Haisch and M.~Misiak, \emph{{Four-Loop Anomalous Dimensions for
  Radiative Flavour-Changing Decays}},
  \href{http://dx.doi.org/10.1088/1126-6708/2007/03/008}{\emph{JHEP} {\bf 03}
  (2007) 008}, [\href{http://arxiv.org/abs/hep-ph/0612329}{{\tt
  hep-ph/0612329}}].

\bibitem{Hermann:2012fc}
T.~Hermann, M.~Misiak and M.~Steinhauser, \emph{{$\bar{B}\to X_s \gamma$ in the
  Two Higgs Doublet Model up to Next-to-Next-to-Leading Order in QCD}},
  \href{http://dx.doi.org/10.1007/JHEP11(2012)036}{\emph{JHEP} {\bf 11} (2012)
  036}, [\href{http://arxiv.org/abs/1208.2788}{{\tt 1208.2788}}].

\bibitem{Misiak:2015xwa}
M.~Misiak et~al., \emph{{Updated NNLO QCD predictions for the weak radiative
  B-meson decays}},
  \href{http://dx.doi.org/10.1103/PhysRevLett.114.221801}{\emph{Phys. Rev.
  Lett.} {\bf 114} (2015) 221801}, [\href{http://arxiv.org/abs/1503.01789}{{\tt
  1503.01789}}].

\bibitem{Misiak:2017bgg}
M.~Misiak and M.~Steinhauser, \emph{{Weak radiative decays of the B meson and
  bounds on $M_{H^\pm }$ in the Two-Higgs-Doublet Model}},
  \href{http://dx.doi.org/10.1140/epjc/s10052-017-4776-y}{\emph{Eur. Phys. J.
  C} {\bf 77} (2017) 201}, [\href{http://arxiv.org/abs/1702.04571}{{\tt
  1702.04571}}].

\bibitem{Misiak:2020vlo}
M.~Misiak, A.~Rehman and M.~Steinhauser, \emph{{Towards $ \overline{B}\to
  {X}_s\gamma $ at the NNLO in QCD without interpolation in m$_{c}$}},
  \href{http://dx.doi.org/10.1007/JHEP06(2020)175}{\emph{JHEP} {\bf 06} (2020)
  175}, [\href{http://arxiv.org/abs/2002.01548}{{\tt 2002.01548}}].

\bibitem{Czaja:2023ren}
M.~Czaja, M.~Czakon, T.~Huber, M.~Misiak, M.~Niggetiedt, A.~Rehman et~al.,
  \emph{{The $Q_{1,2}$\textendash{}$Q_7$ interference contributions to $b
  \rightarrow s \gamma $ at ${\mathcal O}(\alpha _{\mathrm s}^2)$ for the
  physical value of $m_c$}},
  \href{http://dx.doi.org/10.1140/epjc/s10052-023-12270-8}{\emph{Eur. Phys. J.
  C} {\bf 83} (2023) 1108}, [\href{http://arxiv.org/abs/2309.14707}{{\tt
  2309.14707}}].

\bibitem{Misiak:2006ab}
M.~Misiak and M.~Steinhauser, \emph{{NNLO QCD corrections to the anti-B $\to$
  X(s) gamma matrix elements using interpolation in m(c)}},
  \href{http://dx.doi.org/10.1016/j.nuclphysb.2006.11.027}{\emph{Nucl. Phys. B}
  {\bf 764} (2007) 62--82}, [\href{http://arxiv.org/abs/hep-ph/0609241}{{\tt
  hep-ph/0609241}}].

\bibitem{HFLAV:2022esi}
Y.~S. Amhis et~al., \emph{{Averages of b-hadron, c-hadron, and
  \ensuremath{\tau}-lepton properties as of 2021}},
  \href{http://dx.doi.org/10.1103/PhysRevD.107.052008}{\emph{Phys. Rev. D} {\bf
  107} (2023) 052008}, [\href{http://arxiv.org/abs/2206.07501}{{\tt
  2206.07501}}].

\bibitem{Belle-II:2022hys}
F.~Abudin\'en et~al., \emph{{Measurement of the photon-energy spectrum in
  inclusive $B\rightarrow X_{s}\gamma$ decays identified using hadronic decays
  of the recoil $B$ meson in 2019-2021 Belle II data}},
  \href{http://arxiv.org/abs/2210.10220}{{\tt 2210.10220}}.

\bibitem{Bahl:2022igd}
H.~Bahl, T.~Biek\"otter, S.~Heinemeyer, C.~Li, S.~Paasch, G.~Weiglein et~al.,
  \emph{{HiggsTools: BSM scalar phenomenology with new versions of HiggsBounds
  and HiggsSignals}},
  \href{http://dx.doi.org/10.1016/j.cpc.2023.108803}{\emph{Comput. Phys.
  Commun.} {\bf 291} (2023) 108803},
  [\href{http://arxiv.org/abs/2210.09332}{{\tt 2210.09332}}].

\bibitem{Bechtle:2008jh}
P.~Bechtle, O.~Brein, S.~Heinemeyer, G.~Weiglein and K.~E. Williams,
  \emph{{HiggsBounds: Confronting Arbitrary Higgs Sectors with Exclusion Bounds
  from LEP and the Tevatron}},
  \href{http://dx.doi.org/10.1016/j.cpc.2009.09.003}{\emph{Comput. Phys.
  Commun.} {\bf 181} (2010) 138--167},
  [\href{http://arxiv.org/abs/0811.4169}{{\tt 0811.4169}}].

\bibitem{Bechtle:2011sb}
P.~Bechtle, O.~Brein, S.~Heinemeyer, G.~Weiglein and K.~E. Williams,
  \emph{{HiggsBounds 2.0.0: Confronting Neutral and Charged Higgs Sector
  Predictions with Exclusion Bounds from LEP and the Tevatron}},
  \href{http://dx.doi.org/10.1016/j.cpc.2011.07.015}{\emph{Comput. Phys.
  Commun.} {\bf 182} (2011) 2605--2631},
  [\href{http://arxiv.org/abs/1102.1898}{{\tt 1102.1898}}].

\bibitem{Bechtle:2012lvg}
P.~Bechtle, O.~Brein, S.~Heinemeyer, O.~St\r{a}l, T.~Stefaniak, G.~Weiglein
  et~al., \emph{{Recent Developments in HiggsBounds and a Preview of
  HiggsSignals}}, \href{http://dx.doi.org/10.22323/1.156.0024}{\emph{PoS} {\bf
  CHARGED2012} (2012) 024}, [\href{http://arxiv.org/abs/1301.2345}{{\tt
  1301.2345}}].

\bibitem{Bechtle:2013wla}
P.~Bechtle, O.~Brein, S.~Heinemeyer, O.~St\r{a}l, T.~Stefaniak, G.~Weiglein
  et~al., \emph{{$\mathsf{HiggsBounds}-4$: Improved Tests of Extended Higgs
  Sectors against Exclusion Bounds from LEP, the Tevatron and the LHC}},
  \href{http://dx.doi.org/10.1140/epjc/s10052-013-2693-2}{\emph{Eur. Phys. J.
  C} {\bf 74} (2014) 2693}, [\href{http://arxiv.org/abs/1311.0055}{{\tt
  1311.0055}}].

\bibitem{Bechtle:2015pma}
P.~Bechtle, S.~Heinemeyer, O.~St\r{a}l, T.~Stefaniak and G.~Weiglein,
  \emph{{Applying Exclusion Likelihoods from LHC Searches to Extended Higgs
  Sectors}}, \href{http://dx.doi.org/10.1140/epjc/s10052-015-3650-z}{\emph{Eur.
  Phys. J. C} {\bf 75} (2015) 421},
  [\href{http://arxiv.org/abs/1507.06706}{{\tt 1507.06706}}].

\bibitem{Bechtle:2020pkv}
P.~Bechtle, D.~Dercks, S.~Heinemeyer, T.~Klingl, T.~Stefaniak, G.~Weiglein
  et~al., \emph{{HiggsBounds-5: Testing Higgs Sectors in the LHC 13 TeV Era}},
  \href{http://dx.doi.org/10.1140/epjc/s10052-020-08557-9}{\emph{Eur. Phys. J.
  C} {\bf 80} (2020) 1211}, [\href{http://arxiv.org/abs/2006.06007}{{\tt
  2006.06007}}].

\bibitem{Bahl:2021yhk}
H.~Bahl, V.~M. Lozano, T.~Stefaniak and J.~Wittbrodt, \emph{{Testing exotic
  scalars with HiggsBounds}},
  \href{http://dx.doi.org/10.1140/epjc/s10052-022-10446-2}{\emph{Eur. Phys. J.
  C} {\bf 82} (2022) 584}, [\href{http://arxiv.org/abs/2109.10366}{{\tt
  2109.10366}}].

\bibitem{Bechtle:2013xfa}
P.~Bechtle, S.~Heinemeyer, O.~St\r{a}l, T.~Stefaniak and G.~Weiglein,
  \emph{{$HiggsSignals$: Confronting arbitrary Higgs sectors with measurements
  at the Tevatron and the LHC}},
  \href{http://dx.doi.org/10.1140/epjc/s10052-013-2711-4}{\emph{Eur. Phys. J.
  C} {\bf 74} (2014) 2711}, [\href{http://arxiv.org/abs/1305.1933}{{\tt
  1305.1933}}].

\bibitem{Stal:2013hwa}
O.~St\r{a}l and T.~Stefaniak, \emph{{Constraining extended Higgs sectors with
  HiggsSignals}}, \href{http://dx.doi.org/10.22323/1.180.0314}{\emph{PoS} {\bf
  EPS-HEP2013} (2013) 314}, [\href{http://arxiv.org/abs/1310.4039}{{\tt
  1310.4039}}].

\bibitem{Bechtle:2014ewa}
P.~Bechtle, S.~Heinemeyer, O.~St\r{a}l, T.~Stefaniak and G.~Weiglein,
  \emph{{Probing the Standard Model with Higgs signal rates from the Tevatron,
  the LHC and a future ILC}},
  \href{http://dx.doi.org/10.1007/JHEP11(2014)039}{\emph{JHEP} {\bf 11} (2014)
  039}, [\href{http://arxiv.org/abs/1403.1582}{{\tt 1403.1582}}].

\bibitem{Bechtle:2020uwn}
P.~Bechtle, S.~Heinemeyer, T.~Klingl, T.~Stefaniak, G.~Weiglein and
  J.~Wittbrodt, \emph{{HiggsSignals-2: Probing new physics with precision Higgs
  measurements in the LHC 13 TeV era}},
  \href{http://dx.doi.org/10.1140/epjc/s10052-021-08942-y}{\emph{Eur. Phys. J.
  C} {\bf 81} (2021) 145}, [\href{http://arxiv.org/abs/2012.09197}{{\tt
  2012.09197}}].

\bibitem{Belanger:2006is}
G.~B\'elanger, F.~Boudjema, A.~Pukhov and A.~Semenov, \emph{{MicrOMEGAs 2.0: A
  Program to calculate the relic density of dark matter in a generic model}},
  \href{http://dx.doi.org/10.1016/j.cpc.2006.11.008}{\emph{Comput. Phys.
  Commun.} {\bf 176} (2007) 367--382},
  [\href{http://arxiv.org/abs/hep-ph/0607059}{{\tt hep-ph/0607059}}].

\bibitem{Belanger:2008sj}
G.~B\'elanger, F.~Boudjema, A.~Pukhov and A.~Semenov, \emph{{Dark matter direct
  detection rate in a generic model with micrOMEGAs 2.2}},
  \href{http://dx.doi.org/10.1016/j.cpc.2008.11.019}{\emph{Comput. Phys.
  Commun.} {\bf 180} (2009) 747--767},
  [\href{http://arxiv.org/abs/0803.2360}{{\tt 0803.2360}}].

\bibitem{Belanger:2010pz}
G.~B\'elanger, F.~Boudjema, A.~Pukhov and A.~Semenov, \emph{{micrOMEGAs: A Tool
  for dark matter studies}},
  \href{http://dx.doi.org/10.1393/ncc/i2010-10591-3}{\emph{Nuovo Cim. C} {\bf
  033N2} (2010) 111--116}, [\href{http://arxiv.org/abs/1005.4133}{{\tt
  1005.4133}}].

\bibitem{Belanger:2013oya}
G.~B\'elanger, F.~Boudjema, A.~Pukhov and A.~Semenov, \emph{{micrOMEGAs\_3: A
  program for calculating dark matter observables}},
  \href{http://dx.doi.org/10.1016/j.cpc.2013.10.016}{\emph{Comput. Phys.
  Commun.} {\bf 185} (2014) 960--985},
  [\href{http://arxiv.org/abs/1305.0237}{{\tt 1305.0237}}].

\bibitem{Belanger:2014vza}
G.~B\'elanger, F.~Boudjema, A.~Pukhov and A.~Semenov, \emph{{micrOMEGAs4.1: two
  dark matter candidates}},
  \href{http://dx.doi.org/10.1016/j.cpc.2015.03.003}{\emph{Comput. Phys.
  Commun.} {\bf 192} (2015) 322--329},
  [\href{http://arxiv.org/abs/1407.6129}{{\tt 1407.6129}}].

\bibitem{Barducci:2016pcb}
D.~Barducci, G.~B\'elanger, J.~Bernon, F.~Boudjema, J.~Da~Silva, S.~Kraml
  et~al., \emph{{Collider limits on new physics within micrOMEGAs\_4.3}},
  \href{http://dx.doi.org/10.1016/j.cpc.2017.08.028}{\emph{Comput. Phys.
  Commun.} {\bf 222} (2018) 327--338},
  [\href{http://arxiv.org/abs/1606.03834}{{\tt 1606.03834}}].

\bibitem{Belanger:2018ccd}
G.~B\'elanger, F.~Boudjema, A.~Goudelis, A.~Pukhov and B.~Zaldivar,
  \emph{{micrOMEGAs5.0 : Freeze-in}},
  \href{http://dx.doi.org/10.1016/j.cpc.2018.04.027}{\emph{Comput. Phys.
  Commun.} {\bf 231} (2018) 173--186},
  [\href{http://arxiv.org/abs/1801.03509}{{\tt 1801.03509}}].

\bibitem{Alguero:2023zol}
G.~Alguero, G.~B\'elanger, F.~Boudjema, S.~Chakraborti, A.~Goudelis, S.~Kraml
  et~al., \emph{{micrOMEGAs 6.0: N-component dark matter}},
  \href{http://dx.doi.org/10.1016/j.cpc.2024.109133}{\emph{Comput. Phys.
  Commun.} {\bf 299} (2024) 109133},
  [\href{http://arxiv.org/abs/2312.14894}{{\tt 2312.14894}}].

\bibitem{Wilczek:1977zn}
F.~Wilczek, \emph{{Decays of Heavy Vector Mesons Into Higgs Particles}},
  \href{http://dx.doi.org/10.1103/PhysRevLett.39.1304}{\emph{Phys. Rev. Lett.}
  {\bf 39} (1977) 1304}.

\bibitem{Georgi:1977gs}
H.~Georgi, S.~Glashow, M.~Machacek and D.~V. Nanopoulos, \emph{{Higgs Bosons
  from Two Gluon Annihilation in Proton Proton Collisions}},
  \href{http://dx.doi.org/10.1103/PhysRevLett.40.692}{\emph{Phys. Rev. Lett.}
  {\bf 40} (1978) 692}.

\bibitem{Ellis:1979jy}
J.~R. Ellis, M.~Gaillard, D.~V. Nanopoulos and C.~T. Sachrajda, \emph{{Is the
  Mass of the Higgs Boson About 10-GeV?}},
  \href{http://dx.doi.org/10.1016/0370-2693(79)91122-5}{\emph{Phys. Lett. B}
  {\bf 83} (1979) 339--344}.

\bibitem{Rizzo:1979mf}
T.~G. Rizzo, \emph{{Gluon Final States in Higgs Boson Decay}},
  \href{http://dx.doi.org/10.1103/PhysRevD.22.178}{\emph{Phys. Rev. D} {\bf 22}
  (1980) 178}.

\bibitem{Ellis:1975ap}
J.~R. Ellis, M.~K. Gaillard and D.~V. Nanopoulos, \emph{{A Phenomenological
  Profile of the Higgs Boson}},
  \href{http://dx.doi.org/10.1016/0550-3213(76)90382-5}{\emph{Nucl. Phys. B}
  {\bf 106} (1976) 292}.

\bibitem{Shifman:1979eb}
M.~A. Shifman, A.~Vainshtein, M.~Voloshin and V.~I. Zakharov, \emph{{Low-Energy
  Theorems for Higgs Boson Couplings to Photons}}, {\emph{Sov. J. Nucl. Phys.}
  {\bf 30} (1979) 711--716}.

\bibitem{ATLAS:2023tkt}
G.~Aad et~al., \emph{{Combination of searches for invisible decays of the Higgs
  boson using 139 fb\ensuremath{-}1 of proton-proton collision data at s=13 TeV
  collected with the ATLAS experiment}},
  \href{http://dx.doi.org/10.1016/j.physletb.2023.137963}{\emph{Phys. Lett. B}
  {\bf 842} (2023) 137963}, [\href{http://arxiv.org/abs/2301.10731}{{\tt
  2301.10731}}].

\bibitem{ATLAS:2022yvh}
G.~Aad et~al., \emph{{Search for invisible Higgs-boson decays in events with
  vector-boson fusion signatures using 139 fb$^{-1}$ of proton-proton data
  recorded by the ATLAS experiment}},
  \href{http://dx.doi.org/10.1007/JHEP08(2022)104}{\emph{JHEP} {\bf 08} (2022)
  104}, [\href{http://arxiv.org/abs/2202.07953}{{\tt 2202.07953}}].

\bibitem{CMS:2022qva}
A.~Tumasyan et~al., \emph{{Search for invisible decays of the Higgs boson
  produced via vector boson fusion in proton-proton collisions at
  s=13\,\,TeV}},
  \href{http://dx.doi.org/10.1103/PhysRevD.105.092007}{\emph{Phys. Rev. D} {\bf
  105} (2022) 092007}, [\href{http://arxiv.org/abs/2201.11585}{{\tt
  2201.11585}}].

\bibitem{Boudjema:1995cb}
F.~Boudjema and E.~Chopin, \emph{{Double Higgs production at the linear
  colliders and the probing of the Higgs selfcoupling}},
  \href{http://dx.doi.org/10.1007/s002880050298}{\emph{Z. Phys. C} {\bf 73}
  (1996) 85--110}, [\href{http://arxiv.org/abs/hep-ph/9507396}{{\tt
  hep-ph/9507396}}].

\bibitem{Belyaev:2012qa}
A.~Belyaev, N.~D. Christensen and A.~Pukhov, \emph{{CalcHEP 3.4 for collider
  physics within and beyond the Standard Model}},
  \href{http://dx.doi.org/10.1016/j.cpc.2013.01.014}{\emph{Comput. Phys.
  Commun.} {\bf 184} (2013) 1729--1769},
  [\href{http://arxiv.org/abs/1207.6082}{{\tt 1207.6082}}].

\bibitem{Fermi-LAT:2015att}
M.~Ackermann et~al., \emph{{Searching for Dark Matter Annihilation from Milky
  Way Dwarf Spheroidal Galaxies with Six Years of Fermi Large Area Telescope
  Data}}, \href{http://dx.doi.org/10.1103/PhysRevLett.115.231301}{\emph{Phys.
  Rev. Lett.} {\bf 115} (2015) 231301},
  [\href{http://arxiv.org/abs/1503.02641}{{\tt 1503.02641}}].

\bibitem{Navarro:1995iw}
J.~F. Navarro, C.~S. Frenk and S.~D.~M. White, \emph{{The Structure of cold
  dark matter halos}}, \href{http://dx.doi.org/10.1086/177173}{\emph{Astrophys.
  J.} {\bf 462} (1996) 563--575},
  [\href{http://arxiv.org/abs/astro-ph/9508025}{{\tt astro-ph/9508025}}].

\bibitem{Navarro:1996gj}
J.~F. Navarro, C.~S. Frenk and S.~D.~M. White, \emph{{A Universal density
  profile from hierarchical clustering}},
  \href{http://dx.doi.org/10.1086/304888}{\emph{Astrophys. J.} {\bf 490} (1997)
  493--508}, [\href{http://arxiv.org/abs/astro-ph/9611107}{{\tt
  astro-ph/9611107}}].

\bibitem{ATLAS:2018gfm}
M.~Aaboud et~al., \emph{{Search for charged Higgs bosons decaying via $H^{\pm}
  \to \tau^{\pm}\nu_{\tau}$ in the $\tau$+jets and $\tau$+lepton final states
  with 36 fb$^{-1}$ of $pp$ collision data recorded at $\sqrt{s} = 13$ TeV with
  the ATLAS experiment}},
  \href{http://dx.doi.org/10.1007/JHEP09(2018)139}{\emph{JHEP} {\bf 09} (2018)
  139}, [\href{http://arxiv.org/abs/1807.07915}{{\tt 1807.07915}}].

\bibitem{CMS:2019bfg}
A.~M. Sirunyan et~al., \emph{{Search for charged Higgs bosons in the H$^{\pm}$
  $\to$ $\tau^{\pm}\nu_\tau$ decay channel in proton-proton collisions at
  $\sqrt{s} =$ 13 TeV}},
  \href{http://dx.doi.org/10.1007/JHEP07(2019)142}{\emph{JHEP} {\bf 07} (2019)
  142}, [\href{http://arxiv.org/abs/1903.04560}{{\tt 1903.04560}}].

\bibitem{CMS:2020imj}
A.~M. Sirunyan et~al., \emph{{Search for charged Higgs bosons decaying into a
  top and a bottom quark in the all-jet final state of pp collisions at $
  \sqrt{s} $ = 13 TeV}},
  \href{http://dx.doi.org/10.1007/JHEP07(2020)126}{\emph{JHEP} {\bf 07} (2020)
  126}, [\href{http://arxiv.org/abs/2001.07763}{{\tt 2001.07763}}].

\bibitem{ATLAS:2021upq}
G.~Aad et~al., \emph{{Search for charged Higgs bosons decaying into a top quark
  and a bottom quark at $ \sqrt{\mathrm{s}} $ = 13 TeV with the ATLAS
  detector}}, \href{http://dx.doi.org/10.1007/JHEP06(2021)145}{\emph{JHEP} {\bf
  06} (2021) 145}, [\href{http://arxiv.org/abs/2102.10076}{{\tt 2102.10076}}].

\bibitem{Jurciukonis:2022oru}
D.~Jur\v{c}iukonis and L.~Lavoura, \emph{{The centers of discrete groups as
  stabilizers of dark matter}},
  \href{http://dx.doi.org/10.1093/ptep/ptad004}{\emph{PTEP} {\bf 2023} (2023)
  023B02}, [\href{http://arxiv.org/abs/2210.12133}{{\tt 2210.12133}}].

\bibitem{Boehm:2003ha}
C.~Boehm, P.~Fayet and J.~Silk, \emph{{Light and heavy dark matter particles}},
  \href{http://dx.doi.org/10.1103/PhysRevD.69.101302}{\emph{Phys. Rev. D} {\bf
  69} (2004) 101302}, [\href{http://arxiv.org/abs/hep-ph/0311143}{{\tt
  hep-ph/0311143}}].

\bibitem{Ma:2006uv}
E.~Ma, \emph{{Supersymmetric Model of Radiative Seesaw Majorana Neutrino
  Masses}}, {\emph{Annales Fond. Broglie} {\bf 31} (2006) 285},
  [\href{http://arxiv.org/abs/hep-ph/0607142}{{\tt hep-ph/0607142}}].

\bibitem{Hur:2007ur}
T.~Hur, H.-S. Lee and S.~Nasri, \emph{{A Supersymmetric U(1)-prime model with
  multiple dark matters}},
  \href{http://dx.doi.org/10.1103/PhysRevD.77.015008}{\emph{Phys. Rev. D} {\bf
  77} (2008) 015008}, [\href{http://arxiv.org/abs/0710.2653}{{\tt 0710.2653}}].

\bibitem{Cao:2007fy}
Q.-H. Cao, E.~Ma, J.~Wudka and C.~P. Yuan, \emph{{Multipartite dark matter}},
  \href{http://arxiv.org/abs/0711.3881}{{\tt 0711.3881}}.

\bibitem{Zurek:2008qg}
K.~M. Zurek, \emph{{Multi-Component Dark Matter}},
  \href{http://dx.doi.org/10.1103/PhysRevD.79.115002}{\emph{Phys. Rev. D} {\bf
  79} (2009) 115002}, [\href{http://arxiv.org/abs/0811.4429}{{\tt 0811.4429}}].

\bibitem{Profumo:2009tb}
S.~Profumo, K.~Sigurdson and L.~Ubaldi, \emph{{Can we discover multi-component
  WIMP dark matter?}},
  \href{http://dx.doi.org/10.1088/1475-7516/2009/12/016}{\emph{JCAP} {\bf 12}
  (2009) 016}, [\href{http://arxiv.org/abs/0907.4374}{{\tt 0907.4374}}].

\bibitem{Batell:2010bp}
B.~Batell, \emph{{Dark Discrete Gauge Symmetries}},
  \href{http://dx.doi.org/10.1103/PhysRevD.83.035006}{\emph{Phys. Rev. D} {\bf
  83} (2011) 035006}, [\href{http://arxiv.org/abs/1007.0045}{{\tt 1007.0045}}].

\bibitem{Liu:2011aa}
Z.-P. Liu, Y.-L. Wu and Y.-F. Zhou, \emph{{Enhancement of dark matter relic
  density from the late time dark matter conversions}},
  \href{http://dx.doi.org/10.1140/epjc/s10052-011-1749-4}{\emph{Eur. Phys. J.
  C} {\bf 71} (2011) 1749}, [\href{http://arxiv.org/abs/1101.4148}{{\tt
  1101.4148}}].

\bibitem{Belanger:2011ww}
G.~B\'elanger and J.-C. Park, \emph{{Assisted freeze-out}},
  \href{http://dx.doi.org/10.1088/1475-7516/2012/03/038}{\emph{JCAP} {\bf 03}
  (2012) 038}, [\href{http://arxiv.org/abs/1112.4491}{{\tt 1112.4491}}].

\bibitem{Belanger:2012vp}
G.~B\'elanger, K.~Kannike, A.~Pukhov and M.~Raidal, \emph{{Impact of
  semi-annihilations on dark matter phenomenology - an example of $Z_N$
  symmetric scalar dark matter}},
  \href{http://dx.doi.org/10.1088/1475-7516/2012/04/010}{\emph{JCAP} {\bf 04}
  (2012) 010}, [\href{http://arxiv.org/abs/1202.2962}{{\tt 1202.2962}}].

\bibitem{Medvedev:2013vsa}
M.~V. Medvedev, \emph{{Cosmological Simulations of Multicomponent Cold Dark
  Matter}}, \href{http://dx.doi.org/10.1103/PhysRevLett.113.071303}{\emph{Phys.
  Rev. Lett.} {\bf 113} (2014) 071303},
  [\href{http://arxiv.org/abs/1305.1307}{{\tt 1305.1307}}].

\bibitem{Esch:2014jpa}
S.~Esch, M.~Klasen and C.~E. Yaguna, \emph{{A minimal model for two-component
  dark matter}}, \href{http://dx.doi.org/10.1007/JHEP09(2014)108}{\emph{JHEP}
  {\bf 09} (2014) 108}, [\href{http://arxiv.org/abs/1406.0617}{{\tt
  1406.0617}}].

\bibitem{Biswas:2015sva}
A.~Biswas, D.~Majumdar and P.~Roy, \emph{{Nonthermal two component dark matter
  model for Fermi-LAT \ensuremath{\gamma}-ray excess and 3.55 keV X-ray line}},
  \href{http://dx.doi.org/10.1007/JHEP04(2015)065}{\emph{JHEP} {\bf 04} (2015)
  065}, [\href{http://arxiv.org/abs/1501.02666}{{\tt 1501.02666}}].

\bibitem{Cai:2015zza}
Y.~Cai and A.~P. Spray, \emph{{Fermionic Semi-Annihilating Dark Matter}},
  \href{http://dx.doi.org/10.1007/JHEP01(2016)087}{\emph{JHEP} {\bf 01} (2016)
  087}, [\href{http://arxiv.org/abs/1509.08481}{{\tt 1509.08481}}].

\bibitem{Arcadi:2016kmk}
G.~Arcadi, C.~Gross, O.~Lebedev, Y.~Mambrini, S.~Pokorski and T.~Toma,
  \emph{{Multicomponent Dark Matter from Gauge Symmetry}},
  \href{http://dx.doi.org/10.1007/JHEP12(2016)081}{\emph{JHEP} {\bf 12} (2016)
  081}, [\href{http://arxiv.org/abs/1611.00365}{{\tt 1611.00365}}].

\bibitem{Ahmed:2017dbb}
A.~Ahmed, M.~Duch, B.~Grzadkowski and M.~Iglicki, \emph{{Multi-Component Dark
  Matter: the vector and fermion case}},
  \href{http://dx.doi.org/10.1140/epjc/s10052-018-6371-2}{\emph{Eur. Phys. J.
  C} {\bf 78} (2018) 905}, [\href{http://arxiv.org/abs/1710.01853}{{\tt
  1710.01853}}].

\bibitem{Bernal:2018aon}
N.~Bernal, D.~Restrepo, C.~Yaguna and O.~Zapata, \emph{{Two-component dark
  matter and a massless neutrino in a new $B-L$ model}},
  \href{http://dx.doi.org/10.1103/PhysRevD.99.015038}{\emph{Phys. Rev. D} {\bf
  99} (2019) 015038}, [\href{http://arxiv.org/abs/1808.03352}{{\tt
  1808.03352}}].

\bibitem{Poulin:2018kap}
A.~Poulin and S.~Godfrey, \emph{{Multicomponent dark matter from a hidden
  gauged SU(3)}},
  \href{http://dx.doi.org/10.1103/PhysRevD.99.076008}{\emph{Phys. Rev. D} {\bf
  99} (2019) 076008}, [\href{http://arxiv.org/abs/1808.04901}{{\tt
  1808.04901}}].

\bibitem{YaserAyazi:2018lrv}
S.~Yaser~Ayazi and A.~Mohamadnejad, \emph{{Scale-Invariant Two Component Dark
  Matter}}, \href{http://dx.doi.org/10.1140/epjc/s10052-019-6651-5}{\emph{Eur.
  Phys. J. C} {\bf 79} (2019) 140},
  [\href{http://arxiv.org/abs/1808.08706}{{\tt 1808.08706}}].

\bibitem{Bhattacharya:2018cgx}
S.~Bhattacharya, P.~Ghosh and N.~Sahu, \emph{{Multipartite Dark Matter with
  Scalars, Fermions and signatures at LHC}},
  \href{http://dx.doi.org/10.1007/JHEP02(2019)059}{\emph{JHEP} {\bf 02} (2019)
  059}, [\href{http://arxiv.org/abs/1809.07474}{{\tt 1809.07474}}].

\bibitem{Elahi:2019jeo}
F.~Elahi and S.~Khatibi, \emph{{Multi-Component Dark Matter in a Non-Abelian
  Dark Sector}},
  \href{http://dx.doi.org/10.1103/PhysRevD.100.015019}{\emph{Phys. Rev. D} {\bf
  100} (2019) 015019}, [\href{http://arxiv.org/abs/1902.04384}{{\tt
  1902.04384}}].

\bibitem{Borah:2019aeq}
D.~Borah, R.~Roshan and A.~Sil, \emph{{Minimal two-component scalar doublet
  dark matter with radiative neutrino mass}},
  \href{http://dx.doi.org/10.1103/PhysRevD.100.055027}{\emph{Phys. Rev. D} {\bf
  100} (2019) 055027}, [\href{http://arxiv.org/abs/1904.04837}{{\tt
  1904.04837}}].

\bibitem{Nanda:2019nqy}
D.~Nanda and D.~Borah, \emph{{Connecting Light Dirac Neutrinos to a
  Multi-component Dark Matter Scenario in Gauged $B-L$ Model}},
  \href{http://dx.doi.org/10.1140/epjc/s10052-020-8122-4}{\emph{Eur. Phys. J.
  C} {\bf 80} (2020) 557}, [\href{http://arxiv.org/abs/1911.04703}{{\tt
  1911.04703}}].

\bibitem{Hall:2019rld}
E.~Hall, T.~Konstandin, R.~McGehee and H.~Murayama, \emph{{Asymmetric matter
  from a dark first-order phase transition}},
  \href{http://dx.doi.org/10.1103/PhysRevD.107.055011}{\emph{Phys. Rev. D} {\bf
  107} (2023) 055011}, [\href{http://arxiv.org/abs/1911.12342}{{\tt
  1911.12342}}].

\bibitem{Betancur:2020fdl}
A.~Betancur, G.~Palacio and A.~Rivera, \emph{{Inert doublet as multicomponent
  dark matter}},
  \href{http://dx.doi.org/10.1016/j.nuclphysb.2020.115276}{\emph{Nucl. Phys. B}
  {\bf 962} (2021) 115276}, [\href{http://arxiv.org/abs/2002.02036}{{\tt
  2002.02036}}].

\bibitem{DuttaBanik:2020jrj}
A.~Dutta~Banik, R.~Roshan and A.~Sil, \emph{{Two component singlet-triplet
  scalar dark matter and electroweak vacuum stability}},
  \href{http://dx.doi.org/10.1103/PhysRevD.103.075001}{\emph{Phys. Rev. D} {\bf
  103} (2021) 075001}, [\href{http://arxiv.org/abs/2009.01262}{{\tt
  2009.01262}}].

\bibitem{Chakrabarty:2021kmr}
N.~Chakrabarty, R.~Roshan and A.~Sil, \emph{{Two-component doublet-triplet
  scalar dark matter stabilizing the electroweak vacuum}},
  \href{http://dx.doi.org/10.1103/PhysRevD.105.115010}{\emph{Phys. Rev. D} {\bf
  105} (2022) 115010}, [\href{http://arxiv.org/abs/2102.06032}{{\tt
  2102.06032}}].

\bibitem{Choi:2021yps}
S.-M. Choi, J.~Kim, P.~Ko and J.~Li, \emph{{A multi-component SIMP model with
  U(1)$_{X}$\textrightarrow{} Z$_{2}$ \texttimes{} Z$_{3}$}},
  \href{http://dx.doi.org/10.1007/JHEP09(2021)028}{\emph{JHEP} {\bf 09} (2021)
  028}, [\href{http://arxiv.org/abs/2103.05956}{{\tt 2103.05956}}].

\bibitem{DiazSaez:2021pmg}
B.~D\'\i{}az~S\'aez, P.~Escalona, S.~Norero and A.~R. Zerwekh, \emph{{Fermion
  singlet dark matter in a pseudoscalar dark matter portal}},
  \href{http://dx.doi.org/10.1007/JHEP10(2021)233}{\emph{JHEP} {\bf 10} (2021)
  233}, [\href{http://arxiv.org/abs/2105.04255}{{\tt 2105.04255}}].

\bibitem{Hall:2021zsk}
E.~Hall, R.~McGehee, H.~Murayama and B.~Suter, \emph{{Asymmetric dark matter
  may not be light}},
  \href{http://dx.doi.org/10.1103/PhysRevD.106.075008}{\emph{Phys. Rev. D} {\bf
  106} (2022) 075008}, [\href{http://arxiv.org/abs/2107.03398}{{\tt
  2107.03398}}].

\bibitem{Mohamadnejad:2021tke}
A.~Mohamadnejad, \emph{{Electroweak phase transition and gravitational waves in
  a two-component dark matter model}},
  \href{http://dx.doi.org/10.1007/JHEP03(2022)188}{\emph{JHEP} {\bf 03} (2022)
  188}, [\href{http://arxiv.org/abs/2111.04342}{{\tt 2111.04342}}].

\bibitem{Yaguna:2021rds}
C.~E. Yaguna and O.~Zapata, \emph{{Fermion and scalar two-component dark matter
  from a Z4 symmetry}},
  \href{http://dx.doi.org/10.1103/PhysRevD.105.095026}{\emph{Phys. Rev. D} {\bf
  105} (2022) 095026}, [\href{http://arxiv.org/abs/2112.07020}{{\tt
  2112.07020}}].

\bibitem{Ho:2022erb}
S.-Y. Ho, P.~Ko and C.-T. Lu, \emph{{Scalar and fermion two-component SIMP dark
  matter with an accidental $\mathbb Z_{4}$ symmetry}},
  \href{http://dx.doi.org/10.1007/JHEP03(2022)005}{\emph{JHEP} {\bf 03} (2022)
  005}, [\href{http://arxiv.org/abs/2201.06856}{{\tt 2201.06856}}].

\bibitem{Das:2022oyx}
A.~Das, S.~Gola, S.~Mandal and N.~Sinha, \emph{{Two-component scalar and
  fermionic dark matter candidates in a generic U(1)X model}},
  \href{http://dx.doi.org/10.1016/j.physletb.2022.137117}{\emph{Phys. Lett. B}
  {\bf 829} (2022) 137117}, [\href{http://arxiv.org/abs/2202.01443}{{\tt
  2202.01443}}].

\bibitem{BasiBeneito:2022qxd}
A.~Bas~i Beneito, J.~Herrero-Garc\'\i{}a and D.~Vatsyayan,
  \emph{{Multi-component dark sectors: symmetries, asymmetries and
  conversions}}, \href{http://dx.doi.org/10.1007/JHEP10(2022)075}{\emph{JHEP}
  {\bf 10} (2022) 075}, [\href{http://arxiv.org/abs/2207.02874}{{\tt
  2207.02874}}].

\bibitem{GAMBIT:2017gge}
P.~Athron et~al., \emph{{Status of the scalar singlet dark matter model}},
  \href{http://dx.doi.org/10.1140/epjc/s10052-017-5113-1}{\emph{Eur. Phys. J.
  C} {\bf 77} (2017) 568}, [\href{http://arxiv.org/abs/1705.07931}{{\tt
  1705.07931}}].

\bibitem{Barger:2008jx}
V.~Barger, P.~Langacker, M.~McCaskey, M.~Ramsey-Musolf and G.~Shaughnessy,
  \emph{{Complex Singlet Extension of the Standard Model}},
  \href{http://dx.doi.org/10.1103/PhysRevD.79.015018}{\emph{Phys. Rev. D} {\bf
  79} (2009) 015018}, [\href{http://arxiv.org/abs/0811.0393}{{\tt 0811.0393}}].

\bibitem{Drozd:2011aa}
A.~Drozd, B.~Grzadkowski and J.~Wudka, \emph{{Multi-Scalar-Singlet Extension of
  the Standard Model - the Case for Dark Matter and an Invisible Higgs Boson}},
  \href{http://dx.doi.org/10.1007/JHEP04(2012)006}{\emph{JHEP} {\bf 04} (2012)
  006}, [\href{http://arxiv.org/abs/1112.2582}{{\tt 1112.2582}}].

\bibitem{Modak:2013jya}
K.~P. Modak, D.~Majumdar and S.~Rakshit, \emph{{A Possible Explanation of Low
  Energy $\gamma$-ray Excess from Galactic Centre and Fermi Bubble by a Dark
  Matter Model with Two Real Scalars}},
  \href{http://dx.doi.org/10.1088/1475-7516/2015/03/011}{\emph{JCAP} {\bf 03}
  (2015) 011}, [\href{http://arxiv.org/abs/1312.7488}{{\tt 1312.7488}}].

\bibitem{Belanger:2014bga}
G.~B\'elanger, K.~Kannike, A.~Pukhov and M.~Raidal, \emph{{Minimal
  semi-annihilating $\mathbb{Z}_N$ scalar dark matter}},
  \href{http://dx.doi.org/10.1088/1475-7516/2014/06/021}{\emph{JCAP} {\bf 06}
  (2014) 021}, [\href{http://arxiv.org/abs/1403.4960}{{\tt 1403.4960}}].

\bibitem{Bhattacharya:2016ysw}
S.~Bhattacharya, P.~Poulose and P.~Ghosh, \emph{{Multipartite Interacting
  Scalar Dark Matter in the light of updated LUX data}},
  \href{http://dx.doi.org/10.1088/1475-7516/2017/04/043}{\emph{JCAP} {\bf 04}
  (2017) 043}, [\href{http://arxiv.org/abs/1607.08461}{{\tt 1607.08461}}].

\bibitem{Bhattacharya:2017fid}
S.~Bhattacharya, P.~Ghosh, T.~N. Maity and T.~S. Ray, \emph{{Mitigating Direct
  Detection Bounds in Non-minimal Higgs Portal Scalar Dark Matter Models}},
  \href{http://dx.doi.org/10.1007/JHEP10(2017)088}{\emph{JHEP} {\bf 10} (2017)
  088}, [\href{http://arxiv.org/abs/1706.04699}{{\tt 1706.04699}}].

\bibitem{Pandey:2017quk}
M.~Pandey, D.~Majumdar and K.~P. Modak, \emph{{Two Component Feebly Interacting
  Massive Particle (FIMP) Dark Matter}},
  \href{http://dx.doi.org/10.1088/1475-7516/2018/06/023}{\emph{JCAP} {\bf 06}
  (2018) 023}, [\href{http://arxiv.org/abs/1709.05955}{{\tt 1709.05955}}].

\bibitem{Belanger:2020hyh}
G.~B\'elanger, A.~Pukhov, C.~E. Yaguna and O.~Zapata, \emph{{The Z$_{5}$ model
  of two-component dark matter}},
  \href{http://dx.doi.org/10.1007/JHEP09(2020)030}{\emph{JHEP} {\bf 09} (2020)
  030}, [\href{http://arxiv.org/abs/2006.14922}{{\tt 2006.14922}}].

\bibitem{Coito:2021fgo}
L.~Coito, C.~Faubel, J.~Herrero-Garcia and A.~Santamaria, \emph{{Dark matter
  from a complex scalar singlet: the role of dark CP and other discrete
  symmetries}}, \href{http://dx.doi.org/10.1007/JHEP11(2021)202}{\emph{JHEP}
  {\bf 11} (2021) 202}, [\href{http://arxiv.org/abs/2106.05289}{{\tt
  2106.05289}}].

\bibitem{Yaguna:2021vhb}
C.~E. Yaguna and O.~Zapata, \emph{{Two-component scalar dark matter in Z$_{2n}$
  scenarios}}, \href{http://dx.doi.org/10.1007/JHEP10(2021)185}{\emph{JHEP}
  {\bf 10} (2021) 185}, [\href{http://arxiv.org/abs/2106.11889}{{\tt
  2106.11889}}].

\bibitem{Belanger:2021lwd}
G.~B\'elanger, A.~Mjallal and A.~Pukhov, \emph{{Two dark matter candidates: The
  case of inert doublet and singlet scalars}},
  \href{http://dx.doi.org/10.1103/PhysRevD.105.035018}{\emph{Phys. Rev. D} {\bf
  105} (2022) 035018}, [\href{http://arxiv.org/abs/2108.08061}{{\tt
  2108.08061}}].

\bibitem{Belanger:2022esk}
G.~B\'elanger, A.~Pukhov, C.~E. Yaguna and O.~Zapata, \emph{{The Z$_{7}$ model
  of three-component scalar dark matter}},
  \href{http://dx.doi.org/10.1007/JHEP03(2023)100}{\emph{JHEP} {\bf 03} (2023)
  100}, [\href{http://arxiv.org/abs/2212.07488}{{\tt 2212.07488}}].

\bibitem{Luhn:2007yr}
C.~Luhn, S.~Nasri and P.~Ramond, \emph{{Simple Finite Non-Abelian Flavor
  Groups}}, \href{http://dx.doi.org/10.1063/1.2823978}{\emph{J. Math. Phys.}
  {\bf 48} (2007) 123519}, [\href{http://arxiv.org/abs/0709.1447}{{\tt
  0709.1447}}].

\bibitem{Zwicky:2009vt}
R.~Zwicky and T.~Fischbacher, \emph{{On discrete Minimal Flavour Violation}},
  \href{http://dx.doi.org/10.1103/PhysRevD.80.076009}{\emph{Phys. Rev. D} {\bf
  80} (2009) 076009}, [\href{http://arxiv.org/abs/0908.4182}{{\tt 0908.4182}}].

\bibitem{Altarelli:2010gt}
G.~Altarelli and F.~Feruglio, \emph{{Discrete Flavor Symmetries and Models of
  Neutrino Mixing}},
  \href{http://dx.doi.org/10.1103/RevModPhys.82.2701}{\emph{Rev. Mod. Phys.}
  {\bf 82} (2010) 2701--2729}, [\href{http://arxiv.org/abs/1002.0211}{{\tt
  1002.0211}}].

\bibitem{Albright:2010ap}
C.~H. Albright, A.~Dueck and W.~Rodejohann, \emph{{Possible Alternatives to
  Tri-bimaximal Mixing}},
  \href{http://dx.doi.org/10.1140/epjc/s10052-010-1492-2}{\emph{Eur. Phys. J.
  C} {\bf 70} (2010) 1099--1110}, [\href{http://arxiv.org/abs/1004.2798}{{\tt
  1004.2798}}].

\bibitem{Parattu:2010cy}
K.~M. Parattu and A.~Wingerter, \emph{{Tribimaximal Mixing From Small Groups}},
  \href{http://dx.doi.org/10.1103/PhysRevD.84.013011}{\emph{Phys. Rev. D} {\bf
  84} (2011) 013011}, [\href{http://arxiv.org/abs/1012.2842}{{\tt 1012.2842}}].

\bibitem{Grimus:2011fk}
W.~Grimus and P.~O. Ludl, \emph{{Finite flavour groups of fermions}},
  \href{http://dx.doi.org/10.1088/1751-8113/45/23/233001}{\emph{J. Phys. A}
  {\bf 45} (2012) 233001}, [\href{http://arxiv.org/abs/1110.6376}{{\tt
  1110.6376}}].

\bibitem{deAdelhartToorop:2011re}
R.~de~Adelhart~Toorop, F.~Feruglio and C.~Hagedorn, \emph{{Finite Modular
  Groups and Lepton Mixing}},
  \href{http://dx.doi.org/10.1016/j.nuclphysb.2012.01.017}{\emph{Nucl. Phys. B}
  {\bf 858} (2012) 437--467}, [\href{http://arxiv.org/abs/1112.1340}{{\tt
  1112.1340}}].

\bibitem{Ferreira:2012ri}
P.~M. Ferreira, W.~Grimus, L.~Lavoura and P.~O. Ludl, \emph{{Maximal CP
  Violation in Lepton Mixing from a Model with Delta(27) flavour Symmetry}},
  \href{http://dx.doi.org/10.1007/JHEP09(2012)128}{\emph{JHEP} {\bf 09} (2012)
  128}, [\href{http://arxiv.org/abs/1206.7072}{{\tt 1206.7072}}].

\bibitem{Feruglio:2012cw}
F.~Feruglio, C.~Hagedorn and R.~Ziegler, \emph{{Lepton Mixing Parameters from
  Discrete and CP Symmetries}},
  \href{http://dx.doi.org/10.1007/JHEP07(2013)027}{\emph{JHEP} {\bf 07} (2013)
  027}, [\href{http://arxiv.org/abs/1211.5560}{{\tt 1211.5560}}].

\bibitem{King:2013eh}
S.~F. King and C.~Luhn, \emph{{Neutrino Mass and Mixing with Discrete
  Symmetry}},
  \href{http://dx.doi.org/10.1088/0034-4885/76/5/056201}{\emph{Rept. Prog.
  Phys.} {\bf 76} (2013) 056201}, [\href{http://arxiv.org/abs/1301.1340}{{\tt
  1301.1340}}].

\bibitem{King:2014nza}
S.~F. King, A.~Merle, S.~Morisi, Y.~Shimizu and M.~Tanimoto, \emph{{Neutrino
  Mass and Mixing: from Theory to Experiment}},
  \href{http://dx.doi.org/10.1088/1367-2630/16/4/045018}{\emph{New J. Phys.}
  {\bf 16} (2014) 045018}, [\href{http://arxiv.org/abs/1402.4271}{{\tt
  1402.4271}}].

\bibitem{Ludl:2014axa}
P.~O. Ludl and W.~Grimus, \emph{{A complete survey of texture zeros in the
  lepton mass matrices}},
  \href{http://dx.doi.org/10.1007/JHEP07(2014)090}{\emph{JHEP} {\bf 07} (2014)
  090}, [\href{http://arxiv.org/abs/1406.3546}{{\tt 1406.3546}}].

\bibitem{King:2017guk}
S.~F. King, \emph{{Unified Models of Neutrinos, Flavour and CP Violation}},
  \href{http://dx.doi.org/10.1016/j.ppnp.2017.01.003}{\emph{Prog. Part. Nucl.
  Phys.} {\bf 94} (2017) 217--256},
  [\href{http://arxiv.org/abs/1701.04413}{{\tt 1701.04413}}].

\bibitem{Kobayashi:2018vbk}
T.~Kobayashi, K.~Tanaka and T.~H. Tatsuishi, \emph{{Neutrino mixing from finite
  modular groups}},
  \href{http://dx.doi.org/10.1103/PhysRevD.98.016004}{\emph{Phys. Rev. D} {\bf
  98} (2018) 016004}, [\href{http://arxiv.org/abs/1803.10391}{{\tt
  1803.10391}}].

\bibitem{Novichkov:2018ovf}
P.~P. Novichkov, J.~T. Penedo, S.~T. Petcov and A.~V. Titov, \emph{{Modular
  S$_{4}$ models of lepton masses and mixing}},
  \href{http://dx.doi.org/10.1007/JHEP04(2019)005}{\emph{JHEP} {\bf 04} (2019)
  005}, [\href{http://arxiv.org/abs/1811.04933}{{\tt 1811.04933}}].

\bibitem{Novichkov:2018nkm}
P.~P. Novichkov, J.~T. Penedo, S.~T. Petcov and A.~V. Titov, \emph{{Modular
  A$_{5}$ symmetry for flavour model building}},
  \href{http://dx.doi.org/10.1007/JHEP04(2019)174}{\emph{JHEP} {\bf 04} (2019)
  174}, [\href{http://arxiv.org/abs/1812.02158}{{\tt 1812.02158}}].

\bibitem{Feruglio:2019ybq}
F.~Feruglio and A.~Romanino, \emph{{Lepton flavor symmetries}},
  \href{http://dx.doi.org/10.1103/RevModPhys.93.015007}{\emph{Rev. Mod. Phys.}
  {\bf 93} (2021) 015007}, [\href{http://arxiv.org/abs/1912.06028}{{\tt
  1912.06028}}].

\bibitem{Xing:2020ijf}
Z.-z. Xing, \emph{{Flavor structures of charged fermions and massive
  neutrinos}},
  \href{http://dx.doi.org/10.1016/j.physrep.2020.02.001}{\emph{Phys. Rept.}
  {\bf 854} (2020) 1--147}, [\href{http://arxiv.org/abs/1909.09610}{{\tt
  1909.09610}}].

\bibitem{Miller:1916}
G.~A. Miller, L.~E. Dickson and H.~F. Blichfeldt, \emph{Theory and applications
  of finite groups}.
\newblock John Wiley \& Sons, 2~ed., 1916.

\bibitem{Fairbairn:1964sga}
W.~M. Fairbairn, T.~Fulton and W.~H. Klink, \emph{{Finite and Disconnected
  Subgroups of SU3 and their Application to the Elementary-Particle Spectrum}},
  \href{http://dx.doi.org/10.1063/1.1704204}{\emph{J. Math. Phys.} {\bf 5}
  (1964) 1038}.

\bibitem{Bovier:1980ga}
A.~Bovier, M.~Luling and D.~Wyler, \emph{{Representations and Clebsch-gordan
  Coefficients of $Z$ Metacyclic Groups}},
  \href{http://dx.doi.org/10.1063/1.525095}{\emph{J. Math. Phys.} {\bf 22}
  (1981) 1536}.

\bibitem{Bovier:1980gc}
A.~Bovier, M.~Luling and D.~Wyler, \emph{{Finite Subgroups of SU(3)}},
  \href{http://dx.doi.org/10.1063/1.525096}{\emph{J. Math. Phys.} {\bf 22}
  (1981) 1543}.

\bibitem{Fairbairn:1982jx}
W.~M. Fairbairn and T.~Fulton, \emph{{SOME COMMENTS ON FINITE SUBGROUPS OF
  SU(3)}}, \href{http://dx.doi.org/10.1063/1.525224}{\emph{J. Math. Phys.} {\bf
  23} (1982) 1747--1748}.

\bibitem{Luhn:2007uq}
C.~Luhn, S.~Nasri and P.~Ramond, \emph{{The Flavor group Delta(3n**2)}},
  \href{http://dx.doi.org/10.1063/1.2734865}{\emph{J. Math. Phys.} {\bf 48}
  (2007) 073501}, [\href{http://arxiv.org/abs/hep-th/0701188}{{\tt
  hep-th/0701188}}].

\bibitem{Escobar:2008vc}
J.~A. Escobar and C.~Luhn, \emph{{The Flavor Group Delta(6n**2)}},
  \href{http://dx.doi.org/10.1063/1.3046563}{\emph{J. Math. Phys.} {\bf 50}
  (2009) 013524}, [\href{http://arxiv.org/abs/0809.0639}{{\tt 0809.0639}}].

\bibitem{Grimus:2010ak}
W.~Grimus and P.~O. Ludl, \emph{{Principal series of finite subgroups of
  SU(3)}}, \href{http://dx.doi.org/10.1088/1751-8113/43/44/445209}{\emph{J.
  Phys. A} {\bf 43} (2010) 445209}, [\href{http://arxiv.org/abs/1006.0098}{{\tt
  1006.0098}}].

\bibitem{Ludl:2010bj}
P.~O. Ludl, \emph{{On the finite subgroups of U(3) of order smaller than 512}},
  \href{http://dx.doi.org/10.1088/1751-8113/43/39/395204}{\emph{J. Phys. A}
  {\bf 43} (2010) 395204}, [\href{http://arxiv.org/abs/1006.1479}{{\tt
  1006.1479}}].

\bibitem{Ludl:2011gn}
P.~O. Ludl, \emph{{Comments on the classification of the finite subgroups of
  SU(3)}}, \href{http://dx.doi.org/10.1088/1751-8113/44/25/255204}{\emph{J.
  Phys. A} {\bf 44} (2011) 255204}, [\href{http://arxiv.org/abs/1101.2308}{{\tt
  1101.2308}}].

\bibitem{Jurciukonis:2017mjp}
D.~Jurciukonis and L.~Lavoura, \emph{{GAP listing of the finite subgroups of
  U(3) of order smaller than 2000}},
  \href{http://dx.doi.org/10.1093/ptep/ptx064}{\emph{PTEP} {\bf 2017} (2017)
  053A03}, [\href{http://arxiv.org/abs/1702.00005}{{\tt 1702.00005}}].

\bibitem{SmallGrp}
H.~U. Besche, B.~Eick, E.~O'Brien and M.~Horn, ``{SmallGrp}, the gap small
  groups library, {V}ersion 1.5.4.'' \href
  {https://gap-packages.github.io/smallgrp/}
  {\texttt{https://gap\texttt{\symbol{45}}packages.github.io/}\discretionary
  {}{}{}\texttt{smallgrp/}}, Jul, 2024.

\bibitem{KunMT}
A.~Kun\v{c}inas, \emph{Properties of $s_3$-symmetric three-higgs-doublet
  models},  Master's thesis.
\newblock
  \href{http://hdl.handle.net/1956/20467}{http://bora.uib.no/handle/1956/20467}.

\bibitem{Goodman:1984dc}
M.~W. Goodman and E.~Witten, \emph{{Detectability of Certain Dark Matter
  Candidates}}, \href{http://dx.doi.org/10.1103/PhysRevD.31.3059}{\emph{Phys.
  Rev. D} {\bf 31} (1985) 3059}.

\bibitem{Boto:2022uwv}
R.~Boto, J.~C. Rom\~ao and J.~P. Silva, \emph{{Bounded from below conditions on
  a class of symmetry constrained 3HDM}},
  \href{http://dx.doi.org/10.1103/PhysRevD.106.115010}{\emph{Phys. Rev. D} {\bf
  106} (2022) 115010}, [\href{http://arxiv.org/abs/2208.01068}{{\tt
  2208.01068}}].

\bibitem{Bento:2022vsb}
M.~P. Bento, J.~C. Rom\~ao and J.~P. Silva, \emph{{Unitarity bounds for all
  symmetry-constrained 3HDMs}},
  \href{http://dx.doi.org/10.1007/JHEP08(2022)273}{\emph{JHEP} {\bf 08} (2022)
  273}, [\href{http://arxiv.org/abs/2204.13130}{{\tt 2204.13130}}].

\end{thebibliography}\endgroup

\newpage
\thispagestyle{empty}
\vspace*{\fill}
\begin{center}
    \large \textit{This page intentionally right blank}
\end{center}
\vspace*{\fill}

\newpage

\end{document}